\documentclass[12pt]{article}

\usepackage{amsmath2000,euscript,array,amssymb,cite} 
\renewcommand{\thesection}{\Roman{section}}
\renewcommand{\theequation}{\arabic{section}.\arabic{equation}}
\def\rsen{\newpage\setcounter{equation}{0}}

\newcommand{\startappendix}{
\setcounter{section}{0}
\renewcommand{\thesection}{\Alph{section}}
\renewcommand{\theequation}{\Alph{section}.\arabic{equation}}}
\newcommand{\Appendix}[1]{
\refstepcounter{section}
\begin{flushleft}
{\large\bf Appendix \thesection: #1}
\end{flushleft}}

\def\elabel#1{\label{#1}}

\def\sfc{\hat\Omega}
\def\com{X}
\def\rmb{{\rm b}}
\def\rmf{{\rm f}}
\newcommand{\Rowspace}{\phantom{$\Big($}}
\def\H{H}
\def\qvac{q^0}
\def\vhiggs{\phi^0}
\def\K{\cal K}
\def\sfc{\hat\Omega}
\def\com{X}
\def\rmb{{\rm b}}
\def\rmf{{\rm f}}
\def\zo{{\sst(0)}}
\def\Pinfty{{\cal P}_\infty}
\def\bbar{\bar b}
\def\two{{(2)}}
\def\CM{{\cal M}}
\def\eighth{\tfrac18}
\def\fourth{\tfrac14}
\def\trN{{\rm tr}_N}
\def\hf{\tfrac12}
\def\CN{{\cal N}}
\def\ii{{\tilde{\rm i}}}
\def\jj{{\tilde{\rm j}}}
\def\kk{{\tilde{\rm k}}}
\def\llr{{\tilde{\rm l}}}
\def\xii{{\rm i}}
\def\xjj{{\rm j}}
\def\xkk{{\rm k}}
\def\xll{{\rm l}}
\def\aD{{\dot\alpha}}

\def\bD{{\dot\beta}}
\def\M{{{\cal M}'}}

\def\N{{\cal N}}
\def\psibar{\bar\psi}

\def\Ubar{\bar U}

\def\TrN{{\rm tr}_N}

\def\Mbar{\bar{\cal M}}
\def\dalpha{{\dot\alpha}}
\def\Deltabar{\bar\Delta}
\def\G{{\cal G}}
\def\dbeta{{\dot\beta}}
\def\gD{{\dot\gamma}}
\def\sqrtwo{\sqrt2}
\def\dD{{\dot\delta}}
\def\eD{{\dot\epsilon}}
\def\wbar{\bar w}

\def\abar{\bar a}
\def\sigmabar{\bar\sigma}
\def\etabar{\bar\eta}

\def\nubar{\bar\nu}

\def\sst{\scriptscriptstyle}
\def\det{{\rm det}}
\def\SU{\text{SU}}
\def\U{\text{U}}
\def\SL{\text{SL}}
\def\SS{\text{S}}

\def\grp{{\EuScript U}}
\newcommand{\BN}{{\boldsymbol{N}}}
\newcommand{\BM}{\boldsymbol{M}}

\newcommand{\Bk}{\boldsymbol{k}}
\newcommand{\BL}{\boldsymbol{L}}

\newcommand{\BX}{\boldsymbol{X}}

\newcommand{\Bomega}{{\boldsymbol{\omega}}}
\newcommand{\BI}{{\boldsymbol{I}}}

\def\J{{\cal J}}
\def\D{{\cal D}}
\def\Dbarslash{\,\,{\raise.15ex\hbox{/}\mkern-12mu {\bar\D}}}
\def\Dslash{\,{\raise.15ex\hbox{/}\mkern-12mu \D}}
\def\delslash{\,{\raise.15ex\hbox{/}\mkern-9mu \partial}}
\def\delbarslash{\,{\raise.15ex\hbox{/}\mkern-9mu {\bar\partial}}}
\def\SU{\text{SU}}
\def\SO{\text{SO}}
\def\O{\text{O}}
\def\Gl{\text{Gl}}
\def\Sp{\text{Sp}}

\def\USp{\text{USp}}
\def\K{{\cal K}}
\def\One{{1}}

\def\ms{{\mathfrak M}}
\def\ns{{\mathfrak N}}
\def\cms{\widehat{\mathfrak M}}
\def\F{{\mathfrak F}}
\def\Q{{\mathfrak Q}}

\def\AA{A}
\def\Q{{\cal Q}}

\def\psibar{\bar\psi}
\def\vhiggs{{\rm v}}

\def\omegabar{\bar\omega}
\def\Kt{\tilde\K}
\def\hf{{\textstyle{1\over2}}}

\def\quarter{{\textstyle{1\over4}}}

\def\eighth{{\textstyle{1\over8}}}
\def\fourth{\quarter}

\def\z{\zeta}
\def\zb{\bar{z}}
\def\th{\theta}
\def\cH{{\cal H}}

\def\ket#1{\left| #1\right\rangle}

\def\VEV#1{\left\langle #1\right\rangle}
\def\Vev#1{\big\langle{#1}\big\rangle}

\newcommand{\PD}[2]{\frac{\partial #1}{\partial #2}}
\newcommand{\MAT}[1]{\begin{pmatrix}#1\end{pmatrix}}

\newcommand{\EQ}[1]{\begin{equation} #1 \end{equation}}
\newcommand{\AL}[1]{\begin{subequations}\begin{align} #1 \end{align}\end{subequations}}
\newcommand{\SP}[1]{\begin{equation}\begin{split} #1 \end{split}\end{equation}}
\newcommand{\ALT}[2]{\begin{subequations}\begin{alignat}{#1} #2
\end{alignat}\end{subequations}}

\begin{document}

\addtolength{\baselineskip}{2pt}
\thispagestyle{empty}

\begin{flushright}
{\tt hep-th/0206063}\\
May 2002
\end{flushright}

\vspace{0.8cm}

\begin{center}
{\scshape\Huge The Calculus of Many Instantons\\}

\vspace{1.2cm}

{\scshape Nick Dorey,$^{1}$ Timothy J.~Hollowood,$^{1}$
Valentin V.~Khoze$^2$ and Michael P.~Mattis$^{3}$}

\vspace{0.15cm}

$^1${\sl Department of Physics, University of Wales Swansea,\\
Swansea, SA2 8PP, UK}\\ {\tt n.dorey@swan.ac.uk,
t.hollowood@swan.ac.uk}\\

\vspace{0.15cm}
$^2${\sl Department of Physics, University of Durham,\\
Durham, DH1 3LE, UK}\\ {\tt valya.khoze@durham.ac.uk}\\

\vspace{0.15cm}
$^3${\sl 121 Fox Meadow Road, Scarsdale, NY10583, USA\\
{\tt mattis@post.harvard.edu}}

\vspace{1.5cm}
{\Large ABSTRACT}
\end{center}
\vspace{0.3cm}
\noindent We describe the modern formalism, ideas and applications of
the instanton
calculus for gauge theories with, and without, 
supersymmetry. Particular emphasis is
put on developing a formalism that can deal with any number of
instantons. This necessitates a thorough review 
of the ADHM construction of instantons with
arbitrary charge and an in-depth analysis of the resulting moduli
space of solutions. We review the construction of the ADHM moduli space as 
a hyper-K\"ahler quotient. We show how the
functional integral in the semi-classical approximation reduces to an integral
over the instanton moduli space in each instanton sector and how the
resulting matrix partition function involves various geometrical
quantities on the instanton moduli space: volume form, connection, curvature,
isometries, {\it etc\/}. One important conclusion is that
this partition function is the dimensional reduction of a
higher-dimensional gauged linear sigma model which naturally leads us
to describe the relation of the instanton calculus to D-branes in
string theory. Along the way we describe powerful 
applications of the calculus of many instantons to supersymmetric
gauge theories including (i) the gluino condensate puzzle in 
${\cal N}=1$ theories (ii) Seiberg-Witten theory in $\N=2$ theories;
and  (iii) the AdS/CFT correspondence in $\N=2$ and $\N=4$ theories. 
Finally, we brielfy review the modifications of the instanton calculus
for a gauge theory defined on a non-commutative spacetime and we also
describe a new method for calculating instanton processes using a form
of localization on the instanton moduli space.

\newpage


\contentsline {section}{\numberline {I}\hspace{0.5cm}Introduction}{7}
\contentsline {subsection}{\numberline {I.1}\hspace{0.5cm}The philosophy of instanton calculations}{11}
\contentsline {section}{\numberline {II}\hspace{0.5cm}Instantons in Pure Gauge Theory}{13}
\contentsline {subsection}{\numberline {II.1}\hspace{0.5cm}Some basic facts}{13}
\contentsline {subsection}{\numberline {II.2}\hspace{0.5cm}Collective coordinates and moduli space}{15}
\contentsline {subsection}{\numberline {II.3}\hspace{0.5cm}General properties of the moduli space of instantons}{19}
\contentsline {subsubsection}{\numberline {II.3.1}\hspace{0.5cm}The moduli space as a complex manifold}{19}
\contentsline {subsection}{\numberline {II.4}\hspace{0.5cm}The ADHM construction of instantons}{21}
\contentsline {subsubsection}{\numberline {II.4.1}\hspace{0.5cm}The ADHM construction as a hyper-K\"ahler quotient}{24}
\contentsline {subsubsection}{\numberline {II.4.2}\hspace{0.5cm}Symmetries and the moduli space}{27}
\contentsline {subsubsection}{\numberline {II.4.3}\hspace{0.5cm}Singular gauge, one instanton, the dilute limit and asymptotics}{28}
\contentsline {subsection}{\numberline {II.5}\hspace{0.5cm}Zero modes and the metric on ${\mathfrak M}_k$}{31}
\contentsline {subsection}{\numberline {II.6}\hspace{0.5cm}Singularities and small instantons}{33}
\contentsline {section}{\numberline {III}\hspace{0.5cm}The Collective Coordinate Integral}{37}
\contentsline {subsection}{\numberline {III.1}\hspace{0.5cm}From the functional to the collective coordinate integral}{37}
\contentsline {subsection}{\numberline {III.2}\hspace{0.5cm}The volume form on the instanton moduli space}{39}
\contentsline {subsubsection}{\numberline {III.2.1}\hspace{0.5cm}Clustering}{41}
\contentsline {subsection}{\numberline {III.3}\hspace{0.5cm}Fluctuation determinants in the instanton background}{41}
\contentsline {section}{\numberline {IV}\hspace{0.5cm}Instantons in Supersymmetric Gauge Theories}{45}
\contentsline {subsection}{\numberline {IV.1}\hspace{0.5cm}Action, supersymmetry and equations-of-motion}{51}
\contentsline {subsection}{\numberline {IV.2}\hspace{0.5cm}The super-instanton at linear order}{53}
\contentsline {subsubsection}{\numberline {IV.2.1}\hspace{0.5cm}Adjoint fermion zero modes}{54}
\contentsline {subsubsection}{\numberline {IV.2.2}\hspace{0.5cm}Grassmann collective coordinates and the hyper-K\"ahler quotient construction}{55}
\contentsline {subsubsection}{\numberline {IV.2.3}\hspace{0.5cm}Supersymmetric and superconformal zero modes}{56}
\contentsline {subsection}{\numberline {IV.3}\hspace{0.5cm}Going beyond linear order: the quasi-instanton}{58}
\contentsline {subsection}{\numberline {IV.4}\hspace{0.5cm}Scalar VEVs and constrained instantons}{61}
\contentsline {subsubsection}{\numberline {IV.4.1}\hspace{0.5cm}Constrained instantons on the Coulomb branch}{62}
\contentsline {subsection}{\numberline {IV.5}\hspace{0.5cm}Collective coordinate supersymmetry}{64}
\contentsline {section}{\numberline {V}\hspace{0.5cm}The Supersymmetric Collective Coordinate Integral}{66}
\contentsline {subsection}{\numberline {V.1}\hspace{0.5cm}The supersymmetric collective coordinate measure}{66}
\contentsline {subsection}{\numberline {V.2}\hspace{0.5cm}The instanton effective action}{69}
\contentsline {subsubsection}{\numberline {V.2.1}\hspace{0.5cm}Geometric interpretation}{71}
\contentsline {subsubsection}{\numberline {V.2.2}\hspace{0.5cm}The size of a constrained instanton}{73}
\contentsline {subsubsection}{\numberline {V.2.3}\hspace{0.5cm}The lifting of zero modes}{74}
\contentsline {subsection}{\numberline {V.3}\hspace{0.5cm}The supersymmetric volume form on ${\mathfrak M}_k$}{75}
\contentsline {subsubsection}{\numberline {V.3.1}\hspace{0.5cm}Supersymmetry}{76}
\contentsline {subsection}{\numberline {V.4}\hspace{0.5cm}From ${\cal N}=4$ to ${\cal N}=0$ via decoupling}{79}
\contentsline {section}{\numberline {VI}\hspace{0.5cm}Generalizations and Miscellany}{82}
\contentsline {subsection}{\numberline {VI.1}\hspace{0.5cm}Solving the ADHM constraints for $N\geq 2k$}{82}
\contentsline {subsection}{\numberline {VI.2}\hspace{0.5cm}The ADHM construction for $\text {Sp}(N)$ and $\text {SO}(N)$}{86}
\contentsline {subsection}{\numberline {VI.3}\hspace{0.5cm}Matter fields and the ADHM construction}{89}
\contentsline {subsubsection}{\numberline {VI.3.1}\hspace{0.5cm}${\cal N}=1$ theories on the Higgs branch}{89}
\contentsline {subsubsection}{\numberline {VI.3.2}\hspace{0.5cm}${\cal N}=2$ theories on the Coulomb branch}{93}
\contentsline {subsection}{\numberline {VI.4}\hspace{0.5cm}Masses}{96}
\contentsline {subsection}{\numberline {VI.5}\hspace{0.5cm}The instanton partition function}{97}
\contentsline {section}{\numberline {VII}\hspace{0.5cm}The Gluino Condensate in ${\cal N}=1$ Theories}{99}
\contentsline {subsection}{\numberline {VII.1}\hspace{0.5cm}A supersymmetric Ward identity}{101}
\contentsline {subsection}{\numberline {VII.2}\hspace{0.5cm}One instanton calculations of the gluino condensate}{103}
\contentsline {subsubsection}{\numberline {VII.2.1}\hspace{0.5cm}Strong coupling}{103}
\contentsline {subsubsection}{\numberline {VII.2.2}\hspace{0.5cm}Weak coupling}{104}
\contentsline {subsection}{\numberline {VII.3}\hspace{0.5cm}Multi-instanton calculations of the gluino condensate}{109}
\contentsline {subsubsection}{\numberline {VII.3.1}\hspace{0.5cm}Strong coupling}{110}
\contentsline {subsubsection}{\numberline {VII.3.2}\hspace{0.5cm}Weak coupling}{113}
\contentsline {subsection}{\numberline {VII.4}\hspace{0.5cm}Clustering in instanton calculations}{117}
\contentsline {section}{\numberline {VIII}\hspace{0.5cm}On the Coulomb Branch of ${\cal N}=2$ Gauge Theories}{119}
\contentsline {subsection}{\numberline {VIII.1}\hspace{0.5cm}Seiberg-Witten theory and the prepotential}{120}
\contentsline {subsection}{\numberline {VIII.2}\hspace{0.5cm}Extracting the prepotential from instantons}{122}
\contentsline {subsection}{\numberline {VIII.3}\hspace{0.5cm}Gauge group $\text {SU}(2)$}{125}
\contentsline {subsubsection}{\numberline {VIII.3.1}\hspace{0.5cm}One instanton}{127}
\contentsline {subsubsection}{\numberline {VIII.3.2}\hspace{0.5cm}Two instantons}{128}
\contentsline {subsection}{\numberline {VIII.4}\hspace{0.5cm}One instanton prepotential in $\text {SU}(N)$}{130}
\contentsline {section}{\numberline {IX}\hspace{0.5cm}Conformal Gauge Theories at Large $N$}{134}
\contentsline {subsection}{\numberline {IX.1}\hspace{0.5cm}The collective coordinate integrals at large $N$}{135}
\contentsline {subsubsection}{\numberline {IX.1.1}\hspace{0.5cm}The ${\cal N}=4$ case}{136}
\contentsline {subsubsection}{\numberline {IX.1.2}\hspace{0.5cm}The ${\cal N}=2$ case}{143}
\contentsline {subsection}{\numberline {IX.2}\hspace{0.5cm}Large-$N$ correlation functions}{145}
\contentsline {subsection}{\numberline {IX.3}\hspace{0.5cm}Instantons and the AdS/CFT correspondence}{148}
\contentsline {subsubsection}{\numberline {IX.3.1}\hspace{0.5cm}The instanton collective coordinate integral}{149}
\contentsline {subsubsection}{\numberline {IX.3.2}\hspace{0.5cm}Correlation functions}{150}
\contentsline {section}{\numberline {X}\hspace{0.5cm}Instantons as Solitons in Higher Dimensions and String Theory}{155}
\contentsline {subsection}{\numberline {X.1}\hspace{0.5cm}Non-supersymmetric instanton branes}{156}
\contentsline {subsubsection}{\numberline {X.1.1}\hspace{0.5cm}The moduli space approximation}{157}
\contentsline {subsection}{\numberline {X.2}\hspace{0.5cm}Supersymmetric instanton branes}{158}
\contentsline {subsubsection}{\numberline {X.2.1}\hspace{0.5cm}Action, supersymmetry and equations-of-motion}{159}
\contentsline {subsubsection}{\numberline {X.2.2}\hspace{0.5cm}The moduli space approximation}{161}
\contentsline {subsubsection}{\numberline {X.2.3}\hspace{0.5cm}The effective action}{161}
\contentsline {subsubsection}{\numberline {X.2.4}\hspace{0.5cm}Supersymmetry}{164}
\contentsline {subsubsection}{\numberline {X.2.5}\hspace{0.5cm}Relation to the instanton calculus}{165}
\contentsline {subsection}{\numberline {X.3}\hspace{0.5cm}Instantons and string theory}{166}
\contentsline {subsubsection}{\numberline {X.3.1}\hspace{0.5cm}The ${\cal N}=4$ instanton calculus}{167}
\contentsline {subsubsection}{\numberline {X.3.2}\hspace{0.5cm}Probing the stringy instanton}{177}
\contentsline {subsubsection}{\numberline {X.3.3}\hspace{0.5cm}The ${\cal N}=2$ instanton calculus}{179}
\contentsline {subsubsection}{\numberline {X.3.4}\hspace{0.5cm}Mass couplings and soft supersymmetry breaking}{182}
\contentsline {section}{\numberline {XI}\hspace{0.5cm}Further Directions}{185}
\contentsline {subsection}{\numberline {XI.1}\hspace{0.5cm}Non-commutative gauge theories and instantons}{185}
\contentsline {subsubsection}{\numberline {XI.1.1}\hspace{0.5cm}ADHM construction on non-commutative ${\mathbb R}^4$}{187}
\contentsline {subsubsection}{\numberline {XI.1.2}\hspace{0.5cm}The prepotential of non-commutative ${\cal N}=2$ gauge theory}{189}
\contentsline {subsection}{\numberline {XI.2}\hspace{0.5cm}Calculating the prepotential by localization}{192}
\contentsline {subsubsection}{\numberline {XI.2.1}\hspace{0.5cm}One instanton}{198}
\contentsline {subsubsection}{\numberline {XI.2.2}\hspace{0.5cm}Two instantons}{199}
\contentsline {section}{\numberline {Appendix A}\hspace{2.3cm}Spinors in Diverse
Dimensions}{206}
\contentsline {section}{\numberline {Appendix B}\hspace{2.3cm}Complex 
Geometry and the Quotient Construction}{209}
\contentsline {section}{\numberline {Appendix C}\hspace{2.3cm}ADHM Algebra}{216}


\section{Introduction}\elabel{sec:S1}

Yang-Mills instantons \cite{BPST} have provided an enduring interest for a
generation of physicists and mathematicians. On the physics side,
instanton configurations give the leading non-perturbative
contributions to the functional integral in the semi-classical approximation
\cite{tHooft}. After the initial disappointment that instantons were
not going to
provide a simple explanation of quark confinement, they have proved
interesting and useful both in
phenomenological models of QCD (see, for example
\cite{CDG,ABC,SSI}) and for describing exact
non-perturbative phenomena in supersymmetric gauge theories (as we
describe in this review). On the
mathematical side, instantons lie at the heart of some important recent
developments in topology and, in particular, of Donaldson's
construction of topological invariants of four-manifolds
(see \cite{DonaldsonKron} and references therein).

The main applications of instantons we will consider in this review
are to supersymmetric gauge theories. Recent developments have provided
an extraordinary web of exact non-perturbative results for these theories,
making them an ideal field-theoretical
laboratory in which to test our ideas about
strongly-coupled gauge dynamics. Highlights include Seiberg duality for
$\N=1$ theories, the Seiberg-Witten solution of $\N=2$ supersymmetric
Yang-Mills
and the AdS/CFT correspondence. Typically supersymmetry
constrains the form of quantum corrections allowing exact
results to be obtained for some special quantities. Despite these
constraints, supersymmetric gauge theories still 
exhibit many interesting physical
phenomena including quark confinement and chiral symmetry breaking.
Yang-Mills instantons have played an important r\^ole in some
of these developments which we will review in the following.

Instantons in supersymmetric gauge theories were studied in detail by
several groups in the 1980's. This early work focused on the
contribution of a single instanton in theories with $\N=1$ supersymmetry.
Impressive results were obtained for the vacuum structure of these theories,
including exact formulae for condensates of chiral operators.
Although precise numerical answers were obtained, interest in these was
limited as there was nothing with which to compare them.\footnote{Note
however, that several puzzles emerged on comparing numerical values
for condensates obtained using different approaches to instanton
calculations. These will be re-examined in Chapter IV}
The new developments motivate the extension of the instanton calculus to
theories with extended supersymmetry and, in particular, to instantons of
arbitrary topological charge. This is the main goal of the
first part of this review. Developments such as Seiberg-Witten theory and the
AdS/CFT correspondence also yield precise predictions for
instanton effects, which can be checked explicitly
using the methods we will develop below. These applications, which are
described in the second half of the review are important for two reasons.
Firstly, they provide as a quantitative test of
conjectural dualities which underlie the recent progress in
supersymmetric gauge theories. Secondly, they also increase our
confidence in weak-coupling instanton calculations and some of the
technology which they require such as Wick rotation to
Euclidean space, constrained instantons, {\it etc\/}.

A semi-classical evaluation of the path integral requires us to find
the complete set of finite-action configurations which minimize the
Euclidean action. In pure Yang-Mills theory,
this was accomplished many years ago in pioneering work by
Atiyah, Drinfeld, Hitchin and Manin (ADHM) \cite{ADHM}. In particular, these
authors found the complete set of self-dual gauge fields of arbitrary
topological charge $k$.
Their construction, which works for arbitrary $\SU(N)$, $\SO(N)$
or $\Sp(N)$ gauge groups (but not for the exceptional groups), reduces the
self-dual Yang-Mills equation to a set of non-linear algebraic equations
(the ``ADHM constraints'')
constraining a matrix of parameters (the ``ADHM data''). After modding out
a residual symmetry group, each solution of the ADHM constraints defines a
gauge-inequivalent, $k$-instanton configuration. The space of such solutions,
which we denote $\ms_{k}$, is also known as the $k$-instanton moduli space.

The space $\ms_{k}$, has many remarkable properties which will play a
central r\^ole in our story. Firstly, apart from having some isolated
singularities, it is a
Riemannian manifold with a natural metric. The singularities are relatively
mild, being of conical type, and have a natural interpretation as points
where instantons shrink to zero size.
In addition, the moduli space carries
families of inequivalent complex structures and, when endowed with its
natural metric,
defines a hyper-K\"ahler manifold. This property is manifest in the
ADHM construction, which is actually an example of a general method for
constructing hyper-K\"ahler spaces known as the {\it hyper-K\"ahler quotient}.
This viewpoint turns out to be very useful and we will find that many of the
structures which emerge in the semi-classical analysis of the path integral
have a nice geometrical interpretations in terms of the hyper-K\"ahler
quotient.

Applying the semi-classical
approximation to a Green's function requires us to replace the fields by their
ADHM values and then integrate over the moduli space $\ms_{k}$.
Constructing the appropriate measure for this integration,
both in pure gauge theory and its supersymmetric generalizations, is
the main problem we address in the first part of the review.
The main obstacle to overcome is the fact the ADHM constraints cannot
generally be
solved except for small values of $k$. Hence, we cannot find an explicit
unconstrained parameterization of the moduli space. However, we will be able to
find explicit formulae for the measure in terms of the ADHM data together
with Lagrange multiplier fields which impose the ADHM equations as
$\delta$-function constraints. 
Under certain circumstances we will be
able to evaluate the the moduli space integrals even for general
$k$. For instance at large-$N$ the ADHM constraints can be solved, at
least on a certain generic region of the moduli space, and this means
that the instanton calculus becomes tractable for arbitrary $k$. In
the final Chapter we shall briefly explain important new developments that
allow one to make progress even at finite $N$ when one is on a Higgs
or Coulomb branch.

The plan of this report is as follows. As a prelude we briefly
review the basic philosophy of instanton calculations.
Chapter \ref{sec:S2} provides a
solid grounding of the ADHM construction of instantons in $\SU(N)$
gauge theory.\footnote{The ADHM construction for the other classical
groups is described separately in \S\ref{app:A7}.}
Firstly, we describe some general features of
instantons. This begins with the notion of zero modes in the instanton
background, collective coordinates and the moduli space of instantons
$\ms_k$. Very general arguments then show that $\ms_k$ is a
hyper-K\"ahler manifold with a natural Hermitian metric. After this, we
describe all aspects of the ADHM construction of $\ms_k$, putting
particular emphasis on the hyper-K\"ahler quotient description.
Most of the material
in Chapter \ref{sec:S2} is necessarily very mathematical. A reader
primarily interested in physical applications of instantons can skip
most of \S\ref{sec:S7}, \S\ref{sec:S10},  \S\ref{sec:S14} and \S\ref{sec:S15}.

Chapter \ref{sec:S16}, shows how the semi-classical limit of the functional
integral reduces to an integral over $\ms_k$ with a volume form which
is derived from the Hermitian metric constructed previously. However,
in the setting of pure gauge theory, integrating out all the
non-zero modes of the quadratic fluctuations around the instanton
yields a non-trivial determinant factor which accounts for the one-loop
effects in perturbation theory in the instanton background.
We briefly describe, using results
reviewed in \cite{OSB}, how the determinants may be evaluated
in the ADHM background, although the resulting expressions are rather
unsatisfactory (involving spacetime integrals). This latter point will
not overly concern us, because in the supersymmetric applications
the non-trivial parts of the determinant factors
cancel between bosons and fermions.

Chapter \ref{sec:S25} describes how the instanton calculus is generalized
in a supersymmetric gauge theory. This is done in the context of
$\N=1$, 2 and 4 supersymmetric theories with no additional matter
fields. (Adding matter fields is considered separately in
\S\ref{app:A6}.) First of all, we explain how to
construct the Euclidean version of the theory by Wick rotation from
Minkowski space and how this inevitably leads to a theory where the
fermion action is not real. This is a fact of life in the
supersymmetric instanton calculus, just as it is when fermionic theories
are investigated on the lattice, but does not lead to any
inconsistencies or pathologies. In an instanton background, there are
fermion zero modes and their associated Grassmann collective
coordinates. We show how these are superpartners of the bosonic
collective coordinates and so in a supersymmetric gauge theory the
moduli space $\ms_k$ is, itself, supersymmetrized. We show how this
fits in with the hyper-K\"ahler quotient approach. In this section, we
explain in detail the concept of the super-instanton. In
particular, in the $\N=4$ theory, the appropriate supersymmetric
instanton solution is only an approximate solution of the
equations-of-motion, but one which captures the leading order
semi-classical behaviour of the functional integral. We also explain
in this section how the addition of VEVs for the scalar fields (for
the $\N=2$ and 4 theories) leads to Affleck's notion of a constrained
instanton. Our point-of-view is that constrained
instantons are---in a sense---only a mild modification of the
conventional instanton: some of the collective coordinates cease to be
exact moduli and the whole effect can be described by the turning on a
potential on $\ms_k$. We call this approximate solution a ``quasi-instanton''.
Chapter \ref{sec:S35} goes to describe the
semi-classical approximation in a supersymmetric gauge theory. In all cases we
can formulate the leading-order semi-classical approximation to
the functional integral as an integral over the supersymmetric version of
$\ms_k$.

Some generalizations, and other important miscellany, are collected in
Chapter \ref{sec:S201}. In particular, we explain how the ADHM constraints
can be solved on generic orbit of the gauge group
when $N\geq2k$; how to extend the ADHM construction to
gauge groups $\SO(N)$ and $\Sp(N)$; how to add matter fields
transforming in the fundamental representation; the effect of
adding masses which break various amounts of supersymmetry; and how to
define the notion of the instanton partition function.

Chapters \ref{sec:S43}-\ref{sec:S39} each describe an
application of the multi-instanton calculus. In Chapter \ref{sec:S43} we
review the two distinct approaches to instanton calculus in $\N=1$ theories
developed by different groups in the 1980's. One approach follows closely the
methodology developed in the first part of the report, working with
constrained instantons in a weakly-coupled Higgs phase. As we review,
this approach yields
explicit agreement with the web of exact results which predict a precise
numerical value for the gluino condensate in $\N=1$ supersymmetric
Yang-Mills. The
second, more controversial approach, attempts to evaluate the gluino
condensate directly in the strongly-coupled confining phase. We cast further
doubt on this method, by showing explicitly that the resulting formulae
violate the clustering property, a general axiom of quantum field theory, and therefore
cannot yield the exact answer. This is just as one might have
expected: instantons are a semi-classical phenomena and as such are
not expected to yield quantitatively exact results in a
strongly-coupled phase.

We then go on in Chapter \ref{sec:S60} to consider instanton contributions to
the low-energy effective action of $\N=2$ gauge theories on their
Coulomb branch. These calculations are very important because they can
be compared with a completely different approach based on the
celebrated theory of Seiberg and Witten. Perfect agreement
is found providing strong evidence in favour of the instanton
calculus---including all this entails like the imaginary time formalism
and the resulting saddle-point approximation of the functional
integral---but also in favour of
Seiberg and Witten's ingenious theory. Several
minor modifications of the original Seiberg-Witten solution for ${\cal N}=2$
theories with matter are also described. These involve
ambiguities in the definition  of parameters appearing in the exact solution.

In Chapter \ref{sec:S39} we consider conformal, or finite, gauge theories
with $\N=2$ or $\N=4$ supersymmetry in their non-abelian Coulomb phase.
Here, we show how various instanton effects for arbitrary instanton
charge can be evaluated exactly in the limit of large $N$. The
large-$N$ instanton calculus that we develop provides substantial
evidence in favour of the remarkable AdS/CFT
correspondence which relates the gauge theories to ten-dimensional
superstring theory on a particular background. In particular we will
describe how, even at weak coupling,
the large-$N$ instanton calculus probes the background
ten-dimensional spacetime geometry of the string theory directly.

In Chapter \ref{sec:S49} we describe how instantons can be embedded
in higher-dimensional theories as brane-like solitons. We show in detail, as
expected from general principles, the
resulting collective coordinate dynamics involves a certain
$\sigma$-model on the brane world-volume with the moduli space of
instantons as the target space. This provides a useful way to understand
the relationship of instantons to D-branes in superstring theory, a
subject we go on to describe. Studying instantons in
the context of string theory provides a powerful way
to derive many aspects of the instanton calculus developed in the previous
sections and in the process
removes much of the cloak of mystery surrounding the ADHM construction.

Finally, in Chapter \ref{sec:S562} we describe two recent
developments in the instanton calculus. Firstly, how the instanton is
modified when the underlying gauge theory is defined on a
non-commutative space. It turns out that the instanton moduli space is
modified in a particularly natural way. Secondly, we describe a new
calculational technique which promises to make instanton effects
tractable even at finite $N$ and for all instanton charge. The key
idea is that in the presence of scalar field VEVs, a potential
develops on the instanton moduli space and the collective coordinate
integrals localize around the critical points of this potential.

Appendix A details our conventions for spinors in
different spacetime dimensions. Appendix B includes an introduction to
hyper-K\"ahler geometry. Appendix C reviews
some useful identities for calculations involving the ADHM data.

There are many other excellent reviews of instanton physics including
\cite{Amati:1988ft,Coleman,ABC,Shifman:1994ee,Shifman:1999mv,Belitsky:2000ws};
however these reviews concentrate on the single instanton.

\subsection{The philosophy of instanton calculations}

Before we begin in earnest,
it is perhaps useful to remind the reader of the basic
philosophy of instanton calculations.
In quantum field theory, all information
about the physical observables ({\it i.e.\/}
the spectrum and the $S$-matrix) can be
obtained by calculating the correlation functions of appropriate operators.
These correlation functions are defined by the Feynman path integral. The
most convenient formulation is one where the correlation functions and
the path integral are analytically continued to Euclidean spacetime.
The path integral was first introduced as a formal generating functional for
perturbation theory. However, thanks to the work of Wilson, it is widely
believed that the Euclidean path integral actually provides a first-principles
non-perturbative definition of quantum field theory. Implementing this
definition in practice involves replacing continuous spacetime
by a finite number of points and leads to the subject of
lattice field theory which is beyond the scope of this article.
Nevertheless, the success of this viewpoint
inspires us to take even the {\it continuum\/} path integral seriously.

In continuum quantum field theory we are generally limited to calculating
at weak coupling.
In four dimensions, the only exceptions are certain
supersymmetric gauge theories which we will discuss further below.
If the continuum path integral makes any sense at all, then the very least
should expect is that it should yield sensible calculable
answers at weak coupling. As mentioned above, the Feynman rules of
ordinary weak-coupling perturbation theory can easily be
derived from the path integral. However,
even at weak coupling, the path integral contains much more
information, including effects which are non-perturbative in the
coupling constant.

In the simplest cases, the generating functional in Euclidean
spacetime takes the schematic form,
\EQ{
Z[J]=\int\, \left[d\Phi\right]\,
\exp\Big(-\frac{1}{g^{2}}S[\Phi] + \int d^4x\,J\Phi\Big) \ ,
\elabel{zj}
}
where $\Phi$ denotes some set of fields with sources $J$ and
$S[\Phi]$ is the Euclidean action which is real and
bounded below. The gauge theories we consider below are characterized
(classically) by a single
dimensionless coupling, denoted $g^{2}$, and one may always re-scale the
fields so that the coupling appears in front of the action as indicated in
\eqref{zj}. Of course, in the quantum theory,
we will certainly need to modify our
discussion to account for the running of the coupling but we will,
for the moment, postpone this discussion.

The basic idea of the semi-classical approximation\footnote{The term
``semi-classical'' originates in the observation that $g^{2}$ appears in the
exponent of \eqref{zj} in exactly the same position as Planck's constant
$\hbar$ would if we restored physical units. Thus the
$g^{2}\rightarrow 0$ limit is identical to the $\hbar \rightarrow 0$
limit.} is that, for small $g^{2}$,
the path integral is dominated by the configurations of lowest
Euclidean action and we may proceed by expanding around these configurations.
The simplest such configurations are the perturbative vacua of the theory
({\it i.e.\/}~minima of the classical
potential) and the corresponding expansion
is just the ordinary loop expansion. However the Euclidean action may have
other minima with finite action. Such configurations are known as instantons
the same logic dictates that we should also expand in fluctuations
around them. For an instanton of finite Euclidean action $S_{\rm cl}$, the
leading semi-classical contribution goes like $\exp(-S_{\rm cl}/g^{2})$ and
expanding in fluctuation leads to corrections which are suppressed by
further powers of $g^{2}$.

As we will see, there are many extra
complications and technical difficulties in carrying out this program in
practice. As mentioned above, for asymptotically free theories of direct
physical interest the coupling $g^{2}$ runs as a function of the energy and
becomes large in the infra-red at the dynamical scale $\Lambda$. In this case,
the fluctuations of the fields in the path integral become large and neither
the instanton nor any other classical field configuration determines the
physics. It seems therefore that instantons have little to tell us directly
about these theories. On the other hand there are two types of theory
we will meet where instanton methods are directly applicable. Firstly,
and most straightforwardly, there are theories like $\N=4$ supersymmetric
Yang-Mills theory where the $\beta$-function vanishes and the
coupling does not run. In this case we may set $g^{2}\ll1$ and safely
apply semi-classical methods. The second type of theory is
asymptotically free but also contains scalar fields which can acquire
a vacuum expectation value (VEV) spontaneously breaking the gauge group.
The Higgs mechanism then cuts off the running of the coupling in the IR.
If the mass scale of the VEV is much large than the dynamical
scale $\Lambda$, the coupling small at all length scales and
semi-classical reasoning is valid.

In the case of a non-zero scalar VEV, we will also have to contend
with problems related to the fact that instantons are no longer true
minima of the action. In fact, both types of theory described above
requires a more complicated
application of the semi-classical method than simply expanding around
exact solutions  of the equations-of-motion.
We will frequently solve the classical equations only approximately
order-by-order in $g^{2}$.
However, the basic philosophy remains the same: as long as the
theory is genuinely weakly coupled, an appropriate
semi-classical approximation to the path
integral should be reliable and we can use it to evaluate the leading
non-perturbative corrections to observables. In the following chapters
we will develop the necessary technology to accomplish this goal.

\rsen\section{Instantons in Pure Gauge Theory}\elabel{sec:S2}

In this Chapter we describe instanton solutions of pure gauge
theory, {\it i.e.\/}~without any additional fields present. To begin
with we introduce the central concept of the moduli space of
instantons and describe how this space has a lot structure: it is a
hyper-K\"ahler manifold with singularities. In \S\ref{sec:S8} we
describe the Atiyah, Drinfeld, Hitchin and Manin construction of
arbitrary instanton solutions and then explain how the ``ADHM
construction'' can be viewed as a particular example of the
hyper-K\"ahler quotient construction. This point-of-view is
necessarily rather mathematical and some of the background required is
reviewed in Appendix \ref{app:A2}.

\subsection{Some basic facts}

We start with pure $\SU(N)$ gauge theory
described by a Euclidean space action
\EQ{
S[A]=-\frac1{2}\int d^4x\,\TrN\,F_{mn}^2+i\theta k\ ,
}
where the field strength
$F_{mn}=\partial_mA_n-\partial_nA_m+g[A_m,A_n]$
and the topological charge is
\EQ{
k=-\frac{g^2}{16\pi^2}\int d^4x\,\TrN\,F_{mn}{}^*F_{mn}\in{\mathbb Z}\
\elabel{topc}
}
with $^*F_{mn}=\tfrac12\epsilon_{mnkl}F_{kl}$. In these conventions,
characteristic of the mathematical instanton literature, the
gauge field $A_m$ is anti-Hermitian and so the covariant derivative
is $D_m=\partial_m+g A_m$.

Instantons are the finite action solutions of the classical
equations-of-motion which consequently satisfy a first-order
equation. From the inequality
\EQ{
\int d^4x\,\TrN\,\big(F_{mn}\pm{}^*F_{mn}\big)^2\leq0\ ,
}
one establishes a lower bound on the real part of the action:
\EQ{
-\frac12\int d^4x\,\TrN\,F_{mn}^2\geq \frac{8\pi^2}{g^2}|k|\ ,
}
with equality when the gauge field satisfies the
the self-dual, for $k>0$, or anti-self-dual, for $k<0$, Yang Mills equations:
\EQ{
{}^*F_{mn}\equiv\tfrac12\epsilon_{mnkl}F_{kl}=\pm F_{mn}\ .
\elabel{selfdual}
}
In our convention, instantons are the self-dual solutions and carry
positive topological (instanton)
charge $k>0$. In contrast, anti-instantons
satisfy the anti-self-dual Yang-Mills equations and carry negative
topological charge $k<0$. Their action is
\EQ{
S=\frac{8\pi^2|k|}{g^2}+ik\theta=\begin{cases}-2\pi ik\tau &k>0\\
-2i\pi k\tau^* & k<0\ ,\end{cases}
}
where we have defined the complex coupling
\EQ{
\tau=\frac{4\pi i}{g^2}+\frac{\theta}{2\pi}\ .
}

In this review we will discuss exclusively instantons, rather than
anti-instantons. The fundamental problem then, and the main subject of
\S\ref{sec:S8}, is to find all the solutions to \eqref{selfdual} with a
$+$ sign. When discussing the instanton calculus it is very convenient
to introduce a {\it quaternionic\/} notation for
four-dimensional Euclidean spacetime. The covering group
of the $\SO(4)$ Euclidean Lorentz group in four dimensions is
$\SU(2)_L\times\SU(2)_R$. A 4-vector
$x_n$ can be rewritten as a $({\bf2},{\bf2})$ of this group
with components
$x_{\alpha\aD}$ (or $\bar x^{\aD\alpha}$).
Here, $\alpha,\aD=1,2$ are spinor indices of
$\SU(2)_L$ and $\SU(2)_R$, respectively. The explicit relation
between the two bases is
\EQ{
x_{\alpha\aD}=x_n\sigma_{n\alpha\aD}\ ,\qquad
\bar x^{\aD\alpha}=x_n\bar\sigma_n^{\aD\alpha}\ ,
\elabel{rae}
}
where $\sigma_{n\alpha\aD}$ are four $2\times2$ matrices
$\sigma_n=(i\vec\tau,1_{\sst[2]\times[2]})$,
where $\tau^c$, $c=1,2,3$ are the three Pauli
matrices. In addition we define the Hermitian conjugate
matrices $\bar\sigma_n\equiv
\sigma_n^\dagger=(-i\vec\tau,1_{\sst[2]\times[2]})$
with components $\bar\sigma_n^{\aD\alpha}$.\footnote{
Notice that
indices are raised and lowered with the
$\epsilon$-tensor defined in Appendix \ref{app:A1}:
$\bar x^{\aD\alpha}=\epsilon^{\alpha\beta}
\epsilon^{\aD\bD}x_{\beta\bD}$.}
As explicit $2\times2$-dimensional matrices
\EQ{
x_{\alpha\aD}=\begin{pmatrix}ix_3+x_4 & ix_1+x_2 \\
ix_1-x_2& -ix_3+x_4\end{pmatrix}\ ,\qquad
\bar x^{\aD\alpha}=\begin{pmatrix}-ix_3+x_4 & -ix_1-x_2 \\
-ix_1+x_2& ix_3+x_4\end{pmatrix}\ .
\elabel{raf}
}
Notice that derivatives are defined in the same way:
\EQ{
\partial_{\alpha\aD}=\sigma_{n\alpha\aD}\partial_n\ ,\qquad
\bar\partial^{\aD\alpha}=\bar\sigma_n^{\aD\alpha}\partial_n\ .
}
But note with this definition
$\partial_{\alpha\aD}\neq\partial/\partial x_{\alpha\aD}$.

We now introduce the Lorentz generators,
\EQ{
\sigma_{mn}=\tfrac14(\sigma_m\bar\sigma_n-\sigma_n\bar\sigma_m)\ ,\qquad
\bar\sigma_{mn}=\tfrac14(\bar\sigma_m\sigma_n-\bar\sigma_n\sigma_m)\ ,
}
which are, respectively, self-dual and anti-self-dual:
\EQ{
\sigma_{mn}=\tfrac12\epsilon_{mnkl}\sigma_{kl}\ ,\qquad
\bar\sigma_{mn}=-\tfrac12\epsilon_{mnkl}\bar\sigma_{kl}\ .
\elabel{sdasd}
}
In terms of these, we
write down, in regular gauge, the explicit one-instanton solution
of the $\SU(2)$ gauge theory,
known as the BPST instanton \cite{BPST}, as
\EQ{
A_n=g^{-1}\frac{2(x-X)_m\sigma_{mn}}
{(x-X)^2+\rho^2}\ .
\label{regbpst}}
Note the identification $\SU(2)$ gauge indices and those of $\SU(2)_{L}$
introduced above, reflecting the fact that the instanton configuration
breaks the product of these groups down to a diagonal subgroup.
The instanton gauge-field above depends on five parameters:
one scale-size $\rho$ and the four-vector instanton position
$X_m$. Performing global
$\SU(2)$ gauge-rotations of the right hand side of \eqref{regbpst}
the total number of free parameters of a single-instanton
solution becomes eight. This is the first example of the emergence
of instanton collective coordinates. We will see that in general a
$k$-instanton solution in the $\SU(N)$ theory will contain $4kN$ collective
coordinates (for the $\SU(2)$ theory and $k=1$ we have $4kN=8$).
It is straightforward to see that the corresponding field-strength
\EQ{
F_{mn} = g^{-1}\frac{4\rho^2\sigma_{mn}}
{\big((x-X)^2+\rho^2\big)^2}\ , }
is self-dual. The anti-instanton is obtained from expressions above
via a substitution $\sigma_{mn} \rightarrow \bar\sigma_{mn}$.

Note that the BPST instanton \eqref{regbpst} is a non-singular expression
which falls off at large distances as $1/x$. This slow fall off would
make it difficult to construct square-integrable quantities
involving instanton gauge fields (see \cite{tHooft} for more detail).
An elegant and straightforward resolution of this technical problem
is to gauge-transform the regular instanton \eqref{regbpst} with a
singular gauge transformation $U(x)=\bar\sigma_m (x-X)_m /|x-X|.$
The resulting
expression for the instanton in singular gauge is
\EQ{
A_n=g^{-1}\frac{2\rho^2(x-X)_m\bar\sigma_{mn}}
{(x-X)^2\big((x-X)^2+\rho^2\big)}\ .
\label{singbpst}}
Note in singular gauge the $\SU(2)$ gauge indices are identified with
$\SU(2)_R$ indices in contrast to the expression in regular gauge.
This expression falls off as $1/x^3$ at large distances which improves
the convergence of various integrals.\footnote{The fast fall off
of the singular gauge is also required for applying the LSZ reduction
formulae to various Green functions in the instanton background.}
Hence, from now on we will always
assume that all the multi-instanton solutions we are dealing with are
written in singular gauge. The price to pay for this is the apparent
singularity of the solution \eqref{singbpst} at the instanton centre $x_m=X_m.$
However, this singularity is not a problem, since it is, by
construction, removable by a gauge transformation.
A rigorous way of dealing with a singular gauge transformation
is to introduce it on a punctured Euclidean space with the singular
point(s) being removed. Then the singular-gauge instanton remains regular
on the punctured space. Note that the punctures contribute to the boundary,
hence the integrals of total derivatives, such as the instanton charge
\eqref{topc}, will receive contributions from these punctures.

\subsection{Collective coordinates and moduli space}\elabel{sec:S3}

One of the key concepts associated to instantons (and more generally
to solitons) is the idea of a {\it moduli space of solutions\/}.
For instantons, this is the space of inequivalent solutions of the self-dual
Yang-Mills equations \eqref{selfdual}. Since we
are dealing with a gauge theory, the word ``inequivalent'' requires
some clarification. It is most convenient to think of
``inequivalent'' as being equivalence up to {\it local\/}
gauge transformations. So, for instance,
solutions differing by {\it global\/} gauge
transformation will be deemed inequivalent to each other. The reason for the
distinction between local and global gauge transformations is
discussed in
\cite{Coleman}. In the present context the main point is simply that the
usual covariant gauge fixing condition does not fix global gauge
transformations and hence we must still integrate over the corresponding
orbits in the path integral. If one is calculating a correlation function of
gauge-invariant operators, this integral simply leads to an additional
factor of the volume of the gauge group.

It will turn out that the moduli space of instantons has a lot of
mathematical structure, but for the
moment, there are two properties that are paramount. Firstly, since
finite action classical solutions of gauge theories on
four-dimensional Euclidean space are classified by
the topological (instanton) charge \eqref{topc}, the moduli space must contain
distinct components describing the inequivalent solutions for
each topological charge $k$. When the gauge group is $\SU(N)$ (or
$\U(N)$ since the abelian factor makes
no difference to the instanton solutions on a commutative Euclidean space)
we will denote the moduli
space of instantons with topological charge $k$ as $\ms_k$. The second
property that these moduli spaces have is that they are
manifolds, a fact which is not {\it a priori\/}
obvious. Strictly speaking, as we shall see later,
they have conical-type singularities (which occur physically
when instantons shrink to zero size) and so we will use the term
manifold in a slightly looser sense to encompass spaces with these
kinds of features. We shall also see that
$\ms_k$ has even more structure: it is a complex manifold of
a very particular type known as a hyper-K\"ahler manifold.

Since the moduli space is a manifold, albeit with
singularities, we can introduce local
coordinates to label its points. The
coordinates on the moduli space label various {\it collective\/}
properties of the gauge field and are therefore called {\it collective
coordinates\/}. So the gauge fields of the instanton
$A_n(x;X)$ depend not only on the coordinates on ${\mathbb R}^4$,
$x_n$, but also on a set of collective coordinates that we
denote $X^{\mu}$, $\mu=1,2,\ldots,{\rm dim}\,\ms_k$.
Some of the collective coordinates have an obvious
physical interpretation; for instance, if we have have a given
instanton solution, then since the solution is localized in
${\mathbb R}^4$ it has a definable notion of centre. We can
obviously translate the whole configuration in ${\mathbb R}^4$ and so
there must be collective coordinates that specify the position of the
centre which we denote as $X_n$, $n=1,2,3,4$. Notice that, by symmetry, the
gauge fields can only depend on these coordinates through the difference
$x_n-X_n$. To reflect this, the moduli space $\ms_k$ is a product
\EQ{
\ms_k={\mathbb R}^4\times\cms_k\ .
\elabel{cmsd}
}
The component $\cms_k$, with the centre factored out,
is known as the {\it centred moduli space\/}.

Notice that the collective coordinates $X_n$ arise because, by its very
nature, a given
instanton solution breaks the translational symmetry of the
theory. Another way to say this is that the subspace
spanned by the collective coordinates $X_n$ around a given point in
the moduli space can be generated by acting on the instanton solution
with the group elements of the broken symmetry. In this
case the symmetry in question are translations in spacetime. So if
$T_{X}=-X_n\partial/\partial x_n$
are the generators of translations, then
\EQ{
A_n(x;X,\ldots)=e^{T_{X}}A_n(x;0,\ldots)=A_n(x-X;0,\ldots)\ .
}
This is all quite trivial for the translational collective
coordinates; however, it illustrates a general principle: there are
collective coordinates associated to all the symmetries of the gauge
theory that are broken by a given instanton solution. However, not all
symmetries of the gauge theory give inequivalent collective
coordinates and not all collective coordinates correspond to broken
symmetries: indeed typically the majority do not.
Another way to phrase this is that the symmetries of
the classical equations-of-motion are realized as symmetries of the
moduli space and different symmetries may sweep out the same
subspace of the moduli space while some symmetries may act trivially
on (leave invariant) the moduli space. Finally,
there may be some directions in the moduli space,
and typically the majority, that are not related to any symmetry.

We now describe the symmetries of our theory.
First of all, we have spacetime symmetries including the Poincar\'e
symmetry of four-dimensional Euclidean space. However,
classical gauge theory is actually invariant under the larger group
that includes conformal transformations. In all, this group has fifteen
generators which includes four translations, six rotations, four
special conformal transformations generated by
\EQ{
\delta x_n=2x_n(x\cdot \epsilon)-\epsilon_nx^2\ ,
}
where $\epsilon_n$ is an infinitesimal four-vector, and dilatations
$\delta x_n=\alpha x_n$. In quaternionic language where the vector $x_n$ is
represented as the $2\times2$ matrix $x_{\alpha\aD}$
in \eqref{raf}, the action of the whole conformal group can be written
elegantly as
\EQ{
x\to x'=(Ax+B)(Cx+D)^{-1}\ ,\qquad \det\MAT{A&B\\ C&D}=1\ ,
\elabel{cgr}
}
where $A$, $B$, $C$ and $D$ are quaternions.
Notice that there are fifteen variables in the transformation \eqref{cgr},
matching the dimension of the conformal group.
As well as these spacetime symmetries,
there are global gauge transformations in our chosen gauge group $\SU(N)$.
Some of these symmetries will, like translations, be broken by an
instanton solution, and consequently be realized non-trivially
on the moduli space.

Another key concept that is related to idea of a collective coordinate
is the notion of a {\it zero mode\/}. Suppose that $A_n(x)$ is an
instanton solution. Consider some small fluctuation $A_n(x)+\delta
A_n(x)$ around this
solution which is also a solution of the self-dual Yang-Mills
equations \eqref{selfdual}. To linear order
\EQ{
{\cal D}_m\delta A_n-{\cal D}_n\delta
A_m=\epsilon_{mnkl}{\cal D}_k\delta A_l\ .
\elabel{zme}
}
There are actually three independent equations here,
manifested by writing them in quaternionic form:
\EQ{
\vec\tau^\aD{}_{\bD}\Dbarslash^{\bD\alpha}\delta A_{\alpha\aD}=0\ ,
\elabel{nonp}
}
where $\vec\tau$ are the three Pauli matrices and we define the
covariant derivatives in the quaternionic basis as
\EQ{
\Dslash\equiv\sigma_n{\cal D}_n\ ,\qquad\Dbarslash
\equiv\bar\sigma_n{\cal D}_n\
.
}
We must also fix the gauge in some way to weed out the variations that
are just local gauge transformations rather than genuine physical
variations of the instanton solution. This is conveniently achieved
by demanding that all fluctuations, including the zero modes,
are orthogonal to gauge transformations. Here, orthogonality is
defined in a functional sense by means of the inner product on
adjoint-valued (anti-Hermitian) variations
\EQ{
\langle\delta A_n,\delta'A_n\rangle=-2\int
d^4x\ \trN\,\delta A_n(x)\delta'A_n(x)\ .
}
Implicitly, we are discussing fluctuations which are square-integrable
with respect to this inner product.
So if $\delta A_n$ is orthogonal to a gauge transformation,
$\int d^4x\,\TrN\,{\cal D}_n\Omega\,\delta A_n=0$, then by integrating-by-parts
\EQ{
{\cal D}_n\delta A_n=0
\ .
\elabel{gchoice}
}
Again it is quite instructive to write this in quaternionic form:
\EQ{
\Dbarslash^{\aD\alpha}\delta A_{\alpha\aD}=0\ .
\elabel{nono}
}
The conditions for a zero mode, \eqref{nonp} and \eqref{nono}, can
then be written as a single quaternion equation:
\EQ{
\Dbarslash^{\aD\alpha}\delta A_{\alpha\bD}=0\ .
\elabel{ttyy}
}
We recognize this as the covariant Weyl equation for a Weyl spinor
$\psi_\alpha=\delta A_{\alpha\bD}$ in the
instanton background. Due to the free $\bD$ index each gauge
zero mode actually corresponds to two independent solutions of the
Weyl equation.

The fluctuations $\delta A_n(x)$ satisfying \eqref{zme} and
\eqref{gchoice}, or \eqref{ttyy}, are then
``zero modes'' in the sense that they represent physical fluctuations in field
space which do not change the value of the action. Non-zero mode
fluctuations necessarily increase the action of the instanton
solution. To quadratic order around the instanton solution, one finds
\SP{
S[A]&=-2\pi ik\tau+
\int d^4x\,\TrN\,\delta A_m({\cal D}^2\delta_{mn}+2gF_{mn})\delta
A_n+\cdots\\
&=-2\pi ik\tau-
\frac1{2}\int d^4x\ \TrN\,\delta\bar
A^{\aD\alpha}\Delta^{(+)}{}_\alpha{}^\beta\delta A_{\beta\aD}
+\cdots\ ,
\elabel{ggss}
}
where we have defined the fluctuation operator $\Delta^{(+)}$, and, for
future use, a companion $\Delta^{(-)}$:
\AL{
\Delta^{(+)}&\equiv-\Dslash\Dbarslash
=-1_{\sst[2]\times[2]}{\cal D}^2
-gF_{mn}\sigma_{mn}\ ,\elabel{rraa}\\
\Delta^{(-)}&\equiv-\Dbarslash\Dslash
=-1_{\sst[2]\times[2]}{\cal D}^2
-gF_{mn}\bar\sigma_{mn}\ .
\elabel{rrbb}}
The equality between the two expressions in \eqref{ggss} is
established by using the fact that $\sigma_{mn}$ is a projector onto
self-dual tensors \eqref{projsasd}. In addition, in an instanton (as
opposed to an anti-instanton) background, the second term in
$\Delta^{(-)}$ vanishes; hence
\EQ{
\Delta^{(-)}
=-1_{\sst[2]\times[2]}{\cal D}^2\ .
\elabel{ddww}
}
The two operators $\Delta^{(\pm)}$
play an important r\^ole in governing the behaviour of
fluctuations around the instanton in
the semi-classical approximation. A
important result is that in an instanton (self-dual) background
$\Delta^{(+)}$ has zero modes, whereas, by virtue of
\eqref{ddww}, $\Delta^{(-)}$ is a
positive semi-definite operator and therefore has no normalizable zero modes.

The zero modes are associated to the collective coordinates in the
following way. Consider the space of solutions
$A_n(x;X)$ where $x_n$ are the Euclidean spacetime coordinates and $X^\mu$
are the collective coordinates.
The derivative of the gauge field with respect to a
collective coordinate, $\partial A_n/\partial X^\mu$, is guaranteed to
satisfy the zero mode equation \eqref{zme}:
in other words it is a potential zero mode. In order to ensure that it is
a genuine zero mode, we have to
satisfy the gauge condition \eqref{gchoice}.
The way to achieve this
is to notice that \eqref{zme} is trivially satisfied by a gauge
transformation $\delta
A_n=D_n\Omega$, for some function $\Omega$ in the Lie algebra of the
gauge group. We can then consider a linear combination of the
derivative by the collective coordinate and a compensating gauge
transformation:
\EQ{
\delta_{\mu}A_n(x;X)\equiv\PD{A_n(x;X)}{X^{\mu}}-{\cal D}_n\Omega_{\mu}(x)\ .
\elabel{hhff}
}
The parameter of the gauge transformation $\Omega_{\mu}(x)$
is chosen in order
that \eqref{gchoice} is satisfied by $\delta_{\mu} A_n$; hence
\EQ{
{\cal D}_n\Big(\PD{A_n}{X^{\mu}}\Big)={\cal D}^2\Omega_{\mu}\ .
\elabel{edr}
}
The quantity $\delta_{\mu}A_n$ is then the ``genuine'' zero mode associated
to $X^{\mu}$. For the case of instantons, as long as we work in a
special gauge known as {\it singular gauge\/} (as in Eq.~\eqref{singbpst}),
there are no subtleties
associated with normalizability and $\delta_\mu A_n$ are all square integrable.
This rather convenient choice of gauge will be described
in \S\ref{sec:S12}. Subsequently in \S\ref{sec:S14} we will show in
singular gauge that
$\delta_\mu A_n$ is ${\cal O}(x^{-3})$ for large $x$.

We have seen that all collective coordinates are associated to zero modes.
However, when gauge fields couple to other fields,
it can happen that instanton zero modes are no longer associated to collective
coordinates because they can fail to integrate from solutions of the
linearized problem to solutions of the full coupled equations. Often we find
that zero modes of the linearized problem are lifted by higher order
interactions or by external interactions (for instance when scalar
fields have VEVs). In these circumstances it is still useful to
introduce the notion of {\it quasi-collective coordinates\/}
which are lifted by a non-trivial effective action. The
quasi-collective coordinates give rise to the notion of a
{\it quasi-instanton\/} which only satisfies the classical equation-of-motion
up to a certain order in the coupling.

\subsection{General properties of the moduli space of
instantons}\elabel{sec:S4}

The moduli space of instanton solutions plays a central r\^ole in our
story so it is important to describe in some detail how one arrives at
a description of it. The story is necessarily rather mathematical,
but the final answer, known as the ADHM construction, after Atiyah,
Drinfeld, Hitchin and Manin \cite{ADHM} is a great mathematical
achievement which we will review in \S\ref{sec:S8}. The
ADHM moduli space has a lot
of structure that can be deduced from general considerations.
For example, the dimension of $\ms_k$
can be obtained by simply counting the
number of zero modes at a point in the moduli space using the
Atiyah-Singer Index Theorem \cite{AtiyahSinger} (for applications to
instantons see \cite{BCGW}). The result is that $\ms_k$ has real
dimension $4kN$.

\subsubsection{The moduli space as a complex manifold}\elabel{sec:S7}

We have argued that the moduli space $\ms_k$ is a
space with dimension $4kN$. In fact it is a Riemannian
manifold endowed with a natural metric defined as the
functional inner product of the zero modes (in singular
gauge),\footnote{The factor of $g^2$ is inserted here so that the
metric is independent of the coupling.}
\EQ{
g_{\mu\nu}(X)=-
2g^2\int d^4x\ \TrN\,\delta_{\mu} A_n(x;X)\delta_{\nu}A_n(x;X)\ .
\elabel{ccmet}
}
This metric plays an important r\^ole in the theory since, as we
establish in \S\ref{sec:S16}, it defines
the volume form on $\ms_k$ that arises from changing variables in the
path integral from the gauge field to the collective
coordinates. However, we shall find that various other quantities that
are derived from the metric, like the connection and curvature, also
have an important r\^ole to play in the instanton calculus.

In fact there is more structure
to the moduli space than simply the existence of a metric. It turns
out that it is also a complex manifold of a very particular kind;
namely a hyper-K\"ahler space (with singularities).
A short review of some relevant aspects of such
spaces is provided in Appendix \ref{app:A2}. Fundamentally, these
spaces admit 3 linearly
independent complex structures $\BI^{(c)}$, $c=1,2,3$, that satisfy the
algebra
\EQ{
\BI^{(c)}\BI^{(d)}
=-\delta^{cd}+\epsilon^{cde}\BI^{(e)}\ .
\elabel{csa}
}
Often we will represent them as a three-vector $\vec\BI$.
The key idea for constructing the triplet of complex structures on
$\ms_k$ arises from noticing that Euclidean spacetime
is itself hyper-K\"ahler. The
three complex structures of ${\mathbb R}^4$
can be chosen so that $I^c_{mn}=-\bar\eta^c_{mn}$, where $\bar\eta^c_{mn}$
is a  't Hooft $\eta$-symbol defined in Appendix \ref{app:A1}.
In the quaternion basis
$x_{\aD\alpha}$
\EQ{
(\vec\BI\cdot x)_{\alpha\aD}=
ix_{\alpha\bD}\vec\tau^{\bD}{}_{\aD}
\ ,\qquad
(\vec\BI\cdot\bar x)^{\aD\alpha}=
-i\vec\tau^{\aD}{}_\bD\bar x^{\bD\alpha}\ .
}
These descend to give the three complex structures on
$\ms_k$ in the following way. Notice that in the zero mode
equation \eqref{ttyy} $\bD$
is a free index. This means that if $\delta_{\mu}
A_{\alpha\aD}$ is a zero mode then so is
$\delta_{\mu}A_{\alpha\bD}G^\bD{}_{\aD}$, for any constant matrix
$G$. In particular,
if $\delta_{\mu}A_{\alpha\aD}$ is a zero mode then so is
$(\vec\BI\cdot\delta_\mu A)_{\alpha\aD}
=i\delta_\mu A_{\alpha\bD}\vec\tau^{\bD}{}_{\aD}$.
Since the zero modes form a complete set,
there must exist $\vec\BI{}^{\nu}{}_{\mu}$ such that
\EQ{
(\vec\BI\cdot\delta_{\mu}
A)_{\alpha\aD}=\delta_{\nu}A_{\alpha\aD}\vec\BI{}^{\nu}{}_{\mu}\ ,
\elabel{ddq}
}
from which it follows
that the triplet $\vec\BI{}^{\nu}{}_{\mu}$ satisfies
the algebra \eqref{csa}.

At this stage, $\vec\BI^\mu{}_\nu$ are {\it almost\/} complex
structures because
we have not proved that they are integrable \eqref{xmk}. Rather than
prove this directly we will follow the analysis of
Maciocia \cite{Maciocia:1991ph} and construct a  {\it hyper-K\"ahler
potential\/} for
$\ms_k$. This proves that $\ms_k$ is not only hyper-K\"ahler
but also it is a rather special kind for which
each complex structure shares
the same K\"ahler potential (called the hyper-K\"ahler potential).
The expression for the potential is \cite{Maciocia:1991ph}
\EQ{
\chi=-\frac{g^2}{4}\int d^4x\,x^2\,\TrN\,F_{mn}^2\ .
\elabel{hypkp}
}
In order to prove this, we pick out one of the complex structures
$\BI^{(c)}$
of ${\mathbb R}^4$. Choose holomorphic coordinates $(z^i,\bar z^{i})$,
$i=1,2$, with respect to this particular complex structure. For
example, choosing $\BI^{(3)}$ we have $z^1=ix^3+x^4$ and $z^2=ix^1-x^2$.
The complex structure is associated, via
\eqref{ddq}, with a complex structure on $\ms_k$.
We can then choose a set of matching holomorphic coordinates on $\ms_k$
$(Z^\xii,\bar Z^{\xii})$, $\xii=1,\ldots,\tfrac12{\rm dim}\,\ms_k$
for which the complex structure on the moduli space is
\EQ{
\BI^{(c)}=\MAT{i\delta^\xii{}_\xjj&0\\ 0&-i\delta^\xii{}_\xjj}\ .
}
Then,
\EQ{
\Big(\BI^{(c)}\cdot\PD{A}{Z^\xii}\Big)_{\alpha\aD}=i\PD{A_{\alpha\aD}}{Z^\xii}
\ ,\qquad
\Big(\BI^{(c)}\cdot\PD{A}{\bar Z^{\xii}}\Big)_{\alpha\aD}
=-i\PD{A_{\alpha\aD}}{\bar Z^\xii}\
.
}
For example, for $\BI^{(3)}$ we have
\EQ{
\PD{A_{\alpha2}}{Z^\xii}=0\ ,\qquad\PD{A_{\alpha1}}{\bar Z^{\xii}}=0\ .
\elabel{tarez}
}
Furthermore, the variation equation \eqref{zme} implies
that the derivatives above automatically satisfy the
background gauge condition \eqref{nono}. Consequently
\EQ{
\delta_\xii A_n\equiv\PD{A_n}{Z^\xii}\ ,\qquad
\bar\delta_{\xii} A_n\equiv\PD{A_n}{\bar Z^{\xii}}
}
are zero modes directly without the need
for a compensating gauge transformation \eqref{hhff}. This fact turns out to
be crucial. Furthermore,
\EQ{
\frac{\partial^2 A_n}{\partial \bar Z^{\xjj}\partial Z^\xii}=0\ ,
\elabel{ddder}
}
as is clear from \eqref{tarez} for the case of $\BI^{(3)}$.

By explicit calculation, using the zero mode condition \eqref{ttyy}
and \eqref{ddder}, one finds
\EQ{
\frac{\partial^2}{\partial \bar Z^{\xjj}\partial Z^\xii}
\TrN\,F_{mn}^2=\square\,\TrN\,\delta_\xii A_n\bar\delta_{\xjj} A_n
-2\partial_m\partial_n\TrN\,\delta_\xii A_m\,\bar\delta_{\xjj} A_n\ .
}
Using this in \eqref{hypkp} and
integrating-by-parts twice, discarding the surface terms
since the zero modes decay as ${\cal O}(x^{-3})$, we have
\EQ{
\frac{\partial^2\chi}{\partial \bar Z^{\xjj}\partial Z^\xii}=
-2g^2\int d^4x\,\TrN\,\delta_\xii A_n\,\bar\delta_{\xjj}A_n\ .
}
By comparing with
\eqref{ccmet}, we see that the above expression is a component of the
metric on the space of zero modes:
\EQ{
g(X)=\frac{\partial^2\chi}{\partial \bar
Z^{\xjj}\partial Z^\xii}\,dZ^\xii\,d\bar Z^{\xjj}\ .
\elabel{methkp}
}
This proves that $\chi$ is the K\"ahler potential for the complex
structure $\BI^{(c)}$ for $c=3$.
However, $\chi$ manifestly does not depend on the choice
of the index $c=1,2,3$ and so it is a hyper-K\"ahler potential, and by
implication $\ms_k$ is hyper-K\"ahler space.

\subsection{The ADHM construction of instantons}\elabel{sec:S8}

In this section, we describe the construction of instantons due to
Atiyah, Drinfeld, Hitchin and Manin (ADHM) \cite{ADHM} and how this
leads to a description of $\ms_k$ for which the hyper-K\"ahler
property is manifest.
This remarkable construction of ADHM was originally discussed in
Refs.~\cite{Corrigan:1979pr,CGTone,CWS}. Here we follow, with minor
modifications, the $\SU(N)$ formalism of
Refs.~\cite{KMS,MO3}. Our approach here is to
describe the ADHM construction as an ansatz for producing instanton
solutions and we direct the reader to the references above for the
more mathematical technicalities.

The basic object in the ADHM construction
is the $(N+2k)\times 2k$ complex-valued matrix
$\Delta_{\lambda i\aD}$ which is taken to be linear in the
spacetime variable $x_n$:
\EQ{
\Delta_{\lambda i \aD}(x)
= a_{\lambda  i \aD}+
b_{\lambda  i}^{\alpha}x_{\alpha\aD}\ ,\qquad
\bar\Delta^{\aD\lambda}_{i}(x)
 = \abar^{\aD\lambda}_{i} +
\bar{x}^{\aD \alpha} \, \bar b_{i\alpha}^\lambda\ .
\elabel{del}
}
Here, we have introduced ``ADHM indices''
$\lambda,\mu\ldots=1,\ldots,N+2k$ and ``instanton indices''
$i,j,\ldots=1,\ldots,k$ and used the
quaternionic representation of
$x_n$ as in \eqref{rae} and \eqref{raf}. By definition the
conjugate is\footnote{Throughout this, and other sections, an over-bar means
Hermitian conjugation: $\bar\Delta\equiv\Delta^\dagger$. Notice as
usual we have to distinguish between upper and lower spinor indices,
and hence ADHM indices, but not for the other types of index.}
\EQ{
\bar\Delta^{\aD\lambda}_{i}\equiv (\Delta_{\lambda i\aD})^*\ .
\elabel{relp}
}
We will soon
verify by direct calculation that $k$ is
the instanton charge of the solution. As discussed below, the
complex-valued constant matrices $a$ and $b$ in \eqref{del}
constitute a (highly over complete) set of collective
coordinates on $\ms_k$.

Generically, the null-space of the Hermitian conjugate matrix
$\bar\Delta(x)$
is $N$-dimensional, as it has $N$ fewer rows than columns.
The basis vectors for this null-space can be assembled
into an $(N+2k)\times N$ dimensional  complex-valued matrix
$U_{\lambda u}(x)$, $u=1,\ldots,N$:
\EQ{
\bar\Delta^{\aD\lambda}_{i}
U_{\lambda u}
= 0 =
\bar U_{u}^\lambda\Delta_{\lambda i\aD}\ ,
\elabel{uan}
}
where $U$ is orthonormalized according to
\begin{equation}
\bar U_{u}^{\lambda}U_{\lambda v}=\delta_{uv}\ .
\elabel{udef}\end{equation}
The construction requires a non-degeneracy condition: the maps
$\Delta_\aD(x):\,{\mathbb C}^k\rightarrow{\mathbb C}^{N+2k}$ must be injective
while the maps $\bar\Delta^\aD(x):\,{\mathbb C}^{N+2k}\rightarrow{\mathbb
C}^k$ must be surjective. Having said this, we will see in \S\ref{sec:S15},
that points where the non-degeneracy conditions break down have an
interesting physical interpretation.

In turn, the classical ADHM gauge field $A_n$ is constructed from $U$ as
follows. Note first that
for the special case $k=0$,  the antisymmetric
gauge configuration $A_n$  defined by\footnote{Frequently,
where confusion cannot
arise, we do not indicate all the indices. Often only the spinor
indices have to labelled explicitly.}
\begin{equation}(A_n)_{uv}=
g^{-1}\bar U_u^\lambda\partial_n U_{\lambda v}
\elabel{vdef}\end{equation}
is ``pure gauge'' so that it automatically
solves the self-duality equation \eqref{selfdual}
in the vacuum sector. The ADHM ansatz is that Eq.~\eqref{vdef} continues
to give a solution to Eq.~\eqref{selfdual}, even for nonzero $k$. As we
shall see, this requires the additional condition
\begin{equation}
\bar\Delta^{\aD\lambda}_{i}
\Delta_{\lambda j\bD}=\delta^\aD{}_\bD
(f^{-1})_{ij}\ ,
\elabel{dbd}\end{equation}
where $f$ is an arbitrary $x$-dependent $k\times k$ dimensional
Hermitian matrix. Note that the existence of the inverse $f^{-1}$ is
guaranteed by the non-degeneracy condition.

To check the validity of the ADHM ansatz, we first observe
that  Eq.~\eqref{dbd} combined with the null-space condition \eqref{uan}
imply the completeness relation
\EQ{
{\cal P}_{\lambda}{}^{\mu}\equiv
U_{\lambda u}\bar U_{u}^{\mu}=
\delta_{\lambda}^{\mu}-
\Delta_{\lambda i\aD}f_{ij}
\bar\Delta_{j}^{\aD\mu}\ .
\elabel{cmpl}
}
Note that $\cal P$, as defined, is actually a projection operator; the
fact that one can write $\cal P$ in these two equivalent ways turns
out to be a useful trick in ADHM algebra, used pervasively throughout
the instanton calculus.
With the above relations the expression
for the field strength $F_{mn}$ may  be massaged as follows:
\begin{equation}\begin{split}
F_{mn} &\equiv \partial_mA_n-\partial_nA_m + g [A_m, A_n] =
g^{-1}\partial_{[m}(\bar U\partial_{n]}U)
+g^{-1}(\bar U\partial_{\,[m}U)(\bar U\partial_{n]}U)\\&=
g^{-1}\partial_{\,[m}\bar U(1-U\bar U)\partial_{n]}U
= g^{-1}\partial_{\,[m}\bar U\Delta f \bar\Delta\partial_{n]}U\\&=
g^{-1}\bar U\partial_{[m}\Delta f \partial_{n]}\bar\Delta U =
g^{-1}\bar Ub \sigma_{[m}\bar\sigma_{n]}f \bar{b} U  \\&=
4g^{-1}\bar U b \sigma_{mn}f\bar b U\ .
\elabel{sdu}\end{split}\end{equation}
Self-duality of the field strength then follows automatically from the
self-duality property of the tensor
$\sigma_{mn}$ \eqref{sdasd}.

The instanton number of the ADHM configuration can be calculated
directly using a remarkable identity of Osborn \cite{OSB}
for $\TrN\,F^2$ in the ADHM background proved in Appendix~\ref{app:A4}:
\SP{
-\frac{g^2}{16\pi^2}\int d^4x\,\TrN\,F_{mn}^2=
\frac1{16\pi^2}\int d^4x\,\square^2{\rm tr}_k\log\,f\ .
}
Now from \eqref{dbd} one can deduce
the asymptotic form for $f(x)$ for large $x$,
$f(x)\overset{x\to\infty}\longrightarrow x^{-2}1_{\sst[k]\times[k]}$,
and therefore the right-hand side is equal to $k$ as claimed.

Let us analyse the factorization condition \eqref{dbd} in more detail.
Noting that $f_{ij}(x)$ is arbitrary,
one extracts three $x$-independent conditions on $a$ and $b$:
\begin{subequations}
\begin{align}
\bar a^{\aD\lambda}_{i}a_{\lambda  j \bD}
&=  (\tfrac12 \abar a)_{ij} \ \delta^{\aD}_{\ \bD}
\ ,\elabel{conea}\\
\bar a^{\aD\lambda}_{i}b_{\lambda  j}^{\beta}
&=\bar b^{\beta\lambda}_{i} a_{\lambda  j}^\aD\ ,
\elabel{coneb}\\
\bar b_{\alpha i}^{\lambda}b_{\lambda  j}^{\beta}
&=(\tfrac12 \bar b b)^{}_{ij} \, \delta_{\alpha}^{\ \beta}\ .
\elabel{conec}\end{align}
\end{subequations}
These three conditions are generally known as the ``ADHM constraints''
\cite{CGTone,CWS}.\footnote{We should warn the reader that we use the
term ``ADHM constraints'' in a rather more restricted sense after a
canonical choice has been made below.}
They define a set of coupled quadratic conditions
on the matrix elements of $a,$ $\bar a$ $b$ and $\bar b$. Note that
\eqref{coneb} and \eqref{conec} can be combined in the useful form
\EQ{
\bar\Delta^\aD b^\alpha=\bar b^\alpha\Delta^\aD\ .
\elabel{ADHMbi}
}
The fact that the ADHM construction involves non-linear constraints
presents considerable difficulties for practical applications.
However, it turns out, as we shall we shall see in \S\ref{app:A5},
that the ADHM constraints can be resolved in a simple way, at least
generically, when $N\geq2k$.

The elements of the matrices $a$ and $b$ comprise the collective coordinates
for the $k$-instanton gauge configuration. Clearly the number
of independent such elements grows as $k^2$, even after accounting
for the constraints \eqref{conea}-\eqref{conec}. In contrast, the
number of physical collective coordinates should equal $4kN$ which scales
 linearly with $k$. It follows that
$a$ and $b$ constitute a highly redundant set. Much of this redundancy
can be eliminated by noting that the ADHM construction is
unaffected by $x$-independent transformations of the form
\EQ{
\Delta\to \Lambda\Delta\Upsilon^{-1}\ ,\qquad
U\to \Lambda U\ ,\qquad f\to\Upsilon f\Upsilon^\dagger\ ,
\label{tra}
}

provided $\Lambda \in \U(N+2k)$ and $\Upsilon \in {\rm Gl}(k, {\mathbb C})$.
Exploiting these symmetries, one can choose
a representation in which $b$ assumes a simple  canonical
form \cite{CGTone}. Decomposing the index $\lambda=u+i\alpha$
\EQ{
b_{\lambda j}^\beta=b_{(u+i\alpha) j}^\beta =
\MAT{0 \\ \delta_{\alpha}^{\ \beta}
\delta_{ij}}
\ ,\qquad
\bar b_{\beta j}^\lambda= \bar b_{\beta j}^{(u+i\alpha)}=
\big(0 \ \ \delta^{\ \alpha}_{\beta}\, \delta_{ji}
\big)\ .
\elabel{aad}
}
The remaining variables all reside in $a$ and we will split them up in
a way that mirrors the canonical form for $b$:
\EQ{
a_{\lambda  j \aD}=
a_{(u+i\alpha) j \aD}=
\begin{pmatrix}  w_{u j \aD}\\
(a'_{\alpha\aD})^{ }_{ij}\end{pmatrix}_{\phantom{q}}
\ ,\qquad
\bar a^{\aD\lambda}_{j}= \bar a^{\aD(u+i\alpha)}_j=
\big(\bar w^\aD_{j u}\ \ (\bar a^{\prime\aD\alpha})^{}_{ji}
\big)^{\phantom{T} }_{\phantom{q}}\ ,\elabel{aab}
}

With $b$ in the canonical form \eqref{aad},
the third ADHM constraint of \eqref{conec} is
satisfied automatically, while the remaining constraints \eqref{conea}
and \eqref{coneb} boil down to the $k\times k$ matrix equations:
\begin{subequations}
\begin{align}
\vec\tau^\aD{}_{\bD}(\bar a^\bD a_\aD)&= 0\elabel{fconea}\\
(a^{\prime}_n)^\dagger  &= a^{\prime}_n\ .
\elabel{fconeb}\end{align}
\end{subequations}
In \eqref{fconea} there are three separate equations since
we have contracted $\bar a^\bD a_\aD$
with any of the
three Pauli matrices, while in Eq.~\eqref{fconeb} we have decomposed
 $a'_{\alpha\aD}$ and $\bar a^{\prime\aD\alpha}$
in our usual quaternionic basis of spin matrices \eqref{rae}:
\begin{equation}a'_{\alpha \aD}=
a'_n\sigma_{n \alpha\aD}\ , \quad
\bar a^{\prime\aD \alpha} =
a'_n\bar\sigma_n^{\aD \alpha}\ .
\elabel{dec}\end{equation}
In the canonical form \eqref{aad} we also have the useful identity
\EQ{
\bar b_\alpha b^\beta=\delta_\alpha{}^\beta1_{\sst[k]\times[k]}
\elabel{bbident}
}
and the ADHM matrix $f$ takes the form
\EQ{
f=2\big(\bar w^\aD w_\aD+(a'_n+x_n1_{\sst[k]\times[k]})^2\big)^{-1}\ .
\label{fac}
}

Note that the canonical form for $b$ \eqref{aad}
is  preserved by a residual $\U(k)$ subgroup
of the $\U(N+2k)\times\Gl(k,{\mathbb C})$ symmetry group
\eqref{tra}:
\EQ{
\Lambda=\MAT{1_{\sst[N]\times[N]}&0\\ 0&\Xi\,1_{\sst[2]\times[2]}}
\ ,\qquad \Upsilon=\Xi\ ,\qquad \Xi\in\U(k)\ .
}
These residual transformations act non-trivially
on the remaining variables:
\begin{equation}w_{ui\aD} \to w_{\aD}\Xi
\ , \qquad
a'_n\to \Xi^\dagger\,
a'_n\,\Xi\ .
\elabel{restw}\end{equation}

Henceforth, we shall use exclusively the
streamlined version of the ADHM construction obtained by
fixing $b$ as in \eqref{aad}.
The basic variables will be $a_\aD=\{w_\aD,a'_n\}$ where we will
automatically assume \eqref{fconeb} so
that the four $k\times k$ matrices $a'_n$ are defined from the outset
to be Hermitian. The remaining constraints \eqref{fconea}
\EQ{
\vec\tau^\aD{}_\bD
\,\bar a^\bD a_\aD\equiv
\vec\tau^\aD{}_\bD\big(\bar w^\bD w_\aD+\bar
a^{\prime\bD\alpha}a'_{\alpha\aD}\big)=0
\elabel{badhm}
}
will be called the ``ADHM constraints''.

\subsubsection{The ADHM construction as a
hyper-K\"ahler quotient}\elabel{sec:S10}

It follows from the ADHM construction of the solutions of the
self-dual Yang-Mills equations that the moduli space $\ms_k$ is
identified with the
variables $a_\aD$ subject to the
ADHM constraints \eqref{badhm} quotiented by the
residual symmetry group $\U(k)$ \eqref{restw}. Subsequently it was realized
that the ADHM construction was actually an example of a more
general construction that has become known as the hyper-K\"ahler
quotient \cite{Hitchin:1987ea}. This
construction is reviewed in Appendix \ref{app:A2} and provides a way of
constructing a new hyper-K\"ahler space $\ms$ starting from a ``mother''
hyper-K\"ahler space $\tilde\ms$ with suitable isometries.
In the case of the ADHM construction, the mother space $\tilde\ms$ is simply
the Euclidean space ${\mathbb R}^{4k(k+N)}$ with coordinates $a_\aD$
and metric
\EQ{
\tilde g=4\pi^2\big(2d\bar w^\aD_{iu}\,dw_{ui\aD}+d(\bar
a^{\prime\aD\alpha})_{ij}\, d(a'_{\alpha\aD})_{ji}\big)\equiv
8\pi^2{\rm tr}_k\,\big(d\bar w^\aD\,dw_\aD+da'_n\,da'_n\big)\ .
\elabel{qum}
}
The normalization factor here will be justified in \S\ref{sec:S14}.
Flat space with dimension a multiple of four
is trivially hyper-K\"ahler as described in Appendix \ref{app:A2}.
The tangent space of a
hyper-K\"ahler space of real dimension $4n$, admits a distinguished
$\Sp(n)\times\SU(2)$ basis of tangent vectors. For flat space
${\mathbb R}^{4k(N+k)}$, this
symplectic structure is realized by a set of coordinates
$z^{\ii\aD}$, $\ii=1,\ldots,2k(N+k)$ and $\aD=1,2$, in terms of which the
flat metric is
\EQ{
\tilde g=\tilde
\Omega_{\ii\,\jj}\epsilon_{\aD\bD}dz^{\ii\aD}\,dz^{\jj\bD}\ ,
\elabel{flmt}
}
where $\tilde\Omega$ is a symplectic matrix.
In the present case,
where $\tilde\ms={\mathbb R}^{4k(N+k)}$ is identified with $a_\aD$, the
coordinates and matrix $\tilde\Omega$ are
\EQ{
z^{\ii\aD}=\MAT{\bar w^\aD_{iu}\\ (\bar a^{\prime\aD1})_{ij} \\
\epsilon^{\aD\bD}w_{ui\bD} \\
\epsilon^{\aD\bD}(a'_{1\bD})_{ij}}\ ,\qquad
\tilde\Omega_{\ii\,\jj}=4\pi^2\MAT{0 & 0 & 1_{\sst[kN]\times[kN]} & 0\\
0 & 0 & 0 & 1_{\sst[k^2]\times[k^2]}\\ -1_{\sst[kN]\times[kN]} & 0 & 0\\
0 & -1_{\sst[k^2]\times[k^2]} &0 & 0}\ ,
\elabel{jws}
}
where the Roman indices $\ii,\jj,\ldots$ each
run over the set of $2k(N+k)$ composite indices $\{iu,ij,ui,ij\}$.

The ADHM construction
is the hyper-K\"ahler quotient of $\tilde\ms={\mathbb R}^{4k(N+k)}$
by the isometry group $\U(k)$ which acts on the variables
$a_\aD$ as in \eqref{restw}. This action defines a set of tri-holomorphic
Killing vector fields on $\tilde\ms$:
\EQ{
X_r=iT^r_{ij}\bar w^\aD_{ju}\PD{}{\bar w^\aD_{iu}}
-iT^r_{ji}w_{uj\aD}\PD{}{w_{ui\aD}}+i[T^r,a'_n]_{ij}\PD{}{(a'_n)_{ij}}\
,
\elabel{qqp}
}
where $T^r$ is a generator of $\U(k)$ in the fundamental
representation.\footnote{We will choose a normalization ${\rm
tr}_kT^rT^s=\delta^{rs}$.}

There are two main parts to the quotient construction. Firstly,
one restricts to the {\it level set\/} $\ns\subset\tilde\ms$, defined by the
vanishing of the moment maps associated to the isometries, in this
case the $\U(k)$ vector fields
\eqref{qqp}. The expressions for the moment maps are given in
\eqref{mommp}, using \eqref{jws} and \eqref{qqp}, the moment maps are
\EQ{
i\vec\mu^{X_r}=4\pi^2\vec\tau^\bD{}_\aD{\rm tr}_k\,\big(T^r\bar
a^{\aD}a_{\bD}\big)-\vec\zeta^r\ .
\elabel{mommap}
}
The central elements $\vec\zeta^r$ can, in
this case, take values in the Lie algebra of the $\U(1)$ factor of the
gauge group. One notices immediately that the ADHM constraints
\eqref{badhm} are precisely the conditions $\vec\mu^{X_r}=0$ (with
vanishing central element). In other
words, the ADHM constraints explicitly implements the first part of the
quotient construction.
The second part of the quotient construction involves an ordinary
quotient of $\ns$ by the $\U(k)$ action (which is guaranteed to fix
$\ns$). But this quotient, as we have seen, is also an essential
ingredient of the ADHM construction.

The conclusion is that the
ADHM construction realizes the instanton moduli space $\ms_k$ as a
hyper-K\"ahler quotient of flat space by a $\U(k)$ group of
isometries (with vanishing
central elements $\vec\zeta^r=0$). Later in \S\ref{sec:S129}, we shall
describe what the physical interpretation of taking a non-vanishing
$\vec\zeta^r$. The dimension of the quotient space $\ms_k$ is
\EQ{
{\rm dim}\,\ms_k={\rm dim}\,\tilde\ms-4\,{\rm dim}\,\U(k)=4kN\ ,
}
as anticipated by the Index Theorem.

The importance of the hyper-K\"ahler quotient construction is that
geometric properties of $\ms_k$ are inherited from the mother space
$\tilde\ms$ in a rather straightforward way (as described in
Appendix~\ref{app:A2}). For example of particular importance will be
the metric on $\ms_k$ and we will now describe in some detail
how to construct it. First of all, we focus on $T\ns$, the tangent space of
the level set. Locally, as described in Appendix~\ref{app:A2},
this is the subspace of $T\tilde\ms$ orthogonal to
the $3k^2$ vectors $\tilde\BI^{(c)}X_r$, where $\tilde\BI^{(c)}$ are the three
independent complex structures of the mother space $\tilde\ms$.
Then we have the following decomposition
\EQ{
T\ns={\EuScript V}\oplus{\EuScript H}\ ,
}
where the {\it horizontal\/} subspace
${\EuScript H}$ is the subspace orthogonal to the vectors $X_r$ and
the {\it vertical\/} subspace ${\EuScript V}$ is the orthogonal complement.
The tangent space of $\ms_k$ is identified
with the quotient $T\ns/{\EuScript V}$. This means that each $X\in
T\ms$ has a unique lift to
${\EuScript H}$, which, by a slight abuse of notation, we denote by the same
letter. The metric on the quotient $g(X,Y)$ is then identified with
the metric on $\tilde\ms$, $\tilde g(X,Y)$, evaluated on the lifts to
$\EuScript H$.

What we would like to show is that the metric on $\ms_k$ inherited
from the quotient construction is equal to the metric on $\ms_k$
that arises from the functional inner product of zero modes
\eqref{ccmet}. Rather than implement the procedure that we describe
above explicitly, it is more straightforward to use the fact that the
hyper-K\"ahler spaces that we are considering are of a special class
which admit a hyper-K\"ahler potential. We have already determined
in \eqref{hypkp} the form of this potential for the metric arising
from the inner product of zero modes. Is this equal to
the hyper-K\"ahler potential arising from the quotient construction?
First of all, the mother space $\tilde\ms$ trivially admits such a
potential with
\EQ{
\tilde\chi=\tilde
\Omega_{\ii\,\jj}\epsilon_{\aD\bD}z^{\ii\aD}\,z^{\jj\bD}\equiv
8\pi^2{\rm tr}_k\big(\bar w^\aD w_\aD+a'_na'_n\big)\ .
\elabel{hkpadhmm}
}
The hyper-K\"ahler potential on the quotient space $\ms_k$ is then
simply obtained by finding a parameterization of the ADHM variables in
terms of the coordinates $\{X^\mu\}$ on $\ms_k$. In other words
$z^{\ii\aD}(X)$, or $a_\aD(X)$,
which solves the ADHM constraints and for some choice
of gauge slice for the $\U(k)$ action on the level set $\ns$. The
hyper-K\"ahler potential for $\ms_k$ is then obtained by restriction
\EQ{
\chi(X)=\tilde\chi\big(z^{\ii\aD}(X)\big)=
8\pi^2{\rm tr}_k\big(\bar w^\aD(X)w_\aD(X)+a'_n(X)a'_n(X)\big)\ .
\elabel{hkpadhm}
}
We now evaluate \eqref{hypkp} using
Osborn's formula \eqref{oruf} for $\TrN\,F_{mn}^2$, giving
\EQ{
\chi=\frac{g^2}{4}\int d^4x\,x^2\,\square^2{\rm tr}_k\,\log\,f\ .
}
Applying Gauss's
Theorem along with the ADHM form for the matrix $f$ \eqref{fac}
and \eqref{del}, one finds precisely \eqref{hkpadhm}.

\subsubsection{Symmetries and the moduli space}\elabel{sec:S11}

We now show how the symmetries of the classical gauge theory broken by
an instanton solution are realized on the ADHM moduli space.
Consider, first of all, the action of the conformal group on
the instanton solutions. The action of the conformal group is
given most elegantly in the quaternion basis \eqref{cgr}. So acting on
the ADHM variable $\Delta(x)$ we find
\EQ{
\Delta(x';a,b)=\Delta(x;aD+bB,aC+bA)(Cx+D)^{-1}\ .
}
Notice that since the gauge field only depends on $\bar U$ and $U$, defined by
\eqref{uan}, the factor of $(Cx+D)^{-1}$ on the right is redundant.
Hence, the action
of the conformal group on the ADHM variables is
\EQ{
a\to aD+bB\ ,\qquad b\to aC+bA\ .
}
To get the transformation on our canonical basis \eqref{aad} and
\eqref{aab} we have to perform a
compensating transformation of the form \eqref{tra} in order to return
$b$ to its canonical form \eqref{aad}. So there exist transformations of the
form \eqref{tra}, with $\Lambda$ and  $\Upsilon$
dependent on $a$ and the element of the conformal group, that takes
\EQ{
\Lambda(aC+bA)\Upsilon^{-1}=b\ ,
}
where $b$ assumes its canonical form \eqref{aad}.
The resulting action of
the conformal group on the ADHM variable $a$ is then
\EQ{
a\to\Lambda(aD+bB)\Upsilon^{-1}\ .
}
For example, translations act on the ADHM coordinates in the following
way. From
\EQ{
\Delta(x+\epsilon;a,b)=
\Delta(x;a+b\epsilon,b)\ ,
}
we deduce
\EQ{
a'_n\to a'_n+\epsilon_n1_{\sst[k]\times[k]}\ ,\qquad w_\aD\to w_\aD\ .
}
This allows us to identify the coordinates of the centre of the instanton with
the component of $a'_n$ proportional the identity matrix:
\begin{equation}
X_n=-k^{-1}{\rm tr}_k\,a'_n\ .
\elabel{transmo}\end{equation}
Note that these centre-of-mass coordinates do not appear in the ADHM
constraints \eqref{badhm} reflecting of the fact that the moduli
space is the product \eqref{cmsd}.

Global gauge transformation act on the gauge indices $u,v,\ldots$, so only on
the quantities
$w_{ui\aD}$. Generically, $w_{ui\aD}$ constitute a
set of $2k$ complex $N$-vectors. Consequently
if $N\leq2k$ then all global gauge
transformation generically act non-trivially on the ADHM variables
while if $N>2k$ then there is a non-trivial subgroup that leaves the
instanton fixed. This is the {\it stability group\/} of the
instanton. In order to identify it, we follow Bernard's description of
the one instanton moduli space \cite{Bernard} and embed the $k$
instanton solution in an $\SU(2k)$
dimensional subgroup of the gauge group. This involves choosing a
suitable gauge transformation that puts the $N\times2k$ matrix
$w$, with elements $w_{ui\aD}$, into upper-triangular form:
\def\hf{\textstyle{{1\over2}}}
\def\Ovec{{\vec0}}
\begin{equation}
\begin{pmatrix}w_{11}&\cdots&w_{12}\\ \vdots&\ddots&\vdots\\
w_{N1}&\cdots&w_{N,2k}\end{pmatrix}
=\grp\cdot\begin{pmatrix}\xi_{11}&\xi_{12}&\cdots&\xi_{1,2k}\\
0&\xi_{22}&\cdots&\xi_{2,2k}\\
{}&\ddots&\ddots&\vdots\\
\vdots&{}&\ddots&\xi_{2k,2k}\\
{}&{}&{}&0\\  {}&{}&{}&\vdots\\  0&0&\cdots&0\end{pmatrix}\ .
\elabel{uptri}\end{equation}
The $\xi_{ab}$, $a,b=1,\ldots,2k$,
are complex except for the diagonal elements $\xi_{aa}$ which we can choose
to be real. Note the group elements $\grp$ in the $\SU(N-2k)$
subgroup in the lower $(N-2k)\times(N-2k)$ corner leave $\xi$
invariant. Therefore, at least generically $\grp$ is valued in the coset
\EQ{
\grp\in\frac{\SU(N)}{\SU(N-2k)}\ .
\elabel{ncsp}
}
There are $4kN$ independent real parameters in $w$
on the left-hand side of \eqref{uptri},
matching $4k(N-k)$ in $\grp$ and $4k^2$ in
$\xi$, on the right-hand side.

With the parameterization \eqref{uptri}, the instanton solution
has the form
\EQ{
A_n=\grp^\dagger\MAT{(A_n^{\sst\text{inst}})_{\sst[2k]\times[2k]} &
0_{\sst[2k]\times[N-2k]} \\
0_{\sst[N-2k]\times[2k]}& 0_{\sst[N-2k]\times[N-2k]}}\grp\ .
}
Where $A_n^{\sst\text{inst}}$ is the $k$-instanton solution lying,
generically, in
$\SU(2k)\subset\SU(N)$. Generically,
the stability group of the instanton solution
consists of the $\SU(N-2k)$ in the denominator of \eqref{ncsp} along
an additional $\U(1)$ transformation generated by
\EQ{
\lambda=
\MAT{1_{\sst[2k]\times[2k]}&0_{\sst[2k]\times[N-2k]} \\
0_{\sst[N-2k]\times[2k]}& -\tfrac{2k}{N-2k}1_{\sst[N-2k]\times[N-2k]}}\ .
}
The stability group, therefore,
consists of elements $e^{i\lambda\theta}g$, with
$g\in\SU(N-2k)$. Notice that
this is not exactly $\SU(N-2k)\times\U(1)$ because
elements elements $e^{i\lambda\theta_1}$ and $e^{i\lambda\theta_2}$
with $\theta_1-\theta_2=\pi$ differ by an element of
the centre of $\SU(N-2k)$. More precisely, therefore,
the stability group is $\SS\big(\U(N-2k)\times\U(1)\big)\subset\SU(N)$.
To summarize: the non-trivial global
gauge transformations acting on a generic instanton solution are
\EQ{
\SU(N)\ \ \text{for}\ \  N\leq2k\ ,
\qquad \frac{\SU(N)}{\SS\big(\U(N-2k)\times\U(1)\big)}\ \ \text{for}\ \ N>2k\ .
\elabel{goc}
}
For $N>2k$, we
will loosely refer to $\grp$, taking values in \eqref{ncsp}, as the ``gauge
orientation'' of the instanton, even though the true gauge orientation
involves quotienting by the additional $\U(1)$ described above. Notice, the
additional $\U(1)$ can be identified with the $\U(1)$ subgroup of the
$\U(k)$ auxiliary group.

\subsubsection{Singular gauge, one instanton, the dilute limit and
asymptotics}\elabel{sec:S12}

Let us determine the gauge field $A_n$ more explicitly.
This entails solving for $U$, and hence $A_n$ itself via \eqref{vdef},
in terms of $\Delta$.  It is
convenient to make the decomposition:
\begin{equation}U_{\lambda v}=U_{(u+i\alpha)v}=
\begin{pmatrix} V_{uv} \\  (U'_{\alpha})_{iv}\end{pmatrix}
 \ , \quad
\Delta_{\lambda j\aD}=\Delta_{(u+i\alpha)j\aD}=
\begin{pmatrix} w_{uj\aD} \\
(\Delta'_{\alpha\aD})_{ij}\end{pmatrix}\ .
\elabel{dcmp}\end{equation}
Then from the completeness condition \eqref{cmpl} one finds
\begin{equation}
 V=
1_{\sst [N]\times [N]}-
w_\aD f\bar w^\aD\ .
\elabel{cmpag}\end{equation}
For any $V$ that solves this equation, one can find another by
right-multiplying it by an $x$-dependent $\U(N)$ matrix. A specific choice of
$V$ corresponds to fixing the spacetime gauge. The ``singular
gauges'' correspond to taking any one of the $2^N$ choices of
matrix square roots:
\begin{equation}V = (1_{\sst[N]\times[N]} - w_\aD f \wbar^\aD)^{1/2} \ .
\elabel{sga}\end{equation}
Next, $U'$ in \eqref{dcmp} is determined in terms of
$V$ via
\begin{equation}
U' = -\Delta'_\aD f \wbar^\aD \bar V^{-1}
\elabel{prim}\end{equation}
which likewise follows from \eqref{cmpl}.
Equations \eqref{sga} and \eqref{prim} determine $U$ in \eqref{dcmp},
and hence the gauge field $A_n$ via \eqref{vdef}.

We now show how the well-known \eqref{singbpst}
of 't~Hooft for the one instanton solution in $\SU(2)$
(in singular gauge) is reproduced by the ADHM construction.
Adopting the canonical form \eqref{aad}, we set the
instanton number $k=1$, thus dropping
the instanton, $i,j$,
indices. Eq.~\eqref{fconeb} then says that $a'_n$ is a real
4-vector which from the last section we will identify as minus the centre
$-X_n$ of the instanton, as per \eqref{transmo},
\begin{equation}
a'_n \equiv - X_n \in  {\mathbb R}^4\ .
\elabel{xcon}\end{equation}
The ADHM constraint \eqref{badhm} collapses to
\begin{equation}\bar w^\aD_u  w_{u \bD}=
\rho^2\,\delta^\aD_{\ \bD}  \ .
\elabel{wcon}\end{equation}
The parameter $\rho$ will soon be identified with the instanton
scale size. The constraint \eqref{wcon} can be explicitly solved by taking
\EQ{
w
=\rho\,\grp\
\begin{pmatrix}1_{\sst [2]\times [2]}\\
0_{\sst [N-2] \times [2]} \end{pmatrix} \ ,
\elabel{wcon2}
}
which is the decomposition \eqref{uptri} for one instanton and
identifies
$\grp\in\SU(N)$ as the gauge orientation of the
instanton. For one instanton, the ADHM quantity $f$ is a scalar; from
\eqref{fac}
\EQ{
f=\frac1{(x-X)^2+\rho^2}\ .
}
Then, from \eqref{sga} and \eqref{prim},
\SP{
V&=1_{\sst[N]\times[N]}+\frac1{\rho^2}\Big(\sqrt{\frac{(x-X)^2}
{(x-X)^2+\rho^2}}-1\Big)
w_\aD\bar w^\aD\ ,\\
U'&=-\frac{(x-X)_{\alpha\aD}\bar w^\aD}{|x-X|\sqrt{(x-X)^2+\rho^2}}\ .
}
Using \eqref{vdef}, one finds the
expression for the gauge potential in singular gauge
\EQ{
A_n=g^{-1}\frac{2w_\aD(x-X)_m\bar\sigma_{mn}{}^\aD{}_\bD
\bar w^\bD}{(x-X)^2\big((x-X)^2+\rho^2\big)}\ .
}
Comparing with the $\SU(2)$ solution \eqref{singbpst},
this form of the solution manifests the fact that a
single $\SU(N)$ instanton is described by taking
the $\SU(2)$ instanton solution and embedding it in
$\SU(2)\subset\SU(N)$. In this case the
three $\SU(2)$ generators are
\begin{equation}
T^c_{uv}=\rho^{-2}w_{u\aD}
\tau^c{}^\aD_{\ \bD}\bar w_{v}^\bD\ ,\qquad c=1,2,3.
\elabel{embed1}
\end{equation}
The fact that these generators satisfy the $\SU(2)$ algebra is
guaranteed by the ADHM constraints \eqref{wcon}. With the explicit
solution \eqref{wcon2}
\EQ{
A_n = \grp
\begin{pmatrix} A_n^{\sst \SU(2)} & 0 \\ 0 & 0\end{pmatrix}\grp^\dagger
\ ,
\elabel{inss}
}
which manifests the fact that $\grp$ is the gauge orientation of the instanton
taking values in the coset $\SU(N)/\SS(\U(N-2)\times\U(1))$.

In general a multi-instanton configuration cannot be thought of as a
combination of single instantons. However, there are asymptotic
regions of $\ms_k$ where the solutions can be identified as being
composed of well-separated single instantons.
Here, we consider the {\it completely
clustered limit\/} in which the $k$-instanton configuration looks like
$k$ well-separated single instantons. Up to the action of
the auxiliary $\U(k)$ symmetry, the completely clustered limit
is the region of moduli space
where the differences between the diagonal elements of $a'_n$
are much greater than the off-diagonal elements (in a sense that we
make precise below). We can then
identify $X^i_n\equiv-(a'_n)_{ii}$ as the centres of each of the $k$ single
instantons.

To be more specific, it is useful to fix the $\U(k)$ symmetry by
setting to zero the off-diagonal components of $a'_n$ that are
generated by $\U(k)$ adjoint action on the
diagonal matrix ${\rm diag}(-X^1_n,\ldots,-X^k_n)$. The ``gauge choice''
amounts to taking
\EQ{
(a'_n)_{ij}(X^i-X^j)_n=0
\elabel{gukch}
}
and we will denote the $a'$ so constrained by $\tilde a'$.
This leaves
the diagonal symmetry $\U(1)^k$ which will be identified with the
auxiliary symmetry of each of the $k$ single instantons.
In the complete clustering limit, the terms
$(\tilde a^{\prime\aD\alpha})_{ik}(\tilde a'_{\alpha\bD}
)_{kj}$, $k\neq i,j$, can be ignored in the ADHM constraints \eqref{badhm}.
In this case, these off-diagonal constraints are linear in $
(\tilde a'_{\alpha\aD})_{ij}$, for $i\neq j$,
\EQ{
(\bar X^i-\bar X^j)^{\aD\alpha}(\tilde a'_{\alpha\bD})_{ij}+\bar
w^\aD_{iu}w_{uj\bD}\propto\delta^\aD_{\ \bD}\
\elabel{boscl}
}
and can be solved, although we will not require explicit expressions for
the solutions. The
diagonal components of the ADHM constraints \eqref{badhm} in this
limit are then simply
\EQ{
\bar w^\aD_{iu}w_{ui\bD}=\rho_{i}^2\delta^\aD_{\ \bD}\ ,
\elabel{oiladhm}
}
(no sum on $i$)
for arbitrary $\rho_i$. The constraint \eqref{oiladhm} is then the
ADHM constraint of a single instanton and we can therefore identify
$\rho_i$ with the scale size of the $i^{\rm th}$ instanton. Each
instanton is associated with a particular $\SU(2)$ embedding of
$\SU(N)$ defined by the generators
\EQ{
(T^c_i)_{uv}=\rho_i^{-2}w_{ui\aD}
\tau^c{}^\aD{}_{\bD}\bar w_{iv}^\bD\ ,\qquad c=1,2,3,
\elabel{embed2}
}
with no sum on $i$. One can show that the completely clustered limit
is valid when, for each $i\neq j$,
\EQ{
(X^i-X^j)^2\gg\rho_i\rho_j\TrN(T^c_iT^c_j)\ .
}
In other words, the separation between the instantons must be much
greater than the product of the scale sizes times a trace over
generators which measures the overlap of the $\SU(2)$ embeddings of
each of the instantons.

It will be important for many of the applications of the instanton
calculus to describe the
asymptotic fall off of the fields from the centre of the
instanton. The nature of the fall off depends on the gauge
used which for us means singular gauge as described above.
{}From the previous formulae of
this section for arbitrary instanton charge $k$, we find
the leading large-$x$ asymptotic behavior of
several key ADHM quantities, in singular gauge
\eqref{sga}:
\begin{equation}
\Delta_\aD\ \rightarrow\ b^\alpha x_{\alpha\aD}\ ,\qquad f_{ij}\ \rightarrow\
{1\over x^2}\,\delta_{ij}\ , \qquad U'\ \rightarrow\
-\frac{x_{\alpha\aD}}{x^2}\,\wbar^\aD\ ,
\qquad V\ \rightarrow\ 1_{\sst [N] \times
[N]} \elabel{asymadhm}\end{equation}
and in addition, for the gauge field
\EQ{
A_n\ \rightarrow\ g^{-1}\frac{x_m}{x^4}w_\aD\bar\sigma_{mn}{}^\aD{}_\bD\bar
w^\bD\ .
}

\subsection{Zero modes and the metric on $\ms_k$}\elabel{sec:S14}

The ADHM construction yields an explicit solution for the self-dual
gauge field in terms of the (over-complete)
ADHM collective coordinates $a_\aD$. We now show how one can find explicit
expressions for the zero modes. In \S\ref{sec:S3}, we showed, in the
quaternion basis, that zero modes $\delta A_{\alpha\aD}$
are solutions of the covariant Weyl equation $\Dbarslash^{\aD\alpha}
\delta A_{\alpha\bD}=0$. In Appendix \ref{app:A4} (Eq.~\eqref{dzmm}) we
verify that the following linear functions
\EQ{
\Lambda_\alpha(C)\ \overset{\rm def}=\ \bar U Cf\bar b_\alpha U
-\bar U b_\alpha f\bar CU\ ,
\elabel{xxoo}
}
for a constant $(N+2k)\times k$ matrix $C_{\lambda i}$, is a solution of the
covariant Weyl equation in the background of the instanton,
\EQ{
\bar{\cal D}^{\aD\alpha}\Lambda_\alpha(C)=0\ ,
\elabel{bvz}
}
as long as $C$ satisfies the constraints
\AL{
\bar C_{i\lambda}a_{\lambda j\aD}&=-\bar a_{i\aD\lambda}C_{\lambda
j}\elabel{opa}\\
\bar C_{i\lambda}b^\alpha_{\lambda j}&=\bar
b^{\alpha}_{i\lambda}C_{\lambda j}\ .
\elabel{opb}
}
There are $2k(N+2k)$ real degrees-of-freedom in $C$ subject to $4k^2$
constraints \eqref{opa}-\eqref{opb}. Hence, there are $2kN$
independent solutions to the Weyl equation as anticipated by the Index
Theorem.

The question is how these solutions are related to the derivatives of
the gauge field with respect to the collective coordinates? If $X^\mu$
is an arbitrary collective coordinate then we prove in Appendix
\ref{app:A4} (Eq.~\eqref{wwtpt}) that
\EQ{
\PD{A_n}{X^{\mu}}=-g^{-1}
{\cal D}_n\Big(\PD{\bar U}{X^{\mu}}U\Big)+g^{-1}\bar U\PD a{X^{\mu}}
f\bar\sigma_n bU-g^{-1}\bar U b\sigma_n f\PD{\bar a}{X^{\mu}}U\ .
\elabel{eerr}
}
In the quaternion basis we recognize the 2nd and 3rd terms as being
$2g^{-1}\Lambda_\alpha(C_\aD)$ with $C_\aD=\partial a_\aD/\partial X^\mu$:
\EQ{
\PD{A_{\alpha\aD}}{X^{\mu}}=-
g^{-1}{\cal D}_{\alpha\aD}\big(\PD{\bar
U}{X^{\mu}}U\Big)+2g^{-1}\Lambda_\alpha(\partial a_\aD/\partial X^\mu)\ .
\elabel{tmz}
}
The first term in \eqref{tmz},
as its form suggests, is precisely the compensating gauge transformation
\eqref{hhff} needed in order to force the zero mode into background gauge:
\EQ{
\Omega_{\mu}=-g^{-1}\PD{\bar U}{X^{\mu}}U\ .
\elabel{cgt}
}
Hence the expression for the zero modes is given explicitly by the
linear function defined in \eqref{xxoo}
\EQ{
\delta_{\mu}A_{\alpha\aD}=2g^{-1}
\Lambda_\alpha(\partial a_\aD/\partial X^\mu)\equiv
2g^{-1}\bar U\Big(\PD{a_\aD}{X^{\mu}}f\bar b_\alpha
-b_\alpha f\PD{\bar a_\aD}{X^{\mu}}\Big)U\ .
\elabel{zmq}
}

The fact that $C_\aD=\partial a_\aD/\partial X^\mu$ satisfies the
constraints \eqref{opa} and \eqref{opb} is partially taken care of
by taking the
$X^\mu$-derivative of the ADHM constraints
\eqref{conea}-\eqref{coneb}. From the hyper-K\"ahler quotient
perspective, described in \S\ref{sec:S10} and Appendix \ref{app:A2},
the conditions that arise
from taking the $X^\mu$ derivative of \eqref{conea}
are equivalent to the requirement that $\partial a_\aD/\partial X^\mu$ is
orthogonal to the $3k^2$ vectors $\tilde\BI^{(c)}X_r$: {\it
i.e.\/}~$\partial a_\aD/\partial X^\mu$ lies in the
tangent space $T\ns\subset T\tilde\ms$. However, this leaves $k^2$
additional constraints on $\partial
a_\aD/\partial X^\mu$ which have not been
accounted for; namely,
\EQ{
\PD{\bar a^\aD}{X^\mu}a_\aD-\bar a^\aD\PD{a_\aD}{X^\mu}=0\ .
\elabel{extcon}
}
{}From the hyper-K\"ahler quotient
point-of-view, these extra constraints simply require
that the vector $\partial a_\aD/\partial X^\mu$ is orthogonal to the
tri-holomorphic Killings vectors $X_r$, \eqref{qqp}, which generate the $\U(k)$
isometries of $\tilde\ms$. This means that it lies in the horizontal subspace
${\EuScript H}\subset T\ns$.
The horizontal subspace is identified with the
tangent space $T\ms_k$ by a unique lifting procedure. Hence, the zero
modes are naturally associated to tangent vectors to $\ms_k$.

It is remarkable that one can calculate the
explicit functional inner product of the zero modes in singular gauge
by using a
formula attributed in Ref.~\cite{OSB} to Corrigan
\cite{Corrup}. The proof, reviewed in Appendix \ref{app:A4}
(Eq.~\eqref{corrigan}),
was first published in \cite{MO-I}, for gauge
group $\Sp(1)$, and extended to gauge group $\SU(N)$
in \cite{MO3}.
The identity can then be written in terms of the Weyl spinor quantities
$\Lambda(C)$ as\footnote{The relative
minus sign in the second term compared to
Eq.~(2.61) in \cite{MO3} is due to the fact that in that reference we
have written the identity for the fermion zero modes and so ${\cal
M}$ and ${\cal N}$ are Grassmann collective coordinates, whereas here
$C$ and $C'$ are $c$-number-valued.}
\SP{
\int d^4x\,\TrN\,\Lambda(C)\Lambda(C')
=-\frac{\pi^2}{2}{\rm tr}_k\big[\bar C({\cal P}_\infty+1)C'
-\bar C'({\cal P}_\infty+1)C\big]\ .
\elabel{jjqq}
}
Here
\begin{equation}
{\cal P}_\infty =
\underset{x\rightarrow\infty}{\rm lim}{\cal P} = 1-b\bar
b=\MAT{1_{\sst[N]\times[N]}&0_{\sst[N]\times[2k]}\\
0_{\sst[2k]\times[N]}&0_{\sst[2k]\times[2k]}}\ .
\elabel{Pinftydef}
\end{equation}
{}From this identity, and the expression \eqref{zmq}, we can deduce the
expression for the metric on the
space of collective coordinates \eqref{ccmet}
\EQ{
g_{\mu\nu}(X)=2\pi^2{\rm tr}_k\big(\PD{\bar a^\aD}{X^{\mu}}({\cal
P}_\infty+1)\PD{a_\aD}{X^{\nu}}+\PD{\bar a^\aD}{X^{\nu}}({\cal P}_\infty+1)
\PD{a_\aD}{X^{\mu}}\big)\ .
\elabel{htyh}
}
Since, $\partial a_\aD/\partial X^\mu$ are the components of a vector
in $\EuScript H$ (due to the constraints \eqref{opa}-\eqref{opb}),
the metric arising from the functional inner product
of zero modes (in singular gauge) \eqref{htyh} is identical to the
metric on $\ms_k$ induced by the hyper-K\"ahler quotient
construction. This is completely in accord with the argument in
\S\ref{sec:S10} involving the hyper-K\"ahler potential.

Now that we have found the explicit form of the zero modes and
compensating gauge transformations in singular gauge, one can easily
verify that for large $x$
\EQ{
\delta_\mu A_m\thicksim{\cal O}(x^{-3})\ ,\qquad \PD{A_m}{X^\mu}\thicksim
{\cal O}(x^{-3})\ ,\qquad\Omega_\mu\thicksim{\cal O}(x^{-2})\ .
\label{asymbe}
}
These formulae follow from the explicit expressions
\eqref{eerr} and \eqref{cgt} and asymptotic formulae in \eqref{asymadhm}.

\subsection{Singularities and small instantons}\elabel{sec:S15}

We have previously remarked that the instanton moduli space $\ms_k$
fails to be a smooth manifold due to certain singularities. In this
section we will explore these interesting features in more detail.
Before we embark on this analysis, it is important to emphasize that
these singularities are
{\em not\/} evidence of any pathology in the instanton calculus. The
integrals over the $\ms_k$ appearing in the semi-classical
approximation of the functional integral are
perfectly well defined in the vicinity of these singularities
and in this sense the singularities are integrable.

Before we talk about the singularities specifically, let us develop
a picture of the geometry of $\ms_k$. In fact it will be more
convenient, for the most part,
to consider the centered moduli space $\widehat\ms_k$
defined in \eqref{cmsd}.
As we have shown, the ADHM construction
of $\ms_k$, or $\widehat\ms_k$,
is an example of a hyper-K\"ahler quotient
based on the quotient group $\U(k)$ and starting from flat
space. Since the quotient group has an abelian factor the quotient
construction can, in general, involve the parameters $\zeta^c$ of
\eqref{mommap} taking values in the $u(1)$ subalgebra of the Lie
algebra of $\U(k)$. However, the ADHM construction requires
$\zeta^c=0$ and this has some interesting consequences that we now
elaborate.\footnote{The case $\zeta^c\neq0$ corresponds to the ADHM
construction in the gauge theory defined on a non-commutative
spacetime as described in
\S\ref{sec:S129}.} Firstly, the quotient space admits a
{\it dilatation\/} generated by a {\it homothetic Killing
vector\/} which we denote $\kappa$.\footnote{A homothetic Killing vector
is a conformal Killing vector, so
${\cal L}_{\kappa}g=\phi g$, for which $\phi$ is a constant.} To see this,
let us note that the mother space
$\tilde\ms={\mathbb R}^{4n}$ trivially admits a dilatation
generated by the homothetic Killing vector $\tilde
\kappa$ with components
\EQ{
\tilde\kappa^{\ii\aD}=z^{\ii\aD}\ .
}
In addition $\kappa$ is {\it hypersurface orthogonal\/}, {\it i.e.\/}
\EQ{
\tilde\kappa^{\ii\aD}=\partial^{\ii\aD}\tilde\chi\ ,
}
for some function $\tilde\chi$, which we identify with the hyper-K\"ahler
potential \eqref{hkpadhmm}.
When the central terms $\zeta^c=0$ it is easy to see that
$\tilde\kappa$ preserves the level set $\ns$ and hence is the lift to
$\EuScript H$ of a
(hypersurface orthogonal) homothetic Killing vector $\kappa$
on the quotient space. The dilatation arises
as a consequence of
the classical conformal invariance of the gauge theory described
in \ref{sec:S11}.
The existence of the dilatation implies that the
metric on the quotient $\ms_k$ has the form of a cone
\cite{Gibbons:1998xa}. A space of this kind is known as a
{\it hyper-K\"ahler cone\/}. At this point it is convenient to talk
about the centered moduli space $\widehat\ms_k$. Everything
we have said above is equally applicable to
$\widehat\ms_k$ which has hyper-K\"ahler potential $\hat\chi$ equal to
$\chi$ in \eqref{hkpadhm} with the trace
components of $a'_n$ set to zero.
The fact that $\widehat\ms_k$ is a cone means that
$\widehat\ms_k\simeq{\mathbb R}^+\times\F$ with a metric
\EQ{
ds^2=\frac{d\hat\chi^2}{4\hat\chi}+\hat\chi ds_{\F}^2\ .
\elabel{conemet}
}
Since $\F$ is not a sphere, $\widehat\ms_k$ has a conical singularity at
$\hat\chi=0$ of which we will have more to say below.

The quotient space also inherits the obvious $\SU(2)$
isometries (acting in the obvious way on the
$\aD$ indices) from the mother. These isometries on $\tilde\ms$
are generated by three Killing vectors with components
\EQ{
\tilde n^{(c)\ii\aD}=\tilde\BI^{(c)}{}^{\aD}{}_\bD z^{\ii\bD}\equiv
i\tau^{c\aD}{}_\bD z^{\ii\bD}\ .
}
As with the dilatation above, it is
precisely when the central terms $\zeta^c$ vanish that
these symmetries are inherited by the quotient space $\ms_k$. To see
this one simply has to verify that the action preserves the level set
$\mu^c=0$.\footnote{It is worth pointing out that the action is {\it not\/}
tri-holomorphic because the $\SU(2)$ rotates the complex structures.}
Notice that the four vectors $\kappa$ and $n^{(c)}$ are mutually orthogonal.

The existence of the dilatation and
$\SU(2)$ isometries together implies that the centered
moduli space is locally a product $\widehat\ms_k=
[{\mathbb R}^4/{\mathbb Z}_2]\times{\mathfrak Q}$ for some
$4kN-8$ dimensional space $\mathfrak Q$.\footnote{The ${\mathbb
Z}_2$ quotient (acting by inversion)
arises because the centre of $\SU(2)$ lies in the
$\U(k)$ quotient group.} In fact, when
$\widehat\ms_k$ is described as the cone \eqref{conemet} the space
$\mathfrak F$ is a tri-Sasakian manifold, which due to
the structure of $\SU(2)$ isometries described above, has the form
of a non-trivial fibration of $\SO(3)$ over ${\mathfrak Q}$.
One can further show that ${\mathfrak Q}$ is a {\it
quaternionic K\"ahler space\/} (see \cite{Boyer:1998sf,Vandoren:2000qr}
for more details and other references).

Having established the interesting cone geometry of $\ms_k$
we now turn more specifically to its singularities.
It is not hard to pin down the reason for the singularities.
The ADHM construction requires us to quotient the level set $\ns$
by the $\U(k)$ auxiliary symmetry group and this
procedure will introduce orbifold singularities at points
in $\ns$ where $\U(k)$ does not act freely, {\it i.e.\/}~where some
subgroup of $\U(k)$ leaves $a_\aD$, a solution to \eqref{badhm},
fixed. Geometrically, the volume of the gauge orbit through any point
in $\ns$ is proportional to
the determinant of the matrix of inner products of the $\U(k)$
tri-holomorphic Killing vectors $\{X_r\}$:
\EQ{
\Big|{\rm det}_{k^2}\tilde g(X_r,X_s)\Big|^{1/2}\ .
}
The $k^2\times k^2$ matrix with elements $\tilde g(X_r,X_s)$ plays a
ubiquitous r\^ole in the instanton calculus and we introduce the
notation
\EQ{
\tilde g(X_r,X_s)\equiv\BL_{rs}\ \overset{\text{def}}=\
8\pi^2{\rm tr}_k\big(T^r\BL\, T^s\big)\ ,
\elabel{gwww}
}
where the generators $T^r$ of $\U(k)$ were defined previously in
\S\ref{sec:S10}.
Here, using \eqref{qum} and \eqref{qqp},
$\BL$ is an operator on $k\times k$ Hermitian matrices of the form
\SP{
\BL\cdot\Omega&\,\overset{\text{def}}=\,\tfrac12\{\bar w^\aD
w_\aD,\Omega\}+\tfrac12\bar a^{\prime\aD\alpha}a'_{\alpha\aD}\Omega
-\bar a^{\prime\aD\alpha}\Omega a'_{\alpha\aD}
+\tfrac12\Omega\bar a^{\prime\aD\alpha}a'_{\alpha\aD}\ ,\\
&=\tfrac12\{\bar w^\aD w_\aD,\Omega\}+[a'_n,[a'_n,\Omega]]\ .
\elabel{vvxx}
}
At a point where $\U(k)$ does not act freely, the operator $\BL$
develops one, or more, null eigenvectors. The
relevant effect can be seen already at the one instanton level. In this case,
the ADHM constraints are more explicitly \eqref{wcon}.
The auxiliary group $\U(1)$ acts by
phase rotation, $w_\aD\rightarrow e^{i\phi}w_\aD$,
but does not act freely when $w_\aD=0$ which, from \S\ref{sec:S12}, is
the point at which the instanton
has zero scale size $\rho$. This is precisely in accord with our description
of the centered moduli space
$\widehat\ms_k$ above as a cone over the tri-Sasakian space
$\F$. For the single instanton the apex of the cone is the point
$\rho=0$. At the one-instanton level we can be more explicit. In
\eqref{goc} we have identified the action of global gauge
transformations on $\ms_k$. In fact at the one-instanton level
all gauge orbits are equivalent to the generic one \eqref{goc} and therefore
the whole centered moduli space
$\widehat\ms_1$ is simply a cone over the gauge orbit:
\EQ{
\widehat\ms_1\simeq{\mathbb
R}^+\times\frac{\SU(N)}{\SS\big(\U(N-2)\times\U(1)\big)}
\ ,
}
where the variable along the cone is $\hat\chi\propto\rho^2$.
Since, at the one-instanton level,
there are no additional singularities in $\widehat\ms_k$, the
corresponding $\F$ is an example of a homogeneous tri-Sasakian space
\cite{Boyer:1998sf}.

As discussed above,
the centered moduli spaces $\widehat\ms_k$ for instanton charge $k>1$
also have the structure of a cone over a tri-Sasakian space. The
apex of the cone in this case corresponds to the point where $w_\aD=0$
and all the non-trace parts of $a'_n$ vanish. We will shortly
interpret this as the
point in the moduli space where all the instantons have shrunk to zero
size and all lie coincident in ${\mathbb R}^4$. This is the point of
maximal degeneracy where the whole of $\U(k)$ is fixed.
However, there are other singularities of smaller
co-dimension in the moduli space reflecting the fact that the
tri-Sasakian space $\F$ in this case is not homogeneous and has
singularities of its own.
These can be uncovered in the following way. In \S\ref{sec:S12}, we have
described how multi-instanton configurations in various asymptotic
regions of the moduli space can be identified with clusters of smaller
numbers of instantons. In particular, when $w_{ui\aD}\to0$ for some
fixed $i=l$, the ADHM constraints imply that the elements
$(a'_n)_{il},(a'_n)_{li}\to0$, for $i\neq l$. In the limit, the
subgroup $\U(1)\subset\U(k)$ corresponding to
\EQ{
w_{ui\aD}\to w_{ui\aD}e^{i\delta_{il}\phi}\ ,\qquad
(a'_n)_{ij}\to e^{(\delta_{jl}-\delta_{il})\phi} (a'_n)_{ij}
}
does not act freely. This is a limit in the moduli space, where the
$k$-instanton configuration looks like a smooth $(k-1)$-instanton
configuration along with a single instanton that has shrunk to zero
size. The point-like instanton still has a position
$X^l_n=-(a'_n)_{ll}$; hence, in the limit
$\ms_k\rightarrow\ms_{k-1}\times{\mathbb R}^4$. This process can
continue. There are regions where some subgroup $\U(1)^r\subset\U(k)$
does not act freely, which corresponds to a $(k-r)$-instanton
configuration along with $r$ point-like instantons. In this limit the
moduli space is of the form
\EQ{
\ms_k\longrightarrow
\ms_{k-r}\times{\rm Sym}^r{\mathbb R}^4\ .
}
Here, ${\rm Sym}^r{\mathbb R}^4$ is the symmetric product of $r$
points in ${\mathbb R}^4$.\footnote{This is 
the product $({\mathbb R}^4)^r$ modded out
by the group of permutations on $r$ objects arising
from the subgroup of the Weyl group of $\U(k)$ that permutes the $r$ labels
$\{i\}$ for which $w_{ui\aD}\to0$.} Additional
singularities arise when the point-like instantons come together at
the same spacetime point. This describes a situation where a
non-abelian subgroup of $\U(k)$ does not act freely. The maximally
degenerate situation is when the whole of $\U(k)$ is fixed. This occur
at the apex of the hyper-K\"ahler cone and describes a configuration
where all the instantons have shrunk to zero size at the same point in
${\mathbb R}^4$.

In a certain mathematical sense, the moduli space $\ms_k$ excludes the regions
with point-like instantons. It then has a natural compactification
by including the regions with point-like instantons at the boundary
\cite{Uhlenbeck:1982zm} (see also \cite{DonaldsonKron}).
However, in the semi-classical approximation
of the functional integral, the singularities of the
instanton moduli space do not lead to a divergence of the
collective coordinate integral and---for all practical 
purposes---we need not distinguish between $\ms_k$ and its compactification.

\newpage

\rsen\section{The Collective Coordinate Integral}\elabel{sec:S16}

In this section, we describe how instantons contribute to the
functional integral of the field theory in the semi-classical
limit. To start with in \S\ref{sec:S17} we consider the problem of
expanding around an instanton solution in the functional integral. We
follow the approach of Bernard \cite{Bernard} suitably generalized in
an obvious way to $k>1$ (see also the thorough treatment in
Osborn's review \cite{OSB}). In particular, following
\cite{Bernard}, we take advantage
of certain simplifications that occur in singular gauge arising from
the fast fall off of the gauge potential. The goal of this section is
to show that the leading-order semi-classical approximation involves
an integral over the collective coordinates with a measure defined by
the inner product of the corresponding zero modes along with a ratio
of determinants arising from the fluctuations around the instanton.
In \S\ref{sec:S18}
we explain how a concrete expression for the integration measure can
be obtained using the hyper-K\"ahler quotient perspective, although
this was not originally how the measure was found in
Refs.~\cite{measure1,DHKM}. We go on in \S\ref{sec:S24}
to review the successes and
failures of the old instanton literature regarding the fluctuation
determinants in the general $k$ instanton background. We do not develop this
subject any further because the ratio of fluctuations in a
supersymmetric theory is a trivial collective coordinate independent
constant factor.

\subsection{From the functional to the
collective coordinate integral}\elabel{sec:S17}

The semi-classical approximation is a saddle-point method
and, as such, in order to find the leading order behaviour of the
functional integral, we need to consider the fluctuations
around the instanton solution. To this end, we expand
\EQ{
A_n(x)=A_n(x;X)+\delta A_n(x;X)\ ,
\elabel{ioa}
}
where $A_n(x;X)$ is the instanton solution as a function of the
collective coordinates and $\delta A_n(x;X)$ are the fluctuations
which are chosen to satisfy the background gauge condition
\eqref{gchoice}\footnote{In the following all covariant derivatives
are defined with respect to the instanton solution $A_n(x;X)$.}
\EQ{
{\cal D}_n\delta A_n=0\ .
}
Notice that the fluctuations depend implicitly on the collective
coordinates. The action of the theory to quadratic order in the
fluctuations is \eqref{ggss}
\EQ{
S=-2\pi i k\tau-\frac1{2}\int d^4x\,
\TrN\,\delta\bar
A^{\aD\alpha}\Delta^{(+)}{}_\alpha{}^\beta\delta A_{\beta\aD}
+\cdots\ .
\elabel{uuaq}
}
We must also include the gauge-fixing term involving the ghost fields
\EQ{
S_{\rm gf}=2\int d^4x\,\trN\,b\,{\cal D}^2c\ .
}

The fluctuations can be expanded in terms of the eigenfunctions of the
operator $\Delta^{(+)}$ which can be split into the zero modes,
already extensively investigated in \S\ref{sec:S14},
and the non-zero modes:
\EQ{
\delta A_n=\sum_\mu\xi^\mu\delta_\mu A_n+\tilde
A_n\ .
\elabel{lsw}
}
Here, the non-zero mode
fluctuations $\tilde A_n$ are functionally orthogonal to the zero
modes. In the functional integral we now separate out the integrals
over the zero and non-zero modes:
\EQ{
\int [dA_n]=g^{-4kN}\int\Big\{\sqrt{\det\,
g(X)}\ \prod_{\mu}\frac{d\xi^\mu}{\sqrt{2\pi}}\Big\}\,
[d\tilde A_n]
}
Here, $g(X)$ is the metric on the zero modes defined in
\eqref{ccmet}. The term in braces is the integral over the zero mode
subspace. The factors of $g$ in front of the measure
arise from the fact that we included a factor of
$g^2$ in the definition of the metric \eqref{ccmet}.

The non-zero mode fluctuations $\tilde A_n$ and the ghosts can now
be integrated out, producing the usual determinant factors:
\SP{
\int [d\tilde A_n]\,[db]\,[dc]\,\exp\,
\frac1{2}\int d^4x\,
\TrN\,\Big\{\tilde{\bar
A}^{\aD\alpha}\Delta^{(+)}{}_\alpha{}^\beta\tilde A_{\beta\aD}
-4b\,{\cal D}^2c\Big\}
=\frac{\det(-{\cal D}^2)}
{\det'\,\Delta^{(+)}}\ .
\elabel{detsb}
}
As is conventional, the prime on the determinant indicates that the operator
$\Delta^{(+)}$ has zero modes and these must be excluded in the
product over eigenvalues that defines the determinant. So the leading
order expression for the functional integral in the charge-$k$ sector is
\SP{
\frac{e^{2\pi ik\tau}}{g^{4kN}}\int\Big\{\sqrt{\det\,
g(X)}\ \prod_{\mu}\frac{d\xi^\mu}{\sqrt{2\pi}}\Big\}\,\frac{\det(-{\cal D}^2)}
{\det'\,\Delta^{(+)}}\ .
\elabel{fgg}
}
At this stage, \eqref{fgg} is only schematic because in order to
define the determinants rigorously, we must regularize the theory in
some way. We shall consider this problem in \S\ref{sec:S24}.

In the final part of this section we explain how the integrals over
the expansion coefficients of the zero modes $\{\xi^\mu\}$ may be traded
for integrals over the collective coordinates $\{X^\mu\}$. This change
of variables is facilitated by the well-known trick of inserting an
expression for unity in the guise of
\EQ{
1\equiv\int
\ \prod_\mu dX^\mu\ \Big\vert
\det\,\PD{f_\nu}{X^\mu}\Big\vert\
\prod_\nu\delta(f_\nu(X))\ .
}
A judicious choice for the
function $f_\nu(X)$ is the
inner product of the fluctuation \eqref{ioa} with the zero mode
$\delta_\nu A_n$:
\EQ{
f_\nu(X)=
-2g^2\int d^4x\,
\TrN\,\delta A_n\,\delta_\nu A_n=\sum_\mu \xi^\mu g_{\mu\nu}(X)\ ,
}
where we used \eqref{lsw} and the fact that the zero and non-zero modes
are functionally orthogonal, along with the definition of the metric
$g_{\mu\nu}(X)$ as the inner product of zero modes \eqref{ccmet}.

Now we compute the derivative
\EQ{
\PD{f_\nu}{X^\mu}=-2g^2\int d^4x\,\TrN\,\Big\{
\PD{\delta A_n}{X^\mu}\delta_\nu A_n+
\delta A_n\PD{\delta_\nu A_n}{X^\mu}\Big\}\ .
}
Since the total field $A_n(x)$ in Eq.~\eqref{ioa}
does not depend on the collective coordinates, we can replace $\delta
A_n$ by $-A_n(x;X)$ in the first term. Using the fact that
\EQ{
\int d^4x\,\TrN\,{\cal D}_n\Omega_\mu
\delta_\nu A_n=-\int d^4x\,\TrN\,\Omega_\mu{\cal D}_n\delta_\nu
A_n=0\ ,
}
where the surface term vanishes due to the asymptotic form of
$\Omega_\mu$ and $\delta_\nu A_n$ (see \S\ref{sec:S14}),
and the last equality follows from \eqref{gchoice}, we have
\EQ{
-2g^2\int d^4x\,\TrN\,\PD{A_n(x;X)}{X^\mu}\delta_\nu A_n
=g_{\mu\nu}(X)\ .
}
The insertion of unity is then
\EQ{
1\equiv\int \  \prod_\mu dX^\mu\
\Big\vert\det\,\Big(g_{\mu\nu}(X)-2g^2\int d^4x\,\TrN
\delta A_n\PD{\delta_\nu
A_n}{X^\mu}\Big)\Big\vert\
\prod_\nu\delta\Big(\sum_\mu\xi^\mu g_{\mu\nu}(X)\Big)\ .
\elabel{hwq}
}
Plugging this into \eqref{fgg}, we perform the integrals over the expansion
coefficients using the $\delta$-functions. Since $g_{\mu\nu}(X)$ is
invertible the $\delta$-functions enforce $\xi^\mu=0$. Consequently
in the second term of the determinant we can replace $\delta A_n$ by
the non-zero mode part of the fluctuation $\tilde A_n$. This term is
higher order in $g$ and may be dropped to leading order.
Finally we have to leading order in in the charge-$k$ sector
\EQ{
\int [dA_n]\,[db]\,[dc]\,e^{-S[A,b,c]}\Big|_{\text{charge-}k}\
\overset{g\to 0}
\longrightarrow\ \frac{e^{2\pi ik\tau}}{g^{4kN}}\int_{\ms_k}\Bomega\
\cdot\ \frac{\det(-{\cal D}^2)}{\det'\Delta^{(+)}}\ .
\elabel{ffg}
}
Notice that the resulting form of the collective coordinate integral
is proportional to the
canonical volume form on the moduli space $\ms_k$ associated to the
metric $g_{\mu\nu}(X)$:
\EQ{
\int_{\ms_k}\Bomega \equiv \int  \sqrt{\det\,
g(X)}\ \prod_{\mu}\frac{dX^\mu}{\sqrt{2\pi}}\ ,
\elabel{bos}
}
along with a non-trivial function on $\ms_k$ equal to
the determinants of the operators governing the Gaussian
fluctuations of the gauge field and ghosts in the instanton background
\eqref{detsb}.

Finally, when calculating a correlation function $\langle{\cal
O}_1(x^{(1)})\cdots{\cal O}_n(x^{(n)})\rangle$ in the semi-classical
limit, the fields insertions
${\cal O}_i(x^{(i)})$ are replaced by their values in the instanton
background and so, to leading semi-classical order, become
functions of the collective
coordinates.

\subsection{ The volume form on the instanton moduli space}\elabel{sec:S18}

We have shown that the leading order semi-classical approximation of
the functional integral involves an integral over instanton moduli
space with a measure associated to the natural metric inherited from
the inner product of the zero modes \eqref{ccmet}.
At leading order in the saddle-point expansion, the integrand includes
determinants of the fluctuations \eqref{ffg}, but in the present
section we focus on the volume form.

The most direct way to obtain a workable expression for the volume form
on the instanton moduli space is to use the fact that this space can
be described as a hyper-K\"ahler quotient of flat space.
Actually this is not how the volume form was originally
found in Refs.~\cite{KMS,measure1,DHKM}.
In these references, the constraints of supersymmetry, the Index Theorem and
other consistency conditions (most notably ``clustering'' in the
dilute instanton gas limit)
were used to write down the unique expression for the volume
form for a supersymmetric gauge theory. Then,
decoupling the fermions and scalars, by giving them large masses, an
expression for the volume form on $\ms_k$ was derived. Here, we will
not follow this route, rather we shall take as our central theme the
hyper-K\"ahler quotient construction (an approach also adopted in
\cite{Bruzzo:2001di}). Since the quotient space $\ms_k$
inherits a metric from the mother space $\tilde\ms$ it also inherits a
volume form. Recall from \S\ref{sec:S10} and Appendix \ref{app:A2},
there are two parts to the hyper-K\"ahler
quotient. Firstly, one restricts to the level sets $\ns\subset\tilde\ms$
defined by the vanishing of the moment maps $\vec\mu=0$. Then one
performs a quotient by the group $G$ finally giving
$\ms_k=\ns/G$. An expression
for the volume form on $\ms_k$, is be obtained from that of $\tilde\ms$ by
imposing the vanishing of the moment maps by explicit
$\delta$-functions with the appropriate Jacobian.
One then divides by the volume of the
orbits of the $G$-symmetry. This leads in general to the expression
\eqref{measq}. An expression for the volume form on
$\ms_k$ is then obtained by ``gauge fixing'' the $\U(k)$ symmetry.
However, it turns out to be more
convenient to leave the $G$-symmetry ``un-fixed''. This is because the
expressions for all the fields in an instanton background are
$G$-invariant.

For the ADHM construction the mother space is $\tilde\ms={\mathbb
R}^{4k(N+k)}$ parameterized by $a_\aD$ with metric \eqref{qum}
and $G$ is $\U(k)$ acting in \eqref{restw}. The moment maps are the ADHM
constraints \eqref{fconea}. The $\U(k)$ action defines a set of
vector fields on $\tilde\ms$, $X_r$, $r=1,\ldots,k^2$, given in
\eqref{qqp}. The inner-product of these vector fields is given by
\eqref{gwww} which defines the ADHM operator $\BL$ which appears in
the formula for the volume form.
The volume form of the ADHM moduli space in its $\U(k)$-unfixed
form is, from \eqref{measqex},
\EQ{
\int_{\ms_k}\Bomega=\frac{C_k}{{\rm Vol}\,\U(k)}\int\, d^{4k(N+k)} a
\ \big|\det_{k^2}\BL\big|\
\prod_{r=1}^{k^2}\, \prod_{c=1}^3\,
\delta\Big(\tfrac12{\rm
tr}_k\,T^r\big(\tau^c{}^\aD{}_\bD \bar a^\bD
a_\aD\big)\Big)\ .
\elabel{bmes}
}
In the above, we have defined
\EQ{
\int\, d^{4k(N+k)}a\equiv\int\,\prod_{n=1}^4\prod_{r=1}^{k^2} d
(a'_n)^r \,\prod_{i=1}^k\prod_{u=1}^N\prod_{\aD=1}^2
d\bar w^\aD_{iu}\, dw_{\aD ui}\ ,
}
where the integral over the $k\times k$ matrices $a'_n$ and the arguments of
the ADHM constraints are defined with respect to the generators of $\U(k)$
in the fundamental representation, normalized so that
${\rm tr}_k\,T^rT^s=\delta^{rs}$. The volume of the
$\U(k)$ is the constant
\EQ{
{\rm Vol}\,\U(k)=\frac{2^k\pi^{k(k+1)/2}}{\prod_{i=1}^{k-1}i!}\ .
}
The normalization factor $C_k$ can be determined by
taking into account the normalization of the metric \eqref{qum}, giving
\EQ{
C_k=2^{-k(k-1)/2}(2\pi)^{2kN}\ .
\elabel{irrr}
}

Our expression for the volume form on the instanton moduli space can
be compared with the one-instanton expressions in the literature
\cite{tHooft,Bernard}. For $k=1$ the ADHM are resolved as described in
\S\ref{sec:S12}. In fact the necessary
change of variables from $w_{u\aD}$ to the scale size $\rho$ and gauge
orientation $\grp$ is described in \S\ref{app:A5}. In particular, the
(non-supersymmetric) 
one-instanton measure is given by \eqref{sbames} with $\N=0$, $k=1$
and with $W^0\equiv2\rho^2$ and $a'_n=-X_n$. In order to compare with
the literature we integrate over the  
gauge orientation $\grp$ which is conveniently normalized so that 
$\int d\grp=1$. This leaves
\EQ{
\int_{\ms_1}\Bomega=\frac{2^{4N+2}\pi^{4N-2}}{(N-1)!(N-2)!}\int
d^4X\,d\rho\,\rho^{4N-5}\ ,
}
which agrees with the expression for gauge group $\SU(N)$ 
derived in \cite{Bernard}.

\subsubsection{Clustering}\elabel{sec:S22}

The relative normalization constants $C_k$, for different $k$,
can be checked by using the
clustering property of the instanton measure. This requires in
certain regions of moduli space where a $k$ instanton configuration can be
interpreted as well separated $k_1$ and $k_2$ instanton
configurations, $\ms_k$ is approximately
$\ms_{k_1}\times\ms_{k_2}$, and so the volume form factorizes as
\EQ{
\int_{\ms_k}
\Bomega\ \longrightarrow\
\frac{k_1!k_2!}{k!}\int_{\ms_{k_1}}\Bomega\times\int_{
\ms_{k_2}}\Bomega\ .
}
We can determine the overall normalization of the volume form by going
to the completely clustered limit defined in \S\ref{sec:S12}.
In this limit, we can describe the positions of the single
instanton by the diagonal elements of $a'_n$,
$(a'_n)_{ii}=-X^i_n$. Part of the $\U(k)$ symmetry can be fixed by
setting to zero the off-diagonal elements of $a'_n$ that are
generated by $\U(k)$ adjoint action on the
diagonal matrix ${\rm diag}(-X^1_n,\ldots,-X^k_n)$; as in \S\ref{sec:S12} we
will denote the
remaining off-diagonal elements as $\tilde a'_{ij}$.
In the measure, this gauge fixing involves a Jacobian
factor:\footnote{The following formula can easily be derived
from the well-known Jacobian that arises from changing variables from the
elements of a $k\times k$
Hermitian matrix $X$ (defined with respect to our basis $T^r$) to
its eigenvalues $X^i$: $\int
dX=(2\pi)^{k(k-1)/2}/\prod_{i=1}^ki!\cdot\int
dX_i\,\prod_{i<j}(X^i-X^j)^2$. See for example \cite{Mehta}.}
\EQ{
{1\over{\rm Vol}\,\U(k)}\int d^{4k^2}a'\ \longrightarrow\
\frac{2^{k(k-1)/2}}{[{\rm Vol}\,\U(1)]^k}\frac1{k!}\int
\prod_{i=1}^kd^4X^i\,\bigg\{\prod_{i,j=1
\atop i\neq j}^kd^3\tilde a^{\prime}_{ij}
\,|X^i-X^j|\bigg\}\ .
\elabel{diagcov}
}

The off-diagonal ADHM constraints are linear in $\tilde a'_{ij}$
\eqref{boscl} and the corresponding $\delta$-functions in \eqref{bmes}
can be used to integrate out completely the $3k(k-1)$ variables
$\tilde a'_n$. These integrals produce a factor $\prod_{i\neq
j}(X^i-X^j)^{-3}$.
To complete the analysis we note that in the complete clustering
limit
\EQ{
\det\,\BL=\prod_{i=1}^k2\rho_i^2\ \cdot\
\prod_{i\neq j}(X^i-X^j)^2+\cdots
\elabel{gml}
}
where $2\rho_i^2\equiv\BL_i$, the $\BL$-operator of each individual
instanton.

Putting everything together one finds in the complete
clustering limit, with the normalization constant given in \eqref{irrr},
\EQ{
\int_{\ms_k}\Bomega \longrightarrow \frac1{k!}
\underbrace{\int_{\ms_1}\Bomega\times\cdots\times
\int_{\ms_1}\Bomega}_{k\text{ times}}\ ,
\elabel{cclm}
}
as required.

\subsection{Fluctuation determinants in the instanton
background}\elabel{sec:S24}

Determinants of fluctuation operators in the background of a single instanton
were first evaluated by 't Hooft in the classic paper \cite{tHooft}.
In this section, we sketch how the determinants of the
fluctuation operators in a {\it general} ADHM instanton
background can be found, although we omit many of the technical parts
of the calculation, principally because in the
supersymmetric case the determinants over non-zero modes
cancel between the bosonic and fermionic fields \cite{D'Adda:1978ur}.
A complete review of the old instanton literature pertaining to the
fluctuation determinants may be found in Osborn's Review \cite{OSB}.
Ideally, one would want to express the results in terms of
geometric quantities on the instanton moduli space. This was never
achieved and the final answers are quite implicit, involving spacetime
integrals.

The first thing we need to do is to provide properly UV-regularized
definitions of the determinants. We will do this by introducing
Pauli-Villars regulator fields with large masses $\mu_i$ and alternating
``metric'' $e_i$, such that
\EQ{
\sum_{i=1}^\nu e_i=-1\ ,\qquad \sum_{i=1}^\nu e_i\mu_i^{2p}=0\qquad
p=1,\ldots,\nu-1\ .
}
We will also define
\EQ{
\log\mu=-\sum_{i=1}^\nu e_i\log \mu_i\ .
\elabel{yyqq2}
}
For a consistent regularization the number of
regulator fields $\nu$ must exceed three. The
regularized determinants are
\AL{
\log\det'\Delta^{(+)}&={\rm Tr}\,\Big\{\log(\Delta^{(+)}+{\EuScript P}_0)
+\sum_{i=1}^\nu e_i\log(\Delta^{(+)}+\mu_i^2)\Big\}\ ,\elabel{iiaa}\\
\log\det\Delta^{(-)}&={\rm Tr}\,\Big\{\log\,\Delta^{(-)}
+\sum_{i=1}^\nu e_i\log(\Delta^{(-)}+\mu_i^2)\Big\}\ .\elabel{iibb}
}
In \eqref{iiaa}, ${\EuScript P}_0$ is the projector onto the zero mode
subspace of $\Delta^{(+)}$. Since the non-zero eigenvalues of
$\Delta^{(\pm)}$ are identical; hence, we can extract a very simple
expression for the ratio
\EQ{
\frac{\det'\Delta^{(+)}}{\det\,\Delta^{(-)}}=\exp\big(
2kN\sum_{i=1}^\nu e_i \mu_i^2\big)=\mu^{-4kN}\ ,
}
Here, $2kN$ is the number of zero modes of $\Delta^{(+)}$ and $\mu$ is
the overall Pauli-Villars mass scale \eqref{yyqq2}. Using this relation
and the fact that $\det\,\Delta^{(-)}=\{\det(-{\cal D}^2)\}^2$
we can express the fluctuation determinants in \eqref{ffg}
in terms of the determinant of the
covariant Laplacian:
\EQ{
\frac{\det(-{\cal D}^2)}{\det'\Delta^{(+)}}
=\mu^{4kN}\det(-{\cal D}^2)^{-1}\ .
}
We have succeeded in reducing the problem to that of the fluctuation
determinant of a
scalar field transforming in the adjoint representation of the gauge
group. By pooling together results from the old instanton literature
we can find an expression for this determinant. Firstly,
one can construct the determinant for a scalar field transforming in the
fundamental representation of $\SU(N)$ following
Refs.~\cite{OSB,Berg:1979ku,Corrigan:1979di}.
The determinant for an adjoint-valued field
is then related to this by a remarkable formula of Jack
\cite{Jack:1980rn}.

So we begin with a brief sketch of how one determines the
fluctuation determinant of a scalar field transforming in the
fundamental representation. The key idea involves considering the
variation of $\log\det(-{\cal D}^2)_{\text{fund.}}$
by the collective coordinates.\footnote{Here, the subscript reminds us
that the scalar field transforms in the fundamental representation of
$\SU(N)$.} The resulting formula
established in \cite{OSB,Berg:1979ku,Corrigan:1979di} is
\EQ{
\PD{}{X^\mu}\log\det(-{\cal D}^2)_{\text{fund.}}=\frac1{6\pi^2}\int
d^4x\,\TrN\,\delta_\mu A_nJ_n\ ,
}
where $J_n$ is the conserved current
\EQ{
J_n=\bar U\sigma_{n\alpha\aD}b^\alpha f\bar\Delta^\aD b^\beta
f\bar b_\beta U-\bar Ub^\beta f\bar b_\beta\Delta_\aD f \bar
b_\alpha\bar\sigma_n^{\aD\alpha}U
}
and $\delta_\mu A_n$ is the zero mode associated to $X^\mu$. By
integrating this expression one can extract the ratio of the
determinant in the instanton background to the determinant in the
vacuum. The final expression established in \cite{Berg:1979ku} is expressed
solely in terms of the ADHM matrix $f$:
\SP{
\log\Big[\frac{\det(-{\cal D}^2)}{\det(-{\cal D}^2_0)}\Big]_{\text{fund.}}
&=\frac k6\log\mu+p_k+\frac1{48\pi^2}\int d^4x(I_1(x)+I_2(x))\ ,
\elabel{lass}
}
where ${\cal D}_0$ is the covariant derivative in the vacuum
($A_n=0$), $p_k$ is a constant which will be determined shortly and
the integrands are
\AL{
I_1(x)&={\rm tr}_k\big(f\partial_nf^{-1}  f\partial_nf^{-1}
f\partial_mf^{-1}  f\partial_mf^{-1}-20f^2\big)
+\frac{4k}{(1+x^2)^2}\ ,\\
I_2(x)&=\int_0^1 dt\,
\epsilon_{mnkl}{\rm tr}_k\big(\tilde f\partial_t \tilde f^{-1}
\tilde f\partial_m \tilde f^{-1}
 \tilde f\partial_n \tilde f^{-1}
\tilde f\partial_k \tilde f^{-1}  \tilde f\partial_l \tilde
f^{-1}\big)\ .
}
Here, $t$ is an auxiliary variable and $\tilde f(x,t)$ is the $k\times k$
dimensional matrix derived from $f(x)$:
\EQ{
\tilde f^{-1}(x,t)=tf^{-1}(x)+(1-t)(1+x^2)1_{\sst [k]\times[k]}\ .
}

In order to fix the constant $p_k$ we can appeal to a clustering
argument. In the complete clustering limit the $k\times k$ matrix $f$
is diagonal:
\EQ{
f={\rm diag}\big(\ldots,\frac1{(x-X^i)^2+\rho_i^2},\ldots\big)\ ,
}
where $\rho_i$ and $X^i_n$ are the scale size and position of the
$i^{\rm th}$ instanton, respectively. In this limit the integrals
in \eqref{lass} can easily be evaluated. (Actually the integral of
$I_2$ vanishes.) One finds
\SP{
\log\Big[\frac{\det(-{\cal D}^2)}{\det(-{\cal D}^2_0)}\Big]_{\text{fund.}}
&=\sum_{i=1}^k\big(\tfrac16\log(\mu\rho_i)-\tfrac5{18}\big)+p_k\ .
\elabel{kkss}
}
Comparing with the one instanton expression for this determinant
\cite{tHooft} fixes the constant to be
\EQ{
p_k=k\big(\alpha(\tfrac12)+\tfrac5{18})=
k\big(-2\zeta'(-1)-\tfrac16\log2+\tfrac{15}{72}\big)\ .
}
Here, $\alpha(\tfrac12)$ is a constant defined in \cite{tHooft}.

The second part of the problem involves relating the fluctuation
determinant associated to the fundamental representation of the gauge group
to that of the adjoint representation. The way this was achieved
\cite{Jack:1980rn}
relied on the tensor product ADHM formalism developed in
\cite{CGTone}. The
analysis provides the following explicit relation for the
fluctuation determinant of the Laplace operator for an adjoint-valued
field in terms of one for a fundamental-valued field:
\SP{
\log\frac{\det(-{\cal D}^2)}{\det(-{\cal D}^2_0)}&=2N
\log\Big[\frac{\det(-{\cal D}^2)}{\det(-{\cal
D}^2_0)}\Big]_{\text{fund.}}\\
&+\log\det_{k^2}\BL-\frac1{16\pi^2}\int dx^4\,\log\det_k f\
\square^2\log\det_kf+q_k\ .
}
Here, $\BL$ is the ubiquitous ADHM
operator on $k\times k$ matrices that we defined in
\eqref{vvxx}. As above, we will determine the constant $q_k$ by a clustering
argument. In the complete clustering limit, we find
\EQ{
\log\frac{\det(-{\cal D}^2)}{\det(-{\cal
D}^2_0)}=2N\sum_{i=1}^k\big(\tfrac16\log(\mu\rho_i)
+\alpha(\tfrac12)\big)+k(\log2-\tfrac56)+q_k\ .
}
Comparing this with the one instanton expression \cite{tHooft}, we find
\EQ{
q_k=k\big(\alpha(1)-4\alpha(\tfrac12)-\log2+\tfrac56\big)=\tfrac59k\ .
}

\rsen\section{Instantons in Supersymmetric Gauge Theories}\elabel{sec:S25}

In this Chapter we consider instanton configurations in gauge theories with
supersymmetry. There is an intimate
relation between instantons and supersymmetry which can be traced to the
fact that, in any supersymmetric gauge theory,
self-dual gauge fields are invariant under precisely
half the supersymmetry generators. As we will see, this is manifest in the
ADHM construction, which has a very natural supersymmetric generalization.
Specifically we consider the minimal
theories in four dimensions with gauge group $\SU(N)$ and $\N=1$, $2$ and $4$
supersymmetry.
Like the purely bosonic gauge theory considered in previous chapters,
the $\N=1$ and $\N=2$ theories are asymptotically free and the
relevance of instantons in the quantum theory
is not immediately obvious. However, we will begin by focusing on the
classical aspects of these configurations.

Before searching for supersymmetric instantons there is the worrisome
issue of supersymmetry in Euclidean space to discuss.
The nub of the issue is the following: Weyl
spinors in $D=4$ Minkowski space are in a ``real'' representation of
the covering group of the Lorentz group $\overline{\SO}(3,1)$.\footnote{Our
conventions for spinors are described in Appendix \ref{app:A1}. In particular
in Minkowski space our conventions are those of Wess and Bagger.} Consequently
the minimal spinor in $D=4$ is a Majorana spinor, which we can think of as
two Weyl spinors $\lambda$ and $\bar\lambda$ subject to the
reality condition
\EQ{
\bar\lambda^\aD=(\lambda_\alpha)^\dagger\qquad (\aD=\alpha)\ .
\elabel{pso}
}
This ensures, for instance, that the canonical fermion kinetic term
is real. On the contrary, in $D=4$ Euclidean space, Weyl spinors are in
``pseudo real'' representations of $\overline{\SO}(4)$
and one cannot impose the reality condition
\eqref{pso}. On the contrary,
$\overline{\SO}(4)\simeq\SU(2)\times\SU(2)$ so the spinor indices $\alpha$ and
$\aD$ refer to each of the $\SU(2)$'s and they are not mixed under
complex conjugation. In Euclidean space there is no notion of a Majorana
spinor and, apparently, no Euclidean version of the theory with a real
action. This problem may be by-passed in theories with extended supersymmetry
as the Weyl spinors in these theories may be combined in pairs to form
Dirac spinors which do have Euclidean counterparts. However, the
problem is unavoidable for theories with $\N=1$ supersymmetry.

There are several alternative approaches to this issue favoured by different
workers in the field. The most conservative approach as described in
\cite{NSVZ} (See Appendix A of this reference), is to abandon the idea of
constructing a Euclidean quantum field theory with
$\N=1$ supersymmetric as unnecessary. In
particular, our main interest should be calculating Green's functions in the
Minkowski space theory. As in standard perturbative calculations, these
Green's functions are most conveniently calculated by analytic continuation
of the corresponding Minkowski space path integral to Euclidean spacetime.
This is effected by the standard Wick rotation of the time coordinate
$x_{0} \rightarrow -i x_{4}$. The path-integral exponent is then
$-S_E=iS_M$.  The resulting path-integral can
then be evaluated in the saddle-point approximation by expanding in
fluctuations around the minima of $S_{E}$. This procedure yields
finite and well-defined answers for the original Minkowski space Green's
functions. In this context, the fact that the fermionic part of $S_{E}$
is not real is inconsequential: it does not affect the convergence of
the integrals. Some authors argue that
reality of the action is not, in any case, the appropriate condition for
Euclidean space theories. Rather we should impose a modified condition
known as ``reflection positivity'' which is characteristic of fermionic
actions on a spacetime lattice.

In the following we will adopt the conservative viewpoint described above.
However, there are other approaches to instantons which divorce
themselves from the Minkowski space theory and seek to define
the supersymmetric theory directly in Euclidean space
\cite{Belitsky:2000ws,Belitsky:2000ii}. This follows work of
several authors who have shown that it is actually possible
to define Euclidean versions of the theories with both extended supersymmetry
and real actions \cite{Zumino:1977yh,Nicolai:1978vc,
vanNieuwenhuizen:1996tv,Blau:1997pp,Acharya:1998jn}.
These theories have the potentially undesirable feature
of a non-compact $R$-symmetry group
and a scalar field with a negative kinetic term.
The philosophy is therefore slightly
different from our viewpoint---and the
relation with the original Minkowski space theory
is now rather obscure---but the
resulting calculations are essentially identical to those
described below.

In the bulk of this chapter we will discuss the $\N=1$, 2 and 4
theories in a unified formalism. In the remainder of this
introductory section we will introduce some of the key points which
arise, starting with the
$\N=1$ theory which contains the gauge field and a
single species of Weyl fermion $\lambda_{\alpha}$ in the adjoint
representation of the gauge group. Setting the fermion fields to zero to
start with, the ADHM instanton configuration trivially solves the
equations-of-motion. However, must now consider
the fluctuations of the fermions
around this solution. These are governed by the following
covariant Dirac equations,
\AL{
\Dbarslash\lambda^{A}&=0\ ,\elabel{jsa1}\\
\Dslash\bar\lambda_A&=0\ ,\elabel{jsb1}
}
where the covariant derivatives are taken in the adjoint representation and
evaluated in the instanton background of topological charge $k$.
As already discussed in \S\ref{sec:S4},
a standard application of the Atiyah-Singer index theorem shows that
equation \eqref{jsa1} has $2kN$
linearly independent (normalizable)
solutions. These were identified explicitly in
\S\ref{sec:S14} as $\Lambda_\alpha(C_\xii)$, were $\xii=1,\ldots 2kN$
labels the solutions of the constraints \eqref{opa} and \eqref{opb}.
On the contrary, \eqref{jsb1} has no non-trivial solutions in an
instanton background.

These solutions are known as the {\it fermion zero modes\/} of the instanton.
There are two distinct---but ultimately equivalent---approaches to treating
these modes. One way of incorporating these modes is to treat them
perturbatively as fluctuations around the ordinary bosonic instanton.
This expansion is perfectly consistent, but is hard to carry out in practice
beyond the lowest orders.
An alternative approach,
and the one we will adopt here, originates in the work of
Novikov {\it et al\/} \cite{NSVZ}. In this approach, we interpret
the fermionic zero modes as corresponding to a degeneracy of classical
solutions just like zero modes of the
gauge field we met in the previous chapter. More generally,
the standard semi-classical
reasoning suggests that we should look for finite-action
solutions of the full coupled equations-of-motion of the supersymmetric
theory. We will call these configurations {\it super-instantons}.
In the present case they are easy to find: the gauge field takes is
ADHM value, while the fermions solve equations \eqref{jsa1} and \eqref{jsb1}.
Thus the right-handed fermion $\bar{\lambda}_{\dot{\alpha}}$ is zero; this ensures that the fermions do not modify the Yang-Mills equation for the
gauge field. Meanwhile,
the left-handed fermion $\lambda_{\alpha}$ is general linear
superposition of the normalizable zero modes:
\EQ{\lambda_{\alpha}=\sum_{i=1}^{2kN}\, \psi^\xii\,
\Lambda_\alpha(C_\xii)\ .
\elabel{supos}}
As $\lambda_{\alpha}$ is a fermionic field, the $2kN$ coefficients
$\psi^\xii$ are Grassmann variables.
In fact \eqref{supos} can be written more compactly as
$\lambda_\alpha=\Lambda_\alpha(\CM)$, where the Grassmann
quantities $\CM=\psi^\xii C_\xii$ themselves satisfy the constraints
\eqref{opa} and \eqref{opb}. We will call these constraints the
``fermionic ADHM constraints'' since they will turn out to be the
Grassmann superpartners of the ``bosonic'' ADHM constraints
\eqref{badhm}. The solutions to these constraints will then be
parameterized as $\CM(\psi^\xii)$.

The Grassmann variables $\psi^\xii$ are the
fermionic analogs of the collective coordinates which parameterize
the general ADHM solution and we will henceforth refer to them as
{\it Grassmann collective coordinates\/}. While
the bosonic collective coordinates $X^{\mu}$
have the interpretation as intrinsic coordinates on the
moduli space $\ms_{k}$,
their Grassmann counterparts correspond to intrinsic
{\it symplectic tangent vectors\/}
on this manifold (see Appendix \ref{app:A2}). Like the bosonic collective
coordinates they parameterize degenerate minima of the action and we
must integrate over them.
We will work out the exact integration measure in the
following but the important qualitative features follow from the basic rules
of Grassmann integration,
\EQ{
\int\, d\psi^{\xii} =0\ ,\qquad \int\, d\psi^{\xii}\,
\psi^{\xjj}=\delta^{\xii\xjj}\ .
\elabel{berezin}
}
We see that to obtain a non-zero answer, each Grassmann integration in the
measure must be saturated by a single power of the integration variable.
This leads to simple selection rule for fermionic Green's functions in the
$k$ instanton background. Specifically the Green's function,
$\langle \lambda(x_{1}) \lambda(x_{2}) \ldots \lambda(x_{l}) \rangle$
will vanish unless $l=2kN$. This counting can be understood by
noting that $\N=1$ supersymmetric Yang-Mills theory has an anomalous abelian
$R$-symmetry under which the left- and right-handed fermions have charges
$\pm 1$. The exact anomaly in the $R$-symmetry current $j_{\mu}^{R}$ is
determined at one loop as,
\EQ{
\partial_m j_m^{R} = {{Ng^{2}}\over{8\pi^{2}}}\,
{\rm tr}_N\,F_{mn}{}^*F_{mn}
\elabel{anom}
}
Integrating this equation over spacetime we find that conservation of the
corresponding $\U(1)$ charge is
violated by $2k N$ units in the background of topological charge $k$.
which agrees with the selection rule given above. This agrees with the
selection rule described above. The symmetry means that
observables of definite
$R$-charge receive corrections from a single topological sector if at all.
As we will see below, much more interesting behaviour is possible in theories
without an anomalous abelian $R$-symmetry.

In the previous chapter we saw that a subset of the bosonic collective
coordinates can be understood in terms of the action of
symmetries of the theory on the instanton solution. The same is true for
the Grassmann collective coordinates. As mentioned above,
any self-dual configuration of the gauge field is invariant under
half the generators of the supersymmetry algebra. On the other hand,
the other half of the generators act non-trivially on the solution
generating fermion zero modes. The $\N=1$ supersymmetry algebra
has four supercharges and yields two zero modes.
In fact, classical supersymmetric gauge theories in four dimensions are
invariant under a larger superconformal algebra which,
in the $\N=1$ case, includes four additional fermionic generators, two of
which act non-trivially on the instanton.
Broken symmetries thus yield a total of four fermion zero modes.
The argument which guarantees the existence of
these zero modes is very robust (it is essentially Goldstone's
theorem) and holds at all orders in the full quantum theory.
As above, the actual number of
zero modes is $2kN$ which equals four in the minimal case $k=1$, $N=2$.
Thus, the situation for fermionic zero modes matches nicely with the
corresponding counting of bosonic zero modes given in the previous chapter.
Specifically, the bosonic and fermionic
zero modes of a single instanton of gauge group
$\SU(2)$ are all associated with the action of broken symmetry generators.
For higher instanton number and/or larger gauge group there are additional
zero modes which do not correspond to broken symmetries.

Theories with extended supersymmetry or, more generally,
$\N=1$ theories with additional matter necessarily contain scalar fields.
As usual, the scalars can acquire vacuum expectation values (VEVs) which
spontaneously break the gauge symmetry. This possibility is particularly
important for applications of instanton calculus to theories with
asymptotic freedom. As usual
such theories are characterized by logarithmic running of the coupling
which introduces a dynamical scale $\Lambda$.
In the absence of scalar VEVs, the running coupling becomes large in the IR
at mass scales of order $\Lambda$ and we do not expect semi-classical methods
to work. However, if we introduce a scalar VEV
which breaks the gauge group at some scale $v$ then the effective
coupling will not run below this scale. In particular, if the scale of
the VEV is much greater than $\Lambda$,
the running coupling is frozen before it has a chance to become large
and the theory is weakly coupled at all length scales. In these
circumstances we can expect a semi-classical analysis of the path integral
to be reliable.

Introducing scalar fields, with or without VEVs, affects the
instanton calculus in several important ways. In addition to the gauge
couplings of the scalars, supersymmetric Lagrangians necessarily contain
Yukawa couplings between the scalars and fermions.
As above, we are looking for
a super-instanton which solves the full equations of motion of the theory.
A promising starting point is the configuration described above where the
gauge-field takes its ADHM value and the fermions are a general
linear combination of the zero-modes.
The scalars themselves satisfy a covariant Laplace equation in the
gauge field background with a fermion bilinear source term which comes
from the Yukawa coupling and, as we
review below, this equation can be solved explicitly. The solution exhibits
a new feature of the super-instanton: bosonic fields can have pieces which
are bilinear (or of higher even power) in the Grassmann collective
coordinates. A potentially worrying feature is that a complex
scalar field $\phi$ typically acquires a non-zero Grassmann bilinear part,
while its complex conjugate $\phi^{\dagger}$ does not.
There is no inconsistency here, as we can illustrate
by considering the following toy integral:
\EQ{
{\cal I}=\int d^2\phi\,e^{-|\phi|^2+\phi A+\phi^*B}\ ,
}
where $A$ and $B$ are quadratic expressions in Grassmann
variables and we suppose $A\neq B^*$.
Obviously, one should expand in the Grassmann composites
$A$ and $B$, since the series terminates because there are only a
finite number of Grassmann variables, and then do the $\phi$ integral:
\EQ{
{\cal I}=\sum_{m,n=0}^P\frac1{m!n!}A^mB^n\,\int d^2\phi\,\phi^m
\phi^{*n}\,e^{-|\phi|^2}=\pi\sum_{m=0}^P\frac1{m!}(AB)^m=\pi e^{AB}\ .
\label{iresult}
}
However, the same result can be obtained by solving the
``equations-of-motion'' of $\phi$ and $\phi^*$
and shifting the integration variables by the solution. The solutions are
\EQ{
\phi=B\ ,\qquad\phi^*=A\ ,
}
which, since by hypothesis $A\neq B^*$,
violate the reality condition on $\phi$. However,
\EQ{
{\cal I}=\int d^2\phi\,e^{-(\phi-B)(\phi^*-A)+AB}=
e^{AB}\,\int\,d^2\phi\,e^{-|\phi|^2}=\pi e^{AB}
}
reproducing \eqref{iresult}. In this way solving the
equation-of-motion in the instanton background will involve Grassmann
composite terms which violate the reality conditions of the fields,
however, as we have seen with the toy integral above, it is a
convenient book-keeping device.


Even in the absence of scalar VEVs, the configuration described
above is not the end of the story because it
does not necessarily solve the full coupled equations-of-motion.
In particular,
the non-zero Grassmann bilinear part of the scalar field can modify the
equations of motion for the fermions and even for the gauge field itself,
invalidating our starting ansatz for these fields. In general this
modification will be non-trivial unless it is forbidden by the
symmetries of the theory. For theories with an abelian
$R$-symmetry, such as the $\N=2$ theory without scalar VEVs,
the new terms in the equations are zero and our
candidate super-instanton actually solves the full equations of motion.
However, in one of the most important cases,
that of $\N=4$ supersymmetric Yang-Mills,
there is no such symmetry and the modification is unavoidable.
Even in this case the situation is
not as hopeless as it might appear because the offending couplings
are each suppressed by powers of $g^{2}$ at weak coupling.
Fortunately, for our stated purpose
of calculating the leading semi-classical contributions to
Green's functions, it suffices to solve the equations of motion
perturbatively to some order in $g^{2}$. The resulting
configuration differs from being an exact solution by a power of $g^{2}$
and hence we will refer to it as a {\it quasi-instanton\/}. The question of
whether a corresponding {\it exact\/} solution exists is an interesting one.
As we are expanding in Grassmann bilinears, the perturbation
series must truncate at some finite order and one might imagine that the
final result should be an exact solution. However, at least for the $\N=4$
theory we will argue that this is not the case. In fact, we will
exhibit an obstruction to solving the equations beyond next-to-leading order
in this case. In any case, the utility of proceeding beyond the first
few orders in the classical equations is unclear,
because one must also take into account quantum corrections which modify the
equations themselves at the same order in $g^{2}$.

An important property of the quasi-instanton can be exhibited by
calculating its action. The $\N=4$ theory has four species of
Weyl fermions which lie in the fundamental representation of an $SU(4)$
$R$-symmetry. The corresponding Grassmann collective coordinates are
$\psi^{\xii A}$ where, as in the $\N=1$ theory discussed above,
$\xii=1,\ldots,2kN$  and $A=1,2,3,4$ is the $R$-symmetry index.
In the absence of scalar VEVs, the action of our
quasi-instanton of topological charge $k$ is,
\EQ{
\tilde S =
\frac{8\pi^{2}k}{g^{2}}-ik\theta+\frac1{96}\epsilon_{ABCD}
R_{\xii\xjj\xkk\xll}(X)
\psi^{\xii A}\psi^{\xjj B}\psi^{\xkk C}\psi^{\xll D}\ .
\elabel{qinstact}
}
Here, $R(X)$ is the symplectic curvature of the hyper-K\"ahler quotient space
$\ms_k$, which depends explicitly the chosen point in the moduli space.
The action depends explicitly on both
the bosonic and fermionic collective coordinate reflecting the fact that
we are not dealing with an exact solution of the equation of motion.

{}From the point of view of the semi-classical approximation,
the expression \eqref{qinstact} can be thought of as an
``effective action''
for the quasi-zero modes, to order $g^0$ in the coupling, which results from
integrating out the remaining modes of non-zero frequency.
As we will see in the next Chapter, the exponential of this action is an
essential ingredient in the collective coordinate integration measure.
The effective action for the instanton
collective coordinates closely related to the idea of a
world-volume effective action for solitons and other extended object in
higher dimensions. This point of view will be developed substantially
in Chapter \ref{sec:S49}.

The dependence of the action \eqref{qinstact}, on the Grassmann
collective coordinates means that the corresponding fermionic modes are
not exact zero modes. These modes are {\it lifted\/} at order
$g^0$ in the semi-classical expansion.
The zero modes which are generated by the action of fermionic
symmetry generators on the instanton are an important exception.
The same symmetries lead to zero eigenvalues of the curvature tensor which
mean that the action \eqref{qinstact} is independent of the corresponding
Grassmann collective coordinates. The associated
zero modes remain unlifted to all orders in $g^{2}$.
The lifting of fermion zero modes means that the selection rules
determining which Green's functions can receive instanton corrections are
much less restrictive than those of the $\N=1$ theory described above.
Typically this is also related to the absence of an anomalous abelian
$R$-symmetry.
For example, in the $\N=4$ theory with zero scalar VEVs, the total number of
exact fermion zero modes is sixteen (two supersymmetric and two
superconformal modes for each of the four species of Weyl fermion).
This number does not depend on the topological charge and, for example,
sixteen-fermion correlators receive an
infinite series of corrections from all numbers of instantons.

{}Finally we turn to the case where scalar fields develop non-zero
VEVs. In this
case, it is well known that there is no non-trivial instanton solution of
the coupled equations of motion for the gauge field and scalar. Indeed, the
existence of such a solution is forbidden by Derrick's theorem.
The problem is best illustrated by proceeding na\"\i vely in the
$\SU(2)$ theory. If we solve the
covariant Laplace equation for an adjoint scalar $\phi$ with VEV
${\rm diag}(\phi^0,-\phi^0)$
in the background of a single $\SU(2)$ instanton of scale-size $\rho$
and substitute the solution back into the action the
result is,
\EQ{S=\frac{8\pi^{2}}{g^{2}}-i\theta+ 4 \pi^{2}
\rho ^{2}(\phi^0)^2\ .
\elabel{derr}
}
The instanton action depends explicitly on the scale size and may be lowered
continuously to zero by shrinking the instanton. However, it is useful to
note that the second term in the action \eqref{derr} is down by a power of
$g^{2}$ relative to the first.\footnote{Strictly
speaking this is true as long  as we treat
$\rho\phi^0$ as order $g^{0}$. As we must integrate overall
values of $\rho$ this assumption needs to be justified.
In the cases of interest is not hard to show that
that larger values of $\rho\phi^0$ are exponentially suppressed and the
na\"\i ve scaling is correct.}
Thus our candidate configuration is actually a quasi-instanton in the same
sense as in our discussion of the $\N=4$ theory above.
In either case, our
aim is to determine the leading semi-classical behaviour of Green's functions
and it is legitimate to solve the saddle-point conditions order by order in
$g^{2}$.

In the case of non-zero scalar VEVs, this approach was first
developed by Affleck who referred to the corresponding field configurations
as {\it constrained instantons\/}. The formalism of constrained
instantons is rather technical. However, as we explain below, a large
part of it can actually be understood in terms of a modified
instanton action like that given in \eqref{derr}. As mentioned above,
any quasi-instanton will have an action which depends explicitly on the
collective coordinates. In the case of the $\N=4$ theory without scalar
VEVs, the resulting expression, given as \eqref{qinstact} above,
is determined in terms of the symplectic curvature on $\ms_{k}$.
One of our main results is that the effect of introducing scalar VEVs is
simply to introduce an appropriate potential on the moduli
space. Furthermore this potential has a nice interpretation as the
norm squared of a tri-holomorphic Killing vector on $\ms_{k}$ 
(for related results in the context of dyons see \cite{Tong1,Tong2}).

The remainder of this Chapter is organized as follows. In \S\ref{sec:S26},
we present the Minkowski space action, equations-of-motion and
supersymmetric transformations of the minimal gauge theories with
$\N=1$, 2 and
4 supersymmetries in four dimensions. We discuss the analytic
continuation of the theory to Euclidean spacetime. In \S\ref{sec:S27} we
construct the super-instanton at the first non-trivial order in $g^{2}$.
This necessitates solving the adjoint Dirac equation \eqref{jsb1} for the
left-handed fermion in the general ADHM background. The general
solution involves a matrix of Grassmann variables constrained by a linear
equation which generalizes the ADHM constraint equations of the previous
chapter. We introduce intrinsic Grassmann collective coordinates and
identify them as symplectic tangent vectors on $\ms_{k}$.
We discuss the action of supersymmetry on the collective coordinates.

In \S\ref{sec:S31}, we consider the construction of the super-instanton
beyond linear order. This requires solving the
covariant Laplace equation for the
adjoint scalar in the general ADHM background with an appropriate
fermion bilinear source term. We show that this yields an exact super-instanton
for the $\N=2$ theory (without scalar VEVs), but only a
quasi-instanton, in the sense described above, in the $\N=4$ case. We
demonstrate an obstruction to the existence of an exact solution
in this case. In \S\ref{sec:S33}, we review the necessary aspects of the
constrained instanton formalism. Finally, in \S\ref{sec:S32} we
explain how the supersymmetry of the field theory descends to the
collective coordinates.

In the next chapter,
we derive the instanton measure in supersymmetric theories.
We introduce the {\it instanton effective action\/} for these theories and
discuss the consequent lifting of fermion zero modes.
Then we derive an explicit formulae for the appropriate supersymmetric 
volume form on $\ms_{k}$.

\subsection{Action, supersymmetry and equations-of-motion}\elabel{sec:S26}

We start by defining theories in four-dimensional Minkowski space
with $\N=1,2$ and $4$ supersymmetry. In the interests of brevity we will
develop a unified notion that allows us to deal with all these
cases together. To this end, we introduce the fermionic partners of the gauge
field $\lambda^A$ and $\bar\lambda_A$. Here,
$A=1,\ldots,\N$ is an $R$-symmetry index of the supersymmetry. Since
we are working---at least initially---in Minkowski space, these spinors are
subject to the reality conditions
\EQ{
(\lambda_\alpha^A)^\dagger=\bar\lambda_{\aD A}\ ,\qquad
(\bar\lambda^\aD_A)^\dagger=\lambda^{\alpha A}\qquad(\alpha=\aD)\ .
\elabel{rew}
}
In addition, for the theories with extended supersymmetry there are
real scalar fields $\phi_a$, $a=1,\ldots,2(\N-1)$. The
Minkowski space action is\footnote{We remind the reader that our gauge
field is anti-Hermitian rather than Hermitian, otherwise our
conventions in Minkowski space are those of Wess and Bagger \cite{WB}.}
\SP{
S^{\sst{\rm Mink}}=&
\int d^4x\,\TrN\Big\{\tfrac12F_{mn}^2+\frac{i\theta g^2}{16\pi^2}
F_{mn}{}^*F^{mn}
+2i{\cal D}_n\bar\lambda_{A}\bar\sigma^n\lambda^A
-{\cal D}^n\phi_a{\cal D}_n\phi_a
\\
&+g\bar\lambda_{A}\Sigma^{AB}_a[\phi_a,\bar\lambda_B]
+g\lambda^{A}\bar\Sigma_{aAB}[\phi_a,\lambda^B]+
\tfrac12g^2[\phi_a,\phi_b]^2\Big\} .
\elabel{cpta}
}
The terms involving the scalar fields are, of course, absent in the
$\N=1$ theory. The $\Sigma$-matrices are associated to the $\SU(2)$
and $\SU(4)$ $R$-symmetry group of the $\N=2$ and $\N=4$ theories,
respectively. For $\N=2$ we take
\EQ{
\Sigma_a^{AB}=\epsilon^{AB}(i,1)\ ,\qquad
\bar\Sigma_{aAB}=\epsilon_{AB}(-i,1)\ ,
\elabel{llzz}
}
In this case the indices $A,B,\cdots=1,2$ are spinor indices of the
$\SU(2)$ subgroup of the $\U(1)\times\SU(2)$ $R$-symmetry group. SO
in this case, we can raise and lower the indices using the
$\epsilon$-tensor in the usual way following the conventions of \cite{WB}.
For the $\N=4$ case
\begin{equation}\begin{split}
\Sigma_a&=\big(\eta^3,i\bar\eta^3,\eta^2,
i\bar\eta^2,\eta^1,i\bar\eta^1\big)\ ,\\
\bar\Sigma_a&=\big(-\eta^3,i\bar\eta^3,-\eta^2,i\bar\eta^2,-\eta^1,
i\bar\eta^1\big)\ ,
\end{split}\elabel{mmzz}\end{equation}
where $\eta^c$, $\bar\eta^c$, $c=1,2,3$,
are 't~Hooft's $\eta$-symbols defined in Appendix \ref{app:A1}.

The theory \eqref{cpta} is invariant under the on-shell supersymmetry
transformations
\AL{
\delta A_{n}&=-\xi^A\sigma_n
\bar\lambda_A-\bar\xi_A\bar\sigma_n
\lambda^A\ ,\elabel{hha}\\
\delta\lambda^A&=-i\sigma^{mn}\xi^AF_{mn}-
ig\Sigma_{ab}{}^A{}_B\xi^B[\phi_a,\phi_b]+\Sigma_a^{AB}
\sigma^n\bar\xi_B{\cal D}_n\phi_a\
,\elabel{hhb}\\
\delta\bar\lambda_A&=-i\bar
\sigma^{mn}\bar\xi_AF_{mn}-ig\bar\Sigma_{abA}{}^B\bar\xi_B[\phi_a,\phi_b]
+\bar\Sigma_{aAB}\bar\sigma^n\xi^B{\cal D}_n\phi_a\
,\elabel{hhc}\\
\delta\phi_a&=i\xi^A\bar\Sigma_{aAB}\lambda^B+i\bar\xi_A\Sigma_a^{AB}
\bar\lambda_B\ .\elabel{hhd}
}
In the above,
\EQ{
\sigma^{mn}=\tfrac14(\sigma^m\bar\sigma^n-\sigma^n\bar\sigma^m)\ ,\qquad
\bar\sigma^{mn}=\tfrac14(\bar\sigma^m\sigma^n-\bar\sigma^n\sigma^m)\ .
}
and
\EQ{
\Sigma_{ab}=\tfrac14(\Sigma_a\bar\Sigma_b-\Sigma_b\bar\Sigma_a)\ ,\qquad
\bar\Sigma_{ab}=\tfrac14(\bar\Sigma_a\Sigma_b-\bar\Sigma_b\Sigma_a)\ .
}

In order to construct instanton solutions, we now
Wick rotate to Euclidean space. Vector quantities
in Minkowski space $a^n=(a^0,\vec a)$, with
$n=0,1,2,3$, become
$a_n=(\vec a,ia^0)$, with $n=1,2,3,4$, in Euclidean space.
The Euclidean action is then $-i$ times the Minkowski space
action. The exception to this is that we define the Euclidean
$\sigma$-matrices as in \eqref{rae} and \eqref{raf},
So in Minkowski space $\sigma^n=(-1,\vec\tau)$ and
$\bar\sigma^n=(-1,-\vec\tau)$, whereas in Euclidean space
$\sigma_n=(i\vec\tau,1)$ and
$\bar\sigma_n=(-i\vec\tau,1)$. Operationally,
this means that when Wick rotating from Minkowski space to Euclidean
space we should actually replace the Minkowski space
$\sigma$-matrices by $-i$ times the Euclidean space $\sigma$-matrices.
As usual we
treat $\lambda_\alpha$ and $\bar\lambda^\aD$ as independent
spinors, {\it i.e.\/}~independent integration variables in the
functional integral. The Euclidean space action is
\SP{
S=&\int d^4x\,\TrN\Big\{-\tfrac12F_{mn}^2-
\frac{i\theta g^2}{16\pi^2}F_{mn}{}^*F_{mn}
-2{\cal D}_n\bar\lambda_{A}\bar\sigma_n\lambda^A
+{\cal D}_n\phi_a{\cal D}_n\phi_a
\\
&-g\bar\lambda_{A}\Sigma^{AB}_a[\phi_a,\bar\lambda_B]
-g\lambda^{A}\bar\Sigma_{aAB}[\phi_a,\lambda^B]-
\tfrac12g^2[\phi_a,\phi_b]^2\Big\}\ .
\elabel{cptae}
}
As discussed above, the fact that the fermionic terms in this action
are not real will not concern us further.
For the case with $\N=2$ supersymmetry, we can recover the more
conventional presentation of the theory by defining a complex scalar
field
\EQ{
\phi=\phi_1-i\phi_2\
,\qquad \phi^\dagger
=\phi_1+i\phi_2\ .
\elabel{ntsc}
}
and spinors  $\lambda\equiv\lambda^1$ and
$\psi\equiv\lambda^2$. The fields $\Phi=\{\phi/\sqrt2,\psi\}$ form a chiral
multiplet and $V=\{A_m,\lambda\}$ a
vector multiplet of $\N=1$ supersymmetry.
In terms of these variables, the Euclidean space
action of the $\N=2$ theory \eqref{cptae} is
\SP{
S_{\sst\N=2}=&\int d^4x\,\TrN\Big\{-\tfrac12F_{mn}^2-
\frac{i\theta g^2}{16\pi^2}F_{mn}{}^*F_{mn}
-2{\cal D}_n\bar\lambda\bar\sigma_n\lambda-2{\cal D}_n\bar\psi
\bar\sigma_n\psi
+{\cal D}_n\phi^\dagger{\cal D}_n\phi
\\
&+2ig\bar\psi[\phi,\bar\lambda]
+2ig[\phi^\dagger,\lambda]\psi+\tfrac14g^2
[\phi,\phi^\dagger]^2\Big\}\ .
\elabel{cptam}
}
In the following, we prefer the presentation of the theory in
\eqref{cptae} since this will allow us to deal with the
theories with different numbers of supersymmetries in a unified way.

The equations-of-motion following from \eqref{cptae} are
\AL{
{\cal D}_mF_{nm}&=2g[\phi_a,{\cal D}_n\phi_a]+2g
\bar\sigma_{n}\{\lambda^A,\bar\lambda_A\}\ ,\elabel{uta}\\
\Dbarslash\lambda^A&=
g\Sigma^{AB}_a[\phi_a,\bar\lambda_B]\ ,\elabel{utb}\\
\Dslash\bar\lambda_A&=
g\bar\Sigma_{aAB}[\phi_a,\lambda^B]\ ,\elabel{utc}\\
{\cal D}^2\phi_a&=g^2[\phi_b,[\phi_b,\phi_a]]+
g\bar\Sigma_{aAB}\lambda^A\lambda^B
+g\Sigma_a^{AB}\bar\lambda_A\bar\lambda_B\ .\elabel{utd}
}
The supersymmetry transformations in Euclidean space are given by
\eqref{hha}-\eqref{hhd} by replacing the sigma matrices with $-i$
times their Euclidean space versions and by replacing Minkowski space
inner-products with Euclidean ones:
\AL{
\delta A_{n}&=i\xi^A\sigma_n
\bar\lambda_A+i\bar\xi_A\bar\sigma_n
\lambda^A\ ,\elabel{lha}\\
\delta\lambda^A&=i\sigma_{mn}\xi^AF_{mn}-
ig\Sigma_{ab}{}^A{}_B\xi^B[\phi_a,\phi_b]-i\Sigma_a^{AB}
\Dslash\phi_a\bar\xi_B\
,\elabel{lhb}\\
\delta\bar\lambda_A&=i\bar
\sigma_{mn}\bar\xi_AF_{mn}-ig\bar\Sigma_{abA}{}^B\bar\xi_B[\phi_a,\phi_b]
-i\bar\Sigma_{aAB}\Dbarslash\phi_a\xi^B\
,\elabel{lhc}\\
\delta\phi_a&=i\xi^A\bar\Sigma_{aAB}\lambda^B+i\bar\xi_A\Sigma_a^{AB}
\bar\lambda_B\ .\elabel{lhd}
}

\subsection{The super-instanton at linear order}\elabel{sec:S27}

We will now attempt to find {\it super-instanton\/} configurations which
solve the full coupled equations-of-motion \eqref{uta}-\eqref{utd}.
First notice that the original instanton solution of the pure gauge theory
\eqref{vdef} is a solution of the
full equations-of-motion when all other fields are set to zero.
In fact, we can use $A_m(x;X)$ as a starting point to find the
more general solutions where the fermion and scalar fields are non-vanishing.
As explained in the introduction to this chapter we will proceed
perturbatively order by order in the coupling. In this connection note the
explicit powers of $g$ appearing on the right-hand side of
Eqs.~\eqref{uta}-\eqref{utd}.

The first step, following \cite{NSVZ},
is to expand to linear order in the fields
around the bosonic instanton solution.
To the next order, we must therefore solve the
covariant Weyl equations
\AL{
\Dbarslash\lambda^{A}&=0\ ,\elabel{jsa}\\
\Dslash\bar\lambda_A&=0\ ,\elabel{jsb}
}
for the fermions, and the covariant Laplace equation
\EQ{
{\cal D}^2\phi_a=0\ ,
\elabel{gdf}
}
for the scalars. It then remains to be seen whether the
original instanton solution needs to be modified due the source term
on the right-hand side of \eqref{uta}.

A key result follows from the fact that $\Dslash$ has no zero modes in
an instanton (rather than anti-instanton) background. Consequently,
the solution to \eqref{jsb} is $\bar\lambda_A=0$.
To prove this, \eqref{jsb} implies
$\Dbarslash\Dslash\bar\lambda_A=0$. But we can expand the product of
operators using the definition of $\Delta^{(-)}$ in \eqref{rrbb}, so
\EQ{
\Dbarslash\Dslash\bar\lambda_A
\equiv-\Delta^{(-)}\bar\lambda_A=
{\cal D}^2\bar\lambda_A+\bar\sigma_{mn}F_{mn}\bar\lambda_A
\elabel{iioo}
}
and use the fact that in an instanton
background $F_{mn}$ is self-dual. Since
$\bar\sigma_{mn}$ is a projector onto the anti-self-dual part
\eqref{sdasd}, we have
\EQ{
\Dbarslash\Dslash\bar\lambda_A={\cal
D}^2\bar\lambda_A=0\ .
}
But ${\cal D}^2\equiv{\cal D}_n{\cal D}_n$ is a positive
operator. This means ${\cal D}^2\bar\lambda_A=0$ implies
$\bar\lambda_A=0$ and consequently there are no zero modes for the
anti-chiral fermions. The discussion above also implies, for
vanishing VEVs at least, that
the scalar fields vanish also vanish at linear order.

On the contrary, for $\lambda^A$ a similar maneuver to \eqref{iioo} gives
\EQ{
\Dslash\Dbarslash\lambda^A
\equiv-\Delta^{(+)}\lambda^A=
{\cal D}^2\lambda^A+F_{mn}\sigma_{mn}\lambda^A=0\ .
}
In this case the second term does not vanish and so there is no reason
for $\lambda^A$ to vanish. In fact, we should have anticipated this
since, as discussed in
\S\ref{sec:S14}, the Atiyah-Singer index theorem dictates that
the operator $\Dbarslash$
has $2kN$ normalizable zero modes in the $k$ instanton background for
gauge group $\SU(N)$.

\subsubsection{Adjoint fermion zero modes}\elabel{sec:S28}
In this section, we consider in more detail the adjoint-valued
fermion zero
modes in the background of the bosonic instanton solution; in other words, the
solutions of the covariant Weyl
equation \eqref{jsa} in the ADHM background. Due to the linearity of
the equation we can consider a single Weyl fermion
$\lambda$ with $\Dbarslash\lambda=0$. Fortunately, no addition work
need be done since we have already solved this equation in \S\ref{sec:S14}
in the context of the gauge field
zero modes. We can immediately write
down the solutions in terms of the linear functions defined in \eqref{xxoo}:
\begin{equation}
\lambda_\alpha=g^{-1/2}\Lambda_\alpha(\CM)\equiv
g^{-1/2}\Big(\bar U{\cal M} f\bar
b_{\alpha}U- \bar Ub_{\alpha} f \bar{\cal M}U\Big) \ .
\elabel{lam}\end{equation}
(The unconventional
power of $g^{-1/2}$ in the definition
reflects the true $g$-scaling of the fermion zero modes, as will
emerge in due course.)
One difference from the gauge zero modes follows from the
fact that $\lambda$ is a Grassmann quantity; hence,
${\cal M}_{\lambda i}$ and $\bar{\cal M}_{i}^\lambda$ are constant
$(N+2k)\times k$
and $k\times (N+2k)$ matrices of {\it Grassmann\/} collective
coordinates, respectively, which replace the $c$-number
quantities $C_\aD$ and $\bar C_\aD$ in \eqref{xxoo}. In addition, as
indicated, the
Grassmann collective coordinates do {\it not\/}
carry the Weyl spinor index $\aD$.

In order for \eqref{lam} to be a solution of \eqref{jsa}, the
Grassmann collective coordinates must satisfy the constraints \cite{CGTone}
\eqref{hap}
\EQ{
\bar\Delta^\aD\CM+\bar\CM\Delta^\aD=0\ .
\elabel{mdid}
}
Expanding $\bar\Delta$ and $\Delta$ as in \eqref{del} and writing all
the indices explicitly, this becomes
\begin{subequations}
\begin{align}
\bar{\cal M}_{i}^\lambda a_{\lambda j\aD}
 &=- \bar a^\lambda_{i\aD}  {\cal M}_{\lambda j}
\ ,\elabel{zmcona}\\
\bar{\cal M}_{i}^\lambda b_{\lambda j}^\alpha &= \bar b_{i}^{\alpha\lambda}
 {\cal M}_{\lambda j}\ .
\elabel{zmconb}\end{align}
\end{subequations}
In a formal sense discussed
in Ref.~\cite{KMS}, these fermionic constraints are the ``spin-$\tfrac12$''
superpartners of the original ``spin-1'' ADHM constraints
\eqref{conea} and \eqref{coneb},
respectively.  Note further that \eqref{zmconb} is easily solved when
$b$ is in the canonical form
\eqref{aad}.
With the ADHM index decomposition $\lambda=u+i\alpha$, we set
\begin{equation}{\cal M}_{\lambda j} \equiv {\cal M}_{(u+i\alpha) j} =
\begin{pmatrix} \mu_{u j} \\  ({\cal M}^{\prime}_\alpha)_{ij}\end{pmatrix}
\ ,\quad
\bar{\cal M}_{j}^{\lambda}  \equiv  \bar{\cal M}_{j(u+i\alpha)} =
\left( \bar\mu_{j u} \ \ (\bar{\cal M}^{\prime\alpha})_{ji} \right)\ .
\elabel{mrep}\end{equation}
Eq.~\eqref{zmconb} then collapses to
\begin{equation}\bar{\cal M}^{\prime}_\alpha =
{\cal M}^{\prime}_\alpha
\elabel{mctw}\end{equation}
which allows us to eliminate $\bar{\cal M}^{\prime}$ in favor of
${\cal M}^{\prime}$.
In terms of the variables $\mu$, $\bar\mu$ and
${\cal M}'_\alpha$, the ``fermionic ADHM
constraints'' \eqref{zmcona} are
\EQ{
\bar\CM a_\aD+\bar a_\aD\CM\equiv\bar\mu w_\aD+\bar w_\aD\mu+[\M^\alpha,
a'_{\alpha\aD}]=0\ .
\elabel{fadhm}
}
The quantities $\CM=\{\mu,\bar\mu,\CM'_\alpha\}$, subject to
\eqref{fadhm}, are the Grassmann-valued partners of the ADHM variables
$a_\aD=\{w_\aD,a'_n\}$. Later in \S\ref{sec:S32}
we shall see that they are, indeed, related by supersymmetry.

We can now count the number of independent fermion zero modes. There
are $2k(N+k)$ independent Grassmann variables $\CM=\{\mu,\bar\mu,
{\cal M}'\}$, subject to $2k^2$ constraints
\eqref{fadhm}. Hence, there are $2kN$ independent zero modes, in
agreement with the Index Theorem.

\subsubsection{Grassmann collective coordinates and the hyper-K\"ahler
quotient construction}\elabel{sec:S29}

It is interesting to establish the relationship
between these Grassmann collective
coordinates and the hyper-K\"ahler quotient construction. A
hyper-K\"ahler space of dimension $4n$
admits a preferred $\SU(2)\times\Sp(n)$ basis of tangent vectors. In
particular, this leads to the notion of a {\it symplectic tangent
vector\/} \eqref{yrr}. With the relation
between the ADHM data $a_\aD$ and the symplectic variables
$z^{\ii\aD}$
in \eqref{jws}, we can see that the fermionic ADHM constraints
\eqref{fadhm} are precisely the condition that the symplectic tangent
vector to the mother space
\EQ{
\CM^\ii=\MAT{\bar\mu_{iu} \\ (\M^1)_{ij} \\ \mu_{ui} \\ (\CM'_1)_{ij}}\ ,
\elabel{spp}
}
is a symplectic tangent vector to the
hyper-K\"ahler quotient space $\ms_k$. Here, $\ii$ is the an index
that, as earlier in \S\ref{sec:S10},
runs over $\{iu,ij,ui,ij\}$. In order to prove this we must show
that the inner-product
\EQ{
\CM^\ii\tilde\Omega_{\ii\,\jj}X_r^{\jj\aD}=
-4i\pi^2{\rm tr}_k\,T^r\big(\bar\mu w^\aD+\bar w^\aD\mu+[\bar
a^{\prime\aD\alpha},\M_\alpha]\big)\ ,
\elabel{rwfadhm}
}
vanishes, where $X_r$ are the vectors, defined in \eqref{qqp},
that generate the $\U(k)$ action on $\tilde\ms$.
It is easy to see that this condition is
equivalent to the fermionic ADHM constraints \eqref{fadhm}.
To summarize, we have shown that the Grassmann collective coordinates are
Grassmann-valued symplectic tangent vectors to
the instanton moduli space $\ms_k$.

The functional inner product of
the fermion zero modes can be calculated using the same integral
formula that we used to establish the functional inner-product of the
gauge zero modes in \S\ref{sec:S14} and Appendix \ref{app:A4}
(Eq.~\eqref{corrigan}). For two such zero modes
\begin{equation}
\int d^4x\,\TrN\,\Lambda(\CM)\Lambda(\CN)=-{\pi^2\over2}
{\rm tr}_k\big[\bar{\cal M}({\cal P}_\infty+1){\cal N}+
\bar{\cal N}({\cal P}_\infty+1){\cal M}\big]\ .
\elabel{corriganf}
\end{equation}
Notice that the extra minus sign in \eqref{corriganf} relative to \eqref{jjqq}
arises because of the Grassmann-valued nature of the collective coordinates.
This is precisely the inner product of symplectic tangent vectors on $\tilde\ms$:
\EQ{
\tilde\Omega(\CM,\CN)=-4\int  d^4x\,\TrN\,\Lambda(\CM)
\Lambda(\CN)\ .
\elabel{ipstv}}

Just as in the $c$-number sector, where we have $\{X^\mu\}$ as intrinsic
coordinates on $\ms_k$, so we can define intrinsic Grassmann-valued
symplectic tangent vectors on $\ms_k$. To do this we solve the
fermionic ADHM constraints $\CM=\CM(\psi,X)$. Since the fermionic ADHM
constraints are linear in the Grassmann variables, the
parameterization
$\CM(\psi,X)$ will be linear
in the intrinsic Grassmann
coordinates $\psi^\xii$, $\xii=1,\ldots,2kN$. In much the same way
that the metric on $\ms_k$ is induced by that on $\tilde\ms$, so
the inner product of Grassmann-valued symplectic tangent vectors on
$\tilde\ms$, \eqref{ipstv},
then induces a similar inner product on $\ms_k$. So for a
pair of symplectic tangent vectors $\psi$ and $\theta$:
\EQ{
\Omega_{\xii\xjj}(X)\psi^\xii\theta^\xjj=
\tilde\Omega(\CM(\psi,X),\CM(\theta,X))\ .
\elabel{ipint}
}
Here, $\Omega_{\xii\xjj}(X)$ is the symplectic matrix which appears in the
expression for the metric on $\ms_k$ in \eqref{metsp}.

\subsubsection{Supersymmetric and superconformal zero modes}\elabel{sec:S30}

In \S\ref{sec:S11}, we described how those global symmetries of the classical
equations-of-motion which act non-trivially on the instanton
are represented on the moduli
space. The symmetries described there---global gauge
and conformal---will also have an action on
the Grassmann collective coordinates. In addition, we have
supersymmetry, enhanced to the superconformal group, that acts as
symmetries of the classical equations-of-motion. These will act on the
supersymmetrized moduli space \cite{KMS,NSVZ,MO-II}.

A special set of the fermion zero modes can be
identified with the action of supersymmetry transformations on the
bosonic instanton solution. As is evident from \eqref{lhb} and
\eqref{lhc}, supersymmetry transformations on the purely bosonic
instanton turn on the fermionic fields:
\AL{
\lambda^A&=i\sigma_{mn}\xi^AF_{mn}\ ,\elabel{svolam}\\
\bar\lambda_A&=i\bar\sigma_{mn}\bar\xi_AF_{mn}\ .
}
Actually, to be more precise,
in the bosonic instanton background $\bar\sigma_{mn}F_{mn}=0$, so the
anti-chiral fermions are not turned on. In particular, this means that the
bosonic instanton is invariant under half the supersymmetries, namely
the anti-chiral ones $\bar\xi_A$. Correspondingly the
chiral supersymmetry transformations generate fermion zero modes.
Using the expression for the field
strength and fermion zero modes in the ADHM instanton background,
Eqs.~\eqref{sdu} and \eqref{lam},
respectively, we find
\SP{
\lambda^A_\alpha&=4ig^{-1/2}(\sigma_{mn}\xi^A)_\alpha\bar Ub\sigma_{mn}\bar
bfU\\ &=-4ig^{-1}\bar U\Big(b\xi^Af\bar b_\alpha-b_\alpha f\xi^A\bar
b\Big)U\\
&\equiv g^{-1/2}\Lambda_\alpha(-4ib\xi^A)\ .
\elabel{ppll}
}
Here, we have  re-scaled
\EQ{
\xi^A\to g^{1/2}\xi^A
\elabel{ressp}
}
so that the following equations do not have $g$ dependence.
Consequently, the chiral
supersymmetry transformations generate fermionic zero modes
with the Grassmann collective coordinates
\EQ{
\CM_{\lambda i}^A=-4i\xi^{A}_\alpha b_{\lambda i}^\alpha\, ,\qquad
\bar\CM^{\lambda A}_i=-4i\xi^{\alpha A}
\bar b^\lambda_{\alpha i}\ .
\elabel{susymo}
}
These privileged fermion zero modes are known as the ``supersymmetric
zero modes''. There are obviously two such modes for each species of
fermion, hence the total number of supersymmetric modes is equal to
half the number of component supercharges of the gauge theory, {\it
i.e.\/}~$2\N$.

In addition to these supersymmetric zero modes, there are fermionic
zero modes corresponding to broken superconformal invariance.
These transformations can be obtained by generalizing the
supersymmetry transformations
\eqref{lha}-\eqref{lhd} by making the parameters $\xi^A$ local in a
particular way:
\EQ{
\xi^A_\alpha(x)=\xi^A_\alpha-x_{\alpha\aD}\bar\eta^{\aD A}\ ,\qquad
\bar\xi_A^\alpha(x)=\bar\xi^{\alpha}_A+\eta_{\aD A}\bar x^{\aD\alpha}\ .
\elabel{xdep}
}
This defines a basis of both supersymmetric and superconformal
transformations generated by $\{\xi_A,\bar\xi^A\}$ and $\{\eta^A,\bar\eta_A\}$,
respectively. The bosonic instanton breaks half the superconformal
transformations, namely those generated by $\bar\eta^A$. These
transformations generate fermion zero modes in much the same way as
\eqref{ppll}:
\SP{
\lambda^A_\alpha&=-4ig^{-1/2}(\sigma_{mn}x\bar\eta^A)_\alpha
\bar Ub\sigma_{mn}\bar
bfU\\ &=4ig^{-1/2}\bar U\Big(b x\bar\eta^Af\bar b_\alpha-b_\alpha
f\bar\eta^A\bar x\bar
b\Big)U\\
&=-4ig^{-1/2}\bar U\Big(a\bar\eta^Af\bar b_\alpha-b_\alpha
f\bar\eta^A\bar a\Big)U\\
&\equiv g^{-1/2}\Lambda_\alpha(-4ai\bar\eta^A)\ .
\elabel{qqll}
}
where in the second to last equality, we used the ADHM relations
\eqref{uan} which imply $\bar Ubx=-\bar Ua$ and $\bar x\bar bU=-\bar
aU$. Consequently, the associated Grassmann collective
coordinates are
\EQ{
\CM_{\lambda i}=-4ia_{\lambda i\aD}\bar\eta^{\aD}\
,\qquad \bar\CM^{\lambda}_i=
-4i\bar\eta_\aD\bar a^{\aD\lambda}_i\ .
\elabel{suconmo}
}
As for the supersymmetric modes,
there are two such modes for each species of
fermion, hence the total number of ``superconformal modes'' is equal
again to $2\N$.

\subsection{Going beyond linear order: the quasi-instanton}\elabel{sec:S31}

In order to evaluate correlation functions at their first non-vanishing order
in the semi-classical expansion, we will find it necessary
to go beyond the linear order analysis of the previous subsection.
The systematics of the semi-classical expansion can be
deduced from the equations-of-motion \eqref{uta}-\eqref{utd} and the
fact that the fermion zero modes $\lambda^A$ are order
${\cal O}(g^{-1/2})$ and the gauge field is
${\cal O}(g^{-1})$. Schematically, the various fields have the following
$g$-expansions:
\SP{
A_m&= g^{-1}A_m^\zo+gA_m^{\sst(1)}+g^3A_m^{\sst(2)}+\cdots\ ,\\
\lambda^A&= g^{-1/2}\lambda^{\zo A}+g^{3/2}\lambda^{{\sst(1)}A}+
g^{7/2}\lambda^{{\sst(2)}A}+\cdots\ ,\\
\bar\lambda_A&=g^{1/2}\bar\lambda^{\zo}_A+
g^{5/2}\bar\lambda^{\sst(1)}_A+g^{9/2}\bar\lambda_A^{\sst(2)}+\cdots\ ,\\
\phi_a&=g^0\phi_a^\zo+g^2\phi_a^{\sst(1)}+g^4\phi_a^{\sst(2)}+\cdots\ .
\elabel{whg}
}
In the above,
\EQ{
A_m^\zo=\bar U\partial_m U\ ,\qquad\lambda^{\zo A}=\Lambda(\CM^A)\ .
}
Note in the $\N=1$ theory there are  no scalar fields, hence the
leading order instanton solution,
\EQ{
A_m= g^{-1}A_m^\zo\ , \qquad \lambda^A= g^{-1/2}\lambda^{\zo A}\ , \qquad
\bar\lambda_A =0 \ , \label{whgn1}}
is exact. We will also see shortly that in the absence of scalar VEVs the
leading-order instanton configuration of the $\N=2$ theory,
\EQ{
A_m= g^{-1}A_m^\zo\ , \qquad \lambda^A= g^{-1/2}\lambda^{\zo A}\ , \qquad
\bar\lambda_A =0\ , \qquad \phi_a=g^0\phi_a^\zo
 \ , \label{whgn2}}
is again an exact solution. This will no longer be the case when VEVs
are turned on.
The $\N=4$ case is more subtle, and we will treat it carefully.

When the expansions are substituting into the action, the
latter is
\EQ{
S=
\frac{8\pi^2 k}{g^2}+ik\theta+\tilde S\,g^0+{\cal O}(g^2)\ ,
\elabel{expac}
}
where we have defined
\EQ{
\tilde S=\int d^4x\,
\TrN\,\Big\{{\cal D}_n\phi_a^\zo{\cal D}_n\phi_a^\zo
-\lambda^{\zo A}\bar\Sigma_{aAB}[\phi_a^\zo,\lambda^{\zo B}]
\Big\}\ ,
\elabel{yyip}
}
which will play an important r\^ole in what
follows. So if we wish to work to
leading order in $g$, we only need to solve the equations-of-motion to
order $g^0$.\footnote{The higher order effects are, in any case, mixed
non-trivially with the quantum effects from the fluctuations as we
shall see in \S\ref{sec:S35}. The fluctuations actually also
contribute determinant factors at ${\cal O}(g^0)$. However,
since our theory is supersymmetric this is just a constant.}
{}From \eqref{expac}, it follows we need the expression for
the scalar field to order $g^0$. This is given by the solution of
equation-of-motion \eqref{utd} with
only the source bi-linear in the fermion zero modes included:
\EQ{
{\cal D}^2\phi_a=g\bar\Sigma_{aAB}\lambda^{A}\lambda^{B}\ .
\elabel{gzp}
}
Since $\lambda^A$ is ${\cal O}(g^{-1/2})$ the solution is ${\cal O}(g^0)$.
The solution to \eqref{gzp} is found in Appendix \ref{app:A4}
(Eq.~\eqref{uyt}) with
arbitrary VEV. Setting the VEV to zero, the solution has the form
\cite{MO-I,MO3}
\EQ{
\phi_a =-\tfrac 14\bar\Sigma_{aAB}\bar U\CM^Af\bar\CM^BU+\bar
U\MAT{0_{\sst[N]\times[N]} & 0_{\sst[N]\times[2k]}\\
0_{\sst[2k]\times[N]}&\varphi_{a}1_{\sst[2]\times[2]}}U\ ,
\elabel{ggdd}
}
where the $k\times k$ matrices $\varphi_a$ are
\EQ{
\varphi_a=\tfrac 14\bar\Sigma_{aAB}\BL^{-1}\big(\bar\CM^A\CM^B\big)\ .
\elabel{dll}
}
For the case of $\N=2$, one can see that \eqref{ggdd} implies that the
holomorphic field $\phi$ is non-trivial, while $\phi^\dagger$ remains
zero. In other words, we can see that the lack of a reality condition
on the fermions means that the scalar field in the instanton
background also violates its reality condition. The same is true in
the $\N=4$ theory. However, this violation of the reality condition
only occurs in terms are quadratic in the Grassmann collective
coordinates and so will not affect the convergence of the integrals
over the fluctuations of the scalar field (which will still satisfy the
usual reality condition). The fact that the reality properties of
fields are violated in the instanton background by polynomial factors
in the Grassmann collective coordinates will be a constant feature of
the supersymmetric instanton calculus.

For the applications considered below to evaluating correlation
functions at leading semi-classical order,
it turns out that we need go no further in iterating the
equations-of-motion. However, it is instructive to go one step further
by solving for the anti-chiral fermions at order $g^{1/2}$. At this order
it appears from \eqref{utc} that the source term for the
anti-chiral fermion turns on. However, for the $\N=2$ theories there
is a major simplification compared with the $\N=4$ theories. From
\eqref{gzp} we can see that only the components of $\phi_a$ which are
non-trivial are of the form $\phi_a=\phi_{AB}\bar\Sigma_{aAB}$.
But we recall from \eqref{llzz} for the $\N=2$ theories
\EQ{
\bar\Sigma_{aAB}=\epsilon_{AB}\bar\Sigma_a\
,\qquad\bar\Sigma_a=(-i,1)\ .
}
Consequently
\EQ{
\N=2:\qquad
\bar\Sigma_{aAB}\bar\Sigma_{aCD}=\epsilon_{AB}\epsilon_{CD}\bar\Sigma_a
\bar\Sigma_a=0
\elabel{poo}
}
and so the source term for the anti-chiral
fermions vanishes and the anti-chiral fermion
remain zero. This should be contrasted with the situation in the
$\N=4$ theories where one can show from the definition \eqref{mmzz}
\EQ{
\N=4:\qquad\bar\Sigma_{aAB}\bar\Sigma_{aCD}=2\epsilon_{ABCD}\ .
\elabel{poq}
}
Hence, in these theories the source for $\bar\lambda_A$ is
non-vanishing and the anti-chiral fermions are non-trivial.

Another way to see the difference between the $\N=2$ and $\N=4$
theories is to note that the former has an abelian factor in their
$R$-symmetry group under which the fields have the following charges
\EQ{
q(A_n)=0\ ,\qquad q(\lambda)=1\ ,\qquad q(\bar\lambda)=-1\ ,
\qquad q(\phi)=2\ ,\qquad q(\phi^\dagger)=-2\ .
}
In the instanton background, the charge 2 component $\phi$ is non-trivial,
however, it is $\phi^\dagger$, the charge $-2$ component, that
couples in the source for $\bar\lambda$. In the $\N=4$ case there is
no abelian $R$-symmetry to provide a similar selection rule of this kind.
The existence of the abelian $R$-symmetry in the $\N=2$ theory
means that no other sources are turned on
and so there is fully-fledged supersymmetric generalization of
the bosonic instanton solution which depends on
$4kN$ Grassmann collective coordinates.
The solution has $A_n$, $\lambda^A$ and $\phi_a$ all non-trivial, where
$A_n$ is the original gauge field, $\lambda^A$ are the zero modes of the
Weyl equation and $\phi_a$ is the solution \eqref{ggdd}. It is easy to
verify that this ``supersymmetric instanton'' is degenerate with the
original bosonic instanton and so has action $-2\pi i\tau$. In this
case, therefore, the instanton effective action $\tilde S$ in
\eqref{expac} vanishes, as do all terms higher order in $g$.

The picture in the $\N=4$ theory is much more subtle. Na\"\i vely one
would think that a supersymmetric instanton with $8kN$ Grassmann
collective coordinates exists also in this case,
albeit that it is much harder to find. We will argue that this is not the
case. Remarkably, we will find that most of the Grassmann collective
coordinates, exactly the $8kN-16$ of them, are not true collective
coordinates in the sense of parameterizing a space of solutions
of the classical
equations-of-motion. In view of this, we shall call them {\it
quasi-collective
coordinates\/} and associate to them a {\it quasi-instanton\/}.
They correspond to zero modes of the linearized
system that are lifted by interactions. As a consequence, the
action of the theory evaluated on the instanton solution
\eqref{expac}, actually depends non-trivially on the quasi-collective
coordinate modes. The only ``genuine'' collective
coordinates are those protected by symmetries of the theory.
For the $\N=4$ theory these number sixteen in total,
corresponding to the supersymmetry
and superconformal generators which, as explained in
\S\ref{sec:S30}, act non-trivially on
the instanton. These zero modes are protected by symmetries and cannot
be lifted by interactions. Hence the exact instanton solution of the
equations-of-motion of the $\N=4$ theory
contains only sixteen Grassmann
collective coordinates. The expressions for
the multiplet of fields of this solution
can be found by acting on the bosonic instanton
with a series of supersymmetry transformations
\eqref{lha}-\eqref{lhd}, with the $x$-dependent
Grassmann parameters $\xi^A$ in \eqref{xdep} and with the re-scaling
\eqref{ressp}. Iterating this ``sweeping-out'' procedure to fourth
order yields
\SP{
\lambda^A&=ig^{1/2}\sigma_{mn}\xi^A(x)F_{mn}\ ,\\
\phi_a&=-\tfrac12g\bar\Sigma_{aAB}\xi^A(x)\sigma_{mn}\xi^B(x)F_{mn}\ ,\\
\bar\lambda_A&=\tfrac13ig^{3/2}\epsilon_{ABCD}(\Dbarslash
F_{mn})\xi^B(x)\big[\xi^C(x)\sigma_{mn}\xi^D(x)\big]\ ,\\
A_m&=A_m-\tfrac16g^2\epsilon_{ABCD}\big[\xi^A(x)\sigma_{mn}\xi^B(x)\big]
\big[\xi^C(x)\sigma_{kl}\xi^D(x)\big]{\cal D}_nF_{kl}\ .
\label{sweep}
}
However, this exact instanton solution is not the most convenient
starting point for the semi-classical approximation to the functional
integral. Integrating over the quadratic fluctuations in the background
of this exact solution, one would have to introduce the additional
$8kN-16$ fermion zero modes of the Dirac operator. These modes will
couple to the scalar field fluctuation and in order to lift them
one would have to re-sum tree-level perturbative contributions in the
instanton background. A much faster and more elegant way of addressing
this problem is to modify the instanton background by including
in it from the beginning all $8kN$ fermion zero modes. Thus,
as in Ref.~\cite{MO3},
we will always work with the quasi-instanton configuration in the $\N=4$ theory
which is not an exact solution, but does not require perturbation theory
to lift the quasi-zero modes.

The subtlety of the quasi-instanton
of the $\N=4$ theory appears when we try to solve for the
anti-chiral fermions \eqref{utc}. In Appendix \ref{app:A4}
(Eq.~\eqref{psibardef}), we show that the
right-hand side of the \eqref{utc} can be decomposed as
\EQ{
g\bar\Sigma_{aAB}[\phi_a,\lambda^{B}]=g^{1/2}\big(\Dslash
\bar\psi_A+\Lambda(\N_A)\big)\ .
\elabel{aamm}
}
Here, $\N_A$ is a Grassmann odd function of the collective
coordinates which satisfies the fermionic ADHM
constraints \eqref{zmcona} and \eqref{zmconb}
whose form is written down in Appendix \ref{app:A4},
Eq.~\eqref{calNdef}.\footnote{Actually the Appendix considers the more
general case when the scalar fields have VEVs.}
This means that the second term in \eqref{aamm}
is a zero mode of $\Dbarslash$. The expression for
$\psi_A$ in \eqref{aamm} is also given in
Appendix \ref{app:A4}, Eq.~\eqref{psiansatz}.
Notice that the right-hand side of \eqref{aamm} has a
component in the kernel of $\Dbarslash$. However this kernel
is not in the image of $\Dslash$, in other words there is
no $\bar\lambda_A$ such that $\Dslash
\bar\lambda_A=g^{1/2}(\Dslash
\bar\psi_A+\Xi_A)$. Hence it is impossible to solve this
equation-of-motion. The best that can be done is to set
\EQ{
\bar\lambda_A=g^{1/2}\bar\psi_A\ ,
\label{expaf}
}
which gives the leading order ${\cal O}(g^{1/2})$ expression
for the anti-chiral fermion in the instanton background. In particular,
from the expression for $\bar\psi_A$ \eqref{psiansatz}, this solution
contains a cubic dependence on the Grassmann collective coordinates.

We shall see in \S\ref{sec:S35}, the failure to solve the anti-chiral fermion
equation-of-motion
is a symptom of the fact that, all but the sixteen Grassmann collective
coordinates associated to broken supersymmetric and
superconformal invariance, are lifted by interactions and so are only
{\it quasi-\/}collective coordinates. The leading order effect is
captured by the order $g^0$ term in the action \eqref{yyip}.
Even though an exact supersymmetric instanton does not exist in the
$\N=4$ case, dependent on all $4kN$ Grassmann collective
coordinates, the
approximate solution  $\{A_m,\lambda^{
A},\phi_a\}$, the ``quasi-instanton'',
is sufficient to capture semi-classical contributions
to the leading order  in
$g$. The expression for the anti-chiral fermion is then only needed when one
considers correlations functions with explicit insertions of $\bar\lambda_A$.

\subsection{Scalar VEVs and constrained instantons}\elabel{sec:S33}

In this section, we examine instantons in cases where the scalar
fields have non-vanishing vacuum expectation values. From the outset,
we emphasized that
instantons are a semi-classical phenomenon and therefore are only
expected to describe the physics of these theories in a weakly-coupled
phase. The $\N=4$ theory has a weakly coupled regime, obtained simply by
taking $g$---which does not run---to be arbitrarily small.
In this case the theory
is in a non-abelian Coulomb phase and the semi-classical
approximation is reliable. However, the pure $\N=1$ and $\N=2$
theories (with
vanishing VEVs in the latter) are
strongly interacting. These theories can be rendered weakly coupled by
breaking the gauge symmetry via the Higgs mechanism. Either the
gauge symmetry is broken to an
abelian subgroup, yielding a Coulomb phase, or it is broken
completely, yielding a
Higgs phase. In the $\N=1$ theory this can only be achieved by adding
matter fields, a subject that we will pursue in \S\ref{app:A6} and
\S\ref{sec:S43}. The $\N=2$, like
the $\N=4$, theory has adjoint-valued scalar fields, $\phi_a$,
which can develop a VEV driving
the theory into an abelian Coulomb phase. By taking the VEV to be
large the theory is weakly coupled and semi-classical methods can be
rigorously justified. As mentioned in the introduction to this chapter,
scalar VEVs have an unfortunate
side effect on instantons: strictly speaking they no longer exist!
The way to resurrect them and make sense of the theory in the Coulomb
or Higgs phase was worked out some time ago by Affleck \cite{Affleck} (see also
\cite{ADS}). Instantons are replaced by ``constrained instantons'' in a
rather technically demanding formalism. But it turns out that the
constrained instanton formalism is a paper tiger: working to lowest order
in the semi-classical expansion with constrained instantons involves
only a relatively mild generalization of the instanton calculus and,
moreover, one which has a nice geometrical interpretation in the moduli space
picture. Roughly speaking, instantons are no longer true minima of the
action and a potential develops on $\ms_k$:
instantons now have non-trivial action as well as entropy. In this
sense constrained instanton are examples of the more general notion of
a quasi-instanton that we have already seen in the $\N=4$ theory.

\subsubsection{Constrained instantons on the Coulomb
branch}\elabel{sec:S34}

The $\N=2$ and $\N=4$ theories contain two and six real adjoint-valued
scalar fields, respectively. The classical potentials of these theories have
flat directions along which these scalars develop VEVs breaking the gauge
group to its maximal abelian subgroup by the adjoint Higgs
mechanism. In the case of gauge group
$\SU(N)$ the unbroken subgroup is $\U(1)^{N-1}$.
Without loss of generality, we can simultaneously
diagonalize all the components of the VEVs and we label the elements
\EQ{
\phi_a^0={\rm diag}\Big((\phi_a^0)_1,(\phi_a^0)_2,\ldots,(\phi_a^0)_N\Big)\
,
\label{VEVs}
}
with $\sum_{u=1}^N(\phi_a^0)_u=0$ for tracelessness. Here,
and in the following, the superscript ``$0$'' on a scalar field
denotes a vacuum expectation value. Up to Weyl transformations, these VEVs
parameterize a moduli space of inequivalent vacua known as the
Coulomb branch.

After the Higgs
mechanism has done its work,
the diagonal components of all the fields remain
massless whereas the off-diagonal components gain masses
$|(\phi_a^0)_u-(\phi_a^0)_v|$.
The $\N=2$ theory is asymptotically free with dynamical scale $\Lambda$.
The running of the coupling is cut-off in the IR at a scale set by the
masses of the off-diagonal components. Thus, as long as we choose
VEVs such that $|(\phi_a^0)_u-(\phi_a^0)_v|\gg\Lambda$, for all $u,v$,
the theory is weakly coupled at all length scales and
semi-classical (instanton) methods should be reliable.
In contrast, the coupling constant of the $\N=4$ theory does not run and we
may achieve weak coupling simply by setting $g^{2}\ll1$, either on the
Coulomb branch or at the conformal point where the non-abelian
gauge symmetry is restored.

As we mentioned in the introduction to this Chapter, in the background of
scalar VEVs, instantons are no longer exact solutions of
the equations-of-motion due to Derrick's Theorem \cite{Derrick}. The
action can always be lowered by shrinking an instanton, so the
the size of the configuration cannot be a genuine
modulus when the VEVs are turned on.
The way to implement the semi-classical approximation
around instanton quasi-solutions in theories
with symmetry
breaking was developed by Affleck \cite{Affleck,ADS}. For simplicity,
we review this approach
in the context of a single BPST instanton of scale size $\rho$.
The basic idea is to
introduce a new operator, or ``Affleck constraint'', into the action
by means of a Faddeev-Popov
insertion of unity. If this operator is of suitably high dimension
Derrick's theorem is avoided
and instantons stabilize at a fixed scale size $\rho$.
The integration over the
Faddeev-Popov Lagrange multiplier can then be traded off for the
integration over $\rho$.
The now-stable solutions are known as {\it constrained\/} instantons.
Of course, the detailed
shape of the constrained instanton depends in a complicated
way on one's choice of constraint, but certain important features remain
constraint independent, namely:

(i) The short-distance regime, $x\ll1/(g\phi^0)$.\footnote{Here,
$\phi^0$ is the characteristic scale of the VEVs.}
In this regime the equations-of-motion
can be solved perturbatively in $g^2\rho^2\phi^0$;
since ultimately, as seen {\it ex post facto\/} in \S\ref{sec:S38},
the integration over scale size is dominated by $\rho\lesssim
1/\phi^0$ this is tantamount to
perturbation theory in $g$. As the constraints do not enter
into these equations until some high order, the first few
terms in this expansion are robust. In particular,
to leading order in the semi-classical approximation, the
gauge fields and fermions are equal to their BPST expressions
while the
scalar fields undergo a minor modification to take account of the VEV.

(ii) The long-distance
regime, $x\gg1/(g\phi^0)$. The long-distance ``tail'' of the instanton
reflects the Higgs mechanism.
In the model at hand, the instanton component fields which gain a mass
via the Higgs mechanism decay exponentially.
In contrast, the diagonal components of the fields
fall off as powers of $\rho^2/x^2$. It is an important assumption of
the constrained instanton method that to leading order in the
semi-classical approximation,
the long-range behaviour of these massless fields is simply an
extrapolation of the BPST core.\footnote{In principle,
this simple patching of the short and long distance behaviour is
modified at higher order in a way that is dependent on the precise
form of the Affleck constraint.} So for the massless components the
BPST form of the solution is all that one needs to discuss the
instanton on all length scales (at leading order).
However, even for the massive fields
the BPST form often suffices, even though it is not correct at large
distance. The reason is that the error made in using the BPST form
rather than the actual exponential fall off is higher order in the coupling.

The most important conclusion of the constrained instanton method is
that to leading order,
small constrained instantons are well approximated by ordinary BPST
instantons. But
since these are the ones that are favoured by the now size-sensitive
action of the constrained instanton, little error is made by replacing
the constrained instanton by a conventional instanton.
We will apply the same reasoning to the case of arbitrary topological
charge to find that the core of the required
constrained instanton is more or less the ADHM
instanton that we have constructed:
the gauge field assumes its ADHM form, \eqref{vdef}, and the
fermions $\lambda^A$ are the linear combination of zero modes of
$\Dbarslash$, \eqref{lam}. The scalar field continues to obey
\eqref{gzp} but now the boundary condition on $\phi_a$ is that it must
approach the VEV $\phi_a^0$ at large distance from the instanton.
The general solution with VEV \cite{MO-I} is derived in Appendix
\ref{app:A4} (Eq.~\eqref{uyt})
\EQ{
\phi_a=-\tfrac 14\bar\Sigma_{aAB}\bar U\CM^Af\bar\CM^BU+\bar
U\MAT{\phi_a^0 & 0\\
0&\varphi_a1_{\sst[2]\times[2]}}U\ ,
\elabel{ssdd}
}
with
\EQ{
\varphi_a=\BL^{-1}
\Big(\tfrac 14\bar\Sigma_{aAB}\bar\CM^A\CM^B+\bar w^\aD\phi_a^0w_\aD\Big)\ .
\elabel{dxx}
}
That this has the requisite boundary condition \eqref{VEVs} can be verified
using the asymptotic formulae in \S\ref{sec:S12}.

As long as we work to leading order in $g$,
the Affleck constraint does not explicitly appear and furthermore
there is no need to iterate the instanton solution any further.
Just as in the $\N=4$ theory with zero VEV, the
quasi-instanton solution $\{A_m,\lambda^{
A},\phi_a\}$ is sufficient to
capture the leading order semi-classical approximation of the
functional integral, as we describe in \S\ref{sec:S35}.

\subsection{Collective coordinate supersymmetry}\elabel{sec:S32}

In \S\ref{sec:S30},
we saw that chiral supersymmetry transformations on the bosonic
instanton generate fermion zero modes: these are the supersymmetries
that are broken by the purely bosonic instanton. Once, the fermion zero modes
are turned on, the other half of the
supersymmetry generators, the anti-chiral ones
which left unbroken the bosonic instanton, now act non-trivially on
the super-instanton. It turns out, as we uncover in this section, that
the supersymmetry transformations on the super-instanton
can be traded for supersymmetry transformations on the collective
coordinates themselves. In fact the unbroken supersymmetries of the
ADHM background are linearly realized on the bosonic and
fermionic collective coordinates. This is an example of a general feature
of BPS saturated solitons and corresponding higher-dimensional extended
objects
({\it i.e.\/}~branes). In all of these cases the unbroken supersymmetries
are linearly realized in the world-volume theory. Invariance under these
symmetries is an important constraint which must be satisfied by the
super-instanton measure we will construct in the next section.

To find the transformations, we need to consider the
supersymmetry variation
of the gauge field $A_m$ and fermions $\lambda^A$ in the background of
the non-linear supersymmetric instanton solution. By the latter, we
mean the solution with $A_m$ and
$\lambda^A$ equal to their ADHM forms, \eqref{vdef} and \eqref{lam},
and with the
scalar fields given by \eqref{ssdd}. In the $\N=4$ case, or on the
Coulomb branch of the $\N=2$ theory,
this quasi-instanton
configuration, as we have already described in \S\ref{sec:S31}
and \S\ref{sec:S33}, is only an
approximate solution to the equations-of-motion,
but, nevertheless, provides a convenient way of capturing the leading
order semi-classical contribution to
the functional integral. As a symptom of the non-exactness of the
instanton solution in these cases, supersymmetry
transformations on the fields, as well as transforming the
collective coordinates, also turn on components of the fields at a
higher order in the semi-classical expansion. We have already seen
this phenomena as the ``sweeping-out'' procedure that led to the
expressions for $\phi_a$, $\bar\lambda_A$ and $A_m$ in \eqref{sweep}.

We start by considering the supersymmetry variation of the gauge field
\eqref{lha}. Since the anti-chiral fermions vanish, we have
\EQ{
\delta A_{\alpha\aD}=2ig^{-1/2}\bar\xi_{\aD A}\lambda^A_\alpha
\equiv2g^{-1}\Lambda_\alpha\big(i\bar\xi_{\aD A}\CM^A\big)\ .
}
Here, for convenience we have re-scaled
\EQ{
\bar\xi_A\to g^{-1/2}\bar\xi_A\ .
\elabel{resspb}
}
Comparing this with \eqref{eerr}, the variation of the gauge field up
to a local gauge transformation,
under a variation of the $c$-number collective coordinates,
we deduce the simple rule
\EQ{
\delta a_\aD=i\bar\xi_{\aD A}\CM^A\ ,\qquad\delta\bar a^\aD=i
\bar\xi^\aD_A\bar\CM^A\ .
\elabel{uiui}
}
This leaves the transformations of the Grassmann collective
coordinates which are deduced by considering the variation of the
fermions \eqref{lhb}.
We have already shown how the first term lifts to the collective
coordinates in \S\ref{sec:S30} but for consistency and contrary to
\eqref{resspba}, we should re-scale
\EQ{
\xi^A\to g^{1/2}\xi^A\ .
\elabel{resspba}
}
The second term involves an expression which is higher-order in the
VEVs and Grassmann collective
coordinates of ${\cal O}(g^{3/2})$. In other words this term does
not contribute to $\delta\lambda^{\zo A}$
but rather the next term
$\delta\lambda^{{\sst(1)}A}$ in th expansion \eqref{whg}. This is
example of the sweeping-out procedure turning on a higher-order term
in the semi-classical expansion of a field.
Note with vanishing VEVs and $\N<4$ supersymmetry this term vanishes.
The final term is ${\cal O}(g^{-1/2})$ and in Appendix \ref{app:A4}
(Eq.~\eqref{susst}),
we show how it lifts to a variation of the
Grassmann collective coordinates. Putting the two variations, from the
first and third terms, together, we have
\EQ{
\delta\CM^A=-4i\xi^A_\alpha b^\alpha+
2i\Sigma_a^{AB}{\cal C}_{a\aD}\bar\xi_{B}^\aD\ ,\qquad
\delta\bar\CM^A=-4i\xi^{\alpha A}\bar b_\alpha+
2i\Sigma_a^{AB}\bar\xi_{\aD B}\bar{\cal C}_a^\aD\ ,
\elabel{vivi}
}
where
\EQ{
{\cal C}_{a\aD}=\MAT{\phi_a^0& 0\\ 0 &\varphi_a}a_\aD-a_\aD\varphi_a\
,\qquad \bar{\cal C}_{a}^\aD=\bar a^\aD
\MAT{\phi_a^0& 0\\ 0 &\varphi_a}-\varphi_a\bar a^\aD\ ,
\elabel{defbc}
}
where $\varphi_a$ was defined in \eqref{dxx} in the context of the
scalar field equation. Notice that with the re-scaling \eqref{resspb}
and \eqref{resspba},
all $g$-dependence drops out of \eqref{vivi}.

The anti-chiral supersymmetry transformation also
turn on the anti-chiral fermions at ${\cal O}(g^{1/2})$:
\EQ{
\delta\bar\lambda_A=-ig^{1/2}\Sigma_{abA}{}^B\bar\xi_B[\phi_a,\phi_b]\ .
\elabel{sact}
}
Again, this is a symptom of the fact
that the quasi-instanton is not an exact solution
to the equation-of-motion.\footnote{Except for $\N<4$, with
zero VEVs, in which case \eqref{sact} vanishes.} Nevertheless, the
supersymmetry transformations on the collective coordinates
that we have derived in \eqref{uiui} and
\eqref{vivi} will turn out to be symmetries
of the leading-order semi-classical approximation of the functional
integral.

\rsen\section{The Supersymmetric
Collective Coordinate Integral}\elabel{sec:S35}

This Chapter is a companion to the previous one and in it we consider
how the supersymmetric instanton is used to
implement the semi-classical approximation of the functional integral
in the context of a supersymmetric gauge theory. As we have already
seen, there are extra subtleties in the supersymmetric case
arising from the existence of quasi-zero modes. In this
Chapter we show how the procedure that we adopted to solve the
equations-of-motion is precisely in accord with an expansion in
$g^2$. First of all, in \S\ref{sec:S36} we describe how the
semi-classical approximation of the functional integral leads to a
supersymmetrized version of the collective coordinate integral in
which the Grassmann collective coordinates are also integrated
over. We show that, even though the non-trivial parts of the fluctuation
determinants cancel between bosonic and fermionic
fluctuations, there is still, in general, a non-trivial
integrand which arises as a consequence of
quasi-zero modes. In fact the integrand involves the exponential of
the ``instanton
effective action'', a suitably supersymmetrized potential on the
instanton moduli space. In the case of constrained instantons the
instanton effective action penalizes large instantons but, even when
the VEVs vanish
it is a non-trivial function of the Grassmann collective
coordinates in the $\N=4$ theory. The instanton effective action
is constructed and analysed in \S\ref{sec:S38}. In \S\ref{sec:S385} we
find an expression for the supersymmetric volume form on the
instanton moduli space. As in the pure gauge theory, the
hyper-K\"ahler quotient construction plays a central r\^ole here. Finally,
in \S\ref{sec:S386}, we show how the supersymmetry of the underlying
field theory is manifested on the collective coordinates.

\subsection{The supersymmetric collective coordinate
measure}\elabel{sec:S36}

In order to implement the semi-classical approximation we must expand
around the bosonic instanton solution. In this section, we will
initially consider the
case where the VEVs vanish, so avoiding the complications of the
constrained instanton. The gauge field is expanded as in \eqref{ioa}
and \eqref{lsw} where we recall that the fluctuations are split into
the zero mode piece and the component orthogonal to the zero modes;
the latter being denoted
$\tilde A_n$. The fermions are treated in a similar way: we separate
out the zero modes of $\Dbarslash$ and write the chiral fermion as
\EQ{
\lambda^A(x)=g^{-1/2}\lambda^{\zo A}(x;X,\psi^A)+\tilde\lambda^A(x;X,\psi)\ .
\elabel{ttmk}
}
The $2kN$ Grassmann collective coordinates, for each species,
are denoted by $\psi^A$.
Since the anti-chiral fermions and scalar field have no zero modes in
an instanton background we continue
to denote them as $\bar\lambda_A$ and $\phi_a$.

The functional integral over the fermions can be factorized into
integrals over the Grassmann collective coordinates $\psi^A$ and the
non-zero mode fluctuations; schematically\footnote{The factor of
$g^{kN\N}$ arises from the factor of $g^{-1/2}$ included in the
definition of each fermion zero modes in \eqref{lam}.}
\EQ{
\int\prod_{A=1}^\N\, [d\lambda^A]\,[d\bar\lambda_A]=g^{kN\N}
\int \prod_{A=1}^\N\,\bigg\{\frac{\prod_{\xii=1}^{2kN}d\psi^{\xii A}}{
{\rm
Pfaff}\,\tfrac12\Omega(X)}\  [d\tilde\lambda^A]\,[d\bar\lambda_A]\bigg\}\ .
}
Here, $\Omega(X)$ is the anti-symmetric $2kN\times2kN$ matrix defined
by the functional inner-product of the zero-modes \eqref{ipint}:
\EQ{
\int d^4x\,\TrN\,\lambda^{\zo}(x;X,\psi^A)\lambda^{\zo}(x;X,\psi^B)
=-\tfrac14\delta^{AB}\Omega_{\xii\xjj}(X)\psi^{\xii A}\psi^{\xjj B}
}
and the Pfaffian ensures that the integral is invariant under
re-parameterizations of the Grassmann collective coordinates.

We now plug the expansion of the fields into the
action of theory we obtain
\EQ{
S[g^{-1}A^\zo_m+\tilde A_m,g^{-1/2}\lambda^{\zo
A}+\tilde\lambda^A,\bar\lambda^A,\phi_a]
=-2\pi ik\tau+S_{{\rm kin}}+S_{{\rm int}}\ .
}
Here, the second term denotes the
kinetic terms for the non-zero mode fluctuations:\footnote{In the
following, as previously, all covariant derivatives are defined with
respect to the bosonic instanton solution \eqref{vdef}.}
\EQ{
S_{{\rm kin}}=\int d^4x\,
\TrN\,\Big\{-\tfrac12\tilde{\bar
A}^{\aD\alpha}\Delta^{(+)}{}_\alpha{}^\beta\tilde A_{\beta\aD}-
2{\cal D}_n\bar\lambda_{A}\bar\sigma_n\tilde\lambda^A
+{\cal D}_n\phi_a{\cal D}_n\phi_a\Big\}
\elabel{uuip}
}
and the third term includes the interactions between the zero
and non-zero modes:
\EQ{
S_{{\rm int}}=\int d^4x\,\TrN\Big\{
-\lambda^{\zo A}\bar\Sigma_{aAB}[\phi_a,\lambda^{\zo B}]
-2g^{1/2}[\tilde A_n,\bar\lambda_{A}]\bar\sigma_n\lambda^{\zo A}
-2g^{1/2}\lambda^{\zo A}\bar\Sigma_{aAB}[\phi_a,\tilde\lambda^B]+\cdots
\Big\}\ .
\elabel{qqip}
}
The remaining terms, whose presence is indicated by the
ellipsis are higher order in $g$. We
have also not written down any terms involving the zero-mode piece of
the gauge field because, as we saw in \S\ref{sec:S17},
once the integrals of the
expansion parameters of the zero modes, the
$\xi^\mu$ in \eqref{lsw},
are traded for integrals over the collective
coordinates $X^\mu$, at leading order the $\xi^\mu$ are set
to zero.

We can now integrate over the non-zero modes $\{\tilde
A_m,\tilde\lambda^A,\bar\lambda^A,\phi_a\}$, and ghosts. This defines
a kind of effective action on the collective coordinates
$S_{\rm eff}$, which we call the {\it instanton effective action\/}:
\EQ{
e^{-S_{\rm eff}}\ \overset{\text{def}}=\
e^{2\pi ik\tau}
\int [d\tilde A]\,[db]\,[dc]\,[d\tilde\lambda]\,[d\bar\lambda]\,[d\phi]
\,e^{-S_{\rm kin}-S_{\rm int}-S_{\rm gh}}\ .
}
The instanton effective action is only non-vanishing in the $\N=4$
theory (or more generally in the $\N=1$ and $\N=2$ theories with
non-vanishing VEVs) when the supersymmetric instanton is not an exact
solution to the equations of motion.
We can think of \eqref{uuip} and \eqref{qqip} as specifying a set of Feynman
rules which determines $S_{\rm eff}$ perturbatively in $g$.
Only the non-zero mode pieces of the fields actually propagate
and, in particular, the zero modes of the fermions $\lambda^{\zo A}$ are
non-propagating and act as sources. When we integrate out the
fluctuations the second and third terms
in \eqref{qqip} will not contribute at the leading $g^0$ order since
they are order $g^{1/2}$; only the first
term which is linear in the fluctuation of the scalar field is relevant.
We can include the leading order
effect in an efficient way via a shift of
the scalar field by $\phi_a^\zo$, the solution to the equation
\EQ{
{\cal D}^2\phi_a^\zo=\bar\Sigma_{aAB}\lambda^{\zo A}\lambda^{\zo B}\ .
\elabel{gzzp}
}
This equation is identical to
one which we solved in \S\ref{sec:S31} and yields an
expression for $\phi_a$ which is bi-linear in the Grassmann
collective coordinates. So when working to
leading order in $g$, it is convenient to think of the background
configuration of the instanton as being the multiplet
$\{A_m,\lambda^{A},\phi_a\}$, defined in \eqref{vdef}, \eqref{lam} and
\eqref{ssdd}. Notice that as we explained in the introduction to
Chapter \ref{sec:S25}, the solution for the
scalar field has terms bilinear the Grassmann collective coordinates
which violates the reality condition on the field:
$\phi_a^\dagger$ is no longer the Hermitian conjugate of
$\phi_a$. However, as we explained, there is no inconsistency.

Once we have shifted the scalar field, the fluctuations can be
integrated out. At leading order this yields the usual determinant
factors. The gauge field and ghosts have already been dealt with in
\S\ref{sec:S17}, while for the fermions we obtain
\EQ{
{\rm Pfaff}'\,\MAT{0&\Dslash\\ \Dbarslash
&0}\
\overset{\text{def}}=\ \Big|\det'\MAT{0&\Dslash\\ \Dbarslash
&0}^2\Big|^{1/4}=
\big|\det'\,\Delta^{(+)}\cdot\det\,\Delta^{(-)}\big|^{1/4}\
,
\elabel{flll}
}
for each flavour of fermion and
where $\Delta^{(\pm)}$ are defined in \eqref{rraa} and \eqref{rrbb}.
In \eqref{flll}, the prime, as usual,
indicates that $\Delta^{(+)}$ has zero modes and the determinant must be
taken over the subspace orthogonal to the zero mode space.
Finally, the integrals over the scalar field fluctuation $\tilde\phi_a$ simply
gives an additional factor of $[\det(-{\cal
D}^2)]^{1-\N}$. Putting the determinant factors together we have
\EQ{
\Big|\frac{\det(-{\cal D}^2)}
{\det'\,\Delta^{(+)}}\Big|\cdot\big|\det'\,\Delta^{(+)}
\big|^{\N/4}\cdot\big|\det\,\Delta^{(-)}\big|^{\N/4}\cdot\big|\det(-{\cal
D}^2)\big|^{1-\N}=
\Big|\frac{\det'\Delta^{(+)}}{\det\,\Delta^{(-)}}\Big|^{\N/4-1}\ ,
\label{condet}
}
using \eqref{ddww} valid in an instanton background.

There are two leading-order $g^0$ contributions to $S_{\rm eff}$. The
first contribution arises from evaluating the action of the theory on the
multiplet $\{A_m,\lambda^{A},\phi_a\}$. This is simply the
quantity $\tilde S$ defined in \eqref{yyip}. The second contribution
comes from the determinants \eqref{condet}. Putting these together we have
\EQ{
S_{\rm eff}=-2\pi ik\tau+\Big\{\tilde
S+(\N/4-1)\log\big[\det'\Delta^{(+)}/\det\,\Delta^{(-)}\big]\Big\}g^0
+{\cal O}(g^2)
\label{seffint}
}
As we described in \S\ref{sec:S24}, the ratio of determinants in
\eqref{seffint} is simply a power of the Pauli-Villars mass scale:
\EQ{
\frac{\det'\Delta^{(+)}}{\det\,\Delta^{(-)}}=\mu^{-4Nk}\ .
}
This cancelling of the determinants, up to a constant factor, is
perhaps the biggest simplification that occurs for instantons
in a supersymmetric gauge theory \cite{D'Adda:1978ur}.
Recall that $\tilde S$
is an expression quartic in the Grassmann collective coordinates.
In the $\N=1$ theory, there are no scalar fields, so this term cannot
appear. In the $\N=2$ theory with zero VEVs, this term also vanishes
because the
multiplet $\{A_m,\lambda^{A},\phi_a\}$ is an exact
solution to the equations-of-motion. With zero VEVs, it is only in
the $\N=4$ theory that this contribution to
$\tilde S$ is non-vanishing as we shall find by
explicit evaluation in \S\ref{sec:S38}.

Including the bosonic parts of the functional integral as described in
\S\ref{sec:S17}, the final expression for the leading-order
semi-classical approximation of the functional integral in the
charge-$k$ sector is
\EQ{
\Big(\frac\mu g\Big)^{kN(4-\N)}e^{2\pi ik\tau}
\int\,\frac{\sqrt{\det\,g(X)}}
{\big[{\rm Pfaff}\,\tfrac12\Omega\big]^\N}\,
\bigg\{\prod_{\mu=1}^{4kN}\frac{dX^\mu}{\sqrt{2\pi}}\,
\prod_{A=1}^\N\prod_{\xii=1}^{2kN} d\psi^{\xii A}\bigg\}\
e^{-\tilde S(X,\psi)}\ .
\elabel{fhg}
}

So the final expression for the leading-order semi-classical
approximation of the functional integral in the
$k$-instanton sector is simply a multiple of the supersymmetrized
volume form on the instanton moduli space $\ms_k$ with an integrand
involving the instanton effective action $\tilde S$:
\EQ{
\int[dA]\,[d\lambda]\,[d\phi]\,[db]\,[dc]\,
e^{-S}\Big|_{\text{charge-}k}\ \overset{g\to0}\longrightarrow\
\Big(\frac\mu g\Big)^{kN(4-\N)}e^{2\pi ik\tau}
\int_{\ms_k}\Bomega^{\sst(\N)}e^{-\tilde S}\ ,
\elabel{intm}
}
where we have defined an $\N$ supersymmetric volume form on $\ms_k$:
\EQ{
\int_{\ms_k}\Bomega^{\sst(\N)}\ \overset{\text{def}}=\
\int\,\frac{\sqrt{\det\,g(X)}}{\big[{\rm Pfaff}\,\tfrac12\Omega(X)\big]^\N}
\prod_{\mu=1}^{4kN}\frac{dX^\mu}{\sqrt{2\pi}}\,
\prod_{A=1}^\N\prod_{\xii=1}^{2kN} d\psi^{\xii A}\ .
\elabel{susmes}
}
We will refer to the quantity
\EQ{
{\EuScript Z}_k^{\sst(\N)}=\int_{\ms_k}\Bomega^{\sst(\N)}e^{-\tilde
S}\ ,
\elabel{ipfun}
}
as the {\it instanton partition function\/} since it has the form of a
partition function of a zero dimensional field theory. Later in Chapter
\ref{sec:S49} we will see that for $\N>1$ it can be viewed as
the dimensional reduction of the partition function of a
higher-dimensional field theory. Specifically for $\N=2$, respectively
$\N=4$, the field theory is a two-dimensional, respectively
four-dimensional, $\sigma$-model with $\ms_k$ as the target
space. This point-of-view leads very naturally to the relation of the
instanton calculus to the dynamics of D-branes in string theory
described in \S\ref{sec:S100}.

In the pure $\N=1$ and $\N=2$ theories, the pre-factors of the collective
coordinates measures \eqref{intm} can be related to the
renormalization group invariant $\Lambda$-parameters. The point is
that the coupling $g$ must run with the Pauli-Villars mass scale in
such a way that the combination in front of the measure is a
renormalization group invariant. This defines the $\Lambda$-parameters
in the Pauli-Villars scheme:
\EQ{
\Lambda_{\sst\N=1}^{3N}=\mu^{3N}g(\mu)^{-2N}e^{-8\pi^2/g(\mu)^2+i\theta}\
,\qquad\Lambda_{\sst\N=2}^{2N}=\mu^{2N}
e^{-8\pi^2/g(\mu)^2+i\theta}\ .
\elabel{deflp}
}
(Notice that the powers of $g$ in \eqref{intm} and \eqref{deflp} do
not match. The additional powers of $g$ come from the insertions when
one calculates correlations functions.)
On the contrary, the $\N=4$ theory is finite, the coupling does not
run and as a consequence the integration measure \eqref{intm}
is independent of the Pauli-Villars mass scale $\mu$.

\subsection{The instanton effective action}\elabel{sec:S38}

In this section, we evaluate the leading order contribution to the
instanton effective action $\tilde S$ in \eqref{yyip}.
When the scalar fields have a non-trivial VEV, the semi-classical
approximation proceeds via Affleck's
constraint method as explained in \S\ref{sec:S33}.
In this case, the scale sizes of instantons cease to be true
moduli and the instanton effective action will be a non-trivial function on
$\ms_k$ which breaks superconformal invariance.\footnote{In the $\N=4$
case, this is in addition to the non-trivial instanton effective action that is
present even when the VEVs vanish.} However, the now {\it
quasi\/}-collective coordinates are
still to be integrated over in the semi-classical approximation of the
functional integral. Furthermore,
the Affleck constraint, at leading order, does not explicitly appear
and further discussion of it is unnecessary.
To leading order, the net effect of introducing the VEV is to change
the boundary condition on $\phi_a$, as indicated in \eqref{ssdd}
and this will feed into the instanton effective action $\tilde S$ in a
way we now calculate.

With an integration by parts together with the scalar
equation-of-motion, Eq.~\eqref{yyip} may be re-cast as\footnote{In the
following $\{A_m,\lambda^A,\phi_a\}$ take the ADHM expressions
\eqref{vdef}, \eqref{lam} and \eqref{ssdd}.}
\EQ{
\tilde S =\int d^4x\,\Big\{\partial_n
\TrN\,\big(\phi_a{\cal D}_n\phi_a\big)-\tfrac12g\trN\,
\lambda^{A}
\bar\Sigma_{aAB}[\phi_a,\lambda^{B}]\Big\}\ .
\elabel{bbxx}
}
The first term, being a total derivative may
be converted to a surface integral over the sphere
at infinity in spacetime. Since it is gauge invariant we can evaluate it in any
convenient gauge. In particular, in singular gauge
defined in \S\ref{sec:S12}, Eqs.~\eqref{asymadhm} and \eqref{ssdd}
imply, in the limit of large $x$,
\EQ{
\frac{x_n}{x}{\cal
D}_n\phi_a\ \overset{x\to\infty}\longrightarrow\
\frac1{x^3}\big(\tfrac12\bar\Sigma_{aAB}
\mu^A\bar\mu^B+w_\aD\bar w^\aD\phi_a^0+\phi^0_aw_\aD\bar w^\aD
-2w_\aD\varphi_a\bar w^\aD\big)\ .
}
Hence, the first term in \eqref{bbxx} is
\EQ{
4\pi^2{\rm tr}_k\Big[\tfrac14\bar\Sigma_{aAB}\bar\mu^A\phi_a^0\mu^B
+\bar w^\aD\phi_a^0\phi_a^0w_\aD-\bar
w^\aD\phi_a^0w_\aD\varphi_a\Big]\ .
\elabel{gtty}
}
The second term can be evaluated by using the identity
\eqref{psibardef} in Appendix \ref{app:A4}:
\EQ{
\bar\Sigma_{aAB}[\phi_a,\Lambda(\CM^B)]=
\Dslash\bar\psi_A+\Lambda(\N_A)\ .
}
Then
\EQ{
-\tfrac1{2}\int d^4x\,\trN\Lambda(\CM^A)\Big(
\Dslash\bar\psi_A+\Lambda(\CN_{A})\Big)=
-\tfrac1{2}\int d^4x\,\Big(\partial_n\trN\,\Lambda(\CM^A)
\sigma_{n}\bar\psi_A+\trN\,\Lambda(\CM^A)
\Lambda(\N_A)\Big)\ .
\elabel{tyw}
}
In the first term, we have used the fact $\Dbarslash
\lambda^A=0$ to pull the derivative outside the trace. One can
verify that for large $x$, $\bar\psi\sim x^{-2}$ and $\lambda\sim
x^{-3}$; hence, the first term on the right-hand side of \eqref{tyw}
gives a vanishing contribution at infinity and may be dropped. The
second term can be evaluated using the inner-product formula
\eqref{corriganf}:
\SP{
-\tfrac1{2}\int d^4x\,
\trN\,\Lambda(\CM^A)
\Lambda(\N_A)&=-\frac{\pi^2}{4}{\rm tr}_k\big[\bar\CM^A(\Pinfty+1)\N_A+
\bar\N_A(\Pinfty+1)\CM^A\big]\\
&=\pi^2\bar\Sigma_{aAB}{\rm
tr}_k\Big[\bar\mu^A\phi^0_a\mu^B-\bar\CM^A\CM^B\varphi_a\Big]
\ ,
\elabel{htty}
}
where we used the expression for $\N_A$ in \eqref{calNdef}
Appendix \ref{app:A4}:
\EQ{
\N_A=-\bar\Sigma_{aAB}\Big\{\,
\begin{pmatrix}\phi_a^0&0\\0&\varphi_a
\end{pmatrix}{\cal M}^B-{\cal M}^B\varphi_a\,\Big\}
+2\begin{pmatrix}0&0\\0&\G^\dalpha_A\end{pmatrix}a_\dalpha-
2a_\dalpha\G^\dalpha_A\ .
}
Here, $\G^{\aD A}$ is a quantity needed to ensure that $\N_A$ satisfy
the fermionic ADHM constraints \eqref{fadhm}, but which, as one can show by
explicit substitution, does not contribute to \eqref{htty}.

Assembling all the pieces, the leading-order contribution to the
instanton effective action
is
\SP{
\tilde S &=4\pi^2{\rm tr}_k\Big\{\tfrac12\bar\Sigma_{aAB}
\bar\mu^A\phi_a^0\mu^B
+\bar w^\aD\phi_a^0\phi_a^0w_\aD-\varphi_a\BL\varphi_a\Big\}\\
&=4\pi^2{\rm tr}_k\Big\{\tfrac12\bar\Sigma_{aAB}
\bar\mu^A\phi_a^0\mu^B+\bar w^\aD\phi_a^0\phi_a^0w_\aD\\
&-\Big(\tfrac14\bar\Sigma_{aAB}\bar\CM^A\CM^B+\bar
w^\aD\phi_a^0w_\aD\Big)
\BL^{-1}\Big(\tfrac14\bar\Sigma_{aCD}\bar\CM^C\CM^D+\bar
w^\bD\phi_a^0w_\bD\Big)\Big\}\ .
\elabel{yyvv}
}
One can verify explicitly that the expression above is invariant under
the supersymmetry transformations \eqref{uiui} and \eqref{vivi}.
Notice in the $\N=2$ theory when the VEVs vanish $\tilde S =0$.
However, in the $\N=4$ theory with vanishing VEVs, $\tilde S$
remains a non-trivial expression quartic in the Grassmann
collective coordinates:
\EQ{
\tilde S =-\frac{\pi^2}{2}\epsilon_{ABCD}{\rm
tr}_k\big(\bar\CM^A\CM^B\BL^{-1}\bar\CM^C\CM^D\big)\ .
\elabel{rrww}
}
This reflects the fact that in the $\N=4$ theory the super-instanton is not an
exact solution of the equations-of-motion.

\subsubsection{Geometric interpretation}\label{sec:S401}

The terms in the instanton effective \eqref{yyvv} have an elegant
interpretation in terms  of the geometry of $\ms_k$. To start with, the
quartic coupling is
precisely the coupling of the Grassmann-valued symplectic tangent
vectors $\CM^A$ to the symplectic curvature of $\ms_k$:
\EQ{
\tilde S
=\frac1{96}\epsilon_{ABCD}R(\CM^{A},\CM^{B},\CM^{C},\CM^{D})
=\frac1{96}\epsilon_{ABCD}R_{\xii\xjj\xkk\xll}
\psi^{\xii A}\psi^{\xjj B}\psi^{\xkk C}\psi^{\xll D}\ .
}
The second expression is written in terms of the intrinsic coordinates
$\psi^A$.
Here, $R$ is the symplectic curvature of the hyper-K\"ahler quotient space
$\ms_k$. It is a totally symmetric tensor in the $\Sp(n)$
indices. In Appendix \ref{app:A2}
we derive a general formula \eqref{gju} for the
symplectic curvature of a hyper-K\"ahler quotient in terms of the curvature
and connection of the mother space $\tilde\ms$.
Using the relation \eqref{spp} and the
explicit representation of the generators of the $\U(k)$ symmetry
in Eq.~\eqref{qqp}, we find
\EQ{
\CM^{\ii A}(\tilde\Omega T^r)_{\ii\,\jj}\CM^{\jj B}
\equiv-4\pi^2i{\rm tr}_k\,T^r\big(\bar\CM^A\CM^B-\bar\CM^B\CM^A\big)\ .
\elabel{gitup}
}
Hence, using the formula for the symplectic curvature \eqref{gju} (in
terms of the ADHM variables $\CM^{\ii A}=\CM^\ii(\psi^A,X)$) and
the definition of $\BL_{rs}$ in Eq.~\eqref{gwww}, we have
\SP{
&\frac1{96}\epsilon_{ABCD}R_{\xii\xjj\xkk\xll}\psi^{\xii
A}\psi^{\xjj B}\psi^{\xkk C}\psi^{\xll D}\\
&=\frac1{16}\epsilon_{ABCD}\sum_{rs}
\Big[\CM^{\ii A}(\tilde\Omega T^r)_{\ii\,\jj}\CM^{\jj B}
\Big]\BL^{-1}_{rs}
\Big[\CM^{\kk C}(\tilde\Omega T^s)_{\kk\,\llr}\CM^{\llr D}\Big]\\
&=-\frac{\pi^2}{2}\epsilon_{ABCD}{\rm tr}_k\big(\bar\CM^A\CM^B\BL^{-1}
\bar\CM^C\CM^D\big)\ .
}
which is precisely \eqref{rrww}.

The pieces of the instanton effective
action which do not depend on the Grassmann
collective coordinates also have a very
concrete geometrical interpretation in terms of the hyper-K\"ahler
quotient construction. The first observation is that the $\SU(N)$ gauge
symmetry of the theory acts as a group of isometries on the
hyper-K\"ahler quotient space. Firstly, consider the $\SU(N)$ isometries
generated by the Killing vectors
\EQ{
\tilde V=i\bar w^\aD_{iu}\PD{}{\bar w^\aD_{iv}}-i
w_{ui\aD}\PD{}{w_{iv\aD}}
\elabel{defsun1}
}
on the mother space $\tilde\ms$. This group action descend to act as
isometries on the quotient $\ms_k$. To see this, it suffices to notice
that the moment maps are invariant: $\tilde V\mu^{X_r}=0$. In
fact the group action is also holomorphic with respect to each of the
three independent complex structures, ${\cal L}_{\tilde V}\tilde\BI^{(c)}=0$, a
property that can be easily be shown to be 
inherited by the action on the quotient. Consequently
the vector fields generated by the group action are {\it tri-holomorphic\/}.

Now consider a set of such isometries associated to
$\U(1)^{N-1}\subset\U(N)$ picked out by the VEVs of the scalar field:
\EQ{
\tilde V_a=i\bar w^\aD_{iu}(\phi_a^0)_u\PD{}{\bar w^\aD_{iu}}-i(\phi_a^0)_u
w_{ui\aD}\PD{}{w_{iu\aD}}\ .
\elabel{defsun2}
}
This action, as described above is tri-holomorphic on the quotient
space $\ms_k$. The lifts of the
Killing vectors on the quotient $\ms_k$, $V_a$, are equal to projection of the
$\tilde V_a$ to ${\EuScript H}\subset T\ns$.
The bosonic parts of the action \eqref{yyvv} are then equal to a sum
over the inner-products of the vectors $V_a$:
\EQ{
4\pi^2{\rm tr}_k\Big[\bar
w^\aD\phi_a^0\phi_a^0w_\aD-\bar
w^\aD\phi_a^0w_\aD\BL^{-1}\bar
w^\aD\phi_a^0w_\aD\Big]=\tfrac12g_{\mu\nu}(X)V_a^\mu V_a^\nu\ ,
\elabel{xxxv}
}
where $g$ is the metric on the quotient space. As explained in
\S\ref{sec:S10} (and more fully in Appendix \ref{app:A2})
the metric on the quotient space $g(X,Y)$ is equal to
$\tilde g(X,Y)$, the metric on the mother space evaluated on the lifts of $X$
and $Y$. Hence,
\EQ{
g(V_a,V_a)=\tilde g(V_a,V_a)=
\tilde g(\tilde V_a,\tilde V_a)-\tilde g(\tilde V_a^\perp,\tilde V_a^\perp)\ ,
}
where $\tilde V_a^\perp$ is the projection of $\tilde
V_a$ to the vertical subspace $\EuScript V$. 
The $\U(1)^{N-1}$ group action on $\tilde\ms$ is
tri-holomorphic,
hence $\tilde g(\tilde V_a,\BI^{(c)}X_r)=0$, so that $\tilde V_a$ already
lies in the tangent space of the level set $T\ns\subset T\tilde\ms$.
Since the vectors $X_r$, $r=1,\ldots,k^2$, form a basis for the
vertical subspace,
\EQ{
\tilde V^\perp_a=\sum_{rs}X_r\BL_{rs}^{-1}\tilde g(X_s,\tilde V_a)
\elabel{perpv}
}
and so
\EQ{
g(V_a,V_a)=\tilde g(\tilde V_a,\tilde V_a)-
\sum_{rs}\tilde g(\tilde V_a,X_r)\BL^{-1}_{rs}\tilde g(X_s,\tilde V_a)\ .
\elabel{kioo}
}
The matrix of inner-products $\BL_{rs}$ is given in \eqref{gwww}. By
explicit calculation we have
\EQ{
\tilde g(\tilde V_a,\tilde V_a)=8\pi^2{\rm tr}_k\big(\bar
w^\aD\phi_a^0\phi_a^0w_\aD\big)\ ,\qquad
\tilde g(\tilde V_a,X_r)=8\pi^2{\rm tr}_k\big(T^r\bar
w^\aD\phi_a^0w_\aD\big)\ .
\elabel{ipofv}
}
Substituting these expressions in \eqref{kioo} gives \eqref{xxxv}.

The remaining terms in \eqref{yyvv} involving the Grassmann collective
coordinates can also be given a geometric interpretation. First of
all, we recall from \S\ref{sec:S29} that, from a geometric
perspective, the Grassmann collective coordinates are Grassmann-valued
symplectic tangent vectors to the instanton moduli space. The action
of the $\U(1)^{N-1}$ symmetry on the Grassmann collective coordinates
generated by the vector fields $\tilde V_a$ in \eqref{defsun2} is simply
\EQ{
\delta_a\mu^A_{ui}=i(\phi^0_a)_u\mu_{ui}^A\ ,\qquad
\delta_a\bar\mu^A_{iu}=i\bar\mu^A_{iu}(\phi^0_a)_u\ ,\qquad
\delta_a\CM^{\prime A}_{ij\alpha}=0\ ,
\elabel{vargcc}
}
which defines $\delta_a\CM^A$. This variation can be written in a
completely geometric way using the (flat) connection on the mother space:
\EQ{
\delta_a\CM^{\ii A}=\big(\tilde\nabla_{\jj\aD}\,\tilde
V^{\ii\aD}_a\big)\CM^{\jj A}\ .
}
We now show that the Grassmann terms in the instanton effective action
\eqref{yyvv} can be interpreted in terms of the intrinsic geometry of
$\ms_k$ as
\EQ{
2\pi^2\bar\Sigma_{aAB}{\rm tr}_k\Big\{\bar\mu^A\phi^0_a\mu^B-\bar
w^\aD\phi^0_aw_\aD\BL^{-1}\bar\CM^A\CM^B\Big\}
=\tfrac i4\bar\Sigma_{aAB}\Omega_{\xii\xjj}(X)\psi^{\xii
A}\big(\nabla_{\xkk\aD} V_a^{\xjj\aD}\big)\psi^{\xkk B}\ .
\elabel{whatww}
}
This follows by proving
\EQ{
\Omega_{\xii\xjj}(X)\psi^{\xii
A}\big(\nabla_{\xkk\aD} V_a^{\xjj\aD}\big)\psi^{\xkk
B}=\tilde\Omega_{\ii\,\jj}\CM^{\ii A}\big(\tilde\nabla_{\kk\aD}(\tilde
V_a-\tilde V_a^\perp)^{\jj\aD}\big)\CM^{\kk B}
}
and then,
from \eqref{jws}, \eqref{qqp}, \eqref{spp} and \eqref{vargcc},
\SP{
&\bar\Sigma_{aAB}\tilde\Omega_{\ii\,\jj}\CM^{\ii A}\big(\tilde\nabla_{\kk\aD}\tilde
V_a^{\jj\aD}\big)\CM^{\kk B}=-8i\pi^2\bar\Sigma_{aAB}{\rm
tr}_k\,\bar\mu^A\phi^0_a\mu^B\ ,\\
&\bar\Sigma_{aAB}\tilde\Omega_{\ii\,\jj}\CM^{\ii A}\big(\tilde\nabla_{\kk\aD}
\tilde V_a^{\perp\jj\aD}\big)\CM^{\kk B}
=-8i\pi^2\bar\Sigma_{aAB}{\rm
tr}_k\,\bar w^\aD\phi_a^0w_\aD\BL^{-1}\bar\CM^A\CM^B\ .
\elabel{gituq}
}
Putting this together with \eqref{ipofv}, and using \eqref{gwww}, we
have proved \eqref{whatww}.

To summarize, the instanton effective action
\eqref{yyvv} can be written in an elegant way involving only the intrinsic
geometry of $\ms_k$:
\EQ{
\tilde S
=\tfrac1{2}\Big\{g_{\mu\nu}(X)V_a^\mu V_a^\nu+\tfrac
i2\bar\Sigma_{aAB}\Omega_{\xii\xjj}(X)\psi^{\xii A}
\big(\nabla_{\xkk\aD}V_a^{\xjj\aD}\big)\psi^{\xkk B}
+\tfrac1{48}\epsilon_{ABCD}R_{\xii\xjj\xkk\xll}
\psi^{\xii A}\psi^{\xjj B}
\psi^{\xkk C}\psi^{\xll D}\Big\}\ .
\elabel{geoia}
}
We will see in \S\ref{sec:S49} that this effective action can be
obtained from a non-trivial 
dimensional reduction of a $\sigma$-model in two and six
dimensions, for $\N=2$ and $\N=4$, respectively, with the instanton
moduli space $\ms_k$ as target.

\subsubsection{The size of a constrained instanton}\elabel{sec:S37}

Having derived the leading order expression for the instanton
effective action, \eqref{yyvv} or \eqref{geoia},
we can, {\it ex post facto\/}, justify the main assumption of the
constrained instanton method that there is an effective cut-off on
large instanton sizes. First of all, since the bosonic part of the instanton
effective action is simply \eqref{xxxv}, we can identify the minima of $\tilde
S$ as fixed points of the $\U(1)^{N-1}$ action on the instanton moduli
space $\ms_k$. Since the construction of $\ms_k$ involves a quotient by
$\U(k)$, the fixed-point condition is
\EQ{
\big(\phi_a^0\big)_uw_{ui\aD}=w_{uj\aD}(\chi_a)_{ji}\ ,\qquad
[a'_n,\chi_a]=0\ ,
\elabel{fpcond}
}
in addition to the ADHM constraints \eqref{badhm}.
Here, $\chi_a$ are $k\times k$ Hermitian matrices
acting as infinitesimal
compensating transformation in the auxiliary group
$\U(k)$ which can depend on the VEVs $\phi^0_a$. For generic values of
the VEVs the solution of the equations
\eqref{fpcond} are as follows. At least until we impose the ADHM constraints,
each fixed-point set is associated to the partition
\EQ{
k\to k_1+k_2+\cdots+k_N\ ,
\label{partit}
}
up to the $\U(k)$ auxiliary symmetry.
Each $i\in\{1,2,\ldots,k\}$ is then
associated to a given $u$ by a map $u_i$ as follows:
\SP{
&\Big\{\underbrace{1,2,\ldots,k_1}_{u=1},
\underbrace{k_1+1,\ldots,k_1+k_2}_{u=2},\ldots,\\
&\qquad\ldots,\underbrace{k_1+\cdots+k_{u-1}+1,
\ldots,k_1+\cdots+k_u}_u,\ldots,\ldots,
\underbrace{k_1+\cdots+k_{N-1}+1,\ldots,k}_{u=N}\Big\}
}
For a given partition the variables have a block-diagonal form
\EQ{
(\chi_a)_{ij}=-(\phi_a^0)_{u_i}\delta_{ij}\ ,\qquad
w_{ui\aD}\propto \delta_{uu_i}\ ,\qquad
(a'_n)_{ij}\propto\delta_{u_iu_j}\ .
\label{critp}
}
Now we impose the ADHM
constraints \eqref{badhm}. In the $u^{\rm th}$ block the
constraints are of the
form of a set of ADHM constraints for $k_u$ instantons for which $N=1$.
In fact the
$N$ blocks each correspond to the $N$ $\U(1)$ subgroups of the gauge
group picked out by the VEVs.\footnote{Actually this statement is not
quite correct because we are taking a gauge group $\SU(N)$. However,
the ADHM construction actually yields a $\U(N)$ gauge potential, where
the abelian part of the gauge group is pure gauge: in this sense the
statement is correct.}
Taking the trace of the ADHM constraints within the block removes the $a'_n$
dependent terms to leave equations of the form
\EQ{
\vec\tau^{\aD}{}_\bD\sum_{i=k_{u-1}+1}^{k_u}\bar w^{\bD}_{iu}w_{ui\aD}=0\ .
\label{qeer}
}
However, the solution to these equations is
$w_{ui\aD}=0$. Therefore the structure of the partitions collapses to
leave, up to the $\U(k)$ symmetry, $a'_n$ and $\chi_a$ diagonal:
\EQ{
w_{ui\aD}=0\ ,\qquad a'_n=-{\rm diag}\big(X_n^1,\ldots,X_n^k\big)\ .
}
Taking the solution above fixes all of the auxiliary symmetry apart
from permutations; hence, the fixed-point space is simply ${\rm
Sym}^{k}\,{\mathbb R}^4$, the
symmetric product of $k$ points in ${\mathbb R}^4$. So the
fixed-point set describes a configuration where all the instantons has
shrunk down to zero size. It is easy to
verify that the gauge potential on this singular subspace is pure
gauge. As the sizes of the instantons grow, the
effective action favours instantons with sizes
up to a scale $\sim 1/\phi^0$, where
$\phi^0$ is the characteristic scale of the VEVs, after which they
are exponentially suppressed. Therefore, constrained instantons have a
natural cut-off on the their scale size.

\subsubsection{The lifting of zero modes}\label{sec:S384}

Of paramount importance for applications is the process of lifting
fermion zero modes encoded in the instanton effective action. First of
all, consider the case with vanishing VEVs. In this case, the only
genuinely weakly-coupled scenario is the $\N=4$ theory. As we have
emphasized in \S\ref{sec:S31}, only the 8 supersymmetric and
8 superconformal zero modes are exact, the remaining $8kN-16$ are
lifted by interactions and consequently we expect the instanton
effective action $\tilde S$ to depend on all the Grassmann collective
coordinates except \eqref{susymo} and \eqref{suconmo}. It is easily
verified from \eqref{rrww} (using the fermionic ADHM constraints
\eqref{fadhm}) that the latter sixteen variables decouple from
$\tilde S$ as expected. There is another way to phrase this result.
The decoupling of the supersymmetric and superconformal Grassmann
collective coordinates implies that
the symplectic curvature of $\ms_k$ admits four null eigenvectors:
\EQ{
R_{\xii\xjj\xkk\xll}\ell^\xii=0\ .
}
The fact that the symplectic curvature has four null eigenvectors implies
that the holonomy group of $\ms_k$ is reduced from $\Sp(kN)$ to
$\Sp(kN-2)$ (or for $\widehat\ms_k$ from $\Sp(kN-1)$ to $\Sp(kN-2)$)
\cite{Vandoren:2000qr,deWit:2000fp}.

When VEVs are turned on, it is easy to see that the instanton
effective action now lifts the superconformal zero modes via the first
term in \eqref{yyvv}, as one
expects since conformal invariance is broken. Of course the
supersymmetric zero modes remain unlifted because the introduction of 
VEVs does not break supersymmetry.

\subsection{The supersymmetric volume form on $\ms_k$}\elabel{sec:S385}

In this section, we show how to construct a volume form on the
space of collective coordinates of a supersymmetric instanton,
generalizing the result \eqref{bmes} in the pure gauge theory. We shall
adopt an approach based on the hyper-K\"ahler quotient construction,
but we shall find a result that is identical with the original
approach of Refs.~\cite{measure1,DHKM,KMS} which relied on various consistency
conditions, principally supersymmetry and clustering, in order to
construct the volume form.

We have already seen how the Grassmann collective
coordinates arise in the context of
the hyper-K\"ahler quotient construction. The
quantities $\{\mu,\bar\mu,\M_\alpha\}$ can be arranged \eqref{spp} as an
$n$-vector $\CM^\ii$ ($n=2k(N+k)$).
The fermionic ADHM constraints are precisely the conditions for
$\CM^{\ii}$ to be a symplectic tangent vector to the
hyper-K\"ahler quotient space. Using this fact it
straightforward to write down the integration measure for the
Grassmann collective coordinates. As for the volume form itself, the
integration measure on the quotient $\ms_k$ is induced from that on the mother
space $\tilde\ms$. The covariant expression for this latter quantity
is
\EQ{
\int\ \frac{\prod_{\ii=1}^{2n}d\CM^\ii}{\text{Pfaff}\,\tilde\Omega}\ ,
}
where $\tilde\Omega_{\ii\,\jj}$ is the anti-symmetric symplectic tensor
on the mother space \eqref{jws}. We
restrict to the quotient space by inserting explicit Grassmann-valued
$\delta$-functions to impose the
symplectic tangent vector condition \eqref{yrr}.
These $\delta$-functions must be accompanied by a suitable Jacobian
giving
\EQ{
\int\,\frac{\prod_{\ii=1}^{2n}d\CM^\ii}{\text{Pfaff}\,
\tilde\Omega}\,\frac1{J_f}
\,\prod_{r=1}^{\text{dim}\,G}\,\prod_{\aD=1}^2\,
\delta\big(\CM^\ii\tilde\Omega_{\ii\,\jj}X_r^{\jj\aD}\big)\ .
}
The Jacobian $J_f$ is related to the determinant
of $\BL$, \eqref{gwww},
the matrix of inner products of the normal vectors to the
quotient space:\footnote{The following can be deduced from Eq.~\eqref{stvr}.}
\EQ{
J_f=\big\vert\det_{k^2}\,\BL\big\vert\ .
}

For the particular quotient that yields the ADHM construction $\tilde\ms$ is
flat, and $\text{Pfaff}\,\tilde\Omega$ is simply a numerical factor and
we can write the measure for each species of fermion
as
\EQ{
\int d^{2k(N+k)}\CM\,\big\vert\det_{k^2}\,\BL\big|^{-1}
 \,\prod_{r=1}^{k^2}\, \prod_{\aD=1}^{2}\,\delta\big({\rm
tr}_k\,T^r(
\bar{\cal M}a_\aD+\bar a_\aD{\cal M})\big)\ .
\elabel{intsf}
}
where\footnote{Our convention
for integrating a two-component spinor 
$\psi_\alpha$ is $\int d^2\psi\equiv\int d\psi_1\,d\psi_2$.}
\EQ{
\int\,d^{2k(N+k)}\CM\ \overset{\text{def}}=\
\int\,\prod_{r=1}^{k^2}d^2(\M)^r\,\prod_{i=1}^k\prod_{u=1}^N
d\bar\mu_{iu}\,d\mu_{ui}\ .
}

Putting this together with volume form for $\ms_k$, \eqref{bmes},
the collective
coordinate volume form for an arbitrary supersymmetric theory is
\SP{
&\int_{\ms_k}\Bomega^{\sst(\N)}
=\frac{C_k^{\sst(\N)}}{{\rm Vol}\,\U(k)}\int\, d^{4k(N+k)} a \,
\,\prod_{A=1}^\N\,
 d^{2k(N+k)}{\cal M}^{A}\ \big|\det_{k^2}\,\BL\big|^{1-\N}\\
&\qquad\times\,\prod_{r=1}^{k^2}\,\bigg\{
\prod_{c=1}^3\,
\delta\big(\tfrac12{\rm
tr}_k\,T^r(\tau^c{}^\aD{}_\bD \bar a^\bD a_\aD)\big)
\prod_{A=1}^\N\prod_{\aD=1}^2\,\delta\big({\rm
tr}_k\,T^r(
\bar{\cal M}^Aa_\aD+\bar a_\aD{\cal M}^A)\big)\bigg\}\ .
\elabel{fms}
}
The normalization factor $C_k^{\sst(\N)}$ can be determined by a
careful analysis of the 
inner products of the zero modes:
\EQ{
C_k^{\sst(\N)}=2^{-k(k-1)/2+kN(2-\N)}\pi^{2kN(1-\N)}\ .
\elabel{sdl}
}

An independent check of the normalization constants $C_k^{\sst(\N)}$ can
be achieved, as in \S\ref{sec:S22}, by invoking the clustering property of the
instanton integration measure.
In order to apply this argument we have to consider how
the Grassmann integrals behave in the complete clustering limit. In
this limit, the off-diagonal components of the fermionic ADHM
constraints are dominated by the term
\EQ{
(\bar X^i-\bar X^j)^{\aD\alpha}\M_\alpha+\cdots\ .
}
We can therefore use the off-diagonal constraints to saturate the
$(\M_\alpha)_{ij}$, $i\neq j$, integrals. This yields a factor of
\EQ{
\prod_{i\neq j}(X^i-X^j)^2\ ,
}
for each species of fermion. However, each species of fermion is
accompanied by a factor of $|\det_{k^2}\BL|^{-1}$ which clusters as
\eqref{gml}. Hence, what remains in the complete clustering limit are the
diagonal fermionic ADHM constraints which to leading order are the
fermionic ADHM of the individual instantons. Hence, taking into
account the clustering of the purely bosonic parts of the volume form,
as in \S\ref{sec:S22}, we find that the supersymmetric measures clusters
consistently.

\subsubsection{Supersymmetry}\elabel{sec:S386}

In this section, we will verify that the volume form Eq.~\eqref{fms} is
invariant under the supersymmetry transformations acting on the
collective coordinates that we established in \S\ref{sec:S32}.

To begin with, consider the supersymmetry variations of the
bosonic and fermionic ADHM constraints, \eqref{badhm} and
\eqref{fadhm}, since these appear as the argument of the
$\delta$-functions in \eqref{fms}.
First, the $c$-number ADHM constraints \eqref{badhm}:
\EQ{
\delta\big(\vec\tau^\bD{}_\aD\bar a^\aD a_\bD\big)=
\vec\tau^\bD{}_\aD\big(-i\bar\xi^\aD_A\bar\CM^Aa_\bD+i\bar\xi_{\bD
A}\bar
a^\aD\CM^A\big)=-i\vec\tau^\bD{}_\aD\xi^\aD_A\big(\bar\CM^Aa_\bD+\bar
a_\bD\CM^A\big)\ .
}
In other words, the bosonic ADHM constraints transform into a linear
combination of the fermionic ADHM constraints \eqref{fadhm}. Now the
fermionic ADHM constraints themselves:
\SP{
\delta(\bar\CM^Aa_\aD+\bar
a_\aD\CM^A\big)&=4i\xi^A_\alpha\big(b^\alpha a_\aD-\bar a_\aD b^\alpha\big)
+i\bar\xi_{\aD B}\big(\bar\CM^A\CM^B-\bar\CM^B\CM^A\big)+2i\bar\xi_{\bD
B}
\Sigma^{AB}_a\big(\bar{\cal C}^\bD_aa_\aD-\bar a_\aD{\cal
C}^\bD_a\big)\\
&=i\bar\xi_{\aD
B}\big(\bar\CM^A\CM^B-\bar\CM^B\CM^A-2\Sigma^{AB}_a\bar
w^\aD\phi^0_aw_\aD+2\Sigma^{AB}_a\BL\varphi_a\big)\ .
\elabel{jqqj}
}
To prove this, we used \eqref{coneb} to show that first bracket on the
right-hand side of \eqref{jqqj} vanishes. Going from the first line to
the second line involved using the bosonic ADHM constraints
\eqref{badhm} along with the definition of $\BL$ in Eq.~\eqref{vvxx}. The final
expression then vanishes by virtue of the definition of
$\varphi_a$ in Eq.~\eqref{dxx} along with a
$\Sigma$-matrix identity \eqref{smid}.

The fact that the variation of the bosonic (fermionic) ADHM constraints
involves the fermionic (bosonic) ADHM constraints looks promising
because it means that the variation of the product of $\delta$-functions in
\eqref{fms} vanishes to linear order. For the $\N=1$ measure, this is
sufficient to prove supersymmetric invariance. The reason is that the
transformation of $a_\aD$ involves $\CM$ but that of $\CM$ is a
constant. Hence, the super-Jacobian for the transformation of
$\{a_\aD,\CM\}$ vanishes at linear order. Since there are no other
contributions to consider, the $\N=1$ measure is a supersymmetric invariant.

However, the story is more involved with extended
supersymmetry. The reason is that, contrary to the $\N=1$ case,
the variation of
the Grassmann collective coordinates \eqref{vivi} involves the
Grassmann collective coordinates through the dependence of
${\cal C}^\aD_a$ and $\bar{\cal C}^\aD_a$ on $\varphi_a$,
\eqref{defbc}, which in
turn depends on $\CM^A$ via \eqref{dxx}. This means that at linear
order there is
non-trivial super-Jacobian for the transformation of $\{a_\aD,\CM^A\}$.
In order to prove supersymmetric invariance, this
super-Jacobian must cancel the transformation of the remaining
factor
$\big|\text{det}_{k^2}\BL\big|^{1-\N}$ that we have not, hitherto,
considered.

Rather than evaluate the variations of these two quantities,
and then show that they
cancel, we will proceed in more indirect fashion. The idea is to remove
the dependence of $\delta\CM^A$ on the Grassmann collective coordinates by
introducing some auxiliary variables. The obvious candidate is
$\varphi_a$ itself. In order to implement this idea, we need to find out how
$\varphi_a$ transforms. We can answer this by considering the
variation of the definition \eqref{dxx}. Writing this as
\EQ{
\BL\cdot\varphi_a=\tfrac14\bar\Sigma_{aAB}\bar\CM^A\CM^B+\bar
w^\aD\phi^0_aw_\aD\ ,
\elabel{loopy}
}
the variation of the left-hand side is the sum of
$\BL\cdot\delta\varphi_a$ and
\SP{
(\delta\BL)\cdot\varphi_a&=
\tfrac i2\bar\xi^\aD_A\Big(
\big\{\bar\mu^Aw_\aD-\bar w_\aD\mu^A+\CM^{\prime\alpha A}a'_{\alpha\aD}
+a'_{\alpha\aD}\CM^{\prime A},\varphi_a\}
-2\CM^{\prime\alpha A}\varphi_aa'_{\alpha\aD}-2a'_{\alpha\aD}\varphi_a
\CM^{\prime\alpha A}\Big)\\
&=i\bar\xi^\aD_A\Big(
\bar\mu^Aw_\aD\varphi_a+[\CM^{\prime\alpha A}a'_{\alpha\aD},\varphi_a]
-\varphi_a\bar w_\aD\mu^A+[\varphi_a,a'_{\alpha\aD}\CM^{\prime\alpha
A}]\Big)\ ,
\elabel{weea}
}
using the definition of $\BL$ in \eqref{vvxx}.
The last equality follows by using the fermionic ADHM constraints
\eqref{fadhm}. For the variation of the right-hand side of
\eqref{loopy}, one finds
\EQ{
\tfrac i2\bar\Sigma_{aBC}\Sigma_b^{CA}\bar\xi^\aD_A
\big(\bar{\cal C}_{b\aD}\CM^B-\bar\CM^B{\cal C}_{b\aD}\big)
+i\bar\xi^\aD_A\big(\bar\mu^A\phi^0_aw_\aD-\bar
w_\aD\phi^0_a\mu^A\big)\ .
\elabel{weeb}
}
Taking the difference of \eqref{weea} and \eqref{weeb}, one deduces
the variation of $\varphi_a$:
\EQ{
\BL\cdot
\delta\varphi_a=i\Big\{\tfrac12\bar\Sigma_{aBC}\Sigma_b^{CA}-\delta_{ab}
\delta_B^A\Big\}\bar\xi^\aD_A{\cal F}_{b\aD}^B\ ,
\elabel{weec}
}
where we have defined
\EQ{
{\cal F}_{b\aD}^B=
\big(\bar{\cal C}_{b\aD}\CM^B-\bar\CM^B{\cal C}_{b\aD}\big)\ .
}
This looks disappointing because the variation of $\varphi_a$ seems to
depend on itself via ${\cal C}_{b\aD}$ and $\bar{\cal
C}_{b\aD}$. However, notice that the variation of the Grassmann
collective coordinates \eqref{vivi} actually depends not on
$\varphi_a$ directly
but, more precisely, on the combination
$\Sigma_a^{AB}\varphi_a$. From \eqref{weec}, we have
\EQ{
\BL\cdot\delta\big(\Sigma_a^{AB}\varphi_a\big)=i\bar\xi^\aD_C\big(
-\Sigma_a^{BC}{\cal F}^A_{a\aD}+\Sigma_a^{AC}{\cal
F}^B_{a\aD}-\Sigma_a^{AB}{\cal F}^C_{a\aD}\big)\ .
\elabel{weed}
}
Now for $\N=2$ supersymmetry, there is only a single independent quantity
$\Sigma_a^{AB}\varphi_a$ which we can take to be
$\varphi\equiv\Sigma_a^{12}\varphi_a=i\varphi_1+\varphi_2$. In this
case the right-hand side of \eqref{weed} vanishes identically; hence
\EQ{
\delta\varphi=0\ .
}
We now consider the extended set of variables
$\{a_\aD,\CM^A,\varphi\}$, where the latter is subject to its own
``ADHM constraint'' following from its definition \eqref{dxx}:
\EQ{
\BL\cdot\varphi=-\tfrac12(\bar\CM^1\CM^2-\bar\CM^2\CM^1)
+\bar w^\aD\phi^0w_\aD\
,
\elabel{nadhm}
}
where $\phi^0\equiv\Sigma^{12}_a\phi^0_a$. The $\N=2$ measure
\eqref{fms} can then be written in a suggestive way by replacing the
factor of $|\det_{k^2}\,\BL|^{-1}$ with an integral
over the auxiliary variable $\varphi$ with an explicit
$\delta$-function which imposes the new ADHM constraint \eqref{nadhm}:
\EQ{
\big|{\rm det}_{k^2}\BL\big|^{-1}=\int d^{k^2}\varphi\,
\prod_{r=1}^{k^2}\delta\big({\rm
tr}_kT^r(\BL\cdot\varphi+\tfrac12\bar\CM^1\CM^2-\tfrac12\bar\CM^2\CM^1-\bar
w^\aD\phi^0w_\aD)\big)\ .
}
It is then easy to see that, when re-written using the above identity,
the new form of the measure is
supersymmetric. This follows because the variations of
$\{a_\aD,\CM^A,\varphi\}$ are either off-diagonal or vanish;
hence, the super-Jacobian vanishes
to linear order. To complete the proof, the
variations of the ADHM constraints, \eqref{badhm}, \eqref{fadhm} and
\eqref{nadhm} are also either off-diagonal or
vanish,\footnote{Note that the variation
of \eqref{nadhm} vanishes since we imposed it to find the variation of
$\varphi_a$ above.} and so the variation of the
product of $\delta$-functions also vanishes to linear order.

For $\N=4$ supersymmetry, we have to go back to \eqref{weed} and follow the
same logic. To make things simpler, and ultimately
without loss-of-generality, we can focus on particular a
supersymmetry variation, say $\bar\xi^\aD_1$. In that case,
the variation of the Grassmann collective coordinates \eqref{vivi}
involves the three quantities
$\varphi^{A1}\equiv\Sigma_a^{A1}\varphi$,
$A=2,3,4$. In that case, from
\eqref{weed}, we have
\EQ{
\BL\cdot\delta\varphi^{A1}=i\bar\xi^\aD_1\big(
-\Sigma_a^{11}{\cal F}^A_{a\aD}+\Sigma_a^{A1}{\cal
F}^1_{a\aD}-\Sigma_a^{A1}F^1_{a\aD}\big)=0\ .
\elabel{weednf}
}
The new auxiliary variables $\varphi^{A1}$ are subject to their own ``ADHM
constraints'':
\EQ{
\BL\cdot\varphi^{A1}=\tfrac12\big(\bar\CM^1\CM^A-\bar\CM^A\CM^1\big)
+\Sigma_a^{A1}\bar w^\aD\phi^0_aw_\aD\ .
\elabel{adhmnf}
}
So for this
particular supersymmetry variation, we consider the multiplet of variables
$\{a_\aD,\CM^A,\varphi^{A1}\}$. The $\N=4$
collective coordinate integral \eqref{fms} can then be re-cast
by introducing integrals
over the auxiliary variables $\varphi^{A1}$ along with explicit
$\delta$-functions to impose \eqref{adhmnf}:
\EQ{
\big|{\rm det}_{k^2}\BL\big|^{-3}=\int\prod_{A=2}^4\bigg\{
d^{k^2}\varphi^{A1}\,
\prod_{r=1}^{k^2}\delta\big({\rm
tr}_kT^r(\BL\cdot\varphi^{A1}-
\tfrac12\bar\CM^1\CM^A+\tfrac12\bar\CM^A\CM^1-\Sigma_a^{A1}\bar
w^\aD\phi^0w_\aD)\big)\bigg\}\ .
}
It is then straightforward to prove, following the same logic as for
$\N=2$, supersymmetric invariance. Firstly, the variations of
$\{a_\aD,\CM^A,\varphi^{A1}\}$ are either off-diagonal or vanish;
hence, the super-Jacobian vanishes
to linear order. In addition, the
variations of the ADHM constraints, \eqref{badhm}, \eqref{fadhm} and
\eqref{adhmnf} are also off-diagonal or zero. Consequently,
the variation of the
product of $\delta$-functions also vanishes to linear order.
The new twist in the $\N=4$ case is,
for each supersymmetry variation $\bar\xi_A$, we must use a
different set of three
auxiliary variables $\Sigma_a^{AB}\varphi_a$, $B\neq A$ to prove
supersymmetric invariance. Invariance of the measure \eqref{fms}
under a general supersymmetry variation then follows by linearity.

\subsection{From $\N=4$ to $\N=0$ via decoupling}\elabel{sec:S42.6}

An interesting consistency check on our collective coordinate
integrals, Eq.~\eqref{intm} along with \eqref{fms}, which relates the
expressions for different numbers of supersymmetries follows from
renormalization group decoupling. The idea is to take one of the
supersymmetric theories and add mass terms for some of the fields in such
a way that for large masses the massive fields can be ``integrated
out'' and the theory flows in the infra-red to a theory with fewer
supersymmetries. This procedure, when implemented at the level of the
semi-classical approximation, provides a way to relate the instanton
integration measure for different numbers of supersymmetries.

We begin, with the $\N=2$ theory, and give a mass to one of the two
$\N=1$ chiral multiplets. Adding such mass terms is discussed in
\S\ref{app:A8}. For a single flavour of fermion, the leading
order effect, is to introduce \eqref{nonepn}
\EQ{
\tilde S _{\text{mass}}=m\int d^4x \,\TrN\,\lambda^2
=-\frac{m\pi^2}{g}{\rm tr}_k\,\bar\CM(\Pinfty+1)\CM\equiv -\frac
m{4g}\tilde\Omega_{\ii\,\jj}\CM^\ii\CM^\jj\ ,
}
into the instanton effective action,
where the inner product of two Grassmann symplectic tangent vectors was
defined in \eqref{ipstv}.

The integral over the Grassmann collective coordinates
$\CM$ is written down in \eqref{intsf}. It is convenient to re-write the
argument of the fermionic ADHM constraints using \eqref{rwfadhm} and
then including the mass term, we have
\EQ{
{\EuScript I}=\big\vert\det_{k^2}\BL\big|^{-1}\,
\int d^{2k(N+k)}\CM
 \,\prod_{r=1}^{k^2}\,
\prod_{\aD=1}^{2}\,\delta\big(\tfrac1{4\pi^2}\CM^\ii
\tilde\Omega_{\ii\,\jj}X_r^{\jj\aD}\big)\exp\Big(\frac
m{4g}\tilde\Omega_{\ii\,\jj}\CM^\ii\CM^\jj\Big)\ .
\elabel{intsfm}
}

In order to evaluate the integral it is useful to decompose
\EQ{
\CM^\ii=\sigma_{r\aD}X_r^{\ii\aD}+\CM^\perp
\elabel{decf}
}
where $(\CM^{\perp})^\ii\tilde\Omega_{\ii\,\jj}X_r^{\jj\aD}=0$ for all
$r=1,\ldots,k$ and $\aD=1,2$. The quantity $\CM^\perp$ is the
projection of $\CM$ which does not appear in the arguments of the
$\delta$-functions and its integrals must be saturated by bringing
down powers of the mass term. Hence with the decomposition
\eqref{decf}, the integral \eqref{intsfm} factorizes as
\EQ{
{\EuScript I}=\big\vert\det_{k^2}\BL\big|^{-1}\,
\,\frac{\int\prod_{r=1}^{k^2}\prod_{\aD=1}^{2}\,d\sigma_{r\aD}\,
\delta\big(\tfrac1{4\pi^2}\sigma_{s\bD}X_s^{\ii
\bD}\tilde\Omega_{\ii\,\jj}X_r^{\jj\aD}\big)}
{\Big|\det_{2k^2}\tfrac1{4\pi^2}X_r^{\ii\aD}\tilde
\Omega_{\ii\,\jj}X_s^{\jj\bD}\Big|^{1/2}}\ \cdot
\ \int d^{2kN}\CM^\perp\,
\exp\Big(\frac
m{8g}\tilde\Omega_{\ii\,\jj}\CM^{\perp\ii}\CM^{\perp\jj}\Big)\ .
\elabel{inta}
}
The determinant in the denominator is the Jacobian for transforming to
the variables $\sigma_{r\aD}$.
The Grassmann integrals can now be done to give
\EQ{
{\EuScript I}=\big\vert\det_{k^2}\BL\big|^{-1}\,
\cdot
\Big|\det_{2k^2}\tfrac1{4\pi^2}X_r^{\ii\aD}\tilde
\Omega_{\ii\,\jj}X_s^{\jj\bD}\Big|^{1/2}\,
\cdot\,\Big(\frac{2m\pi^2}{g}\Big)^{kN}\ .
\elabel{intaa}
}
The second determinant factor can be related to a more familiar quantity by
noting the following. From Appendix~\ref{app:A2}, we established that
on the level set of the hyper-K\"ahler quotient,
the $\U(k)$-vectors $X_r$ satisfy
$\tilde g(X_r,\tilde\BI^{(c)}X_s)=0$. In terms of components
\EQ{
\epsilon_{\aD\bD}X_r^{\ii\aD}\tilde
\Omega_{\ii\,\jj}X_s^{\jj\gD}\vec\tau^\bD{}_\gD=0\
\quad\text{implies}\quad
X_r^{\ii\aD}\tilde\Omega_{\ii\,\jj}X_s^{\jj\bD}\propto\epsilon^{\aD\bD}\ .
}
Hence, from the definition of $\BL$ is \eqref{gwww} and the explicit
relation $\tilde g(X_r,X_s)=\epsilon_{\aD\bD}
X_r^{\ii\aD}\tilde\Omega_{\ii\,\jj}X_s^{\jj\bD}$, we have
\EQ{
\Big|\det_{2k^2}\tfrac1{4\pi^2}X_r^{\ii\aD}\tilde\Omega_{\ii\,\jj}
X_s^{\jj\bD}\Big|=
\Big|\det_{k^2}\tfrac1{8\pi^2}\tfrac12\epsilon_{\aD\bD}
X_r^{\ii\aD}\tilde\Omega_{\ii\,\jj}X_s^{\jj\bD}\Big|^2\equiv
\big|\det_{k^2}\BL\big|^2\ .
\label{stvr}
}
Consequently, the two determinants in \eqref{intaa} cancel to leave
a purely numerical factor
\EQ{
{\EuScript I}=\Big(\frac{2m\pi^2}{g}\Big)^{kN}\ .
\elabel{intbb}
}

Using this expression for the integral \eqref{intsfm}, we find that
after decoupling the $\N=2$ collective coordinate integral
gives the $\N=1$ collective coordinate integral
with the following relation between the Pauli-Villars
mass scales:
\EQ{
\mu_{\sst\N=1}^{3N}=m^{N}\mu_{\sst\N=2}^{2N}\ ,
}
Alternatively, we can phrase the result in terms of the
renormalization group $\Lambda$-parameters defined in \eqref{deflp}:
\EQ{
\Lambda_{\sst\N=1}=\big(g^2m\big)^N\Lambda_{\sst\N=2} \ .
\elabel{rgrto}
}
Therefore, the decoupling limit is $g^2m\to\infty$ and $\Lambda_{\sst\N=2}\to0$
in such a way that $\Lambda_{\sst\N=1}$ is fixed.

Now we turn to the relation between the $\N=4$ and $\N=2$ theory. We
start with the $\N=4$ theory and add an $\N=2$ preserving mass term
for a pair of $\N=1$ chiral multiplets. This results in a contribution
\eqref{nonepn} to the instanton effective action:
\EQ{
\tilde S _{\text{mass}}=m \int d^4x
\,\TrN\,\big(\lambda^3\lambda^3+\lambda^4\lambda^4\big)
=-\frac{m\pi^2}{g}{\rm
tr}_k\,\big[\bar\CM^3(\Pinfty+1)\CM^3+\bar\CM^4
(\Pinfty+1)\CM^4\big]
}
We can now integrate out the two flavours of
Grassmann collective coordinates $\CM^3$ and $\CM^4$ as above to
verify that the collective coordinate integrals are related if
\EQ{
\mu_{\sst\N=2}^{2N}=m^{2N}\ .
}
This means that
\EQ{
\Lambda_{\sst\N=2}=m^{2N}e^{-8\pi^2/g^2+i\theta}
\elabel{rgrft}
}
and so the decoupling limit is $m\to\infty$ and $g^2\to0$, in such a
way that $\Lambda_{\sst\N=2}$ is fixed.

We remark that \eqref{rgrto} and \eqref{rgrft} are
consistent with the standard prescriptions in the literature
for the renormalization group matching of a low-energy and a
high-energy theory \cite{weinb,FP}. The absence of numerical factors on the
right-hand side of these relations reflects the absence of threshold
corrections in the Pauli-Villars scheme.

Finally, we can take the $\N=1$ theory and decouple the gluino to flow
at low energies to the non-supersymmetric theory. In this limit, the
$\N=1$ collective coordinate integral gives the extrapolation of
\eqref{fms} to $\N=0$ with the relation
\EQ{
\mu_{\sst\N=0}^{4N}=m^N\mu_{\sst\N=1}^{3N}\ .
}
However, in the non-supersymmetric theory there is also a non-trivial
fluctuation determinant described in \S\ref{sec:S24} which is not, of
course, reproduced.

\rsen\section{Generalizations and Miscellany}\label{sec:S201}

In this section we describe some important generalizations of the
instanton calculus and other results. In
\S\ref{app:A5}, we describe how the ADHM constraints can be solved when
$N\geq2k$. This will be useful in applications
where a large-$N$ limit is involved, as in \S\ref{sec:S43} and
\S\ref{sec:S39}. Then, in \S\ref{app:A7}, we explain how to generalize
the instanton calculus to theories with gauge
groups $\Sp(N)$ and $\SO(N)$. This is useful in the application to
$\N=2$ theories described in
\S\ref{sec:S60} involving
gauge group $\SU(2)$ because it turns out that the instanton calculus
of $\Sp(1)$ ($\simeq\SU(2)$) is actually more economical and convenient
than the $\SU(2)$-as-an-example-of-$\SU(N)$ formalism developed previously.
Moving on, in \S\ref{app:A6}, we describe how to include matter fields
in the calculus. This is important for the applications in
\S\ref{sec:S43} and \S\ref{sec:S50}. The question of how to modify the
instanton calculus when fields have masses is considered in
\S\ref{app:A8}. Finally in \S\ref{sec:N1}
we consider the instanton partition function in more detail. In
particular we show how it can be ``linearized'' by introducing various
auxiliary variables including Lagrange multipliers for the bosonic and
fermionic ADHM constraints. We also show how the integrals over the
overall position coordinate and its superpartners can be separated out,
leading to the notion of the centred instanton partition function.

\subsection{Solving the ADHM constraints for $N\geq2k$}\elabel{app:A5}

It turns out that when $N\geq2k$ the ADHM constraints can be solved in
a certain generic region of the moduli space by a change of
variables \cite{MO3}. The idea is to introduce the gauge invariant
$\U(N)$-invariant bi-linears
\EQ{
\big(W^\dalpha{}_\bD\big)_{ij}=\bar w_{iu}^\dalpha
 \,w_{uj\dbeta}
\elabel{defbw}
}
which can be incorporated into four $k\times k$ matrices
\EQ{
W^0={\rm tr}_2\,W,\quad W^c={\rm tr}_2(\tau^cW)\ .
\elabel{bosbi}
}
When $N\geq2k$, these variables are independent and the
change of variables from the $4kN$ real variables
$\{w_\aD,\bar w^\aD\}$, to the  bi-linear variables
$\{W^0,W^c\}$ and the gauge orientation
$\grp$ (defined in \S\ref{sec:S11})
is invertible, at least when the gauge orbit is generic as in Eq.~\eqref{goc}.
In changing variables, one must also, of course,
specify the integration domain for the new variables. In particular, since the
$W$ variables are the inner products of vectors they will be constrained
by various triangle inequalities.\footnote{Fortunately, these technicalities
will not be relevant in our applications at large-$N$. In this case,
steepest-descent methods apply and
the saddle-point values of the $W$'s obtained in \S\ref{sec:S39} and
\S\ref{sec:S43} satisfy all such triangle inequalities by inspection.
In addition, we will show that on the saddle-points the $k$ instantons
inhabit $k$ commuting $\SU(2)$ subgroups of the the gauge
group. Hence, manifestly, they lie on a generic orbit of the gauge
group \eqref{goc}.}

The reason why this change of variables is useful, is because the ADHM
constraints \eqref{badhm} are linear in $W^c$; indeed they can be
re-cast as
\EQ{
W^c=-a'_ma'_n{\rm tr}_2\big(\tau^{c}
\bar\sigma_m\sigma_{n}\big)\ .
\elabel{linad}
}
The final description of the subspace of the moduli space on a generic
orbit of the gauge group for $N\geq2k$
is then in terms of the five
$k\times k$ Hermitian matrices, $W^0$ and $a'_n$, in addition to
$\grp$. However, as long as one is in a phase where the gauge
symmetry is not broken then physical quantities are gauge invariant
and one can readily integrate over the gauge orientation $\grp$.

In order to exploit this linearizing change-of-variables, we must
determine how the instanton integration measure transforms in going
from $\{w_\aD,\bar w^\aD\}$
to the bi-linears $\{W^0,W^c\}$
in Eq.~\eqref{bosbi} and the gauge orientation $\grp$. In what follows
it will be useful to think of $w_\aD$ as an $N\times K$-dimensional
matrix $w$, where $K=2k$, and to introduce the composite index $a\equiv
i\aD=1,\ldots,K$, so that the elements of $w$ are $w_{ua}$.
A suitable $\SU(N)$ gauge transformation $\grp$ puts the matrix $w$ into
upper-triangular form, as in  \eqref{uptri}.
The $\xi_{ab}$
are complex except for the diagonal elements $\xi_{aa}$ which we can choose
to be real.
While \it a priori \rm the $\xi_{ab}$ are only defined for $1\leq a\le b\le
K,$ it is convenient to extend them to $1\leq b<a\le K$ as well, by defining
$\xi_{ab}=(\xi_{ba})^*$.  In terms of these extended variables,
the gauge-invariant bi-linear $W$ defined in \eqref{defbw} is the
matrix 
\begin{equation}
W =w^\dagger\,w=
\begin{pmatrix}\xi_{11}&0&\cdots&0\\
\xi_{21}&\xi_{22}&{}&\vdots\\
\vdots&\vdots&\ddots&0\\
\xi_{K1}&\xi_{K2}&\cdots&\xi_{KK}\end{pmatrix}
\begin{pmatrix}\xi_{11}&\xi_{21}&\cdots&\xi_{K1}\\
0&\xi_{22}&\cdots&\xi_{K2}\\
\vdots&{}&\ddots&\vdots\\
0&\cdots&0&\xi_{KK}\end{pmatrix}.
\elabel{Wequals}\end{equation}
Note that there are as many real degrees-of-freedom in the $\{W_{ab}\}$ as
in the $\{\xi_{ab}\}.$ From Eq.~\eqref{Wequals} it follows that
\begin{equation}
\det_KW=\Big(\prod_{a=1}^K\xi_{aa}\Big)^2\ .
\elabel{detWdef}\end{equation}

When calculating the Jacobian, it is useful to pass
through an intermediate change-of-variables involving $\xi$
rather than the $W$. The Jacobian for this is
\EQ{
\int\,d^{K^2}W=2^{K}\int\,d^{K^2}\xi\,\prod_{a=1}^K
\xi_{aa}^{2K-2a+1}\ .
\elabel{intjac}
}
This can be proved by induction.
For $K=1$ one has simply $W_{11}=\xi_{11}^2$.
{}From Eq.~\eqref{Wequals} we can also easily relate the Jacobian for $K$ to
that for $K-1$; one finds
\begin{equation}\begin{split}
{\partial\big(\{W_{ab}\}\big)\over\partial\big(\{\xi_{ab}\}\big)}\bigg|_K^{}
&=
{\partial W_{KK}\over\partial\xi_{KK}}
\Big(\prod_{a=1}^{K-1}{\partial W_{aK}\over\partial\xi_{aK}}
{\partial W_{Ka}\over\partial\xi_{Ka}}\Big)
{\partial\big(\{W_{ab}\}\big)\over\partial\big(\{\xi_{ab}\}\big)}\bigg|_{K-1}^{}
\\ &=
2\xi_{KK}\Big(\prod_{a=1}^{K-1}\xi_{aa}^2
\,\Big)
{\partial\big(\{W_{ab}\}\big)\over\partial\big(\{\xi_{ab}\}\big)}
\bigg|_{K-1}^{}\ .
\elabel{inductcomp}
\end{split}\end{equation}
The result \eqref{intjac} follows by induction.

Next, we calculate the Jacobian for the change of variables from $w$
to $\{\xi,\grp\}$. To this end, define $u^a\equiv\grp_{ua}$
to be the $a^{\rm th}$ column of $\grp$, $a=1,\ldots,K$. Since $\grp\subset
\SU(N)$
\begin{equation}
(u^\dagger)^a\cdot u^{b}=\delta^{ab}\ .
\end{equation}
The $K$ $N$-vectors $u^a$ then provide the well-known parameterization of the
coset \eqref{ncsp}
as a product of spheres \cite{GILMORE}. To see this, $u^1$ is a unit
vector in an $N$-dimensional complex space and consequently parameterizes
$S^{2N-1}$. The second vector $u^2$ is also a unit vector, but one which is
orthogonal to $u^1$, and consequently parameterizes
$S^{2N-3}$. Continuing this chain of argument, we see that the
vectors $\{u^a\}$ parameterize the product of spheres
\begin{equation}
{\SU(N)\over \SU(N-K)}
\simeq S^{2N-1}\times S^{2N-3}\times\cdots\times S^{2N-2K+1}\ .
\end{equation}

{}From \eqref{uptri} we can read off the expansion of the elements of
$w$ in terms of the elements of the vectors $u^a$:
\begin{equation}
w_{ua}=\sum_{b=1}^a\xi_{ba}u^b_u\ .
\end{equation}
The Jacobian can be
determined through the following iterative process. First we start
with $w_{u1}=\xi_{11}u^1_u$, whose measure can be written in polar
coordinates as
\EQ{
\int\prod_{u=1}^N dw_{u1}dw_{u1}^*=
2^N\int \xi_{11}^{2N-1}\,d\xi_{11}\,d^{2N-1}\hat\Omega_1\ .
}
Here $d^{2N-1}\hat\Omega_1$ is the usual measure for the solid angles on
$S^{2N-1}$ parameterized by $u^1$. Continuing the process on the next vector
$w_{u2}=\xi_{12}u^1_u+\xi_{22}u^2_u$, we have
\EQ{
\int\prod_{u=1}^N dw_{u2}dw_{u2}^{*}=2^{N-1}\int d\xi_{12}d\xi_{12}^*
\,\xi_{22}^{2N-3}\,d\xi_{22}\,d^{2N-3}\hat\Omega_2\ .
}
In general
\EQ{
\int\prod_{u=1}^N\, dw_{ua}\,dw_{ua}^*=2^{N-a-1}
\int\Big\{\prod_{b=1}^{a-1}\,d\xi_{ba}\,d\xi_{ba}^*\Big\}
\,\xi_{aa}^{2N-2a+1}\,d\xi_{aa}\,d^{2N-2a+1}\hat\Omega_a\ .
}
where $\hat\Omega_a$ is parameterized by $u^a$. Hence
\EQ{
\int d^{2KN}w=
2^{NK-K(K-1)/2}\int\Big\{
\prod_{a=1}^K\xi_{aa}^{2N-2a+1}
d\xi_{aa}\,d^{2N-2a+1}\hat\Omega_a\Big\}\,
\Big\{\prod_{a<b}d\xi_{ab}\,d\xi^{*}_{ab}\Big\}\ .
\elabel{numanss}
}
Using \eqref{detWdef} and \eqref{intjac}, we obtain
\begin{equation}
\int d^{2KN}w=
2^{NK-K(K+1)/2}\int\,\big|\det_{K}
W\big|^{N-K}\,d^{K^2}W\,\Big\{\prod_{a=1}^Kd^{2N-2a+1}\hat\Omega_a\Big\}\ .
\end{equation}

Re-introducing $k=K/2$, we have
\begin{equation}
\int d^{2kN}w\, d^{2kN}\bar w=A_k\int\,\big|{\rm
det}_{2k}W\big|^{N-2k}\,
d^{k^2}W^0\prod_{c=1,2,3}d^{k^2}W^c\,d^{4k(N-k)}\grp\  .
\elabel{E36.1}\end{equation}
The integral over the $2k\times 2k$
matrix $W$ has been written as four separate integrals over the
$k\times k$ matrices $W^0$ and $W^c$, defined in \eqref{bosbi}, with
respect to the basis $\{T^r\}$, defined in
\S\ref{sec:S18}.\footnote{This
accounts for an additional factor of
$2^{-2k^2}$.} For convenience we
have defined a unit normalized measure on the coset space:
\EQ{
\int d^{4k(N-k)}\grp\ \overset{\text{def}}=\ \frac1{\prod_{a=1}^{2k}{\rm
Vol}\,
S^{2(N-a)+1}}\int\ \prod_{a=1}^{2k}d^{2N-2a+1}\hat\Omega_a
\elabel{defgrp}
}
and this fixes the normalization constant to be
\EQ{
A_k= 2^{2kN-4k^2-k}\,\prod_{a=1}^{2k}{\rm Vol}\,S^{2(N-a)+1}=
{2^{2kN-4k^2+k}\pi^{2kN-2k^2+k}\over\prod_{a=1}^{2k}(N-a)!}\ .
\elabel{E37.1}
}

In the expression \eqref{bmes} for
volume form on $\ms_k$,
the $\delta$-functions imposing the
ADHM constraints simply soak up the integrals over $W^c$ (giving rise to the
numerical factor of $2^{3k^2}$ from the $\tfrac12$'s in the arguments of
the $\delta$-functions) to leave
\EQ{
\int_{\ms_k}\Bomega=\frac{2^{3k^2}A_k\,C_k}
{{\rm Vol}\,\U(k)}\int\, d^{4k^2} a' \,
d^{k^2}W^0\,d^{4k(N-k)}\grp\,
\big|\det_{k^2}\,\BL\big|\,\big|\det_{2k}W\big|^{N-2k}\ .
\elabel{bames}
}

The simplifications for $N\geq2k$
also extend to the Grassmann sector.
The trick is to find a change of variables in the
Grassmann sector which mirrors that which we have just described for
the $c$-number collective coordinates. To this end, let us identify the
super-partners of the collective coordinates associated
to global gauge transformations on the instanton solution.
Infinitesimally, the latter are the
subset of $\delta w$ which preserve the gauge invariant variables
$W$, {\it i.e.\/}, which satisfy
\begin{equation}
\bar w^\aD_{iu}\delta w_{uj\bD}+\delta\bar w^\aD_{iu}w_{uj\bD}=0\, .
\elabel{stwbw}
\end{equation}
Under a supersymmetry transformation \eqref{uiui} one has
\begin{equation}
\delta w_{ui\aD}=i\bar\xi_{\aD A}\mu^A_{ui}\ , \qquad \delta
\bar w^\aD_{iu}=-i\bar\mu_{iu}^A\bar\xi^{\aD}_A\ .
\elabel{susytr}
\end{equation}
Inserting Eq.~\eqref{susytr} into Eq.~\eqref{stwbw} produces the
gauge-invariant conditions
\begin{equation}
\bar\xi_{\bD A}\bar w^\aD_{iu}\mu_{uj}^A+\bar\xi^{\aD
}_A\bar\mu^A_{iu}w_{uj\bD}=0\,
\elabel{stwbe1}
\end{equation}
or equivalently,
\begin{equation}
 \bar w_{iu}^\aD\mu_{uj}^A=0 \quad\text{and}\quad
\bar\mu_{iu}^Aw^{}_{uj\aD}=0\ .
\elabel{stwbe}\end{equation}
To satisfy these constraints, it is convenient to decompose $\mu^A$ as
follows:
\begin{equation}
\mu_{iu}^A=w_{uj\aD}(\zeta^{\aD
A})_{ji}+\nu_{iu}^A,\qquad
\bar\mu_{iu}^A=(\bar\zeta^{
A}_\aD)_{ij}\bar w_{ju}^\aD+\bar\nu_{iu}^A\ ,
\elabel{E44}\end{equation}
where $\nu^A$ lies in the orthogonal subspace to $w$:
\begin{equation}
\bar w_{iu}^\aD\nu_{uj}^A=0,\qquad \bar\nu_{iu}^Aw^{}_{uj\aD}=0\ .
\elabel{E45}\end{equation}
The superpartners of the bosonic coset coordinates $\grp$
are then precisely the variables
$\{\nu^A,\bar\nu^A\}$.\footnote{It is worth
mentioning that although the coset coordinates correspond to bosonic zero
modes which are generated by Lagrangian symmetries, this is not true
of their Grassmann partners.} Notice in the case of a single
instanton, the variables $\bar\zeta^A=\zeta^A$ are precisely the
Grassmann collective coordinates associated to superconformal
transformations.

We now turn to the Grassmann part of the instanton measure \eqref{fms}.
As explained above, the superpartners of the global gauge collective
coordinates $\grp$
are the Grassmann variables $\{\nu^A,\bar\nu^A\}$ defined
in Eqs.~\eqref{E44}-\eqref{E45}. Since these coordinates are
orthogonal to the $\bar w$ and $w$ vectors, respectively, it is easy
to see from \eqref{zmcona} that they do not appear in the fermionic
ADHM constraints. The Jacobian for the change of variables from
the original Grassmann coordinates $\{\mu^A,\bar\mu^A\}$
to $\{\zeta^A,\bar\zeta^A,\nu^A,\bar\nu^A\}$, is, for each
value of $A$,\footnote{We define the integrals over the $k\times k$
matrices $\zeta^{\aD A}$ and $\bar\zeta^{\aD A}$ with respect to the
Hermitian basis $T^r$ introduced in \S\ref{sec:S10}.}
\begin{equation}
{\partial\big(\{\mu^A,\bar\mu^A\}\big)\over
\partial\big(\{\zeta^A,\bar\zeta^A,\nu^A,\bar\nu^A\}\big)}
=\big|\det_{2k}\, W\big|^{-k}\ .
\elabel{jacznu}\end{equation}

As in the bosonic sector, the change of variables allows us to
integrate out the Grassmann-valued $\delta$-functions in the measure
\eqref{fms}
trivially. The arguments of the $\delta$-functions are the
fermionic ADHM constraints which, in terms of the new variables, are
\eqref{zmcona} are
\begin{equation}
\bar\zeta_{\bD}^AW_{\ \,\aD}^{\bD}+W_{\aD\bD}\zeta^{\bD A}+
\big[{\cal M}^{\prime\alpha A},a'_{\alpha\aD}\big]=0\ .
\elabel{E46}\end{equation}
These equations can be used to eliminate the $2k^2$ variables
$\bar\zeta_\aD^A$, for each $A$.
The relevant integral is simply
\begin{equation}
\int d^{2k^2}\bar\zeta^A\,
\prod_{r=1}^{k^2}\prod_{\aD=1}^2\delta
\Big({\rm tr}_k\,T^r(\bar\zeta_{\bD}^AW_{\ \,\aD}^{\bD}+W_{\aD\bD}
\zeta^{\bD A}+
\big[{\cal M}^{\prime\alpha A},a'_{\alpha\aD}\big])\Big)
=\big|\det_{2k}\, W\big|^{k}.
\end{equation}
Notice that the factor on the right-hand side conveniently cancels the
Jacobian of \eqref{jacznu}.

The expression for the supersymmetric volume form on the instanton
moduli space in the case $N\geq2k$ is
\SP{
\int_{\ms_k}\Bomega^{\sst(\N)}&=\frac{2^{3k^2}C_k^{\sst(\N)}A_k}
{{\rm Vol}\,\U(k)}\int\, d^{4k^2} a' \,
d^{k^2}W^0\,d^{4k(N-k)}\grp\
\prod_{A=1}^\N\Big\{
d^{k(N-2k)}\nu^A\,d^{k(N-2k)}\bar\nu^A\,d^{2k^2}\zeta^A\,d^{2k^2}\CM^{\prime
A}\Big\}\\ &\times
\big|\det_{k^2}\,\BL\big|^{1-\N}
\,\big|\det_{2k}W\big|^{N-2k}\ .
\elabel{sbames}
}
This expression for the measure is the starting point for two
applications of the multi-instanton calculus at large $N$ that we
describe in \S\ref{sec:S43} and \S\ref{sec:S39}.

\subsection{The ADHM construction for $\Sp(N)$ and $\SO(N)$}\elabel{app:A7}

The ADHM formalism \cite{ADHM} for constructing instanton solutions was
adapted for dealing with any of the classical gauge groups
in the early instanton literature.
The method adopted was to consider the construction for
one of the series of classical groups, {\it e.g.\/}~symplectic groups in
Ref.~\cite{Corrigan:1978ce} and
orthogonal groups in Ref.~\cite{Christ:1978jy}, and
then embed the other two series
in this series. Our approach will be no different, although we will
start from the unitary series. The advantage of this is that all the
previous formulae that we have established in the $\SU(N)$ case
can easily be adapted to describe the other gauge groups.

In order to construct instanton
solutions for gauge theories with $\Sp(N)$ and $\SO(N)$ gauge groups
we use the embeddings\footnote{For the
orthogonal groups we restrict $N\geq4$.}
\begin{equation}
\Sp(N)\subset \SU(2N)\ ,\qquad \SO(N)\subset \SU(N)\ ,
\end{equation}
to extract the ADHM formalism for these groups in terms of the
$\SU(N)$ ADHM construction. The surprising feature of the resulting formalism
is that the auxiliary group, $\U(k)$ in the $\SU(N)$
case and denoted generally as $\H(k)$ at instanton number $k$,
is {\it not\/} in the same series as the gauge group
$G$. Table 6.1 shows the
auxiliary groups and defines the quantities $N'$ and $k'$
allowing us to present a unified treatment of $\Sp(N)$ and $\SO(N)$.

\begin{center}
\begin{tabular}{cccc}
\hline
\Rowspace $G$ & $\SU(N)$ & $\Sp(N)$ & $\SO(N)$\\
\Rowspace $\H(k)$ & $\U(k)$ & $\O(k)$ & $\Sp(k)$\\
\Rowspace $ N'$ & $N$ & $2N$ & $N$\\
\Rowspace $ k'$ & $k$ & $k$ & $2k$\\
\hline
\end{tabular}\\
\vspace{0.5cm}
Table 6.1: Gauge and associated auxiliary groups
\end{center}

To describe the other classical groups we start with the theory with
gauge group $\SU( N')$ at instanton instanton number
$k'$. Instanton solutions in the $\Sp(N)$ and $\SO(N)$
theories follow by simply imposing certain reality conditions on the
ADHM construction of the $\SU( N')$ theory which ensures that the
gauge potential lies in the appropriate $sp(N)$ and $so(N)$
subalgebra of $su(N')$.
In order to deal with both the $\Sp(N)$ and $\SO(N)$ case at the same time
it is useful to define the notion of a generalized transpose operation
denoted $t$ which acts either on gauge or instanton indices.
Specifically, on $\Sp(n)$ group indices $t$ acts
as a symplectic transpose, {\it i.e.\/}~on a column vector
$v$, $v^t=v^TJ^T$, where $J$ is $2n\times2n$ the symplectic matrix
\begin{equation}
J=\begin{pmatrix} 0 & 1 \\ -1 & 0\end{pmatrix}\ ;
\end{equation}
while on $\O(n)$ group indices $t$ is a conventional transpose
$t\equiv T$. The adjoint representations of both groups
are Hermitian $t$-anti-symmetric matrices and have dimensions $n(2n+1)$
and $n(n-1)/2$, respectively. Hermitian $t$-symmetric matrices
correspond to the {\it anti-symmetric\/} representation of $\Sp(n)$,
with dimension $n(2n-1)$ and the symmetric representation of $\SO(n)$,
with dimensions $n(n+1)/2$.

The additional reality conditions on the ADHM variables are
\begin{equation}
\bar w^\aD=
\epsilon^{\aD\bD}(w_\bD)^t\ ,\qquad
(a'_{\alpha\aD})^t=a'_{\alpha\aD}\ .
\elabel{breality}\end{equation}
These reality conditions are only preserved by the subgroup
$\H(k)\subset\U(k')$ of the auxiliary symmetry group of the $\SU(N')$ theory.
The matrices $a'_n$ are Hermitian and, by \eqref{breality},
$t$-symmetric, {\it i.e.\/}~real
symmetric in the case of auxiliary group $\O(k)$, and
symplectic anti-symmetric in the case of auxiliary group $\Sp(k)$.
It is easy to verify that the
ADHM constraints \eqref{badhm} themselves are anti-Hermitian
$t$-anti-symmetric, in other words $\H(k)$ adjoint-valued.
It is straightforward to show that these reality conditions are
precisely what is required to render the gauge field \eqref{vdef}
$t$-anti-symmetric, in other words to restrict it
to an $sp(N)$ and $so(N)$ subalgebra of $su(N')$, respectively.

The Grassmann collective coordinates are subject to a similar set of reality
conditions:
\begin{equation}
\bar\mu=\mu^t\ ,\qquad
\qquad(\CM^{\prime}_\alpha)^t=\CM^{\prime}_\alpha\ .
\elabel{freality}\end{equation}
So ${\cal M}^{\prime}_\alpha$ is $t$-symmetric. The fermionic ADHM
constraints \eqref{fadhm} are, like their bosonic counterparts
$t$-anti-symmetric.\footnote{In proving this it
is useful to notice that $(w_\aD^t)^t=-w_\aD$
and $(\mu^t)^t=-\mu$.}

We can now count the number of $c$-number
and Grassmann collective coordinates. For both $\Sp(N)$ and $\SO(N)$ at
instanton number $k$ there are
$4kN$ real independent $w$ variables, taking into account
the reality conditions. The number
of $a'_n$ variables is
$4\times k(k+1)/2$ and $4\times k(2k-1)$, for $\Sp(N)$ and $\SO(N)$,
respectively. The physical moduli space is then the space of these
variables modulo the three $\H(k)$-valued
ADHM constraints \eqref{badhm} and auxiliary $\H(k)$ symmetry.
Hence the dimension of the physical moduli space is
$4k(N+1)$ and $4k(N-2)$, for $\Sp(N)$ and $\SO(N)$, respectively. This
agrees with the counting via the Index Theorem.
The counting of the Grassmann sector of the physical moduli space
goes as follows. There are $2kN$ real
degrees-of-freedom in $\mu$ and $2\times k(k+1)/2$
and $2\times k(2k-1)$, in ${\cal
M}^{\prime}_\alpha$, for $\Sp(N)$ and $\SO(N)$, respectively.
The ADHM constraints then impose $2\times k(k-1)/2$
and $2\times k(2k+1)$ conditions, for $\Sp(N)$ and $\SO(N)$,
respectively. Hence there are $2k(N+1)$ and $2k(N-2)$ real physical
Grassmann collective coordinates for $\Sp(N)$ and $\SO(N)$, respectively.
Again this agrees with the counting via the Index Theorem.

As an example, consider the $\Sp(1)$ theory. Since
$\Sp(1)\simeq\SU(2)$, this should have the same content as
the $\SU(2)$ ADHM
construction that we have described in \S\ref{sec:S8}; however, we
will find that the $\Sp(1)$ description is more economical in that
there are fewer ADHM variables subject to fewer constraints for a given
instanton number. This makes the $\Sp(1)$ formalism particularly attractive in
certain applications (for instance in \S\ref{sec:S60}).
In this case, the reality conditions \eqref{breality} are explicitly
\EQ{
w_{ui\aD}^*=\epsilon^{\aD\bD}J_{uv}w_{ui\bD}\ ,\qquad
(a'_n)_{ij}=(a'_n)_{ji}\ .
}
Given that $a'_n$ are Hermitian, the second condition implies that
$a'_n$ are real symmetric $k\times k$ matrices.
We now write the gauge
group indices as $\alpha=1,2$ rather than $u=1,2$. In this case the
first condition becomes
\EQ{
w_{i\alpha\aD}^*=\epsilon^{\aD\bD}\epsilon^{\alpha\beta}w_{i\beta\bD}\ .
}
This means that $w_i$ are quaternions of the form \eqref{raf}, so
$w_{i\alpha\aD}=w_{in}\sigma_{n\alpha\aD}$ for real $w_{in}$.
The fact that the gauge indices are also labelled by $\alpha$ means
that the ADHM variable $a_\aD$ can be written as a quaternion:
\EQ{
a_{\alpha\aD}=\MAT{w_{\alpha\aD}\\ a'_{\alpha\aD}}\ .
\label{giiy}
}
At instanton number $k$, there are $2k(k+3)$ variables $a_{\alpha\aD}$
subject to $3k(k-1)/2$ ADHM constraints and $k(k-1)/2$ symmetries to
give the dimension of $\ms_k$ as $8k$. This compares with the
$\SU(2)$ description, where there are $4k(k+2)$ variables $a_\aD$
subject to $3k^2$ ADHM constraints and $k^2$ symmetries. Clearly the
$\Sp(1)$ formalism is more economical. For example, at the
one-instanton level there are no ADHM constraints in the $\Sp(1)$
formalism compared with three in the $\SU(2)$ formalism.

In general all the formulae that we have established in the instanton
calculus carry through to the $\Sp(N)$ and $\SO(N)$ cases without
change. The only difference is the ADHM variables are subject to the
reality conditions \eqref{breality} and \eqref{freality}. In
particular, the collective coordinate integration measure and the
instanton effective action have the same form established in
\S\ref{sec:S35}. Only the normalization constant for
the integral over the instanton moduli space $C^{\sst(\N)}$ is changed.

\subsection{Matter fields and the ADHM construction}\elabel{app:A6}

In this section we consider various aspects of the
instanton calculus in supersymmetric
theories with matter fields. We will only consider matter in the
fundamental representation of the gauge group (although the ADHM
formalism can be extended to any higher representation by the tensor
product formalism developed in \cite{CGTone}).
There are two distinct
applications: to $\N=1$ theories on the Higgs branch and $\N=2$
theories on the Coulomb branch.

\subsubsection{$\N=1$ theories on the Higgs branch}

Let us start by considering
a theory with the $\N=1$ vector multiplet coupled to a
a single fundamental chiral multiplet $Q=(q,\chi)$, where
$q$ is the scalar (Higgs) field and
$\chi$ is the Weyl fermion partner (Higgsino). The Euclidean space
action for the matter fields (with no super-potential) is
\EQ{
S_{\text{matter}}=\int d^4x\,\Big\{{\cal D}_nq^\dagger{\cal D}_nq-{\cal
D}_n\bar\chi\bar\sigma_n\chi-\sqrt2ig\bar\chi\bar\lambda q+\sqrt2ig
q^\dagger\lambda\chi+\tfrac14g^2(q^\dagger q)^2\Big\}\ .
\elabel{actmat}
}
On the Higgs branch,
the scalar field will
have an arbitrary VEV $q^0$ and the instantons become constrained. As
previously we can capture the leading-order behaviour in the
semi-classical limit by an appropriate approximate instanton
solution. As on the Coulomb branch of the $\N=2$ theories discussed in
\S\ref{sec:S33}, the anti-chiral fermions are zero for the
approximate instanton: $\bar\lambda=\bar\chi=0$. To leading
order, therefore, the fundamental-valued
chiral fermion $\chi$ satisfies the source free Weyl
equation in the ADHM instanton background
\EQ{
\Dbarslash\chi=0\ .
\elabel{covWeyl}
}
The fundamental fermion zero modes were originally constructed in
\cite{CGTone}.
In our language, they read:
\EQ{
\chi_{\alpha}=
g^{-1/2}\bar{U}b_{\alpha}f{\cal K}\ ,
\elabel{fund}
}
where ${\cal K}_i$ are $k$ new Grassmann collective
coordinates. In the case a fundamental fermion there are no analogues
of the
fermionic ADHM constraints for the new Grassmann coordinates $\K$. So
there are $k$ independent zero modes which agrees with counting via the Index
Theorem. The proof that \eqref{fund} satisfies \eqref{covWeyl} is a
straightforward exercise in ADHM algebra, using \eqref{cmpl},
\eqref{silid} and \eqref{fderivs}:
\SP{
\Dbarslash^{\aD\alpha}\chi_\alpha=&g^{-1/2}\bar\sigma_n^{\aD\alpha}
\big(\partial_n\bar
U \Delta_\bD f\bar\Delta^\bD b_\alpha f+\bar
Ub_\alpha\partial_nf\big)\K\\
=&g^{-1/2}\bar\sigma_n^{\aD\alpha}\bar U\big(-b_\alpha f\bar\Delta^\bD
b^\beta\sigma_{n\beta\bD}f-b^\beta\sigma_{n\beta\bD}f\bar\Delta^\bD
b_\alpha f\big)\K=0\ .
}

On the other hand, to leading order the Higgs field $q$
satisfies an inhomogeneous covariant Laplace equation:
\EQ{
{\cal D}^2 q=\sqrtwo ig\lambda\chi
\elabel{squarkeq}
}
together with the VEV boundary conditions
\EQ{
q\  \overset{x\rightarrow\infty}=
\ \qvac\ ,
\elabel{vevbcon}
}
where $\qvac=(q^0_1,\ldots,q^0_N)$
denotes the fundamental VEV. The right-hand side of
Eq.~\eqref{squarkeq} is the product of the classical configurations
\eqref{lam} and \eqref{fund},
respectively. The general solution to Eqs.~\eqref{squarkeq}-\eqref{vevbcon}
is
\EQ{
q=\bar{U}\MAT{\qvac_{\sst[N]} \\ 0_{\sst[2k]}}-
\tfrac i{2\sqrt2}\bar U\CM f\K\ .
\elabel{squarkeqm}
}
The proof is another straightforward exercise in ADHM algebra (very similar to
that in Appendix \ref{app:A4} Eq.~\eqref{uyt}).
Firstly, similar to \eqref{gdd},
\EQ{
{\cal D}^2(\bar U\J)=-4\bar Ub^\alpha f\bar b_\alpha\J
+\bar U\partial^2\J -2\bar Ub^\alpha
f\sigma_{n\alpha\aD}\bar\Delta^\aD
\partial_n\J\ .
\elabel{mgdd}
}
Now take $\J=-\tfrac i4\CM f\K$ and compare with
\EQ{
g\lambda\chi=\bar U\big(\CM f\bar b^\alpha-
b^\alpha f\bar\CM\big){\cal P}b_\alpha f\K\ .
\elabel{mbxx}
}
Using the differentiation formulae \eqref{ddf} and \eqref{fderivs}
along with  $\bar\Delta_\aD\CM=-\bar
\CM\Delta_\aD$ and $\bar b_\alpha\CM=\bar\CM b_\alpha$, one finds that
the 2nd term in \eqref{mgdd} matches the 1st term in \eqref{mbxx},
while the sum of the 1st and 3rd terms in \eqref{mgdd} matches the 2nd
term in \eqref{mbxx}.

As usual in the supersymmetric instanton calculus,
$q^\dagger$ ceases to be the conjugate of $q$ in the
instanton background, since it satisfies ${\cal D}^2q^\dagger=0$ to
leading order. Consequently
\EQ{
q^\dagger=\MAT{(\qvac)^\dagger_{\sst[N]} & 0_{\sst[2k]}}U\ .
\elabel{squarkeqd}
}
Notice that $q^\dagger$ fails to be the conjugate of $q$ due to the
presence of terms bi-linear in the Grassmann collective coordinates in
the latter. This is a---by now familiar---symptom
of working in Euclidean space.

To capture the leading order behaviour we have to take the approximate
instanton solution, including the matter field solutions discussed
above, and substitute them into the Euclidean action. The gauge field term in
the component Lagrangian yields $-2\pi i\tau$ as always.
Following the method of Refs.~\cite{MO-II,DKMn4},
the two other relevant terms of the
action, namely the Higgs kinetic term and the Yukawa interaction
involving the chiral fermions 
are turned into a surface term with an integration by parts in the former
together with the Euler-Lagrange equation \eqref{squarkeq}
for the fundamental scalar:
\EQ{
\tilde S=\int d^4x\,\Big\{{\cal D}_nq^\dagger{\cal D}_nq+i\sqrt2
q^\dagger\lambda\chi\Big\}=\int d^4x\,\Big\{
\partial_n\big(q^\dagger{\cal D}_nq\big)
+q^\dagger\big(-{\cal D}^2q+\sqrt2ig\lambda\chi\big)\Big\}
\ .
}
Using Gauss's Theorem, the contribution to the action
may then be extracted from the asymptotic fall-off at infinity
\EQ{
\frac{x_n}x\D_n q\ \overset{x\to\infty}\longrightarrow\
{1\over 2 x^3}
\Big(
w_{\dalpha} \bar w^{\dalpha} \qvac
 + \tfrac i{\sqrt2}\,\mu\K\Big)
\elabel{funasym}
}
and hence to lowest order the instanton effective action is
\EQ{
\tilde S=
\pi^2\, \Big\{(\qvac)^\dagger w_\aD \bar w^\aD\,\qvac
+\tfrac i{\sqrt2}(\qvac)^\dagger\mu\K\Big\}\ .
\elabel{funaction}
}
This $k$-instanton formula, although written in ADHM collective coordinates,
is nevertheless easily compared with the one-instanton
expression for the action with $\SU(2)$ gauge symmetry
found in Ref.~\cite{NSVZ}. The first
term in parentheses is equivalent to $\sum_i\,|q^0|^2\,\rho_i^2,$
summed over the $k$ different instantons, where $\rho_i$ is the
scale size of the $i^{\rm th}$ instanton. Also the second term in parentheses
is the fermion bi-linear necessary to promote this $\rho_i^2$ to
the supersymmetric invariant scale size constructed in \cite{NSVZ}.
Independent of one's choice of collective coordinates,
the presence of the VEV in the instanton effective
action \eqref{funaction} gives a natural
cut-off to the integrations over instanton scale sizes \cite{tHooft},
leading to an infrared-safe application of instanton calculus.

The expressions given above may be immediately extended to phenomenologically
more interesting models with $N_F$ fundamental
flavors of Dirac fermions (known as a hypermultiplet). In this
case the gauge multiplet is minimally coupled to $2N_F$ chiral superfields
$Q_f$ and $\tilde Q_f,$ $\,1\le f\le N_F,\,$ where $Q_f$ transforms in
the $\BN$  and $\tilde Q_f$ in the $\bar{\BN}$
representation of the gauge group. We will take $Q$ ($\tilde Q$)
to be an $N\times N_F$ ($N_F\times N$), but for greater clarity we
will often write the flavour indices explicitly.
The action for the matter fields is
\SP{
S_{\text{matter}}=&\int d^4x\,\Big\{{\cal D}_nq^\dagger{\cal D}_nq+
{\cal D}_n\tilde q{\cal D}_n\tilde q^\dagger-{\cal
D}_n\bar\chi\bar\sigma_n\chi+
\tilde\chi\sigma_n{\cal
D}_n\bar{\tilde\chi}\\
&-\sqrt2ig\bar\chi\bar\lambda q+i\sqrt2
q^\dagger\lambda\chi+\sqrt2ig\tilde q\bar\lambda\bar{\tilde\chi}-\sqrt2ig
\tilde\chi\lambda\tilde q^\dagger+\tfrac14g^2(q^\dagger q-\tilde q\tilde
q^\dagger)^2\Big\}\ .
\elabel{actmatp}
}
The classical moduli space of the theory in the Higgs phase
is given by \cite{IS,Shifman}:
\EQ{
\qvac_{uf}= \MAT{
{\vhiggs}_1 & 0 & \ldots & 0  \\
0 & {\vhiggs}_2 & \ldots & 0  \\
\vdots & \vdots & \ddots  & \vdots \\
0 & 0 & \ldots &{\vhiggs}_f\\
\vdots & \vdots &\ddots & \vdots\\
0 & 0 & \ldots & 0}\ ,\qquad
\tilde q^0_{fu} = \MAT{
\tilde\vhiggs_1 & 0 & \ldots & 0 & \ldots & 0 \\
0 & \tilde\vhiggs_2 & \ldots & 0 & \ldots & 0 \\
\vdots & \vdots & \ddots  & \vdots  & \ddots  & \vdots \\
0 & 0 & \ldots & \tilde\vhiggs_f & \ldots & 0 } \ .
\elabel{qvev}
}
The VEV matrices   in Eq.~\eqref{qvev} correspond to the cases
$N_F \leq N$. The cases $N_F > N$ are similar except that the VEV
matrices have extra rows of zeroes rather than columns, or vice-versa.
These VEVs are not all independent;
the D-flatness condition requires that for each value of $f$,
\EQ{
|v_f|^2 =\begin{cases}
 |\tilde v_f|^2+ a^2 &
N_F \geq N\ ,\\
|\tilde v_f|^2 &
N_F < N\ , \end{cases}
\elabel{dflg}
}
where $a^2$ is an arbitrary constant, independent of the color index $u$.

Now Eqs.~\eqref{fund} and \eqref{squarkeq} generalize to\footnote{Here,
and in the following, one should not confuse the flavour subscript $f$
with the ADHM quantity $f$.}
\EQ{
\chi_{\alpha f}=
g^{-1/2}\bar{U}b_{\alpha}f{\cal K}_f
\ ,\qquad
\tilde{\chi}_{f\alpha} =g^{-1/2}
\tilde{\cal K}_{f} f\bar b_{\alpha} U
\elabel{fundredux}
}
and
\EQ{
q_{f}=\bar{U}\MAT{ \qvac_f \\ 0} -
\tfrac i{2\sqrt2}\bar U
\CM f\K_{f}  \ , \qquad
\tilde{q}_{f}=\MAT{
\tilde q^0_{f} &0}U +
\tfrac i{2\sqrt2}
\tilde\K_{f}f
\bar\CM U\ ,
\elabel{squarkdredux}
}
respectively. The leading contribution to the instanton effective action can be
worked out in a completely analogous way to \eqref{funaction}, yielding
\EQ{
\tilde S= \pi^2\sum_{f=1}^{N_F}
\Big\{q_{f}^{0\dagger}
w_{\aD}\bar w^{\aD}q^0_f
+\tfrac i{\sqrt2}q_f^{0\dagger}\mu\K_{f}
+\tilde q^0_f
w_{\aD}\bar w^{\aD}\tilde q^{0\dagger}
-\tfrac i{\sqrt2}\tilde\K_{f}\bar\mu\tilde q^{0\dagger}_f
\Big\}\ .
\elabel{funactionredux}
}
The $\N=1$
supersymmetry transformation
properties of the ADHM variables were constructed in \S\ref{sec:S32}.
To check the invariance of the expression \eqref{funactionredux}, it is
necessary as well to derive the transformation properties for the
Grassmann collective coordinates $\K$ and $\tilde\K$ associated with the
fundamental fermions. As with the other collective coordinates, this
may be straightforwardly accomplished by equating ``active'' and ``passive''
supersymmetry transformations on the Higgsinos $\chi$ and $\tilde\chi$.
In this way one obtains:
\EQ{
\delta\K_{f}=
- 2 \sqrt2 \bar\xi_\dalpha \wbar^\aD
\qvac_{f}  \ ,\qquad
\delta\tilde\K_{f}= - 2 \sqrt2 \tilde q^0_{f}w_{\aD}
\bar\xi^\aD\ .
\elabel{deltaKnz}
}
It is now easily checked that the action \eqref{funactionredux} is invariant
under the supersymmetry transformations \eqref{uiui}, \eqref{vivi}
and  \eqref{deltaKnz}.

It remains to determine how one integrates over the new Grassmann
collective coordinates in the semi-classical approximation of the
functional integral. As for the adjoint-valued fields, the non-zero
mode fluctuation determinants
cancel, up to a power of the Pauli-Villars mass scale $\mu^{-2kN_F}$. The
measure for integrating over the matter zero modes follows from the
inner-product formula
\EQ{
\int d^4x\,\sum_{f=1}^{N_F}\tilde\chi_f\chi_f
=g^{-1}\int d^4x\,\sum_{f=1}^{N_F}
\tilde\K_f f\bar b_\alpha{\cal P}b^\alpha\K_f=-\frac1{4g}\int
d^4x\,\sum_{f=1}^{N_F}
\tilde\K_f\,\square\,f\K_f=\frac{\pi^2}{g}\sum_{f=1}^{N_F}\,\tilde\K_f\K_f\ ,
\elabel{ipform}
}
employing \eqref{ddf}. The integration
measure for the matter Grassmann collective coordinates is then
\EQ{
\pi^{-2kN_F}\mu^{-kN_F}g^{kN_F}\int \,d^{kN_F}{\cal
K}\,d^{kN_F}\tilde{\cal K}\ .
\elabel{msci}
}
For convenience, we define the combination
\EQ{
\int_{\ms_k}\Bomega^{\sst(\N,N_F)}=
\pi^{-2kN_F}\int_{\ms_k}\Bomega^{\sst(\N)}\,\cdot\,d^{kN_F}{\cal
K}\,d^{kN_F}\tilde{\cal K}
\label{mattmeas}
}
and then a generalization of the instanton partition function \eqref{ipfun}
\EQ{
{\EuScript Z}_k^{\sst(\N,N_F)}=\int_{\ms_k}\Bomega^{\sst(\N,N_F)}e^{-\tilde
S}
\elabel{ipfunh}
}
to include the matter fields.

We will see that the
$\N=1$ action \eqref{funactionredux} possesses two simplifying
properties that the $\N=2$
action, to be discussed below, does not.
First, Eq.~\eqref{funactionredux}
 has the form of a disconnected sum of $k$ single instantons;
with the choice of these ADHM coordinates there is no interaction between them.
Second, the only gaugino modes that are lifted ({\it i.e.\/}, that appear
in the action) are those associated
with the top elements $\mu$ and $\bar\mu$ of the collective coordinate
matrices
$\CM$ and $\bar\CM$. This leaves ${\cal O}(k)$ unlifted gaugino modes
after one implements the fermionic
constraints \eqref{fadhm}. This counting contrasts
sharply with the $\N=2$ theories in which the number of unlifted modes
is independent of the winding number $k$.
Saturating each of these unlifted modes with an anti-Higgsino as per
Affleck, Dine and Seiberg \cite{ADS} one sees that
unlike the $\N=2$ theory, here the sectors of
different topological number cannot interfere with one another, since the
corresponding Green's functions
are distinguished by different (anti-)fermion content.

\subsubsection{$\N=2$ theories on the Coulomb branch}\elabel{sec:S91}

Now we consider $\N=2$ theories with
$N_F$ flavors of fundamental hypermultiplets.
Each such hypermultiplet comprises
a pair of $\N=1$ chiral multiplets, $Q_{f}$ and $\tilde{Q}_{f}$, with
the same conventions for component fields as in the $\N=1$ case
discussed in the last section.
In $\N=1$ language, these matter fields couple to the gauge multiplet
via a superpotential,
\EQ{
W =\sqrt2g\sum_{f=1}^{N_{F}}\tilde{Q}_{f}\Phi
Q_{f}\ .
\elabel{superp}
}
In component form, the matter fields have the Euclidean action
\SP{
S_{\text{matter}}=&
\int d^4x\,\Big\{{\cal D}_nq^\dagger{\cal D}_nq+
{\cal D}_n\tilde q{\cal D}_n\tilde q^\dagger-{\cal
D}_n\bar\chi\bar\sigma_n\chi+
\tilde\chi\sigma_n{\cal
D}_n\bar{\tilde\chi}\\
&-\sqrt2ig\bar\chi\bar\lambda q+i\sqrt2
q^\dagger\lambda\chi+\sqrt2ig\tilde q\bar\lambda\bar{\tilde\chi}-\sqrt2ig
\tilde\chi\lambda\tilde q^\dagger-\sqrt2g\tilde\chi\psi q-\sqrt2g\tilde
q\psi\chi\\
&\qquad\qquad
-g\tilde\chi\phi\chi-\sqrt2gq^\dagger\bar\psi\bar{\tilde\chi}
-\sqrt2g\bar\chi\bar\psi\tilde
q^\dagger-g\bar\chi\phi^\dagger\bar{\tilde
\chi}\Big\}+S_{\text{scalar}}\ ,
\elabel{actmatpn}
}
where $S_{\text{scalar}}$ are the interaction terms between the scalar
fields whose explicit form we do not need.
In what follows, we will restrict our attention to the Coulomb branch of
the $\N=2$ theory where the hypermultiplet squarks do not acquire VEVs.
The classical component fields $\chi_f,$ $\tilde\chi_f,$ $q_f$ and $\tilde
q_f$ are still given by Eqs.~\eqref{fundredux}-\eqref{squarkdredux},
except that on the Coulomb branch
the first terms on the right-hand sides of Eq.~\eqref{squarkdredux} are
zero. The adjoint-valued fermions have their usual ADHM form
$\lambda=g^{-1/2}\Lambda(\CM^1)$ and $\psi=g^{-1/2}
\Lambda(\CM^2)$. To leading order, the
scalar field $\phi$
satisfies the same equation-of-motion as in the pure $\N=2$ gauge
theory and so the solution follows from Eqs.~\eqref{ssdd} and
\eqref{dxx} (along with definition of $\phi$ in
\eqref{ntsc}):\footnote{In the following when discussing the $\N=2$
theory, we use the $\epsilon$ tensor to raise and lower $\SU(2)$
$R$-symmetry spinor indices in the usual way following the conventions
of \cite{WB}.}
\EQ{
\phi=
\tfrac i{2}\bar U\CM^A f\bar\CM_AU+\bar
U\MAT{\phi^0& 0\\ 0&\varphi1_{\sst[2]\times[2]}}U\ ,
}
where
\EQ{
\varphi=
\BL^{-1}\Big(-\tfrac i{2}\bar\CM^A\CM_A+\bar
w^\aD\phi^0w_\aD\Big)\ .
\elabel{dvpoo}
}
On the other hand the anti-holomorphic component now satisfies the
inhomogeneous equation
\EQ{
{\cal D}^2\phi^\dagger=-g\chi\tilde\chi\ .
}
The solution of this equation is readily shown to be
\EQ{
\phi^\dagger=\bar U\MAT{\phi^{0\dagger} & 0\\ 0&
\varphi^\dagger1_{\sst[2]\times[2]}}U\ ,
\label{aholsf}
}
where
\EQ{
\varphi^\dagger=\BL^{-1}\Big(-\tfrac 1{4}\sum_{f=1}^{N_F}
\K_f\tilde\K_f+\bar w^\aD\phi^{0\dagger} w_\aD\Big)\ .
}
Now we consider the matter fields themselves.
The new feature in the $\N=2$ theory is that the fields
$q^\dagger$ and $\tilde q^\dagger$ now satisfy the non-trivial
equations\footnote{As usual, to
leading order where we set the anti-chiral fermions to zero and
ignore the potential terms for the scalar fields.}
\EQ{
{\cal D}^2q^\dagger=-\sqrt2g\tilde\chi\psi\ ,\qquad
{\cal D}^2\tilde q^\dagger=-\sqrt2g\psi\chi
}
with solutions
\EQ{
q^\dagger_f=\tfrac1{2\sqrt2}\tilde\K_f f\bar\CM^2 U\ ,\qquad
\tilde q^\dagger_f=\tfrac1{2\sqrt2}\bar U\CM^2 f\K_f\ .
}

As usual, only the kinetic terms of the scalar fields and the Yukawa
interactions involving the fermions (rather than the anti-chiral fermions)
contribute at leading order to the instanton effective action:
\SP{
\tilde S=&\int d^4x\,\Big\{{\cal D}_n\phi^\dagger{\cal D}_n\phi+
{\cal D}_nq^\dagger{\cal D}_nq+
{\cal D}_n\tilde q{\cal D}_n\tilde q^\dagger
+2ig[\phi^\dagger,\lambda]\psi\\&\qquad+\sqrt2ig
q^\dagger\lambda\chi-\sqrt2ig
\tilde\chi\lambda\tilde q^\dagger-\sqrt2g\tilde\chi\psi q-\sqrt2g\tilde
q\psi\chi-g\tilde\chi\phi\chi\Big\}\ .
\elabel{poop}
}
This can be simplified by integrating the kinetic terms by parts
and using the equations-of-motion for the scalar fields:
\EQ{
\tilde S=
\int d^4x\,\Big\{\partial_n(\phi^\dagger{\cal D}_n\phi)+
\partial_n(q^\dagger{\cal D}_nq)+
\partial_n(\tilde q{\cal D}_n\tilde q^\dagger)
-g\tilde\chi(\sqrt2i\lambda\tilde q^\dagger+\sqrt2\psi q
+\phi\chi)\Big\}\ ,
\elabel{poopb}
}
The first three terms are converted into surface terms on a large
sphere at infinity. Only $\phi$ has a VEV, so only the first
term is none vanishing. Using
\EQ{
\frac{x_n}{x}{\cal D}_n\phi\ \overset{x\to\infty}\longrightarrow\
\frac1{x^3}\Big\{2i\mu^A\bar\mu_A+w_\aD\bar
w^\aD\phi^0+\phi^0w_\aD\bar w^\aD
-2w_\aD\varphi\bar w^\aD\Big\}
}
we find the contribution to $\tilde S$:
\EQ{
4\pi^2{\rm tr}_k\Big\{-\tfrac
i{2}\bar\mu^A\phi^{0\dagger}\mu_A
+\bar w^\aD\big|\phi^0\big|^2w_\aD
-\varphi\bar w^\aD\phi^{0\dagger} w_\aD\Big\}\ .
\elabel{ktcon}
}

The contribution from the remaining Yukawa interactions can be
evaluated in a similar way to the manipulations required to evaluate
the instanton effective action in \S\ref{sec:S38} (see also
Appendix~\ref{app:A4}
Eq.~\eqref{psibardef} and \cite{AOYAMA}). The strategy is to write
\EQ{
\sqrt2i\lambda\tilde q^\dagger+\sqrt2\psi q
+\phi\chi=g^{-1/2}\big(\Dslash\bar\Upsilon+\Theta\big)\ ,
\elabel{poopc}
}
where $\Theta$ is a fundamental zero mode, $\Dbarslash\Theta=0$.
We now verify that the solution for $\bar\Upsilon$ and $\Theta$ is
\EQ{
\bar\Upsilon^\aD_f=-\tfrac i{4}
\bar U\CM^A f\bar\Delta^\aD\CM_A
f\K_f+
\bar U\MAT{\phi^0&0\\ 0&\varphi}\Delta^\aD f\K_f
}
and
\EQ{
\Theta_f=\bar Ub_\alpha f\varphi\K_f\ .
}
First of all, using \eqref{cmpl}, the left-hand side of \eqref{poopc} is
\SP{
g^{1/2}\Big(\sqrt2i\lambda\tilde q^\dagger+\sqrt2\psi q
+\phi\chi\Big)=
&\tfrac i{2}\bar U\Big\{\CM^A fb_\alpha{\cal
P}\CM_A-
b_\alpha
f\bar\CM^A{\cal P}\CM_A\\
&+\CM^A f\bar\CM_A{\cal P}b_\alpha\Big\}f\K_f+
\bar U\MAT{\phi^0&0\\ 0&\varphi}{\cal P}b_\alpha f\K_f\ .
\elabel{puupa}
}
Then using the formula
\EQ{
{\cal D}_{\alpha\aD}(\bar U{\cal J})=\bar U\partial_{\alpha\aD}{\cal
J}-2\bar Ub_\alpha f\bar\Delta_\aD{\cal J}\ ,
}
we have
\SP{
\Dslash_{\alpha\aD}\bar\Upsilon^\aD&=
\tfrac i{2}\epsilon_{AB}\bar U\Big\{b_\alpha f\bar\CM^A(1-{\cal P})\CM_A
+\CM^A f\bar b_\alpha{\cal P}\CM_A+\CM^A f\bar\CM_A{\cal
P}b_\alpha\Big\}f\K_f \\
&+\bar U\bigg\{\MAT{\phi^0&0\\ 0&\varphi}(
{\cal P}-1)b_\alpha+
b_\alpha f\bar\Delta^\aD\MAT{\phi^0&0\\ 0&\varphi}\Delta_\aD\bigg\} f\K_f\ .
\elabel{puupb}
}
In the last term we can now use the identity \eqref{funiwl}, along with
the expression for $\varphi$ in \eqref{dvpoo}:
\EQ{
\bar\Delta^\aD\MAT{\phi^0&0\\ 0&\varphi}\Delta_\aD=-\tfrac
i{2}\bar\CM^A\CM_A+\{\varphi,f^{-1}\}\ .
}
This simplifies \eqref{puupb}
\SP{
\Dslash_{\alpha\aD}\bar\Upsilon^\aD&=
\tfrac i{2}\bar U\Big\{-b_\alpha f\bar\CM^A{\cal P}\CM_A
+\CM^A f\bar b_\alpha{\cal P}\CM_A+\CM^A f\bar\CM_A{\cal P}b_\alpha
\Big\}f\K_f\\ &\qquad\qquad
+\bar U\MAT{\phi^0&0\\ 0&\varphi}
{\cal P}b_\alpha f\K_f+\bar Ub_\alpha f\varphi\K_f\ .
\elabel{puupc}
}
Now one can see that apart from the last term, which
is the fundamental zero mode
$\Theta$, this is precisely \eqref{puupa}. So the contribution to
$\tilde S$ from the Yukawa terms is given by
\EQ{
-g^{1/2}\int d^4x\,\tilde\chi\Theta\ ,
}
which can be evaluated using the inner-product formula
\eqref{ipform}:
\EQ{
-\pi^2\sum_{f=1}^{N_F}\tilde\K_f\varphi\K_f=
\frac{\pi^2}{g^2}
{\rm tr}_k\Big[\Big(\sum_{f=1}^{N_F}\K_f\tilde\K_f\Big)\BL^{-1}\big(
-\tfrac i2\bar\CM^A\CM_A
+\bar w^\aD\phi^0w_\aD\big)\Big]\ .
}
Summing the result with \eqref{ktcon} gives the leading order expression
for the instanton effective action:
\SP{
\tilde S&=4\pi^2{\rm tr}_k\Bigg\{-\tfrac
i2\bar\mu^A\phi^{0\dagger}\mu_A+\bar
w^\aD\big|\phi^0\big|^2w_\aD\\
&+\Big(\tfrac 1{4}\sum_{f=1}^{N_F}
\K_f\tilde\K_f-\bar w^\aD\phi^{0\dagger} w_\aD\Big)\BL^{-1}\Big(
-\tfrac i{2}\bar\CM^A\CM_A+\bar w^\aD\phi^0w_\aD\Big)\Bigg\}\ .
\elabel{ieant}
}
As with the $\N=1$ action \eqref{funactionredux}, one can check that
this expression is a supersymmetric invariant. Notice on the Coulomb branch,
Eq.~\eqref{deltaKnz} collapses to
\EQ{
\delta\K_f = \delta\tilde\K_f=0\ .
\elabel{deltaKzero}
}

We can also add $\N=2$ preserving masses for the hypermultiplets as
explained in \S\ref{app:A8} below.
Finally, the collective coordinate integration measure for the
Grassmann collective coordinates is given by the same expression,
Eq.~\eqref{msci}, as in the $\N=1$ theory.

\subsection{Masses}\elabel{app:A8}

In certain circumstances one wants to add masses for some of the
fields. Here we assess the affects on the instanton calculus. It turns
out that the effect will be simple to incorporate when working to
leading order in the semi-classical approximation.

The interesting cases involve $\N=1$ preserving mass terms in
either $\N=2$ or $\N=4$ theories.
The simplest case consists of a mass term for a fundamental hypermultiplet
transforming in the $(\BN,\bar\BN)$ as described
in \S\ref{app:A6}. When Wick rotated to Euclidean space,
the conventional mass term in components is
\EQ{
S_{\text{mass}}=\int
d^4x\,\Big\{m\tilde\chi\chi+m^*\bar\chi
\bar{\tilde\chi}
+2|m|^2q^\dagger q+2|m|^2\tilde q\tilde q^\dagger\Big\}\ .
\elabel{masst}
}
In principle, these mass terms contribute new terms to the
equations-of-motion which will affect the instanton solutions
themselves. In applications one is typically
interested quantities that are
known to have holomorphic dependence on the mass. We then expect that
the semi-classical approximation respects this
dependence.\footnote{It should be possible to prove this by showing 
that the relevant
Ward identity is respected. Such a proof would follow the lines of
\S\ref{sec:S44} which considers
a different Ward identity.} Hence for a holomorphic quantity,
we can treat $m$ and $m^*$ as independent variables 
and then set $m^*=0$. The mass
terms then only affect the equations-of-motion of the anti-chiral fermions
$\bar\chi$ and $\bar{\tilde\chi}$ and the supersymmetric
instanton remains unaffected to leading order. Hence,
we simply have to evaluate the first term in Eq.~\eqref{masst} in the
background of the supersymmetric instanton
and add the resulting expression to the instanton
effective action $\tilde S$. The contribution can be determined
using the inner-product formula \eqref{ipform}. For $N_F$ such hypermultiplets
\EQ{
\tilde S_{\text{mass}}=\frac{\pi^2}{g}\sum_{f=1}^{N_F}m_f\tilde\K_f\K_f\ .
\elabel{simm}
}

Now we turn to masses for adjoint chiral superfields. 
As discussed above only the mass
term for the chiral fermions is relevant. In the $\N=4$ theory
the most general mass term for the chiral fermions is of the form
\EQ{
S_{\rm mass}=\int d^4x\,m_{AB}\TrN\,\lambda^A\lambda^B\ .
\label{genmss}
}
where $m_{AB}$ is a symmetric matrix which in a suitable basis is diagonal
$m_{AB}={\rm diag}(m_1,m_2,m_3,m_4)$. In the general case, all the
supersymmetry is broken, while if one, respectively two, of the
eigenvalues vanish then the coupling arise from mass terms that
preserve $\N=1$, respectively $\N=2$, supersymmetry.
To find the effect on the instanton effective action we substitute
the ADHM form $\lambda^A=g^{-1/2}\Lambda(\CM^A)$ and evaluate the
integral using
Corrigan's inner-product formula \eqref{corriganf}:
\EQ{
\tilde S_{\text{mass}}=-\frac{m_{AB}\pi^2}{g}{\rm
tr}_k\,\bar\CM^A(\Pinfty+1)\CM^B=
-\frac{m_{AB}\pi^2}{g}{\rm
tr}_k\,\big[2\bar\mu^A\mu^B+\CM^{\prime\alpha A}\CM^{\prime B}_\alpha\big]\ .
\elabel{nonepn}
}

\subsection{The instanton partition
function}\label{sec:N1}

For many applications and for later conceptual developments
it is useful to re-formulate the collective
coordinate integral, or instanton partition function \eqref{ipfun},
by introducing some auxiliary variables which have the effect of the
``linearizing'' problem in a sense to be described.
To start with one introduces
Lagrange multipliers for the
bosonic and fermionic ADHM $\delta$-function constraints and also, for
$\N>1$, other additional auxiliary variables. In Chapter
\ref{sec:S49} we will see for $\N>1$
this linearized form of the collective
coordinate integral has an important relation to higher-dimensional
field theories. This point-of-view will also be the starting point for
understanding the relation of the instanton calculus to D-branes in
string theory as described in \S\ref{sec:S100}.

In the linearized formalism one introduces auxiliary variables in
the form of: $\chi_a$ a $2(\N-1)$-vector of Hermitian $k\times
k$ matrices; $\vec D$,
a 3-vector of Hermitian $k\times k$ matrices; and $k\times k$ matrices
of Grassmann superpartners $\bar\psi^\aD_A$, $A=1,\ldots,\N$.
The instanton partition function \eqref{ipfun} can be written as
\EQ{
{\EuScript Z}_k^{\sst(\N)}=
\frac{2^{2(2-\N)k^2}\pi^{(2-3\N)k^2}
C_k^{\sst(\N)}}{{\rm Vol}\,\U(k)}\int\, d^{4k(N+k)} a \,
\,d^{3k^2}D\,d^{2(\N-1)k^2}\chi\,\prod_{A=1}^\N\,
 d^{2k(N+k)}{\cal M}^{A}\,d^{2k^2}\bar\psi_A\,
e^{-\tilde S}\ ,
\elabel{linvolf}
}
where the instanton effective action is
\EQ{
\tilde S=4\pi^2{\rm tr}_k\Big\{\chi_a\BL\chi_a+\tfrac
12\bar\Sigma_{aAB}\CM^A\CM^B\chi_a\Big\}
+\tilde S_{\text{L.m.}}\ ,
\elabel{ieal}
}
where
\EQ{
\tilde S_{\text{L.m.}}=-4i\pi^2{\rm
tr}_k\Big\{\bar\psi_A^\aD\big(\bar\CM^A
a_\aD+\bar a_\aD\CM^A\big)+\vec D\cdot\vec
\tau^{c\aD}{}_\bD\bar a^\bD a_\aD\Big\}\ .
}
The previous form of the collective coordinate integral \eqref{fms} is
recovered by integrating out the auxiliary variables
$\{\chi_a,\vec D,\bar\psi^\aD_A\}$. Specifically, integrating out
the Lagrange multipliers $\vec D$
and $\bar\psi_A$ yield the $\delta$-functions in \eqref{fms} imposing
the ADHM constraints and their Grassmann analogues, \eqref{badhm} and
\eqref{fadhm}. The Gaussian integrals over $\chi_a$ yield the
appropriate power of $\det_{k^2}\BL$ in \eqref{fms} as well as the
quadrilinear couplings of Grassmann collective coordinates of the
$\N=4$ theory \eqref{rrww}.

One of the advantages of the linearized form of the partition function
is that it is straightforward to incorporate scalar VEVs on
the Coulomb branch in the $\N=2$ and $\N=4$ theories.
One simply generalizes the following couplings in
the instanton effective action $\tilde S$:
\SP{
&w_\aD\chi_a\to w_\aD\chi_a+\phi^0_aw_\aD\ ,\qquad
\chi_a\bar w^\aD\to \chi_a\bar w^\aD+\bar w^\aD\phi^0_a\ ,\\
&\mu^A\chi_a\to\mu^A\chi_a+\phi^0_a\mu^A\ ,\qquad
\chi_a\bar\mu^A\to \chi_a
\bar\mu^A+\bar\mu^A\phi_a^0\ .
\elabel{addvev}
}
The effective instanton action is now explicitly
\EQ{
\tilde S=4\pi^2{\rm tr}_k\Big\{\,
\big|w_\aD\chi_a+
\phi^0_a w_\aD\big|^2
-[\chi_a,a'_n]^2
+\tfrac 12\bar\Sigma_{aAB}\bar\mu^A(\mu^B\chi_a+\phi^0_a\mu^B)
+\tfrac 12\bar\Sigma_{aAB}\CM^{\prime
A}\CM^{\prime B}\chi_a\Big\}+\tilde S_{\text{L.m.}}\ ,
\elabel{withvevs}
}
where we used the definition of $\BL$ in \eqref{vvxx}.
The supersymmetry transformations on the linearized system are
\ALT{2}{
\delta a'_{\alpha\aD}&=  i\bar{\xi}_{\dot{\alpha}A}
{\cal M}^{\prime A}_{\alpha}\ ,&\qquad
 \delta{\cal M}^{\prime A}_{\alpha}& =  -2i\Sigma_a^{AB}\bar\xi^{\aD}_B
[a'_{\alpha\aD},\chi_a]\ ,\label{susyl1}\\
\delta w_\aD&= i\bar\xi_{\aD A} \mu^A\ ,&\qquad
\delta \mu^A&=  -2i\Sigma_a^{AB}
\bar\xi^{\aD}_B\big(w_\aD\chi_a+\phi^0_a w_\aD\big)\ ,
\label{susyl2}\\
\delta\chi_a&=-\Sigma_a^{AB}\bar\xi_{\aD A}\bar\psi^\aD_B\ ,
&\qquad \delta \bar\psi_{A}^\aD &=
2\bar\Sigma_{abA}{}^B
[\chi_a,\chi_b]\bar\xi_{B}^\aD-i\vec D\cdot\vec\tau^{\aD}{}_\bD
\bar\xi_{A}^\bD\\
\delta\vec D&=
-i\vec\tau^{\aD}{}_\bD\Sigma_a^{AB}\bar\xi_{\aD
 B}[\bar\psi^\bD_A,\chi_a]\ .&&
\label{susyl3}
}
On integrating out the auxiliary variables
$\{\chi_a,\vec D,\bar\psi_A\}$, these transformations reduce to those
constructed in \S\ref{sec:S32}.

In the $\N=2$ theory it rather simple to incorporate fundamental
hypermultiplets. The instanton effective action,
generalizing \eqref{withvevs}, is
\SP{
\tilde S&=4\pi^2{\rm tr}_k\Big\{\,
\big|w_\aD\chi_a+
\phi^0_a w_\aD\big|^2
-[\chi_a,a'_n]^2
+\tfrac i2
\bar\mu^A(\mu_A\chi^\dagger+\phi^{0\dagger}\mu_A)\\
&\qquad+\tfrac i2\CM^{\prime
A}\CM'_A\chi^\dagger+\tfrac14\sum_{f=1}^{N_F}\K_f\tilde\K_f(
\chi-g^{-1}m_f)\Big\}+\tilde S_{\text{L.m.}}\ ,
\label{iean2}
}
where we have defined 
\EQ{
\phi^0=\phi^0_1-i\phi^0_2\ ,\qquad
\chi=\chi_1-i\chi_2
\label{defcchi}
}
and we have allowed for arbitrary hypermultiplet masses \eqref{simm}.
Integrating out $\chi_a$ gives the instanton effective action \eqref{ieant}.
The supersymmetry transformations \eqref{susyl1}-\eqref{susyl3}
are then augmented with
\eqref{deltaKzero}.

In applications of the instanton calculus it is useful to define
the notion of the ``centred instanton partition function''. This is
defined in terms of an integral over the centred instanton moduli
space $\cms_k$, \eqref{cmsd}, where the overall position coordinates and their
superpartners, the Grassmann collective coordinates for the
supersymmetries broken by the bosonic instanton solution (see \S\ref{sec:S30}),
\EQ{
X_n=-k^{-1}{\rm tr}_k\,a'_n\ ,\qquad\xi^A=\tfrac i4k^{-1}{\rm tr}_k\,
\CM^{\prime A}\ ,
}
have been factored off. First of all, from the expression for the
metric \eqref{qum} we have
\EQ{
ds^2_{\ms_k}=8\pi^2k\,dX_n\,dX_n+ds^2_{\widehat\ms_k}\ .
}
Therefore taking account of the normalization of the measure in the
non-supersymmetric theory
\eqref{bos}, we have
\EQ{
\int_{\ms_k}\Bomega=
\int\, (4\pi k)^2d^4X\ \cdot\
\int_{\widehat\ms_k}\Bomega\ .
\label{sepouta}
}
Now we consider the supersymmetric integral. The inner-product of the
supersymmetric zero modes \eqref{susymo} is from
\eqref{corriganf}
\EQ{
\int d^4x\,\TrN\,\Lambda(-4i\xi^A1_{\sst[k]\times[k]})\Lambda
(-4i\xi^B1_{\sst[k]\times[k]})=16\pi^2k\,\xi^A\xi^B\ .
}
This means
\EQ{
\int_{\ms_k}\Bomega^{\sst(\N,N_F)}=
\int\, (4\pi k)^2d^4X\,\prod_{A=1}^{\N}(32\pi^2 k)^{-1}\,d^2\xi^A\,\cdot\,
\int_{\widehat\ms_k}\Bomega^{\sst(\N,N_F)}\ .
\label{sepout}
}

Since the instanton effective action is always independent of $X_n$
and $\xi^A$, we can define the {\it centred instanton partition
function}, generalizing \eqref{ipfunh} as
\EQ{
\widehat{\EuScript Z}_k^{\sst(\N,N_F)}=
\int_{\widehat\ms_k}\Bomega^{\sst(\N,N_F)}e^{-\tilde
S}\ .
\elabel{cipfun}
}

\newpage

\rsen\section{The Gluino Condensate in $\N=1$ Theories}\elabel{sec:S43}

Pure $\N=1$ gauge theory is in some respects the simplest
supersymmetric gauge theory; however, a puzzle arose in the
mid-1980's over the numerical
value of the gluino condensate.
In this section, we will describe how the
puzzle---described in
detail below---can be resolved using the calculus of many instantons
and, in particular, by exploiting the
simplifications that occur in the large-$N$ limit.

It is well established that $\N=1$ supersymmetric gauge theory has $N$
physically equivalent vacua, as dictated by the Witten Index,
which differ by the phase of the gluino
condensate. By dimensional analysis one expects
\begin{equation}
\VEV{\tfrac{g^2}{16\pi^2}\TrN \lambda^2}=c\ \Lambda^3e^{2\pi iu/N}\ ,\qquad
u=1,\ldots,N\ ,
\elabel{expect}\end{equation}
where $\Lambda$ is the dynamical scale in the theory
while $c$ is a numerical
constant. In particular, by a powerful supersymmetric
non-renormalization theorem there can be no
perturbative corrections to the form \eqref{expect} since this would
be a series in $g$ and hence would
not be holomorphic in $\Lambda$.\footnote{In this context,
$\Lambda$ is a complex quantity carrying with it the phase
$e^{i\theta/3N}$.}  This suggests that if
$c$ can be calculated to leading semi-classical order then it will be exact.
Remarkably, there are two approaches in the
literature for calculating the gluino condensate
each superficially appearing to be exact
but which differ in their predictions of the
constant $c$. This disagreement is especially vexing in light of the
fact that {\it both\/}
involve the use of a single supersymmetric instanton. The first,
we shall call the ``strong-coupling instanton'' (SCI)
approach
\cite{Amati:1988ft,NSVZ,Rossi:1984bu,Amati:1985uz,Fuchs:1986ft},
while the second, we call the ``weak-coupling instanton'' (WCI)
approach \cite{ADS,Fuchs:1986ft,Novikov:1985ic,Shifman:1988ia}.

In the SCI
approach one calculates the instanton contributions to the correlation
functions of $g^2\TrN\lambda^2$ directly in the
strongly-coupled confining phase of the theory. Strictly speaking, it is
therefore not a semi-classical calculation
even though it uses instantons and ultimately
we will conclude that this is why the SCI approach is flawed.
In an instanton background, each insertion
of $g^2\TrN\lambda^2$ is quadratic in the Grassmann collective
coordinates and since the charge-$k$ instanton has $2kN$ fermion zero modes,
the latter can only contribute to the $kN$-point function.\footnote{This can
also be deduced from the relation between $\Lambda$ and the running
coupling \eqref{deflp} $\Lambda^{3N}\sim e^{2\pi i\tau(\mu)}$.}
Furthermore, by a Ward identity, which we prove in
\S\ref{sec:S44} is
respected in the instanton approximation, the $kN$-point correlation
function is independent of the $kN$ spacetime insertion points
$\{x^{(i)}\}$. The one-instanton calculation of the $N$-point
function, reviewed in
\S\ref{sec:S45}, can be done exactly and yields the result
\begin{equation}
\VEV{\tfrac{g^2}{16\pi^2}\TrN\lambda^2(x^{(1)})\times\cdots\times
\tfrac{g^2}{16\pi^2}\TrN\lambda^2(x^{(N)})}
={2^N\over(N-1)!\,(3N-1)}\,\Lambda^{3N}\ .
\elabel{SCIans}\end{equation}
In order to extract $\Vev{g^2
\TrN\lambda^2}$ from the correlator
\eqref{SCIans}, one then invokes cluster decomposition: taking
$|x^{(i)}-x^{(j)}|\gg\Lambda^{-1}$, and
remembering the independence of the correlator on $\{x^{(i)}\}$,
one simply replaces the left-hand side of Eq.~\eqref{SCIans} by
$\Vev{\tfrac{g^2}{16\pi^2}
\TrN\lambda^2}^N.$ Taking the $N^{\rm th}$ root,
the net result, based on this one-instanton calculation, reads:
\begin{equation}
\VEV{\tfrac{g^2}{16\pi^2}\TrN\lambda^2}_{\rm SCI}=
 {2\over\big[(N-1)!\,(3N-1)\big]^{1/N}}\ \Lambda^{3}\,e^{2\pi
iu/N}\ ,
\elabel{SCIansb}\end{equation}
where $u=1,\ldots,N$ indexes the $N$ vacua
of the $\SU(N)$ theory, and reflects the ambiguity in taking the
$N^{\rm th}$ root. In retrospect---as argued in
Refs.~\cite{Amati:1988ft,Amati:1985uz}---the reason why
the na\"\i ve instanton calculation of
$\Vev{\tfrac{g^2}{16\pi^2}\TrN\lambda^2}$ gives zero is that the $N$ vacua
are being averaged over with equal weight in the instanton approximation
and the phases cancel.\footnote{The same argument implies that
only the $kN$-point functions are non-vanishing, matching the selection
rule arising from counting fermion zero modes in the $k$-instanton background.}

In contrast, in the WCI approach, one modifies the pure gauge theory by adding
matter superfields in such a way that the gauge symmetry is either entirely
broken, or only an abelian subgroup remains, after the Higgs
mechanism. In this way the theory can be rendered weakly coupled in
the infra-red and
semi-classical approaches should be reliable.
Here, we choose the former option by adding $N_F=N-1$
matter hypermultiplets in the fundamental representation of the
gauge group. In the Higgs phase, all the $n$-point functions
of $g^2\TrN\lambda^2$ receive contributions from
constrained instantons. In particular, based on fermion zero-mode counting,
$k$-instantons now contribute uniquely to the $k$-point function. We will
argue
in \S\ref{sec:S44} that, just as for the SCI approach, the supersymmetric Ward
identity guaranteeing independence of the result on the insertion
points, is respected in the instanton approximation. In
particular, the one instanton contribution to the one-point function
can be performed exactly, as reviewed in \S\ref{sec:S46}.
The non-renormalization theorems of
$\N=1$ supersymmetric gauge theory
then permit the analytic continuation of the
answer into the confining phase by decoupling
the extraneous matter fields. This is achieved by giving them a mass
$m$, and taking the joint
limit $m\rightarrow\infty$ and $\Lambda_{\sst(N_F)}\rightarrow0$ in the
manner dictated by renormalization group
decoupling. Matching the exact one-instanton calculation
onto the effective low-energy theory without matter gives:
\begin{equation}
\VEV{\tfrac{g^2}{16\pi^2}\TrN\lambda^2}_{\rm WCI}=
\Lambda^{3}e^{2\pi iu/N}\  .
\elabel{WCIans}\end{equation}
Note that the renormalization group
decoupling procedure forces the low-energy theory
into one of the $N$ degenerate vacua. The puzzle now reveals itself as
the mismatch between \eqref{SCIansb} and \eqref{WCIans}.

In this Chapter, following Ref.~\cite{Hollowood:2000qn},
we review this old controversy, using the
many-instanton calculus that we have developed in previous sections. In
particular, we shall calculate the $k$-instanton contribution to the
$kN$-point correlator, in the SCI approach, and the $k$-point
function, in the WCI approach, in the large-$N$ limit where the
solution of the ADHM constraints
described in \S\ref{app:A5} in conjunction with
saddle-point methods simplify the instanton calculus.
In a nutshell, our results cast serious doubt
on the validity of the SCI calculations of the
condensate. Specifically, we will demonstrate that in the SCI approach
cluster decomposition is violated at leading order in $1/N$:
\begin{equation}
\VEV{\tfrac{g^2}{16\pi^2}\TrN\lambda^2(x^{(1)})\times\cdots\times
\tfrac{g^2}{16\pi^2}\TrN\lambda^2(x^{(kN)})}_{\rm SCI}
\ \overset{N\to\infty}\neq\  \VEV{\tfrac{g^2}{16\pi^2}
\TrN\lambda^2}^k_{\rm SCI}\ .
\elabel{viocl}\end{equation}
On the contrary, we will present a new calculation showing that 
the WCI approach
is perfectly consistent with clustering at leading order
in $1/N$. But we shall go further and argue that the WCI approach is
also consistent with clustering for finite $N$.
The important implications of this
observation are as follows. Since cluster decomposition is an
essential requirement of quantum field theories (with very mild
assumptions that are certainly met by supersymmetric Yang-Mills),
the exact quantum
correlators must have this property. The fact that cluster
decomposition is violated by the
instanton-saturated SCI correlators means that---contrary to
claims in the literature---the SCI approximation is only giving {\it
part\/} of the full answer. Since the SCI correlators obey
supersymmetric perturbative non-renormalization theorems
\cite{Novikov:1985ic}, it necessarily follows that additional
$non$-perturbative configurations must be contributing to the correlators.
On the contrary we will argue that the WCI approach, which is a
genuine semi-classical technique, is
consistent with cluster decomposition. We believe that this conclusion
is perfectly natural and simply underlines the fact that instantons
are a semi-classical phenomenon and should only be trusted at weak coupling.
We should add that there are
other non-instanton approaches to calculating the gluino condensate,
for instance, via softly-broken $\N=2$ gauge theory
\cite{Hollowood:2000qn} solved using the
theory of Seiberg and Witten \cite{SeibWitt}, or
from monopoles acting as instantons in the
three-dimensional gauge theory that arises after compactification on a circle 
\cite{Glumag,Davies:2000nw}. These alternative approaches all 
agree with the WCI answer.

\subsection{A supersymmetric Ward identity}\elabel{sec:S44}

A fundamental property of correlation functions involving insertions of lowest
components of gauge invariant chiral superfields, of which
$\TrN\lambda^2$ is an example, is that they are
independent of the insertion points
\cite{NSVZ,Rossi:1984bu}. It is of paramount
importance for our subsequent arguments that this property is preserved
within both the SCI and WCI approaches. To start with, let 
us briefly review the field theoretic 
argument. For any gauge invariant chiral superfield
$\Phi=A(x)+\sqrt2\theta\psi(x)+\cdots$ one can show
\EQ{
{\partial\over\partial x_n}A(x)=\tfrac i4\bar\sigma_n^{\aD\alpha}
\bar\delta_\aD\psi_\alpha(x)\ .
\elabel{derivact}
}
Here, we have defined $\bar\delta_\aD$ via the supersymmetry variation $\delta
=\xi^\alpha\delta_\alpha+\bar\xi_\aD\bar\delta^\aD$. In particular,
\EQ{
\frac{\partial}{\partial x_n}\TrN\,\lambda^2
=\bar\delta_\aD\big(\bar\sigma^{\aD\alpha}_m{\rm
tr}_N\,F_{mn}\lambda_\alpha\big)\ .
\elabel{dxll}
}
Now consider the derivative of a correlator of lowest-component fields
with respect to one of the insertion points:
\EQ{
\PD{}{x_n^{(l)}}\VEV{A_1(x^{(1)})\times\cdots\times
A_p(x^{(p)})}=\tfrac i4\bar\sigma_n^{\aD\alpha}
\VEV{A_1(x^{(1)})\cdots \bar\delta_\aD\psi_\alpha(x^{(l)})\cdots
A_p(x^{(p)})}\ .
\elabel{lrmm}
}
The supersymmetry variation can then be commuted through the operators
to the left and right, since for a lowest component of a gauge
invariant chiral multiplet $\bar\delta_\aD A_{j}(x)=0$. Furthermore, the
vacuum is a supersymmetry invariant so the right-hand side of 
\eqref{lrmm} vanishes and
the correlation function is independent of the insertion point $\{x^{(l)}$.

The question which we now address is whether the
supersymmetric Ward identity described above is respected in either
the SCI or WCI approaches. In an $\N=1$ supersymmetric theory, the
supersymmetry
transformation $\bar\delta_\aD$ lifts to the collective coordinates as
\eqref{uiui} and \eqref{vivi}: $\bar\delta^\aD
a_\bD=i\delta^\aD{}_\bD\CM$ and $\bar\delta^\aD\CM=0$. Using
\eqref{dxll}, this means
\EQ{
\frac{\partial}{\partial x_n}\TrN\,\lambda^2(x)
=i\Big(\mu_{ui}\PD{}{w_{ui\aD}}+
\bar\mu_{iu}\PD{}{\bar
w_{iu\aD}}+\CM'_{ij\alpha}\PD{}{a'_{ij\alpha\aD}}\Big)
\sigma_{m\alpha\aD}\TrN\,F_{mn}\lambda^\alpha
\elabel{derlam}
}
and, in addition, $\bar\delta^\aD\TrN\,\lambda^2=0$.
In the instanton background,
therefore, the $x^{(l)}$ derivative of the multi-point correlator
of $g^2\TrN\lambda^2$ is equal to an integral
over $\ms_k$ of a total derivative.
We can use Gauss's Theorem to write this as an integral over the
boundary of $\ms_k$.
The only possible contributions can come from the
large sphere at infinity or the small spheres surrounding one of the
insertion points. To judge whether the contributions are
non-vanishing we need to determine the asymptotic behaviour of the integrand.

First of all, consider the SCI instanton approach, for which the
relevant collective coordinate measure is \eqref{fms} with $\N=1$.
Consider the contribution from the sphere of large radius $R$. By this
we mean where the ADHM variables scale as
\EQ{
{\rm tr}_k\big(\bar w^\aD w_\aD+a'_na'_n\big)\ \overset{R\to\infty}
\thicksim\ R^2\ .
}
The relevant asymptotic behaviour we need is
\EQ{
\TrN\,\lambda^2\thicksim R^{-4}\ ,\qquad
\TrN\,F_{mn}\lambda\thicksim R^{-4}\ ,
\elabel{lra}
}
so, collectively, the $kN$ operator insertions scale as $R^{-4kN}$.
To complete the analysis we have to determine the scaling of the
supersymmetric volume form \eqref{fms}.
There are two sources of $R$ dependence: firstly there are $4k(N+k)-1$
integrals over the $c$-number collective coordinates
on the boundary, giving $R^{4k(N+k)-1}$, and
secondly the
the $\delta$-functions for the bosonic and fermionic ADHM constraints scale as
$R^{-4k^2}$.\footnote{The $3k^2$ bosonic ADHM constraints are
quadratic in the bosonic ADHM variables
and so scale as $R^{-6k^2}$ while the $2k^2$ fermionic ADHM
constraints are linear in the bosonic variables and so scale as
$R^{2k^2}$, giving $R^{-4k^2}$.} Overall, therefore,
the volume form on the boundary scales as $R^{4kN-1}$.
Putting this together with the scaling of the operator insertions,
we see that the contribution from the large sphere
scales as $R^{4kN-1}\times R^{-4kN}=1/R$ and therefore vanishes as
$R\to\infty$.
Now consider the behaviour on a small sphere of radius $R$ around
one of the insertion
points $x^{(j)}$:
\EQ{
{\rm tr}_k\big(\bar w^\aD w_\aD+(a'_n+x^{(j)}_n1_{\sst[k]\times[k]})
(a'_n+x^{(j)}_n1_{\sst[k]\times[k]})\big)\ \overset{R\to0}
\thicksim\ R^2\ .
}

 On this sphere,
the insertion at $x^{(j)}$ scales as
$R^{-4}$, while the other insertions at other points
remain finite.\footnote{Note that this argument is only valid if the
insertion points are distinct. We shall find out by explicit
calculation that ambiguities can arise if too many insertions are made
at the same point.} The collective coordinate integral
over this boundary scales as
$R^{4kN-1}$. Therefore, as long as we avoid making all the insertions
at the same point, there is a vanishing contribution as $R\to0$.
Since there are no contributions from the boundaries, we conclude that the
integral vanishes and the supersymmetric Ward identity is respected
within the SCI approach.

Now we turn to the same considerations in the 
WCI approach. In this case, $k$ instantons
contribute to the $k$-point function rather than the $kN$-point
function. To leading order in the semi-classical expansion, even though
the instantons are now constrained, the insertions
take their ADHM form and so satisfy \eqref{derlam}. Hence, we use the
same logic to write the $x^{(l)}$-derivative of the $k$-point function with
respect to one of the insertion points as a total derivative on
$\ms_k$. However, there are some new subtleties. Firstly,
since we are on the Higgs branch, there is a non-trivial instanton
effective action \eqref{funactionredux} characteristic of a
constrained instanton calculation. Since it is a supersymmetric
invariant though, we can pull $\bar\delta^\aD$ past $e^{-\tilde S}$,
as well as the other insertions,
in order to apply Gauss's Theorem. However,
the instanton effective action \eqref{funactionredux} will
modifies the asymptotic
behaviour on the large sphere of radius $R$. The bosonic terms in
$\tilde S$ effectively prevent $4k(N-1)$ of the variables
$\{w_{ui\aD},\bar
w^\aD_{iu}\}$ from becoming large on the boundary manifesting the fast
that the instantons are constrained and there is a cut off on their
size. Taking this into account, the measure now scales effectively as
$R^{4k-1}$ on the sphere, rather than the
$R^{4kN-1}$ of the SCI approach. Each of the insertions (which are equal to
their ADHM expressions to leading order) still scale as \eqref{lra} and
so the insertions, together, scale as $R^{4k}$. Overall, measure and
insertions scale as $1/R$ and so there is a vanishing
contribution as $R\to\infty$. Likewise it is straightforward to show
that there are no contributions
from around the insertion points themselves. The point is that in this
case the instantons are small and so the constraining plays no r\^ole
and the analysis is identical to the SCI case above. Consequently 
the $k$-point correlations functions are independent of the insertions 
points in the WCI approach.

\subsection{One instanton calculations of the gluino condensate}
\elabel{sec:S44.5}

In the following subsections,
we review the one-instanton SCI and WCI calculations
of the gluino condensate.

\subsubsection{Strong coupling}\elabel{sec:S45}

The strong-coupling one-instanton calculation
was done originally for gauge group
$\SU(2)$ in  \cite{NSVZ} and then
extended to the
$\SU(N)$ theories in \cite{Amati:1985uz} (see also the very comprehensive
review articles \cite{Amati:1988ft,Shifman:1999mv}).

To begin with the supersymmetric volume form on the instanton moduli
space is given in \eqref{fms}.
However, since at the one-instanton level
$N\geq 2k\equiv 2$, we find it
more convenient to use the expression given in \eqref{sbames} where the
ADHM $\delta$-functions have been resolved:
\begin{equation}
\int_{\ms_1}\Bomega^{\sst(1)}
={2^{3N}\pi^{2N-2}\over(N-1)!(N-2)!}\int\rho^{4N-8}\,
d^4X\, d\rho^2\,d^{4(N-1)}\grp\, d^2\CM'\, d^2\zeta\,
d^{(N-2)}\nu\, d^{(N-2)}\bar\nu\ .
\elabel{oivfg}\end{equation}
Here, $d^{4(N-1)}\grp$ is the unit normalized ($\int
d^{4(N-1)}\grp=1$) volume form for the gauge orientation of the
instanton. We have identified
$X_n=-a'_n$, the position of
the instanton, $\rho=\sqrt{W^0/2}$, the scale size,
and $\{\CM',\zeta\}$ as the Grassmann collective
coordinates associated to broken supersymmetric and
superconformal invariance. In particular, recall from \S\ref{app:A5},
the definitions
\EQ{
\mu=w_\aD\zeta^\aD+\nu\ ,\qquad\bar\mu=\zeta_\aD\bar w^\aD+\bar\nu\ .
}
By explicit evaluation, using \eqref{lam} and
the one-instanton formulae of \S\ref{sec:S12}, the
gluino insertions in the one-instanton
background are
\EQ{
g^2\TrN\,\lambda^2(x)
=\frac{4g\rho^2\bar\nu\nu}{\big((x-X)^2+\rho^2\big)^3}
+\frac{6g\rho^4\big(\CM^{\prime\alpha}+\zeta_\aD
(\bar x-\bar X)^{\aD\alpha}\big)
\big(\CM^{\prime}_\alpha-(x-X)_{\alpha\aD}\zeta^\aD
\big)}{\big((x-X)^2+\rho^2\big)^4}\ .
\elabel{oneil}
}
We then insert $\prod_{i=1}^N\tfrac{g^2}{16\pi^2}\TrN\lambda^2(x^{(i)})$
into the collective coordinate integral. Since in the confining phase
there is no symmetry breaking and
the insertions are gauge invariant there is no dependence
on the gauge orientation of the instanton and one can
simply integrate over the associated collective coordinates
$\int d^{4(N-1)}\grp=1$. Next, let us carry out the Grassmann integrations.
Obviously the $\zeta$ and $\CM'$ Grassmann integrals have to be
saturated from the insertions at two points
$\{x^{(i)},x^{(j)}\}$ chosen from amongst the $N$.
After integrating over $\zeta$ and $\CM'$, the contribution from this pair is
\begin{equation}
{36\rho^8\,(x^{(i)}-x^{(j)})^2\over ((x^{(i)}-X)^2
+\rho^2)^4((x^{(j)}-X)^2+\rho^2)^4}\ .
\elabel{adding}\end{equation}
Now we take advantage of the fact that
the $N$-point function is independent of the $x^{(i)}$,
to choose the insertion points for maximum simplicity of
the algebra. The simplest conceivable such choice, $x^{(i)}=0$ for all
$i,$ turns out to give an ill-defined answer of the form
``$0\times\infty$'' (the zero coming from the Grassmann integrations
as follows from Eq.~\eqref{adding}, and the infinity from divergences
in the $\rho^2$ integration due to coincident poles).  In order to
sidestep this ambiguity, one chooses instead:
\begin{equation}
x^{(1)}=\cdots=x^{(N-1)}=0\ ,\qquad x^{(N)}=x\ .
\elabel{insertionone}\end{equation}
This choice is the simplest
which gives a well-defined answer with no ``$0\times\infty$''
ambiguity. More ambitiously, one can still perform the calculation
even if all the insertion points are taken to be arbitrary
\cite{Amati:1988ft,Amati:1985uz}; however, we find it convenient
to take the minimal resolution provided by
\eqref{insertionone}. Due to the $(x^{(i)}-x^{(j)})^2$ factor in
Eq.~\eqref{adding}, it follows that the pair of insertions
$\{x^{(i)},x^{(j)}\}$ responsible for the $\{\CM',\zeta\}$ integrations must
include the point $x^{(N)}=x$; there are $N-1$ possible such pairs,
giving
\begin{equation}
{36(N-1)\rho^8\,x^2\over
((x-X)^2+\rho^2)^4\,(X^2+\rho^2)^4}
\elabel{addingb}\end{equation}
for these contributions. The remaining Grassmann integrations over
$\{\nu,\nubar\}$ are saturated by the $N-1$ insertions at $x^{(i)}=0,$ and give
\begin{equation}
(N-2)!\,\left[{4\rho^2\over
(X^2+\rho^2)^3}\right]^{N-2}\ .
\elabel{nuresult}\end{equation}

Combining the denominators in Eqs.~\eqref{addingb}-\eqref{nuresult}
with a Feynman parameter $\alpha$,
\begin{equation}
{1\over(X^2+\rho^2)^{3N-2}}\,
{1\over((x-X)^2+\rho^2)^4}=
{(3N+1)!\over3!\,(3N-3)!}\int_0^1d\alpha\,{\alpha^3(1-\alpha)^{3N-3}\over\big(
(\alpha x-X)^2+\alpha(1-\alpha)x^2+\rho^2\big)^{3N+2}}
\elabel{alphadef}\end{equation}
and performing the $X$ integrals yields:
\SP{
&\VEV{\tfrac{g^2}{16\pi^2}\TrN\lambda^2(x^{(1)})\times\cdots\times
\tfrac{g^2}{16\pi^2}\TrN\lambda^2(x^{(N)})}\\
&\qquad\qquad={3(3N-2)\,2^{N-1}\mu^{3N}e^{2\pi i\tau}\over g^{2N}(N-2)!}x^2\int_0^1d\alpha
\int_0^\infty d\rho^2\,(\rho^2)^{2N-4}
\frac{\rho^{6N-4}\alpha^3(1-\alpha)^{3N-3}}{\big(
\rho^2+\alpha(1-\alpha)x^2\big)^{3N}}\\&\qquad\qquad=
{3(3N-2)\,2^{N-1}\,
\mu^{3N}e^{2\pi i\tau}\over g^{2N}
(N-2)!}\int_0^1d\alpha\,\alpha^2(1-\alpha)^{3N-4}
\\&\qquad\qquad={2^{N}\,\Lambda^{3N}\over(N-1)!\,(3N-1)}\ ,
\elabel{finalone}
}
in agreement with Eqs.~\eqref{SCIans}-\eqref{SCIansb}. In the final
expression, we have used \eqref{deflp} to remove any dependence on the
running coupling in favour of the $\Lambda$-parameter.

\subsubsection{Weak coupling}\elabel{sec:S46}

Next, let us review the weak coupling instanton calculation of the
gluino condensate. These kinds of calculation were originally done in
\cite{ADS,Fuchs:1986ft,Novikov:1985ic,Shifman:1988ia}
and reviewed in \cite{Shifman:1999mv}.
The strategy here is to add sufficient matter
fields in order to completely break the gauge group by the Higgs
mechanism. For matter fields transforming in the fundamental
representation, this means we need add $N_F=N-1$ hypermultiplets,
that is $N-1$ chiral
multiplets in both the $\BN$ and $\bar\BN$ representations.
On the Higgs branch, the scalar fields of the chiral multiplets $q_f$ and
$\tilde q_f$, $f=1,\ldots,N_F$ gain a VEV. For the case $N_F=N-1$ we
can choose the VEVs as in \eqref{qvev}.
For large values of the VEVs, the theory is weakly coupled and
semi-classical methods should be reliable. In contrast with the SCI
calculation, in the weakly-coupled Higgs
phase $\Vev{g^2\TrN\lambda^2}$
receives a non-zero contribution directly at the
one-instanton level. Decoupling the extraneous matter and matching to the
low-energy pure gauge theory is then accomplished using standard
renormalization group prescriptions. What is absolutely crucial is
that, due to holomorphy,
the value of the the gluino condensate is not renormalized as the
matter is decoupled and in this way a result
calculated at weak coupling can yield a result also valid at strong coupling.

We now calculate the instanton contribution to the
one-point function $\Vev{\tfrac{g^2}{16\pi^2}\TrN\lambda^2}$
in the theory with $N_F=N-1$. The
instanton calculus in theories with fundamental
matter fields is described in \S\ref{app:A6}.
Since the
scalar fields in the matter sector have VEVs the calculation involves
constrained instantons. As we have seen, to leading order (which in the
present context is exact) the effect of constraining the
instantons appears at the level of the instanton effective action which becomes
a non-trivial function of the collective coordinates as is evident in 
\eqref{funactionredux}.

The existence of the $N_F$ fundamental hypermultiplets leads to new
Grassmann collective coordinates $\{{\cal K},\tilde{\cal K}\}$
as described in \S\ref{app:A6}. The new feature of the
semi-classical limit are
integrals over these new collective coordinates
Eq.~\eqref{msci}. To leading order in the semi-classical approximation,
the instanton measure is
obtained by amalgamating \eqref{intm}, for
$\N=1$, with \eqref{msci}:
\EQ{
\mu^{k(3N-N_F)}g^{k(N_F-3N)}e^{2\pi ik\tau}
\int_{\ms_k}\Bomega^{\sst(\N=1,N_F)}\,e^{-\tilde S}\ ,
\label{colin1}
}
where the volume form $\Bomega^{\sst(\N,N_F)}$ is defined in \eqref{mattmeas}.

In order to calculate the gluino condensate, we
must insert into \eqref{colin1} (with $N_F=N-1$)
the one instanton expression for
$\tfrac{g^2}{16\pi^2}\TrN\lambda^2$.
Note that in contrast to the SCI approach, in this WCI calculation a single
instanton contributes directly to the one-point function. However,
since the matter fields have non-vanishing VEVs the
instantons are constrained as described in \S\ref{sec:S33}.
This seems to preclude an
evaluation of the condensate because we do not actually know the exact
profile of $\lambda$ in the constrained instanton
background. However, we do know the profile of $\lambda$ in the core
of the instanton where it is well approximated by its ADHM form
\eqref{lam}. Now suppose
we simply use the ADHM expression for $\tfrac{g^2}{16\pi^2}\TrN\lambda^2$
(written in \eqref{oneil}) have we any right to expect this to be a
a good approximation? At first it appears not because
the ADHM profile behaves differently in the tail of the instanton
where it has a power-law
fall off rather than the exponential fall off of the constrained
instanton. However, this would only be a problem for large instantons
which are in any case suppressed by the
instanton effective action \eqref{oista}. Consequently
the error made in
substituting the ADHM profile for the insertion will be higher
order in $g$ and as long as we are only after the leading order
expression, the ADHM profile will suffice.
Here is the rub: we know that the condensate cannot receive
any corrections in $g$ by holomorphy
and so the result we obtain will therefore be exact.

The insertion $\tfrac{g^2}{16\pi^2}\TrN\lambda^2$ in the background of
one instanton is given in \eqref{oneil} and is quadratic in Grassmann
collective coordinates. Since in the measure \eqref{colin1} includes
an integral over the two Grassmann collective coordinates associated
to broken supersymmetry transformations $\CM^{\prime}$ and
the instanton effective action \eqref{oista} is independent of them,
the insertion must saturate these integrals:
\EQ{
\int d^2\CM'\,\TrN\,\lambda^2(x)=g^{-1}
\frac{6\rho^4}{\big((x-X)^2+\rho^2\big)^4}\ .
}
The instanton effective action is likewise independent of the position
of the instanton $X_n$ and so the integrals over these collective
coordinates give
\EQ{
\int d^4X\,\frac{1}{\big((x-X)^2+\rho^2\big)^4}=\frac{\pi^2}{6\rho^4}\ .
\elabel{Xintci}
}
What remains is then a supersymmetrized integral over the centred
moduli space $\widehat\ms_1$. In fact
the gluino condensate in the case $N_F=N-1$ can be written simply
in terms of the centred instanton partition function \eqref{cipfun} as
\EQ{
\VEV{\tfrac{g^2}{16\pi^2}\TrN\lambda^2}
=\tfrac12\mu^{2N+1}g^{-2N}e^{2\pi i\tau}
\widehat{\EuScript Z}_1^{\sst(\N=1,N_F=N-1)}\ .
\label{oipfn1}
}

At the one-instanton level, since $N\geq2k\equiv2$,
we can use results from \S\ref{app:A5}
to resolve the bosonic ADHM constraints. When this has been done
the centred one-instanton partition function is
\SP{
\widehat{\EuScript Z}_1^{\sst(\N=1,N_F=N-1)}&=
{2^{3N-3}\pi^{2N-2-2N_F}\over
(N-1)!(N-2)!}\int\rho^{4N-8}\, d\rho^2\,d^{4(N-1)}\grp\\
&\qquad\times\, d^N\bar\mu\,d^N\mu
\,d^{N_F}{\cal K}\,d^{N_F}\tilde{\cal
K}\ \prod_{\aD=1}^2\delta\big(\bar w_\aD\mu+\bar\mu w_\aD\big)\
e^{-\tilde S}\ ,
\elabel{cioi}
}
where the the instanton effective action is
\eqref{funactionredux} (with the choice of VEVs \eqref{qvev})
\EQ{
\tilde S= 2\pi^2\rho^2\sum_{f=1}^{N-1}
|v_f|^2\big(|\grp_{f1}|^2+|\grp_{f2}|^2\big)
+\frac{i\pi^2}{\sqrt2}
\sum_{f=1}^{N-1}\big(v_f^*\K_f\mu_f-\tilde
v_f^*\bar\mu_f\tilde\K_f\big)\ .
\elabel{oista}
}
Here, we have used the one-instanton form for $w_\aD$ in terms of the
gauge orientation $\grp$ in Eq.~\eqref{wcon2}.
Notice that the form of the measure that we will use \eqref{cioi} is
rather schizophrenic, since we have chosen to resolve the bosonic ADHM
constraints as in \S\ref{app:A5},
but not the fermionic ones, as indicated by
the remaining Grassmann $\delta$-functions in \eqref{cioi}.

We now evaluate the partition function explicitly.
First of all,
the $\{\K_f,\tilde\K_f\}$ and $\{\mu_u,\bar\mu_u\}$, $u\neq N$,
integrals are saturated by pulling down powers of
the Yukawa coupling terms
in the instanton effective action \eqref{oista}. This yields a constant term
\EQ{
\Big(\frac{\pi^2}{\sqrt2}\Big)^{2(N-1)}\prod_{f=1}^{N-1}(v_f\tilde v_f)^*\ .
\elabel{kkres}
}
The integrals over the two remaining Grassmann variables
$\{\mu_N,\bar\mu_N\}$ are saturated by the Grassmann $\delta$-functions:
\EQ{
\int d\mu_N\,d\bar\mu_N\,\prod_{\aD=1}^2\delta\big(\bar w_{N\aD}\mu_N
+\bar\mu_Nw_{N\aD}\big)=\rho^2\big(|\grp_{N1}|^2+|\grp_{N2}|^2\big)\ .
\elabel{munmun}
}

Putting all these factor together, we have the remaining bosonic
integrals to perform
\SP{
\widehat{\EuScript Z}_1^{\sst(\N=1,N_F=N-1)}=
&=\frac{\pi^{4N-8}2^{2N-1}}{(N-1)!(N-2)!}
\prod_{f=1}^{N-1}v_f^*\tilde v_f^*\,\int
d^4X\,d\rho^2\,d^{4(N-1)}\grp\\
&\times\,\rho^{4N-6}
\big(|\grp_{N1}|^2+|\grp_{N2}|^2\big)
\,\exp\big[-2\pi^2\rho^2\sum_{f=1}^{N-1}
|v_f|^2\big(|\grp_{f1}|^2+|\grp_{f2}|^2\big)
\big]\ .
}
The integral over the scale size is
\EQ{
\int_0^\infty d\rho^2\,\rho^{4N-6}\exp\big[-2\pi^2\rho^2
\sum_{f=1}^{N-1}
v_f^2\big(|\grp_{f1}|^2+|\grp_{f2}|^2\big)
\big]=
\frac{(2N-3)!}{\Big(2\pi^2\sum_{f=1}^{N-1}
v_f^2\big(|\grp_{f1}|^2+|\grp_{f2}|^2\big)\Big)^{2N-2}}
\ .
}
This leaves the remaining integral over the gauge orientation $\grp$
which can be done by using the formulae in
Ref.~\cite{Fuchs:1987ae}:\footnote{Alternatively, making the choice
$v_f=\tilde v_f\equiv v$ yields an elementary integral.}
\EQ{
\int
d^{4(N-1)}\grp\,\frac{|\grp_{N1}|^2+|\grp_{N2}|^2}{\Big[
\sum_{f=1}^{N-1}
v_f^2\big(|\grp_{f1}|^2+|\grp_{f2}|^2\big)\Big]^{2N-2}}
=\frac{(N-1)!(N-2)!}{(2N-3)!}\prod_{f=1}^{N-1}|v_f|^{-4}\ .
\elabel{intcos}
}
Collecting the results, we have
\EQ{
\widehat{\EuScript Z}_1^{\sst(\N=1,N_F=N-1)}=2\prod_{f=1}^{N-1}(v_f\tilde
v_f)^{-1}
}
and hence
\EQ{
\VEV{\tfrac{g^2}{16\pi^2}\TrN\lambda^2}
=\mu^{2N+1}g^{-2N}e^{2\pi i\tau}\prod_{f=1}^{N-1}(v_f\tilde
v_f)^{-1}\ .
\elabel{wcigc}
}

The result \eqref{wcigc} is expressed in terms of bare quantities and
involves the Pauli-Villars mass-scale $\mu$. We need to re-express it
in terms of renormalized quantities. Firstly, the bare VEVs $v_f$ and
$\tilde v_f$ should be
replaced by $Z_f^{1/2}v_f^{\text{ren.}}$ and $Z_f^{1/2}\tilde
v_f^{\text{ren.}}$,
respectively, where
$Z_f$ is the usual multiplicative wavefunction renormalization factor.
{}From now on we shall drop the ``ren.'' superscripts
on the renormalized VEVs with
the understanding that, henceforth, all quantities are understood to
be renormalized.
The coupling constant $g$ and multiplicative renormalization factors $Z_f$ run
with the Pauli-Villars mass scale $\mu$ in such a
way that the remaining factors are equal to a
certain power of the strong coupling scale of the theory in
the Pauli-Villars scheme:
\EQ{
\Lambda^{2N+1}_{\sst(N-1)}=e^{-8\pi^2/g(\mu)^2+i\theta}
\mu^{2N+1}g(\mu)^{-2N}\prod_{f=1}^{N-1}Z_f(\mu)^{-1}\ ,
\elabel{grrea}
}
generalizing the $N_F=0$ relation \eqref{deflp}.
Note that due to holomorphy
this relation must be exact. By differentiating with respect
to $\mu$ this yields an exact expression for the $\beta$-function.
It is an example of the more general relation for arbitrary
$N_F$
\EQ{
\Lambda^{3N-N_F}_{\sst(N_F)}=e^{-8\pi^2/g(\mu)^2+i\theta}
\mu^{3N-N_F}g(\mu)^{-2N}\prod_{f=1}^{N_F}Z_f(\mu)^{-1}\ ,
\elabel{grre}
}
which upon differentiation leads to the
famous Novikov-Shifman-Vainstein-Zakharov $\beta$-function of
supersymmetric QCD \cite{Shifman:1999mv,Novikov:1983uc}:
\EQ{
\mu\frac{\partial\alpha}{\partial\mu}=-\frac{\alpha^2}{2\pi}\cdot
\frac{3N-\sum_{f=1}^{N_F}(1-\gamma_f)}
{1-\frac{N\alpha}{2\pi}}\ ,\qquad\alpha\equiv\frac{g^2}{4\pi}\ ,\qquad
\gamma_f\equiv-\frac{d\ln Z_f}{d\ln\mu}\,
\elabel{betafm}
}
The final result for the gluino condensate is
\EQ{
\VEV{\tfrac{g^2}{16\pi^2}\TrN\lambda^2}
=\Lambda_{\sst(N-1)}^{2N+1}\prod_{f=1}^{N-1}(v_f\tilde
v_f)^{-1}\ .
\elabel{fwcigc}
}

We still have to relate the value of the gluino condensate
\eqref{fwcigc} in the Higgs phase to that in the confining phase. In
other words, we have to track the value of the gluino condensate as we
decouple the matter fields. In order to achieve this in a controlled
way, it is useful to re-interpret \eqref{fwcigc} in terms of a coupling
in the low energy effective superpotential $W_{\text{eff}}$ for the
matter fields in the Higgs phase. In order to determine the
superpotential we exploit the general functional identity (see for
example \cite{Peskin:1997qi}):
\begin{equation}
\VEV{\tfrac{g^2}{16\pi^2}\trN\lambda^2}
=b_0^{-1}\Lambda\PD{}{\Lambda}\VEV{W_{\text{eff}}}\ .
\elabel{fcnlid}\end{equation}
Here, $\Lambda$ is the appropriate strong coupling scale and $b_0$ is
the first coefficient of the $\beta$-function, equal to $3N-N_F=2N+1$
in the present context. Since, the expression for the superpotential
must be invariant under gauge and global
symmetries, it can be uniquely determined:
\EQ{
W_{\text{eff}}=\frac{\Lambda_{\sst(N-1)}^{2N+1}}{\text{det}\,\tilde
QQ}\ .
\elabel{adssp}
}
This is the famous Affleck-Dine-Seiberg (ADS) superpotential for
$\N=1$ supersymmetric $\SU(N)$ QCD. (See Ref.~\cite{ADS} for the $\SU(2)$
case and Ref.~\cite{Cordes} for the $\SU(N)$
calculation.)
Note we have arrived at the ADS superpotential by a somewhat
different route to Refs.~\cite{ADS,Cordes}. In those references, the form of
the superpotential was deduced by first noticing that after the
Higgs mechanism, there is a combination of the matter fields which
remains classically massless. The ADS superpotential then implies that
these classically massless fields receive a non-perturbative mass
through a one-instanton effect. This mass, and hence the form of the
superpotential itself, can be deduced by calculating the long distance
behaviour of the two-point function of the massless component of the
anti-Higgsinos $\bar\chi$ and $\bar{\tilde\chi}$.
The computation involves using a single constrained
instanton but to leading order we can replace the massless component
of $\bar\chi$ and $\bar{\tilde\chi}$ by their
value in the ADHM instanton background.\footnote{Notice that this
involves iterating the constrained instanton to the required order
where the anti-chiral fermions are non-zero.} The net result of this
calculation is entirely consistent with our approach via the gluino
condensate.

As it stands,
the ADS superpotential implies that the vacuum runs away away to
infinity. One way to stabilize the theory is to add a small mass
term. Using the ubiquitous holomorphy argument, the mass cannot get
renormalized and will appear directly in the effective superpotential:
\EQ{
W_{\text{eff}}=\frac{\Lambda_{\sst(N-1)}^{2N+1}}{\text{det}\,\tilde
QQ}+\sum_{f=1}^{N-1}m_f\tilde Q_fQ_f\ .
\elabel{madssp}
}
The superpotential \eqref{madssp} now has well defined
critical points at
\EQ{
m_fv_f\tilde
v_f=\frac{\Lambda_{\sst(N-1)}^{2N+1}}{\prod_{f'=1}^{N-1}v_{f'}\tilde
v_{f'}}\ .
\elabel{massvev}
}
It is easy to see that there are $N$
solutions where $v_f\tilde v_f$ differ by $N^{\rm th}$ roots of
unity. At these critical points the values of the gluino condensate
obtained by using \eqref{fcnlid} are
\EQ{
\VEV{\tfrac{g^2}{16\pi^2}\TrN\lambda^2}=\Big(\Lambda^{2N+1}_{\sst(N-1)}
\prod_{f=1}^{N-1}m_f\Big)^{1/N}\ ,
\elabel{hvads}
}
which exhibits the $N$-fold degeneracy explicitly via the choice of the
$N^{\rm th}$ root.

The result \eqref{hvads} is justified in the weakly-coupled Higgs
phase where the
VEVs are large (or masses small)
compared with $\Lambda_{\sst(N-1)}$. However the powerful
Ward Identities of supersymmetric theory allow us to extrapolate the
result into the regime of small VEV (large masses)
and strong coupling. The point is
that since $\TrN\,\lambda^2$ is gauge-invariant and the lowest
component of a chiral superfield, its VEV must be a holomorphic
expression in the coupling constants of the theory. In the present
context this means $\Lambda_{\sst(N-1)}$ and the masses. This implies that the
result \eqref{hvads}
cannot be subject to any perturbative corrections since a power series
in $g$ would translate into an expression
which could not be holomorphic in $\Lambda_{\sst(N-1)}$. Hence,
\eqref{hvads} must also be valid for small VEVs, or, correspondingly
from \eqref{massvev}, large masses. This is precisely the limit in
which the matter fields are decoupled and the theory should
flow into the pure $\N=1$ supersymmetric gauge theory. In
particular the $N$ supersymmetric vacua that we see at weak coupling
are continuously connected with the $N$ supersymmetric vacua of the pure gauge
theory. In order to track the value of the gluino condensate from weak to
strong coupling
we simply have to match the $\Lambda$-parameters of the two
theories. The correct renormalization group matching in this case is
\cite{FP,Weinberg}
\EQ{
\Lambda^{3N}=\Lambda_{\sst(N-1)}^{2N+1}\prod_{f=1}^{N-1}m_f\ .
\elabel{rgme}
}
Finally putting \eqref{hvads} together with \eqref{rgme}, we deduce
\EQ{
\VEV{\tfrac{g^2}{16\pi^2}\TrN\lambda^2}=\Lambda^3e^{2\pi iu/N}\ ,
\elabel{hvscc}
}
for $u=1,\ldots,N$ which is the expression in \eqref{WCIans}.

\subsection{Multi-instanton calculations of the gluino
condensate}\elabel{sec:S46.5}

In this section, we calculate the contributions of arbitrary numbers
of instantons to the gluino condensate in both the SCI and WCI
approaches at large $N$ where saddle-point methods are available to
simplify the instanton calculus. 

\subsubsection{Strong coupling}\elabel{sec:S47}

In \S\ref{app:A5}, we derived a form for the multi-instanton
volume form in a supersymmetric gauge theory
with $N\geq2k$. In the confining phase there is symmetry breaking and
all insertions are gauge
invariant, hence, we can immediately integrate over the gauge
orientation: $\int d^{4k(N-k)}\grp=1$. This gauge invariant form
for the multi-instanton volume form is now particularly useful for
taking a large-$N$ limit.

In order to calculate the $kN$-point correlation function, we must
insert into the measure the expression for
$\tfrac{g^2}{16\pi^2}\TrN\lambda^2$
evaluated in the $k$ ADHM instanton background.
At the multi-instanton level we find it useful to use the identity
\eqref{express2} to express the insertions as
\EQ{
\TrN\,\lambda^2=\frac{1}{4g}\,\square\,{\rm tr}_k
\bar\CM({\cal P}+1)\CM f\ .
\elabel{reexlam}
}
As proved earlier, the correlation function is expected to be
independent of the insertion points and so,
as in the one-instanton sector in \S\ref{sec:S44},
the $\{x^{(l)}\}$ can therefore be chosen for maximum simplicity of
the algebra. The simplest conceivable choice, $x^{(l)}=0$ for all $l$,
results in an ill-defined answer of the form ``$0\times\infty$'' (the
zero coming from unsaturated Grassmann integrations, and the infinity
from divergences in the bosonic integrations due to coincident poles);
we have already noted this fact in the one-instanton sector in
\S\ref{sec:S45}. The simplest choice of the $\{x^{(l)}\}$
that avoids this problem, generalizing \eqref{insertionone}, is
\begin{equation}\begin{split}
&x^{(1)}=\cdots=x^{(kN-k^2)}=0\
,\\&x^{(kN-k^2+1)}=\cdots=x^{(kN)}\equiv x
\elabel{simple}\end{split}\end{equation}
which we adopt for the remainder of this section.\footnote{As a
nontrivial check on the Ward identity, we have also numerically integrated
the large-$N$ correlator for insertions other than Eq.~\eqref{simple}, and
verified the constancy of the answer presented below.}

In the large-$N$ limit, there is a large preponderance of insertions
\eqref{simple} at $x^{(l)}=0,$ and the resulting factor of
$(\TrN\lambda^2(0))^{k(N-k)},$ taken together with the Jacobian factor
$|\det_{2k}W|^{N-2k}$ from the measure \eqref{sbames}, dominate the integral
and can be treated in a saddle-point approximation. Below we will carry
out this saddle-point
evaluation in full detail, but we can already quite easily understand
the source of the linear dependence on $k$ in our final result.
The argument goes as follows:

(i) Let us imagine carrying
out all the Grassmann integrations in the problem. The remaining
large-$N$ integrand will then have the form
$\exp\big(-N\Gamma+{\cal O}(\log
N)\big)$ where $\Gamma$ might be termed the ``effective
large-$N$ instanton action.'' The large-$N$ saddle-points are then the
stationary points of $\Gamma$ with respect to the $c$-number collective
coordinates. By Lorentz symmetry, $\Gamma$ can only depend on
the four $k\times k$  matrices $a'_n$ through
even powers of $a'_n$. (This is because the bulk of the insertions
have been chosen to be
at $x^{(l)}=0$; otherwise one could form the Lorentz scalar
$x^{}_na'_n$ and so
have odd powers of $a'_n$.) It follows that the ansatz
\begin{subequations}
\begin{align}
&a'_n=0\ ,\quad n=1,2,3,4\ ,\elabel{spansatza}\\&W^c=0\ ,\quad
c=1,2,3 \elabel{spansatzb}
\end{align}\end{subequations}
is automatically a stationary point of $\Gamma$ with respect to these
collective coordinates. (Note that \eqref{spansatzb} follows automatically
from \eqref{spansatza} by virtue of the ADHM constraints
\eqref{linad}.)
It will actually turn out that, once one assumes these saddle-point
values, $\Gamma$ is independent of the remaining collective
coordinate matrix $W^0$; furthermore we will verify  that this
saddle-point is actually a  minimum of $\Gamma$.

(ii) Having anticipated the saddle-point
\eqref{spansatza}-\eqref{spansatzb} using these elementary
symmetry considerations, let us back up to a stage in the analysis
prior to the Grassmann integration, and proceed a little more carefully.
Evaluating the insertions $\TrN\lambda^2(x^{(l)})$ on this
saddle-point, one easily verifies that the $\zeta$ modes vanish when $x^{(l)}=0$;
consequently the $\zeta$ integrations must be saturated entirely from
the $k^2$ insertions at $x^{(l)}=x.$  This leaves the $\CM',$ $\nu$ and
$\nubar$ integrations to be saturated purely from the insertions at
$x^{(l)}=0$. Moreover, because $\CM'$ carries a Weyl spinor index $\alpha$ whereas
$\nu$ and $\nubar$ do not, the $\TrN\lambda^2(0)$ insertions
depend on these Grassmann coordinates only through  bi-linears of the form
$\nubar\times \nu$ or $\CM'\times\CM'$; there are no cross terms.

(iii) Performing all the Grassmann integrations then automatically generates
a combinatoric factor
\begin{equation}
(k^2)!\,(k^2)!\,(kN-2k^2)!\,\begin{pmatrix}kN-k^2\\ k^2\end{pmatrix}\ .
\elabel{combfact}\end{equation}
Here the first three factors account for the indistinguishable
bi-linear insertions of the $\zeta,$ $\CM',$ and $\{\nu,\nubar\}$ modes,
respectively, while the final factor counts the ways of
selecting the
$k^2$ bi-linears in
 $\CM'$ from  the $kN-k^2$ insertions at $x^{(i)}=0.$
Multiplying these combinatoric factors together, as well as the
normalization constants from then instanton measure, and taking the
$(kN)^{\rm th}$ root yields, in the large-$N$ limit:
\begin{equation}
\big[
A_k(C_1^{\sst(1)})^k\,(k^2)!\,(kN-k^2)!\big]^{1/kN}\
\overset{N\to\infty}\longrightarrow\
2^{3}\pi^{2}eN^{-1}k\Lambda^3+{\cal O}(N^{-2})\ .
\elabel{kNroot}\end{equation}
Remarkably, apart from a
factor of four, this back-of-the-envelope analysis precisely accounts
for the leading term in the $1/N$ in the final answer,
Eq.~\eqref{Gnfinal} below. Note that
most of the remaining contributions to the saddle-point analysis,
which involve a specific convergent bosonic integral derived below, as
well as the factor $2^{3k^2}/{\rm Vol\,}\U(k)$ from
Eq.~\eqref{sbames}, reduce
to unity when the $(kN)^{\rm th}$ root is taken in the large-$N$
limit; the missing factor of four will simply come from the leading
saddle-point evaluation of the bosonic integrand.

Here are the details of the large-$N$ calculation of the $kN$-point
correlation function.
First, we find it convenient to partially fix the auxiliary $\U(k)$
symmetry by taking
a basis where $W^0$ is diagonal:
\begin{equation}W^0={\rm diag}\big(
2\rho_1^2,\ldots,2\rho_k^2\big)\ .
\elabel{Wdiag}\end{equation}
As the notation implies, in the dilute instanton gas limit $\rho_i$
can be identified with the scale size of the $i^{\rm th}$ instanton in
the $k$-instanton sector (see \S\ref{sec:S27}). The appropriate
gauge fixing involves a Jacobian:
\begin{equation}
{1\over{\rm Vol}\,\U(k)}\int d^{k^2}W^0\ \longrightarrow\ {2^{3k(k-1)/2}\pi^{-k}\over k!}
\int_0^\infty\,
\prod_{i=1}^kd\rho_i^2\,\prod_{i<j=1}^k(\rho_i^2-\rho_j^2)^2\ .
\elabel{diagcv}\end{equation}
For $k=1$ one has, of course, $\int
dW^0\rightarrow 2\int_0^\infty d\rho^2.$

Now let us consider the Grassmann integrations, beginning with the
$\zeta$ variables. We assume the saddle-point conditions
\eqref{spansatza}-\eqref{spansatzb}, in which case
\begin{equation}
\Delta=\begin{pmatrix}w\\ x\cdot1_{\sst [k]\times[k]}^{}\end{pmatrix}\ ,\qquad
f={\rm diag}\big(\tfrac1{x^2+\rho_1^2},\ldots,\tfrac1
{x^2+\rho_k^2}\big)
\elabel{spmore}\end{equation}
and from Eq.~\eqref{reexlam},
\begin{equation}
\TrN\lambda^2(x)=\sum_{i,j=1}^k(\zeta_\dalpha)_{ij}
(\zeta^\dalpha)_{ji}\,F_{ij}(x)+\cdots\ ,
\elabel{zetastuff}\end{equation}
where
\begin{equation}
F_{ij}(x)=
\frac1{4g}\square\,{x^4\over(x^2+\rho_i^2)(x^2+\rho_j^2)}
\elabel{Fijdef}\end{equation}
and the terms omitted in Eq.~\eqref{zetastuff} represent dependence on the
other Grassmann collective coordinates
$\{\CM',\nu,\nubar\}$. It is obvious from
Eq.~\eqref{Fijdef} that $F_{ij}(0)=0$, so that the $\zeta$ must be
entirely saturated from the $k^2$ insertions at $x^{(i)}=x$ as claimed
above. Performing the $\zeta$ integrations then yields
\EQ{
(k^2)!\,\prod_{i,j=1}^kF_{ij}(x)\ .
\elabel{zetaans}
}

Next we consider the insertions at $x^{(l)}=0$. Focusing on the $\CM'$ modes
first, one finds from Eq.~\eqref{reexlam}:
\EQ{
\TrN\lambda^2(0)=
\frac2g\sum_{i,j=1}^k(\CM^{\prime\alpha})_{ij}(\CM'_{\alpha})_{ji}\,
(\rho_i^{-4}+\rho_j^{-4}+\rho_i^{-2}\rho_j^{-2})\
 +\ \cdots\ ,
\elabel{Mpstuff}
}
omitting the $\nu\times\nubar$ terms. Hence the $\CM'$ integrations yield
\begin{equation}
\frac{2^{k^2}}{g^{k^2}}\,\begin{pmatrix}kN-k^2\\ k^2\end{pmatrix}\,(k^2)!\ \prod_{i,j=1}^k
(\rho_i^{-4}+\rho_j^{-4}+\rho_i^{-2}\rho_j^{-2})\ ,
\elabel{Mpans}\end{equation}
where the combinatoric factors in \eqref{Mpans} (as well as in \eqref{zetaans})
have been explained previously.\footnote{One can easily check that
these large-$N$ formulae are consistent with the explicit 1-instanton
calculation presented in \S\ref{sec:S45} which is exact in $N$.}

Finally we turn to the $\{\nu,\nubar\}$ integrations. Since (unlike
$\zeta$ and $\CM'$) the number of $\nu$ and $\nubar$ variables
grows with $N$ as $kN-2k^2,$ it does not suffice merely to plug in the
saddle-point values \eqref{spansatza}-\eqref{spansatzb} and
\eqref{spmore}. One must also calculate
the Gaussian determinant about the saddle-point, which provides an
${\cal O}(N^0)$ multiplicative contribution to the answer. Accordingly we
expand about \eqref{spansatza}-\eqref{spansatzb}
to quadratic order in the $a'_n$. From \eqref{reexlam},
the $\nu\times\nubar$
contribution to $\TrN\lambda^2(0)$ has the form
\begin{equation}
-\frac1{2g}{\rm tr}_k\,\nubar\nu\,\square\, f{\Big|}_{x=0}=
\frac2g{\rm tr}_k\,\nubar\nu f\bar b_\alpha{\cal P}b^\alpha f
{\Big|}_{x=0}
\elabel{nunubar}\end{equation}
as follows from the differentiation formula \eqref{ddf}.
Performing the $\{\nu,\nubar\}$ integrations therefore gives
\begin{equation}\begin{split}
&(kN-2k^2)!\,\exp\Big\{(N-2k){\rm tr}_k\log\big(2f\bar b_\alpha
{\cal P}b^\alpha
f\big){\Big|}_{x=0}\,\Big\}
\\ &=
(kN-2k^2)!\,\exp\Big\{(N-2k)\big(\log\det_k16(W^0)^{-2}-
\tfrac32\sum_{i,j=1}^k(a'_n)_{ij}(a'_n)_{ji}\,
(\rho_i^{-2}+\rho_j^{-2})\,
+\, {\cal O}(a'_n)^4\big)\,\Big\}\ .
\elabel{nuans}\end{split}\end{equation}
The negative sign in front of the quadratic term in $a'_n$ confirms
that our saddle-point \eqref{spansatza}-\eqref{spansatzb} is in fact a minimum of the action.
Combining this expression with the measure factor in Eq.~\eqref{sbames}, namely
\EQ{
\big|\det_{2k}W\big|^{N-2k}=  \exp\big((N-2k)\log\det_{2k}W\big)
=\exp\Big\{(N-2k)\big(\log\det_k(\tfrac12W^0)^{2}\,+\,{\cal O}(a'_n)^4\,
\big)\Big\}\ ,
\elabel{measfact}
}
and performing the Gaussian integrations over $a'_n$, yields:
\begin{equation}
2^{2k(N-2k)}\,(kN-2k^2)!\,\prod_{i,j=1}^k
\left({2\pi\over3N(\rho_i^{-2}+\rho_j^{-2})}\right)^2
+ \cdots\ ,
\elabel{nuansb}\end{equation}
where the omitted terms are suppressed by powers of $1/N$.

Finally one combines Eqs.~\eqref{sbames},
\eqref{diagcov}, \eqref{zetaans}, \eqref{Mpans} and
\eqref{nuansb} (and the definition of the $\Lambda$-parameter in
\eqref{deflp}) to obtain the leading-order result for the
correlator:
\SP{
&\VEV{\tfrac{g^2}{16\pi^2}\TrN\lambda^2(x^{(1)})
\times\cdots\times\tfrac{g^2}{16\pi^2}\TrN\lambda^2(x^{(kN)})}\\
&\qquad\qquad\qquad
\overset{N\to\infty}\longrightarrow\ {2^{kN+k^2-k+1/2}\,\pi^{-k+1/2}\,
e^{kN}\,(k^2)!\,k^{kN-k^2+1/2}
\,\Lambda^{3kN}\over3^{2k^2}\,N^{kN+k^2-1/2}\,k!}{\EuScript I}_k\ ,
\elabel{Gnfinal}
}
where ${\EuScript I}_k$ is the convergent integral
\begin{equation}
{\EuScript I}_k=
\int_0^\infty
\,\prod_{i=1}^kd\rho_i^2\,\prod_{i,j\atop i\neq j}
^k|\rho_i^2-\rho_j^2|\cdot\prod_{i,j=1}^k
F_{ij}(x)\,\big(1-(\rho_j/\rho_i+
\rho_i/\rho_j)^{-2}\big)\ .
\elabel{Ikdef}\end{equation}
Note that ${\EuScript I}_k$ is independent of $x$ as a simple re-scaling
argument confirms.
For the simple case $k=1$, ${\EuScript I}_1=\tfrac32$ and the
expression \eqref{Gnfinal}
agrees---as it must---with the large-$N$ limit of the one-instanton
result in Eq.~\eqref{finalone}.

\subsubsection{Weak coupling}\elabel{sec:S48}

In this section, we turn to multi-instanton effects in the WCI approach.
We apply the same large-$N$ saddle-point methods to
extract the large-$N$ behaviour of the $k$-instanton contribution
to the $k$-point of $\tfrac{g^2}{16\pi^2}
\trN\lambda^2$.
In order to make the calculation more manageable, we will choose the
VEVs from the outset to be
$v_1=\cdots=v_{N-1}=\tilde v_1=\cdots=\tilde v_{N-1}\equiv v$. In this
case, the bosonic part of the instanton effective action
\eqref{funactionredux} can be written as
\EQ{
\tilde S_{\rm b}=\pi^2v^2{\rm tr}_{2k}(W-\xi^\dagger
uu^\dagger\xi)\ .
\elabel{miab}
}
Here, $W$ and $\xi$ are the $2k\times2k$ matrix defined in
\eqref{defbw} and \eqref{uptri}, respectively, and
the complex $2k$ vector $u_{i\aD}$ is defined in terms of the gauge
orientation via
\begin{equation}
u=\big(\grp_{1N},\ldots,\grp_{2k,N}\big)\ .
\end{equation}

Notice that the correlation function only depends on the gauge
orientation through the vector $u$ in the instanton effective action
\eqref{miab}. So we need an expression for the reduced
measure $\int d\grp=\int d^{2k}u\,f(u)$. We now pause to evaluate the
function $f(u)$. The complex $2k$-vector
$u$ is composed of the first $2k$ components of the
complex unit $N$-vector $U=(\grp_{1N},\ldots,\grp_{NN})$. Let $\tilde
u=(\grp_{2k+1,N},\ldots,\grp_{NN})$ be the complementary complex
$(N-2k)$-vector: $U=u+\tilde u$.
It is clear that the measure on the unit vector $U$
inherited from the group measure $\int d\grp$ is proportional to
\begin{equation}
\int d^{2N}U\,\delta(|U|^2-1)
=\int d^{4k}u\,d^{2(N-2k)}\tilde u\,\delta(|u|^2+|\tilde
u|^2-1)\ .
\elabel{hmeas}\end{equation}
The measure on $u$ is obtained by
integrating over $\tilde u$ giving (up to constant)
\begin{equation}
\int_{|u|\leq1} d^{4k}u\,
(1-|u|^2)^{N-2k-1}\ .
\elabel{hqmeas}\end{equation}
Actually \eqref{hqmeas} is still too general for our needs. It
will be sufficient for our purposes to know the measure on the
$k$ quantities
$z_i=\bar u^\aD_iu_{i\aD}$ (with no sum on $i$). The
measure on the $\{z_i\}$ follows easily from \eqref{hqmeas}.
Finally, taking into account the normalization \eqref{defgrp}, we have
the required expression for the reduced measure of the gauge orientation:
\begin{equation}
\int d^{4k(N-k)}\grp=
{\Gamma(N)\over2\Gamma(N-2k)}\int_0^1\,\Big\{\prod_{i=1}^kdz_i\,z_i\Big\}\,
\big(1-\sum_{i=1}^kz_i\big)^{N-2k-1}\ ,
\elabel{mhmeas}\end{equation}
with the constraint $\sum_{i=1}^kz_i\leq1$.

We now have all the ingredients needed to start a large-$N$
analysis. In this limit, we will evaluate the collective coordinate
integral by a method of steepest descent. As
usual in such a saddle-point analysis,
one exponentiates any terms in the collective coordinate integral
raised to the power $N$, and includes them,
along with the bosonic terms in the instanton effective action, in a
large-$N$ effective action. There are two contributions besides the
instanton effective action \eqref{miab}. The first, is the factor of
$|\det_{2k}W|^N$ in the measure
\eqref{sbames} and the second the factor of $(1-|u|^2)^N$ in
the measure for the gauge orientation
\eqref{hqmeas}. Placing all these contributions up in exponent the defines a
saddle-point action:
\begin{equation}
\Gamma=-N\log\det_{2k}\big(\tfrac12W^01_{\sst[2]\times[2]}
-2a'_ma'_n\bar\sigma_{mn}\big)+\pi^2v^2{\rm tr}_{2k}(W-\bar\xi
u\bar u\xi)-N\log(1-|u|^2)\ .
\end{equation}
Now we find the extrema of $\Gamma$ with respect to the
independent variables $a'_n$, $W^0$ and $u_\aD$.

Rather than write down the saddle-point equations and proceed to solve
them, we will simply
write down an ansatz for the solution and then show
{\it ex post facto\/} from a fluctuation analysis that it is
indeed an extremum. Up to the auxiliary $\U(k)$ symmetry, our ansatz is
\begin{subequations}
\begin{align}
W^0&={N\over\pi^2v^2}1_{\sst[k]\times[k]}\ ,\elabel{spsa}\\
a'_n&=-{\rm diag}(X^1_n,\ldots,X^k_n)\ ,\elabel{spsb}\\
u_{i\aD}&=0\ .\elabel{spsc}
\end{align}
\end{subequations}
These values imply
\begin{equation}
\xi=\sqrt{N\over2\pi^2v^2}1_{\sst[2k]\times[2k]}\ ,\qquad
W={N\over2\pi^2v^2}1_{\sst[2k]\times[2k]}\ .
\end{equation}
The solution has a simple physical interpretation. It represents $k$
instantons in the dilute-gas limit
with positions $X^i_n$, all with the same
scale size $\sqrt{N/(2\pi^2v^2)}$
inhabiting $k$ mutually commuting $\SU(2)$ subgroups of the gauge
group\footnote{The generators of the $\SU(2)$ embeddings of the $k$
instantons in the dilute limit are given in \eqref{embed2}. The fact
that the commutators of these generators for different instantons
vanishes follows from the fact that the solution
above has $W^c=0$.} orthogonal to the gauge orientation of the VEV
(since $u_{i\aD}=0$). The fact that the instantons live in commuting
subgroups of the gauge group implies that they
are non-interacting and the solution is dilute-gas-like.

We now turn to an analysis of the fluctuations around the solution
\eqref{spsa}-\eqref{spsc}.
The explicit expression for the expansion of action around the
saddle-point to quadratic order is
\begin{equation}\begin{split}
&\Gamma^{(2)}=2kN(1+\log(2\pi^2v^2/N))
+{(2\pi^2v^2)^2\over4N}\sum_{i,j=1}^k\delta W^0_{ij}\delta W^0_{ji}\\
&\qquad\qquad+{(2\pi^2v^2)^2\over N}\sum_{i,j=1\atop i\neq
j}^k(X^i-X^j)^2(a'_n)_{ij}
(a'_n)_{ji}+{N\over4}\sum_{i=1}^kz_i^2\ ,
\elabel{expq}\end{split}\end{equation}
where we have removed the cross terms between fluctuations involving
$\delta W^0$ with $u$ and $a'_n$ with $u$ by appropriate shifts
in $\delta W^0$ and $a'_n$. At this point, we
find it convenient to fix the $\U(k)$ symmetry by making the ``gauge
choice'' \eqref{gukch} for the fluctuations $a'_n$ and then denote the
remaining fluctuations $\tilde a'_n$.

To leading order in the $1/N$ expansion, we substitute the
saddle-point expressions for the variables \eqref{spsa}-\eqref{spsb}
into the insertions. Due to the dilute-gas nature of the saddle-point
solution, at leading order, each insertion $\tfrac{g^2}{16\pi^2}
\TrN\lambda^2$ will be a sum over $k$
one-instanton expressions involving the one-instanton collective
coordinates
$\{\rho_i,X^i_n,u_{i\aD},\CM'_{ii},\mu_{ui},\bar\mu_{ui}\}$ for a
fixed $i\in\{1,\ldots,k\}$. In particular, the only
dependence on the Grassmann collective coordinates $\CM'_{ii}$ is
through the insertions because at leading order they decouple from the
fermionic ADHM constraints. Therefore using the
one-instanton
expression \eqref{oneil}, we can replace each insertion with
\EQ{
\frac{g^2}{16\pi^2}\trN\lambda^2(x)\ \longrightarrow\
\sum_{i=1}^k{2^{-5}3
\pi^{-6}v^{-4}gN^2
\over\big((x-X^i)^2+N/(2\pi^2v^2)\big)^4}
(\CM^{\prime\alpha})_{ii}(\CM^{\prime}_\alpha)_{ii}\ .
\elabel{inser}
}
Even at this
intermediate stage in the calculation we can draw an important
conclusion. Since, to leading order, the
insertions do not involve any ``off-diagonal''
collective coordinates which communicate between the individual
instantons of the saddle-point solution, the result for the
correlation function is destined to
cluster correctly. Nevertheless, it is interesting to show how this
happens explicitly.

As we have noted, the leading order expression for the
insertions are independent of the
fluctuations $\delta W^0_{ij}$ and $\tilde a'_n$; hence,
we can immediately integrate them out:
\begin{equation}\begin{split}
&\int d^{k^2}\delta W^0\,d^{3k(k-1)}\tilde a'\exp\bigg\{-
{(2\pi^2v^2)^2\over4N}\sum_{i,j=1}^k\delta W^0_{ij}\delta W^0_{ji}-
{(2\pi^2v^2)^2\over N}\sum_{i,j=1\atop i\neq j}^k(X^i-
X^j)^2(a'_n)_{ij}
(a'_n)_{ji}\bigg\}\\
&\qquad\qquad\qquad=\left({N\pi\over\pi^4v^4}\right)^{k^2/2}
\left({N\pi\over4\pi^4v^4}\right)^{4k(k-1)/2}\prod_{i,j=1\atop i\neq
j}^k(X^i-X^j)^{-3}\
.\elabel{flucti}
\end{split}\end{equation}

We have succeeded in integrating out all the off-diagonal $c$-number
collective coordinates. Now for the Grassmann sector. The off-diagonal
elements of $\CM'_\alpha$ do not appear in the instanton effective action
\eqref{funactionredux} and, at leading order in $1/N$, in the
insertions \eqref{inser}; hence, the integrals over these quantities must be
saturated by the fermionic ADHM constraints. Since at the saddle point
$a'_n$ is diagonal, the integrals are
\EQ{
\int\,\prod_{i,j=1\atop i\neq j}^k\Big\{d^2\CM'_{ij}\
\prod_{\aD=1}^2\,\delta\big((\bar X^i-\bar
X^j)^{\aD\alpha}(\CM'_\alpha)_{ij}\big)\Big\}
=\prod_{i,j=1\atop i\neq j}^k(X^i-X^j)^2\ .
\elabel{ioodf}
}
This conveniently cancels against similar factors in \eqref{diagcov} and
\eqref{flucti}.

The integrals that remain are over the positions of the instantons
$X^i_n$, the gauge orientation coordinates $z_i$ and
the Grassmann collective coordinates $\{\mu_{iu},\bar\mu_{iu},
(\CM'_\alpha)_{ii},\K,\tilde\K\}$. The integrals over the
remaining Grassmann collective coordinates proceeds as for the exact
one-instanton calculation in \S\ref{sec:S46}: the $\{\K,\tilde\K\}$
and $\{\mu_{ui},\bar\mu_{iu}\}$ integrals are saturated by bringing
down terms from the instanton effective action
\eqref{funactionredux}, as in \eqref{kkres} for each $i$.
The integrals over the remaining variables
$\{\mu_{Ni},\bar\mu_{iN}\}$ are then saturated by the diagonal
fermionic ADHM constraints, as in \eqref{munmun}, for each $i$.

Putting all the pieces together,
we have the leading order $1/N$ contribution
\SP{
&\VEV{\tfrac{g^2}{16\pi^2}\TrN\lambda^2(x^{(1)})\times\cdots\times
\tfrac{g^2}{16\pi^2}\TrN\lambda^2(x^{(k)})}
\ \overset{N\to\infty}\longrightarrow\
{3^k\mu^{k(2N+1)}g^{-2kN}e^{2\pi
ik\tau}v^{-2kN}N^{3k/2}\over2^k\pi^{9k/2} k!}\\
&\qquad\qquad\qquad
\times\sum_{\text{perms}\atop\{l_i\}\,\text{of}\,\{1,\ldots,k\}}
\prod_{i=1}^k\bigg\{\int
d^4X^i\frac{1}{\big((x^{(l_i)}-X^i)^2+N/(2\pi^2v^2)\big)^4}
\int_0^\infty dz_i\,z_i^2e^{-Nz_i^2/4}\bigg\}\ .
}
Performing the remaining integrals
\begin{equation}
\VEV{\tfrac{g^2}{16\pi^2}\TrN\lambda^2(x^{(1)})\times\cdots\times
\tfrac{g^2}{16\pi^2}\TrN\lambda^2(x^{(k)})}\
\overset{N\to\infty}\longrightarrow\
\Big(\mu^{2N+1}g^{2N}e^{2\pi i\tau}v^{-2(N-1)}\Big)^k
\end{equation}
After renormalizing the VEV and matching to the pure gauge theory in
the way described the end of \S\ref{sec:S46}, our leading-order
large-$N$ result is
\EQ{
\VEV{\tfrac{g^2}{16\pi^2}\TrN\lambda^2(x^{(1)})\times\cdots\times
\tfrac{g^2}{16\pi^2}\TrN\lambda^2(x^{(k)})}
\ \overset{N\to\infty}\longrightarrow\
\Lambda^{3k}\ .
\elabel{reswcik}
}
As noted above above, clustering is satisfied at large $N$.

\subsection{Clustering in instanton calculations}\elabel{sec:S485}

Taking the SCI calculations in \S\ref{sec:S45} and \S\ref{sec:S47},
compare the $kN^{\rm th}$-root of \eqref{Gnfinal} with the
large-$N$ limit of the $N^{\rm th}$-root of the
one-instanton result \eqref{finalone}:
\EQ{
{2e\over
N}k\Lambda^3+{\cal O}(N^{-2})
\qquad\text{verses}\qquad{2e\over N}\Lambda^3+{\cal O}(N^{-2})\ .
\elabel{mism}
}
The implication is that the clustering property of the field theory is not
respected by approximating the functional integral with an integral
over instantons in the confining phase.
In Ref.~\cite{Hollowood:2000qn}, it was further
shown that the existence of the Shifman-Kovner vacuum designed to
reconcile the SCI and WCI at the one-instanton level, cannot account
for the mismatch \eqref{mism} for general $k$.
We interpret the result to mean that instantons in the confining phase
by themselves cannot account for all of the gluino
condensate and that additional non-perturbative configurations must
necessarily contribute.
This conclusion is also supported by numerical calculations of the
two-instanton contribution to four-point function for gauge group
$\SU(2)$ which were performed in \cite{Hollowood:2000qn}.
 
In contrast the WCI approach is consistent with clustering,
at least at large $N$: compare \eqref{hvscc} with \eqref{reswcik}.
But we can argue further that consistency with clustering is
guaranteed in the WCI approach, even at finite $N$. The argument
proceeds as follows. Suppose one calculated the
$k$-instanton contribution to the $k$-point function
in the regime where $|x^{(i)}-x^{(j)}|\gg g/v$, where $v$ is the
characteristic scale of the VEVs. In other words, the separation
between the insertions is much greater than the effective cut-off on
the instanton scale size imposed by the instanton effective action
\eqref{funactionredux}. Since each insertion can saturate the
integrals over two Grassmann collective coordinates, the dominant
contribution to the integral comes from a region where one of the $k$
instantons lies in the vicinity of each of the $k$ insertion points.
Choosing the insertion points in this way selects a region of
$\ms_k$ in which the $k$ instanton configuration is completely clustered,
as described in \S\ref{sec:S12}. So to leading order we can ignore any
interactions between them. The collective coordinate integration
measure then clusters as in \eqref{cclm}. The factor of $1/k!$ is then
cancelled by the $k!$ ways of pairing the $k$ instantons and $k$
insertion points. Hence, to leading order, for these large separations
we have
\EQ{
\VEV{\tfrac{g^2}{16\pi^2}\TrN\lambda^2(x^{(1)})\times\cdots\times
\tfrac{g^2}{16\pi^2}\TrN\lambda^2(x^{(k)})}_{k\text{-inst}}=
\VEV{\tfrac{g^2}{16\pi^2}\TrN\lambda^2}_{1\text{-inst}}^k\ .
\elabel{wcc}
}
But we have proved in \S\ref{sec:S44} that the correlation functions are
independent of the insertion points and so \eqref{wcc} is an exact statement.

Now that we have established that the SCI approach is inconsistent
with clustering it is appropriate for us to make some comments
regarding this methodology. The usual justification put forward for
this method can be paraphrased as
follows \cite{Amati:1988ft,NSVZ,Amati:1985uz}.
The correlation functions of $\TrN\,\lambda^2$ are independent of the
insertion points and so we can consider the configuration where the
insertions are very close compared with the scale of strong coupling
effects $\Lambda$. In this case only small instanton
configurations should contribute to the integral over the instanton
moduli space. If this is true then we should be able in invoke
asymptotic freedom, namely the fact that the integral over the
instanton moduli space involves the
the instanton action factor $\exp(-8\pi^2/g^2)$, where $g$ runs with
the characteristic scale size of the instanton, to argue that the
calculation is reliable for insertions which are arbitrarily close. The fact
that the correlation functions are independent of the insertions then
allows us to continue to large separations. Unfortunately this
argument is potentially
flawed because, as in is clear from \eqref{finalone}, the integral over the
instanton scale size is {\em not\/} cut-off exponentially by the separations
between the insertions but only by a power law and so large instantons
are not adequately suppressed. This behaviour should
at least question the use of asymptotic freedom as an argument for
exactness. 

Although we have only chosen to discuss the gluino condensate
the analysis can be generalized to any correlation function
involving 
gauge-invariant lowest-components of chiral superfields. For example, in the
theories with fundamental hypermultiplets, the gauge invariant
composite $\tilde q_f q_{f'}$ is also gauge invariant and the lowest
component of a chiral superfield. So correlation functions of this
insertion, along with $g^2\TrN\lambda^2$ will, by the Ward
identity, be independent of the
insertion points and holomorphic in the couplings. The SCI verses WCI
mismatch that we have witnessed for the gluino condensate extends to
these correlation functions and the same conclusion applies: only the
WCI approaches are reliable. The mismatch is particularly
striking in the theory with $N_F=N-1$ hypermultiplets. In this case
the correlation function
\EQ{
{\cal G}=
\VEV{\tfrac{g^2}{16\pi^2}\TrN\lambda^2(x^{(1)})\tilde q_{f_1}q_{f'_1}(x^{(2)})
\times\cdots\times\tilde q_{f_{N-1}}q_{f'_{N-1}}(x^{(N)})}
}
receives a contribution from a single instanton in both the SCI and
WCI approaches. In the SCI methodology one takes the Higgs to be zero
in the bare theory so that the instantons are not constrained. VEVs
for the scalar fields then arise from instanton effects. On the
contrary, in the WCI methodology the Higgs VEV is included in the bare
theory and the instantons are constrained. 
The SCI approach yields the result \cite{Amati:1988ft}
\EQ{
{\cal G}_{\rm
SCI}=\frac{\Lambda_{\sst(N-1)}^{2N+1}}{N!}\epsilon_{\{f_l\},\{f'_l\}}\
,
\elabel{ollia}
}
where $\epsilon_{\{f_l\},\{f'_l\}}$
equals 1 if all the $f_l$ are distinct and $\{f_l\}\equiv\{f'_l\}$,
otherwise it is 0. The WCI result is very easy to establish given the
calculation of $\Vev{\tfrac{g^2}{16\pi^2}\lambda^2}$ in
\S\ref{sec:S46}. This is because the
insertions of the matter fields can be replaced by their VEVs $\tilde
q_fq_{f'}\to\tilde v_fv_{f}\delta_{ff'}$. Multiplying \eqref{wcigc}
by the product of VEVs one finds
\EQ{
{\cal G}_{\rm
WCI}=\Lambda_{\sst(N-1)}^{2N+1}\prod_{l=1}^{N-1}\delta_{f_lf'_l}\
.
\elabel{ollib}
}
What makes the mismatch between \eqref{ollia} and \eqref{ollib}
particularly striking is that the integral over the collective
coordinates that yields the WCI result is formally just the limit of
the SCI integral with the VEVs set to zero: $v_f=\tilde v_f=0$ as 
emphasized in the context of gauge group $\SU(2)$ in
Ref.~\cite{Ritz:2000mq}. So the collective coordinate integral is
discontinuous at this point. These considerations show that the SCI approach
yields an incorrect result even when applied in a weakly-coupled Higgs phase.

\rsen\section{On the Coulomb Branch of 
$\N=2$ Gauge Theories}\elabel{sec:S60}

Theories with $\N=2$ supersymmetry have an adjoint-valued scalar field
and consequently a Coulomb branch where the gauge symmetry is broken
to the maximal abelian subgroup. For large values of the VEVs the
theory is weakly-coupled and constrained instanton methods should be
reliable. Moreover certain holomorphic quantities, and in particular
the {\it prepotential\/} described below, are protected against
quantum corrections (beyond one-loop)
and leading-order semi-classical methods should be
exact. On the Coulomb branch of an $\N=2$
theory there is a completely different approach to calculating the
prepotential based on the
theory of Seiberg and Witten \cite{SeibWitt,SWtwo}. This
remarkable theory predicts in a rather implicit way
the exact form of the prepotential which
can then be expanded in the semi-classical regime yielding the sum
of a tree-level
and one-loop contribution plus a series of instanton terms of
arbitrary charge. In general closed-form expressions for higher
instanton numbers are not known, however what has been established are
recursion relations which
give the higher instanton coefficients, in terms of those of lower
instanton number. The situation gives us an unprecedented
laboratory for testing semi-classical instanton methods and all that
entails: imaginary-time formalism, Euclidean path integral, {\it etc\/}.
Actually the result cuts both ways: not only can we
quantitatively
test semi-classical instanton methods in gauge theory, but, in addition,
we can subject the ingenious theory of Seiberg Witten to
stringent tests.\footnote{We take it as
unreasonably perverse that the instanton
calculus and Seiberg-Witten theory could both be wrong while being
consistent with one another. In any case there are other tests of
Seiberg-Witten theory based, for example, on the spectrum of BPS
states \cite{mybps}.}

After introducing the notion of the low-energy effective action and
the prepotential, we
then go on to show how the instanton contributions at each instanton
number are the centred instanton partition function introduced in
\S\ref{sec:N1}.
We then go on to describe the explicit evaluation of this partition
function for the particular case of gauge group $\SU(2)$ with $N_F$ flavours of
hypermultiplets in the fundamental representation for instanton
number $k=1$ and $k=2$. This calculation relies on the fact that the
gauge group $\SU(2)$ is also $\Sp(1)$ and the ADHM construction based
on $\Sp(N)$, for $N=1$, is more economical than that for $\SU(N)$, for
$N=2$. We then present the
one-instanton calculation for general unitary groups. The results we
obtain are in precise agreement with
Seiberg-Witten theory for any number of flavours. Nevertheless,
the cases with $N_F=2N$ are
rather special since there are non-trivial dictionary issues to
resolve.

We shall conclude this overture with a brief guide to literature
which studied instanton effects in Seiberg-Witten models.
First instanton tests of pure ${\cal N}=2$ supersymmetric $\SU(2)$ theories
were performed in \cite{FP} at a one-instanton level and in \cite{MO-I}
at a two-instanton level. Two-instanton contributions to the prepotential
in $\SU(2)$ theories
with $N_F$ fundamental hypermultiplets were calculated in
\cite{MO-II,AOYAMA,test2} and the general expression for the $k$-instanton
contribution to the prepotential as an integral over the ADHM
moduli space was derived in \cite{MO-II}. The relation of Matone
\cite{Matone:1995rx} between the prepotential
and the condensate $u_2$ in $\SU(2)$ was tested at a two-instanton level in
\cite{Fucito:1997ua}
and derived to all orders in instantons in \cite{dkmmatone}.
$\SU(N)$ gauge theories with and without matter fields were studied
at one-instanton level in \cite{KMS,Ito:1996qj,Ito:1997fd}.
In all these cases exact results of Seiberg-Witten and their generalizations
\cite{Klemm:1995qs,Argyres:1995xh,Argyres:1995wt,Hanany:1995na,Minahan:1996er,D'Hoker:1997nv}
for the low-energy effective action
were reproduced exactly for $N_F<2N$. The case of $N_F=2N$ was considered
in \cite{KMS,Dorey:1997bn} where it was pointed out that the mismatch
between instanton calculations and the proposed exact solutions
 for $N_F=2N$
arises due to a finite (perturbative and non-perturbative)
renormalization of the coupling constant of the low-energy effective theory.
This effect has to be incorporated into exact solutions which then
agree with instanton results. For a recent careful treatment of these
issues see \cite{Argyres:2000ty}.
In addition to these effects,
explicit instanton calculations are also necessary in order to fix
the dictionary between the quantum moduli used in constructing exact solutions
and the gauge-invariant VEVs $u_n$, see \cite{KMS,
DKMn4,AOYAMA,Ito:1997fd,Dorey:1997bn,Argyres:2000ty,Slater:1997zn,Pomeroy}
for more detail.
A completely new technique for evaluating the instanton contributions
to the prepotential has been pioneered in \cite{local,locn4} leading
to the first calculations of instanton effects for all instanton number
(beyond the large-$N$ calculations reported in Chapters \ref{sec:S43}
and \ref{sec:S39}) and the first test of Seiberg-Witten theory to all
orders in the instanton expansion. We will
briefly describe this new formalism in \S\ref{sec:N2}.

\subsection{Seiberg-Witten theory and the
prepotential}\label{sec:V2}

Theories with $\N=2$ supersymmetry and gauge group
$\SU(N)$ on the Coulomb branch
have an $N-1$ complex dimensional vacuum moduli space parameterized by
the VEVs \eqref{VEVs} for the complex scalar
field $\phi$.
At a generic point on this classical moduli space the gauge group is
broken to its maximal abelian subgroup $\U(1)^{N-1}\subset\SU(N)$ and
the theory is in a Coulomb phase. In these circumstances it is
possible to describe the long-distance behaviour of the
theory in terms of a low-energy
effective action. This can be written in terms of abelian
$\N=1$ superfields\footnote{For each of these superfields
we have the traceless condition $\sum_{u=1}^NW_{\alpha
u}=\sum_{u=1}^N\Phi_u=0$.} $W_{\alpha u}=(A_{mu},\lambda_u)$
and $\Phi_u=(\phi_u,\psi_u)$: 
\EQ{
S_{\rm eff}  = \frac1{4\pi}\int d^4x\,
{\rm Im}\Big\{\tfrac12\tau_{uv}(\Phi)\,W^{\alpha}_u
W_{\alpha v} \Big\vert_{\theta^2} + \Phi_{{\rm D}u}(\Phi) \ \Phi_u^\dagger
\Big\vert_{\theta^2\bar\theta^2}\Big\}
\ .
\label{prept}}
where the VEVs of the scalar components of $\Phi_u$ are identified
with $g\phi^0_u/\sqrt2$ of the field of the microscopic theory.
This low-energy effective action is uniquely determined
by the holomorphic prepotential ${\cal F}(\Phi)$ \cite{SeibWitt} as
\EQ{
\Phi_{{\rm D}u} \equiv {\partial {\cal F} \over \partial \Phi_u} \ ,
\qquad
\tau_{uv} \equiv
{\partial^2 {\cal F} \over \partial \Phi_u \partial \Phi_v}
\ .
\label{dtaud}}
Equation \eqref{prept} implies that the matrix of
complexified coupling constants
of the low-energy $\U(1)^{N-1}$ theory has components
$\tau_{uv}$ that depend on the scalar VEVs.

The electric-magnetic duality of the $\U(1)^{N-1}$ theory, uncovered
by Seiberg and Witten \cite{SeibWitt,SWtwo},
identifies $\Phi_{{\rm D}u}$ with the magnetic dual of
the original (electric) matter superfield $\Phi_u$.
The vacuum expectation values of the scalar component of $\Phi_{{\rm D}u}$
are $g\phi_{{\rm D}u}^0/\sqrt2$, with $\sum_{u=1}^N\phi_{{\rm D}u}^0=0$,
which provide an alternative
parameterization of the vacuum moduli space. Both parameterizations,
$\{\phi^0_u\}$ and $\{\phi^0_{{\rm D}u}\}$,
are not valid globally on the moduli
space. Such a global parameterization is provided \cite{SeibWitt,SWtwo}
by the gauge-invariant condensates $\{u_n\}$ $n=2,\ldots,N$:
\EQ{
u_n =2^{-n/2}g^n\VEV{\TrN\,\phi^n}\ .
\label{udcls}
}
On a patch of the moduli space where $\{\phi^0_u\}$ are a
local coordinate system we have $\phi^0_u=\phi^0_u(u_n)$ while on the patch
where the dual variables $\{\phi^0_{{\rm D}u}\}$ are a local coordinate system
we have $\phi^0_{{\rm D}u}=\phi^0_{{\rm D}u}(u_n)$.
The analysis of Seiberg and Witten \cite{SeibWitt,SWtwo},
and its generalizations, explicitly determine these
functions in terms of an auxiliary Riemann surface
defined by a family of hyper-elliptic curves.

The effective action specifies the long-range behaviour, compared
with the scale $1/(g\phi^0)$, of a series of
correlation functions. In particular, we will focus on
a four-point anti-chiral fermion correlator which involves the fourth
derivative of the prepotential. In Euclidean space
\SP{
&\langle \bar{\lambda}^{\dot{\alpha}}_{u_1}(x^{(1)})
\bar{\lambda}^{\dot{\beta}}_{u_2}(x^{(2)})
\bar{\psi}^{\dot{\gamma}}_{u_3}
(x^{(3)})\bar{\psi}^{\dot{\delta}}_{u_4}(x^{(4)})\rangle\\
&=
\frac{1}{2\pi i}\,\frac{\partial^{4}{\cal F}}
{\partial\phi^0_{u_1}\partial\phi^0_{u_2}
\partial\phi^0_{u_3}\partial\phi^0_{u_4}} \,
\int\, d^{4}X\,
\bar{\rm S}^{\aD\alpha}(x^{(1)},X)
\bar{\rm S}^{\bD}{}_\beta(x^{(2)},X)
\bar{\rm S}^{\gD\gamma}(x^{(3)},X)
\bar{\rm S}^{\dD}{}_\gamma(x^{(4)},X)\ ,
\label{foureffc}
}
where
$\bar{\rm S}(x,X)$ is the free anti-Weyl spinor propagator
\EQ{
\bar{\rm S}(x,X)=\frac1{4\pi^2}\delbarslash\frac1{(x-X)^2}\ .
\label{spinprop}
}

The restrictions imposed by holomorphy, renormalization
group invariance and the anomaly imply that the prepotential  has a
weak coupling expansion consisting of perturbative piece that is
one-loop exact and a sum over instanton contributions which are exact
to leading order in the semi-classical expansion. For $\SU(N)$ with
$N_F<2N$ fundamental hypermultiplets
\EQ{
{\cal F}={\cal F}_{\rm pert}+\frac1{2\pi i}\sum_{k=1}^\infty{\cal
F}_k\Lambda^{k(2N-N_F)}\ .
\elabel{ppseries}
}
In the special case $N_F=2N$, for which the $\beta$-function vanishes,
\EQ{
{\cal F}\Big|_{N_F=2N}=\tfrac{g^2}2\tau\sum_{u=1}^N(\phi^0_u)^2+
\frac1{2\pi i}\sum_{k=1}^\infty{\cal
F}_ke^{2\pi ik\tau}\ .
\elabel{preci}
}
In general, the coefficients ${\cal F}_k$ depend on the VEVs and
hypermultiplet masses.

The theory of Seiberg and Witten, extended from the original
setting for $\SU(2)$ \cite{SeibWitt}, to $\SU(N)$, $N>2$, in
Refs.~\cite{Klemm:1995qs,Argyres:1995xh,Argyres:1995wt,Hanany:1995na,Minahan:1996er},
determines
the exact form of the prepotential and in particular the $k$-instanton
coefficients ${\cal F}_k$. The construction involves an auxiliary algebraic
curve, or Riemann surface,
the {\it Seiberg-Witten curve\/}.\footnote{In a nut-shell, the
matrix of coupling $\tau_{uv}$ is identified with the period matrix of
the curve in a suitable basis.} Once the curve has been
identified, the coefficients can, in principle, be extracted. In
particular, explicit expression for low instanton number have been
found in this way
\cite{Ito:1996qj,D'Hoker:1997nv}. For arbitrary instanton
number, explicit expressions are hard to obtain, however, recursion
relations have been established that relate instanton contributions at
charge $k$ to those of lower charge
\cite{Matone:1995rx,Edelstein:1999sp,Edelstein:1999dd,Edelstein:2000xk,Chan:2000gj}.
These recursion relations are only
valid in the asymptotically-free theories with $N_F<2N$. The finite
theory with $N_F=2N$ is rather special since there are various
re-definitions of the physical quantities that have to be
taken into account: see Ref.~\cite{Argyres:2000ty}.

Here, we use the
notation of \cite{D'Hoker:1997nv} to write down the $k=1$ and $k=2$ contributions
for $N_F<2N$:
\AL{
{\cal F}_1&=\sum_{u=1}^NS_u(\phi^0_u)\ ,
\label{onei}\\
{\cal F}_2&=
\sum_{u,v=1\atop(u\neq v)}^N\frac{S_u(\phi^0_u)S_v(\phi^0_v)}{
(\phi^0_u-\phi^0_v)^2}+
\tfrac14\sum_{u=1}^NS_u(\phi^0_u)\frac{\partial^2S_u(\phi^0_u)}
{\partial(\phi^0_u)^2}\ ,
\label{twoi}
}
where we have used the function
\EQ{
S_u(x)\equiv \prod_{{v=1\atop(\neq u)}}^N
\frac1{(x-\phi^0_v)^2}\prod_{f=1}^{N_F}(m_f+x)\ .
\label{defs}
}

\subsection{Extracting the prepotential from instantons}\label{sec:V3}

There are two distinct---but ultimately equivalent---ways of
determining the instanton contributions to the prepotential.
The first involves calculating the leading
semi-classical contribution to
Green's functions whose long-distance limit can be matched
with the low-energy effective action \eqref{prept}. Here, we will
concentrate on the four-point anti-chiral fermion correlator
\eqref{foureffc} which involves the fourth derivative of the
prepotential. However, other correlators which we do not consider
explicitly
determine the second derivative of the prepotential in an analogous way.
The second approach involves calculating the instanton
contribution to the condensate $u_2$ defined in \eqref{udcls} and then
relating this to the prepotential by use of a renormalization group
equation which yields the derivative of the prepotential with respect
to $\log\,\Lambda$ (for $N_F<2N$)
\cite{Matone:1995rx,Eguchi:1996jh,D'Hoker:1997ph,Sonnenschein:1996hv}.
Alternatively, we will take the view that the instanton calculation of
the condensate provides an independent way to establish the
renormalization group relation.

For the first approach we turn our focus to the instanton contributions to the
four-point anti-chiral fermion correlator \eqref{foureffc}. The first point
to make is that since there are non-trivial scalar VEVs the instantons
are constrained as described in  \S\ref{sec:S34}.
It will transpire that the integral over the 4-vector $X$ in
\eqref{foureffc} arises in the instanton context
from the integral over the centre of the instanton
\eqref{transmo}. Consequently, in
order to extract the long-distance behaviour of the correlator
we should insert the corresponding
behaviour of the anti-chiral fermions $\bar\lambda_A
=(\bar\lambda,\bar\psi)$ far from the core of the constrained
instanton. Recall from our discussion in \S\ref{sec:S34}, that
in the tail of the constrained instanton
fields decay exponentially except for the
components which remain massless after the Higgs mechanism. These are
the fields valued in the $\U(1)^{N-1}$ subgroup picked out by the VEV
$\phi^0$; in other words the diagonal elements of the
$N\times N$ matrices $\bar\lambda_A$.
It is precisely these components of the anti-chiral fermions that
we need in order calculate the long-distance behaviour of the
correlator. This is fortunate because, as we argued in \S\ref{sec:S34},
to leading order in the semi-classical expansion these massless
components are simply given by their ADHM expressions.
Moreover, we will only need the components of the
anti-chiral fermions that depend on the four supersymmetric Grassmann
collective coordinates $\xi^A$ \eqref{susymo}, since these are the only
Grassmann variables whose integrals are not saturated by the instanton
effective action \eqref{ieant}. It will turn out that each insertion
of $\bar\lambda_A$ is then linear in $\xi^B$ and therefore the
insertion of the four anti-chiral fermions precisely saturates the integrals
over the four Grassmann variables associated to the
unlifted supersymmetric zero modes. In principle, we can extract the
terms that we need from the solution for the anti-chiral fermions in
\eqref{expaf} (with $\bar\psi_A$ given in \eqref{psiansatz} and
\eqref{psionedef}-\eqref{psithreedef}).
However, it more straightforward to use the sweeping-out
procedure described in \S\ref{sec:S31}. 
The supersymmetry transformation \eqref{lhc}
(re-scaled by $g^{1/2}$) in the
super-instanton background gives the required dependence on $\xi^A$:
\EQ{
\bar\lambda_A=-ig^{1/2}\bar\Sigma_{aAB}\Dbarslash\phi_a\xi^B=
-g^{1/2}\Dbarslash\phi^\dagger\xi_A
\ ,
\label{jjblob}
}
where in the latter expression we used the relations \eqref{ntsc}.
The expression for the anti-holomorphic component of the scalar field
is given in \eqref{aholsf} (it is important to remember that in an
instanton background $\phi^\dagger$ is {\it not\/} the conjugate of
$\phi$). We then expand the diagonal components of
$\Dbarslash\phi^\dagger$  (these are the massless components)
for large distances from the instanton core
using the asymptotic formulae in \S\ref{sec:S12}:
\EQ{
{}(\Dbarslash\phi^\dagger)_{uu}=-
\delbarslash\frac1{(x-X)^2}
w_{u\aD}\bigg\{\phi^{0\dagger}_u1_{\sst[k]\times[k]} +
\BL^{-1}\Big(\bar w^\bD\phi^{0\dagger}
w_\bD-\tfrac 1{4}\sum_{f=1}^{N_F}
\K_f\tilde\K_f\Big)\bigg\}\bar w^\aD_u\ .
}
Plugging this into \eqref{jjblob}, and using the form of the instanton
effective action \eqref{ieant}, we are able to 
establish the following simple relation between the 
long-distance behaviour of the anti-chiral fermion and the instanton
effective action:\footnote{Here, $\bar\lambda_{u}
\equiv(\bar\lambda)_{uu}$ the diagonal elements of $\bar\lambda$.}
\EQ{
\bar\lambda_{Au}^{\aD}=
2\sqrt{g}\,\bar{\rm S}^{\aD\alpha}(x,X)\epsilon_{AB}\xi^A_\alpha
\PD{\tilde S}{\phi^0_u}+\cdots\ .
\label{jjins}
}
The ellipsis represent terms depending on other Grassmann collective
coordinate which will not be required to determine the long-range
behaviour of the correlator in question.

The $k$-instanton contribution to the correlation function \eqref{foureffc}
is evaluated in the usual way by making
insertions of the long-distance component
\eqref{jjins} into the collective coordinate integral:
\SP{
&\VEV{\bar{\lambda}^{\dot{\alpha}}_{u_1}(x_{1})
\bar{\lambda}^{\dot{\beta}}_{u_2}(x_{2})
\bar{\psi}^{\dot{\gamma}}_{u_3}
(x_{3})\bar{\psi}^{\dot{\delta}}_{u_4}(x_{4})}_k\\
&= \left(\frac\mu g\right)^{k(2N-N_F)}e^{2\pi
ik\tau}\int_{\ms_k}\Bomega^{\sst(\N=2,N_F)}
\,e^{-\tilde S}\,\bar{\lambda}^{\aD}_{u_1}(x^{(1)})
\bar{\lambda}^{\bD}_{u_2}(x^{(2)})
\bar{\psi}^{\gD}_{u_3}
(x^{(3)})\bar{\psi}^{\dD}_{u_4}(x^{(4)})
\ .
\label{res}
}
Here, $\Bomega^{\sst(\N=2,N_F)}$ is the supersymmetric volume form on the
instanton moduli space in a theory with $N_F$ hypermultiplets
defined in \eqref{fms} and \eqref{mattmeas}.
The instanton effective action $\tilde S$
is precisely \eqref{ieant} along with the hypermultiplet mass term
\eqref{simm}.
Following \cite{MO-II,dkmmatone}, we substitute the expression
\eqref{jjins} into the right-hand side and perform the
integrals over the four
supersymmetric Grassmann variables $\xi^A$. This leaves
\SP{
&\VEV{\bar{\lambda}^{\dot{\alpha}}(x^{(1)})
\bar{\lambda}^{\dot{\beta}}(x^{(2)})
\bar{\psi}^{\dot{\gamma}}
(x^{(3)})\bar{\psi}^{\dot{\delta}}(x_{4})}_k= \\
&\qquad\qquad\frac1{4\pi^2}g^2\left(\frac\mu g\right)^{k(2N-N_F)}e^{2\pi
ik\tau}
\frac{\partial^{4}}{\partial
\phi^0_{u_1}\partial\phi^0_{u_2}\partial\phi^0_{u_3}\partial\phi^0_{u_4}}
\int_{\widehat\ms_k}\Bomega^{\sst(\N=2,N_F)}
\,e^{-\tilde S}\\
&\qquad\qquad\times\,\int\, d^{4}X\,
\bar{\rm S}^{\aD\alpha}(x^{(1)},X)
\bar{\rm S}^{\bD}{}_\alpha(x^{(2)},X)
\bar{\rm S}^{\gD\gamma}(x^{(3)},X)
\bar{\rm S}^{\dot{\delta}}{}_\gamma(x^{(4)},X)\ .
\label{greeniesc}
}
The integral over the centred moduli space $\widehat\ms_k$ is
precisely the {\it centred instanton partition function\/} $\widehat{\EuScript
Z}_k^{\sst(\N=2,N_F)}$ defined in \eqref{cipfun}.
Note in \eqref{greeniesc} the linearity of
$\tilde S$ in $\phi^0_u$, apparent in \eqref{ieant}, has been used to pull the
$\phi^0_u$ derivatives outside the collective coordinate integral.

Comparing the semi-classical expression \eqref{greeniesc} with its exact
counterpart \eqref{foureffc}, we deduce the following expression for
the $k$-instanton expansion coefficient of the prepotential in
\eqref{ppseries}:
\EQ{
{\cal F}_k=g^{-k(2N-N_F)+2}
\widehat{\EuScript Z}_k^{\sst(\N=2,N_F)}\ .
\label{ndeqb}
}
One might have thought that we could add to this relation any function
whose fourth derivative with respect to the VEVs vanishes. Actually,
by considering other correlators, one can establish \eqref{ndeqb} up
to functions whose second derivative with respect to the VEVs
vanishes. However, since there are no possible linear functions of the
VEVs, \eqref{ndeqb} must be true up to an undetermined constant which does
not affect the physics because the low-energy effective action only
depends on derivatives of ${\cal F}$.
Note that in order to write the instanton contributions in the form \eqref{ppseries}
we have substituted the expression for the $\Lambda$-parameter of
the theory in terms of the running coupling:
\EQ{
\Lambda_{\sst(N_F)}^{2N-N_F}\equiv\mu^{2N-N_F}e^{2\pi i\tau}\ .
\label{defn2l}
}
{}For the special case when $N_f=2N$, where the $\beta$-function
vanishes, the factor $\Lambda_{\sst(N_{F})}^{2N-N_{F}}$ should
be replaced by $q=e^{2i\pi\tau}$.

The other way to calculate the instanton expansion of the prepotential
is to use the renormalization group equation for the prepotential
first established by Matone in the context
of pure $\SU(2)$ theory \cite{Matone:1995rx} and then generalized to
$\SU(N)$ in Refs.~\cite{Eguchi:1996jh,D'Hoker:1997ph,Sonnenschein:1996hv}.
Rather than write down the relevant equation we shall proceed to prove
a version of the renormalization
group equation using
the instanton calculus, generalizing the $\SU(2)$ calculation of
\cite{dkmmatone}. One may then check that our instanton version is
consistent with those in the literature cited above.
Our relation will also be valid in the finite theory where $N_F=2N$.
The approach is rather different from that just follows. Instead of
calculating the instanton contribution to the long-range behaviour of
a correlator, one calculates the instanton contributions to
the condensate $u_2$.

To begin, we establish the
form of the insertion $\TrN\,\phi^2$ in the constrained
instanton background. Just
as in the calculation of the gluino condensate in the $\N=1$ theories
in the Higgs phase described in \S\ref{sec:S46} and \S\ref{sec:S48},
to leading order we may replace the insertion by its value in the
unconstrained ADHM instanton background. This is because any error
incurred is necessarily of a higher order in $g$. Just as for the 4-point
anti-chiral fermion correlator \eqref{foureffc},
the insertion must saturate the integrals over
the four supersymmetric Grassmann variables $\xi^A$. Since, as is
evident from
\eqref{ssdd}, the scalar field is quadratic in the Grassmann collective
coordinates so the composite $\TrN\,\phi^2$ is indeed quartic in the
Grassmann collective coordinates. The dependence on $\xi^A$ may be
obtained by the sweeping out procedure using the supersymmetry
transformation \eqref{lhd}:
\EQ{
\phi_a=-\tfrac12g\bar\Sigma_{aAB}\xi^A\sigma_{mn}\xi^B F_{mn}+\cdots\ .
}
Therefore, using \eqref{ntsc},
$\phi=\phi_1-i\phi_2$, and the identity
\EQ{
\xi\sigma_{mn}\eta\,\xi\sigma_{pq}\eta=-\tfrac18(\xi\xi)(\eta\eta)\ ,
}
for arbitrary spinors $\xi$ and $\eta$,
we have
\EQ{
\tfrac12g^2\TrN\,\phi^2(x)=
\tfrac12g^4(\xi^1\xi^1)(\xi^2\xi^2)\TrN\,F_{mn}^2(x)+\cdots\ .
\label{insu2}
}
We now insert \eqref{insu2} into the collective coordinate
integral. Separating out the variables $\{X,\xi^A\}$ as in
\eqref{sepout} one can then trivially integrate over the Grassmann
variables $\xi^A$. The integral over the 4-vector $X$ is also trivial
because by translational symmetry it may be traded for an integral
over the insertion point $x$ yielding
$\int d^4x\,\TrN F_{mn}^2(x)$ which is the integral that gives the
instanton charge \eqref{topc}. Hence, the
$k$-instanton contribution to the condensate is
\EQ{
u_2\big|_k=k\Lambda^{k(2N-N_F)}_{\sst(N_F)}g^{-k(2N-N_F)+2}
\,\widehat{\EuScript Z}_k^{\sst(\N=2,N_F)}\ .
\label{kfune}
}
Notice that the instanton contribution to the condensate $u_2$ is
proportional to the centred instanton partition function just as for
the expression for the gluino condensate \eqref{oipfn1} in the
$\N=1$ theory. Of course in the $\N=2$ theory the condensate receives
contributions from all instanton numbers whereas one can easily show
that in the $\N=1$ theory the centred instanton partition function
vanishes for $k>1$.

Comparing \eqref{kfune} with our formula for the prepotential \eqref{ndeq}
we obtain a version of the renormalization group equation
in the form of Ref.~\cite{dkmmatone}.
The result is most easy to state in terms of the $k$-instanton contribution:
\EQ{
u_2\big|_k=k\Lambda^{k(2N-N_F)}{\cal F}_k\ .
\label{matone}
}
This expression relates only the non-perturbative
contributions to both quantities. The rest, however,
is easy to determine via direct perturbative calculation.
In addition to instantons,
the prepotential receives only 1-loop perturbative
contributions, while $u_2$ receives no perturbative
corrections to its classical value.
It is remarkable that we were able to derive \eqref{matone} using
instanton calculus, but without actually having to integrate
over the instanton moduli space. Note that the result is
equally valid in the finite theory $N_F=2N$ on replacing $\Lambda^{2N-N_F}\to
e^{2\pi i\tau}$ and it is also valid for arbitrary hypermultiplet masses.

At this point we find it useful to perform various re-scalings of the
variables by powers of the coupling $g$ in order that there are is no
explicit $g$ dependence in $\widehat{\EuScript Z}_k$. The required
re-scalings are
\EQ{
\phi^0\to g^{-1}\phi^0\ ,\qquad a_\aD\to ga_\aD\ ,\qquad
\CM^A\to g^{1/2}\CM^A\ ,\qquad
\K\to g^{1/2}\K\ ,\qquad\tilde\K\to g^{1/2}\tilde\K\ .
\elabel{nicer}
}
Under these re-scalings, the
centred instanton partition function scales as
\EQ{
\widehat{\EuScript Z}_k^{\sst(\N=2,N_F)}\to g^{k(2N-N_F)+2}\widehat{\EuScript
Z}_k^{\sst(\N=2,N_F)}\ .
}
In particular there is no $g$ dependence in the re-scaled instanton
effective action action and the re-scaled version of
\eqref{ndeqb} is simply
\EQ{
{\cal F}_k=\widehat{\EuScript Z}_k^{\sst(\N=2,N_F)}\ ,
\label{ndeq}
}
where all $g$ dependence has disappeared (as it should).

\subsection{Gauge group $\SU(2)$}\elabel{sec:S84}

In this section, we specialize to gauge group $\SU(2)$ and compute the
one and two instanton contributions to the prepotential in order to
compare with the exact theory of Seiberg and Witten.
For $\SU(2)$, there is a single
VEV $\phi^0\equiv\phi^0_1/2=-\phi^0_2/2$. The predictions for the one-
and two-instanton coefficients are obtained from \eqref{onei} by
setting $N=2$.

Since the gauge group $\SU(2)$ is isomorphic to $\Sp(1)$ one can proceed with
either of the two formalisms; however, as described in \S\ref{app:A7},
the ADHM construction for $\Sp(1)$ is more economical in the sense that
for a given $k$
their are fewer ADHM variables and constraints compared with the
$\SU(2)$ formalism. Hence, the $\Sp(1)$ formalism
is better suited for a direct calculation of the prepotential.

Almost all of the formulae that we
established in the $\SU(N)$ instanton calculus case carry over to
$\Sp(1)$ by simply imposing the extra restrictions
\eqref{breality} and \eqref{freality} on $a_\aD$ and $\CM^A$,
respectively. Recall, once we have replaced the gauge index
$u=1,2$ by a spinor index $\alpha=1,2$, the ADHM variables
$w_{\aD ui}$ become quaternions $w_{\alpha\aD i}$. From \eqref{giiy}
\EQ{
a=\MAT{w\\
a'}
\ ,
\label{bcanonical}
}
where $w$ is a $k$-vector of quaternions and
$a'$ is a $k\times k$ symmetric matrix whose elements are
quaternions. In \eqref{bcanonical} and in much of the following
the quaternion indices are understood.
Products of quaternions are defined as $2\times 2$
matrix multiplication. We also define the quaternion inner product
\EQ{
x\cdot y=\tfrac14{\rm tr}_2(\bar xy+\bar yx)\ ,\qquad
|x|^2\equiv x\cdot x=x_nx_n\ .
}
For the Grassmann collective coordinates
\EQ{
\CM^A=\MAT{\mu^A\\ \CM^{\prime A}}\ ,
}
for $A=1,2$, where $\mu^A$ and $\CM^{\prime A}$ are Weyl
spinor-valued $k$-vectors and $k\times k$ symmetric matrices,
respectively.

The collective coordinate integral will involve the volume form on the
instanton moduli space which is given by an obvious translation of the
general $\SU(N)$ formula \eqref{fms}:
\SP{
&\int_{\ms_k}\Bomega^{\sst(\N=2)}
={C^{\sst(\N=2)}_k
\over{\rm Vol}\,\O(k)}\,\int\,\prod_{i=1}^k\,
d^4w_i\,\prod_{i\leq j}d^4a'_{ij}\,\Big\{\prod_{A=1}^2
\prod_{i=1}^kd^2\mu^A_i\,\prod_{i\leq
j=1}^kd^2\CM_{ij}^A\Big\}\\
&\times\big|\det\,\BL\big|^{-1}\,\prod_{i<j=1}^k\bigg\{
\prod_{c=1}^3\delta\big(\tfrac12{\rm
tr}_k\,\tau^c{}^\aD{}_\bD((\bar a^\bD a_\aD)_{ij}-
(\bar a^\bD a_\aD)_{ji})\big)
\prod_{A=1}^2\prod_{\aD=1}^2\,\delta\big(
(\bar{\cal M}^Aa_\aD)_{ij}-(\bar{\cal M}^Aa_\aD)_{ji}\big)\bigg\}\ .
\label{sponem}
}
In this expression, integrals over a quaternion $w$
are defined as $\int d^4w\equiv \int dw_1\,dw_2\,dw_3\,dw_4$.
Note in the one-instanton sector the $\delta$-functions are absent
(unlike the $\SU(2)$ formalism). In this formalism $\BL$ is now a
linear operator on the space of $k\times k$
antisymmetric matrices but still defined as in \eqref{vvxx}. Finally,
the normalization constant is fixed by using
clustering and comparison with the one-instanton collective coordinate
integral in \cite{tHooft}:
\EQ{
\label{Conepequals}
C^{\sst(\N=2)}_k=2^{5k-k^2}\pi^{-4k}\ .
}

The leading order expression for the
instanton effective action of the $\N=2$ theory with $N_F$
fundamental hypermultiplets can be read of the expression in the
$\SU(N)$ theory \eqref{ieant}. In addition, we add the mass terms
\eqref{simm}. The collective coordinate integral also includes integrals
over the hypermultiplet
collective coordinates $\{\K,\tilde\K\}$ in \eqref{msci}.
Consider the complete expression for the collective coordinate
integral in the chiral limit, $m_f=0$.
In this limit, for fixed flavor index $f$,
the Grassmann measure \eqref{msci} is obviously even or odd under the
discrete symmetry
\EQ{
\K_{if}\ \leftrightarrow\ \Kt_{fi}\ ,
\label{discretion}
}
depending on whether $k$ itself is even or odd.
On the other hand, the instanton effective action \eqref{ieant}
(without the mass terms) is always even under this symmetry.
Therefore, for $N_F>0,$ only the even-instanton sectors $k=0,2,\ldots$
can contribute in the chiral limit\footnote{The
absence of a one-instanton contribution is quite easy
to see since $\{\K_f,\tilde\K_f\}$ completely decouple from the instanton
effective action \eqref{ieant} and consequently their
integrals remain unsaturated.}
(recall that when $N_F=0$ all instanton numbers contribute).
This selection rule was already noted
by Seiberg and Witten in Sec.~3 of \cite{SWtwo} so it is satisfying
to see it arising naturally in the instanton calculus.
Of course it is violated once the masses are non-zero,
since $\K_{if}\Kt_{fi}$ is odd under the symmetry.

\subsubsection{One instanton}\elabel{sec:S842}

Recall that in order to calculate instanton contributions to the
prepotential we have to calculate the centred instanton partition
function \eqref{ndeq}. We begin with the one-instanton case.
In ADHM language, the bosonic and fermionic parameters of a single
$\N=2$ super-instanton are contained in three $2\times 1$ matrices of
unconstrained parameters:
\EQ{
a=\MAT{w \\ -X}\ ,\qquad
{\cal M}^A=\MAT{\mu^A \\ -4i\xi^A}\ .
\label{oneI}
}
In addition, there are $2N_{F}$ Grassmann variables $\{{\cal K}_f,
\tilde{\cal K}_f\}$ which parameterize the fundamental zero modes.
The centred-instanton volume form is extracted from
\eqref{sponem}:
\EQ{
\int_{\widehat\ms_1} \Bomega^{\sst(\N=2,N_F)}
=\frac{2^{3}}
{\pi^{4+2N_{F}}}\int\,d^{4}w\,\prod_{A=1}^2d^{2}\mu^A\,
\prod_{f=1}^{N_F}d{\cal K}_f\,d\tilde{\cal K}_f\ ,
\label{oneImeasure}
}
where the instanton effective action for $k=1$ is easily deduce from
\eqref{ieant} and \eqref{simm}:
\EQ{
\tilde S=8\pi^{2}|w|^{2}
|\phi^0|^2
-2i\pi^{2}\mu^A\phi^{0\dagger}\mu_A+\pi^{2}
\sum_{f=1}^{N_{F}}m_f{\cal K}_f\tilde{\cal K}_f\ .
\label{oneIaction}
}
Notice that the only dependence on
${\cal K}_f$ and $\tilde{\cal K}_f$ comes exclusively from the mass
term at the one-instanton level. The corresponding Grassmann
integrations are easily saturated by bringing down appropriate powers
of this term from the exponent. As
expected from the discrete symmetry \eqref{discretion}, the result is
non-zero only when all the  $m_f$ are non-vanishing.
The remaining integrals are easily performed, yielding
\EQ{
\label{oneIprepot}
\widehat{\EuScript Z}_1^{\sst(\N=2,N_F)}\Big|_{N=2}=\frac2{(\phi^0)^2}
\prod_{f=1}^{N_{F}}m_{f}\ .
}
Using \eqref{ndeq}, this gives ${\cal F}_1$ which
is in agreement with \eqref{onei} for $N_F<4$, up to
irrelevant VEV-independent constants for the cases $N_F=2,3$.

\subsubsection{Two instantons}\elabel{sec:S843}

Next we turn to the more calculationally intensive
2-instanton contribution. The parameters of the $k=2$
ADHM super-instanton are contained in the following $3\times 2$ matrices:
\EQ{
a= \begin{pmatrix}w_1&w_2\cr -X+a_3&a_1\cr a_1&-X-a_3
\end{pmatrix}\ ,\qquad
\CM^A=\begin{pmatrix}\mu_{1}^A&\mu_{2}^A\\
-4i\xi^A+\CM^{\prime A}_{3}&\CM^{\prime A}_{1}\\
\CM^{\prime A}_{1}&-4i\xi^{A}-\CM^{\prime A}_{3}\end{pmatrix}\ .
}
Each element of $a$ is a quaternion and of $\CM^A$ a Weyl spinor.
In addition, there are now $4N_{F}$ fundamental zero modes
parameterized
by the Grassmann numbers ${\cal K}_{if}$ and $\tilde{\cal
K}_{fi}$. We also define the following frequently
occurring combinations of these collective coordinates:
\SP{
L &=|w_{1}|^{2}+|w_{2}|^{2}\ ,\qquad
H =|w_{1}|^{2}+|w_{2}|^{2} + 4|a_{1}|^{2}+4|a_{3}|^{2} \ ,\\
\Omega&=w_1\bar w_2-w_2\bar w_1\ ,\qquad
\omega =\tfrac12\phi^0{\rm tr}_2\big(
\bar{w}_{2}\tau^3w_{1}-\bar{w}_{1}\tau^3 w_{2}\big)\ ,\\
Y &=-\mu^A_{1}\mu_{2A}-2
{\cal M}^{\prime A}_{3}{\cal M}^{\prime}_{1A}\ ,\qquad
Z =\sum_{f=1}^{N_{F}}\big({\cal K}_{1f}\tilde{\cal K}_{f2}-
{\cal K}_{2f}\tilde{\cal K}_{f1}\big)\ .
\label{useful}
}
For $k=2$ $\BL$ is just multiplication by the
quantity $H$.
In terms of these variables the instanton effective action
\eqref{ieant} (with the mass term \eqref{simm} and re-scalings
\eqref{nicer}) is written
\SP{
\tilde S&=
8\pi^2L|\phi^0|^2-2\pi^2i\phi^{0\dagger}
\big(\mu^A_1\mu_{1A}-\mu^A_2\mu_{2A}\big)\\
&-{8\pi^2\over H}\big(\bar\omega-\tfrac18Z\big)
\big(\omega-\tfrac i2Y\big)+\pi^2\sum_{f=1}^{N_F}
m_f\big(\K_{1f}\tilde\K_{f1}+\K_{2f}\tilde\K_{f2}\big)\ .
}
We can now proceed to evaluate the centred instanton partition function.
To start with, we can explicitly solve the bosonic and fermionic ADHM
constraints. This is conveniently done by
eliminating the off-diagonal elements $a_{1}$ and ${\cal M}^{\prime A}_{1}$,
as follows:
\EQ{
a_1={1\over4|a_3|^2}\,a_3(\wbar_2w_1-\wbar_1w_2)\ ,
\qquad\CM^{\prime A}_1={1\over2|a_3|^2}\,a_3\,\big(2\abar_1
\CM^{\prime A}_3+\wbar_2\mu^A_1
-\wbar_1\mu^A_2\,\big)\ , \label{aonedefb}\ .
}

We can now explicitly integrate out the $\delta$-functions in the
expression for the centred $k=2$ instanton volume form \eqref{sponem}:
\EQ{
\int_{\widehat\ms_k}\Bomega^{\sst(\N=2)}
=\frac{C^{\sst(\N=2)}_2}{2^{10}}
\,\int\,d^4a'_1\,d^4w_1\,d^4w_2\,
\Big\{\prod_{A=1}^2d^2\CM^{\prime A}_3\,d^2\mu^A_1\,d^2\mu^A_2\Big\}\,
 \frac{\big||a_{3}|^{2}-|a_{1}|^{2} \big|}{H}\ .
\label{defmua}
}
The bosonic parts of
the 2-instanton collective coordinate integral
was originally derived in Refs.~\cite{OSB,GMO,Mansfield:1981sk} by directly
changing variables in the path integral.

We now have a series of integrals to perform. First of all,
performing the Grassmann integrals over the
parameters of the adjoint zero modes gives \cite{MO-II}
\SP{
&\int\, \Big\{\prod_{A=1}^2d^{2}{\cal M}^{\prime A}_{3}\,d^{2}\mu^A_{1}\,
d^{2}\mu^A_{2}\Big\}
\,\exp\Big\{2\pi^2i\big(\phi^{0\dagger}
(\mu^A_1\mu_{1A}-\mu^A_2\mu_{2A})
+{Y\over H}\,\big(\omegabar-\tfrac18Z\big)\big)\Big\}\\
&=
-\left(\frac{16\pi^{6}(\bar{\omega}-\tfrac18Z)}
{|a_{3}|^{2}H}
\right)^{2}\Big\{2^{-8}(\phi^{0\dagger})^{4}|\Omega|^{2}
+\frac{L}{8H}
(\phi^0)^{\dagger2}
(\bar{\omega}-\tfrac18Z)\bar{\omega}\\
& \qquad\qquad\qquad+\tfrac14H^{-2}(\bar{\omega}
-\tfrac18Z)^{2}
\big(\quarter(\phi^{0\dagger})^{2}
(L^{2}-|\Omega|^{2})+\bar{\omega}^{2}\big)\Big\}\ .
\label{lift}
}
This is the generalization to $N_F>0$ of the Yukawa determinant
given in Eq.~(8.13) \hbox{of \cite{MO-I}.}
The next step is to integrate over the Grassmann  collective
coordinates $\{\K,\tilde\K\}$ using the identity
\SP{
&\int\,d^{2N_F}\K\,d^{2N_F}\tilde\K\, G(Z)\exp\Big(-\pi^2\sum_{f=1}^{N_F}
m_f\big(\K_{1f}\tilde\K_{f1}+\K_{2f}\tilde\K_{f2}\big)\Big)\\
&\qquad\qquad\qquad =\sum_{l=0}^{N_{F}}
\frac{M^{({N_{F}})}_{N_{F}-l}}{\pi^{4l}}
\frac{\partial^{2l}G}{\partial Z^{2l}}\Big|_{Z=0}\ ,
\label{littlest}
}
where
\EQ{
M^{\sst(N_{F})}_l\ \overset{\text{def}}=\
\sum_{f_1<f_2<\cdots<f_l=1}^{N_F}m_{f_1}^2m_{f_2}^2\cdots
m_{f_l}^2\qquad (M^{\sst(N_F)}_0=1\ ,\quad M_{l}^{\sst(N_F)}=0\quad l<0)\ .
\label{defml}
}

Finally we turn to the remaining integration over the bosonic moduli.
Following \cite{MO-I}, it is convenient to change variables in the
bosonic measure from $\{a_{3},w_{1},w_{2}\}$ to the new
set  $\{H,L,\Omega\}$ defined in \eqref{useful}. The relevant Jacobians are
\AL{
&\int_{-\infty}^\infty d^4a_3\
\frac{\big||a_3|^2-|a_1|^2\big|}{|a_3|^4}\ \longrightarrow\
{\pi^2\over2}\,\int_{L+2|\Omega|}^\infty\ dH\ ,\label{covone}\\
&\int_{-\infty}^\infty\,d^4w_1\,d^4w_2\
\longrightarrow\
{\pi^3\over8}\int_0^\infty dL\ \int_{|\Omega|\le L}d^3\Omega\ .
\label{covtwo}
}
The numerator and denominator in the left-hand side of \eqref{covone} are
supplied by \eqref{defmua} and \eqref{lift}, respectively.
In addition, we introduce re-scaled variables $\Omega=L\Omega'$, $H=LH'$,
and $\omega=L\omega'$. The integral over $L$ is now trivial.
Finally we switch to spherical polar coordinates,
\EQ{
\int d^3\Omega'\ \longrightarrow\ 2\pi\int_{-1}^1d(\cos
\theta)\int_0^1|\Omega'|^2d|\Omega'|\ ,\label{polarcoor}
}
where the polar angle is defined by $|\omega'|=
\hf|\Omega'||\phi^0|\cos\theta.$ This leaves an
ordinary 3-dimensional integral over the remaining
variables $H,$ $\cos\theta$ and $|\Omega'|$ which is the precise
analog of Eq.~(8.19) in \cite{MO-I}. Performing this
integral with the help of a standard symbolic manipulation routine
gives results which can be summed up by the formula
\EQ{
{\cal F}_k\Big|_{N=2,N_F}=
\frac{5}{(\phi^0)^6}M_{N_F}^{\sst(N_F)}
-\frac{3}{4(\phi^0)^4}M_{N_F-1}^{\sst(N_F)}
\frac1{16(\phi^0)^2}M_{N_F-2}^{\sst(N_F)}
-\frac5{2^63^3}M_{N_F-3}^{\sst(N_F)}+\frac{7(\phi^0)^2}{2^83^5}
M_{N_F-4}^{\sst(N_F)}\ ,
\elabel{bigres}
}
where the coefficients $M_l^{\sst(N_F)}$ are defined in \eqref{defml}.
For $N_F<4$ these expressions are identical to the predictions from
the Seiberg-Witten curve \eqref{twoi}, up to a physically unimportant
additive constant in the case $N_F=3$.

The situation in the $N_F=4$ theory is rather more subtle. To our
knowledge the two-instanton prediction from Seiberg-Witten theory 
with arbitrary hypermultiplet
masses has not been determined in the literature. However, in the
massless case ones expects that the prepotential is classically exact:
in other words given by the first term in \eqref{preci}. The result
\eqref{bigres} for $N_F=4$ with vanishes masses
\EQ{
{\cal F}_2\Big|_{N_F=4,m_f=0}=\frac{7(\phi^0)^2}{2^83^5}
}
is obviously in contradiction to this. The resolution of this
discrepancy is explained in Refs.~\cite{MO-II,Dorey:1997bn}. 
The point is that the
Seiberg-Witten curve is parameterized by an effective coupling
$\tau_{\text{eff}}$ rather than by the microscopic coupling $\tau$. The
two definitions differ by an infinite series of even-charge 
instanton corrections:
\EQ{
\tau_{\text{eff}}=\tau+\sum_{k=2,4,\ldots}
c_ke^{2\pi ik\tau}\ .
}
The two-instanton computation above shows that
\EQ{
c_2=\frac1{i\pi}\cdot\frac7{2^73^5}\ .
}

\subsection{One instanton prepotential in $\SU(N)$}\elabel{sec:N3}

In this section we perform, following \cite{KMS}, 
the explicit evaluation of the 
centred instanton partition function
for $k=1$ in an theory with arbitrary gauge group $\SU(N)$. It is best
to start from the linearized formulation described in
\S\ref{sec:N1}. The instanton effective action for $k=1$ is, from
\eqref{iean2} (with factors of $g$ removed by the re-scalings \eqref{nicer}):
\EQ{
\tilde S=4\pi^2\Big\{\,
\big|w_{u\aD}\chi+
\phi^0w_{u\aD}\big|^2
+\tfrac i2
\bar\mu^A_u(\mu_{uA}\chi^*+\phi^{0*}_u\mu_{uA})
+\tfrac14\sum_{f=1}^{N_F}\K_f\tilde\K_f(
\chi-m_f)\Big\}+\tilde S_{\text{L.m.}}\ ,
\label{iean2k1}
}
where the Lagrange multiplier terms for the ADHM constraints are
\EQ{
\tilde S_{\text{L.m.}}=-4i\pi^2
\Big\{\bar\psi_A^\aD\big(\bar\mu^A_u
w_{u\aD}+\bar w_{u\aD}\mu^A_u\big)+
\vec D\cdot\vec\tau^{\aD}{}_\bD\bar w^\bD_u w_{u\aD}\Big\}\ .
}
Note that the quantity $\chi$ for $k=1$ is just a complex variable
rather than being a matrix. The linear shifts
\EQ{
\mu^A_u\to\mu^A_u-\frac{2w_{u\aD}}{\alpha_u^*}\bar\psi^{\aD A}\
,\qquad
\bar\mu^A_u\to\bar\mu^A_u+\frac{2\bar w_{u\aD}}{\alpha_u^*}
\bar\psi^{\aD A}\ .
}
eliminate the linear terms of these variables in the action. The
Grassmann integrals over $\{\mu^A_u,\bar\mu^A_u\}$ then bring down the
factors
\EQ{
\prod_{u=1}^N\Big(2\pi^2\alpha_u^*\Big)^2\ ,
}
where we have defined
\EQ{
\alpha_u=\chi+\phi^0_u\
,\qquad\alpha^*=\chi^*+\phi_u^{0*}\ .
}
The Grassmann integrals over the matter field coordinates
$\{\K_f,\tilde\K_f\}$ are simply evaluated:
\EQ{
\int d^{N_F}\K\,d^{N_F}\tilde\K\,\exp\Big(-\pi^2\sum_{f=1}^{N_F}
\K_f\tilde\K_f(\chi-m_f)\Big)=\pi^{2N_F}\prod_{f=1}^N(m_f-\chi)\ .
}
The integrals over $\{w_{u\aD},\bar w^\aD_u\}$ are Gaussian and are
accomplished using the identity
\EQ{
\int d^{2N}w\,d^{2N}\bar w\,\exp\Big(-4\pi^2A_u\bar
w^\aD_uw_{u\aD}+4i\pi^2\vec B_u\cdot\vec\tau^{\aD}
{}_\bD\bar w^\bD_uw_{u\aD}\Big)=(2\pi)^{-2N}
\prod_{u=1}^N\frac{1}{A_u^2+\vec B_u^2}\ .
}
All that remains are integrals over the auxiliary
variables $\{\chi,\vec D,\bar\psi_A\}$:
\EQ{
\widehat{\EuScript Z}_{k}^{\sst(\N=2,N_F)}=\frac1{(2\pi)^3}\int d^2\chi\,d^3D\,
\prod_{A=1}^2d^2\bar\psi_A\,
\prod_{u=1}^N\frac{\alpha_u^{*2}}{|\alpha_u|^4+(\vec D+\vec\Xi_u)^2}
\prod_{f=1}^{N_F}(m_f-\chi)\ ,
}
where $\vec\Xi_u$ is the Grassmann bilinear
\EQ{
\vec\Xi_u=\big(\alpha_u^*\big)^{-1}\bar\psi^A_\aD\vec\tau^{\aD}{}_\bD
\bar\psi_A^\bD\ .
}
The integrals over $\bar\psi_A$ must be saturated with two insertions
of $\Xi$:
\EQ{
\int\prod_{A=1}^2d^2\bar\psi_A\Xi^c_u\Xi^d_v=-8\frac{\delta^{cd}}
{\alpha^*_u \alpha^*_v}\ ,
}
leading to the identity
\EQ{
\int\prod_{A=1}^2d^2\bar\psi_A\,F(\Xi)
=-4\sum_{u,v=1}^N\frac1{\alpha^*_u \alpha^*_v}\,\frac{\partial^2F(\Xi)}{
\partial\Xi^c_u\partial\Xi^c_v}\Big|_{\Xi=0}
=-\frac4{\vec D^2}\frac{\partial^2F(\Xi=0)}{\partial\chi^{*2}}\ ,
}
where, in our case,
\EQ{
F(\Xi)=\prod_{u=1}^N
\frac{\alpha_u^{*2}}{|\alpha_u|^4+(\vec D+\vec\Xi_u)^2}\ .
}

We are now in a position to integrate out the Lagrange
multipliers $\vec D$. The non-trivial part of the integral
is easily performed by a standard contour integration in the
variable $|\vec D|\equiv\sqrt{\vec D\cdot\vec D}$
extended to run from $-\infty$ to $+\infty$:
\EQ{
\int\frac{d^3D}{\vec D^2}\prod_{u=1}^N\frac{\alpha^{*2}_u}{
|\alpha_u|^4+\vec D^2}=
2\pi^2\sum_{u=1}^N\frac{\alpha_u^*}{\alpha_u}\prod_{v=1\atop(\neq
u)}^N
\frac{\alpha^{*2}_v}{|\alpha_v|^4-|\alpha_u|^4}\ .
\label{jyy}
}
Finally, it only remains to integrate over $\chi$:
\EQ{
\widehat{\EuScript Z}_k^{\sst(\N=2,N_F)}=-\pi^{-1}\int d^2\chi\,
\frac{\partial^2}{\partial\chi^{*2}}f_1(\chi,\chi^*)f_2(\chi)
\ ,
}
where we have defined the two quantities
\EQ{
f_1(\chi,\chi^*)=
\sum_{u=1}^N\frac{\alpha_u^*}{\alpha_u}\prod_{v=1\atop(\neq
u)}^N
\frac{\alpha^{*2}_v}{|\alpha_v|^4-|\alpha_u|^4}\ ,\qquad
f_2(\chi)=\prod_{f=1}^{N_F}(m_f-\chi)\ .
\elabel{easyw}
}
The resulting integral over the $\chi$-plane can be evaluated by
Stoke's Theorem. There are two kinds of boundary to
consider: on the sphere at infinity and around the singularities
of the integrand. First the singularities. Contrary to appearances
the integrand is completely regular at $|\alpha_u|^2-|\alpha_v|^2$ due
to the cancellation between the $u^{\rm th}$ and $v^{\rm
th}$ terms in the sum. However, there are $N$ singularities at
$\alpha_u=0$, {\it i.e.\/}~$\chi=-\phi^0_u$, 
following from the fact that for $z=x+iy$
\EQ{
\PD{}{z^*}\frac1{z}=\pi\delta(x)\delta(y)\ .
}
Introducing polar coordinates
in the vicinity of the point $\chi=-\phi^0_u$, $\alpha_u=re^{i\theta}$,
we take a boundary in the form of a small circle of radius
$r\to0$. The resulting contribution is then
\EQ{
\lim_{r\to0}\frac1{4\pi}\Big(2+r\PD{}{r}\Big)\int_0^{2\pi}
d\theta\,e^{2i\theta}f_1(r,\theta)f_2(re^{i\theta})\ .
\elabel{consi}
}
Applying this formula in the vicinity of $\chi=-\phi^0_u$ we find a
contribution to the centred instanton partition function of
\EQ{
\prod_{v=1\atop(\neq
u)}^N\frac1{(\phi^0_u-\phi^0_v)^2}\prod_{f=1}^{N_F}(m_f+\phi^0_u)\ .
\label{contsing}
}

We can determine the contribution from the sphere at infinity by once again
introducing polar coordinates. Then in a similar fashion to
\eqref{consi} the contribution is
\EQ{
-\lim_{r\to\infty}\frac1{4\pi}\Big(2+r\PD{}{r}\Big)\int_0^{2\pi}
d\theta\,e^{2i\theta}f_1(r,\theta)f_2(re^{i\theta})\ .
\elabel{michw}
}
It is easy to establish the following asymptotic forms for large $r$
\EQ{
f_1(r,\theta)\thicksim\frac{e^{-2iN\theta}}{r^{2(N-1)}}\ ,\qquad
f_2(re^{i\theta})\thicksim e^{iN_F\theta}r^{N_F}\ .
\elabel{bkol}
}
This means that the contribution from the
sphere at infinity is only non-vanishing when $N_F\geq2(N-1)$, {\it
i.e.\/}~in the three cases $N_F=2N-2,2N-1,2N$. Taking
into account the selection rule arising from the integration over $\theta$,
there are only two relevant terms in the asymptotic expansion
of $f_1(r,\theta)$, namely
\EQ{
f_1(r,\theta)\thicksim
\alpha_1\frac{e^{-2iN\theta}}{r^{2(N-1)}}+\alpha_2
\frac{e^{-2i(N+1)\theta}}{r^{2N}}\sum_{u=1}^N(\phi^0_u)^2
\elabel{bkolb}
}
where we have defined the two constants
\EQ{
\alpha_1=2^{3-2N}\MAT{2N-3\\ N-1}\ ,\qquad\alpha_2=
2^{-2N}\MAT{2N\\
N-1}\ .
\label{defaot}
}
All other terms are either sub-leading in $1/r$ or have the wrong
$\theta$ dependence (taking account of the fact that $f_2$ is a
polynomial in $e^{i\theta}$ along with the factor of $e^{2i\theta}$ in
\eqref{michw}).
Hence, the contribution to the centred instanton partition function
from the sphere at infinity is
\EQ{
{\cal S}_{k=1}^{\sst(N_F)}=-\begin{cases}0 & N_F<2N-2\ ,\\ \alpha_1 &
N_F=2N-2\ ,\\
\alpha_1\sum_{f=1}^Nm_f & N_F=2N-1\ ,\\
\alpha_1\sum_{f,f'=1\atop(f<f')}^{N_F}
m_fm_{f'}+\alpha_2\sum_{u=1}^N(\phi^0_u)^2 & N_F=2N\ .\end{cases}
\label{contsung}
}
All-in-all, the $k=1$ contribution to the centred partition function,
and hence the coefficient ${\cal F}_1$ of the prepotential, is the sum
of \eqref{contsing} and \eqref{contsung}:
\EQ{
{\cal F}_1\equiv\widehat{\EuScript Z}_k^{\sst(\N=2,N_F)}=
\sum_{u=1}^N\prod_{v=1\atop(\neq
u)}^N\frac1{(\phi^0_v-\phi^0_u)^2}\,\prod_{f=1}^{N_F}
(m_f+\phi^0_u)+{\cal S}_1^{\sst(N_F)}\ .
\elabel{bfoi}
}

This calculation illustrates the complexity of integrating
over the instanton moduli space even at the one instanton
level. However, if one follows the method in detail an interesting
intuitive picture emerges. First recall that the final integral over $\chi$ was
evaluated by using Stoke's Theorem: there are only contributions from
an isolated set of $N$ points along
with a contribution from the sphere at infinity. We now show that
these contributions arise at the 
critical points of the instanton effective action
\eqref{iean2k1}. The latter correspond to the vanishing 
\EQ{
w_\aD\chi+\phi^0w_\aD=0\ .
}
So there is a branch consisting of $N$ solutions, labelled by $u$, where
\EQ{
\chi=-\phi^0_u\ ,\qquad w_{v\aD}\propto\delta_{uv}\ ,
}
{\it i.e.\/}~precisely at the positions of the isolated contributions
to the instanton effective action. However, it is apparent that
these are critical points of
$\tilde S$ only on a branch where the
instanton has non-zero scale size: $\rho^2=\tfrac12\bar w^\aD
w_\aD>0$. There is another branch of solutions when $\rho=0$,
{\it i.e.\/}~$w_\aD=0$ where physically
the instanton has shrunk to zero size. This
second branch is associated with the contribution from the
integral over the sphere at
infinity in $\chi$-space. To see this notice that when $\chi$ is
eliminated via its ``equation-of-motion'' one has
$\chi\sim\rho^{-1}$: so $\rho\to0$ does indeed correspond to the
circle at infinity in $\chi$-space. These observations are intended as
anecdotal evidence in favour of some kind of localization on the
moduli space of instantons. This is subject that we will pursue in
earnest in \S\ref{sec:N2}.

\newpage

\rsen\section{Conformal Gauge Theories at Large $N$}\elabel{sec:S39}

Instanton calculations are only reliable at weak coupling which can be
achieved in a Higgs or Coulomb phase, giving rise to the
applications reported in Chapters \ref{sec:S50} and \ref{sec:S43}. However,
theories can be weakly-coupled without invoking the Higgs mechanism if
they are finite, or conformal. In this case the gauge coupling does
not run with scale and weak coupling prevails for small $g$. The two
main examples that we discuss here are the $\N=4$ theory and the
$\N=2$ theory with $N_F=2N$ hypermultiplets in the fundamental
representation of the gauge group. We shall see that these two
examples have some interesting features in common and, in particular,
the calculation of
instanton effects in both cases considerably simplify in the large
$N$ limit (see Refs.~\cite{MO3,LETT,Hollowood:1999sk}). The most
striking feature of this limit is that the instanton measure
concentrates on a subspace of the moduli space, which, in the ${\cal
N}=4$ case, is essentially $AdS_{5}\times S^{5}$.
Of course this phenomenon is directly related to the AdS/CFT
correspondence, which we discuss in \S\ref{sec:S320}.
The approach of Refs.~\cite{MO3,LETT,Hollowood:1999sk} that
we will describe has been generalized to other finite gauge theories.
Ref.~\cite{Hollowood:1999ev} considered the $\N=4$ theory with gauge groups
$\Sp(N)$ and $\SO(N)$; Refs.~\cite{Gava:2000ky,Hollowood:1999nq}
considered finite $\N=2$ gauge theories with gauge group $\Sp(N)$; and
Ref.~\cite{Hollowood:2000bm} considered finite $\N=2$ theories with
product gauge group $\SU(N)^k$ and hypermultiplets in bi-fundamental
representations of the gauge group---the ``quiver'' models.

The main common feature of finite gauge theories
is the fact, discussed in \S\ref{sec:S31}, that the
supersymmetric instanton with the full set of fermion zero modes turned on
is {\it not\/} an exact solution of the
equations-of-motion. Put another way, most of the
fermion zero modes that appear at linear order around the instanton
background are lifted by Yukawa interactions as evidenced by
the leading order expression for the
instanton effective action \eqref{yyvv}
or \eqref{ieant} (with $\phi_a^0=0$ in the present context).
In both cases, this effective action involves a
quadrilinear coupling of the Grassmann collective coordinates. Of course,
in both cases, the supersymmetric and superconformal zero modes are
protected by symmetries
and remain unlifted by the instanton action. The pattern of lifting
dictated by the Grassmann quadrilinear implies that instantons of all
charges $k$ will contribute to a class of correlation functions whose
insertions evaluated in the instanton background are responsible for
saturating the
integrals over a finite number of unlifted modes. In both
cases, the number of unlifted fermion modes is sixteen. This is
directly the
number of supersymmetric and superconformal modes in the $\N=4$ theory
while in the $\N=2$ theory it includes the supersymmetric and
superconformal modes---numbering eight---along with eight additional
fundamental fermion zero modes.

The aim of this section is show how instanton contributions to
these kinds of correlation
functions can be calculated in the
large-$N$ limit. They will turn out to have remarkable properties.
The only dependence on the instanton charge appears in an overall
numerical pre-factor:
\EQ{
\VEV{{\cal O}_1(x^{(1)})\times\cdots\times{\cal
O}_n(x^{(n)})}_{\text{inst.}}
\ \overset{N\to\infty}\longrightarrow\
\Big(\sum_{k=1}^\infty k^nc_k\,e^{2\pi
ik\tau}\Big)f(x^{(1)},\ldots,x^{(n)})\ ,
\elabel{corrf}
}
for some function $f$ and coefficients $c_k$ independent of the
correlator in question. At first sight,
the limit \eqref{corrf} looks absurd. How could it be that the
only dependence on the instanton number $k$ is via an overall
multiplicative factor? The
reason for the simple dependence on $k$ rests on certain very
special properties of the large-$N$ limit of the instanton
calculus. Intuitively what happens in the large-$N$ limit is that
integral over the $k$-instanton moduli space is dominated by
configurations of $k$ single instantons occupying $k$ commuting
$\SU(2)$ subgroups of the gauge group which are therefore totally
non-interacting. In this
sense the dominant configurations are dilute-gas-like. But, contrary to
the dilute gas, the dominant configuration
also has all the instantons lying at the same spacetime point
$X_n$ and having the same scale size $\rho$. This accounts for the fact
that the functional dependence on the insertion points is the same for
all instanton number. We will show that
the $k$-dependent numerical factors $c_k$ are related to
interesting matrix integrals; namely, the partition functions of
dimensionally reduced gauge theories.

In addition, when taking the large-$N$
limit, it is convenient to introduce auxiliary bosonic collective
coordinates to bi-linearize the Grassmann quadrilinear in the instanton
effective action. In the large-$N$ limit these additional variables
are confined to a sphere: $S^5$ for $\N=4$ theory \cite{MO3}
and $S^1$ for $\N=2$ \cite{Hollowood:1999sk}.
The appearance of these auxiliary
collective coordinate is especially interesting in the
light of the AdS/CFT correspondence \cite{MAL,Aharony:2000ti}
relating the $\N=4$ theory to Type IIB string theory on $AdS_5\times S^5$
as discussed in \S\ref{sec:S320}.

\subsection{The collective coordinate integrals
at large $N$}\elabel{sec:S41}

We begin our analysis by establishing an expression for the
instanton partition function in the
large-$N$ limit. There are two relevant cases: $\N=4$ and $\N=2$
with $N_F=2N$ hypermultiplets. The expressions for the collective
coordinate integral \eqref{intm} is obtained from \eqref{fms}, with
$\tilde S$ equal to \eqref{yyvv}, in the $\N=4$ theory and
\eqref{ieant}, for the $\N=2$ case.\footnote{The VEVs are to set to
zero in both cases to preserve conformal invariance in the present
applications.}
Since we are interested in pursuing a large-$N$ limit, we can use the
results described in \S\ref{app:A5} to resolve the ADHM constraints
and therefore use the explicit version for the
supersymmetric volume form in \eqref{sbames}.
Note that the requirement $N\geq2k$ will certainly be met in the
large-$N$ limit (for fixed $k$). Since we are working in a non-abelian
Coulomb phase all the VEV vanish and the gauge symmetry remains
unbroken. In this case, we can immediately
integrate over the gauge orientation of
the instanton: $\int d^{4k(N-k)}\grp=1$.

The key to taking a large-$N$ limit is to bi-linearize
the Grassmann quadrilinear effective action \eqref{rrww} or
\eqref{ieant}, by introducing some auxiliary variables. This kind of
transformation is a well-known tool for analyzing the large-$N$ limit
of field theories with four-fermion interactions, like the Gross-Neveu
and Thirring models \cite{GN}. In fact we have already introduced the
necessary auxiliary variables in the form of the $k\times k$ matrices
$\chi_a$ in \S\ref{sec:N1}. In particularly the linearized instanton
effective actions \eqref{ieal} and \eqref{iean2} involve only
Grassmann bilinears.

In the $\N=2$ theory the identity we use is
\SP{
&\big|\det_{k^2}\BL\big|^{-1}\exp\,\Big\{\frac{i\pi^2}{2}\sum_{f=1}^{N_F}{\rm
tr}_k\,\K_f
\tilde\K_f\BL^{-1}\bar\CM^A\CM_A\Big\}\\
 &=2^{2k^2}\pi^{k^2}\int
d^{2k^2}\chi\exp\Big\{-4\pi^2\big({\rm tr}_k\,\chi_a\BL
\chi_a+\tfrac i2{\rm
tr}_k\,\bar\CM^A\CM_A\chi^\dagger+\tfrac14\sum_{f=1}^{N_F}{\rm
tr}_k\,\K_f\tilde\K_f\chi\big)\Big\}\ ,
\elabel{bilin}
}
whilst in the $\N=4$ case
\SP{
&\big|\det_{k^2}\BL\big|^{-3}\exp\Big\{
\frac{\pi^2}{2}\epsilon_{ABCD}{\rm
tr}_k\big(\bar\CM^A\CM^B\BL^{-1}\bar\CM^C\CM^D\big)\Big\}\\
&=2^{6k^2}\pi^{3k^2}\int
d^{6k^2}\chi\ \exp\Big\{-4\pi^2\big({\rm tr}_k\,\chi_a\BL
\chi_a+\tfrac12\bar\Sigma_{aAB}{\rm
tr}_k\,\bar\CM^A\CM^B\chi_a\big)\Big\}\ .
\elabel{E53}
}
Notice in both cases that the appropriate factor of
$|\det_{k^2}\BL|$ is already present in the supersymmetric
volume form on $\ms_k$ \eqref{fms} (or \eqref{sbames}).

\subsubsection{The $\N=4$ case}\elabel{sec:S512}

Before we proceed, and with the large-$N$ limit in mind,
it is useful to re-scale
\EQ{
\chi_a\to\frac{\sqrt N}{2\pi}\chi_a\ .
}
It is also convenient to define
\EQ{
\chi_{AB}=\tfrac 1{\sqrt8}\bar\Sigma_{aAB}\chi_a\
,\qquad\chi_a=-\tfrac1{\sqrt2}\Sigma_a^{AB}\chi_{AB}
\elabel{E54.1}
}
which satisfies a 
pseudo-reality condition following from the Hermiticity of $\chi_a$:
\EQ{
\chi_{AB}^\dagger=\tfrac12\epsilon^{ABCD}\chi_{CD}\ ,
\elabel{E54}
}
where $\dagger$ only acts on the instanton indices rather than the
$\SU(4)$ $R$-symmetry indices.

After
integrating over the gauge orientation and using the
identity \eqref{E53}, the instanton partition function is
\SP{
&{\EuScript Z}_k^{\sst(\N=4)}
=\frac{2^{3k^2}N^{3k^2}C_k^{\sst(\N=4)}A_k}
{\pi^{3k^2}
{\rm Vol}\,\U(k)}\int\, d^{4k^2} a' \,
d^{k^2}W^0\,d^{6k^2}\chi
\prod_{A=1}^4\Big\{
d^{k(N-2k)}\nu^A\,d^{k(N-2k)}\bar\nu^A\,d^{2k^2}\zeta^A\,d^{2k^2}\CM^{\prime
A}\Big\}\\ &\times
\,\big|\det_{2k}W\big|^{N-2k}\,e^{-N{\rm tr}_k\,\chi_a\BL
\chi_a+\sqrt{8N}\pi {\rm
tr}_k\,\bar\CM^A\CM^B\chi_{AB}} .
\elabel{sbames2}
}
In this expression, and in the following, $W$ is the $2k\times2k$
matrix defined in \eqref{defbw}, where the ADHM constraints
\eqref{linad} are imposed; hence,
\EQ{
W=\tfrac12W^01_{\sst[2]\times[2]}
-2a'_ma'_n\bar\sigma_{mn}\ .
}

The next stage
in the large-$N$ program is to
to integrate out the Grassmann variables
$\{\bar\nu^A,\nu^A\}$
by pulling down powers of the second term in the
exponential in \eqref{E53}:
\EQ{
\int\bigg\{\prod_{A=1}^4d^{k(N-2k)}\nu^A\, d^{k(N-2k)}\bar\nu^A\bigg\}\
\exp\sqrt{8N}\pi
{\rm tr}_k
\,\bar\nu^A\nu^B\chi_{AB}=\big(8N\pi^2
\big)^{2k(N-2k)} \left|{\rm det}_{4k}\chi\right|^{N-2k}\ ,
\elabel{intoutnu}
}
where the determinant is over the $4k\times4k$ dimensional matrix with
elements $(\chi_{AB})_{ij}$.
One might think that it is rather premature to integrate out the
Grassmann variables $\{\bar\nu^A,\nu^A\}$ since they could be saturated by
insertions made into the functional integral. This is possible,
but we shall find later a very simple prescription for including this kind of
dependence when working at large $N$.

After integrating out the subset of Grassmann variables as above,
we can collect together all
the non-constant
terms in the collective coordinate integral which are raised to the power
$N$ as $\exp-N\Gamma$. The large-$N$ ``effective action''
is\footnote{In the following
we translate back and forth as convenient between the
antisymmetric tensor representation $\chi_{AB}$ and the $\SO(6)_R$
vector representation $\chi_a$, $a=1,\ldots,6$ \eqref{E54.1}.}
\begin{equation}
\Gamma
=-\log{\rm det}_{2k}\,W
-\log{\rm det}_{4k}\chi +{\rm
tr}_k\,\chi_a\BL\chi_a\, .
\elabel{E57}\end{equation}
This expression
involves the $11k^2$ bosonic variables comprising the eleven
independent $k\times k$ Hermitian matrices $W^0$, $a'_n$ and
$\chi_a$. The remaining components $W^c$, $c=1,2,3,$ are eliminated
in favor of the $a'_n$ via the ADHM constraints \eqref{linad}.  The
action is invariant under the $\U(k)$ symmetry \eqref{restw} which
acts by adjoint action on all the variables.

With $N$ factored out of the exponent, the measure is in a form which
is amenable to a saddle-point treatment as $N\rightarrow\infty$. The
critical points of $\Gamma$ satisfy the equations
\begin{subequations}
\begin{align}
\epsilon^{ABCD}\left(\BL\cdot\chi_{AB}\right) \chi_{CE}\ &=\
\tfrac12\delta^D_E \,1_{\sst [k]\times[k]}\ ,\elabel{E58}\\
\chi_a\chi_a\ &=\ \tfrac12(W^{-1})^0\ ,\elabel{E59}\\
[\chi_a,[\chi_a,a'_n]]\ &=\ {\rm tr}_2(\tau^c\bar\sigma_{nm})
[(W^{-1})^c,a'_m]\ .
\elabel{E60}
\end{align}
\end{subequations}
These are obtained by varying $\Gamma$ with respect to the matrix
elements of $\chi$, $W^0$ and $a'_n$, respectively, and rewriting
``log$\,$det'' as ``tr$\,$log.''  We have defined the $k\times k$
matrices
\begin{equation}
(W^{-1})^0=
{\rm tr}_2\,W^{-1}\ ,\qquad
(W^{-1})^c=
{\rm tr}_2\,(\tau^cW^{-1})\ .
\elabel{E60.1}\end{equation}

The general solution to these coupled saddle-point equations is easily
found. It has the simple property that all the quantities are diagonal
in instanton indices:
\begin{subequations}\begin{align}
W^0\ &=\ {\rm diag}\big(2\rho_1^2,\ldots,2\rho_k^2\big)\ ,
\elabel{E61a}\\ \chi_a\ &=\ {\rm
diag}\big(\rho_1^{-1}\sfc^1_a,\ldots,\rho_k^{-1}\sfc^k_{a}\big)\
,\elabel{E61b} \\ a'_n\ &=\ {\rm
diag}\big(-\com^1_n,\ldots,-\com^k_n\big)\ ,
\elabel{E61c}\end{align}\end{subequations}
up to adjoint
action by the $\U(k)$ auxiliary 
symmetry. For quantities $\sfc_a^{i}$ parameterize $k$ unit
six-vectors, or points in $S^5$,
\begin{equation}
\sfc^i_{a}\sfc^i_{a}=1\qquad(\text{no sum on }i)\, ,
\elabel{E62}\end{equation}
where the radius of the $i^{\rm th}$ $S^5$ factor is $\rho_i^{-1}$.

A simple picture of this leading-order saddle-point solution emerges:
it can be thought of as $k$ independent copies of a large-$N$ one-instanton
saddle-point solution, where the $i^{\rm th}$ large-$N$
instanton is parameterized by the triple $\{X_n^i,\rho_i,\hat\Omega_a^i\}$.
Additional insight into this solution emerges
from considering the $\SU(2)$ generators $(T^c_i)^{}_{uv}$
describing the embedding of the $i^{\rm th}$ instanton inside
$\SU(N)$. From Eqs.~\eqref{embed2}, \eqref{bosbi} and \eqref{E61a}
one derives the commutation relations
\begin{equation}
[\,T^a_i\,,\,T^b_j\,]\ =\ 2i\delta_{ij}\,\epsilon_{abc}\,T^c_i
\elabel{relcomm}
\end{equation}
so that at the saddle-point, thanks to the Kronecker-$\delta,$ the $k$
individual instantons lie in $k$ mutually commuting $\SU(2)$
subgroups. Actually this feature follows intuitively from large-$N$ statistics
alone,\footnote{Consider the analogous problem of $k$ randomly
oriented vectors in ${\mathbb R}^N$ in the limit $N\rightarrow\infty;$
clearly the dot products of these vectors tend to zero simply due to
statistics.} and has nothing to do with either the existence of
supersymmetry or with the details of the ADHM construction.
Another important property is that
the effective action \eqref{E57} evaluated on these saddle-point solutions is
zero; hence there is no exponential dependence on $N$ in the final
result. Finally we should make the technical point that, thanks to
the diagonal structure of these solutions, they are automatically consistent
with the triangle inequalities on the bosonic bi-linear $W$ discussed in
the paragraph following Eq.~\eqref{ncsp}; hence we never need to
specify more explicitly the integration limits on the $W$ variables.

The saddle-point solution depends on the moduli of each of the $k$
large-$N$ instantons. However, not all these $10k$ parameters are
the genuine moduli, or flat directions of $\Gamma$,
because, as we shall see, they are ``lifted'' by terms
beyond quadratic order in the expansion of $\Gamma$. In fact the
only genuine moduli are the 10 overall large-$N$ coordinates. The
correct large-$N$ behaviour is obtained \cite{MO3}
by expanding $\Gamma$ around
the \it maximally degenerate \rm saddle-point solution
\begin{equation}
W^0=2\rho^2\,1_{\sst[k]\times[k]},\qquad
\chi_a=\rho^{-1}\sfc_a\,1_{\sst[k]\times[k]},\qquad
a'_n=-\com_n\,1_{\sst[k]\times[k]}\ ,
\elabel{specsol}
\end{equation}
which corresponds to the $k$ large-$N$
instantons living at a common point $\{\com_n,\sfc_a,\rho\}$.
(From the ADHM constraint \eqref{linad} it follows
that the remaining components of $W$ vanish: $W^c=0$ for $c=1,2,3.$)
This degenerate solution, unlike Eqs.~\eqref{E61a}-\eqref{E61c}, is
invariant under the residual $\U(k)$. With the instantons sitting on
top of one another, it looks like the complete opposite of the dilute
instanton gas limit; however the instantons still live in $k$ mutually
commuting $\SU(2)$ subgroups of $\SU(N)$ as per Eq.~\eqref{relcomm},
which is a dilute-gas-like feature.

In order to expand about this special solution, one first needs to
factor out the integrals over the parameters $\{\com_n,\rho,\sfc_a\}$.
This is done in the following way:
for each $k\times k$ matrix, we  introduce a basis of traceless
Hermitian matrices $\hat T^r$, $r=1,\ldots,k^2-1$, normalized by ${\rm
tr}_k\,\hat T^r\hat T^s=\delta^{rs}$.  For each $k\times k$ matrix $v$
we separate out the ``scalar'' component $v_0$ by taking
\begin{equation}
v=v_01_{\sst [k]\times[k]}+\hat v^r\hat T^r\, .
\end{equation}
The change of variables from the $\{T^r\}$ basis used in
\eqref{sbames2} to $\{1_{\sst
[k]\times[k]},\hat T^r\}$ involves a Jacobian
\begin{equation}
\int d^{k^2}v=k^{\pm1/2}\int dv_0\,d^{k^2-1}\hat v,
\end{equation}
where $\pm1$ refers to $c$-number and Grassmann quantities,
respectively. For the
moment we continue to focus on the bosonic variables, which are
decomposed as follows:
\begin{subequations}
\begin{align}
a'_n&=-X_n1_{\sst [k]\times[k]}+\hat a'_n\ ,\elabel{asplitdef}\\
\chi_a&=\rho^{-1}\sfc_a1_{\sst [k]\times[k]}+\hat\chi_a\ .\elabel{chisplitdef}
\end{align}
\end{subequations}
By definition the traceless matrix variables $\hat a'_n$ and
$\hat\chi_a$ are the fluctuating fields.

Inserting Eqs.~\eqref{asplitdef}-\eqref{chisplitdef} into
Eq.~\eqref{E57} and Taylor expanding is a tedious though
straightforward exercise. It is necessary to expand
to fourth order in the fluctuating fields around the solution parameterized
by the ten exact moduli.
The expansion of the determinant terms in \eqref{E57} is facilitated
by first writing ``$\log\det$'' as ``${\rm tr}\log$'' and then
expanding the logarithm:
\SP{
{\rm tr}_{2k}W
&=2k\log\rho^2+{1\over\rho^2}{\rm tr}_k(\delta
W^0)-{1\over4\rho^4}{\rm tr}_k(\delta
W^0)^2\\ &\qquad\qquad
+{1\over12\rho^6}{\rm tr}_k(\delta W^0)^3-{1\over32\rho^8}{\rm
tr}_k(\delta W^0)^4+{1\over2\rho^4}{\rm
tr}_k\,\big[\hat a'_n,\hat a'_m\big]^2+\cdots
\elabel{expdw}
}
and
\begin{equation}
{\rm tr}_{4k}\log\chi
=-2k\log(8\rho^2)-2^5\rho^2{\rm
tr}_{4k}\,\big(\sfc^*\hat\chi\big)^2
+{2^9\rho^3\over3}{\rm tr}_{4k}\,\big(\sfc^*\hat\chi\big)^3-
2^{10}\rho^4{\rm tr}_{4k}\,\big(\sfc^*\hat\chi\big)^4+\cdots\ .
\elabel{expdchi}
\end{equation}
In these expansions we have  dropped fifth- and
higher-order terms in the fluctuating fields.
Here we are anticipating the fact that these terms are not
needed at leading order in $1/N$. To see this
we can re-scale the integration variables in a standard way which shows that
the fluctuations around the maximally degenerate saddle point are of
order $N^{-1/4}$. The higher order terms in the exponent therefore yield
subleading contributions in the $1/N$ expansion. In particular, this
is true for the diagonal components of $\hat{a}'_{n}$ and
$\hat{\chi}_{a}$ which correspond to the moduli of the generic saddle-point
solution discussed earlier in this Section. This shows that our large-$N$
expansion around the maximally degenerate saddle-point is self-consistent.
In Eq.~\eqref{expdchi}, and in
subsequent equations, we move back and forth as convenient between the
6-vector and the antisymmetric tensor representations of $\sfc$ and
$\chi$ using Eq.~\eqref{E54.1}. In particular, the $\SO(6)$
orthonormality condition $\sfc\cdot\sfc=1$ becomes, in $4\times4$
matrix language,
\begin{equation}
\sfc\sfc^*=-\tfrac181_{\sst[4]\times[4]}\ ,\quad\hbox{or}\quad
\sfc^{-1}=-8\sfc^*\ ,
\elabel{resu2}
\end{equation}
which has been implemented in Eq.~\eqref{expdchi}.

Next we need a systematic method for re-expressing the traces over
$4k\times4k$ matrices in Eq.~\eqref{expdchi} as traces over $k\times k$
matrices. We will exploit the following ``moves'':\footnote{In the
following, we should emphasize that $\dagger$ only acts on instanton
indices, as per the reality condition \eqref{E54},
and {\it not\/} on  $\SU(4)$ matrix indices.}
\begin{subequations}
\begin{align}
\sfc^*\hat\chi&=-\hat\chi^\dagger\sfc-\tfrac14(\sfc\cdot\hat\chi)
1_{\sst[4]\times[4]}\ ,\elabel{resu1}\\
{\rm tr}_4\,E^\dagger F&={\rm tr}_4\,F^\dagger E=-\tfrac12(E\cdot F)\ ,\elabel{resu3}
\\ {\rm tr}_4\,E^\dagger FG^\dagger H&=\tfrac1{16}\big(E_aF_aG_bH_b-E_aF_bG_aH_b
+E_aF_bG_bH_a\big)\
.\elabel{resu4}
\end{align}
\end{subequations}
On the left-hand sides of Eq.~\eqref{resu3}-\eqref{resu4}, the
$4k\times4k$ matrices $\{E,F,G,H\}$ are assumed to be antisymmetric in
$\SU(4)_R$ indices and subject to the usual conditions
\eqref{E54}-\eqref{E54.1}; the identity \eqref{resu1} follows from a
double application of Eq.~\eqref{E54}.  Using
Eqs.~\eqref{resu2}-\eqref{resu4} in an iterative fashion, it is then
easy to derive the following trace identities:
\begin{subequations}
\begin{align}
{\rm tr}_{4k}\,\big(\sfc^*\hat\chi\big)^2\ &=\
-{\rm
tr}_{4k}\,\hat\chi^\dagger\sfc\sfc^*\hat\chi-\tfrac14{\rm
tr}_k\big(\sfc\cdot\hat\chi\,
{\rm tr}_4(\sfc^*\hat\chi)\big)\notag\\
&=
{1\over2^3}{\rm tr}_k\,
\big(\sfc\cdot\hat\chi\big)^2-{1\over2^4}
{\rm tr}_k\,\hat\chi\cdot\hat\chi\, ,\elabel{ident1}\\
{\rm tr}_{4k}\,\big(\sfc^*\hat\chi\big)^3\ &=\
\tfrac18{\rm
tr}_{4k}(\hat\chi^\dagger\hat\chi\sfc^*\hat\chi)+\tfrac1{64}{\rm
tr}_k\,(\sfc\cdot\hat\chi)^2
\hat\chi\cdot\hat\chi-\tfrac1{32}{\rm
tr}_k\,(\sfc\cdot\hat\chi)^3\notag\\
&=\
-{1\over2^5}{\rm tr}_k\,
\big(\sfc\cdot\hat\chi\big)^3
+{3\over2^7}{\rm tr}_k\,\hat\chi\cdot\hat\chi\, \sfc\cdot
\hat\chi\, ,\elabel{ident2}\\
{\rm tr}_{4k}\,\big(\sfc^*\hat\chi\big)^4\ &=\
{\rm tr}_{4k}\big(
\tfrac1{64}\hat\chi^\dagger\hat\chi\hat\chi^\dagger\hat\chi-\tfrac1{32}
\hat\chi^\dagger\hat\chi(\sfc\cdot\hat\chi)\sfc^*\hat\chi
\notag\\ &\qquad\qquad\qquad\qquad\qquad
-\tfrac1{32}(\sfc\cdot\hat\chi)\sfc^*\hat\chi\hat\chi^\dagger\hat\chi
+\tfrac1{16}(\sfc\cdot\hat\chi)\sfc^*\hat\chi(\sfc\cdot\hat\chi)\sfc^*
\hat\chi\big)\notag\\
&=\
{1\over2^7}{\rm tr}_k\,\big(\sfc\cdot\hat\chi\big)^4-
{1\over2^7}{\rm tr}_k\,\big(\sfc\cdot\hat\chi\big)^2
\hat\chi\cdot\hat\chi
+{1\over2^{9}}{\rm tr}_k\,\big(\hat\chi\cdot\hat\chi\big)^2
-{1\over2^{10}}{\rm tr}_k\,\hat\chi_a\hat\chi_b
\hat\chi_a\hat\chi_b\, .\elabel{ident3}
\end{align}
\end{subequations}
As before, on the left-hand side of these formulae, $4k\times4k$
 matrix multiplication is implied, whereas on the right-hand side, all
 $\SO(6)$ indices are saturated in standard 6-vector inner products,
 leaving the traces over $k\times k$ matrices.

{}From Eqs.~\eqref{E57}, \eqref{expdw}, \eqref{expdchi}, and
\eqref{ident1}-\eqref{ident3}, one obtains for the bosonic part of the
expansion of the action $\Gamma$:
\begin{equation}
\Gamma_{\rm b}=\Gamma^{(2)}+\Gamma^{(3)}+\Gamma^{(4)}+\cdots
\elabel{sumSb}
\end{equation}
where the quadratic, cubic and quartic actions are now given entirely
as $k$-dimensional (rather than $2k$ or $4k$-dimensional) traces:
\begin{subequations}
\begin{align}
\Gamma^{(2)}&={\rm tr}_k\,\varphi^2\  ,\qquad
\varphi=2\rho \sfc\cdot\hat\chi+{1\over2\rho^2}\delta W^0\, ,\elabel{S2def}
\\
\Gamma^{(3)}&=-{1\over12\rho^6}{\rm tr}_k\,(\delta W^0)^3+4\rho^3
{\rm tr}_k\,\sfc\cdot\hat\chi\,\hat\chi\cdot\hat\chi-{16\rho^3\over3}
{\rm tr}_k\,(\sfc\cdot\hat\chi)^3+{\rm tr}_k\,\delta
W^0\hat\chi\cdot\hat\chi
\notag\\
 &=2\rho^2\,{\rm tr}_k\,\varphi\big(\hat\chi\cdot\hat\chi
-4(\sfc\cdot\hat\chi)^2\big)+\cdots\ ,\elabel{S3con}
\\
\Gamma^{(4)}&=
-{1\over2\rho^4}{\rm
tr}_k\,\big[\hat a'_n,\hat a'_m\big]^2+
{1\over32\rho^8}{\rm
tr}_k(\delta W^0)^4-
{\rm tr}_k\,\big[\hat\chi_a,\hat a'_n\big]\big[\hat\chi_a
,\hat a'_n\big]
+8\rho^4{\rm tr}_k\,\big(\sfc\cdot\hat\chi\big)^4\notag\\ &\qquad\qquad\qquad-
8\rho^4{\rm tr}_k\,\big(\sfc\cdot\hat\chi\big)^2
\hat\chi\cdot\hat\chi
+2\rho^4{\rm tr}_k\,\big(\hat\chi\cdot\hat\chi\big)^2
-\rho^4{\rm tr}_k\,\hat\chi_a\hat\chi_b\hat\chi_a\hat\chi_b\notag
\\
 &=
-{1\over2\rho^4}{\rm
tr}_k\,\big[\hat a'_n,\hat a'_m\big]^2
-8\rho^4{\rm tr}_k\,\big(\sfc\cdot\hat\chi\big)^2
\hat\chi\cdot\hat\chi
+2\rho^4{\rm tr}_k\,\big(\hat\chi\cdot\hat\chi\big)^2\notag\\
&\qquad\qquad\qquad+16\rho^4{\rm tr}_k\,\big(\sfc\cdot\hat\chi\big)^4-
\rho^4{\rm tr}_k\,\hat\chi_a\hat\chi_b\hat\chi_a\hat\chi_b
-{\rm tr}_k\,\big[\hat\chi_a,\hat a'_n\big]\big[\hat\chi_a
,\hat a'_n\big]+\cdots\ .\elabel{S4con}
\end{align}
\end{subequations}
Notice that only $k^2$ fluctuations, denoted $\varphi,$ are actually lifted at
quadratic order. This, in turn, implies that certain terms in
$\Gamma^{(3)}$ and $\Gamma^{(4)}$
are subleading, and can be omitted. Specifically,
the omitted terms in the final rewrites in Eqs.~\eqref{S3con}-\eqref{S4con}
contain, respectively, two or more, and one or more, factors of
the quadratically lifted $\varphi$ modes, and consequently are suppressed in
$1/N$ (as a simple re-scaling argument again confirms).

Now let us perform the elementary Gaussian integration over the
$\varphi$'s. Changing integration variables in Eq.~\eqref{sbames2} from
$d^{k^2}W^0$ to $d^{k^2}\varphi$ using Eq.~\eqref{S2def}, one finds:
\begin{equation}
\int d^{k^2}W^0\,e^{-N(\Gamma^{(2)}+\Gamma^{(3)})}\ =\ \Big({4\pi\rho^4\over
N}\Big)^{k^2/2}\, e^{-N\Gamma^{(4)\prime}}
\elabel{S4'def}
\end{equation}
where the new induced quartic coupling reads
\begin{equation}
\Gamma^{(4)\prime}\ =\
-\rho^4{\rm tr}_k\,
\big(\hat\chi\cdot\hat\chi-4(\sfc\cdot\hat\chi)^2\big)^2\ .
\elabel{S4'def1}
\end{equation}
Combining $\Gamma^{(4)\prime}$ with the original quartic coupling
\eqref{S4con} gives for the
effective bosonic small-fluctuations action:
\begin{equation} \Gamma_{ \rm b}=
-{1\over2}{\rm
tr}_k\left(\rho^{-4}[\hat a'_n,\hat a'_m]^2+2[\hat\chi_a,\hat a'_n]^2
+\rho^4[\hat\chi_a,\hat\chi_b]^2\right)\, .
\elabel{Sbanswer}
\end{equation}
Remarkably, all dependence on the unit vector $\sfc_a$ has canceled out.

Notice that apart from the absence of derivative terms, the expression
\eqref{Sbanswer} looks like a Yang-Mills field strength for
the gauge group $\SU(k)$! We can make this explicit by introducing a
ten-dimensional vector field $A_M$,
\begin{equation}
A_M=N^{1/4}\big(\rho^{-1}\hat a'_n,\rho\hat\chi_a\big)\ ,\quad
M=1,\ldots,10\ ,
\elabel{amudef}\end{equation}
in terms of which
\begin{equation}
N\Gamma_\rmb(A_M)\ =\ -{1\over2}{\rm tr}_k\,\left[A_M,A_N\right]^2 \ .
\elabel{actymb}
\end{equation}
We recognize this as precisely the action of ten-dimensional $\SU(k)$
gauge theory, reduced to $0$ dimensions, {\it i.e.\/}~with all derivatives
set to zero.

Now let us turn to the Grassmann collective coordinates.
Since $\N=4$ supersymmetry in four
dimensions descends from $\N=1$ supersymmetry in ten dimensions, and
since all our saddle-point manipulations commute with supersymmetry,
we expect to find the $\N=1$ supersymmetric completion of the ten-dimensional
dimensionally-reduced action \eqref{actymb}, namely
\begin{equation}
N\Gamma_\rmf(A_M,\Psi)=-i{\rm
tr}_k\,\bar\Psi\Gamma_M\left[A_M,\Psi\right]\ ,
\elabel{actymf}
\end{equation}
where $\Psi$ is a ten-dimensional Majorana-Weyl spinor, and
$\Gamma_M$ is an element of the ten-dimensional Clifford algebra. To
see how this comes about, we first separate out from the fermionic
collective coordinates the exact zero modes, in analogy
to Eqs.~\eqref{asplitdef}-\eqref{chisplitdef}:
\begin{subequations}
\begin{align}
\CM^{\prime A}_\alpha&=-4i\xi^A_\alpha1_{\sst
[k]\times[k]}-4i
a'_{\alpha\aD}\bar\eta^{\aD A}+\hat\CM^{\prime A}_\alpha\ ,\elabel{sosm}\\
\zeta^{\aD A}&=-4i\bar\eta^{\aD A}1_{\sst [k]\times[k]}+\hat\zeta^{\aD
A}\ .
\end{align}
\end{subequations}
Here, $\xi^A_\alpha$ and $\bar\eta^{\aD A}$ are the
supersymmetric and superconformal fermion
modes \eqref{susymo}-\eqref{suconmo}.
Expanding the remaining part of the
fermion coupling in the exponent of \eqref{E53} around the
special solution \eqref{specsol} and using the relations \eqref{linad}
 and \eqref{S2def}, we find
\begin{equation}\begin{split}
 N\Gamma_{\rm f}=
\big(8\pi^2 N\big)^{1/2}\,
{\rm tr}_k\Big[&
\big(\varphi
-2\rho(\sfc\cdot\hat\chi)\big)\rho\sfc_{AB}
\hat \zeta^{\aD A}\hat \zeta_\aD^B+\rho^{-1}\sfc_{AB}\big[\hat a'_{\alpha\aD},
\hat {\cal M}^{\prime\alpha A}\big]\hat \zeta^{\aD B}\\
&\qquad\qquad\qquad+\hat\chi_{AB}\big(\rho^2\hat
\zeta^{\aD A}\hat \zeta_\aD^B+\hat {\cal M}^{\prime
\alpha A}\hat {\cal M}^{\prime B}_\alpha\big)\Big] \ .
\elabel{fermc}\end{split}\end{equation}
If we now define the
ten-dimensional Majorana-Weyl fermion field $\Psi$
\begin{equation}
\Psi=\sqrt{\pi\over2}\,N^{1/8}\Big[\rho^{-1/2}\MAT{0\\1}\otimes
\MAT{1\\ 0}\hat{\cal M}^{\prime A}_\alpha+
\rho^{1/2}\MAT{0\\1}\otimes\MAT{0\\ 1}\hat\zeta^{\aD A}\big]\ ,
\elabel{psidef}
\end{equation}
and the $\Gamma_M$ matrices according to
Eq.~\eqref{rotca} below, we do in fact
recover the simple form \eqref{actymf}.  In moving from
Eq.~\eqref{fermc} to Eq.~\eqref{actymf} we have dropped the term
depending on $\varphi$; since $\varphi$ is a quadratically lifted bosonic
mode its contribution is suppressed in large $N$ compared to the other
couplings in Eq.~\eqref{fermc}, as a simple re-scaling argument
confirms. The representation of the $\SO(10)$ $\Gamma$-matrices is constructed
via the decomposition $\SO(10)\supset\SO(6)\times\SO(4)$
in the following way. Firstly,
We define an $\SO(6)$ rotation matrix $R$,
$RR^T=1$, such that $\sfc'_a=R_{ab}\sfc_b$ lies entirely along, say, the first
direction, {\it i.e.\/}~$\sfc'_a\propto\delta_{a1}$.
In the new basis, we have a representation of the $\SO(6)$ Clifford algebra
$\Gamma'_a=R_{ab}\Gamma_b$.
In the rotated basis, we can construct a representation of the
$\SO(10)$ Clifford algebra as follows:
\begin{equation}
\Gamma'_M=\Big\{\Gamma'_1\otimes\gamma_n\, ,\,
\Gamma'_a\otimes\big(\delta_{a1}\gamma_5+(1-\delta_{a1})
1_{\sst[4]\times[4]}\big)\Big\}\, ,
\elabel{rotca}\end{equation}
where $n=1,\ldots,4$ and $a=1,\ldots,6$.
The representation of the $\SO(10)$ Clifford algebra that we need
is then found by un-doing the rotation on the six-dimensional subspace:
\begin{equation}
\Gamma_M=\Big\{\Gamma'_1\otimes\gamma_n\, ,\,
(R^{-1})_{ab}\Gamma'_b\otimes\big(\delta_{b1}\gamma_5+(1-\delta_{b1})
1_{\sst[4]\times[4]}\big)\Big\}\, .
\elabel{rotcam}\end{equation}
One can then verify that \eqref{psidef} is Majorana-Weyl with respect
to this basis. Note that unlike the bosonic sector, the $\sfc_a$
dependence of the fermionic action does not actually disappear;
it is simply subsumed into the representation
of the $\SO(10)$ $\Gamma$-matrices.

Finally our effective large-$N$ $k$-instanton
partition function has the form
\EQ{
{\EuScript Z}_k^{\sst(\N=4)}\
\overset{N\rightarrow\infty}\longrightarrow\
\frac{\sqrt N}{
k^32^{17k^2/2-k/2+25}\pi^{9k^2/2+9}{\rm Vol}\,\U(k)}
\int\rho^{-5}d\rho\,d^4\com\, d^5\sfc\prod_{A=1}^4d^2\xi^A
d^2\bar\eta^A\cdot\widehat{\EuScript Z}_{\SU(k)}^{\sst(d=10)}\ ,
\elabel{hello}
}
where $\widehat{\EuScript Z}_{\SU(k)}^{\sst(d=10)}$
is the partition function of an ${\cal N}=1$
supersymmetric $\SU(k)$ gauge theory in ten dimensions dimensionally reduced
to zero dimensions:
\begin{equation}\begin{split}
\widehat{\EuScript Z}_{\SU(k)}^{\sst(d=10)}&=\int_{\SU(k)}\,
d^{10}A\, d^{16}\Psi\,e^{-S(A_\mu,\Psi)}\ ,\\
S(A_M,\Psi)\ &=\ N(\Gamma_{\rm b}+\Gamma_{\rm f})\ =\
-{1\over2}{\rm tr}_k\,\left[A_M,A_N\right]^2-i{\rm
tr}_k\,\bar\Psi\Gamma_M\left[A_M,\Psi\right]\ .
\elabel{sukpart}
\end{split}
\end{equation}
Notice that the rest of the measure, up to numerical factors, is
independent of the instanton number $k$.
When integrating expressions which are independent of the $\SU(k)$
degrees-of-freedom, $\widehat{\EuScript Z}_{\SU(k)}^{\sst(d=10)}$ is
simply an overall constant
factor. A calculation of Ref.~\cite{MNS} revealed that $\widehat{\EuScript
Z}_{\SU(k)}^{\sst(d=10)}$
is proportional to $\sum_{d|k}d^{-2}$, a sum over the positive integer
divisors $d$ of $k$. However, the constant of proportionality was
fixed definitively in Ref.~\cite{KNS} to give\footnote{In comparing to
Ref.~\cite{KNS}, it is important to note that our convention
for the normalization of the generators is  ${\rm tr}_k\hat
T^r\hat T^s=\delta^{rs}$, rather than $\tfrac12\delta^{rs}$ in
 Ref.~\cite{KNS}.}
\begin{equation}
\widehat{\EuScript Z}_{\SU(k)}^{\sst(d=10)}=\frac{2^{17k^2/2+k/2-9}\pi^{5k^2+k/2-7/2}k^{-1/2}}
{\prod_{i=1}^{k-1}i!}
\sum_{d\vert k}{1\over d^2}\ .
\elabel{parte}\end{equation}

In summary, the
effective large-$N$, semi-classically leading-order,
collective coordinate measure has the following
simple form \cite{MO3}:
\EQ{
e^{2\pi ik\tau}{\EuScript Z}_k^{\sst(\N=4)}\
\overset{N\rightarrow\infty}\longrightarrow\
{\sqrt N\over
2^{33}\pi^{27/2}}\,k^{-7/2}e^{2\pi ik
\tau}\Big\{\sum_{d\vert k}{1\over d^2}\Big\}
\int\,
{d^4\com\,d\rho\over\rho^5}\, d^5\sfc\prod_{A=1}^4d^2\xi^A
d^2\bar\eta^A\ .
\elabel{endexp}
}

\subsubsection{The $\N=2$ case}

We now follow essentially the same steps to deduce a form for the
large-$N$ collective coordinates integral in the finite $\N=2$
theory following Ref.~\cite{Hollowood:1999sk}.
In view of the large-$N$ limit to come, we first re-scale
\EQ{
\chi\to\frac{\sqrt N}{2\pi}\chi\ .
}
Performing the integration over the $\nu^A$'s and $\bar\nu^A$'s gives
\begin{equation}
\int\prod_{A=1,2}d^{k(N-2k)}\nu^A\, d^{k(N-2k)}\bar\nu^A\,
\exp\Big[-\sqrt{N}\pi{\rm tr}_k
\,\chi^\dagger\bar\nu^A\nu_A\Big]=\big(N\pi^2
\big)^{k(N-2k)}\,
\big|{\rm det}_{k}\chi^\dagger\big|^{2(N-2k)}\ .
\elabel{ntoutnu}\end{equation}
Similarly, integrating out the matter Grassmann collective coordinates
\begin{equation}
\int d^{2kN}\K\,d^{2kN}\tilde\K\,\exp\Big[\sqrt{\frac N4}\pi
\sum_{f=1}^{N_F=2N}{\rm tr}_k\,\K_f\tilde\K_f\chi\Big]=
\Big(\frac{N\pi^2}4\Big)^{kN} \big|{\rm det}_{k}\chi\big|^{2N}\ .
\elabel{ntoutk}\end{equation}

Once again, we collect all the terms to the power $N$ to give the
large-$N$ ``effective action''
\begin{equation}
\Gamma
=-\log{\rm det}_{2k}W
-2\log{\rm det}_{k}\chi\chi^\dagger+{\rm
tr}_k\,\chi_a\BL\chi_a\ .
\elabel{E57n}\end{equation}

We now turn to the solution of the large-$N$ saddle-point
equations. These are the coupled Euler-Lagrange equations that come
from extremizing $\Gamma$ with respect to $\{a'_n,W^0,\chi_a\}$.
The analysis of these coupled equations is
virtually identical to the $\N=4$ case in \S\ref{sec:S512}; hence
we suppress the calculational details. In particular, the dominant
configuration is the maximally degenerate solution \eqref{specsol}

where now $\sfc_a$ is a unit 2-vector. It is convenient to parameterize
$\sfc_a$ by the phase angle $\phi$:
\begin{equation}
\sfc_1+i\sfc_2=e^{i\phi}\ .
\end{equation}

The next stage of the analysis is to expand the effective action in
the fluctuations out to sufficient order to ensure the convergence of
the integrals over the fluctuations. Fortunately, the analysis need
not be repeated because our effective action \eqref{E57n} is, up to a
constant, simply the $\N=4$ effective action \eqref{E57} with the
replacements $\chi_{12}\to\chi/\sqrt8$,
$\chi_{34}\to\chi^\dagger/\sqrt8$ and all other components of
$\chi_{AB}$ set to zero. The fluctuations in $W^0$ are integrated out
at Gaussian order to leave integrals over the remaining fluctuations
in $\hat a'_n$ and $\hat\chi_a$, the traceless parts, that are lifted
at quartic order; the terms beyond quartic order are then formally
suppressed by (fractional) powers of $1/N$, and may be dropped. The
``action'' for these quartic fluctuations is precisely the action for
six-dimensional Yang-Mills theory, as in \eqref{amudef} and
\eqref{actymb} but where the indices $M=1,\ldots,6$.

Not surprisingly the coupling to the Grassmann collective coordinates
completes the
six-dimensional theory to a supersymmetric gauge theory
in six dimensions dimensionally reduced to $0$ dimensions. The
six-dimensional theory has eight independent supersymmetries and is
therefore an $\N=(1,0)$ theory in six dimensions.
The eight component fermion field of the $\N=(1,0)$ theory has components
\begin{equation}
\Psi=\sqrt{\pi\over2}\,N^{1/8}e^{-i\phi/2}\big(\rho^{-1/2}
\hat{\cal M}^{\prime A}_\alpha\,,\,
\rho^{1/2}\hat\zeta^{\aD A}\big)\ ,
\elabel{npsidef}
\end{equation}

Following the same steps as in the $\N=4$ theory, the
effective large-$N$ $k$-instanton
measure, for the leading order semi-classical approximation of the
functional integral, has the form
\SP{
&e^{2\pi ik\tau}{\EuScript Z}_k^{\sst(\N=2,N_F=2N)}
\
\overset{N\rightarrow\infty}\longrightarrow\\
& {N^{1/2}e^{2\pi ik\tau}\over
k2^{9k^2/2-k/2+12}\,\pi^{5k^2/2+4}\,{\rm Vol}\,\U(k)}
\int\rho^{-5}d\rho\,d^4\com\, d\phi\,e^{4i\phi}\,\prod_{A=1}^2d^2\xi^A
d^2\bar\eta^A\cdot\widehat{\EuScript Z}_{\SU(k)}^{\sst(d=6)}\ ,
\elabel{nhello}
}
where $\hat{\EuScript Z}_k$ is the partition function of the ${\cal N}=(1,0)$
supersymmetric $\SU(k)$ gauge theory in six dimensions dimensionally reduced
to zero dimensions:
\begin{equation}\begin{split}
\widehat{\EuScript Z}_{\SU(k)}^{\sst(d=6)}&=\int_{\SU(k)}\,
d^{6}A\, d^{8}\Psi\,e^{-S(A_\mu,\Psi)}\ ,\\
S(A_M,\Psi)\ &=\ N(\Gamma_{\rm b}+\Gamma_{\rm f})\ =\
-{1\over2}{\rm tr}_k\,\left[A_M,A_N\right]^2-i{\rm
tr}_k\,\bar\Psi\Gamma_M\left[A_M,\Psi\right]\ .
\elabel{nsukpart}
\end{split}
\end{equation}
When integrating expressions which are independent of the $\SU(k)$
degrees-of-freedom, $\widehat{\EuScript Z}_{\SU(k)}^{\sst(d=6)}$
is simply an overall constant
factor that was evaluated in \cite{KNS}. In our notation\footnote{We
have written the result in a way which allows an easy comparison with
\cite{KNS}. The factors of $\sqrt{2\pi}$ and
$\sqrt2$ arise, respectively, from the
difference in the definition of the bosonic integrals and the normalization
of the generators: we have ${\rm tr}_k\,T^rT^s=\delta^{rs}$ rather
than $\tfrac12\delta^{rs}$. The
remaining factors are the result of \cite{KNS}.}
\begin{equation}
\hat{\EuScript
Z}_{\SU(k)}^{\sst(d=6)}=
\big(\sqrt{2\pi}\big)^{6(k^2-1)}\big(\sqrt2\big)^{(8-6)(k^2-1)}\cdot
{2^{k(k+1)/2}\pi^{(k-1)/2}\over2\sqrt
k\prod_{i=1}^{k-1}i!}\cdot{1\over k^2}\ .
\elabel{partn2}\end{equation}

In summary, the effective large-$N$ collective
coordinate measure has the following
simple form \cite{Hollowood:1999sk}:
\EQ{
e^{2\pi ik\tau}{\EuScript Z}_k^{\sst(\N=2,N_F=2N)}
\ \overset{N\rightarrow\infty}\longrightarrow\ {\sqrt N\over
2^{17}\pi^{15/2}}\,k^{-7/2}e^{2\pi ik
\tau}
\,\int\,
{d^4\com\,d\rho\over\rho^5}\, d\phi\,e^{4i\phi}\,\prod_{A=1}^2d^2\xi^A
d^2\bar\eta^A\ .
\elabel{nendexp}
}
This has a remarkable similarity to the form of
the $\N=4$ measure \eqref{endexp}. Apart from the
differences in the overall numerical factors, the integral
over $S^5$ is replaced by $S^1$ and the $\N=4$ measure
involves, in addition, the
factor $\sum_{d|k}d^{-2}$, the sum over the integer divisors of
$k$. Notice that the $\sqrt N$ dependence and factor of $k^{-7/2}$ is the same in both cases.

The appearance of the $e^{4i\phi}$ phase in
\eqref{nendexp} implies a selection
rule in order that correlation functions are non-vanishing. This is
a relic of fermion zero mode counting.
{}From \eqref{bilin}, we see each insertion of an adjoint fermion
Grassmann collective coordinate (other than those associated to broken
supersymmetry and superconformal invariance which drop out from the
couplings in \eqref{bilin}) implies an
insertion of $e^{i\phi/2}$ and each insertion of a fundamental fermion
Grassmann collective coordinate
implies an insertion of $e^{-i\phi/2}$. The selection rule implies
that the difference between the latter and former is 8.
Therefore the simplest
kind of non-vanishing correlation functions would involve
insertions that saturate the 8 integrals over $\xi^A$ and $\bar\eta^A$
and, in addition, integrals over
8 of the fundamental fermion Grassmann collective coordinates
$\{\K,\tilde\K\}$.

\subsection{Large-$N$ correlation functions}\elabel{sec:S42}

Having established the form for the collective coordinate integral in
the large-$N$ limit, we can now calculate
correlation functions of various composite operators.  We are
primarily interested in those correlators that receive contributions
from all instanton numbers. From the form of the integrals
\eqref{hello} and \eqref{nhello},
we can now identify those these as the ones for which the leading
order (in $1/N$) expression for the insertion is independent of the
$\SU(k)$ variables. Hence the only non-trivial dependence is on
$\{X_n,\rho,\sfc_a\}$ and
$\{\xi^A,\bar\eta^A\}$.\footnote{We will describe how dependence on
$\sfc_a$ in the insertion
arises in due course.} In this case, we can use the reduced form of
the integrals \eqref{endexp} and \eqref{nendexp}.
For these special correlation functions the only dependence on the
instanton charge will be through an
overall multiplicative factor as advertised in \eqref{corrf}.

Before considering specific insertions, we can make some useful
general statements.
At leading order in $1/N$, we can replace each operator
insertion with its classical $k$-instanton saddle-point value. Since the
saddle-point solution \eqref{specsol} is relatively simple, this
observation greatly streamlines the form of the operator
insertions. We will restrict our attention to operators
consisting of a single trace on the gauge group index of a product
of adjoint scalars, fermions and field strengths. Each of these three adjoint
quantities is of the type $\Ubar\BX U$ where $U$ is the ADHM quantity
defined in \eqref{uan} and $\BX$ is some
matrix of ADHM variables; consequently the insertions have the form
\begin{equation}
{\cal O}(x)=\TrN\big[\bar U\BX_1U\bar U\BX_2U\cdots\bar U\BX_pU\big]=
{\rm tr}_{N+2k}\,\big[{\cal P}\BX_1{\cal P}\BX_2{\cal P} \cdots{\cal P}\BX_p\big]\ ,
\elabel{insop}\end{equation}
where ${\cal P}=U\bar U$ is the projection operator  \eqref{cmpl}.
It is easily checked that
at the saddle-point, the bosonic ADHM quantities $f$, $a'_n$, $\BL$
and ${\cal P}$ collapse to
\SP{
f&\rightarrow{}{1\over (x-X)^2+\rho^2}1_{\sst[k]\times[k]}\ , \quad
a'_n\rightarrow{}-X_n1_{\sst[k]\times[k]}\  ,\quad
\BL\rightarrow{}2\rho^2\ ,
\\
{\cal P}&=\frac1{(x-X)^2+\rho^2}\MAT{
((x-X)^2+\rho^2)1_{\sst[N]\times[N]}-w_\aD\bar w^\aD & -
w_\aD (\bar x-\bar X)^{\aD\alpha}
\\ -(x-X)_{\alpha\aD}\bar w^\aD & \rho^2
1_{\sst[2k]\times[2k]}}\ .\elabel{spuub}
}
One can verify that deviations from these saddle-point values
are suppressed by powers of $N^{-1/4}$.

The analogous replacement prescription for the fermionic ADHM
quantities is, in general, somewhat trickier.
Recall the expansions \eqref{sosm} for
the Grassmann collective coordinates.
In analogy with the bosonic quantities
\eqref{specsol}-\eqref{spuub}, it is useful to think of the unlifted
variables $\xi^A$ and $\etabar^A$ themselves as the
arising from a  saddle-point evaluation:
\begin{equation}
\CM_\alpha^{\prime A}\rightarrow-4i\xi_\alpha^A1_{\sst[k]\times[k]}\, ,\qquad
\zeta^{\aD A}\rightarrow-4i\bar\eta^{\aD A}1_{\sst[k]\times[k]}\ .
\elabel{spvamz}
\end{equation}
Indeed, the scaling \eqref{psidef} implies that the remaining variables
denoted $\hat\CM^{\prime A}$ and $\hat\zeta^A$ in \eqref{sosm} are
subleading compared to $\xi^A$ and $\etabar^A$ by a factor of
$N^{-1/8}.$ There remain the modes $\nu^A$ and $\bar\nu^A$, which are
distinct from the others in that they carry an $\SU(N)$ index
$u$. From their coupling to $\chi_a$ in
Eq.~\eqref{intoutnu}, one
sees that each $\bar\nu^A\nu^B$ pair in an insertion, for a fixed,
un-summed value of the index $u$, costs a factor of $N^{1/2}$;
however, summing on $u$ (as required by gauge invariance) then turns
this $N^{-1/2}$ suppression into an $N^{1/2}$ enhancement. In other
words, $\nu^A$ and $\bar\nu^A$ factors in the insertions should each
be thought of as being enhanced by $N^{1/4}$. The large-$N$ rule of
thumb for choosing Grassmann collective coordinates
in correlators is now clear:
the insertions should saturate as many of the
$\{\nu^A,\bar\nu^A\}$ integrals as possible,
with the proviso that the integrals over the
$\{\xi^A,\etabar^A\}$ must be saturated.

Notice in order to consider insertions of
$\{\nu^A,\bar\nu^A\}$ and $\{\K,\tilde\K\}$ we must return to
return to the collective coordinates before these variables have been
integrated out. We take the $\N=4$ case first and show,
to leading order in $1/N$,
the gauge invariant combination $\bar\nu^A\nu^B$
in the effective integration measure \eqref{endexp} is replaced by
\begin{equation}
\bar\nu^A\nu^B\rightarrow -{\rho N^{1/2}\over\pi}\Sigma_a^{AB}\sfc_a
1_{\sst[k]\times[k]}\ ,
\elabel{replace}\end{equation}
where the $N^{1/2}$ dependence has already been noted.
To this end, consider a general insertion with a string of such
combinations
$\bar\nu^{A_1}\nu^{B_1}\otimes\cdots\otimes\bar\nu^{A_p}\nu^{B_p}$.
We must insert this expression into the measure {\it before\/} the
$\nu$ integrals have been performed.
Then performing the $\nu$ integrals as in \eqref{intoutnu}
in the presence of the insertion leads to a modified expression
involving factors of $\chi^{-1}$ which
can be derived by considering
\begin{equation}\begin{split}
&{\partial\over\partial \chi^t_{A_1B_1}}\otimes
\cdots\otimes{\partial\over\partial \chi^t_{A_pB_p}}
\big|{\rm det}_{4k}\chi\big|^{N-2k}\\
&\qquad\qquad\qquad=N^p\big|{\rm det}_{ 4k}\chi\big|^{N-2k}
(\chi^{-1})_{B_1A_1}\otimes
\cdots\otimes(\chi^{-1})_{B_pA_p}+\cdots\ ,
\end{split}\end{equation}
where $t$ (transpose) acts on instanton indices and
the ellipsis represent terms of lower order in $1/N$. This shows
that, after performing the $\nu$ integrals,
a term of the form $\bar\nu^A\nu^B$ is
replaced by $-(\sqrt N/\sqrt8\pi)(\chi^{-1})_{BA}$, to leading
order. Now we replace
$\chi$ with its saddle-point
value \eqref{specsol} and use \eqref{E54.1} to give \eqref{replace}.
The replacement elucidates the mysterious appearance of the variables
$\sfc_a$.

In the $\N=2$ theory, examination of Eq.~\eqref{bilin}
yields a similar replacement
\EQ{
\bar\nu^A\nu^B\longrightarrow\frac{g\rho\sqrt{2iN}}{\pi}
\epsilon^{AB}e^{i\phi}1_{\sst[k]\times[k]}\ .
}
This leaves the matter Grassmann collective coordinates
$\{\K,\tilde\K\}$. Again after examining \eqref{bilin},
for each flavour $f=1,\ldots,2N$, we have the
replacement
\EQ{
\K_f\tilde\K_f\longrightarrow\frac{\rho\sqrt{2i}}{\pi\sqrt{N}}e^{-i\phi}
1_{\sst[k]\times[k]}\ .
\elabel{matrep}
}

We now consider some examples. First of all, in
the $\N=4$ theory there are a class of correlation functions of type
described above motivated by the AdS/CFT correspondence (discussed in
\S\ref{sec:S320}).
The simplest involves the 16-point correlation
function of the fermionic composite operator
\EQ{
\Lambda^A=g^2\,\sigma_{mn}\TrN\,
F_{mn}\,\lambda^A\ .
\elabel{e4}
}
Since, to leading order in $g$,
each insertion is linear in Grassmann collective coordinates, we
only need the
dependence on the collective coordinates
$\{\xi^A,\bar\eta^A\}$, since the other Grassmann coordinates must be
lifted by the instanton effective action. In this case to find the
dependence, we can use the supersymmetric sweeping out procedure
defined in \S\ref{sec:S31} 
with the $x$-dependent variation parameter \eqref{xdep}
giving\footnote{The extra factor of $g^{1/2}$ follows form the
re-scaling \eqref{ressp}.}
\EQ{
\Lambda^A=ig^{5/2}\sigma_{mn}\sigma_{kl}\xi^A(x)\,\TrN\,F_{mn}F_{kl}+\cdots\ .
=-\tfrac i4g^{5/2}\xi^A(x)\,\TrN\,F_{mn}^2+\cdots\ .
}
The ellipsis represent the dependence on the other Grassmann
collective coordinates that are not needed. Next we can use the identity
\eqref{oruf} and finally insert the saddle-point expression for $f$ in
Eq.~\eqref{spuub}:
\EQ{
\Lambda^A=
24ikg^{1/2}\xi^A(x)\frac{\rho^4}{\big((x-X)^2+\rho^2\big)^4}+\cdots\ .
\elabel{lamins}
}
Then using Eqs.~\eqref{endexp} and \eqref{lamins},
the large-$N$ $k$-instanton contribution to the 16-point correlator
is precisely \cite{MO3}
\SP{
&\VEV{\Lambda_{\alpha_1}^{A_1}(x^{(1)})\times\cdots\times
\Lambda_{\alpha_{16}}^{A_{16}}(x^{(16)})}_{\rm inst}
\ \overset{N\to\infty}\longrightarrow\
{3^{16}2^{15}g^8\sqrt N\over\pi^{27/2}}\Big\{\sum_{k=1}^\infty
k^{25/2}e^{2\pi
ik\tau}\sum_{d\vert k}{1\over d^2}\Big\}\\
&\times\int {d^4\com d\rho\over\rho^5}\,\prod_{A=1}^4d^2\xi^A\,d^2\etabar^A\,
\,\prod_{l=1}^{16}\frac{\xi^{A_l}_{\alpha_l}(x^{(l)})
\rho^4}{\big((x^{(l)}-X)^2+\rho^2\big)^4}\ .
\elabel{finalans}
}
For application to the AdS/CFT correspondence it is unnecessary to
perform the remaining integrals.

The correlator \eqref{finalans} has the distinguishing property that
the functional dependence on the insertion points given by the
integral term can be reproduced by a single instanton calculation with
gauge group $\SU(2)$. The reason is clear: only the dependence of the
insertions on the supersymmetric and superconformal Grassmann
coordinates was required. In particular the saddle-point evaluation of
the relevant part of the insertion \eqref{lamins} is identical
to the single instanton in $\SU(2)$. So at large $N$ the dependence on
$k$ and $N$ only appears in an overall multiplicative factor. In fact
there is a whole family of related correlators considered in
Ref.~\cite{BGKR} for which the same property holds and a calculation for a
single instanton in $\SU(2)$ is sufficient to obtain the functional
dependence on the insertion points. We have now explained the puzzle
of why the analysis of Ref.~\cite{BGKR} involving a single instanton in
$\SU(2)$ was able to capture effects that ultimately, via the AdS/CFT
correspondence should have been valid at large-$N$ only.

There are more general correlation functions involving insertions with
non-trivial $\SO(6)$ $R$-transformation properties for which a calculation
in $\SU(2)$ would not suffice. For example consider the insertion of
the composite field
\EQ{
{\cal O}_{a_1\cdots
a_p}(x)=\TrN\,\big[\phi_{a_1}\cdots\phi_{a_p}F_{mn}^2\big]\ ,
}
whose form is motivated by the AdS/CFT correspondence. We now consider
this evaluated at the saddle point. In general, ${\cal O}$
depends on the Grassmann variables $\{\mu^A,\bar\mu^A\}$, as well as
the supersymmetric and superconformal variables. The resulting
expressions are rather cumbersome so we will assume that the integrals
over the supersymmetric and superconformal variables are saturated by
other insertions, for example, by sixteen insertions of $\Lambda^A$. In
this case we only need the dependence of ${\cal O}$ on the
$\{\nu^A,\bar\nu^A\}$:
\EQ{
{\cal O}_{a_1\cdots a_p}(x)\thicksim
\frac{\rho^{4}}{\big((x-X)^2+\rho^2\big)^{p+4}}
\bar\Sigma_{a_1A_1B_1}\cdots\bar\Sigma_{a_pA_pB_p}
{\rm tr}_k\big[\bar\nu^{A_1}\nu^{B_1}\cdots\bar\nu^{A_p}\nu^{B_p}\big]\ .
}
At leading order in $1/N$ we need use the replacement \eqref{replace}
to obtain
\EQ{
{\cal O}_{a_1\cdots a_p}(x)\thicksim
kN^{p/2}\frac{\rho^{4+p}}{\big((x-X)^2+\rho^2\big)^{p+4}}
\sfc_{a_1}\cdots\sfc_{a_p}\ .
}
In this case, we see that the insertion depends explicitly on the
$S^5$ coordinate $\sfc_a$.

In the $\N=2$ theory, a simple correlator involves 8 insertions
of $\Lambda^A$, as in \eqref{e4} where now $A=1,2$,
and 4 insertions of the gauge-invariant operator
\EQ{
Q_f(x)=g^2\TrN\,\chi_f\,\tilde\chi_f\ ,
}
which is quadratic in the matter Grassmann collective coordinates.
Evaluating this on the saddle-point solution using the expression
\eqref{fundredux} along with the leading-order replacement
\eqref{matrep}, we have
\EQ{
Q_f(x)=\frac{gk\sqrt{8i}}{\pi\sqrt N}\frac{\rho^2}{
\big((x-X)^2+\rho^2\big)^3}e^{-i\phi}+\cdots\ .
}
Notice that the 4 insertions of $Q_f$ involve a factor of
$e^{-4i\phi}$ which cancels the factor of $e^{4i\phi}$ in
Eq.~\eqref{nendexp}, so the integral over $\phi$ yields the
constant $2\pi$. Hence,
\SP{
&\VEV{\Lambda_{\alpha_1}^{A_1}(x^{(1)})\times\cdots\times
\Lambda_{\alpha_8}^{A_8}
(x^{(8)})Q_{f_1}(x^{(9)})\times\cdots\times Q_{f_4}(x^{(12)})}_{\rm inst}\
\overset{N\to\infty}\longrightarrow\
 {3^82^{14}g^{8}\over
\pi^{21/2}N^{3/2}}\,\Big\{\sum_{k=1}^\infty
k^{17/2}e^{2\pi
ik\tau}\Big\}\\
&\qquad\qquad
\times\,\int {d^4\com d\rho\over\rho^5}\,\prod_{A=1}^2d^2\xi^A\,d^2\etabar^A\,
\,\prod_{l=1}^{8}\frac{\xi^{A_l}_{\alpha_l}(x^{(l)})
\rho^4}{\big((x^{(l)}-X)^2+\rho^2\big)^4}\,\prod_{l=1}^{4}
\frac{\rho^2}{\big((x^{(l+8)}-X)^2+\rho^2\big)^3}
\ .
\elabel{nfinalans}
}

\subsection{Instantons and the AdS/CFT correspondence}\label{sec:S320}

The formalism that we have used to calculate the
instanton contributions to certain correlation functions at leading
order in $1/N$ is particularly interesting in light of the AdS/CFT
correspondence. The basic example of the AdS/CFT correspondence,
due to Maldacena \cite{MAL} (see also the comprehensive review
\cite{Aharony:2000ti}), states that the ${\cal N}=4$
supersymmetric gauge theory is equivalent to
Type IIB string theory on an $AdS_{5}\times S^{5}$ background.
In particular, the gauge coupling $g$ and vacuum angle $\theta$
of the four-dimensional theory are given in terms of the string
parameters by
\begin{equation}
g=\sqrt{4\pi g_{st}}=\sqrt{4\pi e^\phi}\ ,\qquad \theta=2\pi
C^{(0)}\ .
\elabel{corresp}
\end{equation}
Here, $g_{st}$ is the string coupling while $\phi$ and
$C^{(0)}$ are the
expectation values of the dilaton and Ramond-Ramond scalar,
respectively, of Type IIB string
theory. Also $N$ appears explicitly, through the relation
\begin{equation}
{L^2\over\alpha'} = \sqrt{g^2\,N}
\elabel{alphapdef}
\end{equation}
 where
$(\alpha')^{-1}$ is the string tension and $L$ is the radius of both
the $AdS_5$ and $S^5$ factors of the background.

\subsubsection{The instanton collective coordinate integral}

Even before we consider application of the AdS/CFT
correspondence to correlation functions,
and how it applies to instantons, we can already see in
the form of the effective large-$N$ collective coordinate
integral \eqref{endexp}
a remarkable relation to $AdS_5\times S^5$. The $c$-number integrals
are precisely the volume form on this space, where the $AdS_5$ part
has the metric
\EQ{
ds^2=\rho^{-2}(dX_n^2+d\rho^2)
}
so that $\rho$ is the radial variable. This fact alone
strongly suggests that Yang-Mills instantons are identified in the
large-$N$ limit with D-instantons in the Type IIB string theory.
Some relevant aspects of D-branes in
Type II string theory are summarized in \S\ref{sec:S100}. In
particular, the contribution of $k$ D-instantons to low energy
correlators is governed by the $\U(k)$ matrix integral \eqref{ukpfn}.
For the most part, the $\SU(k)$ part of the integral can be factored
off as an overall numerical constant; hence,
\begin{equation}
{\EuScript Z}_{k}\thicksim
\frac{e^{-2\pi k(e^{-\phi}+iC^{(0)})}}{{\rm Vol}\,\U(k)}\,
\int\,d^{10}X\,d^{16}\Theta\,\cdot\,\hat{\EuScript Z}_k\ ,
\elabel{ukpfnm}
\end{equation}
where $\hat{\EuScript Z}_k$ is the same $\SU(k)$ matrix integral that we
defined in \eqref{nsukpart} and $X_M$ and $\Theta$ are the abelian
components of the fields. This D-instanton
collective coordinate integral is appropriate to the case of flat
ten-dimensional space.
However, the results of Banks and Green \cite{BG} imply that
since $AdS_5\times S^5$ is conformally flat a similar expression should
apply to this background with the bosonic abelian integrals replaced by the
appropriate volume form of $AdS_5\times S^5$. Being careful with
the pre-factor, one has the D-instanton collective coordinate
integral\cite{BGKR,MO3}
\EQ{
{\EuScript Z}_k\Big|_{AdS_5\times
S^5}\thicksim
(\alpha')^{-1}(ke^{\phi})^{-7/2}\,e^{-2\pi k(e^{-\phi}+iC^{(0)})}\,
\sum_{d|k}\frac1{d^2}\,\int\,
\frac{d\rho\,d^4X}{\rho^{5}}\, d^5\sfc\, d^{16}\Theta\ .
\elabel{int}
}
Using the relation between the couplings of the string theory and the
gauge theory in \eqref{corresp} this is precisely the leading order
collective coordinate integral in the large-$N$ limit of the $\N=4$
instanton calculus \eqref{endexp}.\footnote{There appears to be
a mismatch of $g^8$, however, this is due to our normalization of the
Grassmann collective coordinates. Re-scaling the supersymmetric and
superconformal collective coordinates, $\xi^A$ and $\bar\eta^A$,
by $g^{1/2}$ produces the missing factor of $g^8$.}
This equivalence of the large-$N$
Yang-Mills instanton measure and the D-instanton measure on
$AdS_5\times S^5$ is rather stunning evidence for the AdS/CFT
correspondence. It is very satisfying that instantons at large $N$
seem to probe
the $AdS_5\times S^5$ geometry directly.
Although, as discussed below, this relation implies the existence
of an as yet un-proved non-renormalization theorem for a certain class of
correlation functions (like those discussed in \S\ref{sec:S42})
protecting them against perturbative corrections in $g^2N$. In what
follows, we shall make this correspondence between large-$N$
instanton and D-instanton effects even more convincing by considering
these correlation functions.

In \S\ref{sec:S100} we will establish
a relation between Yang-Mills instantons and
D-instantons which is apparently rather different from that we have
just described. We will show in \S\ref{sec:S100} how
$k$ D-instantons in the presence of $N$ D3-branes
are precisely identified with $k$ Yang-Mills
instantons in $\U(N)$ gauge theory describing the collective dynamics
of the D3-branes. This is true in the
limit $\alpha'\to0$ for fixed $g_{st}$ (in other
words fixed coupling $g$ on the D3-branes) in which bulk supergravity
modes decouple from the world-volume theory of the D3-branes.
The question is how this description in terms of D-instantons moving in the
background of D3-branes relates to the description of D-instantons
moving in the $AdS_5\times S^5$ background (but with no D3-branes)
established above? The answer is interesting because it illuminates
some basic
features of the AdS/CFT correspondence. In the limit of large
$N$ (with $g^2N$ large and small $g$) the background of D3-branes is
replaced by its near-horizon geometry, namely $AdS_5\times S^5$. In a
certain sense the D3-branes disappear to be replaced by non-trivial
geometry. We can see this happening explicitly with the D-instanton/D3-brane
system. The presence of the D3-branes in the D-instanton matrix theory
is signaled by the fundamental hypermultiplet variables
$\{w_\aD,\mu^A\}$, and their conjugates, describing open strings
stretched between the D-instantons and D3-branes. At large $N$, these
degrees-of-freedom can be integrated out in the way described in
\S\ref{sec:S41} to yield an effective collective coordinate integral
\eqref{hello} which exhibits the $AdS_5\times S^5$ geometry
explicitly. So the analogue of taking the near horizon geometry is the
process of integrating out the degrees-of-freedom of open strings
stretched between the D-instantons and D3-branes and taking the
large-$N$ limit. The remaining puzzle
is that this should be done at weak coupling, {\it i.e.\/}~small
$g^2N$, whereas the dual supergravity region of the AdS/CFT
correspondence is valid at large $g^2N$. Yet again, 
this strongly suggests that some non-renormalization theorem in $g^2N$ is a
work: something we shall comment on in the next section.

\subsubsection{Correlation functions}

A more precise statement of the AdS/CFT correspondence was
presented in Refs.~\cite{WIT150,GKP}.
The ${\cal N}=4$ gauge theory is to be thought of living
on the four-dimensional
boundary of $AdS_{5}$. In particular, each chiral primary operator
${\cal O}$ in the boundary conformal field theory is identified with a
particular Kaluza-Klein mode of the supergravity fields which we
denote as $\Phi_{\cal O}$.
In general it is not known how to solve string theory on an $AdS$
background. However, the AdS/CFT correspondence is still useful
because in a certain limit we can approximate the full string theory
by its supergravity low-energy limit. This requires weak coupling
(small $g$) but, in addition, the length scale $L$ must be
large compared with the string length scale
$\sqrt{\alpha'}$. This latter requirement is met when the 't~Hooft
coupling $g^2N$ of the gauge theory is large and conventional
perturbation theory breaks down. In this limit \cite{WIT150,GKP},
the generating function for the correlation
functions of ${\cal O}$ is then given in terms of the supergravity
action $S_{\rm IIB}[\phi_{\cal O}]$ according to
\begin{equation}
\left\langle \exp\,\int\, d^{4}x \, J_{\cal O}(x){\cal O}(x)\,
\right \rangle = \exp-S_{\rm IIB}\left[\Phi_{\cal O};
J\right]\, .
\elabel{corr}
\end{equation}
The IIB action on the right-hand side of the equation is evaluated on
a configuration
which solves the classical field equations subject to the condition
$\Phi_{\cal O}(x)=J_{\cal O}(x)$ on the four-dimensional boundary.

In most
applications considered so far, the relation \eqref{corr} has
primarily been applied at the level of classical supergravity, which
corresponds to $N\rightarrow \infty$, with $g^{2}N$ fixed and large,
in the boundary theory
\cite{WIT150,GKP,HW,AFGJ,FMMR,EFS,LMRS,CNSS,LT,MV,BGut,AF1,SOLOD,GHEZ}.
However, the full equivalence of IIB
superstrings on $AdS_{5}\times S^{5}$ and ${\cal N}=4$ supersymmetric
Yang-Mills theory  conjectured
in \cite{MAL} suggests that \eqref{corr} should hold more generally,
with quantum and stringy corrections to the classical supergravity
action corresponding to $g^2$ and $1/(g^{2}N)$ corrections in the ${\cal
N}=4$ theory, respectively. The particular comparison that concerns us
here is between Yang-Mills instanton contributions to the correlators
of ${\cal O}$ generated by the left-hand side of \eqref{corr} and
D-instanton corrections to the IIB effective action on the right-hand
side.

Let us describe these D-instanton effects in more detail.
Before we look at some specific correlators, let us consider in more
detail the effects of D-instantons in the string theory \cite{BG}
(closely following the more detailed treatment in \cite{BGKR}). In
\cite{GG1}, Green and Gutperle conjectured an exact form for certain
non-perturbative (in $g_{st}$) corrections to certain terms in the
Type IIB supergravity effective action. In the present application,
where the string theory is compactified on $AdS_5\times S^5$, it is
important for the overall consistency of the Banks-Green prediction
that the non-perturbative terms in the effective action do not alter
the $AdS_5\times S^5$ background, since the latter is conformally flat
\cite{BG}.  In particular, at leading order beyond the
Einstein-Hilbert term in the derivative expansion, the IIB effective
action is expected to contain a totally antisymmetric 16-dilatino
effective vertex of the form \cite{GGK,KP2}
\begin{equation}
(\alpha')^{-1}\int d^{10}x\,\sqrt{\det g}\,e^{-\phi/2}\,
f_{16}(\tau,\bar\tau)\,\Lambda^{16}+ \hbox{H.c.}
\elabel{effvert}\end{equation}
Here $\Lambda$ is a complex chiral $\SO(9,1)$ spinor, and $f_{16}$ is a certain
weight $(12,-12)$ modular form under $\SL(2,{\mathbb Z})$. At the same
order in the derivative expansion there are other terms related to
\eqref{effvert} by supersymmetry and involving other modular forms
\cite{GGK,KP2,KP1} and, in particular, $f_n(\tau,\bar\tau)$, with $n=4$ and 8.
The modular symmetry is precisely $S$-duality of the Type IIB superstring, and
although this does not completely determine the modular
forms $f_n$, for the $n=4$ term, Green and Gutperle
\cite{GG1} made the following conjecture, later proved in
Ref.~\cite{GSI} and generalized to $n\neq4$ in \cite{KP2,KP1}:
\begin{equation}
f_n(\tau,\bar\tau)=({\rm Im}\tau)^{3/2}
\sum_{(p,q)\neq(0,0)}(p+q\bar\tau)^{n-11/2}(p+q\tau)^{-n+5/2}.
\end{equation}
These rather arcane expressions turn out to have the right modular
properties, {\it i.e.\/}~weight $(n-4,-n+4)$, and also have very suggestive
weak coupling expansions \cite{GG1,GGK,KP2,GG2}:
\begin{equation}e^{-\phi/2}f_n=
32\pi^2\zeta(3)g^{-4}-{2\pi^2\over3(9-2n)(7-2n)}+\sum_{k=1}^\infty{\cal
G}_{k,n}\
, \elabel{fexpand}\end{equation}
where
\begin{equation}\begin{split}
{\cal G}_{k,n}=\left({8\pi^2 k\over
g^2}\right)^{n-7/2}\Big\{\sum_{d|k}{1\over d^2}\Big\}&
\Big[e^{-(8\pi^2/g^2-i\theta)k} \sum_{j=0}^\infty
c_{4-n,j-n+4}\left({g^2\over8\pi^2k}\right)^{j} \\
&+e^{-(8\pi^2/g^2+i\theta)k} \sum_{j=0}^\infty
c_{n-4,j+n-4}\left({g^2\over8\pi^2k}\right)^{j+2n-8}\Big]
\elabel{Gdef}\end{split}\end{equation}
and the numerical coefficients are
\begin{equation}
c_{n,r}={(-1)^n\sqrt{8\pi}\,\Gamma(3/2)\Gamma(r-1/2)\over2^r\Gamma(r-n+1)
\Gamma(n+3/2)\Gamma(-r-1/2)}\ .
\elabel{defcs}
\end{equation}
As previously,
the summation over $d$ in \eqref{Gdef} runs over the integral
divisors of $k$.  Notice that, having taken into account the
conjectured correspondence \eqref{corresp} to the couplings of
four-dimensional Yang-Mills theory, the series \eqref{fexpand} has
the structure of a semi-classical expansion: the first two terms
correspond to the tree and one-loop pieces, while the sum on $k$ is
interpretable as a sum on Yang-Mills instanton number, the first and
second terms in the square bracket being instantons and
anti-instantons, respectively (as dictated by the $\theta$
dependence). Each of these terms includes a perturbative expansion
around the instantons, although notice that the leading order
anti-instanton contributions are suppressed by a factor of $g^{4n-16}$
(so not suppressed for $n=4$) relative to the leading order instanton
contributions.

Let us focus on the leading
semi-classical contributions to the $f_n$; by this we mean, for each
value of the topological number $k$, the leading-order contribution in
$g^2$. For $f_{16}$ and $f_8$ the leading semi-classical contributions
come from instantons only and have the form
\begin{equation}
e^{-\phi/2}f_n\Big\vert_{k\hbox{-}\rm instanton}=
{\rm const}\cdot\left({k\over g^2}\right)^{n-7/2}
e^{2\pi ik\tau}\sum_{d|k}{1\over d^2}\ ,\elabel{foksh}
\end{equation}
neglecting $g^2$ corrections.  For the special case of $f_4$ there is
an identical anti-instanton contribution with
$i\tau\rightarrow-i\bar\tau$. Comparing with \eqref{int}, we see that
the terms in
the square bracket in \eqref{Gdef}, which are non-perturbative in the
string coupling, are interpreted as being due to D-instantons.

{}From the effective vertex \eqref{effvert} one can construct Green's functions
$G_{16}(x^{(1)},\ldots,x^{(16)})$ for sixteen
dilatinos $\Lambda(x^{(l)}),$ $1\leq l\leq 16,$ which live on the boundary
of $AdS_5$:
\begin{equation}
G_{16}=\langle\,\Lambda(x^{(1)})\times\cdots\times
\Lambda(x^{(16)})\rangle\thicksim
(\alpha')^{-1}
\,e^{-\phi/2}\,f_{16}\,t_{16}\int{d^4\com\,d{\rho}\over{\rho}^5}\,
\prod_{l=1}^{16}\,K_{7/2}^F(\com,{\rho};x^{(l)},0)
\elabel{grnfcn}\end{equation}
suppressing spinor indices.
Here $K_{7/2}^F$ is the bulk-to-boundary propagator for a spin-$\hf$ Dirac
fermion of mass $m=-\tfrac32L^{-1}$ and scaling dimension $\Delta=\tfrac72\,$
\cite{WIT150,GKP,FMMR,HS1}:
\begin{equation}K_{7/2}^F(\com,{\rho};x,0)=
K_{4}(\com,{\rho};x,0)\,\big(\rho^{1/2}\gamma_5-\rho^{-1/2}(x-\com)_n
\gamma^n\big)
\elabel{KFdef}\end{equation}
with
\begin{equation}K_{4}(\com,{\rho};x,0)={{\rho}^4\over\big({\rho}^2+
(x-\com)^2\big)^4}\ .
\elabel{Kfourdef}\end{equation}
In these expressions the $x^{(l)}$ are four-dimensional spacetime
coordinates for the boundary of $AdS_5$ while $\rho$ is the fifth,
radial, coordinate.  The quantity $t_{16}$ in Eq.~\eqref{grnfcn} is
(in the notation of Ref.~\cite{BGKR}) a 16-index antisymmetric invariant
tensor which enforces Fermi statistics and ensures, {\it inter alia\/},
that precisely 8 factors of $\rho^{1/2}\gamma_5$ and 8 factors of
$\rho^{-1/2}\gamma^n$ are picked out in the product over $K^F_{7/2}$.
It is not difficult to see that the form of \eqref{grnfcn} matches
precisely the large-$N$ instanton contribution to the 16-point correlator
in the gauge theory \eqref{finalans}. In particular, the dilatino
corresponds, on the field theory side, to the fermionic
composite operator defined in
\eqref{e4} \cite{BGKR}. This shows that not only does the
form of the D-instanton collective coordinate integral match the
large-$N$ instanton collective coordinate integral, but, in addition,
the various bulk-to-boundary propagators on the supergravity side
arise in the field theory as operator insertions evaluated in the large-$N$
instanton background.
In the case of $G_{16}$, and
related correlators, because of the nature of the saddle-point
solution, the insertions have the form of $k$ times the same insertion in the
theory with gauge group $\SU(2)$ at the one-instanton level (precisely
the case considered in \cite{BGKR}).

Of course this equivalence between D-instanton
and Yang-Mills instanton effects
in correlation functions extends to the other cases, $G_8$ and $G_4$,
related by supersymmetry.
Note that there is no explicit dependence in these expressions on
the coordinates on $S^5$; in particular, the propagator does not
depend on them. This is because $G_{16}$, $G_8$ and $G_4$ are
correlators of operators whose supergravity associates are constant on
$S^5$. Generalizations involving non-trivial dependence on $S^5$ are
possible \cite{MO3}.

As emphasized above, the comparison between the Yang-Mills and
supergravity descriptions can be quantitative if and
only if there exists a non-renormalization theorem that allows one to
relate the small $g^2N$ to the large $g^2N$ behaviour of chiral
correlators such as $G_n$, as has been suggested in
Refs.~\cite{DKMV}.  In the absence of such a theorem the best one
can hope for is that qualitative features of the agreement persist
beyond leading order while the exact numerical factor in each
instanton sector does not, in analogy with the mismatch in the
numerical pre-factor between weak and strong coupling results for
black-hole entropy \cite{GKP}.  In our view, however, our
results provide strong evidence in favor of such a non-renormalization
theorem for the correlators $G_n$, for the following reason.  Consider
the planar diagram corrections to the leading semi-classical
({\it i.e.\/}~$g^2N\rightarrow0$)
result for, say, $G_{16}$, Eq.~\eqref{finalans}.
In principle, these would not only modify the above result by an
infinite series in $g^2N,$ but also, at each order in this expansion,
and independently for each value of $k$, they would produce a
different function of spacetime. The fact that the leading
semi-classical form for $G_{16}$ that we obtain is not only
$k$-independent, but already reproduces the spacetime dependence of the
D-instanton/supergravity prediction exactly, strongly suggests that such
diagrammatic corrections (planar and otherwise) must  vanish.
(Note there are necessary subleading corrections, both in $1/N$
and in $g^2$, to our leading semi-classical results.) By evaluating
the possible form of non-perturbative corrections on the string theory
side of the correspondence, it has been
argued that the non-renormalization theorem is rather natural
\cite{Gopakumar:1999xh}; however, there is still no purely field
theoretic proof.

It is interesting to generalize the relation of large-$N$
instantons in other conformal
gauge theories to the D-instantons via the more general AdS/CFT
correspondences. This has been done for the finite $\N=2$ theories with
product gauge group $\SU(N)^k$ in \cite{Hollowood:2000bm}, the
$\N=4$ $\Sp(N)$ and $\SO(N)$ theories in \cite{Hollowood:1999ev} and
for the finite $\N=2$ $\Sp(N)$ theories in
\cite{Gava:2000ky,Hollowood:1999nq}.

The finite $\N=2$ theory with
gauge group $\SU(N)$ and $2N$ fundamental hypermultiplets has not
yet been
discussed in this respect. However, it is not difficult to find a
set-up which can be used to discuss this case. Here we sketch the details.
The idea is to consider
the finite $\N=2$ theory with product gauge group $\SU(N)\times\SU(N)$
and matter in bi-fundamental representations of the gauge group. This
theory can be realized, as explained in \S\ref{sec:S93},
on the world-volume of $N$ D3-branes lying
transverse to the orbifold ${\mathbb R}^2\times({\mathbb R}^4/{\mathbb
Z}_2)$, where ${\mathbb Z}_2$ acts by inversion. The
near horizon geometry of the D3-branes is $AdS_5\times
S^5/{\mathbb Z}_2$, where ${\mathbb Z}_2$ acts on $S^5$, realized as
$x_a^2=1$ by taking $x^3,x^4,x^5,x^6\to-x^3,-x^4,-x^5,-x^6$, leaving
$x^1$ and $x^2$ fixed. So the AdS/CFT correspondence in this case
involves Type IIB
string theory on the $AdS_5\times S^5/{\mathbb Z}_2$
background. Notice that the ${\mathbb Z}_2$ leaves an $S^1\subset S^5$
fixed and so in this case the dual geometry is not smooth.
Now consider instanton effects in one of the $\SU(N)$
gauge group factors. At leading order in such an instanton background,
the adjoint-valued fields
of the other gauge remain zero and so the instanton effects in
question are identical to the $\SU(N)$ theory with $2N$
hypermultiplets (the latter arising from the bi-fundamental
hypermultiplets of the original theory).
On the string theory side, these instanton effects of charge $k$
will be related to $k$ fractional D-instantons. These
fractional D-instantons are to be though of as D-strings which are
wrapped around the non-trivial $S^2$ of the smooth resolution of the
orbifold ${\mathbb R}^4/{\mathbb Z}_2$ in the limit that the $S^2$
cycle collapses. In the near-horizon geometry these fractional
instantons are consequently stuck at the orbifold singularity of
$S^1\subset S^5/{\mathbb Z}_2$. So the fractional D-instanton
collective coordinate integral involves an integral over the six
dimensional space $AdS_5\times
S^1$ rather than ten-dimensions. Furthermore, this measure includes,
in analogy with \eqref{ukpfnm}, a
partition function of six-dimensional $\SU(k)$ gauge theory
dimensionally reduced to zero dimensions.
This is precisely what is seen in the large-$N$ instanton
collective coordinate integral \eqref{nendexp}. It would be
interesting to extend the analysis to particular
correlation functions in order to place
the relationship on the same footing as the $\N=4$ case.

\rsen\section{Instantons as Solitons in Higher Dimensions and
String Theory}\elabel{sec:S49}

Instantons are classical solutions of Euclidean gauge theories in
four-dimensional spacetime with finite
action. They are consequently localized in {\it spacetime\/} rather
than space. However, it will turn
out to be rewarding to think about instantons in gauge
theories in dimensions greater than four. In these other
contexts, if the higher-dimensional theories are defined in Minkowski
space then the instanton solutions have an interpretation as solitons. For
example, consider five-dimensional gauge theories. We can easily take
an instanton solution of four-dimensional gauge theory and embed it in
the five-dimensional theory, by identifying the coordinates
$x_n$, $n=1,2,3,4$, on
which the instanton solution depends,
with the Euclidean subspace of the five-dimensional Minkowski
space\footnote{In Minkowski space we choose a metric
$\eta_{MN}={\rm diag}(-1,1,\ldots,1)$.}
coordinates $y^N$, $N=0,1,2,3,4$, with $y_n\equiv x_n$, $n=1,2,3,4$. The
resulting solution of the five-dimensional Yang-Mills equations is
then independent of the time coordinate $y^0$.
The solution, which had finite
action in four dimensions, now has finite energy in five dimensions
and so is interpreted as a lump localized in space but evolving in time. In
other words, the instanton is a particle-like soliton in five
dimensions. Adding an extra dimension yields a much richer system:
instantons in four-dimensional gauge theories have no dynamics, but
when lifted to five dimensions they can move in the four dimensional
space and have complicated dynamics.

We can continue this process of embedding instantons in gauge theories
of even higher dimension. In six dimensions, the embedded instanton
solutions are independent of two coordinates and so, thinking of one
of them as a time direction, means the solution is extended in a
single space direction; in other words, it is a string-like
``defect''. In seven dimensions, one gets a membrane and so on.
Let us introduce some modern terminology, arising originally in
the context of supergravity and string theory, which affords us a
certain brevity in talking about these more exotic
possibilities. Consider gauge theories in $D$-dimensional
Minkowski space. An instanton embedded in this theory
will be independent of $D-4$ coordinates (one being the time). This
means that the solution will be extended in $D-5$ spatial
dimensions. Such a configuration is called a ``$(D-5)$-brane''.
The case with $D=4$, describes a solution which is localized in both
space and time dimensions and this is precisely the
situation for the original instanton of four-dimensional gauge
theory. This case is therefore a ``$(-1)$-brane''.
For $D=5$, the solution is localized in space, but not time,
and so the instanton lump-like soliton of five-dimensional
gauge theory is a $0$-brane.
It may seem strange to think about instanton solutions in higher
dimensional gauge theories but one of the major lessons of string
theory and its generalizations is that it is compulsory to think in
arbitrary numbers of dimensions and there are big advantages in doing so.

We will show how the instanton moduli space plays an important
r\^ole in describing the dynamics of instanton-branes in
higher-dimensional gauge theories. In \S\ref{sec:S51}
we begin by discussing the
case of the pure gauge instanton. Supersymmetric generalizations will
be considered in \S\ref{sec:S52}. It is worth pointing out that much of
the analysis for instantons is very closely related to the problem of
describing the semi-classical behaviour of monopoles in supersymmetric
gauge theories described in
Refs.~\cite{Gauntlett:1994sh,Bak:2000vd,Gauntlett:2001ks},
and references therein.
In particular the description of instanton lumps in
five-dimensional gauge theory with either
4 or 8 supercharges (corresponding to $\N=2$ and $\N=4$ supersymmetry
in four dimensions) is closely related to the description of
monopoles in four-dimensional gauge
theories, with $\N=2$ and $\N=4$ supersymmetry, respectively, since
both involve a quantum mechanical $\sigma$-model on the
appropriate moduli space. In both cases, the instanton and monopole
moduli space are both hyper-K\"ahler spaces and so the structure of
the $\sigma$-models is identical. The main difference is that the
$\sigma$-model in the instanton case admits a straightforward linear
realization via the ADHM hyper-K\"ahler quotient construction while in
the monopole case the analogous quotient construction based on
Nahm's equations is somewhat more complicated.

After we have described the dynamics of instanton branes, we will, in
\S\ref{sec:S100}
describe how such objects appear naturally in string theory. In this
context they correspond to D-branes dissolved within other
higher-dimensional D-branes. We will show how solving for the
low energy dynamics of these configurations of D-branes
actually leads directly to the instanton calculus. Not only is
the ADHM construction obtained directly, but also the leading order
expression for the collective coordinate integral for the $\N=4$ and
$\N=2$ supersymmetric theories. We also briefly describe how the
actual profile of the instanton can be obtained by using a suitable
probe brane.

\subsection{Non-supersymmetric instanton branes}\elabel{sec:S50}

We begin with the case of pure gauge theory.
The key idea, which originated with Manton \cite{Manton:1982mp} and was
elaborated by Ward in the context of a $2+1$-dimensional model
\cite{Ward:1985ij}, is that the dynamics of slowly moving
solitons can be approximated by assuming that the evolution is
adiabatic in the moduli space of classical solutions. In many cases
involving lumps the time evolution of the system is
governed by the geodesic motion on the moduli space with respect to
the Levi-Civita connection induced by the metric that arises as the
inner-product of the zero modes \eqref{ccmet}. The
intuitive idea that lies behind this approximation
is that, for sufficiently low velocities, the non-zero modes
of the fields are only very weakly excited and
therefore it is consistent to ignore these modes to leading order.

In our case we start with $D>4$-dimensional gauge theory in Minkowski
space with coordinates $y^M$, $M=0,1,\ldots,D-1$
having the conventional action\footnote{In this and following
sections, we will choose a normalization for the fields where the
coupling constant appears outside the action. So the field strength,
for instance, has no factor of $g$ in front of the commutator:
$F_{MN}=\partial_MA_N-\partial_NA_M+[A_M,A_N]$.}
\EQ{
S=\frac1{2g_D^2}\int d^Dy\,{\rm Tr}_N\,F_{MN}F^{MN}\ .
}
Here, $g_D$ is the gauge coupling in
$D$ dimensions which has dimensions of $[L]^{(D-4)/2}$.

We now consider how to embed an instanton in the $D$-dimensional
theory. To this end, we decompose $D$-dimensional Minkowski space into
$D-4$-dimensional Minkowski space and four-dimensional Euclidean
space, by taking $y^M=(\xi^a,x_m)$, $a=0,\ldots,p$, where $p=D-5$,
and $m=1,2,3,4$.
We now embed the instanton solution in the subspace $x_m$ by taking
\EQ{
A_M(y)=(0,A_m(x;X))\ ,
}
where $A_m(x;X)$ is the instanton solution constructed in
\S\ref{sec:S2} (with coupling $g$ set to 1). Recall that
$X^{\mu}$ denote the collective coordinates of $\ms_k$.
This ansatz obviously satisfies the equations-of-motion of the
$D$-dimensional theory and represents an object extended in
$p=D-5$-space dimensions, {\it i.e.\/}~a $p$-brane.

\subsubsection{The moduli space approximation}\elabel{sec:S51}

Manton's moduli space
approximation amounts to modelling the dynamics of the $p$-brane
by allowing dependence on the transverse coordinates $\xi^a$
(so including time $\xi^0$) to enter implicitly through the
collective coordinates. Of course, the
original classical solution with $\xi^a$-dependent collective
coordinates $A_N(0,x;X(\xi))$ does {\it not\/} satisfy the
classical equations-of-motion. The idea is that for sufficiently
slowly varying $X^{\mu}(\xi)$ it is {\it almost\/} a solution.
The point can be
illustrated with the position coordinates of the centre of the
instanton $X_n$. In this case we know the
exact solution for motion in time $t\equiv\xi^0$
with constant velocity because we can Lorentz-boost the static solution.
The boosted solution only approximates
$\AA_N(0,x;X_n=v_nt)$ if the velocity $|v|\ll 1$.

The moduli space approximation is an expansion in powers
of $\xi^a$ derivatives. In order to find the leading order effective
dynamics we have to
substitute the instanton solution with $\xi^a$-dependent collective
coordinates into the action. This then yields an
effective action for the collective coordinates now interpreted as
fields on the $p+1$-dimensional world-volume of the $p$-brane:
$X^\mu(\xi)$. In order that the effective action is at least
quadratic order in
the $\xi^a$ derivatives, we must ensure that the
equations-of-motion are satisfied to linear order in the
$\xi^a$ derivatives. In components $(A^a,A_n)$, the equations-of-motion,
${\cal D}^NF_{MN}=0$, are
\AL{
&{\cal D}_mF_{mn}+{\cal D}^a(\partial_aA_n-{\cal D}_nA_a)=0\
,\elabel{uiia}\\
&{\cal D}_n({\cal D}_nA_a-\partial_aA_n)+{\cal D}^bF_{ab}=0\ .
\elabel{uiib}
}
Now we substitute in the $\xi$-dependent instanton solution
$A_n=A_n(x;X(\xi))$. To linear order in $\xi^a$ we can ignore the
second term in \eqref{uiia} and the second term in \eqref{uiib}; this leaves
\EQ{
\big(\PD{A_n}{X^\mu}\partial_aX^\mu-{\cal D}_nA_a\big)=0\ ,
\elabel{hqp}
}
for $A_a$. Notice that this equation is very similar to the background
gauge condition for an instanton zero mode \eqref{edr}. Indeed, if we
take\footnote{The following discussion and its supersymmetric
generalization is motivated by the treatment of the analogous monopole
problem in Refs.~\cite{Gauntlett:1994sh,Gauntlett:2001ks}.}
\EQ{
A_a=\Omega_\mu\partial_a X^\mu\ ,
}
where $\Omega_\mu$ is the compensating gauge transformation
associated to the collective coordinate $X^\mu$, then
\eqref{hqp} is satisfied by virtue of \eqref{edr}. Since $A_a$ is linear
in $\partial_aX^\mu$, it is then trivial to see that the ansatz
\EQ{
A_N=\Big(\Omega_\mu\partial_aX^\mu(\xi),A_n(x;X(\xi))\Big)\ ,
\elabel{tyx}
}
satisfies the equations-of-motion \eqref{uiia}-\eqref{uiib}
to linear order in $\xi^a$-derivatives. In
making this ansatz, we have fixed the gauge symmetry of
the theory in a rather unconventional way. We are working in the
gauge where $A_0=\Omega_\mu\partial_0X^\mu$.
Of course we could always return to a more familiar gauge, say $A_0=0$,
by performing a gauge transformation on the ansatz.

Since the $\xi^a$-dependent expression \eqref{tyx} is only an
solution of the equations-of-motion to linear order in
$\xi^a$ derivatives, it will now contribute non-trivially to the
action at quadratic order:
\SP{
S^{(2)}&=\frac1{g_D^2}\int
d^{p+1}\xi\,d^4x\,\trN\,\delta_{\mu}A_n(x;X(\xi))
\partial^aX^{\mu}(\xi)\,\delta_{\nu}A_n(x;X(\xi))\partial_a X^{\nu}(\xi)\\
&=-\frac{1}{2g_D^2}\int d^{p+1}\xi\,
g_{\mu\nu}(X)\partial^aX^{\mu}\partial_aX^{\nu}\ ,
\elabel{sml}
}
where $g_{\mu\nu}(X)$ is the metric tensor on the instanton moduli
space defined in \eqref{ccmet}.

The effective collective coordinate dynamics embodied in \eqref{sml}
has the form of a $\sigma$-model in $p+1$-dimensions whose target space
is the moduli space $\ms_k$. The classical
collective coordinate dynamics follows from the equation-of-motion
\EQ{
\partial^a\partial_a
X^{\mu}+\tfrac12g^{\mu\nu}\PD{g_{\rho\sigma}}{X^{\nu}}\partial^a
X^{\rho}\partial_aX^{\sigma}=0\ .
}
For the case with $D=5$, where the instanton is a lump in five
dimensions, this is simply the equation
for geodesics in $\ms_k$.

\subsection{Supersymmetric instanton branes}\elabel{sec:S52}

In this section, we will investigate
how the description of instanton-branes extends to the supersymmetric
theories. Necessarily the theories must have extended supersymmetry
from the four-dimensional perspective, because the $\N=1$ theory
cannot be obtained by dimensional reduction from higher dimensions.
Since theories with the same number of supercharges
are related by dimensional reduction, we loose no loss of generality
by considering the theories in their maximal dimension: $D=10$ for
the theory having $16$ supersymmetries ($\N=4$ in $D=4$)
and $D=6$ for the theory having $8$ supersymmetries ($\N=2$ in $D=4$).
In the last section we have shown how the moduli space approximation
involves a $p+1=D-4$-dimensional
$\sigma$-model with $\ms_k$ as target. In the supersymmetric
case, we expect the moduli space approximation to lead to a
supersymmetric $\sigma$-model with the appropriate number of
supersymmetries. Since an instanton breaks half the supersymmetries of
the parent theory (see \S\ref{sec:S30}), we expect this $\sigma$-model to
manifest half the supersymmetries; {\it i.e.\/}~have 8 supercharges,
in the $p=5$ case,
and 4 supercharges in the $p=1$ case. Furthermore, the supersymmetry
in both cases is chiral, being $\N=(0,1)$ in six spacetime dimensions and
$\N=(0,4)$ in two spacetime dimensions, respectively.
It is a well-established fact that $\sigma$-models with the relevant
kinds of supersymmetry only exist in these dimensions if the target
space is hyper-k\"ahler \cite{Alvarez-Gaume:1981hm,Howe:1987qv,Howe:1988cj}.
In some respects, the very fact that
instantons can be embedded as branes in the higher-dimensional
supersymmetric gauge theories ``proves'' that the instanton moduli
space $\ms_k$ has to be hyper-K\"ahler.

What is interesting about these $\sigma$-models is that on Wick
rotation and complete dimensional reduction to zero dimensions
their partition functions reproduce
the leading-order semi-classical expression for the
collective coordinate integral of the instanton calculus. This includes both
the volume form on the instanton moduli space as well as the
non-trivial instanton effective action. In fact the $\sigma$-models
have both a non-linear realization, where the target space is directly
$\ms_k$, but also, importantly, a {\it linear\/} realization. In this
second formulation there is an auxiliary supermultiplet involving a
non-dynamical $\U(k)$ gauge field. It turns out that this linear formulation
is directly related to the ADHM construction, where the bosonic and fermionic
ADHM constraints arise from integrating out Lagrange multiplier
fields. However, there is more. When one dimensionally
reduces these $\sigma$-models there are certain kinds of
deformation which can be added in form of very special potentials
\cite{Bak:2000vd,Alvarez-Gaume:1983ab,Gauntlett:2001bd}. These
potentials naturally arise
in the instanton calculus when one moves onto the Coulomb branch of the
original theory and correspond to the adjoint-valued VEVs described in
\S\ref{sec:S38}. In the 2-dimensional $\sigma$-model
example, we also have the freedom to add
other fields to the $\sigma$-model
which correspond in the $\N=2$ instanton calculus to the Grassmann collective
coordinates of hypermultiplet matter fields.
We will subsequently show how these $\sigma$-models arise very
naturally in the context of D-branes in string theory.

\subsubsection{Action, supersymmetry and equations-of-motion}\elabel{sec:S53}

The actions of the $D=6$ and $D=10$ theories with 8 and 16
supercharges, respectively, can both be written as
\EQ{
S=\frac1{g^2_D}\int d^Dy\,\TrN\,\bigg\{\tfrac12F_{MN}F^{MN}-i\bar\Psi
\Gamma^M{\cal D}_M\Psi\bigg\}\ .
\elabel{hdgt}
}
The theory is invariant under the supersymmetry transformations
\AL{
\delta A_N&=-\bar\Xi\,\Gamma_N\Psi\ ,\elabel{sstoa}\\
\delta\Psi&=i\Gamma^{MN}\Xi\, F_{MN}\ .\elabel{sstob}
}
As in the purely bosonic case,
$y^M=(\xi^a,x_m)$, where $\xi^a$, $a=0,1,\ldots,p$,
are $p+1=D-4$-dimensional Minkowski space coordinates and $x_m$,
$m=1,2,3,4$, are four-dimensional Euclidean coordinates.

The spectrum of fields in both case consists of a gauge field
$A_N=(A_a,A_n)$, with $a=0,1$ in $D=6$ and $a=0,1,\ldots,5$ in $D=10$,
and the minimal spinor $\Psi$. (Our conventions for spinors are
described in Appendix \ref{app:A1}.)
In $D=6$ this is a Weyl, while in $D=10$
it is a Majorana-Weyl, spinor.
In both cases, it is convenient to
represent the $\Gamma$-matrices using a tensor product notation that
reflects the subgroups of the Lorentz group, $\SO(4)\times\SO(1,1)$ and
$\SO(4)\times\SO(5,1)$, respectively:
\EQ{
\Gamma_N=\Big\{
\Gamma_a\otimes\gamma_5,\One\otimes\gamma_n\Big\}\ ,
\elabel{cliftd}
}
where $\gamma_n$ are the $\SO(4)$ $\gamma$-matrices, \eqref{fdc}, and
$\Gamma_a$ are the $\SO(1,1)$ and $\SO(5,1)$
$\Gamma$-matrices, in the two cases, respectively. In both cases we write
\EQ{
\Gamma_a=\MAT{0&\Sigma_a\\ \bar\Sigma_a&0}\ ,
}
where, for $\SO(1,1)$, $\Sigma_a=(1,1)$ and
$\bar\Sigma_a=(-1,1)$ and for $\SO(5,1)$ the $\Sigma$-matrices are
defined in \eqref{minksm}.

In both $D=6$ and 10, in the tensor product notation \eqref{cliftd},
a Weyl spinor can be written
\EQ{
\Psi=\MAT{1\\0}\otimes\MAT{1\\0}\lambda_\alpha+\MAT{0\\
1}\otimes\MAT{0\\1}\bar\lambda^\aD\ ,
\elabel{spc}
}
where the $\alpha$ and $\aD$ are four-dimensional Euclidean space
spinor indices. In the $D=6$ case, $\Psi$ is a pseudo-real spinor and
$\lambda_\aD$ and $\bar\lambda^\aD$ are
independent complex quantities. It is convenient in this case,
to introduce the notion of a two-component
symplectic real spinor (see Appendix \ref{app:A1}). In terms of
the component spinors, we define
$\lambda_\aD^A$ and $\bar\lambda^\aD_A$, $A=1,2$, via
\EQ{
\lambda^1_\alpha=\tfrac 1{\sqrt2}\epsilon_{\alpha\beta}(\lambda_\beta)^\dagger\
,\qquad \lambda^2_\alpha=\tfrac1{\sqrt2}\lambda_\alpha\ ,\qquad
\bar\lambda^\aD_1=\tfrac 1{\sqrt2}\bar\lambda^\aD\
,\qquad \bar\lambda^\aD_2=-\tfrac 1{\sqrt2}\epsilon^{\aD\bD}
(\bar\lambda^\bD)^\dagger\ .
\elabel{soo}
}
The reason for choosing these definitions will emerge shortly. To
accompany the two-component spinors we also define the $\Sigma$-matrices
\EQ{
\Sigma^{AB}_a=\Sigma_a\epsilon^{AB}\ ,\qquad
\bar\Sigma_{aAB}=\bar\Sigma_a\epsilon_{AB}\ .
}
By virtue of \eqref{soo}, the two-component spinors
satisfy the pseudo-reality conditions
\EQ{
(\lambda^A)^\dagger=\bar\Sigma^0_{AB}\lambda^B\ ,\qquad
(\bar\lambda_A)^\dagger=\Sigma^{0AB}\bar\lambda_B\ .
\elabel{rew2}
}

In the $D=10$ case, $\Psi$ is
subjected to a Majorana spinor condition, which means in
terms of the component spinors, $\lambda^A$ and
$\bar\lambda_A$, $A=1,2,3,4$, that
\EQ{
(\lambda^A)^\dagger=\bar\Sigma^0_{AB}\lambda^B\ ,\qquad
(\bar\lambda_A)^\dagger=\Sigma^{0AB}\bar\lambda_B\ .
\elabel{reww}
}
Notice that this is identical in notion to the reality condition for
the symplectic real spinor in $D=6$ case \eqref{rew2} (and explains our
choice of definition).

In terms of the spinors $\lambda$ and $\bar\lambda$, the action
\eqref{hdgt} is
\EQ{
S=
\frac1{g_D^2}\int d^Dy\,\TrN\,\bigg\{\tfrac12F_{MN}F^{MN}
+2{\cal D}_n\bar\lambda_{A}
\bar\sigma_n\lambda^A
-i\bar\lambda_{A}\Sigma^{aAB}{\cal D}_a\bar\lambda_B-i
\lambda^{A}\bar\Sigma_{AB}^a{\cal D}_a\lambda^B\bigg\}\ .
\elabel{actded}
}
Our definitions have been chosen so that for $\xi$-independent field
configurations, the Lagrangian density is identical to {\it minus\/}
that in \eqref{cpta} with the relation $A_a=i\phi_a$ and with the metric on the
$a$-indices changed from a Euclidean to Minkowski signature. The
subgroups of the Lorentz group,
$\SO(1,1)$ and $\SO(5,1)$, respectively, are identified, after the
signature change, with part of the $R$-symmetry group of the
four-dimensional theory, namely
$\SO(2)$ and $\SO(6)$, respectively. The other difference is the
reality condition on the spinors: compare \eqref{rew2} or \eqref{reww}
with the situation in the four-dimensional context where the
spinors $\lambda^A$ and $\bar\lambda_A$ are independently real.

The equations-of-motion which follow form \eqref{actded} are
\AL{
{\cal D}_mF_{mn}+{\cal D}^aF_{an}&=2
\bar\sigma_{n}\{\lambda^A,\bar\lambda_A\}\ ,\elabel{uuta}\\
\Dbarslash\lambda^A&=
-i\Sigma^{aAB}{\cal D}_a\bar\lambda_B\ ,\elabel{uutb}\\
\Dslash\bar\lambda_A&=
-i\bar\Sigma^a_{AB}{\cal D}_a\lambda^B\ ,\elabel{uutc}\\
{\cal D}_nF_{na}+{\cal D}^bF_{ab}
&=i\bar\Sigma_{aAB}\lambda^A\lambda^B
+i\Sigma_a^{AB}\bar\lambda_A\bar\lambda_B\ .\elabel{uutd}
}

\subsubsection{The moduli space approximation}\elabel{sec:S54}

The case of pure gauge theory the effective moduli space dynamics
involved an expansion in $\xi_a$ derivatives. When there are
fermion fields are involved, we also allow the associated Grassmann
collective coordinates to depend on $\xi^a$. The moduli space
approximation is then an
expansion in $n=n_{\partial}+\tfrac12 n_f$: the
number of $\xi^a$ derivative plus half the number of Grassmann
collective coordinates. The lowest non-trivial term in the effective
action are those terms of order $n=2$ and in order to derive them we
must solve the equations-of-motions up to order
$n=1$.\footnote{Note that the possible
cross-terms between fields of lower order, $n=0$ and $n=1/2$, and
fields of higher order, $n=3/2$ and $n=2$, which potentially could
contribute to the effective action at orders $\leq2$, actually vanish
by the equations-of-motion.}

The equations to order $n=0$ are solved, as in \S\ref{sec:S50},
by embedding the instanton solution as
$A_N=(0,A_n(x;X(\xi)))$. At the next order, $n=\tfrac12$, the fermions
$\lambda^A$ satisfy the covariant Weyl equation \eqref{jsa} with the
solution $\lambda^A=\Lambda(\CM^A)$, where $\CM^A$ are the Grassmann
collective coordinates. In order to extend the
notion of the moduli space approximation to the supersymmetric theories
we should also
allow the Grassmann collective coordinates $\CM^A$ to depend on $\xi^a$.

At order $n=1$, we have the following equation (generalizing \eqref{hqp})):
\EQ{
{\cal D}_n\Big(\PD{A_n}{X^\mu}\partial_aX^\mu-
{\cal D}_nA_a\Big)=-i\bar\Sigma_{aAB}\lambda^A\lambda^B\ .
}
As in the case of pure gauge theory, we can use the background gauge
condition on the zero modes of $A_n$ to solve these equations for
$A_a$. The new ingredient is the presence of the source bi-linear in
fermion zero modes on the right-hand side. However, as
the equation is linear in $A_a$ we can take the linear combination
\EQ{
A_a(x;X(\xi),\CM^A(\xi))
=\Omega_\mu(x;X(\xi))\partial_a X^\mu(\xi)+i\phi_a(x;X(\xi),\CM^A(\xi))\ ,
\elabel{valaa}
}
where the Hermitian field $\phi_a$ satisfies
\EQ{
{\cal D}^2
\phi_a=\bar\Sigma_{aAB}\lambda^A\lambda^B\ .
}
Fortunately, we have already solved the covariant
Laplace equation in the ADHM background
a bi-fermion source (see \S\ref{sec:S31} and Appendix \ref{app:A4}
Eq.~\eqref{uyt}).

\subsubsection{The effective action}\elabel{sec:S55}

We now substitute our solution into the action of the
theory and extract the terms of order $n=2$. At this order,
the terms which contribute to the effective action are
\EQ{
S^{(2)}=
\frac1{g_D^2}\int d^Dy\,\TrN\,\Big\{F_{na}F^{na}
-i\lambda^{A}\bar\Sigma_{AB}^a{\cal D}_a
\lambda^{B}\Big\}\ .
\elabel{wee}
}
We now evaluate this expression. First of all, we have
\EQ{
F_{na}={\cal D}_nA_a-\partial_a A_n
=i{\cal D}_n\phi_a-\delta_\mu A_n\partial_a X^\mu
\elabel{zbb}
}
and, writing $\lambda^B=\Lambda(\CM^B)$,
\SP{
\lambda^{A}\bar\Sigma_{AB}^a{\cal D}_a
\lambda^{B}&=\lambda^{\alpha A}\bar\Sigma_{AB}^a
\Big\{\Lambda_\alpha(\partial_a\CM^B)+
\PD{\lambda_\alpha^B}{X^\mu}\partial_aX^\mu-
\Lambda_\alpha\big(\PD{\CM^B}{X^\mu}\big)\partial_aX^\mu
+[A_a,\lambda^B]\Big\}\\
&=\lambda^{\alpha A}\bar\Sigma_{AB}^a
\Big\{\Lambda_\alpha(\partial_a\CM^B)+
\Big(\PD{\lambda_\alpha^B}{X^\mu}+[\Omega_\mu,\lambda_\alpha^B]
-\Lambda_\alpha\big(\PD{\CM^B}{X^\mu}\big)\Big)\partial_aX^\mu
+i[\phi_a,\lambda^B_\alpha]\Big\}\ .
\elabel{zaa}
}
Consequently, using the gauge condition \eqref{gchoice},
the action \eqref{wee} becomes
\SP{
S^{(2)}&=
\frac1{g_D^2}\int d^Dy\,\TrN
\,\bigg\{\delta_\mu A_n\partial^aX^\mu\delta_\nu A_n\partial_aX^\nu
-i\lambda^{A}\bar\Sigma_{AB}^a\Lambda(\partial_a\CM^B)\\
&-i\lambda^{A}\bar\Sigma_{AB}^a
\Big(\PD{\lambda^B}{X^\mu}+[\Omega_\mu,\lambda^B]
-\Lambda\big(\PD{\CM^B}{X^\mu}\big)
\Big)\partial_aX^\mu
-{\cal D}_n\phi^a{\cal D}_n\phi_a+\lambda^{A}\bar\Sigma_{AB}^a
[\phi_a,\lambda^B]\bigg\}\ .
\elabel{wef}
}
The first two terms can be evaluated using the inner-product formulae
\eqref{ccmet} and \eqref{corriganf}. The third term can be evaluated
using the identity \eqref{toproo} established in Appendix~\ref{app:A4}:
\EQ{
\PD{\lambda^A}{X^\mu}+[\Omega_\mu,\lambda^A]=\Dslash\bar\varrho^A_\mu
+\Lambda\big(\PD{\CM^A}{X^\mu}\big) \ ,
}
where
\EQ{
\bar\varrho^{\aD A}_\mu=\tfrac14\bar U\PD{a^\aD}{X^\mu}f\bar{\cal M}^A
U\ .
\elabel{vxxk}
}
Therefore the third term in \eqref{wef} involves the integral
\EQ{
\int d^4x\,\trN\,\lambda^{A}\Dslash
\bar\varrho^{B}_\mu\ .
\elabel{sillj}
}
We can now use the fact that $\Dbarslash\lambda^A=0$ to
write this as
\EQ{
\int d^4x\,\partial_{\alpha\aD}\big(\trN\,
\lambda^{\alpha A}\bar\varrho^{\aD B}\big)\ .
}
Using the asymptotic formulae
in \S\ref{sec:S12}, one sees that the surface term at infinity
vanishes and so there is no contribution from the third term in
\eqref{wef}. The final two
terms \eqref{wef} can be evaluated following
an identical analysis to the construction of the instanton effective action in
\S\ref{sec:S38}. Notice that this
term vanishes in the $p=1$ case.\footnote{Note
the fact that we are in Minkowski
space does not affect the result since
$\big(\bar\Sigma^a_{AB}\bar\Sigma_{aCD}\big)_{\rm Mink}=
\big(\bar\Sigma_{aAB}\bar\Sigma_{aCD}\big)_{\rm Eucl}=2\epsilon_{ABCD}$.}

Putting everything together we have
\SP{
S^{(2)}=\frac{1}{g_D^2}\int d^{p+1}\xi\,&
{\rm tr}_k\bigg\{-\tfrac1{2}g_{\mu\nu}(X)\partial^aX^\mu\partial_aX^\nu
-2i\pi^2\bar\Sigma_{AB}^a\bar\mu^A\partial_a\mu^B
-i\pi^2\bar\Sigma_{AB}^a\CM^{\prime A}\partial_a\CM^{\prime B}\\
&\qquad\qquad+\tfrac12\pi^2\epsilon_{ABCD}\bar\CM^A
\CM^B\BL^{-1}\bar\CM^C\CM^D\bigg\}\ .
\elabel{jsl}
}
This expression is somewhat schizophrenic because the bosonic fields $X^\mu$
are the intrinsic coordinates while the fermionic fields $\CM^A$ are
the ADHM variables, subject to the fermionic ADHM constraints.
In order to unify things we can procedure in two alternative
ways, either writing everything in terms of intrinsic variables to
$\ms_k$ or in
terms of the ADHM variables with the ADHM constraints explicitly
imposed. Both viewpoints are worth developing.

First the intrinsic expression. In \S\ref{sec:S29}, we introduced
$\psi^{\xii A}$, $i=1,\ldots,2kN$, the intrinsic Grassmann-valued symplectic
tangent vectors to $\ms_k$. In \eqref{jsl}, the terms quadratic in the fermions
are written in terms of the intrinsic objects $\psi^A$ as
\EQ{
i\pi^2\bar\Sigma_{AB}^a\big(2\bar\mu^A\partial_a\mu^B
+\CM^{\prime A}\partial_a\CM^{\prime B}\big)=
4\bar\Sigma^a_{AB}\tilde\Omega\Big(\CM(\psi^A,X),
\CM(\partial_a\psi^B,X)+\PD{\CM^B}{X^\mu}\partial_aX^\mu\Big)
\elabel{resab}
}
using the linearity of $\CM^A$ on $\psi^A$. The first term can be
written in terms of the intrinsic symplectic matrix $\Omega_{ij}(X)$
defined in \eqref{ipint}. The relevant geometrical quantity
corresponding to the second term is
the symplectic spin connection on $\ms_k$. In Appendix \ref{app:A2}
we explain in the context of the hyper-K\"ahler quotient construction
how the connection $\nabla$ on the quotient space is inherited from
the connection $\tilde\nabla$ on the mother space $\tilde\ms$ by
orthogonal projection to $\EuScript H$. In a completely analogous way
the spin connection for symplectic tangent vectors on $\ms_k$ is
obtained from that of $\tilde\ms$ by projection via the fermionic ADHM
constraints. So since $\CM^A$ are subject to the fermionic ADHM constraints
\EQ{
\tilde\Omega(\CM^A,\nabla_X\CM^B)=\tilde\Omega(\CM^A,\tilde\nabla_X\CM^B)\ .
}
Then since $\tilde\ms$ is flat, we can identify the second term in
\eqref{resab} with
\EQ{
4\bar\Sigma^a_{AB}\Omega_{\xii\xjj}(X)\psi^{\xii A}
\,\omega_\mu{}^\xjj{}_\xkk\partial_aX^\mu\psi^{\xkk B}\ ,
}
where $\omega_\mu{}^\xii{}_\xjj$ are the elements of the symplectic spin
connection.

The final term, as we have explained in \S\ref{sec:S38}, involves a
coupling to the symplectic curvature of $\ms_k$. Writing all the terms
using in the intrinsic variables $X^\mu$ and $\psi^{\xii A}$, we have
\SP{
S^{(2)}=-\frac{1}{2g_D^2}\int d^{p+1}\xi\,&\Big\{
g_{\mu\nu}(X)\partial^aX^\mu\partial_a X^\mu
+\tfrac i2\bar\Sigma_{aAB}\Omega_{\xii\xjj}(X)
\psi^{\xii A}
(\partial_a\delta^\xjj{}_\xkk
+\omega_\mu{}^{\xjj}{}_{\xkk}\partial_a
X^\mu)\psi^{\xkk B}\\
&\qquad\qquad+\tfrac1{48}R_{\xii\xjj\xkk\xll}(X)\epsilon_{ABCD}\psi^{\xii
A}\psi^{\xjj B}\psi^{\xkk C}\psi^{\xll D}\Big\}
\elabel{wer}
}
In both cases $p=1,5$, this is the action of a non-linear $\sigma$-model
with $\ms_k$ as target with, as we shall show in \S\ref{sec:S101},
$\N=(0,4)$ or $\N=(0,1)$
supersymmetry, respectively. For references on
two-dimensional $\N=(0,4)$ theories
see \cite{Howe:1987qv,Howe:1988cj}, while for
six-dimensional $\N=(0,1)$
theories see \cite{Howe:1983fr,Sierra:1983cc} and references therein.

We now follow the opposite philosophy and attempt to write the
effective action \eqref{jsl} entirely in terms of the ADHM variables
$a_\aD$ and $\CM^A$.
In this form, the bosonic, as well as the fermionic,
ADHM constraints are now implicit.
We claim that the correct expression is
\SP{
S^{(2)}&=\frac{\pi^2}{g_D^2}\int d^{p+1}\xi\,
{\rm tr}_k\Big\{-4\partial^a\bar
w^\aD\partial_aw_\aD-4\partial^aa'_n\partial_aa'_n
-2i\bar\Sigma_{AB}^a\bar\mu^A\partial_a\mu^B
-i\bar\Sigma_{AB}^a\CM^{\prime\alpha A}\partial_a\CM^{\prime B}_\alpha\\
&+\tfrac14\big[\bar\Sigma^a_{AB}\bar\CM^A
\CM^B+2i(\bar a^\aD\partial^a a_\aD-\partial^a\bar a^\aD a_\aD)\big]
\BL^{-1}\big[\bar\Sigma_{aCD}\bar\CM^C\CM^D+2i(\bar a^\bD\partial_a
a_\bD-\partial_a\bar a^\bD a_\bD)\big]\Big\}\ .
\elabel{jslm}
}
We must show that this reduces to \eqref{jsl} when we substitute
$a_\aD=a_\aD(X)$. Firstly, we have to recall the expression
\eqref{htyh} for the metric on the quotient space in terms of the
ADHM variables $a_\aD(X)$. This accounts for the first two terms in
\eqref{jslm}. Most of the fermionic terms in \eqref{jslm} are already present
in \eqref{jsl} in particular the fermion quadrilinear term is equal to
that in \eqref{jsl} by the $\Sigma$-matrix identity \eqref{smid}.
We seem to have the additional terms dependent on
$\bar a^\aD\partial^a a_\aD-\partial^a\bar a^\aD a_\aD$, however this vanishes
when we substitute $a_\aD(X^\mu)$ since
\EQ{
\bar a^\aD\,\partial_a a_\aD-\partial_a\bar a^\aD\, a_\aD=\Big(
\bar a^\aD\PD{a_\aD}{X^\mu}-\PD{\bar
a^\aD}{X^\mu}a_\aD\Big)\partial_aX^\mu=0\ ,
}
by virtue of the constraint \eqref{extcon}.

The form \eqref{jslm} is motivated by the fact that one can introduce
a non-dynamical $\U(k)$ (Hermitian) gauge field $\chi_a$ coupled to
the fields via the covariant derivatives
\SP{
&{\cal D}_aw_\aD=\partial_aw_\aD-iw_\aD\chi_a\ ,\qquad
{\cal D}_a\bar w^\aD=\partial_a\bar w^\aD+i\chi_a\bar w^\aD\ ,\qquad
{\cal D}_aa'_n=\partial_aa'_n+i[\chi_a,a'_n]\ ,\\
&{\cal D}_a\mu^A=\partial_a\mu^A-i\mu^A\chi_a\ ,\qquad
{\cal D}_a\bar\mu^A=\partial_a\bar\mu^A+i\chi_a
\bar\mu^A\ ,\qquad
{\cal D}_a\CM^{\prime A}=\partial_a\CM^{\prime
A}+i[\chi_a,\CM^{\prime A}]\ .
}
The action \eqref{jslm} can then be written in the form of a gauge linear
$\sigma$-model
\EQ{
S^{(2)}=-\frac{4\pi^2}{g_D^2}\int d^{p+1}\xi\,
{\rm tr}_k\Big\{{\cal D}^a\bar
w^\aD{\cal D}_aw_\aD+{\cal D}^aa'_n{\cal D}_aa'_n
+\tfrac i2\bar\Sigma_{AB}^a\bar\mu^A{\cal D}_a\mu^B
+\tfrac i4\bar\Sigma_{AB}^a\CM^{\prime A}{\cal D}_a\CM^{\prime
B}\Big\}\ .
\elabel{jslmm}
}
Remember that the bosonic and fermionic ADHM constraints are imposed
implicitly. This can be achieved directly
by introducing, as in \S\ref{sec:N1},
bosonic and fermionic Lagrange multipliers in
the form of $k\times k$ Hermitian matrix fields,
$\vec D$ and $\bar\psi^\aD_A$, respectively:
\EQ{
S_{\text{L.m.}}=\frac{4\pi^2i}{g_D^2}\int d^{p+1}\xi\,{\rm tr}_k\Big\{
\vec D\cdot\vec
\tau^{\aD}{}_\bD\bar a^\bD a_\aD+\bar\psi^\aD_A(\bar\CM^Aa_\aD+\bar
a_\aD\CM^A)\Big\}\ .
\label{lagmult}
}
The non-dynamical gauge field, along with the Lagrange
multipliers then form a vector multiplet
$\{\chi_a,\vec D,\bar\psi_A\}$ of the supersymmetry appropriate to the $p=1$
and $p=5$ cases.

\subsubsection{Supersymmetry}\label{sec:S101}

Just as the supersymmetry of the parent theory is inherited by the
collective coordinates of instanton, we expect the same to be true for
the instanton branes. The only difference now is that the collective
coordinates are now fields and so the supersymmetry transformations
will involve $\xi^a$ derivatives.

{}From \eqref{sstoa}-\eqref{sstob}, one finds
\AL{
\delta
A_m&=i\xi^A\sigma_m\bar\lambda_A+i\bar\xi_A\bar\sigma_m\lambda^A\
,\elabel{huffa}\\
\delta\lambda^A&=i\Sigma^{ab}{}^A{}_B\xi^BF_{ab}+i\sigma_{mn}\xi^AF_{mn}+
\Sigma^{aAB}\sigma_m\bar\xi_BF_{am}\ .\elabel{huffb}
}
These transformations are closely related to \eqref{lha} and
\eqref{lhb}, respectively. (Indeed, by removing $\xi^a$ derivatives
and replacing $A_a$ by $i\phi_a$ they are identical.) Now we let
$A_m$ and $\lambda^A$ take their ADHM values with $\xi^a$-dependent
collective coordinates and $A_a$ as in \eqref{valaa}. The variations
lift to variations of the collective coordinates in an almost identical way
to those derived in \S\ref{sec:S32}. The only difference, is an extra
contribution from the final term in the variation of
$\lambda^A$. Using \eqref{valaa}, this term is
\EQ{
-i\Sigma^{aAB}\Dslash\phi_a\bar\xi_B+\Sigma^{aAB}
\big(\partial_aA_n-\Omega_\mu\partial_aX^\mu\big)
\sigma_n\bar\xi^A\ .
}
The first term, here, is identical to the final term in \eqref{lhb}
and so is already accounted by the analysis in \S\ref{sec:S32}. It
is the second term which is new. Since
\EQ{
\partial_aA_n-\Omega_\mu\partial_aX^\mu=\PD{A_n}{X^\mu}\partial_aX^\mu
-\Omega_\mu\partial_aX^\mu\equiv\delta_\mu A_n
}
we can use the explicit expression for the zero modes \eqref{zmq} to
write the final term as
\EQ{
2\Sigma^{aAB}\bar\xi^\aD_B\Lambda\Big(\PD{a_\aD}{
X^\mu}\Big)\partial_aX^\mu\ .
}
So there is an extra $\xi^a$-derivative term, relative to \eqref{vivi},
in the transformation of the Grassmann collective coordinates:
\SP{
\delta\CM^A&=-4i\xi^A_\alpha b^\alpha+
2i\Sigma^{aAB}{\cal
C}_{a\aD}\bar\xi_{B}^\aD+2\Sigma^{aAB}\partial_aa_\aD\bar\xi_B^\aD
\ ,\\
\delta\bar\CM^A&=-4i\xi^{\alpha A}\bar b_\alpha+
2i\Sigma^{aAB}\bar\xi_{\aD B}\bar{\cal C}_a^\aD+2\Sigma^{aAB}
\partial_a\bar a^\aD\bar\xi_{\aD B}\ .
\elabel{vivim}
}
These transformations are symmetries of the effective action
\eqref{jsl}. In counting the number of supersymmetries, we do not
include the trivial shifts of the fermions generated by $\xi^A$ and this
means that the effective theory of the instanton branes has half the
number of supersymmetries of the parent theory: so 8 for the
six-dimensional ($p=5$)
theory and 4 for the two-dimensional ($p=1$) theory.
In both cases the supersymmetries
are all anti-chiral the hence, using the usual convention, are denoted as
$\N=(0,1)$ and $\N=(0,4)$, for $p=5$ and $p=1$, respectively.

\subsubsection{Relation to the instanton calculus}\elabel{sec:S93}

The instanton calculus for $\N=2$ and $\N=4$ supersymmetry can be
obtained as a particular limit of the $\sigma$-model described above.
This limit involves performing a Wick rotation of the $p+1$-dimensional
world-volume of the brane and then a dimensional reduction to zero
dimensions. In this limit, the field theory reduces to a matrix theory
and the only part of the ``dynamics'' that remains, in the case with
$p=5$ (8
supercharges) is the four-fermion
coupling in \eqref{wer}. This four-fermion coupling then reproduces the
quadrilinear coupling of the Grassmann collective coordinates
described in \S\ref{sec:S38}. On top of this,
the collective coordinate integral of the instanton calculus is
obtained directly from the Wick rotated partition function of the dimensionally
reduced $\sigma$-model. This is most easily seen
in the gauged linear $\sigma$-model description of Eq.~\eqref{jslmm}.
On dimensional reduction the partition function of the $\sigma$ simply
gives the instanton partition in the linearized version constructed in
\S\ref{sec:N1}. One can see that the auxiliary variables
$\{\chi_a,\vec D,\bar\psi_A\}$ introduced in \S\ref{sec:N1} arise from
the dimensional reduction of
the fields of the vector multiplet of the $\sigma$-model.

It is interesting to ask whether the constrained instanton formalism
describing the Coulomb branch of the $\N=2$ and $\N=4$ theories can be
re-produced in this way. To answer this question, we use the fact that
for our six-dimensional
$\sigma$-model with 8 supercharges, relevant to the $\N=4$
case, it is possible when dimensionally reducing to add a
potential which do not break supersymmetry \cite{Alvarez-Gaume:1983ab}.
This potential term has the form of the inner product of a
tri-holomorphic Killing vector field on the hyper-K\"ahler target space.
In fact the most general type of potential for an 8 supercharge
$\sigma$-model in $6-l$-dimensions involves the sum of the
inner-products of $l$ {\it commuting\/} tri-holomorphic Killing vector
fields \cite{Bak:2000vd,Gauntlett:2001bd}.
At the level of the non-linear $\sigma$-model the introduction
of this potential can be described by non-trivial dimensional
reduction of the Scherk-Schwarz kind \cite{Scherk:1979zr}. 
At the level of the linear
$\sigma$-model we can describe it as gauging an additional $\U(1)^l$
symmetry, over and above $\U(k)$, corresponding to the
action of the tri-holomorphic Killing vector fields, and then
giving VEVs to the $l$ components of the six-dimensional gauge field
$\chi_a$ that lie in the dimensionally reduced directions.

In the present setting of the instanton calculus, $\ms_k$ admits a
$\SU(N)$ tri-holomorphic action corresponding to global gauge transformation
in the original $\SU(N)$ gauge theory. The action of these
transformations on the bosonic ADHM variables was described in \S\ref{sec:S12}.
Under global gauge transformations $a'_n$ and
$\CM^{\prime A}$ are invariant, but
\EQ{
w_\aD \to \grp w_\aD\ ,\qquad \mu^A\to \grp \mu^A\ .
}
Commuting actions can be obtained by considering global gauge
transformations in the Cartan subgroup $\U(1)^{N-1}\subset\SU(N)$ generated by
diagonal matrices. We now gauge this symmetry and reduce to zero
dimensions. This amounts to replacing the covariant derivatives in
\eqref{jslmm} by
\SP{
&{\cal D}_aw_\aD\to-iw_\aD\chi_a-i\phi^0_aw_\aD\ ,\qquad
{\cal D}_a\bar w^\aD\to i\chi_a\bar w^\aD+i\bar w^\aD\phi^0_a\ ,\qquad
{\cal D}_aa'_n\to-i[a'_n,\chi_a]\ ,\\
&{\cal D}_a\mu^A\to-i\mu^A\chi_a-i\phi^0_a\mu^A\ ,\qquad
{\cal D}_a\bar\mu^A\to i\chi_a
\bar\mu^A+i\bar\mu^A\phi_a^0\ ,\qquad
{\cal D}_a\CM^{\prime A}\to-i[\CM^{\prime A},\chi_a]\ .
\elabel{repcd}
}
where $\phi_a^0$ are six diagonal $N\times N$ matrices. It is easy to
see that after integrating out the gauge field $\chi_a$ one is left
with the instanton effective action describing constrained instantons
on the Coulomb branch of $\N=4$ gauge theory \eqref{withvevs}.

In a similar way, the constrained instanton calculus of the $\N=2$
theory can be obtained by gauging the same $\U(1)^{N-1}$ symmetry and
dimensionally reducing to zero dimensions
the two-dimensional $\sigma$-model with $\N=(0,4)$ supersymmetry. In
this case, there is another kind of generalization  which plays a
r\^ole in the instanton calculus. Since the $\N=(0,4)$ supersymmetry
is purely chiral it is possible to fermions of the opposite chirality
which are singlet under the supersymmetry. This is precisely what
is required to describe the instanton calculus
when the $\N=2$ gauge theory involves additional matter
hypermultiplets.
As described in \S\ref{sec:S91}, instantons have
additional fermion zero modes and there are new Grassmann collective
coordinates $\{\K_f,\tilde\K_f\}$. Consequently the
$\sigma$-model describing the collective coordinate dynamics will have
additional fermionic fields describing the fluctuations of the hypermultiplet
Grassmann collective coordinates. These are incorporated in
the two-dimensional $\sigma$-model dynamics in precisely the same way
as in description of
monopole dynamics in an $\N=2$ theory with hypermultiplets described
in \cite{Gauntlett:2001ks} (although in this reference the
$\sigma$-model is in one dimension, {\it i.e.\/},~is a quantum
mechanical system).

\subsection{Instantons and string theory}\elabel{sec:S100}

The most remarkable and unexpected developments of the instanton
calculus has come with the realization that the ADHM formalism
arises naturally in the context of string theory. The point is that
supersymmetric
instanton branes, as previously described in \S\ref{sec:S52},
arise when D-branes are ``absorbed'', in a way to be made
precise, on other D-branes. In certain respects,
the string theory point-of-view provides an ``explanation'' for the
rather mysterious ADHM construction in the sense that the ADHM
variables, constraints and the internal $\U(k)$ symmetry have a
simple interpretation in terms of a conventional (but auxiliary)
gauge theory with gauge group $\U(k)$ (see
Refs.~\cite{W1,D2,D1}). Moreover,
the connection is very explicit: not only can the ADHM gauge potential
be derived \cite{Witten:1995tz} but the $\sigma$-model describing the
dynamics of instanton branes is obtained in a direct way
\cite{MO3}. Both are obtained in a certain decoupling limit where
stringy effects can be neglected.
In this way one ``derives''
the leading-order semi-classical expression for the collective
coordinate integral of the original gauge theory. Some aspects of the
relation between Yang-Mills and D-instantons, particularly for $k=1$,
are also discussed in Refs.~\cite{Barbon:1998ak,Barbon:1999fx}.

\subsubsection{The $\N=4$ instanton calculus}\elabel{sec:S70}

We begin by describing how the $\N=4$ instanton calculus can be
recovered from string theory.
The basic idea is that the ${\cal N}=4$
theory with gauge group $\U(N)$ arises as the collective dynamics of
$N$ D3-branes in Type IIB string theory and,
according to \cite{D2,D2}, D-instantons located on
the D3-branes are equivalent to Yang-Mills instantons in the
collective coordinate
world-volume gauge theory. More generally, D$p$-branes located on a
stack of D$(p+4)$-branes appear as instanton $p$-branes in the
world-volume theory of the higher-dimensional branes.
So our attention will focus a general
system of $k$ D$p$-branes and $N$ D$(p+4)$-branes.

We begin by briefly reviewing some basic facts about D-branes in Type II
string theory \cite{P}. There are two distinct ways to think of a
D$p$-brane. Firstly, it can be viewed as a brane-like soliton of the Type
II supergravity in ten dimensions, the low energy limit of the string
theory. In this respect, we expect that in an appropriate limit its
behaviour should be captured by a generalization of Manton's
moduli space dynamics:
in this case some $(p+1)$-dimensional field theory on the world volume.
Secondly, a D$p$-brane can be described
as a $p$-dimensional hyper-plane on which
open strings can end. The connection between these two points-of-view
is that the massless states in the open string
spectrum in the presence of the D$p$-brane
give rise to massless fields which propagate on the $p+1$-dimensional
world volume of the D-brane and these massless modes are
identified with the collective coordinates of the moduli space
approximation.

Specifically, the massless modes or collective coordinates
of a single D$p$-brane come
from a simple dimensional reduction to $p+1$ dimensions of a
$\U(1)$ vector multiplet of ${\cal N}=1$ supersymmetry in
ten dimensions. This supersymmetric theory, which has 16 supercharges,
consequently describes the moduli space dynamics of the D-brane. The
action is \eqref{hdgt} with $D=10$ dimensionally reduced to $p+1$ dimensions.
After dimensional reduction, the ten-dimensional gauge field
$A_N$, $N=0,1,2,\ldots 9$, yields a $p+1$ dimensional gauge field $A_n$,
$n=0,\ldots,p$ and $9-p$ scalars $\phi_a$, $a=1,\ldots,9-p$. In the
collective coordinate interpretation,
the scalars specify the location of the D-brane in the $9-p$ dimensions
transverse to its world volume.  The $d=10$ multiplet also includes a
Majorana-Weyl fermion $\Psi$. The sixteen independent components of $\Psi$
correspond to the sixteen fermion zero modes of the D-brane. This number
reflects the fact that the D-brane is a BPS configuration which breaks
half of the 32 supersymmetries of the Type II theory.

Remarkably, the collective dynamics of a system of
$N$ parallel D$p$-branes is obtained by
simply ``non-abelianizing'' the $\U(1)$ gauge group thereby replacing it with
$\U(N)$. At low energies, the adjoint-valued scalar fields can acquire
VEVs breaking the gauge group to $\U(1)^N$ describing a configuration
of D$p$-branes separated in the transverse directions.
However,
when two or more D-branes coincide, additional states corresponding
to open strings stretched between the two branes become massless
leading to enhanced gauge symmetry
\cite{W135}. In the maximal case, where
all $N$ D$p$-branes coincide, the unbroken gauge group is $\U(N)$.
The low-energy effective action for the
world-volume theory can be obtained from
dimensional reduction of ten-dimensional super-Yang-Mills theory
\eqref{hdgt} (with $D=10$) with gauge group $\U(N)$.
Dimensional reduction to $p+1$ dimensions proceeds by setting all
spacetime derivatives in the reduced directions to zero.
As in the case of a single D$p$-brane, the ten-dimensional gauge field
yields a $p+1$-dimensional, but now $\U(N)$-valued,
gauge field and $9-p$ real adjoint scalars.
Configurations with some or all of the D-branes separated
in spacetime corresponds to the Coulomb branch of the world-volume
gauge theory. In terms of string theory parameters, the Yang-Mills coupling
constant in $p+1$ dimensions is identified as
\EQ{
g^{2}_{p+1}=2(2\pi)^{p-2}g_{\rm st}
\alpha^{\prime(p-3)/2}\ ,
\elabel{coupli}
}
where $g_{\rm st}$ is the string coupling
constant, $\alpha'$ is the inverse string tension.

In the case of
$N$ parallel D3-branes, the low-energy effective theory on
the brane world-volume is four-dimensional
${\cal N}=4$ supersymmetric Yang-Mills theory
with gauge group $\U(N)$ and coupling constant $g^2=4\pi g_{\rm st}$.
In this case we can also introduce a world-volume vacuum angle
$\theta=2\pi C^{(0)}$ where $C^{(0)}$ is the
VEV of the Ramond-Ramond scalar of the IIB theory.
In general, the four-dimensional
fields which propagate on the brane also have couplings
to the ten-dimensional graviton and the other bulk closed string modes.
To decouple the four-dimensional theory from these bulk modes
it is necessary to take
the limit $\alpha'\rightarrow 0$ with
with $g$ ({\it i.e.\/}~$g_{\rm st}$) held fixed and small.
This gives a weakly coupled gauge theory on the D3-brane.

A D-instanton (or D$(-1)$-brane) corresponds to the extreme case where
the dimensional reduction is complete and
the world-volume is a single point in Euclidean spacetime.\footnote{So
before dimensional reduction, we must Wick rotate the $D=10$ action
\eqref{hdgt} to Euclidean space.} Correspondingly, rather
than having finite mass or tension, a single D-instanton has finite action
\EQ{
2\pi(g_{\rm st}^{-1}-iC^{(0)})\equiv-2\pi i\tau\ ,
}
using the relations \eqref{coupli}. Hence, a D-instanton carries the
same action as a Yang-Mills instanton in the world volume gauge theory
of $N$ D3-branes described above.
{}From our discussion above, the collective coordinates
of a charge-$k$ D-instanton correspond to a $\U(k)$ vector multiplet of
ten-dimensional $\N=1$ supersymmetric gauge theory dimensionally
reduced to zero spacetime dimensions.
As we dimensionally reduce to zero dimensions,
the fields become both $c$-number
and Grassmann matrix degrees-of-freedom.
In addition to a constant part equal to $-2\pi ik\tau$,
the action of a charge-$k$ D-instanton also depends on the
collective coordinates via the (Wick rotated)
dimensional reduction of \eqref{hdgt} (with $D=10$):
\begin{equation}
S=-\frac1{2g_0^2}{\rm tr}_{k}\,[A_M,A_N]^{2}
+\frac i{g_0^2}{\rm tr}_k\,\bar{\Psi}\Gamma_M[A_M,\Psi]\ .
\elabel{zerodaction}
\end{equation}
In addition to the manifest $\SO(10)$ symmetry under ten-dimensional
rotations, the action is trivially invariant under translations of the
form $A_M\rightarrow A_M + x_M1_{\sst [k]\times[k]}$.
Hence $k^{-1}{\rm tr}_{k}\,A_M$, which corresponds to the abelian factor of
the $\U(k)$ gauge group, is identified with the position of the centre
of the charge $k$ D-instanton in ${\mathbb R}^{10}$.

The action \eqref{zerodaction} inherits the supersymmetries
\eqref{sstoa}-\eqref{sstob} along with linear shifts in the Grassmann
collective coordinates:
\begin{equation}\begin{split}
\delta A_M &= -\bar{\Xi}\Gamma_M \Psi\ ,\\
\delta \Psi&=  i\Gamma_{MN}\Xi[A_M,A_N]+
1_{\sst [k]\times [k]}\,\varepsilon\ .
\elabel{susy}
\end{split}\end{equation}
The sixteen components of the Majorana-Weyl
$\SO(10)$ spinor $\varepsilon$
correspond to the sixteen zero modes of the D-instanton configuration
generated by the action of the $D=10$ supercharges. Like the bosonic
translation modes, these modes live in
the abelian factor of the corresponding $\U(k)$ field,
$k^{-1}{\rm tr}_{k}\,\Psi$.
In contrast, the
Majorana-Weyl spinor $\Xi$ parameterizes the sixteen
supersymmetries left unbroken by the D-instanton.

In ordinary field theory, as we have seen,
instantons yield non-perturbative corrections
to correlation functions via their saddle-point contribution to the
Euclidean path integral. In the semi-classical limit, the
path-integral in each topological sector reduces to an ordinary integral
over the instanton moduli space \eqref{intm}. The extent to
which similar ideas apply to D-instantons is less clear, in part because
string theory lacks a second-quantized formalism analogous to the path
integral. Despite this, there is considerable evidence that
D-instanton contributions to string theory amplitudes also reduce
to integrals over collective coordinates at weak
string coupling \cite{GG1}. In this case the relevant collective coordinates
are the components the ten-dimensional $\U(k)$ gauge field
$A_M$ and their superpartners $\Psi$ which in the dimensional
reduction end up as matrices. According to Green and
Gutperle \cite{GG1,GG2,GG3,GG4}, the charge-$k$ D-instanton contributions
to the low-energy correlators of the IIB theory are consequently
governed by the partition function
\begin{equation}
{\EuScript Z}_{k} = \frac{e^{-2\pi k(e^{-\phi}+iC^{(0)})}}{{\rm Vol}\,\U(k)}\,
\int\,d^{10}A\,d^{16}\Psi\, e^{-S}\ .
\elabel{ukpfn}
\end{equation}
This partition function can be thought of as the collective coordinate
integration measure for $k$ D-instantons. In particular, the leading-order
semi-classical contribution of $k$ D-instantons to the
correlators of the low-energy supergravity fields is obtained by
inserting into \eqref{ukpfn} the classical value of each field.
The collective coordinate action \eqref{zerodaction} does not depend on the
$\U(1)$ components of the fields $A_M$ and $\Psi$.
Hence, to obtain a non-zero answer,
the inserted fields must, at the very least,
involve a product of at least sixteen fermions to
saturate the corresponding Grassmann integrations. As in field theory
instanton calculations, the resulting amplitudes can be interpreted in
terms of an instanton-induced vertex in the low-energy effective action.
The spacetime position of the D-instanton, $X_M=k^{-1}{\rm tr}_kA_N$,
is interpreted as the
location of the vertex. In particular, the work of Green and Gutperle
\cite{GG1} has
focused on a term of the form $R^{4}$ in the IIB effective action
(here $R$ is the ten-dimensional curvature tensor) and its supersymmetric
completion.

So far we have only considered D-instantons in the IIB theory in a flat
ten-dimensional background and in the absence of other branes. In order
to make contact with four-dimensional gauge theory
we need to understand how these
ideas apply to D-instantons in the presence of D3-branes. In particular
we wish to determine how the D-instanton collective coordinate
integral \eqref{ukpfn}
is modified by the introduction of $N$ parallel D3-branes. Conversely,
in the absence of D-instantons, the theory
on the four-dimensional world volume of the D3-branes is ${\cal N}=4$
Yang-Mills with gauge group $\U(N)$.
Hence a related question is how the D-instantons
appear from the point-of-view of the four-dimensional world-volume
theory of the D3
branes. In fact, the brane configuration considered here is a special
case of a system which has been studied intensively
involving $k$ D$p$-branes
in the presence of $N$ D$(p+4)$-branes, with all branes parallel.
As we will review below, the lower-dimensional branes
corresponds to Yang-Mills instantons in the world-volume gauge theory of the
higher-dimensional branes \cite{D2}.
We begin by reviewing the maximal case $p=5$, which was first
considered (in the context of Type I string theory) by Witten \cite{W1}.
The cases with $p<5$ then follow by straightforward dimensional
reduction.

We start by considering a theory of $k$ parallel D5-branes (in Type IIB
string theory) in isolation.
As above, the world-volume theory is obtained by dimensional reduction
of ten-dimensional ${\cal N}=1$  Yang-Mills theory with gauge group
$\U(k)$. The resulting theory in six dimensions has two Weyl supercharges
of opposite chirality and, therefore,
conventionally has ${\cal N}=(1,1)$
supersymmetry.\footnote{Some convenient facts about six-dimensional
supersymmetry
are reviewed in \cite{CHS}. (See page 67 in particular.)} After dimensional
reduction, the $\SO(10)$ Lorentz group of the Minkowski theory in ten
dimensions is broken to $H=\SO(5,1)\times\SO(4)$. The $\SO(5,1)$ factor is
the Lorentz group of the six-dimensional theory while the $\SO(4)$ is
an $R$-symmetry. The ten-dimensional gauge field $A_M$ splits up into
an adjoint scalar in the vector representation of $\SO(4)$ and
a six dimensional gauge field.
Explicitly we set
\EQ{
A_M=i\Big(\chi_a,\tfrac 1{2\pi\alpha'}a'_{n}\Big)\ ,\qquad
a=1,\ldots,6;\quad n=1,\ldots,4\ .
}
Since, in our conventions, $A_M$ is anti-Hermitian, both $a'_n$ and
$\chi_a$ are Hermitian. The factor of $(2\pi\alpha')^{-1}$ has been
inserted so that $a'_n$ will subsequently be
identified with the quantity of the same name in the instanton calculus.

In order to describe the fermion content of the theory consider the
covering group of $H$, $\bar{H}=\SU(4)\times\SU(2)_{L}\times\SU(2)_{R}$.
We introduce spinor indices $A=1,2,3,4$ and $\alpha,\aD=1,2$ corresponding to
each factor. As mentioned above, a ten-dimensional Majorana-Weyl spinor
is decomposed under $\bar H$ as
\begin{equation}
{\bf 16}\rightarrow
({\bf 4},{\bf 2},{\bf 1})\oplus (\bar{\bf{4}},{\bf 1},{\bf 2})\ ,
\elabel{repn}
\end{equation}
so contains two Weyl spinors of opposite chirality in six dimensions.
With the representation of the ten-dimensional
Clifford algebra as in \eqref{cliftd} a Majorana-Weyl fermion, as in
\eqref{spc}, can be written
\EQ{
\Psi=\frac1{4\pi\alpha'}\MAT{1\\0}\otimes\MAT{1\\0}\CM^{\prime
A}_\alpha
+\MAT{0\\
1}\otimes\MAT{0\\1}\bar\psi^\aD_A\ ,
\elabel{spc2}
}

Altogether, the fields
$(\chi_a,a'_n, {\cal M}^{\prime A}_\alpha, \bar\psi^\aD_A)$ form a vector
multiplet of ${\cal N}=(1,1)$ supersymmetry in six dimensions.
In terms of an ${\cal N}=(0,1)$ subalgebra, the ${\cal N}=(1,1)$
vector multiplet splits up into an ${\cal N}=(0,1)$ vector multiplet
containing $\{\chi_a,\bar\psi^\aD_A\}$
and an adjoint hypermultiplet containing $\{a'_n,
{\cal M}^{\prime A}_\alpha\}$. The action of the ${\cal N}=(1,1)$
theory is deduced from \eqref{actded}  (with $g_D\to g_6$):
\begin{equation}
S=\frac{1}{g_6^2}\Big(S_{\rm gauge}+\frac1{4\pi^2\alpha^{\prime2}}
S_{\rm matter}^{({\rm a})}\Big)
\equiv
\frac{4\pi^2}{g_{p+5}^2}\Big(4\pi^2
\alpha^{\prime2}S_{\rm gauge}+S_{\rm matter}^{({\rm a})}\Big)\ ,
\elabel{de6}
\end{equation}
(in the present case $p=5$) where
\begin{equation}
S_{\rm gauge}=\int\, d^{6}\xi \,{\rm tr}_{k}\Big\{\tfrac{1}{2}F^{2}_{ab}
-i\Sigma^{aAB}\bar\psi_{A}{\cal D}_a\bar\psi_B+
\tfrac{1}{2}D^{2}_{mn}\Big\}
\elabel{sgauge}
\end{equation}
and
\EQ{
S_{\rm matter}^{\rm (a)} =
\int\, d^{6}\xi \,{\rm tr}_{k}
\Big\{{\cal D}^aa'_n{\cal D}_aa'_n
-\tfrac i4\bar\Sigma^a_{AB}{\cal M}^{A}
{\cal D}_a{\cal M}^{\prime B}
-i[\CM^{\prime\alpha A},a'_{\alpha\aD}]\bar\psi^\aD_A+i
\vec D\cdot\vec\tau^{\aD}{}_\bD
\bar a^{\prime\bD\alpha}a'_{\alpha\aD}\Big\}\ .
\elabel{sa}
}
Here, the covariant derivatives are for adjoint-valued fields: ${\cal
D}_aa'_n=
\partial_aa'_n+i[\chi_a,a'_n]$, {\it etc\/}.
For later convenience
we have introduced a real anti-self dual auxiliary field for the vector
multiplet, $D_{mn}=-^*D_{mn})$, transforming in the adjoint
of $\SU(2)_{R}$. Since $D_{mn}$ is anti-self-dual we can write
\EQ{
D_{mn}={\rm tr}_2(\vec\tau\bar\sigma_{mn})\cdot\vec D\ ,
}
which defines the 3-vector $\vec D$.
We have also used the relation $g_{p+5}^2=(2\pi)^4\alpha^{\prime2}g_{p+1}^2$.

Following \cite{D2}, the next step is to introduce $N$ D9-branes of the
Type IIB theory whose world-volume fills the ten-dimensional
spacetime.\footnote{In fact a
IIB background with non-vanishing D9-brane charge
suffers from inconsistencies at the quantum level.
This is not relevant here because the D9-branes in
question are just a starting point for a classical dimensional
reduction.} These will give rise on dimensional reduction
to the $N$ D3-branes in the
$p=-1$ case on which we will eventually focus. The world volume
theory of the D9-branes in isolation ({\it i.e.\/}~in
the absence of the D5-branes)
is simply ten-dimensional $\U(N)$ $\N=1$
supersymmetric gauge theory. As explained
by Douglas \cite{D2}, the effective action for this system contains a
coupling between the 2-form field strength $F$ of the
world-volume gauge field and a
6-form field $C^{(6)}$
which comes from the Ramond-Ramond sector of the Type IIB theory.
This coupling has the form,
\begin{equation}
\int\, C^{(6)}\wedge F\wedge F,
\elabel{douglas}
\end{equation}
where the integration is over the ten-dimensional world-volume of the
D9-branes. The same six-form field $C^{(6)}$ also couples minimally
to the Ramond-Ramond charge carried by D5-branes.  Hence a configuration
of the $\U(N)$ gauge fields
with non-zero second Chern class, $F\wedge F$, acts
as a source for D5-brane charge. More concretely, if the D9-brane
gauge field is chosen to be independent of six of the world-volume
dimensions and an ordinary Yang-Mills instanton is embedded in the
remaining four dimensions, then the resulting configuration has the same
charge density as a single D5-brane. Both objects are also BPS saturated and
therefore they also have the same tension. Further confirmation of the
identification of D5-branes on a D9-brane as instantons was found in
\cite{D2} where the gauge-field background due to a Type I D5-brane
was shown to be self-dual via its coupling to the world-volume of a D1-brane
probe.

As described above, D5-branes appear as BPS saturated instanton
configurations on the D9-brane which break half
of the supersymmetries of the world-volume theory. Conversely,
the presence of D9-branes also breaks half
the supersymmetries of the D5-brane world-volume theory described by
the action \eqref{de6}. Specifically, the ${\cal N}=(1,1)$ supersymmetry
of the six-dimensional theory is broken down to the ${\cal N}=(0,1)$
subalgebra described above equation \eqref{de6}. To explain how this happens
we recall that open strings stretched between branes give rise to fields
which propagate on the D-brane world-volume. So far we have only included
the adjoint representation fields which arise from strings stretching between
pairs of D5-branes. As our configuration now includes both D5- and
D9-branes there is the additional possibility of states corresponding to
strings with one end on each of the two different types of brane.
As the D5-brane and D9-brane ends of the string carry $\U(k)$ and $\U(N)$
Chan-Paton indices respectively, the resulting states transform in the
$(\Bk,\BN)$ representation of $\U(k)\times\U(N)$.

In fact the additional degrees-of-freedom transform
as hypermultiplets of ${\cal N}=(0,1)$ supersymmetry
in six dimensions \cite{W1}.
As these hypermultiplets cannot be combined to form
multiplets of ${\cal N}=(1,1)$ supersymmetry, the residual supersymmetry
of the six dimensional theory is ${\cal N}=(0,1)$ as claimed above.
Each hypermultiplet contains two complex scalars
$w_{ui\dot{\alpha}}$, $\aD=1,2$. Here, as previously, $i$ and $u$
are fundamental representation
indices of $\U(k)$ and $\U(N)$ respectively. The fact that hypermultiplet
scalars transform as doublets of the $\SU(2)$ $R$-symmetry is familiar
from ${\cal N}=2$ theories in four dimensions. Each hypermultiplet also
contains a pair of complex Weyl spinors, $\mu_{ui}^{A}$ and
$\bar{\mu}_{iu}^{A}$. The six-dimensional action for the hypermultiplets
can be deduced from the action \eqref{sa} of the $\{a'_n,\CM^A\}$
hypermultiplet:
\EQ{
\frac{4\pi^2}{g_{p+5}^2}S_{\rm matter}^{\rm (f)} =
\frac{4\pi^2}{g_{p+5}^2}\int\, d^{6}\xi \,{\rm tr}_k\Big\{
-{\cal D}^a\bar{w}^\aD{\cal D}_{a}
w_\aD-\tfrac i2\bar\Sigma^a_{AB}\bar\mu^A{\cal D}_a\mu^B
-i\big(\bar\mu^Aw_\aD+\bar w_\aD\mu^A\big)\bar\psi^\aD_A
+i\vec D\cdot\vec\tau^{\aD}{}_\bD\bar w^\bD w_\aD\Big\}\ .
\elabel{sf}
}
The scalar and fermion kinetic terms in the above
action include the  fundamental representation
covariant derivative, ${\cal D}_{a}w=\partial_{a}w-iw\chi_a$, {\it etc\/}.
The remaining two terms in \eqref{sf} are the
fundamental representation versions
of the Yukawa coupling and D-terms which appear in \eqref{sa}. The complete
action of the six-dimensional theory is then the amalgam of \eqref{sgauge},
\eqref{sa} and \eqref{sf}:
\begin{equation}
S=\frac{4\pi^2}{g_{p+5}^2}\Big(4\pi^2
\alpha^{\prime2}S_{\rm gauge}+S_{\rm matter}^{({\rm a})}+S_{\rm
matter}^{({\rm f})}\Big)\ .
\elabel{de66}
\end{equation}
The various fields of the six dimensional theory and their transformation
properties under $\U(k)\times \U(N)\times \bar{H}$ are collected in
Table 10.1.

\begin{center}
\begin{tabular}{cccccc}
\hline
\Rowspace & $\U(k)$ &  $\U(N) $ & $ \SU(4) $ & $ \SU(2)_{L} $ &
$ \SU(2)_{R} $ \\
\Rowspace $\chi$ & {\bf adj} & ${\bf 1}$ & ${\bf 6}$ & ${\bf 1}$ & ${\bf 1}$ \\
\Rowspace $\bar\psi$ & {\bf adj} & ${\bf 1}$ & $\bar{\bf{4}}$ & ${\bf 1}$ & ${\bf 2}$ \\
\Rowspace $D$ & {\bf adj} & ${\bf 1}$ & ${\bf 1}$ & ${\bf 1}$ & ${\bf 3}$ \\
\Rowspace $a'$ & {\bf adj} & ${\bf 1}$ & ${\bf 1}$ & ${\bf 2}$ & ${\bf 2}$ \\
\Rowspace ${\cal M}'$ & {\bf adj} & ${\bf 1}$ & ${\bf 4}$ & ${\bf 2}$ & ${\bf 1}$ \\
\Rowspace $w$ & $\Bk$ & $\BN$ &  ${\bf 1}$ & ${\bf 1}$ & ${\bf 2}$ \\
\Rowspace $\bar w$ & $\bar\Bk$ & $\bar\BN$ & ${\bf 1}$ & ${\bf 1}$ & ${\bf 2}$ \\
\Rowspace$ \mu$ & $\Bk$ & $\BN$ & ${\bf 4}$ & ${\bf 1}$ & ${\bf 1}$ \\
\Rowspace $\bar{\mu}$ & $\bar\Bk$ & $\bar\BN$ & ${\bf 4}$ & ${\bf 1}$ & ${\bf 1}$ \\
\hline
\end{tabular}

\vspace{0.5cm}
Table 10.1: Transformation properties of the fields
\end{center}

The ${\cal N}=(0,1)$ supersymmetry transformations for
the theory can be deduced from the
supersymmetry transformations of ten-dimensional Yang-Mills theory
in Eqs.~\eqref{sstoa} and \eqref{sstob}.
The ${\cal N}=(0,1)$ supersymmetry of the six-dimensional action
\eqref{sgauge}, \eqref{sa} and \eqref{sf} is then
obtained as the subalgebra of this ten-dimensional ${\cal N}=1$
supersymmetry by taking
\begin{equation}
\Xi=-i
\begin{pmatrix} 0 \\ 1\end{pmatrix}\,\otimes\,
\begin{pmatrix} 0  \\ 1\end{pmatrix}\bar{\xi}^\aD_{A}\ .
\elabel{cov2}
\end{equation}
This yields the transformations
\AL{
\delta a'_{\alpha\aD} =  i\bar{\xi}_{\dot{\alpha}A}
{\cal M}^{\prime A}_{\alpha}\ ,&\qquad
 \delta{\cal M}^{\prime A}_{\alpha} =  2\Sigma^{aAB}\bar\xi^\aD_B
{\cal D}_{a}a'_{\alpha\aD}\ ,\elabel{susy12}\\
\delta \chi_{a} =  i\Sigma_a^{AB}\bar{\xi}_{A}
\bar\psi_B\ ,&\qquad
\delta \bar\psi_{A} =
\bar\Sigma^{ab}{}_A{}^BF_{ab}\bar\xi_{B}-i
\bar\sigma_{mn}D_{mn}\bar\xi_A\ .
\elabel{susy22}
}
The transformations act on the fundamental hypermultiplets
$\{w_\aD,\mu^A\}$ in an analogous
way to the action on the adjoint hypermultiplet $\{a'_n,\CM^{\prime A}\}$:
\EQ{
\delta w_\aD = i\bar\xi_{\aD A} \mu^A\ ,\qquad
\delta \mu^A=  2\Sigma^{aAB}\bar\xi^\aD_B{\cal D}_aw_\aD\ .
\elabel{susy3}
}

The reader will have noticed that we have chosen our notation so
that each hypermultiplet field has a counterpart, denoted by
the same letter, in the $\N=4$ instanton calculus.
In particular the various indices on these fields correspond
with those on the corresponding ADHM variable.
The physical reason for this correspondence is
simple: the six-dimensional fields are the collective
coordinates of the D5-branes. As the D5-branes
are equivalent to
Yang-Mills instantons, the vacuum moduli space of the $\U(k)$ gauge theory
on the D5-branes should coincide with the $\N=4$ supersymmetric
$k$-instanton moduli space
described by the ADHM construction. As the only scalar fields in the
six-dimensional theory lie in ${\cal N}=(0,1)$ hypermultiplets, the relevant
vacuum moduli space is conventionally referred to as a Higgs branch.
Precisely how the proposed equivalence arises was explained in
\cite{W1}. The Higgs branch is described by the vanishing of the
$D$-term. The $D$-term equation-of-motion which follows from varying
the action is
\EQ{
\alpha^{\prime2}\vec D=\frac i{16\pi^2}
\vec\tau^{\aD}{}_\bD\Big(
\bar w^\bD w_\aD+\bar a^{\prime\bD\alpha}a'_{\alpha\aD}\Big)\ .
}
So the condition for a supersymmetric vacuum, $\vec D=0$, is equivalent
to the ADHM constraints \eqref{badhm}. Since
global $\U(k)$ gauge transformations relate equivalent vacua, we see that the
Higgs branch of theory yields precisely the ADHM construction
of the instanton moduli space $\ms_k$ in the form of a hyper-K\"ahler quotient.
This is a particular example of the more general fact
that the Higgs branch of a gauge
theory with 8 supercharges is a hyper-K\"ahler quotient
(for example see
Refs.~\cite{Alvarez-Gaume:1981hm,Fre:1996dw,Antoniadis:1997ra}). The
general construction involves a gauge group $G$ and hypermultiplets
transforming in some representation of $G$. The canonical kinetic
terms of the hypermultiplet scalars $z^{\ii\aD}$
defines the metric on the mother
space \eqref{mothmet} and the gauge group action defines the vector
fields $X_r$ in \eqref{defxr}.
The auxiliary fields $\vec D$ are the moment maps \eqref{mommp}.
Setting $\vec D=0$ and dividing by global gauge
transformations gives the hyper-K\"ahler
quotient construction.\footnote{Note the $\zeta^c$ in \eqref{mommp}
are precisely Fayet-Illiopolos terms for any abelian factors of $G$.}

At a generic point
on the Higgs branch, the $\U(k)$ gauge symmetry is completely broken
and at low energies the vector multiplet can be integrated out
leading to a six-dimensional
$\N=(0,1)$ supersymmetric $\sigma$-model with $\ms_k$ as target
space. At leading order, the low-energy effective action
is obtained by simply removing
$S_{\rm gauge}$, the kinetic terms of the vector multiplet,
from the action \eqref{de66}. In this
limit, the vector multiplet fields $\{\chi_a,\vec D,\bar\psi_A\}$
become auxiliary and the resulting theory is precisely the
linear $\sigma$-model describing the moduli space dynamics of
instanton branes in ten-dimensional $\N=1$ gauge theory constructed in
\S\ref{sec:S52}. Note in this low-energy limit,
$\vec D$ and $\bar\psi_A$ become the
Lagrange multipliers \eqref{lagmult}
for the $c$-number and Grassmann ADHM constraints,
\eqref{badhm} and \eqref{fadhm}, respectively, as . Moreover, $\chi_a$
is the auxiliary $\U(k)$ gauge field in the linear action
\eqref{jslmm} (also, up to a re-scaling,
the auxiliary variables that we used to bi-linearize
the action in \S\ref{sec:S55}). At higher energies, or for small VEVs,
where non-abelian subgroups are restored,
the stringy corrections involving the kinetic terms for the vector multiplet
become increasingly important and the simple moduli space picture
breaks down.

Starting from the configuration of D5- and D9-branes described above,
the general case of parallel D$p$ and D$(p+4)$-branes with $p<5$ can be
obtained by dimensional reduction on the brane world-volumes. In this
case the vacuum structure of the gauge theory describing the brane
configuration is richer. The reason is that on dimensional reduction
$5-p$ of the components of the gauge field $\chi_{\hat a}$, $\hat
a=p+1,\ldots,5$,  become adjoint scalars and
can develop VEVs. These scalars describe the freedom for the
D$p$-branes to move off into the $5-p$ dimensions transverse to both
the D$p$ and D$(p+4)$-branes. In addition to
the ADHM constraints \eqref{badhm}, the
classical equations for a supersymmetric vacuum now include
\EQ{
[\chi_{\hat a},\chi_{\hat b}]=
w_\aD\chi_{\hat a}=[a'_n,\chi_{\hat a}]=0\ .
\elabel{vaceq}
}
The classical vacuum moduli space consists of distinct branches. First of all,
there is a Coulomb branch where $\chi_{\hat a}$ and $a'_n$ are
diagonal with $w_\aD=0$ and on which the gauge group is broken to $\U(1)^k$.
This corresponds to situation where the D$p$-branes are located
(generically) in the bulk away from the D$(p+4)$-branes. The diagonal
elements of $\chi_{\hat a}$ give the positions of the $k$ D$p$-brane
transverse to the D$(p+4)$-branes while the diagonal elements of
$a'_n$ gives the positions of the D$p$-branes along the world-volume
of the D$(p+4)$-branes. When a given element $(\chi_a)_{ii}$ goes
to zero, the corresponding D$p$-brane
touches the world-volume of the D3-branes. In
this case, it is clear from \eqref{vaceq} that it is
then possible for $w_{ui\aD}$ (for the given value of $i$) to become
non-vanishing. The D$p$-brane is then ``absorbed'' onto the
D$(p+4)$-branes. On this new branch of the vacuum moduli space the
D$p$-brane becomes a genuine Yang-Mills instanton with a non-zero scale
size $\rho_i^2=\tfrac12\bar w^\aD_{iu}w_{ui\aD}$ (no sum on $i$). Other kinds
of ``mixed'' branches arise when more of the $k$ D$p$-branes are
absorbed into the D$(p+4)$-branes. The Higgs branch described the
situation where
all the D$p$-branes have been absorbed into the D$(p+4)$-branes
and so $\chi_{\hat a}=0$. Notice that the vacuum moduli space for
on the Higgs branch with $p<5$ continues to be the instanton moduli
space $\ms_k$.
In the reverse sense, when an instanton
shrinks down to zero size, {\it i.e.\/}~becomes an ideal
instanton in the gauge theory on the world-volume of the
D$(p+4)$-branes, it becomes a
D$p$-brane which can move off into the bulk. These transitions
move through the points where the different branches of the vacuum
moduli space touch.
As long as $p>1$, so that the world-volume theory on the
D$p$-branes is more than two-dimensional, we expect the qualitative
picture of distinct branches to be valid, at least semi-classically.
However, the points at which the different branches touch correspond to
situations where a non-abelian subgroup of the gauge group is
restored. The theory here will be strongly coupled and so we expect
the moduli space description to break down. These are regimes where we
cannot ignore the kinetic terms of the vector multiplet and
stringy corrections are expected to be important.
The situations with
$p\leq 1$ are rather different since there can be no symmetry breaking
in this case. However, the picture of the low-energy dynamics being
described by $\sigma$-models on the appropriate moduli spaces is still
expected to be valid.

In particular
the case of D0/D4-branes in the Type IIA theory has been studied
extensively because of its application as a light-cone matrix model of the
$\N=(2,0)$ theory in six dimensions \cite{A}.
In this case, the D0-branes correspond to solitons in the
$4+1$ dimensional gauge theory on the D4-brane world-volume. These solitons
are just four-dimensional Yang-Mills instantons thought of as
static finite-energy configurations in five dimensions.  At weak coupling, we
expect the dynamics of these solitons to be correctly
described by supersymmetric quantum mechanics
on the instanton moduli space. Hence, it is not surprising
that precisely this quantum mechanical system is obtained in \cite{A}
by dimensionally reducing the six-dimensional ${\cal N}=(0,1)$ theory
described above down to a single time dimension.
On adding another world-volume dimension, we obtain the D1/D5 system
discussed by Witten in \cite{W2}. The D1-branes are now strings in
a six dimensional Yang-Mills theory and, in the decoupling limit,
their world-sheet dynamics
is described by a two-dimensional ${\cal N}=(4,4)$ non-linear
$\sigma$-model with the ADHM moduli space as the target manifold.
This $\sigma$-model has kinetic terms for the coordinates on the target
and their superpartners and as usual the supersymmetric completion of
the action involves a four-fermion term which couples to the Riemann tensor
of the target. If we reduce this action back to one dimension by discarding
spatial derivatives we obtain the quantum mechanics of \cite{A}.

In order to arrive at a description of D-instantons in the presence of
D3-branes we must dimensionally reduce the world-volume theory of the
D5-branes all the way to zero dimensions.
However, in order to recover the instanton calculus, we must
first perform a Wick rotation in the world-volume of the
D5-branes. Vector quantities, including the $\Sigma$-matrices, in
Minkowski space $v^a=(v^0,\vec v)$, $a=0,\ldots,5$,
become $v_a=(\vec v,iv^0)$, $a=1,\ldots,6$ and the
Euclidean action is $S^{\rm Eucl}=-iS^{\rm Mink}$.
After dimensional reduction,
the symmetry group $\U(k)\times \U(N)\times\bar{H}$ of the
six-dimensional system now has a simple interpretation:
$\U(k)$ is the auxiliary symmetry of the ADHM
construction, $\U(N)$ is the gauge group of the D3-brane theory,
$\SU(4)$ is the $R$-symmetry
group of the ${\cal N}=4$ supersymmetry algebra
and $\overline{\SO}(4)=\SU(2)_{L}\times \SU(2)_{R}$
is the four-dimensional Lorentz group.

We will now write down the
collective-coordinate integral which determines the leading semi-classical
contribution of $k$ D-instantons to correlation functions of the low-energy
fields of the IIB theory in the presence of $N$ D3-branes. From
the above discussion, the appropriate generalization of \eqref{ukpfn},
is obtained by Wick rotating and then
dimensionally reducing the partition function
of the six-dimensional theory. The resulting matrix theory has a
partition function\footnote{In this section, we shall not keep a
careful track of the overall normalization of the partition function
which, of course, is important for calculating physical quantities.}
\begin{equation}
{\EuScript Z}_{k} = \frac{1}{{\rm Vol}\,\U(k)}\,
\int d^{6}\chi\, d^{8}\lambda\, d^{3}D\,d^{4}a'\,d^{8}{\cal M}'\,
d^{2}w\,d^2\bar w\, d^{4}\mu\, d^{4}\bar{\mu}\,
e^{-S}\ ,
\elabel{final1}
\end{equation}
where the action is deduced from
\eqref{sgauge}, \eqref{sa} and \eqref{sf}
reduced to zero dimensions:
\EQ{
S=\frac{4\pi^2}{g^2}\big(4\pi^2
\alpha^{\prime2}S_G+S_K+S_D\big)\ ,
\elabel{mattact}
}
where $g\equiv g_4$ and
\AL{
S_{G} & ={\rm tr}_{k}\Big\{\tfrac12[\chi_a,\chi_b]^2
-\Sigma_{a}^{AB}\bar\psi_A[\chi_a,\bar\psi_B]
-\tfrac12D_{mn}^2\Big\}\ ,\elabel{p=-1actiona} \\
S_{K} & = {\rm tr}_{k}\Big\{-[\chi_a,a'_{n}]^2
+\chi_a\bar{w}^\aD
w_{\aD}\chi_a -\tfrac14\bar\Sigma_{aAB}
{\cal M}^{\prime \alpha A}[\chi_a,
{\cal M}^{\prime B}_{\alpha}]+\tfrac12\bar\Sigma_{aAB}
\bar{\mu}^{A}\mu^B\chi_a\Big\}\ ,\elabel{p=-1actionb} \\
S_{D} & = {\rm tr}_k\Big\{-i\vec D\cdot\vec\tau^{\aD}{}_\bD
\big(\bar w^\bD w_\aD+\bar a^{\prime\bD\alpha}a'_{\alpha\aD}\big)
+i\big(\bar\mu^Aw_\aD+\bar w_\aD\mu^A+
[\CM^{\prime A},a'_{\alpha\aD}]\big)\bar\psi^\aD_A\Big\}\ .
\elabel{p=-1actionc}
}
Note that $S_{G}$ arises
from the dimensional reduction of $S_{\rm gauge}$ and $S_{K}+S_{D}$ comes
from the dimensional reduction of $S_{\rm matter}^{\rm (a)}+S_{\rm
matter}^{\rm (f)}$.
Specifically, $S_{K}$ contains the six-dimensional gauge couplings of the
hypermultiplets while $S_{D}$ contains the Yukawa couplings and D-terms.

Semi-classical correlation functions of the light fields can be
calculated by replacing each field with its value in the D-instanton
background and performing the collective coordinate integrations
with measure \eqref{final1}. In the case of the low-energy gauge fields on
the D3-brane, the relevant classical configuration is simply the charge-$k$
Yang-Mills instanton specified by the ADHM data which appears explicitly
in the action \eqref{p=-1actiona}-\eqref{p=-1actionc}. Note that the
collective coordinate integral \eqref{final1}
depends explicitly on the string length-scale $\alpha'$ only via the
zero dimensional remnant of the kinetic terms of the vector multiplet.
As a consequence correlation functions which
include fields inserted at distinct spacetime points $x^{(i)}$ and $x^{(j)}$
will have a non-trivial expansion in powers of
$\sqrt{\alpha'}/ |x^{(i)}-x^{(j)}|$. In order to decouple the world-volume
gauge theory from gravity we must take the limit $\alpha'\rightarrow 0$
keeping the four-dimensional coupling $g$ fixed. In this limit,
the coupling of $S_K+S_D$ is held fixed while the remnant of the
kinetic term of the vector multiplet, $S_G$, decouples from the action.
In this limit, the collective coordinate integral
\eqref{final1} becomes
\begin{equation}
{\EuScript Z}_{k}  =  \frac{1}{{\rm Vol}\,\U(k)}\,
\int d^{6}\chi d^{8}\lambda d^{3}D\,d^{4}a'\,d^{8}{\cal M}'\,
d^{4}w\, d^{4}\mu\, d^{4}\bar{\mu}\,
\exp\left(-S_{K}-S_{D}\right) \ .
\elabel{final}
\end{equation}

We can now make contact with the $\N=4$ instanton calculus. Note that $S_K$
can be written more compactly as
\EQ{
S_{K}=\frac{4\pi^2}{g^2}{\rm tr}_k\,\chi_a\BL\chi_a
+\frac{2\pi^2}{g^2}\bar\Sigma_{aAB}
{\rm tr}_k\,\bar\CM^{A}\CM^B\chi_a\ .
\elabel{atisl}
}
We now recognize the partition function \eqref{final} as being the
instanton partition function in its linearized form \eqref{linvolf}
with $S_K+S_D$ being the instanton effective action $\tilde S$ in
\eqref{ieal}.

We have therefore recovered the leading-order semi-classical
expression for the collective coordinate integral for instantons in
the $\N=4$ supersymmetric theory.
In addition the action $S_{K}+S_{D}$ is invariant under eight supercharges
which are inherited from the ${\cal N}=(0,1)$ theory in six
dimensions. One can check that after Wick rotation and
dimensional reduction, the supersymmetry transformations
\eqref{susy12} and \eqref{susy3} (up to appropriate re-scaling by $g$)
match the collective
coordinate supersymmetries of the ${\cal N}=4$ ADHM instanton calculus
written down in \eqref{uiui} and \eqref{vivi}, or
\eqref{susyl1}-\eqref{susyl3}.
There are also eight additional fermionic symmetries, corresponding to
$\xi^A$ in \eqref{vivi}, which
only appear after taking the decoupling limit $\alpha'\rightarrow 0$.
These correspond to the half of the
superconformal transformations of the ${\cal N}=4$ theory which leave
the instanton invariant.

One can ask how constrained instantons appear in this
context. Constrained instantons arise when the $\N=4$ theory is on its
Coulomb branch realized by separating the $N$ D3-branes in
the six-dimensional transverse space. This obviously changes the
lengths of string stretched between the D-instantons and
D3-branes and has the effect of adding mass terms for the fundamental
hypermultiplet fields.
At the level of the matrix theory action \eqref{repcd} the
relevant effect can be introduced by the replacements \eqref{addvev}.
Taking the decoupling limit and integrating out the vector multiplet
one recovers the instanton effective action \eqref{yyvv} for $\N=4$
constrained instantons.

\subsubsection{Probing the stringy instanton}

We have seen in the preceding section that many of the features
of the instanton calculus are reproduced by considering
the dynamics of D$p$-branes
lying inside D$(p+4)$-branes. It is interesting to ask whether
the actual ADHM form for the gauge potential \eqref{vdef} can be
obtained in this string theory context. The answer is affirmative
once an appropriate probe is identified. The concept
of a probe which is able to ``feel'' the instanton background was
first described by Witten \cite{Witten:1995tz} in the context of Type I
string theory. The idea was generalized to the Type II theories needed
for the present discussion by Douglas in \cite{D2}. The appropriate
probe turns out to be a D-brane as well. In the context of the D5/D9-brane
system described in the last section, the probe is a single D1-brane
(or D-string) whose world-volume lies parallel to the other
branes. The D-string ``feels'' the D5/D9-brane background since the
fields of the D-string world-sheet theory include a subset that arise
from open strings stretched between the D-string and the other
higher-dimensional branes. The whole configuration is like a Russian doll:
the D5/D9 and D1/D5 sub-systems are both examples of our general
D$p$/D$(p+4)$ system described in \S\ref{sec:S70}. In the composite system,
the ADHM variables appear both as fields of the D5-branes and as couplings of
the world-sheet theory of the D-string. The conditions for the
resulting world-sheet
theory to be $\N=(0,4)$ supersymmetric are precisely the ADHM
constraints \eqref{badhm}.

In more detail, we now analyse the dynamics of a D-string in the
background of $k$ D5-branes and $N$ D9-branes. The system breaks the
ten-dimensional Lorentz group to
$\SO(1,1)\times\SO(4)_1\times\SO(4)_2$. Here, $\SO(4)_1$ describes the
directions transverse to the D5-branes denoted by vector index
$n=1,2,3,4$. From earlier,
$\overline{\SO}(4)_1\simeq\SU(2)_L\times\SU(2)_R$, where the latter
correspond to the spinor indices $\alpha$ and $\aD$. The other factors
$\SO(1,1)\times\SO(4)_2$ are a subgroup $\SO(5,1)$, the Lorentz group of
the D5-branes' world-volume. We will
denote the corresponding spinor indices of
$\overline{\SO}(4)_2\simeq\SU(2)_A\times\SU(2)_
B$ by $\delta$ and $\dD$, respectively. Under
the decomposition $\SO(5,1)\supset\SO(1,1)\times\SO(4)_2$
\EQ{
{\bf4}\rightarrow({\bf2},{\bf1})_1+({\bf1},{\bf2})_{-1}\ ,\qquad
\bar{\bf4}\rightarrow({\bf2},{\bf1})_{-1}+({\bf1},{\bf2})_1\ ,\qquad
{\bf 6}\rightarrow({\bf2},{\bf2})_0+({\bf1},{\bf1})_{2}+({\bf1},{\bf1})_{-2}\ .
}
Here, the subscripts indicate the $\SO(1,1)$ chirality.

Let us ignore, for the moment, the presence of the D9-branes. The
D-string/D5-brane system is an example of the D$p$/D$(p+4)$-brane
system studied in the last section. Recall that
the D-string world-sheet theory can best be derived as the dimensional
reduction of the D5/D9-brane system, with one D5-brane and $k$
D9-branes. Obviously, we will have to use a different notation to
describe the fields of the D-string theory as well as re-assigning
indices appropriately. The action of the theory is
the dimensional reduction to two dimensions of \eqref{de66} with the following
replacements. Firstly, there is a $\U(1)$ gauge field and fermions
\EQ{
\chi_a\to(A_\pm,x_n)\ ,\qquad\bar\psi^\aD_A\to(\bar\zeta^\dD_\alpha,
\bar\zeta^\dD_\aD)
}
where $A_\pm$ are the light-cone components of the two-dimensional
abelian gauge field. Note that $x_n$ denotes, in this context, a field in
the world-sheet theory rather than a spacetime coordinate. Since the
world-sheet theory is abelian the adjoint
hypermultiplet previously denoted $\{a'_n,\CM^{\prime A}_\alpha\}$
decouples and we can safely ignore it. Finally, the fundamental
hypermultiplets describing string stretched between the D-string and
D5-branes are described by the replacements
\EQ{
w_{ui\aD}\to\phi_{i\dD}\
,\qquad\mu^A\to(\theta^\alpha_i,\theta^\aD_i)\ ,\qquad
\bar w^\aD_{iu}\to\bar\phi^\dD_{i}\
,\qquad\bar\mu^A\to(\bar\theta^\alpha_i,\bar\theta^\aD_i)\ .
}
The world-sheet theory describing the D-string in the presence of the
D5-branes has conventional kinetic terms along with a
potential\footnote{In the following expressions we do not keep careful
track of the coupling constants and normalizations.}
\EQ{
{\EuScript L}_{\rm pot}=
-\bar\phi^\dD_i\phi_{i\dD}x_nx_n+\sum_{c=1}^3\big(\tau^{c\dD}{}_\eD
\,\bar\phi^\eD_i\,\phi_{i\dD}\big)^2\ .
\elabel{pot1}
}
Here, the second term comes from the D-terms which can be viewed
as the ADHM constraint for the D-string viewed as
a single instanton inside the $\U(k)$ gauge theory on the D5-branes.
There are also Yukawa couplings
\EQ{
{\EuScript L}_{\rm Yuk}=\bar\theta^\alpha_i
x_{\alpha\aD}\theta^\aD_i+\bar\theta_{i\aD}\bar
x^{\aD\alpha}\theta_{i\alpha}
+\big(\bar\zeta^\dD_\alpha\bar\theta^\alpha_i+\bar\zeta^\dD_\aD
\bar\theta^\aD_i\big)\phi_{i\dD}+ \big(\bar\zeta^\dD_\alpha
\theta^\alpha_i+\bar\zeta^\dD_\aD
\theta^\aD_i\big)\bar\phi_{i\dD}\ .
\elabel{yuk1}
}
The first two terms are the fermionic ADHM constraints for the
Grassmann collective coordinates of the single
$\U(k)$ instanton.

At the moment we have assumed all the D5-branes are coincident. In
reality the D5-branes will be separated (in addition to being
``thickened out'') in a
way described by the ADHM matrices $a'_n$. By translational symmetry,
the effect can be introduced into the world-sheet theory by replacing
$x_n$ by the matrix quantity
\EQ{
x_n\longrightarrow
x_n1_{\sst[k]\times[k]}+a'_n\ .
}
In this way the ADHM
variables $a'_n$ appear as couplings in the world-sheet theory.

Now we must consider the effect of the D9-branes. First of all, there
are new fields in the world-sheet theory describing open strings
stretched between the D-string and the D9-branes. Since there are no
additional moduli and the new fields are fermionic, $\epsilon_{u}$ and
$\bar\epsilon_u$. The second effect of the D9-branes is that in the
configuration we are interested in, they absorb the D5-branes and the
latter thicken out in a way parameterized by the ADHM variables
$w_\aD$. These variables will, like $a'_n$, appear as couplings in the
world-sheet theory. Taking into account the D9-branes leads to a
modified potential which has a very suggestive form:
\EQ{
{\EuScript L}_{\rm pot}=
-\bar\phi^\dD_i\bar\Delta^{\aD\lambda}_i\Delta_{\lambda
j\aD}\phi_{j\dD}
+\sum_{c=1}^3\big(\tau^{c\dD}{}_\eD
\,\bar\phi^\eD_i\,\phi_{i\dD}\big)^2\ .
\elabel{pot2}
}
Here, $\Delta$ and $\bar\Delta$ are the ADHM quantities introduced in
\S\ref{sec:S8} (with the canonical choice \eqref{aad})
but where now $x_n$ is viewed as a field rather than as
a spacetime coordinate. Note \eqref{pot2} subsumes
\eqref{pot1}. The Yukawa coupling have a similarly suggestive form:
\EQ{
{\EuScript L}_{\rm Yuk}=\bar\Upsilon^\lambda
\Delta_{\lambda i\aD}\theta^\aD_i+\bar\theta_{i\aD}\bar
\Delta^{\aD\lambda}_{i}\Upsilon_\lambda
+\big(\bar\zeta^\dD_\alpha\bar\theta^\alpha_i+\bar\zeta^\dD_\aD
\bar\theta^\aD_i\big)\phi_{i\dD}+ \big(\bar\zeta^\dD_\alpha
\theta^\alpha_i+\bar\zeta^\dD_\aD
\theta^\aD_i\big)\bar\phi_{i\dD}\ ,
\elabel{yuk2}
}
where we have defined the composites\footnote{Recall, from
\S\ref{sec:S8}, $\lambda$ is the ADHM composite index $u+i\alpha$.}
\EQ{
\Upsilon_\lambda=\MAT{\epsilon_u\\ \theta_{i\alpha}}\ ,\qquad
\bar\Upsilon^\lambda=\MAT{\bar\epsilon_u & \bar\theta^\alpha_{i}}\ .
}
Note \eqref{yuk2} subsumes \eqref{yuk1}.

As is often the case, the conditions that the world-sheet theory is
$\N=(0,4)$ invariant is identical to the equations-of-motion in the
target space: in this case the D-flatness conditions in the D5-brane
theory. In other words, as argued in \cite{Witten:1995tz},
the conditions for extended $\N=(0,4)$
supersymmetry are the ADHM constraints \eqref{badhm} on
$\{w_\aD,a'_n\}$ viewed as coupling in the world-sheet theory.

For generic ADHM data, the potential \eqref{pot2} gives a mass to
the fields $\phi$. These fields can then be
integrated out leading to a (gauged) $\sigma$-model. At the classical
level, we can simply set $\phi=0$. The fields $x_n$
are massless and the resulting $\sigma$-model
involves the flat metric on ${\mathbb R}^4$. The fermionic sector is
more interesting. The first two Yukawa couplings in \eqref{yuk2} give
masses to $\theta^\aD_i$, $\bar\theta_{i\aD}$ and a $2k$-dimensional
subspace of both $\Upsilon_\lambda$ and $\bar\Upsilon^\lambda$. What
we are interested in are the remaining massless fermionic
degrees-of-freedom. To order to identify these modes
we need bases for the null spaces of $\Delta$ and $\bar\Delta$. But
these are provided by the ADHM quantities $\bar U$ and $U$, respectively, as
is apparent in \eqref{uan}. Hence, the massless modes
$\{\rho_u,\bar\rho_u\}$ are picked out by
\EQ{
\Upsilon_\lambda=U_{\lambda u}\rho_u\ ,\qquad
\bar\Upsilon^\lambda=\bar\rho_u\bar U_{u}^\lambda\ .
}
The kinetic term for the massless modes then follows by substitution
into the kinetic term for $\Upsilon$:
\EQ{
\bar\Upsilon^\lambda\partial_-\Upsilon_\lambda
\longrightarrow \bar\rho_u\big(\delta_{uv}\partial_-+(\partial_-x_n)
(A_n)_{uv}\big)\rho_v\ ,
}
with
\EQ{
(A_n)_{uv}(x)=\bar U_{u}^\lambda\PD{U_{\lambda v}}{x_n}\ .
}
This is precisely the ADHM expression for the gauge potential
\eqref{vdef} as a function of the field $x_n$ (with $g\to1$).
Of course the expression that we have derived for the gauge potential
is only valid in the classical limit and there will be stringy
corrections to the instanton profile on a scale set by $\sqrt{\alpha'}$.

\subsubsection{The $\N=2$ instanton calculus}\elabel{sec:S94}

There are a number of ways to obtain four-dimensional
$\N=2$ theories as the collective dynamics of branes in string
theory. The first observation is that the $\N=2$ theory has 8
supercharges and may be obtained from the dimensional reduction of a
six-dimensional theory with $\N=(0,1)$ supersymmetry. Just as in the
case of the $\N=4$ theory it is useful to realize the $\N=2$ theory
in its maximal dimension. In this case, an instanton will be a 1-brane
soliton as described in \S\ref{sec:S52}.

One way to realize the six-dimensional theory
is to consider the collective dynamics of D5-branes embedded in a spacetime
where the four transverse dimensions are orbifold ${\mathbb
R}^4/{\mathbb Z}_2$. The collective dynamics of the D5-branes can be
deduced in the following way
\cite{Douglas:1996sw,Douglas:1997xg,Diaconescu:1998br}.
First of all, suppose we start with a set of
$\tilde N$ parallel D5-branes in flat ten-dimensional Minkowski
space. From the discussion in \S\ref{sec:S70}, the
world-volume theory is simply the dimensional reduction to six of
$\N=1$ supersymmetric
$\U(\tilde N)$ gauge theory in ten dimensions. This theory has
16 supercharges, {\it i.e.\/}~has $\N=(1,1)$ supersymmetry
in six dimensions. The theory
has an $\SU(2)_A\times\SU(2)_B$ $R$-symmetry which occurs because of the
$\SO(4)$ rotational symmetry of the transverse space. Now we replace
the transverse space with the orbifold ${\mathbb R}^4/{\mathbb Z}_2$,
where the ${\mathbb Z}_2$ is chosen to act
as the centre of $\SU(2)_A$.
The resulting world-volume theory of the D5-branes is
deduced by a process of projection in the following way. Notice that
the $\sigma_R={\mathbb Z}_2$ acts on the fields via their $R$-symmetry
indices. The next part of the procedure
involves embedding the ${\mathbb
Z}_2$ group action as $\sigma_{\U(\tilde N)}$
in the gauge group. There are different ways to do
this. However, up to conjugation we can take
\EQ{
\sigma_{\U(\tilde N)}=\MAT{1_{\sst[N]\times[N]}&0\\
0&-1_{\sst[M]\times[M]}}\ ,
}
with $\tilde N=N+M$. The resulting theory is then obtained by taking
the fields and action of the original $\N=(1,1)$ theory and
projecting out by hand all the fields which are not invariant under
the simultaneous transformation by $\sigma_R\sigma_{\U(N)}$.
The resulting theory has gauge group $\U(N)\times\U(M)$, the
supersymmetry is reduced to $\N=(0,1)$
and there are two hypermultiplets in the bi-fundamental representation
of the gauge group, {\it i.e.\/}~each having fields in the
$(\BN,\bar{\BM})+(\bar{\BN},\BM)$.

Now we consider instantons. In the six-dimensional theory the
instantons correspond to D1-branes lying inside the D5-brane, the same
situation as in \S\ref{sec:S70}, but now lying transverse to the
orbifold rather than ${\mathbb R}^4$.
The collective dynamics of the D1-branes in this configuration can be deduced
from the D$p$/D$(p+4)$-brane system discussed in \S\ref{sec:S70}
(with $\tilde N$ D$(p+4)$-branes)
and then performing the same kind of projection on the D1-brane
world-sheet theory that we did for the D5-brane world-volume theory above.
Firstly, one embeds ${\mathbb Z}_2$ both in the
$R$-symmetry group and in the gauge group, but now the gauge group is
pertains to the D1-branes. Firstly the $R$-symmetry. In terms of
the variables of the instanton calculus, the ${\mathbb Z}_2$ is
embedded as $\sigma_R$
in the group $\overline{\SO}(6)\simeq\SU(4)$ that arose from the
world-volume Lorentz symmetry of the D5-branes in the D5/D9-brane system.
On dimensional reduction to the D1/D5-brane system $\overline{\SO}(6)$
is broken to $\SO(1,1)\times\SU(2)_A\times\SU(2)_B$ and $\sigma_R$
acts as the centre of $\SU(2)_A$. For instance, $\chi_a$ which
transforms in the vector of $\SO(6)$, breaks up into
\EQ{
\sigma_R(\chi_a)=\begin{cases}\chi_a\qquad & a=0,1\\-\chi_a &
a=2,3,4,5\ .\end{cases}
\elabel{tchi}
}
Here, the invariant components are the two-dimensional $\U(k)$ gauge field on
the D1-brane world-sheet. For spinor quantities like $\bar\psi_A$ and
$\CM^{\prime A}$, we have
\EQ{
\sigma_R(\bar\psi_A)=\begin{cases}\bar\psi^A\qquad &A=1,2\\
-\bar\psi_A &A=3,4\ ,\end{cases}\qquad
\sigma_R(\CM^{\prime A})=\begin{cases}\CM^{\prime A}\qquad &A=1,2\\
-\CM^{\prime A} &A=3,4\end{cases}
\elabel{tlamm}
}
and similarly for $\mu^A$ and $\bar\mu^A$. The embedding ${\mathbb Z}_2$ in the
$\U(k)$ gauge group of the D1-brane world-volume theory
determines the instanton charge of the
configuration. For instance
\EQ{
\sigma_{\U(k)}=\MAT{1_{\sst[k_1]\times[k_1]}&0\\
0&-1_{\sst[k_2]\times[k_2]}}\ ,
}
with $k=k_1+k_2$ describes an instanton configuration
with charges $k_1$ and $k_2$
with respect to the $\U(N)$ and $\U(M)$ factors of the gauge group. In
order to have an instanton which lives solely in the
$\U(N)$ factor we must set $k_2=0$ so that $k_1\equiv k$ and in this
case $\sigma_{\U(k)}=1_{\sst[k]\times[k]}$. From the point-of-view of
this instanton configuration the $\U(M)$ gauge group plays the r\^ole
of a spectator since, to leading order, the $\U(M)$-adjoint fields are
zero and the $\U(M)$ symmetry is effectively a global
symmetry. Consequently on dimensional reduction of the D5-branes
to four dimensions, the resulting theory is effectively an $\N=2$
theory with $N_F=2M$ fundamental hypermultiplets.\footnote{Note only
$N_F$ even theories can be obtained in this way.}

The world-volume theory on the D1-branes is then obtained from
the action of the D5-branes
in the D5/D9-brane system, that is \eqref{de6},
dimensionally reduced to two dimensions and
by removing
by hand any fields that are not invariant under a simultaneous
transformation $\sigma_R\sigma_{\U(k)}\sigma_{\U(\tilde N)}$. The latter
transformation has to be included because the fundamental
hypermultiplets carry $\U(\tilde N)$ gauge indices.
In our case, recall that
$\sigma_{\U(k)}=1_{\sst[k]\times[k]}$.
The fields that remain are the following.
Firstly, from \eqref{tchi}, only the components $\chi_a$,
$a=0,1$ remain. This is a $\U(k)$ gauge field in two dimensions. For
the other adjoint-valued fields,
the bosonic quantities $a'_n$ survive,
since these are invariant under $\sigma_R$.
For the fermions only the components $\bar\psi_A$ and
$\CM^{\prime A}$, with $A=1,2$ remain. The situation is slightly more subtle
for the fundamental hypermultiplets. Each index $\tilde u$ can now run over
$N+M$ values which split into two set: $u=1,\ldots,N$ and
$u'=1,\ldots,M$. Then, for example,
\EQ{
\sigma_{\U(\tilde N)}(w_{ui\aD})=w_{ui\aD}\ ,\qquad
\sigma_{\U(\tilde N)}(w_{u'i\aD})=-w_{u'i\aD}
}
and similarly for the other hypermultiplet fields.
Therefore, the fields
$w_{ui\aD}$ survive along with $\mu^A_{ui}$ and $\bar\mu^A_{iu}$, for
$A=1,2$. However, in addition the fermionic fields $\mu^A_{u'i}$ and
$\mu^A_{iu'}$, with $A=3,4$, are odd with respect to both $\sigma_R$
and $\sigma_{\U(\tilde N)}$ and so survive.

We now show that on dimensional reduction to zero dimensions and in
the decoupling limit, the
partition function of the resulting matrix theory is the leading order
expression for the collective coordinate integral of the $\N=2$ theory
with $N_F=2M$ hypermultiplets. The details are almost identical to the
$\N=4$ case described in \S\ref{sec:S70}. In particular, as previously, in
the decoupling limit, the $\vec D$ and $\bar\psi_A$ ($A=1,2$) become
Lagrange multipliers for the bosonic and fermionic ADHM constraints.
The main difference is that the
the dimensionally reduced (Wick rotated) action \eqref{atisl} is subject to the
projection described above yielding\footnote{It is useful to recall
that in Euclidean
space $\bar\Sigma_1=\eta^3$ and $\bar\Sigma_2=i\bar\eta^3$.}
\EQ{
S_{K}=\frac{4\pi^2}{g^2}\Big\{
{\rm tr}_k\,\chi_a\BL\chi_a
-\tfrac12(\chi_1-i\chi_2)
\big(\bar\CM^1_u\CM^2_u-\bar\CM^2_u\CM^1_u\big)
-\tfrac12(\chi_1+i\chi_2)
\big(\bar\CM^3_{u'}\CM^4_{u'}-\bar\CM^4_{u'}\CM^3_{u'}\big)\Big\}\ .
\elabel{atisl2}
}
In this expression a sum over $a=1,2$ is implied.
The integral over $\chi_a$ yields the factor of $|\det_{k^2}\BL|^{-1}$ in
\eqref{fms} for $\N=2$. While the action becomes
\EQ{
\tilde
S=-\frac{\pi^2}{g^2}{\rm tr}_k\,
\big(\bar\CM^1_u\CM^2_u-\bar\CM^2_u\CM^2_u\big)\BL^{-1}
\big(\bar\CM^3_{u'}\CM^4_{u'}-\bar\CM^4_{u'}\CM^3_{u'}\big)\ .
}
This is precisely equal to \eqref{ieant} after a suitable re-scaling by
$g$, with the VEVs set to zero and with the relations
\EQ{
\K_f=(\bar\CM^3_{u'},\bar\CM^4_{u'})\ ,\qquad
\tilde\K_f=i(\CM^3_{u'},-\CM^4_{u'})\ ,
}
where $f=1,\ldots,N_F=2M$. Just as in the $\N=4$ case, the instanton
effective action on the Coulomb branch of the $\N=2$ theory can
easily be obtained by separating the D5-branes in the two dimensions
transverse to their world-volume orthogonal to the orbifold.

\subsubsection{Mass couplings and soft supersymmetry
breaking}\label{sec:S141.6}

In \S\ref{app:A6}, we described how the instanton calculus was
modified when mass terms were added to the field theory breaking
supersymmetry successively from $\N=4$, through $\N=2$ and $\N=1$ to
$\N=0$. Using the response of the instanton to field theory
masses, we were able to relate in \S\ref{sec:S42.6} the collective
coordinate integrals with different numbers of supersymmetries by a
process of decoupling and renormalization group matching. In this
section, we describe how the effects of mass terms may be realized in
the brane description.

First of all, let us establish at the level of
the $\N=4$ field theory, the form of the mass deformation. It is
useful at this stage to introduce the language of $\N=1$ superfields.
The $\N=4$ theory consists of an $\N=1$ vector
multiplet along with three adjoint-valued chiral multiplets
$\Phi_\xii=\{\varphi_\xii/\sqrt2,\lambda_\xii\}$,
$\xii=1,2,3$. The relation between these fields and the fields that we
introduced in \S\ref{sec:S26} can be chosen as
\EQ{
\varphi_1=i\phi_5+\phi_6\ ,\qquad
\varphi_2=i\phi_3+\phi_4\ ,\qquad
\varphi_3=i\phi_1+\phi_2
\elabel{compb}
}
and
\EQ{
\lambda_\xii=\lambda^A\ ,\qquad\bar\lambda_\xii=\bar\lambda_A\qquad
A=\xii=1,2,3\ .
}
With this choice the vector multiplet contains
$\{A_m,\lambda\equiv\lambda^4\}$.

In terms of $\N=1$ superfields the action of the $\N=4$ theory (in
Minkowski space) can be written
\EQ{
S=\frac1{g^2}\int d^4x\,\TrN\Big\{\tfrac{g^2}{8\pi}{\rm Im}\,\tau W^\alpha
W_\alpha\Big|_{\theta^2}
+2\Phi^\dagger_\xii
e^{V}\Phi_\xii\Big|_{\theta^2\bar\theta^2}
+\tfrac2
3\epsilon_{\xii\xjj\xkk}\Phi_\xii\Phi_\xjj\Phi_\xkk\Big|_{\theta^2}
+\tfrac23\epsilon_{\xii\xjj\xkk}\Phi_\xii^\dagger\Phi_\xjj^\dagger
\Phi_\xkk^\dagger\Big|_{\bar\theta^2}\Big\}\ .
}
The most general mass deformation that preserves
$\N=1$ supersymmetry can be obtained by adding
\EQ{
S_{\rm mass}=\frac1
{g^2}\int dx^4\,\TrN\Big\{m_\xii\Phi_\xii^2\Big|_{\theta^2}+
m_\xii^*\Phi_\xii^{\dagger2}\Big|_{\bar\theta^2}\Big\}\ .
}
In terms of component fields (in Euclidean space)
\EQ{
S_{\rm mass}=\frac1{g^2}\int
d^4x\,\TrN\Big\{m_\xii\epsilon_{\xii\xjj\xkk}\varphi_\xii
\varphi_\xjj^\dagger\varphi_\xkk^\dagger-m_\xii^*
\epsilon_{\xii\xjj\xkk}\varphi_\xii^\dagger
\varphi_\xjj\varphi_\xkk+|m_\xii|^2|\varphi_\xii|^2+m_\xii\lambda_\xii
\lambda_\xii+m_\xii^*\bar\lambda_\xii\bar\lambda_\xii\Big\}\ .
\elabel{commass}
}
It is interesting to establish the $\SO(6)$ R-symmetry properties of
the mass terms above. In general a mass term for the
fermions can be written
\EQ{
\frac1{g^2}\int d^4x\,\TrN\,\Big\{m_{AB}\lambda^A\lambda^B+m^{AB}\bar\lambda_A
\bar\lambda_B\Big\}\,
}
where the symmetric matrix
$m_{AB}$ ($m^{AB}\equiv
m_{AB}^*$) transforms in the $\bar{\bf10}$ ($\bf10$) of
$\SO(6)$. The mass matrix can be chosen to be
diagonal: $m_{AB}={\rm diag}(m_1,m_2,m_3,m_4)$. In order to preserve
at least $\N=1$ supersymmetry,
at least one of the mass eigenvalues must vanish. Choosing $m_4=0$
gives the fermion mass terms in \eqref{commass}. The mass
matrix can also be written as a anti-symmetric rank-3 tensor $T_{abc}$
of $\SO(6)$. The relation between the two bases is provided by the
$\Sigma$-matrices:
\EQ{
T_{abc}\thicksim
m_{AB}\big(\Sigma_{[a}\bar\Sigma_b\Sigma_{c]}\big)^{AB}+
m^{AB}\big(\bar\Sigma_{[a}\Sigma_b\bar\Sigma_{c]}\big)_{AB}\
.
\elabel{tendef}
}
With some choice of normalization, the tensor $T$ is associated to
the following cubic coupling of the scalar fields:
\EQ{
T_{abc}\TrN\,\phi_a\phi_b\phi_c=\epsilon_{\xii\xjj\xkk}\TrN\,\Big\{
m_\xii\varphi_\xii\varphi_\xjj^\dagger\varphi_\xkk^\dagger-m_\xii^*
\varphi_\xii^\dagger
\varphi_\xjj\varphi_\xkk+m_4\varphi_\xii
\varphi_\xjj\varphi_\xkk-m_4^*\varphi_\xii^\dagger
\varphi_\xjj^\dagger\varphi_\xkk^\dagger\Big\}\ .
}
When $m_4=0$, this is precisely the
bosonic part linear in the masses
of the $\N=1$ preserving deformation in \eqref{commass}.

The question before us is how to introduce the mass deformation
when the $\N=4$ theory is realized as the
the collective dynamics of $N$ coincident D3-branes in Type IIB string
theory?
A D$p$-brane carries charge which couples directly to the
Ramond-Ramond $p+1$-form potential $C^{(p+1)}$.
However, D$p$-branes also couple to
other background fields in the string theory.
The most general couplings of the Ramond-Ramond potentials to a
collection of D$p$-branes occurs through the Chern-Simons action whose
form, in the case of multiple branes, was established by Myers
\cite{Myers:1999ps}. In particular, D3-branes can carry D5-brane
dipole moment which is induced by a
coupling to the $C^{(6)}$ potential of the form\footnote{In the
following $0123$ refers to the world-volume directions.}
\EQ{
\frac1{g^2}\int d^4x\,\TrN\,\Big\{\phi_a\phi_b C^{(6)}_{0123ab}(\phi)\Big\}\ .
}
Consider the case where the associated 7-form
field strength $F^{(7)}=dC^{(6)}$ is constant. Then the coupling above
is simply equal to
\EQ{
\frac1{3g^2}\int d^4x\,F^{(7)}_{0123abc}\TrN\,\phi_a\phi_b\phi_c=
\frac 1{3g^2}
\int d^4x\,\big(*_6F^{(3)}\big)_{abc}\TrN\,\phi_a\phi_b\phi_c\ ,
}
using $F^{(7)}=*F^{(3)}$.
Here, $*_6$ is Hodge duality in the six-dimensional transverse
space. Comparison with the mass deformation \eqref{commass} shows that the
background Ramond-Ramond field produces the bosonic
terms linear in the masses if
\EQ{
\big(*_6F^{(3)})_{abc}=6T_{abc}
\elabel{ftot}
}
and $T$ is the tensor defined in \eqref{tendef}.

In this way we have reproduced the bosonic mass coupling
\eqref{commass} using a suitable background Ramond-Ramond
potential. However, as one would expect due to supersymmetry,
the background field also couples to the fermions. The analogous
example of D0-branes coupling to $C^{(3)}$ has
been investigated in \cite{Millar:2000ib}. By using T-duality we can
extract the analogous coupling for D3-branes. The result is most easily written
by recalling that the fermionic fields on D3-branes can be obtained by
a dimensional reduction of a ten-dimensional Majorana-Weyl
spinor. Using the representation of the Clifford algebra as in
\eqref{cliftd} and the decomposition of $\Psi$ in terms of
$\{\lambda^A,\bar\lambda_A\}$ in \eqref{spc},
the coupling to the background $C^{(6)}$ potential can be written in
the form
\EQ{
\frac1{g^2}
\int d^4x\,F^{(7)}_{0123abc}\TrN\,\bar\Psi\Gamma_a\Gamma_b\Gamma_c\Psi\ .
}
Using \eqref{tendef} and \eqref{ftot} this can be written as
\EQ{
\frac1{g^2}\int d^4x\,\TrN\,\Big\{m_{AB}\lambda^A\lambda^B+m^{AB}\bar\lambda_A
\bar\lambda_B\Big\}\ ,
}
precisely the fermion mass coupling in \eqref{commass}.

Now that we have established how to introduce the mass coupling in the
world-volume theory of the D3-branes, we can now consider the effect
of the same background field in the D-instanton theory. Recall that the
relevant effect for the D3-branes could be described as a D5-brane
dipole moment coupling to $C^{(6)}$. But non-trivial $C^{(6)}$ also
implies a non-trivial $C^{(2)}$ background since $dC^{(2)}=*dC^{(6)}$. A
D-instanton will carry D1-dipole moment and so a coupling to the
$C^{(2)}$ background field. The couplings can be deduced from the
D3-$C^{(6)}$ coupling established above by
successive T-dualities reducing the D3-branes to D-instantons. The leads
to the following couplings in the D-instanton action:
\EQ{
\frac1{3g_0^2}F^{(3)}_{abc}{\rm tr}_k\,\chi_a\chi_b\chi_c
+\frac1{g_0^2}F^{(3)}_{abc}{\rm tr}_k\,\bar\Psi\Gamma_a\Gamma_b\Gamma_c\Psi\ ,
\elabel{dicou}
}
where $\Psi$ is the 16-component fermion defined in \eqref{spc2} and
$\Gamma_a$ are components of the ten-dimensional $\Gamma$-matrices
\eqref{cliftd} in the six-dimensional space transverse to the D3-branes.
The field strength can be deduced by taking the dual of \eqref{ftot}:
\EQ{
F^{(3)}=6\tilde T\ ,
}
where $\tilde T$ is equal to the form of $T$
with $m_{AB}$ replaced by $-m_{AB}$.
The couplings \eqref{dicou} are then simply
\EQ{
\frac{\pi^2}{g^2}{\rm tr}_k\,\Big\{-m_{AB}\CM^{\prime A}\CM^{\prime B}
+(4\pi\alpha')^2\big(2\tilde T_{abc}\chi_a\chi_b\chi_c+m^{AB}\bar\psi_A
\bar\psi_B\big)\Big\}\ .
\elabel{yesm}
}
Notice that the couplings involving the vector multiplet
$\{\chi_a,\bar\psi_A\}$
do not survive in the decoupling limit $\alpha'\to0$ (fixed
$g$). In this limit, the only surviving term is the mass coupling
for $\CM^{\prime A}$ which is identical to that in \eqref{nonepn}
(after the re-scaling $\CM^A\to g^{1/2}\CM^A$).
In order to completely reproduce \eqref{nonepn},
the hypermultiplet fermions $\{\mu^A,\bar\mu^A\}$, arising from open
string stretched between the D-instantons and D3-branes must also
couple to $C^{(2)}$ in an analogous way dictated by their $\SO(6)$
transformation properties.

\newpage

\rsen\section{Further Directions}\label{sec:S562}

What we have said in the preceding Chapters is far from the end of
the story of the calculus of many instantons. There are two main
developments that we turn to in this final Chapter. The first involves
the fate of instantons in a rather esoteric generalization of the
underlying gauge theory, or rather on the spacetime on which it is
defined. The idea is to define the theory on a
non-commutative version of ${\mathbb R}^4$. This non-commutative space
is characterized by the fact that the spacetime position coordinates
$x_m$ no longer commute. The motivation for consider such a
generalization comes from string theory. We have already described how
gauge theories can arise as the collective dynamics of D-branes in
string theory and turning on certain background fields can make the
resulting world-volume non-commutative. What is remarkable is that the
non-commutativity of spacetime does little violence to instantons. In
particular the moduli space is still a commutative space. In fact it
is a very simple deformation of $\ms_k$ obtained by taking non-zero
central terms for the $\U(1)\subset\U(k)$ factor in the hyper-K\"ahler
quotient construction. In particular the deformed space, which we
denote $\ms^{(\zeta)}_k$ is still hyper-K\"ahler and rather remarkably
it is a smooth resolution of the original space: instantons can no
longer shrink down to zero size due to the non-commutativity.
In addition, instantons now become non-trivial in theories with an
abelian gauge group like QED.

The second development we describe in \S\ref{sec:N2}
is a new way to calculate instanton
effects beyond a single instanton, or two for $\SU(2)$,
when scalar fields have VEVs. As we have seen, one must use
the constrained instanton formalism of Affleck. To leading order in
the semi-classical approximation, we have seen that the effect of the
VEVs is to turn on a non-trivial instanton effective action which acts
as a potential on the instanton moduli space. We will show in the
context of the calculation of the prepotential in $\N=2$ theories that
the resulting integrals over the instanton moduli space have the
remarkable property that they localize around the critical points of
the instanton effective action.
This is guaranteed by the existence of a nilpotent
fermionic symmetry, or BRST operator \cite{Dorey:2001zq}. In fact the
critical points correspond to configurations where all the instantons
shrink down to zero size. The singular nature of this configuration
may be regularized by making the spacetime non-commutative since this
removes the singularities of the instanton moduli space.
Using this technology we will be able to rather swiftly,
following \cite{local},
re-derive the one-instanton contribution to the
prepotential in $\SU(N)$ that we calculated in \S\ref{sec:N3} and then
extend the method to two instantons. Very recently this technique of
localization has for the first time enabled a calculation of instanton
effects to all orders in the instanton charge for finite $N$
\cite{locn4}. The relevant theory is the $\N=4$ theory softly broken
to $\N=2$ by giving mass $m$ to two of the three adjoint-valued chiral
multiplets and the physical quantity that can be calculated to all
orders in the instanton expansion are the ${\cal O}(m^4)$ terms in the
prepotential. Clearly the new technique of localization looks very
promising and we refer the reader to \cite{local,locn4} for more details.

\subsection{Non-commutative gauge theories and instantons}\label{sec:S129}

It is not at all obvious but the rather bizarre generalization of
gauge theories defined on a
non-commutative background spacetime has a rather pleasant and simple
effect on instantons. Non-commutativity is also of special interest
in the framework of string theory and D-branes
\cite{DHull,Chu:1999qz,Schomerus:1999ug,SWnc} (see Ref.~\cite{DNrev}
for a review). Here we will briefly examine non-commutativity from the
instanton perspective.

We have already seen in \S\ref{sec:S15} that the moduli space $\ms_k$
of instantons in ordinary commutative gauge theories fails to be a
smooth manifold due to
conical singularities arising when the auxiliary symmetry group
$\U(k)$ does not act freely. Physically these are points
where individual instantons shrink to zero size, {\it i.e.\/}~$w_{iu\aD}=0$
for a given $i\in\{1,\ldots,k\}$.
There is a natural way to resolve, or blow up, the singularities of
$\ms_{k}\to\ms_{k}^{(\zeta)}$ preserving the
hyper-K\"ahler structure described by Nakajima
\cite{Nakajima:1993jg,Nakajima:1996ka}. The important observation is that the
quotient group $\U(k)$ has an abelian factor and, as we explain in
Appendix \ref{app:A2}, one has the freedom to add to each of the three
moment maps a constant term in the Lie algebra of any abelian
factor. In the context of the ADHM construction this freedom
involves modifying the ADHM constraints \eqref{fconea}
by adding a term proportional to
the identity $k\times k$ matrix to the right-hand side as in \eqref{mommap}:
\EQ{
\vec\tau^\aD{}_{\bD}\,\bar a^\bD a_\aD=\vec\zeta_{\sst(+)}
1_{\sst[k]\times[k]}\ .
\label{madhm}
}
Now consider the clustering limit in which the $i^{\rm th}$ instanton is
well separated from the remaining. In this limit,
the effective ADHM constraints of
the single instanton are modified from \eqref{oiladhm} to
\EQ{
\bar w^\aD_{iu}
w_{ui\bD}=\rho_i^2\delta^\aD{}_\bD+\tfrac12\vec\tau^\aD{}_\bD
\cdot\zeta_{\sst(+)}\ ,
}
where we recall that $\rho_i$ is a measure of the size of the
instanton. Without-loss-of-generality,
suppose $\vec\zeta_{\sst(+)}\propto=(0,0,1)$, then
it is easy to see that
\EQ{
\rho_i^2\geq \tfrac12|\vec\zeta_{\sst(+)}|\ .
}
Therefore in the presence of the central term, an instanton can no
longer shrink to zero size and the singularity in the instanton moduli
space corresponding to that process is smoothed out. For instance, the
$\U(1)\subset\U(k)$ corresponding to phase rotation of $w_{ui\aD}$ for
fixed $i$, no longer a fixed point. In fact, one can show that the
whole of the auxiliary group $\U(k)$ no longer has any fixed points on
$\ms_k^{(\zeta)}$.

The question is what physical effect can introduce
the central term in the ADHM constraints. Remarkably,
we will see that precisely this
smoothed instanton moduli space $\ms_{k}^{(\zeta)}$, defined by \eqref{madhm},
arises in the ADHM construction of instantons in gauge theories formulated
on a spacetime with non-commuting coordinates \cite{NS}:
\EQ{
[x_m, x_n]=i\theta_{mn}\ .
\label{crxdef}}
The 3-vector $\vec\zeta_{\sst(+)}$  is precisely the anti-self-dual projection
of the non-commutativity parameter $\theta_{mn}$. We also define the
self-dual projection which appears in the ADHM constraints of anti-instantons:
\EQ{
\zeta_{\sst(+)}^c \equiv \bar{\eta}^c_{mn} \theta_{mn}\ ,
 \qquad
\zeta_{\sst(-)}^c \equiv \eta^c_{mn}\theta_{mn} \ ,
\qquad c=1,2,3 \ . \label{zedefs}
}
where $\eta^c_{mn}$ and
$\bar\eta^c_{mn}$ are 't~Hooft $\eta$-symbols (see Appendix \ref{app:A1}).

It is quite remarkable that non-commutative instantons are non-trivial
even for an abelian $\U(1)$ gauge group. In this case, the deformed instanton
moduli space $\ms^{(\zeta)}_k$ is simply a resolution of the space
${\rm Sym}^k\,{\mathbb R}^4$.\footnote{This is the $k$-fold symmetric
product of ${\mathbb R}^4$ which arises from solving the ADHM
constraints \eqref{badhm} for $N=1$. The solution is $w_{i\aD}=0$ and
$a'_n=-{\rm diag}(X^1_n,\ldots,X^k_n)$, $X^i\in{\mathbb R}^4$,
which fixes all the $\U(k)$
symmetry apart from permutations of the positions $X^i_n$. Modding out
by the permutations gives the $k$-fold symmetric product of ${\mathbb
R}^4$. This space is singular whenever two, or more, of the point-like
instantons come together, since then the group of permutations does
not act freely.}
In fact, the case of an abelian gauge group, considered by Nekrasov
and Schwarz in Ref.~\cite{NS},  provided
the first examples of explicit non-commutative instanton configurations.
Various aspects of the ADHM construction on non-commutative spaces were
further developed and clarified in
Refs.~\cite{Furuuchi123,LTY,KLY12,Nekrasov:2000zz,Schwarz:2001ru,Chu:2001cx,Hamanaka:2001dr}.
Explicit examples of instanton solutions
in $\U(N)$ gauge theories with space-space as well as space-time
non-commutativity were constructed and analysed
in \cite{Chu:2001cx}. In general, for $|\vec\zeta|
>0,$ the moduli
spaces of these solutions contain no singularities. However,
for semi-classical functional integral applications it is
important to ensure that the instanton gauge field itself is also
non-singular---or to be more precise
is gauge equivalent to a non-singular configuration---for
all values of the argument \cite{Chu:2001cx}.
The supersymmetric collective-coordinate measure for non-commutative
instantons was constructed and the one-instanton partition function calculated
in Ref.~\cite{Dorey:2001zq}.
Based on these results, instanton contributions to
the prepotential of the non-commutative $\N=2$ pure gauge theory
were determined in \cite{Hollowood:2001ng} and found to be
equivalent to the corresponding commutative contributions confirming
the hypothesis that the non-commutative version of the theory is
described by the same Seiberg-Witten curve
\cite{Hollowood:2001ng,Armoni:2001br}.

\subsubsection{ADHM construction on non-commutative ${\mathbb R}^4$}

We will work in flat Euclidean spacetime ${\mathbb R}^4$ with
non-commutative coordinates $x_m$ which satisfy the commutation
relations \eqref{crxdef} where
$\theta_{mn}$ is
an antisymmetric real constant matrix. Using Euclidean spacetime
rotations, $\theta_{mn}$ can be always brought to the form
\EQ{
\theta_{mn}=
\begin{pmatrix} 0 & \theta_{12} & 0 & 0 \\
-\theta_{12}& 0 & 0 & 0 \\ 0 & 0 & 0 &\theta_{34}\\
 0 & 0 & -\theta_{34}& 0 \end{pmatrix} \ .
\label{thgen}}
In terms of complex coordinates
\SP{
z_1 &= x_2+ i x_1 \ , \quad  \zb_1 = x_2- i x_1 \ , \\
z_2 &=  x_4+ ix_3 \ , \quad  \zb_2 = x_4- ix_3 \ ,
\label{zzbdef}}
the commutation relations \eqref{crxdef} take the form
\SP{
[z_1,\zb_1] &= - 2\th_{12} \ , \quad [z_i,z_j] = 0 \ , \\
[z_2,\zb_2] &= - 2\th_{34}  \ , \quad [z_i,\zb_{j\neq i}]=0 \ ,
\label{crss}}
where $i,j =1,2$.
Besides the usual commutative case, there are two important cases to consider:

{\bf 1.} When either $\theta_{12}$ or $\theta_{34}$ vanishes,
the matrix $\theta_{mn}$ is of rank-two. This case corresponds to
the direct product of the ordinary commutative 2-dimensional space with
the non-commutative 2-dimensional space,
${\mathbb R}_{\rm NC}^2\times {\mathbb R}^2$.
For definiteness we set here $\theta_{34}=0$
and introduce the notation $\theta_{12}\equiv -\zeta/2$ in such a way  that
\EQ{[z_1,\zb_1] = - \zeta \ , \quad [z_2,\zb_2] = 0\ ,
\quad [z_i,z_j] = 0 \ .
\label{crnc2}}
Physical applications
of this situation involve gauge theories defined on a background with
non-commutative space but commutative time.

{\bf 2.} A rank-four matrix $\theta_{mn}$ (with $\theta_{12}\neq 0$ and
$\theta_{34}\neq 0$) generates the non-commutative Euclidean spacetime
${\mathbb R}_{\rm NC}^4={\mathbb R}_{\rm NC}^2\times {\mathbb R}_{\rm NC}^2$.
The corresponding world-volume gauge theory has
non-commutative (Euclidean) time. Since both components of $\theta$ are
non-vanishing, they can be made equal,
$\theta_{12}=\theta_{34}\equiv -\zeta/4,$
with appropriate re-scalings of the four
coordinates $x_m$ and, if necessary, a parity transformation.
Equations \eqref{crss} become
\EQ{
[z_i,\zb_j] = - \frac{\zeta}{2} \delta_{ij} \ , \quad [z_i,z_j] = 0\ .
\label{crsd}}
In fact, the conditions $\theta_{12}=\pm\th_{34}$ corresponds to
(anti-)self-duality: ${\hf}\epsilon^{mnkl}\theta_{kl}=\pm\theta_{mn}$.

A Hilbert space representation for the non-commutative geometry \eqref{crnc2}
or \eqref{crsd} can be easily constructed by using complex variables
\eqref{zzbdef} and realizing
$z$ and $\zb$ as creation and annihilation operators in the Fock space
for simple harmonic oscillators (SHO). The fields in a non-commutative
gauge theory are described by functions
of $z_1,\zb_1,z_2,\zb_2$. In the case of
${\mathbb R}_{\rm NC}^2\times{\mathbb R}^2$,
the arguments $z_2$ and $\zb_2$ are ordinary c-number coordinates, while
$z_1$ and $\zb_1$ are the creation and annihilation operators of a single
SHO:
\EQ{
z_1 \ket{n}
= \sqrt{\z}\; \sqrt{n+1} \ket{n}\ , \quad
\zb_1 \ket{n}
= \sqrt{\z}\; \sqrt{n} \ket{n-1} \ .
}
The non-commutative spacetime
${\mathbb R}_{\rm NC}^4={\mathbb R}_{\rm NC}^2\times{\mathbb R}_{\rm NC}^2$
requires two oscillators. The SHO Fock space
$\cH$ is spanned by the basis $\ket{n_1,n_2}$ with $ n_1,n_2 \geq 0 $:
\SP{
&z_1 \ket{n_1,n_2}
= \sqrt{\frac{\z}{2}}\; \sqrt{n_1+1} \ket{n_1+1,n_2}\ , \quad
z_2 \ket{n_1,n_2}
= \sqrt{\frac{\z}{2}}\; \sqrt{n_2+1} \ket{n_1,n_2+1}\ ,\\
&\zb_1 \ket{n_1,n_2}
= \sqrt{\frac{\z}{2}}\; \sqrt{n_1} \ket{n_1-1,n_2}\ , \quad
\zb_2 \ket{n_1,n_2}
= \sqrt{\frac{\z}{2}}\; \sqrt{n_2} \ket{n_1,n_2-1}\ .
}
The integral on ${\mathbb R}_{\rm NC}^4$ is defined by the operator trace,
\EQ{
\int d^4 x\ * = (2\pi)^2 \sqrt{\det\,
\th} \; \mbox{Tr}\ * = \Big(\frac{\z \pi}{2}\Big)^2 \;
\mbox{Tr}\ *\ .
}

The ADHM construction of the instanton
now proceeds exactly as in Section \S\ref{sec:S8} but one has to keep
an eye on the ordering of operators. We assume the canonical form
\eqref{aad}, \eqref{aab} for the ADHM matrices and analyse
the requirements imposed by the factorization condition \eqref{dbd}.
They amount to the modified ADHM constraints
({\it cf\/}. \eqref{badhm})
\EQ{
\vec\tau^{\aD}{}_{\bD}\,\bar a^\bD a_\aD
=\vec\zeta_{\sst(+)}1_{\sst[k]\times[k]}\ .
\label{fconeam}
}
The three conditions \eqref{fconeam}
are the modified ADHM constraints for the instanton.
When $\vec\zeta_{\sst(+)}=0$ Eqs.~\eqref{fconeam} give the standard
commutative ADHM constraints \eqref{fconea}.
When non-commutativity is present, the
constraints
are modified by the anti-self-dual component of $\th$.
Thus, the ADHM constraints for the instanton in a self-dual-$\th$ background
on  ${\mathbb R}_{\rm NC}^4$ are equal to those of commutative
${\mathbb R}^4$.
However, the constraints for the instanton in non-commutative space
${\mathbb R}_{\rm NC}^2\times {\mathbb R}^2$ are always modified since
$\theta$ cannot be self-dual. The constraints are
\EQ{
\tau^{c\,\aD}{}_{\bD}\,\bar a^\bD a_\aD
=\delta^{c3} \zeta1_{\sst[k]\times[k]}\ .
\label{sdincs}
}

The constraints for an anti-instanton follow from solving
the same factorization condition \eqref{dbd} with the
matrix $\Delta=a +b\bar{x}.$
In this case the ADHM constraints are modified by
the self-dual component of $\th$:
\EQ{
\vec\tau^{\aD}{}_{\bD}\,\bar a^\bD a_\aD
=\vec\zeta_{\sst(-)}1_{\sst[k]\times[k]}\ .
\label{asdca}
}

The explicit one-instanton and one-anti-instanton solutions for
$\U(N)$ gauge theory were constructed in \cite{Chu:2001cx} for space-space,
${\mathbb R}_{\rm NC}^2\times {\mathbb R}^2,$ and spacetime,
${\mathbb R}_{\rm NC}^4,$ non-commutativity by resolving the corresponding
(modified) ADHM constraints and solving the completeness relation
\eqref{cmpl}. As always, one has to distinguish between two types
of singularities: the singularities of the instanton field $A_m(x)$
as the function of the argument, and the singularities arising
for certain values of the collective coordinates interpreted
as singularities on
the moduli space. Instanton configurations can be determined in singular
or regular gauges; in singular gauge all the singularities of instanton
configurations as functions of $x_m$
are simply gauge artifacts and can be gauged away.

Instantons with space-space non-commutativity arise from
\eqref{sdincs} where $\zeta\neq 0$ and their moduli space contains no
singularities. Instantons with spacetime non-commutativity
contain no singularities on the moduli space unless the the
self-duality
of $\theta$ coincides with the self-duality of the instanton field-strength
such that the ADHM constraints \eqref{fconeam} and \eqref{asdca}
collapse to \eqref{fconea}.

\subsubsection{The prepotential of non-commutative
$\N=2$ gauge theory}\label{sec:S1112}

We can now consider the $\N=2$ supersymmetric $\U(N)$ gauge theory
formulated in non-commutative space. On the Coulomb branch, the
gauge group is broken to $\U(1)_{\rm c} \times\U(1)^{N-1},$ where
$\U(1)_{\rm c}$ is the overall $\U(1)$ factor of the non-commutative
$\U(N)$ gauge group which decouples in the infrared due to the
infra-red/ultra-violet
mixing as explained in Ref.~\cite{Hollowood:2001ng}. Our
goal is to determine the low-energy dynamics of the $\U(1)^{N-1}$ factor.

In a similar way to the ordinary
commutative case reviewed in \S\ref{sec:S60},
the corresponding low-energy effective action
is determined by the Seiberg-Witten prepotential
${\cal F}$ as explained in
Refs.~\cite{Hollowood:2001ng,Armoni:2001br}. In particular, note
there are no commutative ``star-products''
in the low energy effective action since we are concerned
only with the leading-order terms in the derivative expansion of the
effective action. As in the commutative theory,
instantons and anti-instantons contribute to various correlation
functions. By considering the instanton contributions to these
correlation functions one can relate the instanton coefficients of the
prepotential to the centred instanton partition function. We shall
assume that the re-scalings \eqref{nicer} have been performed and so
the relation is identical to \eqref{ndeq}. Although we did not
consider the anti-instanton contributions in Chapter \ref{sec:S60},
one can easily derive an anti-instanton version of
\eqref{ndeq} which involves the complex conjugate of the
prepotential. Summarizing, we have for $k>0$
\EQ{
{\cal F}_k=\widehat{\EuScript Z}_{+
k}^{\sst(\N=2,N_F)}\ ,\qquad{\cal F}^*_k=\widehat{\EuScript Z}_{-
k}^{\sst(\N=2,N_F)}\ .
\label{pqra}
}
Here, $\widehat{\EuScript Z}_{\pm k}^{\sst(\N=2,N_F)}$ are the
instanton and anti-instanton centred instanton partition functions of
the non-commutative theory and defined as integrals over the moduli
spaces $\widehat\ms_{\pm k}^{(\zeta)}$:
\EQ{
\widehat{\EuScript
Z}_{\pm k}^{\sst(\N=2,N_F)}=\int_{\widehat\ms_{\pm k}^{(\zeta)}}
\Bomega^{\sst(\N=2,N_F)}e^{-\tilde S}\ ,
}
where, the supersymmetric volume form over the resolved centred instanton
moduli space is explicitly
\SP{
&\int_{\widehat\ms_{\pm k}^{(\zeta)}}\Bomega^{\sst(\N)}
=2^{\N-2}\pi^{2(\N-1)}
\frac{C_k^{\sst(\N)}}{{\rm Vol}\,\U(k)}\int\, d^{4k(N+k)-4} \hat a \,
\,\prod_{A=1}^\N\,
 d^{2k(N+k)-2}\hat{\cal M}^{A}\ \big|\det_{k^2}\,\BL\big|^{1-\N}\\
&\qquad\times\,\prod_{r=1}^{k^2}\,\bigg\{
\prod_{c=1}^3\,
\delta\big(\tfrac12{\rm
tr}_k\,T^r(\tau^c{}^\aD{}_\bD \bar a^\bD
a_\aD-\zeta^c_{\sst(\pm)}1_{\sst[k]\times[k]})\big)
\prod_{A=1}^\N\prod_{\aD=1}^2\,\delta\big({\rm
tr}_k\,T^r(
\bar{\cal M}^Aa_\aD+\bar a_\aD{\cal M}^A)\big)\bigg\}\ .
\elabel{nfmsc}
}
The expression is identical to the expression for the supersymmetric
volume form of $\widehat\ms_k$ defined in
\S\ref{sec:N1} apart from the fact that the bosonic ADHM
constraints are modified appropriately to include the central terms.

{}From \eqref{nfmsc} we conclude that the centred instanton partition
function can only depend on $\vec\zeta_{\sst(+)}$ while the
centred anti-instanton function can only depend on
$\vec\zeta_{\sst(-)}$. Given \eqref{pqra}, the aforementioned
dependences are very restrictive: the prepotential cannot depend on
$\vec\zeta$ and therefore should be identical to that in the
commutative theory. This was the hypothesis that was made in
Ref.~\cite{Hollowood:2001ng} for the theory with no hypermultiplets.
In the next section, we will develop a localization
formalism in which we can prove
rigorously that the centred (anti-)instanton partition function cannot
depend smoothly on $\vec\zeta_{\sst(\pm)}$. There are possible
discontinuities when $\vec\zeta_{\sst(\pm)}=0$, since at this points
the (anti-)instanton moduli space becomes singular and the
localization formalism breaks down.
However, we can already test this hypothesis at the one-instanton level by
re-doing the calculation of \S\ref{sec:N3} in the context of the
non-commutative theory. It is simple to establish the modifications
that arise from including the non-commutativity parameter
$\vec\zeta$. The instanton effective action \eqref{iean2k1} has the
additional coupling
\EQ{
4\pi^2i\vec D\cdot\vec\zeta_{\sst(+)}\ .
}
This modifies the $\vec D$ integral in \eqref{jyy}
\EQ{
\int\frac{d^3D}{\vec D^2}e^{-4\pi^2i\vec D\cdot\vec\zeta_{\sst(+)}}
\prod_{u=1}^N\frac{\alpha^{*2}_u}{
|\alpha_u|^4+\vec D^2}\ .
}
The angular integrals over $\vec D=(|\vec D|,\theta,\phi)$ are now non-trivial:
\EQ{
\int d(\cos\theta)d\theta e^{-4\pi^2i\vec D\cdot\vec\zeta_{\sst(+)}}
=\frac{\sin(4\pi^2|\vec D||\vec
\zeta_{\sst(+)}|)}{\pi|\vec D||\vec\zeta_{\sst(+)}|}\
.
}
We then extend the integral over $|\vec D|$ from $-\infty$ to
$+\infty$ and perform it by standard contour integration yielding
\EQ{
\frac1{2|\vec\zeta_{\sst(+)}|}
\Big(\prod_{u=1\atop(\neq
u)}^N\frac1{\alpha_u^2}-\sum_{u=1}^N\frac{1}{\alpha_u^2}
e^{-4\pi^2|\vec\zeta_{\sst(+)}|\alpha_u^2}
\prod_{v=1\atop(\neq
u)}^N
\frac{\alpha^{*2}_v}{|\alpha_v|^4-|\alpha_u|^4}\Big)\ .
\label{jyym}
}
compared with the right-hand side of \eqref{jyy}. The behaviour of the
modified integrand in the vicinity of one of the $N$ singularities
$\chi=-\phi_u^0$, {\it i.e.\/}~$\alpha_u=0$, is
\EQ{
2\pi^2\frac{\alpha_u^*}{\alpha_u}\prod_{v=1\atop(\neq
u)}^N
\frac{\alpha^{*2}_v}{|\alpha_v|^4-|\alpha_u|^4}+\cdots
}
and so unmodified from the commutative calculation. The contribution
to the centred instanton partition function from the $N$ singularities
is therefore unchanged from \eqref{contsing}. However, the
contribution from the sphere at infinity in $\chi$-space is modified
since in the non-commutative case the asymptotic behaviour of the
integrand is changed. In fact contrary to \eqref{bkol} we now have
\EQ{
f_1(r,\theta)\thicksim \frac{e^{-2iN\theta}}{r^{2N}}
}
and consequently the contribution from the sphere at infinity vanishes
because either the integrand $f_1(r,\theta)f_2(re^{i\theta})$
falls off too fast, for $N_F<2N$, or due
to a vanishing angular integral, for $N_F=2N$.
Hence, in the non-commutative theory the result for the one-instanton
coefficient of the prepotential is
\EQ{
{\cal F}_1^{\text{nc}}=
\sum_{u=1}^N\prod_{v=1\atop(\neq
u)}^N\frac1{(\phi^0_v-\phi^0_u)^2}\,\prod_{f=1}^{N_F}
(m_f+\phi^0_u)\ .
\elabel{bfoim}
}
compared with \eqref{bfoi} in the commutative case.

Notice that the additional contribution denoted ${\cal
S}_1^{\sst(N_F)}$ which we identified as arising form the vicinity of
the singularity on the instanton moduli space, is missing from the
non-commutative result, as might have been expected since the
singularity of $\widehat\ms_1$ has been resolved. At the one-instanton
level, we see that the commutative and non-commutative results are
equivalent up to an unimportant constant for $N_F<2N$. This is
entirely consistent with the hypothesis that commutative and
non-commutative theories have the same $\SU(N)$
low-energy effective action. However, the finite theory with
$N_F=2N$ does have the apparently physically significant difference
due to the term proportional to $\sum_{u=1}^N(\phi_u^0)^2$ that arises
in the commutative theory \eqref{bfoi} but which is missing from
\eqref{bfoim}. This difference, however, is not physically relevant
since it can be explained by a re-definition of the coupling between
the commutative and non-commutative theories. This follows from the
expression for the prepotential in \eqref{preci}. If the coupling in
the non-commutative theory is related to that in the commutative
theory by a series of instanton couplings:
\EQ{
\tau^{\text nc}=\tau+\sum_{k=1}^\infty c_ke^{2\pi ik\tau}\ ,
\label{matco}
}
then $\SU(N)$ part of the
the low-energy of the non-commutative theory is equal to that of the
commutative theory. At the one-instanton level we have
\EQ{
c_1=\frac{\alpha_2}{\pi i}
\label{conett}
}
and $\alpha_2$ is the coefficient in \eqref{defaot}.
Matching the non-commutative and commutative theories requires for
higher instanton number a relation of the form
\EQ{
{\cal F}_k^{\text{nc}}={\cal F}_k-2\pi id_k\sum_{u=1}^N(\phi_u^0)^2\ ,
\label{funnyrel}
}
up to irrelevant (mass-dependent)
constants, where $d_k$ are determined in terms of the $c_k$.
It is difficult to prove this relation
rigorously. However, in the next section we prove that ${\cal
F}_k^{\text{nc}}$ cannot depend smoothly on $\vec\zeta_{\sst(+)}$,
hence, the difference ${\cal F}_k^{\text{nc}}-{\cal F}_k$ cannot depend
on $\vec\zeta_{\sst(+)}$. Then assuming that the contribution to the
difference arises from the sphere at infinity in
$\chi$-space,\footnote{This should follow from the localization
technology developed in the next section.}
generalizing the situation at the one-instanton level, the result can
be argued to be polynomial in the masses and VEVs. Given that the
prepotential has mass-dimension 2, this leaves \eqref{funnyrel} as the only
possible VEV dependence.

It should not have escaped the readers notice that something
unexpected happens in the finite theory when the non-commutativity
parameters are chosen so that either $\vec\zeta_{\sst(+)}=0$ or
$\vec\zeta_{\sst(-)}=0$. In the former case, the instantons are those
of the non-commutative theory, described by the smooth moduli space
$\ms_1^{(\zeta)}$, while the anti-instantons are
conventional commutative ones, described by the singular space
$\ms_1$. In this case, even at the
one-instanton level \eqref{pqra} is violated since the anti-instantons
receive the contribution ${\cal S}_1^{(\sst(N_F=2N)}$
from the singularity on the instanton moduli space, while the
anti-instantons do not; hence
\EQ{
\widehat{\EuScript Z}_{+
1}^{\sst(\N=2,N_F)}\neq \Big(\widehat{\EuScript Z}_{-
1}^{\sst(\N=2,N_F)}\Big)^*\ .
}
In addition, a re-definition of the form \eqref{matco}
cannot reconcile the commutative and non-commutative theories. It is
not clear how one should describe
what is really happening at these points in the non-commutativity
parameter space and whether the resulting behaviour of the
low-energy effective action is physically acceptable.

\subsection{Calculating the prepotential by localization}\label{sec:N2}

The major applications of the many instanton calculus that we have
hitherto reported (the two-instanton contribution to the $\SU(2)$
prepotential in \S\ref{sec:S842} excepted) have
involved a large-$N$ limit. Clearly one would like
techniques to calculate instanton effects for any
$N$ and $k$. The problem is obvious: the integrals
over the instanton moduli space are just too complicated to be done by
brute force beyond instanton number one (or two in the special case of
gauge group $\SU(2)$). Even the integral over the one-instanton moduli
space required to obtain the first instanton coefficient of the
prepotential with $\SU(N)$ gauge group, undertaken in \S\ref{sec:N3}
was far from elementary.

However there is hope: there is interesting mathematical structure
underlying the integrals over the instanton moduli space that define
the instanton partition function and hence the instanton coefficients
of the prepotential.
They are related to integrals that give topological invariants; for example, in
the $\N=4$ case the integral is precisely the
Gauss-Bonnet-Chern integral that---at least on a compact
manifold---gives the Euler characteristic.\footnote{We do not discuss
the mass-deformed $\N=4$ to $\N=2$ theory in this review: see
\cite{Dorey:2001zq}.}
When the theory is on the Coulomb branch
instantons are constrained and there is a non-trivial potential on the
instanton moduli space caused by the non-trivial instanton effective action.
This acts as a Morse potential and the
integrals localize around various subspaces of the instanton
moduli space around which the quadratic approximation is exact. This
kind of localization is a generalization of the Duistermaat-Heckman
Integration Formula that arises in various supersymmetric contexts
(see, for example,
\cite{Blau:1993pm,Niemi:1994ej,Blau:1995rs,Schwarz:1997dg,berline} and
references therein). The obvious problem to
using this theory is the fact that the instanton moduli
space is non-compact and has singularities when instantons shrink to zero size.
The problem with the singularities can be
be alleviated by considering the resolution of the instanton moduli
space $\ms^{(\zeta)}_k$ relevant to a non-commutative theory as described in
\S\ref{sec:S129}. It will turn out that the integrals over the
instanton moduli space in the non-commutative version of the theory on
the Higgs or Coulomb branch is amenable to a form of
localization. This will allow us to calculate the two-instanton
contribution to the prepotential in an $\SU(N)$ (or rather $\U(N)$
since in the non-commutative version the additional abelian factor
is inevitably included) theory. In fact the localization method
greatly simplifies the integrals even at the $k>2$ level and we expect
that further progress may be possible.

In retrospect it should not be a surprise that the instanton partition
function localizes in the non-commutative theory. We will show that
the subspaces of $\widehat\ms_k^{(\zeta)}$ on which the instanton
partition function localizes correspond to products of
(non-commutative) $\U(1)$ moduli spaces
$\widehat\ms_{k}^{(\zeta)}\big|_{N=1}$.
In fact the localization
submanifolds are associated to each partition of $k\to
k_1+\cdots+k_N$, $k_u\geq0$, and have the form
\EQ{
\widehat\ms_{k}^{(\zeta)}\longrightarrow
\bigcup_{\text{partitions}\atop k_1,\ldots,k_N}
\frac{\ms^{(\zeta)}_{k_1}\big|_{N=1}\times
\cdots\times\ms^{(\zeta)}_{k_N}\big|_{N=1}}{{\mathbb R}^4}\ ,
\label{gfps}
}
where the quotient is by the overall centre of the
instanton. This is a manifestation of the
fact that in the non-commutative theory there are exact---no longer
constrained---instanton solutions on the Coulomb branch. The reason is
that the non-commutativity prevents instantons shrinking to zero size
and there are now stable solutions of infinite size and the
consequences of Derrick's theorem are avoided. The fact that these
solutions are exact even in the presence of a VEV
follows because the non-trivial fields only take
values in the $\U(1)^N$ subgroup of the $\U(N)$ gauge group and so
these fields commute with the VEVs. The partitions correspond to the
instanton charge $k_u$ in each of the $N$ $\U(1)$ subgroups.
It is then an issue as how one can relate the prepotential in the
non-commutative theory to that in the commutative theory. We have
already argued in \S\ref{sec:S1112} that they must essentially be the
same. Our calculations at the one- and two-instanton level confirm
this hypothesis, up to an interesting re-parameterization of the
coupling in the theory with $N_F=2N$.

Before proceeding with the more convenient formalism
(from the point-of-view of performing actual
computations) based on the linearized instanton partition
function constructed in \S\ref{sec:N1}, it is useful to consider
the problem in terms of the intrinsic geometry of
$\ms_k$. We first note in the $\N=2$ theory,
the number of Grassmann collective coordinates (we consider only the
case $N_F=0$ for the moment) equals the dimension of
$\ms_k$; namely,
$4kN$. We have already seen in \S\ref{sec:S29}
that the Grassmann collective coordinates $\psi^A$,
for each species $A$, can be thought of as
the components of Grassmann-valued symplectic
tangent vectors to $\ms_k$. In the $\N=2$ case, there is also an
isomorphism between them and the symplectic basis of one forms
\EQ{
\psi^{\xii A}\leftrightarrow h^{\xii\aD}\ ,\qquad\aD\equiv A\ .
\label{gripa}
}
Here, the one forms $h^{\xii\aD}$ manifest the $\Sp(n)\times\SU(2)$
structure of the (co-)tangent space of $T\ms_k$ and
are related to the inverse Vielbeins via
$h^{\xii\aD}=h^{\xii\aD}{}_\mu dX^\mu$ (see Appendix~\ref{app:A2}).
Note that the isomorphism \eqref{gripa} preserves the structure of
differential calculus: both Grassmann variables and one forms
anti-commute. In addition, the $\N=2$ supersymmetric volume form on
$\ms_k$ \eqref{susmes} can be re-interpreted in terms
of differential forms:
\SP{
&\int_{\ms_k}\Bomega^{\sst(\N=2)}{\cal F}(X,\psi^{\xii A})=
\pi^{-2kN}
\int_{\ms_k}\Big\{ \bigwedge_{\xii=1}^{2kN}\bigwedge_{\aD=1}^2
h^{\xii\aD}\Big\}
\,\Big\{\prod_{\xii=1}^{2kN}\prod_{A=1}^2 \PD{}{\psi^{\xii A}}\Big\}\,{\cal
F}(X,\psi^{\xii A})\\
&=\pi^{-2kN}\int_{\ms_k}{\cal F}(X,h^{\xii\aD})\qquad
(\aD=A)\ ,
}
where the function ${\cal F}$ is expanded in wedge powers of the
forms $h^{\xii\aD}$ until a top form is obtained which can then be
integrated .

Using the isomorphism between the Grassmann collective coordinates and
one forms, the instanton effective action \eqref{geoia} for $\N=2$ (so
the final curvature coupling is absent) can be written succinctly as
the inhomogeneous differential form
\EQ{
\tilde S=-\frac14d_V\big(V^\dagger_\mu dX^\mu\big) .
\label{amaza}
}
Here, $V$ and $V^\dagger$ are the (anti-)holomorphic components
(defined as in \eqref{ntsc}) of the (commuting)
tri-holomorphic vector fields on $\ms_k$ denoted $V_a$, $a=1,2$,
introduced in \S\ref{sec:S401}, associated to the VEVs.
The other quantity in \eqref{amaza} is an {\it equivariant\/} exterior
derivative\footnote{Here, $\imath_V$ implies contraction with the vector $V$.}
\EQ{
d_V\equiv d-2\imath_V\ .
}
One can easily verify that
\EQ{
d_V^2=2{\cal L}_V\ ,
}
the Lie derivative with respect to $V$. Hence, $d_V$ is nilpotent
on $\SU(N)$-invariant differential forms. The centred instanton
partition function can then be written in form notation as
\EQ{
\widehat{\EuScript
Z}_k^{\sst(\N=2)}=\pi^{-2kN}\int_{\widehat\ms_k}\exp\Big(-\tfrac14
d_V(V^\dagger_\mu dX^\mu)\Big)\ ,
\label{suggv}
}
which means that the terms must be pulled down from the instanton
effective action in order to make a top form on the centred moduli
space $\widehat\ms_k$ (of degree $4kN-4$).

The final form \eqref{suggv} is precisely the kind that can
localize. To see this consider the more general integral
\EQ{
\widehat{\EuScript
Z}_k^{\sst(\N=2)}(\lambda)
=\pi^{-2kN}\int_{\widehat\ms_k}\exp\Big(-\lambda^{-1}
d_V(V^\dagger_\mu dX^\mu)\Big)\ .
\label{suggvg}
}
We then have
\EQ{
\PD{\widehat{\EuScript
Z}_k^{\sst(\N=2)}(\lambda)}{\lambda}=\pi^{-2kN}
\int_{\widehat\ms_k}\lambda^{-2}d_V\left\{V^\dagger_\mu dX^\mu\,
\exp\Big(-\lambda^{-1}d_V(V^\dagger_\mu dX^\mu)\Big)\right\}
}
using the fact that $d_V^2$ is nilpotent on $\SU(N)$-invariant
quantities. But since the volume form is
$\SU(N)$ invariant the integral---under favourable conditions---vanishes
and so $\widehat{\EuScript Z}_k^{\sst(\N=2)}(\lambda)$
is independent of $\lambda$. Hence, it can be evaluated
in the limit $\lambda\to0$ where the integral is dominated by the
saddle-point approximation around the critical points of the vector
field $V_a$. In the present case, the fixed-point set has been identified
in \S\ref{sec:S37} with the configurations where all the instantons
have shrunk to zero size. The potential problem is that the theory of
localization is most easy to apply to situations involving
compact spaces without
boundary. In the case at hand, the instanton moduli space is obviously
not compact and has conical singularities when instanton shrink to
zero sized. As we have already intimated, the
way to alleviate the problems caused by the singularities
is to consider the analogous problem in the
theory defined on a non-commutative spacetime.
As explained in \S\ref{sec:S129}, the deformed instanton moduli space
$\ms^{(\zeta)}_k$ is then a smooth
resolution of $\ms_k$: instantons can no longer shrink to zero size
and the corresponding canonical singularities are smoothed over. In
this case, as we shall find below, the technique of localization can
be used to considerably simplify the computation of the centred instanton
partition function. In particular the critical-point sets become
smooth manifolds of a very suggestive form.

The problem with formulating localization in terms of the intrinsic
geometry of the instanton moduli space is that, due to the presence of
the ADHM constraints, we do not have an explicit description of the
intrinsic geometry. It will prove much more convenient to work in
terms of the linearized formalism described in \S\ref{sec:N1}. It will
also prove easier to formulate the theory of localization in terms of
Grassmann variables instead of differential forms. The expression for
the centred instanton partition is \eqref{cipfun}.\footnote{We also
perform the re-scaling \eqref{nicer} so that there are no factors of
$g$ in the instanton effective action \eqref{iean2}.}

In the language of Grassmann variables,
the translation of the covariant derivative is a fermionic symmetry $\Q$, or
BRST operator, which is nilpotent---at least up to symmetries. The
fermionic symmetry that we need to prove the localization properties
of the integrals is defined by picking out a particular supersymmetry
transformation. These latter symmetries, acting on the variables of
the linearized formulation, are given in \eqref{susy22} and \eqref{deltaKzero}.
From the supersymmetry transformations we can define corresponding
supercharges via
$\delta=\xi_{\aD A}Q^{\aD A}$. The fermionic symmetry that has the
appropriate properties is then defined as the
combination\footnote{Note that a BRST-type operator was first
constructed in the context of the $\N=2$ instanton calculus
in Refs.~\cite{Fucito:2001ha,Bellisai:2000tn,Bellisai:2000bc}. (In
particular, the latter reference is most closely related to the
approach that we adopt in this paper.) In particular,
these references emphasize the relation with the
topologically twisted version of the original gauge theory. However, these
references did not go on to use the existence of $\Q$ to develop a
calculational technique based on localization. A nilpotent fermionic
symmetry was also constructed in the context of the $\N=4$ instanton
calculus in Ref.~\cite{Dorey:2001zq} where localization was first
proposed as a method to calculate, in this case the $\N=4$, instanton
partition function. It was then shown in \cite{Fucito:2001ha} that the
$\Q$ operator in the $\N=2$ theory could be obtained by the
orbifolding procedure described in \S\ref{sec:S94} in the present article.
Some recent papers \cite{Flume:2001nc} have also
considered the $\Q$-operator and the $\N=2$ instanton calculus, although
they use the equivalent language of differential forms described above.
These references then go some way toward interpreting ${\cal
F}_k$ as a topological intersection number.}
\EQ{
\Q=\epsilon_{\aD A}Q^{\aD A}\ .
}
Notice that the definition of $\Q$ mixes up spacetime and $R$-symmetry
indices as is characteristic of topological twisting.
The action of $\Q$ on the variables is
\ALT{2}{
\Q\, w_\aD&=i\epsilon_{\aD A}\mu^A\ ,&\qquad\Q\,\mu^A&=-2\epsilon^{\aD A}(
w_\aD\chi+\phi^0 w_\aD)\\
\Q\, a'_{\alpha\aD}&=i\epsilon_{\aD A}\CM^{\prime A}_\alpha\ ,&\qquad
\Q\,\CM^{\prime A}_\alpha&=-2\epsilon^{\aD A}[a'_{\alpha\aD},\chi]\\
\Q\,\chi&=0\ ,&\qquad\Q\,\chi^\dagger&=2i\delta^A{}_\aD\bar\psi^\aD_A\ ,\\
\Q\,\bar\psi^\aD_A&=\tfrac 12\delta^\aD{}_A[\chi^\dagger,\chi]
-i\vec D\cdot\vec\tau^{\aD}{}_\bD\delta^\bD{}_A\ ,&\qquad\Q\vec
D&=\delta_\aD{}^A\vec\tau^{\aD}{}_\bD[\bar\psi^\bD_A,\chi]\ ,\\
\Q\,\K&=\Q\,\tilde\K=0\ .&&
}
It is straightforward to show that $\Q$ is nilpotent up to an infinitesimal
$\U(k)\times\SU(N)$ transformation generated by $\chi$ and $\phi$.
For example,
\EQ{
\Q^2\,w_\aD=2i(w_\aD\chi+\phi^0w_\aD)\ .
}
In terms of $\Q$,
the instanton effective action \eqref{iean2} assumes the form
\EQ{
\tilde S=\Q\,\Xi+\Gamma\ ,
\label{specs}
}
with
\EQ{
\Xi=4\pi^2{\rm tr}_k\Big\{\tfrac12\epsilon_{\aD A}\bar
w^\aD\big(\mu^A\chi^\dagger
+\phi^{0\dagger}\mu^A\big)+\tfrac14\epsilon_{\aD A}\bar
a^{\prime\aD\alpha}[\CM^{\prime
A}_\alpha,\chi^\dagger]
+\delta^{A}{}_\aD\bar\psi^\bD_A\big(\bar a^\aD
a_\bD-\tfrac12\vec\zeta_{\sst(+)}\cdot\vec\tau^{\aD}{}_\bD
\big)\Big\}
\elabel{bitia}
}
and
\EQ{
\Gamma=-\pi^2\sum_{f=1}^{N_F}{\rm tr}_k\big(
(m_f-\chi)\K_f\tilde\K_f\big)\ .
}
Note that $\Q\Gamma=0$ so that the instanton effective action is
``$\Q$-closed'': $\Q\,\tilde S=0$. Notice in \eqref{bitia}, to
anticipate what follows, we have
allowed for a non-trivial non-commutativity parameter $\vec\zeta_{\sst(+)}$.

Notice that both the action $\tilde S$ and the integration measure
$\Bomega^{\sst(\N=2,N_F)}$ are $\Q$-invariant. The localization
argument argument proceeds as before. Consider the
general integral
\EQ{
\widehat{\EuScript Z}_k^{\sst(\N=2,N_F)}(\lambda)
=\int_{\widehat\ms_{k}}\Bomega^{\sst(\N=2,N_F)}\,\exp\big(
-\lambda^{-1}\Q\,\Xi-\Gamma\big)\ .
\label{suggvgg}
}
We then have
\EQ{
\PD{\widehat{\EuScript Z}_k^{\sst(\N=2,N_F)}(\lambda)
}{\lambda}=
\lambda^{-2}\int_{\widehat{\ms_k}}\Bomega^{\sst(\N=2,N_F)}\,\Q
\Big\{\Xi\,\exp\big(-\lambda^{-1}\Q\,\Xi-\Gamma\big)\Big\}\ ,
\label{varet}
}
using the fact that $\Q^2\,\Xi=\Q\Gamma=0$.
Since the volume form is $\Q$-invariant
the right-hand side of \eqref{varet}
vanishes. Consequently, $\widehat{\EuScript Z}_k^{\sst(\N=2,N_F)}(\lambda)$
is independent of $\lambda$ and can, therefore, be evaluated
in the limit $\lambda\to0$ where the integral is dominated by
the critical points
of $\Q\,\Xi$. Since the result is independent of $\lambda$,
under favourable circumstances---which will be shown to hold
in the present application---the Gaussian approximation is exact.

Note that the quantity $\Xi$ depends on $\vec\zeta_{\sst(+)}$ and the
anti-holomorphic components of the VEV. Hence, the derivative of the
instanton effective action with respect to either $\vec\zeta_{\sst(+)}$ or
$\phi^{0\dagger}$ is $\Q$-exact and so the instanton partition
function cannot depend on these parameters. On the contrary, the
holomorphic components of the VEVs enter through the action of $\Q$
itself, while the hypermultiplet masses enter through $\Gamma$, and
so the centred instanton partition function {\it can\/} depend on
these parameters. In particular, from what we have said above, we have
proved that the centred instanton partition function cannot depend
smoothly on the non-commutativity parameter $\vec\zeta_{\sst(+)}$. Of course,
there will be a discontinuity when $\vec\zeta_{\sst(+)}=0$
when singularities appear on the instanton moduli space and the
localization argument breaks down.

Following the logic of localization we should investigate the
critical points of $\Q\,\Xi$. For the moment we suppose
$\vec\zeta_{\sst(+)}=0$. The terms to minimize are, from
\eqref{jslmm},
\EQ{
\big|w_\aD\chi_a+\phi^0w_\aD\big|^2
-[\chi_a,a'_n]^2\ .
}
Notice that this is positive semi-definite and the critical points are
simply the zeros. Hence
\EQ{
w_\aD\chi_a+\phi_a w_\aD=[\chi_a,a'_n]=0\ .
}
Notice that these equations are identical to those in
\eqref{fpcond}. Hence we can identify the critical-points of $Q\Xi$
which the fixed points of the tri-holomorphic vector fields
generated by the VEVs. The fixed-point set corresponds to the singular
subspace ${\rm Sym}^k\,{\mathbb R}^4$ where all the instanton have
shrunk down to zero size and the resulting gauge potential is pure gauge.

In order to avoid the singular nature of point-like
instantons, Ref.~\cite{local} suggested a regularization based on the
smooth resolution of the instanton moduli space first described in
purely geometrical terms without reference to the gauge theory by
Nakajima in Ref.~\cite{Nakajima:1993jg}. Of course, as discussed in
\S\ref{sec:S129}, subsequently
it was realized that this smooth resolution of the
instanton moduli space  $\ms_k^{(\zeta)}$
arises naturally when the theory is defined on
a non-commutative spacetime \cite{NS}.
What happens is that the existence of the
central term in the ADHM constraints \eqref{madhm} prevents
instantons shrinking to zero size. The solutions are associated to the
partitions \eqref{partit} and have the form \eqref{critp}. The ADHM
constraints within the $u^{\rm th}$ block are now the ADHM constraints
are now those of a non-commutative gauge theory with a $\U(1)$
gauge group. Taking the
trace of the ADHM constraints within the block we have
\EQ{
\vec\tau^{\aD}{}_\bD\sum_{i=k_{u-1}+1}^{k_u}\bar
w^{\bD}_{iu}w_{ui\aD}=k_u\vec\zeta_{\sst(+)}\ .
}
Now, in contrast to \eqref{qeer},
in the presence of the central term, $w_{ui\aD}$ must be non-trivial.
The critical-point set associated to the partition \eqref{partit} is
consequently a product of non-commutative $\U(1)$ instanton moduli
spaces:
\EQ{
\frac{\ms^{(\zeta)}_{k_1}\big|_{N=1}\times\cdots\times
\ms^{(\zeta)}_{k_N}\big|_{N=1}}{{\mathbb R}^4}\ .
}
Recall from the discussion in \S\ref{sec:S129} that
$\ms_{k}^{(\zeta)}\big|_{N=1}$ is a
smooth resolution of the symmetric product ${\rm Sym}^k{\mathbb R}^4$.

In fact
in the non-commutative theory the consequences of Derrick's Theorem
are avoided and there are now exact non-singular solutions to the
equations-of-motion even in the presence of VEVs. These exact
solutions were called ``topicons''
in \cite{locn4} since they have a contribution to the action which is
localized and, unlike an instanton, they have no size modulus.
Actually there are $N$ flavours of
topicon, associated the $N$ block in \eqref{partit},
obtained by embedding the spacetime non-commutative $\U(1)$
instanton solutions
in each of $N$ unbroken abelian factors of the gauge group.
So the following picture emerges. For instanton charge $k$, the exact
instanton solutions come as a disjoint union of spaces
associated to the inequivalent partitions \eqref{partit}
where each $k_u$ corresponds to each of $N$ $\U(1)$
subgroups picked out by the VEV. Hence, the space of exact solutions,
or ``moduli space of topicons'', lying within the larger instanton
moduli space is of the form
\EQ{
\widehat\ms_{k}\ \overset{\text{resolve}}\longrightarrow\
\widehat\ms_{k}^{(\zeta)}\ \overset{\text{exact}}\supset\
\widehat\ms_{k}^{(\zeta)}\Big|_{\text{topicon}}=
\bigcup_{\text{partitions}\atop
k_1+\cdots+k_N}\frac{\ms^{(\zeta)}_{k_1}\big|_{N=1}\times
\cdots\times\ms^{(\zeta)}_{k_N}\big|_{N=1}}{{\mathbb R}^4}\ .
\label{partitions}
}

\subsubsection{One instanton}

We will now begin using localization to evaluate
the one-instanton contribution to the prepotential of the
$\N=2$ supersymmetric $\SU(N)$ gauge theory with $N_F$ hypermultiplets.
The instanton effective action \eqref{iean2} has $N$
critical points, labelled by $v\in\{1,2,\ldots,N\}$, at which \eqref{critp}
\EQ{
\chi_a=-(\phi_a^0)_v\ ,\qquad w_{u\aD}\propto\delta_{uv}\ .
\label{critpn}
}
Note that $a'_n=0$ in the one-instanton sector.
Without-loss-of-generality, we choose our
non-commutativity parameters
\EQ{
\zeta_{\sst(+)}^1=\zeta_{\sst(+)}^2=0\ ,\qquad\zeta_{\sst(+)}^3\equiv\zeta>0\ .
\label{ncom}
}
In this case the ADHM constraints
\eqref{madhm} are solved on the critical submanifold with
\EQ{
w_{u\aD}=\sqrt{\zeta}e^{i\theta}\delta_{uv}\delta_{\aD1}\ ,
}
for an arbitrary phase angle $\theta$. The integrals over $w_{v\aD}$ are
then partially annulled by the three $\delta$-functions in
\eqref{bmes} that impose the ADHM constraints, leaving a trivial
integral over the phase angle $\theta$:
\EQ{
\int d^2w_v\,d^2\bar
w_v\,\prod_{c=1}^3\delta\big(\tfrac12\tau^{c\aD}{}_\bD(
\bar w_v^\bD w_{v\aD}-\zeta\delta^{c3})\big)
=8\pi\zeta^{-1}\ .
\label{yuppa}
}
Correspondingly,
the $\delta$-functions for the Grassmann ADHM constraints saturate
the integrals over $\{\mu^A_v,\bar\mu^A_v\}$:
\EQ{
\int\,d\mu^A_v\,d\bar\mu^A_v\,\prod_{\aD=1}^2\delta\big(
\bar w_{v\aD}\mu^A_v+w_{v\aD}\bar\mu^A_v\big)=\zeta\ ,
\label{yuppb}
}
for each $A=1,2$.
The remaining variables, $\{w_{u\aD},\mu^A_{u},\bar\mu^A_u\}$, $u\neq
v$, as well as $\{\K_{if},\tilde\K_{fi}\}$,
are all treated as Gaussian fluctuations around the critical point.
To this order, the instanton effective action \eqref{iean2} is
\EQ{
\tilde S=4\pi^2\bigg\{\zeta\chi_a^2+\sum_{{u=1\atop(\neq v)}}^N
\Big((\phi_a)_{vu}^2\big|w_{u\aD}\big|^2
+\tfrac
i{2}\phi^{0\dagger}_{vu}
\bar\mu^A_u\mu_{uA}\Big)
-\tfrac 14\sum_{f=1}^{N_F}(m_f+\phi^0_v)
\K_f\tilde\K_f\bigg\}+\cdots\ ,
}
where $(\phi^0_a)_{uv}\equiv(\phi^0_a)_u-(\phi^0_a)_v$.
The integrals are easily done. Note that the integral over $\chi_a$
yields a factor of $\zeta^{-1}$ which cancels against the factors of
$\zeta$ arising from \eqref{yuppa} and \eqref{yuppb} so the final
result is, as expected, independent of $\zeta$. Summing over the $N$
critical-point sets
gives the centred one-instanton partition function
\EQ{
{\cal F}_1^{\text{nc}}\equiv\widehat{\EuScript Z}_1^{\sst(\N=2,N_F)}=
\sum_{v=1}^N\Big\{\prod_{{u=1\atop(\neq
v)}}^N\frac{1}{(\phi_{v}^0-\phi_u^0)^2}\,
\prod_{f=1}^{N_F}(m_f+\phi^0_v)\Big\}\ .
\label{locoi}
}
Notice that the resulting expression is holomorphic in the VEVs and
independent of $\zeta$ as required.
The result should be compared with the
brute-force calculation of the one-instanton contribution in
\eqref{bfoi}. The expression is entirely consistent with \eqref{locoi}
for
$N_F<2N$.\footnote{The constant factor ${\cal S}_1^{\sst(N_F)}$ for
$N_F=2N-2$ and $2N-1$ does not affect the low energy effective action
which depends only on derivatives of the prepotential with respect to
the VEVs.}
We can also compare \eqref{locoi} with the predictions of
Seiberg-Witten theory written down in \eqref{onei}. The expression
agree for all for $N_F<2N$. Recall that the Seiberg-Witten
predictions as written are only valid for $N_F<2N$; however, our
expression \eqref{locoi} is simply the obviously extrapolation of
\eqref{onei} to $N_F=2N$.

The case with $N_F=2N$, which corresponds to the finite theory, is
rather special. Notice that the expression in the non-commutative
theory misses the contribution ${\cal S}_1^{\sst(N_F=2N)}$ of
\eqref{bfoi} which is quadratic in the VEVs. We have already
identified this term as arising from the singularity of the
one-instanton moduli space $\widehat\ms_1$. However, we have already
argued at the end of \S\ref{sec:S1112}
that the mismatch is not physically relevant since it can be
accommodated by a non-trivial mapping between the couplings of the
commutative and non-commutative theories.

\subsubsection{Two instantons}

We now turn to the situation for $k=2$.
There are two kinds critical-point corresponding to two topicons of
the same, or of different, flavour, respectively.
We now evaluate these two contribution separately. As in the
one-instanton sector we choose the non-commutativity parameters as in
\eqref{ncom}.

{\it Two topicons of different flavour}

For two topicons of the different flavour, the
ADHM constraints are solved with
\EQ{
w_{ui\aD}=\sqrt\zeta e^{i\theta_i}\delta_{uu_i}\delta_{\aD1}
\ ,\qquad
a'_n=\MAT{Y_n&0\\ 0&-Y_n}\ .
\label{diagsol}
}
The two phase angles $\theta_i$, $i=1,2$, are not genuine moduli since they can
be separately rotated by transformations in the $\U(2)$
auxiliary group. The variables $Y_n$ are the genuine moduli
representing the relative positions of the two topicons.
The corresponding solution of the fermionic ADHM constraints
\eqref{fadhm} on the critical-point set is
\EQ{
\mu^A=\bar\mu^A=0\ ,\qquad\CM^{\prime A}_\alpha=\MAT{\rho^A_\alpha&0
\\ 0&-\rho^A_\alpha}\ ,
\label{fdiagsol}
}
where $\rho^A_\alpha$ are the superpartners of $Y_n$. Notice in
this case the critical-point set is non-compact since the separation
$Y_n\in{\mathbb R}^4$. The issue of whether the integral over $Y_n$
converges is rather delicate. By explicit calculation we shall find
that the integral is indeed convergent.

We now proceed to evaluate the contribution to the
centred instanton partition function from the critical-point set.
First we expand around the critical values \eqref{diagsol} and
\eqref{fdiagsol}. It is convenient to use the following notation for
the fluctuations
\EQ{
\delta a'_n=\MAT{0&Z_n\\ Z_n^*&0}\ ,\qquad\delta\CM^{\prime A}_\alpha=
\MAT{0&\sigma^A_\alpha
\\ \varepsilon^A_\alpha&0}
}
and to make the shift
\EQ{
\chi_a\to\chi_a-\MAT{(\phi_a^0)_{u_1}&0\\ 0&(\phi_a^0)_{u_2}}\ ,
}
so that $\chi_a=0$ on the critical submanifold.
We then integrate over the Lagrange multipliers $\vec D$ and
$\bar\psi_A^\aD$ which impose the ADHM constraints \eqref{badhm} and
\eqref{fadhm}. The diagonal components of the constraints (in $i,j$
indices) are the ADHM constraints of the two single $\U(1)$ instantons.
The off-diagonal components vanish on the critical-point set and must
therefore be expanded to linear order in the fluctuations.
For the bosonic variables we have
\AL{
\sqrt\zeta e^{-i\theta_1}(w_{u_12})_2+\sqrt\zeta
e^{i\theta_2}(w_{u_21})_2^*+4i\bar\eta^1_{mn}
Y_mZ_n&=0\ ,\label{ladhm3}\\
-i\sqrt\zeta e^{-i\theta_1}(w_{u_12})_2+i\sqrt\zeta
e^{i\theta_2}(w_{u_21})_2^*+4i\bar\eta^2_{mn}Y_mZ_n&=0\ ,\\
\sqrt\zeta e^{-i\theta_1}(w_{u_12})_1+\sqrt\zeta
e^{i\theta_2}(w_{u_21})_1^*+4i\bar\eta^3_{mn}Y_mZ_n&=0\ ,\label{ladhm1}
}
where $\bar\eta^c_{mn}=\tfrac1{2i}{\rm
tr}_2(\tau^c\bar\sigma_m\sigma_n)$ are 't~Hooft's
$\eta$-symbols defined in Appendix~\ref{app:A1}.
Similarly in the Grassmann sector
\AL{
\sqrt\zeta e^{i\theta_2}\bar\mu^A_{1u_2}+2(\rho^{\alpha
A}Z_{\alpha1}-\sigma^{\alpha A}Y_{\alpha1})&=0\ ,\label{oo1}\\
\sqrt\zeta e^{-i\theta_1}\mu^A_{u_12}+2(\rho^{\alpha
A}Z_{\alpha2}-\sigma^{\alpha A}Y_{\alpha2})&=0\ ,\\
\sqrt\zeta e^{i\theta_1}\bar\mu^A_{2u_1}+2(\varepsilon^{\alpha
A}Y_{\alpha1}-\rho^{\alpha A}Z^*_{\alpha1})&=0\ ,\\
\sqrt\zeta e^{-i\theta_2}\mu^A_{u_21}+2(\varepsilon^{\alpha
A}Y_{\alpha2}-\rho^{\alpha A}Z^*_{\alpha2})&=0\ ,
\label{oo4}
}
where $Y_{\alpha\aD}=Y_n\sigma_{\alpha\aD n}$, {\it etc\/}.
These equations correspond to a set of
linear relations between the fluctuations.
It is convenient to define

\EQ{
(w_{u_12})_1=e^{i\theta_1}(\xi+\lambda)\ ,\qquad(w_{u_21})_1^*=
e^{-i\theta_2}(-\xi+\lambda)\ ,
}
so that the fluctuations $\xi$ drops out from \eqref{ladhm1}.
We can use \eqref{ladhm3}-\eqref{ladhm1} to solve for
$(w_{u_12})_2$, $(w_{u_21})_2$ and $\lambda$, and \eqref{oo1}-\eqref{oo4} to
solve for
$\mu^A_{1u_2}$, $\mu^A_{2u_1}$, $\bar\mu^A_{u_12}$ and $\bar\mu^A_{u_21}$.
We then use the $\U(2)$ symmetry to fix (i)
the fluctuation $Z_n$ to be orthogonal
to $Y_n$, $Z_nY_n=0$; and (ii) $\theta_i=0$. The Jacobian for
the first part of this gauge fixing is
\EQ{
\frac1{{\rm Vol}\,\U(2)}\int d^{12}a'\to\frac{16}{\pi^2}
\int d^4Y\,d^3Z\,d^3Z^*\,Y^2\ .
}

Now we turn to expanding
the instanton effective action \eqref{iean2}. First the bosonic
pieces. To Gaussian order around the critical point
\EQ{
\tilde S_{\rm b}=\tilde S^{(1)}_{\rm b}+\tilde S^{(2)}_{\rm b}+\cdots\ ,
}
where
\EQ{
\frac1{4\pi^2}\tilde S^{(1)}_{\rm
b}=\zeta\big((\chi_a)_{11}^2+(\chi_a)_{22}^2\big)+8Y^2
\big|(\chi_a)_{12}\big|^2+
2\big|(\phi_a^0)_{u_1u_2}\xi+\sqrt\zeta(\chi_a)_{12}
\big|^2
+2(\phi_a^0)_{u_1u_2}^2\big(1+4\zeta^{-1}Y^2\big)|Z|^2
\label{act1}
}
and
\EQ{
\frac1{4\pi^2}\tilde S^{(2)}_{\rm
b}=\sum_{i=1}^2
\sum_{{u=1\atop(\neq u_1,u_2)}}^N
(\phi^0_a)_{uu_i}^2\big|w_{ui\aD}\big|^2\ .
}
In order to simplify the integration over the fluctuations, it is
convenient to shift
\EQ{
\xi\to
\xi-\sqrt\zeta
\frac{(\phi_a^0)_{u_1u_2}(\chi_a)_{12}}{(\phi_a)^2_{u_1u_2}}
}
and define the orthogonal decomposition
\EQ{
\chi_a=\chi^\parallel_a+\chi_a^\perp\ ,\qquad
\chi_a^\perp(\phi^0_a)_{u_1u_2}=0\ .
}
After having done this \eqref{act1} becomes
\SP{
\frac1{4\pi^2}\tilde S^{(1)}_{\rm
b}&=\zeta\big((\chi_a)_{11}^2+
(\chi_a)_{22}^2\big)+2\zeta\big(1+4\zeta^{-1}Y^2\big)
|(\chi^\perp_a)_{12}|^2+8Y^2
\big|(\chi^\parallel_a)_{12}\big|^2
\\
&+2(\phi_a^0)_{u_1u_2}^2\Big(|\xi|^2+\big(1+4\zeta^{-1}Y^2\big)|Z|^2\Big)
}

To Gaussian order the Grassmann parts of the
instanton effective action \eqref{iean2} are
\EQ{
\tilde S_{\rm f}=\tilde S_{\rm f}^{(1)}+\tilde S_{\rm f}^{(2)}+\cdots\ ,
}
where
\SP{
\frac1{4\pi^2}\tilde S_{\rm f}^{(1)}&=
-\tfrac i2\phi^{0\dagger}_{u_1u_2}\big(1+4\zeta^{-1}Y^2\big)
\sigma^{\alpha A}\varepsilon_{\alpha A}+
i\rho^{\alpha A}\big(2\zeta^{-1}(\phi_{u_1u_2}^0)^\dagger
Z_{\alpha\aD}\bar
Y^{\aD\beta}+\chi_{12}^\dagger\delta_\alpha{}^\beta\big)
\varepsilon_{\beta A}\\
&+i\sigma^{\alpha A}\big(2\zeta^{-1}(\phi_{u_1u_2}^0)^\dagger
Y_{\alpha\aD}\bar
Z^{*\aD\beta}-\chi_{21}^\dagger\delta_\alpha{}^\beta\big)\rho_{\beta A}
-2i\zeta^{-1}\phi_{u_1u_2}^{0\dagger}\rho^{\alpha A}Z_{\alpha\aD}\bar
Z^{*\aD\beta}\rho_{\beta A}
}
and
\EQ{
\frac1{4\pi^2}\tilde S_{\rm f}^{(2)}=
\tfrac i{2}\sum_{i=1}^2\sum_{{u=1\atop(\neq u_1,u_2)}}^N
\phi^{0\dagger}_{uu_i}\bar\mu^A_{iu}\mu_{uiA}
-\tfrac14\sum_{i=1}^2\sum_{f=1}^{N_F}
(m_f+\phi^0_{u_i})\K_{if}\tilde\K_{fi}\ .
}
By shifting the fluctuations $\sigma^A$ and $\varepsilon^A$ by the
appropriate amounts of $\rho^A$, we can complete the square yielding
\SP{
\frac1{4\pi^2}\tilde S^{(1)}_{\rm f}&=
-\tfrac i2(\phi^0_{u_1u_2})^\dagger\big(1+4\zeta^{-1}Y^2\big)
\sigma^{\alpha A}\varepsilon_{\alpha A}\\
&-2i\zeta^{-1}(1+4\zeta^{-1}Y^2)^{-1}\rho^{\alpha
A}\Big(\phi^{0\dagger}_{u_1u_2}
Z_{\alpha\aD}\bar Z^{*\aD\beta}+2\chi_{12}^\dagger
Y_{\alpha\aD}\bar Z^{*\aD\beta}+2\chi_{21}^\dagger Z_{\alpha\aD}
\bar Y^{\aD\beta}\Big)\rho_{\beta A}\ .
}

Before we proceed, let us remind ourselves that only the variables $Y_n$
and $\rho^A_\alpha$ are facets of the critical-point set, the
remaining variables are all fluctuations.
The contribution to the centred
instanton partition function from the critical-point set is then
proportional to
\SP{
&
\int d^4Y\,d\xi\,d\xi^*\,d^3Z\,d^3Z^*\,d^{8}\chi_a\,
d^4\rho\,d^4\sigma\,d^4\varepsilon\\ &\times\
\prod_{i=1}^2\bigg\{\prod_{u=1\atop(\neq u_1,u_2)}^Nd^2w_{ui}\,
d^2\bar w_{iu}\,d^2\mu_{ui}\,d^2\bar\mu_{iu}\,
\prod_{f=1}^{N_F}d\K_{if}\,d\tilde\K_{fi}\bigg\}\ Y^2\
\exp(-\tilde S_{\rm b}^{(1)}-\tilde S_{\rm b}^{(2)}-
\tilde S_{\rm f}^{(1)}-\tilde S_{\rm f}^{(2)})\
.
}
The integrals over the Grassmann variables
$\{\sigma^A_\alpha,\varepsilon^A_\alpha,\rho^A_\alpha\}$ are saturated by
pulling down terms from $S_{\rm
f}^{(1)}$ yielding the factors
\EQ{
(\phi_{u_1u_2}^{0\dagger})^{4}
\zeta^{-2}(1+4\zeta^{-1}Y^2)^2\Big(4Y^2(\chi_{21}^\dagger
Z-\chi_{12}^\dagger Z^*)^2+(\phi^{0\dagger}_{u_1u_2})^{2}(Z^2Z^{*2}-(Z\cdot
Z^*)^2)\Big)\ .
}
The integrals over the remaining Grassmann variables
$\{\mu_{ui}^A,\bar\mu_{iu}^A,\K_{if},\tilde\K_{fi}\}$, $u\neq u_1,u_2$,
are saturated
by pulling down terms from $S_{\rm f}^{(2)}$ giving rise to
\EQ{
\prod_{i=1}^2\prod_{{u=1\atop(\neq
u_1,u_2)}}^N(\phi^\dagger_{uu_i})^2\prod_{f=1}^{N_F}
(m_f+\phi_{u_1})(m_f+\phi_{u_2})\ .
}

The $\{Z,\xi,\chi_a\}$ integrals are
\SP{
&\int d\xi\,d\xi^*\,
d^3Z\,d^3Z^*\,d^8\chi_a\,\Big(4Y^2(\chi_{21}^\dagger
Z-\chi_{12}^\dagger Z^*)^2+2(\phi^0_{u_1u_2})^{\dagger2}(Z^2Z^{*2}-(Z\cdot
Z^*)^2)\Big)\,e^{-S^{(1)}}\ ,
}
which yields the non-trivial factor
\EQ{
\frac{(\phi^{0\dagger}_{u_1u_2})^{2}}
{\zeta^3(\phi_a^0)_{u_1u_2}^{12}Y^2(1+4\zeta^{-1}Y^2)^6}
}
while those over $w_{ui\aD}$, $u\neq u_1,u_2$, give a factor
\EQ{
\prod_{i=1}^2\prod_{u=1\atop(\neq
u_1,u_2)}^N\frac1{(\phi^0_a)_{uu_i}^4}\ .
}

Finally all that remains is to integrate over the relative position of
the instantons. As promised the integral is convergent:
\EQ{
\int d^4Y\,\frac{\zeta^2}{(\zeta+4Y^2)^4}=\frac{\pi^2}{96}\ .
}
Putting all the pieces together with the correct numerical factors
gives the final contribution of the critical-point set to the
centred instanton partition function
\EQ{
\frac2{(\phi^0_{u_1u_2})^6}\prod_{i=1}^2\prod_{{u=1\atop(\neq
u_1,u_2)}}^N\frac1{(\phi^0_{uu_i})^2}\prod_{f=1}^{N_F}
(m_f+\phi^0_{u_1})(m_f+\phi^0_{u_2})\ .
}
Notice that the result is holomorphic in the VEVs as required.
Summing over the $\tfrac12N(N-1)$ critical-points for topicons of
different flavours, we have
the following contribution
\EQ{
\sum_{u,v=1\atop
(u\neq v)}^N\frac{S_u(\phi^0_u)S_v(\phi^0_v)}{(\phi^0_u-\phi_v^0)^2}\ ,
\label{res2}
}
where we have written the answer in terms of the functions
\EQ{
S_u(x)\equiv \prod_{{v=1\atop(\neq u)}}^N
\frac1{(x-\phi^0_v)^2}\prod_{f=1}^{N_F}(m_f+x)
\label{defsu}
}
defined in \cite{D'Hoker:1997nv}.

{\it Two topicons of the same flavour}

There are $N$ critical-points of this type, describing two topicons of
the same flavour: $u_1=u_2\equiv
v\in\{1,\ldots,N\}$. On the critical submanifold  $\{w_{vi\aD},a'_n\}$
and $\{\mu^A_{vi},\bar\mu^A_{iv},\CM^{\prime
A}_\alpha\}$ satisfy the ADHM constraints, \eqref{madhm} and
\eqref{fadhm}, respectively, of two
instantons in a non-commutative $\U(1)$ theory. The remaining
variables all vanish and are treated as fluctuations around the
critical-point set.

As previously, it is convenient to
shift the auxiliary variable $\chi_a$ by its critical-point value:
\EQ{
\chi_a\to\chi_a-(\phi^0_a)_v1_{\sst[2]\times[2]}\ .
\label{shift}
}
We now expand in the fluctuations $\{w_{ui\aD},\mu^A_{ui},
\bar\mu^A_{iu}\}$, for $u\neq v$,
as well as $\{\K_{if},\tilde\K_{fi}\}$. Since all the components of the
ADHM constraints are non-trivial at leading order the fluctuations
decouple from the $\delta$-functions in \eqref{bmes} and \eqref{intsf}
which impose the constraints. The fluctuation integrals only involve
the integrand $\exp-\tilde S$, where $\tilde S$
is expanded to Gaussian order around
the critical-point set.
However, it is important, as we shall see below, to leave $\chi_a$
arbitrary rather than set it to its critical-point value; namely,
$\chi_a=0$, after the shift \eqref{shift}. The fluctuation
integrals produce the non-trivial factor
\EQ{
\prod_{u=1\atop(\neq v)}^N
\frac1{\big({\rm det}_2(\chi+\phi^0_{uv}1_{\sst[2]\times[2]})\big)^2}
\prod_{f=1}^{N_F}\det_2\big((m_f+\sqrt2
\phi^0_v)1_{\sst[2]\times[2]}-\chi\big)
=S_v(\phi^0_v-\lambda_1)S_v(\phi^0_v-\lambda_2)\ .
\label{above}
}
Here, $\lambda_i$, $i=1,2$, are the eigenvalues of the $2\times2$
matrix $\chi$ and $S_u(x)$ was defined in \eqref{defsu}.

The remaining integrals involve the supersymmetric volume integral on
$\widehat\ms_{2}^{(\zeta)}\big|_{N=1}$, into which we insert the integrand
\eqref{above} which depends non-trivially on $\chi$. Now by itself
$\int_{\widehat\ms_{2}^{(\zeta)}\big|_{N=1}}\Bomega^{\sst(\N=2)}=0$.
This is clear from the linearized form of the instanton effective
action \eqref{iean2} with $N=1$ and $N_F=0$:
integrals over the Grassmann collective coordinates pull down two elements
of the matrix $\chi^\dagger$ from the
action and since there are no compensating factors of $\chi$ the resulting
integrals over the phases of the elements of
$\chi$ will integrate to zero. This is why we
left $\chi$ arbitrary in \eqref{above} since after expanding in powers
of the eigenvalues $\lambda_i$ it is potentially the quadratic terms
that will give a non-zero result when inserted into
$\int_{\widehat\ms_{2}^{(\zeta)}\big|_{N=1}}\Bomega^{\sst(\N=2)}$. To quadratic order
\eqref{above} is
\EQ{
\tfrac12
S_v(\phi^0_v)\frac{\partial^2S_v(\phi^0_v)}{\partial(\phi^0_v)^2}
(\lambda_1^2+\lambda_2^2)+
\PD{S_v(\phi^0_v)}{\phi^0_v}\PD{S_v(\phi_v)}{\phi^0_v}
\lambda_1\lambda_2\ .
}
So the contribution from
this critical-point set is of the form
\EQ{
{\EuScript I}_1S_v
(\phi^0_v)\frac{\partial^2S_v(\phi^0_v)}{\partial(\phi^0_v)^2}
+
{\EuScript I}_2\PD{S_v(\phi^0_v)}{
\phi^0_v}\PD{S_v(\phi^0_v)}{\phi^0_v}\Big\}\ ,
\label{ress}
}
where the VEV-independent constants ${\EuScript I}_{1,2}$ are given by the
following integrals
\SP{
&{\EuScript
I}_1=\tfrac12\int_{\widehat\ms_{2}^{(\zeta)}\big|_{N=1}}\Bomega^{\sst(\N=2)}
\,(\lambda_1^2+\lambda_2^2)
\equiv\int_{\widehat\ms_{2}^{(\zeta)}\big|_{N=1}}\Bomega^{\sst(\N=2)}\,\big(
\tfrac12({\rm tr}_2\chi)^2-{\rm det}_2\chi\big)\ ,\\
&{\EuScript
I}_2=\int_{\widehat\ms_{2}^{(\zeta)}\big|_{N=1}}\Bomega^{\sst(\N=2)}\,\lambda_1\lambda_2
\equiv\int_{\widehat\ms_{2}^{(\zeta)}\big|_{N=1}}\Bomega^{\sst(\N=2)}
\,{\rm det}_2\chi\ .
\label{defi}
}
We remark that \eqref{ress} is holomorphic in the VEVs as required.

The moduli space $\widehat\ms_{2}^{(\zeta)}\big|_{N=1}$ is the
Eguchi-Hanson manifold \cite{LTY}, a well-known four-dimensional
hyper-K\"ahler space \cite{EGH}. So after all the Grassmann variables
and $\chi_a$ have been integrated out,
we can write ${\EuScript I}_{1,2}$ as integrals over the
Eguchi-Hanson space of a suitable integrand.
The following results were proved in the Appendix of \cite{local}:
\EQ{
{\EuScript I}_1=\frac14\ ,\qquad{\EuScript I}_2=0\ .
}
Hence the final result for the contributions from the $N$ critical
points of this type to the centred instanton partition function is
\EQ{
\tfrac14\sum_{u=1}^N
S_u(\phi_u)\frac{\partial^2S_u(\phi_u)}{\partial(\phi^0_u)^2}\ .
\label{res1}
}

Finally, summing \eqref{res1} and \eqref{res2} we have the
centred two-instanton partition function
\SP{
{\cal F}_2^{\text{nc}}\equiv\widehat{\EuScript Z}_{2}=
\sum_{{u,v=1\atop(u\neq v)}}^N
\frac{S_u(\phi^0_u)S_v(\phi_v^0)}{(\phi^0_u-\phi^0_v)^2}+
\tfrac14\sum_{u=1}^NS_u(\phi^0_u)\frac{\partial^2S_u(\phi^0_u)}
{\partial(\phi^0_u)^2}\ .
\elabel{amaz}
}
For $N<2N_F$ it is astonishing to find exact agreement
with the prediction from Seiberg-Witten theory \eqref{twoi}. As in the
one-instanton sector,
our expression \eqref{amaz} with $N_F=2N$ is simply
the extrapolation of the formula \eqref{twoi}.
We can also compare \eqref{amaz}
with the explicit brute-force integration over the
instanton moduli space for gauge group $\SU(2)$ that we reviewed in
\S\ref{sec:S842}. The agreement is exact for $N_F<4$. As in the
one-instanton sector the case with $N_F=4$ is rather special since we
expect the non-trivial matching of the coupling constants of the
commutative and non-commutative theories displayed in
\eqref{matco}. At the two-instanton level this matching implies
\EQ{
{\cal F}_2={\cal F}_2^{\text{nc}}+2\pi ic_1{\cal
F}_1^{\text{nc}}+i\pi c_2
\sum_{u=1}^N(\phi_u^0)^2\ ,
\label{comp2}
}
modulo constants. Since we have already determined $c_1$ in
\eqref{conett} this matching
condition is non-trivial even for gauge group $\SU(2)$ since the
coefficients depend non-trivially on the masses and the VEVs. Using
\eqref{locoi}, \eqref{amaz} (for $N=2$) and
\eqref{bigres}, all with $N_F=4$, one finds complete consistency with
\EQ{
c_2\Big|_{\SU(2)}=\frac1{2^3\pi
i}\Big(1+\frac7{2^43^5}-\frac{13}{2^4}\Big)\ .
}

The question is whether the localization technique can be extended
beyond two instantons? In fact it already has in the context of the
$\N=4$ theory softly broken to $\N=2$ by mass terms as described in
Ref.~\cite{locn4}. In this reference the leading-order terms in the
mass expansion of the prepotential were calculated to all orders in
the instanton charge and were shown to agree with the predictions of
Seiberg-Witten theory. It is clear that this subject is moving quickly
and it seems possible that one might be able to provide a completely
self-contained instanton proof of Seiberg-Witten theory.

\vspace{2cm}

\begin{center}
$\sst *************************$
\end{center}

\vspace{1cm}

{\bf Acknowledgments}

We would like to thank
Massimo Bianchi, Chong-Sun Chu, Ed Corrigan, Francesco Fucito,
Michael Green, Prem Kumar, Werner Nahm, Hugh Osborn, Misha Shifman,
Gabriele Travaglini, Arkady Vainshtein and Stefan Vandoren for
a wide variety of discussions on the instanton calculus and related subjects.

\newpage

\startappendix

\rsen\Appendix{Spinors in Diverse Dimensions}\elabel{app:A1}

In this appendix we define our conventions for spinors
in various dimensions. The treatment will
be be geared toward the applications needed in the main text.

The Minkowski space metric will be chosen to be
$\eta_{MN}={\rm diag}(-1,1,\ldots,1)$. The $D$-dimensional Minkowski space
Clifford algebra is
\EQ{
\{\Gamma_M,\Gamma_N\}=2\eta_{MN}\ ,
}
with $M,N=0,\ldots,D-1$. The $D$-dimensional
Euclidean space Clifford algebra is
\EQ{
\{\Gamma_M,\Gamma_N\}=2\delta_{MN}\ ,
}
with $M,N=1\,\ldots,D$.

We now define representations of the Clifford algebra in several
(even) dimensions. In all our representations the additional element
of the Clifford algebra is
\EQ{
\Gamma_{D+1}=\MAT{1&0\\ 0&-1}\ ,
}
so that Weyl spinors are of the form
\EQ{
\MAT{\lambda\\0}\ \qquad\text{and}\qquad\ \MAT{0\\ \bar\psi}\ .
}

For $D=2$, we take
\EQ{
\Gamma_1=\begin{pmatrix}0&-i\\ i&0\end{pmatrix}\ ,\qquad
\Gamma_2=\begin{pmatrix}0&1\\ 1&0\end{pmatrix}
}
in Euclidean space and
\EQ{
\Gamma_0=\begin{pmatrix}0&1\\ -1&0\end{pmatrix}\ ,\qquad
\Gamma_1=\begin{pmatrix}0&1\\ 1&0\end{pmatrix}
}
in Minkowski space. Weyl spinors are complex and real, respectively.

In four-dimensional Euclidean space we take
\EQ{
\gamma_n=\begin{pmatrix}0&-i\sigma_n\\
i\bar\sigma_n&0\end{pmatrix}\ ,\qquad n=1,2,3,4\ ,
\elabel{fdc}
}
where $\sigma_n=(i\vec\tau,1)$ and $\bar\sigma_n=(-i\vec\tau,1)$ are the
Euclidean space $\sigma$-matrices.\footnote{Here, $\vec\tau$ are the usual
Pauli matrices.} We will also use the quantities
\SP{
\sigma_{mn}&=\tfrac14(\sigma_m\bar\sigma_n-\sigma_n\bar\sigma_m)\ ,\\
\bar\sigma_{mn}&=\tfrac14(\bar\sigma_m\sigma_n-\bar\sigma_n\sigma_m)\ .
\label{defsmn}
}
Importantly, $\sigma_{mn}$ is self-dual while
$\bar\sigma_{mn}$ is anti-self-dual:
\EQ{
\sigma_{mn}=\tfrac12\epsilon_{mnkl}\sigma_{kl}\ ,\qquad
\bar\sigma_{mn}=-\tfrac12\epsilon_{mnkl}\bar\sigma_{kl}\ .
\label{projsasd}
}
They can be expressed as
\EQ{
\sigma_{mn}=\tfrac12 i \eta^c_{mn}\tau^c\ ,\qquad
\bar\sigma_{mn}=\tfrac12 i \bar\eta^c_{mn}\tau^c\ ,
\label{sigeta}
}
in terms of 't~Hooft's eta symbols \cite{tHooft}
$\eta^c_{mn}$ and
$\bar\eta^c_{mn}$, $a=1,2,3$,  defined below in \eqref{E15.1}.

In Minkowski space, we take
\EQ{
\gamma_n=\begin{pmatrix}0&\sigma_n\\
-\bar\sigma_n&0\end{pmatrix},\qquad n=0,1,2,3\ ,
\elabel{mfdc}
}
where we take the Minkowski space $\sigma$-matrices
$\sigma^n=(-1,\vec\tau)$ and $\bar\sigma^n=(-1,-\vec\tau)$ to agree with
the notation of Wess and Bagger \cite{WB}.

In both Euclidean and Minkowski space, we follow the usual convention
of writing two-component Weyl spinors as
$\psi_\alpha$, $\alpha=1,2$, and $\bar\psi^\aD$, $\aD=1,2$.
A Dirac spinor is
\EQ{
\psi=\MAT{\psi_\alpha\\ \bar\psi^\aD}\ .
\elabel{bxxy}
}
In Euclidean space the Weyl spinors are pseudo real and so
$\psi_\alpha$ and $\bar\psi^\aD$ are independent
quantities. In Minkowski space the Weyl spinors are complex and one
can define a four-component Majorana spinor of the form \eqref{bxxy}
with
\EQ{
\bar\psi^\aD=(\psi_\alpha)^\dagger\ ,\qquad\aD=\alpha\ .
}
In this notation, the $\sigma$-matrices have indices
$\sigma_{n\alpha\aD}$ and $\bar\sigma_n^{\aD\alpha}$. In both
Minkowski and Euclidean space, the indices
are raised and lowered with the anti-symmetric tensor $\epsilon$:
\EQ{
\epsilon_{21}=\epsilon^{12}=1\ ,\qquad\epsilon_{12}=\epsilon^{21}=-1\
,\qquad \epsilon_{11}=\epsilon_{22}=0\ .
}
In addition, we use the summation conventions:
\EQ{
\psi\phi\equiv\psi^\alpha\phi_\alpha\
,\qquad\bar\psi\bar\phi=\bar\psi_\aD\bar\phi^\aD\ .
}

In six-dimensional Euclidean space we take
\begin{equation}
\Gamma_a=\begin{pmatrix}0&\Sigma_a\\
\bar\Sigma_a&0\end{pmatrix}\ ,
\label{sbtt}\end{equation}
where the $\Sigma$-matrices are defined as
\begin{equation}\begin{split}
\Sigma_a&=\big(\eta^3,i\bar\eta^3,\eta^2,
i\bar\eta^2,\eta^1,i\bar\eta^1\big)\ ,\\
\bar\Sigma_a&=\big(-\eta^3,i\bar\eta^3,-\eta^2,i\bar\eta^2,-\eta^1,
i\bar\eta^1\big)\ ,
\end{split}\elabel{sigmatdef}\end{equation}
and in their turn, the $4\times4$-dimensional matrices $\eta^c$ and
$\bar\eta^c$, $c=1,2,3$, are 't~Hooft's eta symbols \cite{tHooft}:
\begin{align}
&\bar\eta_{AB}^c=\eta_{AB}^c=\epsilon_{cAB}\qquad A,B\in\{1,2,3\}, \notag\\
&\bar\eta_{4A}^c=\eta_{A4}^c=\delta_{cA},\elabel{E15.1}\\
&\eta_{AB}^c=-\eta_{BA}^c,\qquad\bar\eta_{AB}^c=-\bar\eta_{BA}^c.\notag
\end{align}

In six-dimensional Minkowski space we can take \eqref{sbtt} with
\SP{
\Sigma_a&=\big(i\eta^3,i\bar\eta^3,\eta^2,
i\eta^2,\eta^1,i\bar\eta^1\big)\ ,\\
\bar\Sigma_a&=\big(-i\eta^3,i\bar\eta^3,-\eta^2,i\bar\eta^2,-\eta^1,
i\bar\eta^1\big)\ ,
\elabel{minksm}
}
with $a=0,1,2,3,4,5$.

The following properties of the $\Sigma$-matrices are valid both in
Euclidean and Minkowski space:
\SP{
&\Sigma_{aAB}\Sigma_{aCD}=\bar\Sigma_{aAB}\bar\Sigma_{aCD}=2\epsilon_{ABCD}\\
&\Sigma_{aAB}\bar\Sigma_{aCD}=-2\delta_{AC}\delta_{BD}+2\delta_{AD}\delta_{BC}\
.
\elabel{smid}
}

A positive (negative) chirality Weyl spinor is written
$\psi^A$ ($\psi_A$), $A=1,2,3,4$.
For consistency the $\Sigma$-matrices have indices
$\Sigma_a^{AB}$ and $\bar\Sigma_{aAB}$.
In Euclidean space the Weyl spinors are complex while in Minkowski
space they are pseudo-real. In the latter case when there are an even
number of such spinors we can
introduce the concept of a ``symplectic-real spinor'' $\psi^A_i$,
$i=1,\ldots,2p$, satisfying the pseudo-reality condition
\EQ{
\psi^A_i=i\Omega_{ij}\bar\eta^1_{AB}(\psi^B_j)^*\ ,
}
where $\Omega_{ij}$ is a (symplectic) matrix with
$\Omega\Omega^*=-1_{\sst[2p]\times[2p]}$.

We can easily build up a representation of Clifford algebras in higher
dimensions from those in lower dimensions.
Suppose, we have Clifford algebras in $p$ and $q$ dimensions
with
Euclidean space generators $\Gamma^{(p)}_n$, $n=1,\ldots,p$,
and $\Gamma_a^{(q)}$, $a=1,\ldots,q$. A
representation of the Clifford algebra in $p+q$ dimensions can then be
constructed by taking the products
\EQ{
\Gamma_N=\Big\{\Gamma^{(p)}_n\otimes
1,\Gamma_{p+1}^{(p)}\otimes\Gamma_a^{(q)}\Big\}\ ,\qquad
N=1,\ldots,p+q\ .
}

\rsen\Appendix{Complex Geometry and the Quotient
Construction}\elabel{app:A2}

Our intention in this appendix is to introduce some basic properties
of complex manifolds\footnote{We will use the term ``manifold'' loosely to
encompass spaces which are generically smooth but may have
certain singularities.} and describe the ``quotient construction''.

A complex manifold $\ms$ admits a complex structure $\BI$, a
linear map of the tangent space to itself, satisfying
$\BI^2=-1$, which is integrable. This latter property means that the {\it
torsion\/}, or Neijenhuis tensor, vanishes, so for any two vectors $X$
and $Y$
\EQ{
[\BI X,\BI Y]-[X,Y]-\BI [X,\BI Y]-\BI[\BI X,Y]=0\ .
\elabel{xmk}
}
Necessarily $\ms$ must be of even dimension, $2n$. A
{\it Hermitian metric\/} $g$ on $\ms$ is invariant under $\BI$, so
\EQ{
g(\BI X,\BI Y)=g(X,Y)\ ,
\elabel{meti}
}
for any two tangent vectors $X$ and $Y$. The {\it fundamental 2-form\/}
$\omega$ is defined by
\EQ{
\omega(X,Y)=g(\BI X,Y)\ .
\elabel{orm}
}
(Notice that $\omega$ is anti-symmetric as a consequence of
\eqref{meti} and $\BI^2=-1$.) We will denote (real) local coordinates
on $\ms$ by $x^{\mu}$, $\mu=1,\ldots,2n$; however,
a complex manifold always admits local
holomorphic coordinates $(z^{\xii} ,\bar z^{\xii})$,
$\xii=1,\ldots,n$, for which
\EQ{
\BI=\begin{pmatrix} i\delta^{\xii} _{\ \xjj}&0\\ 0&-i\delta^{\xii}{}_{
\xjj}\end{pmatrix}\ .
}
In this basis, the Hermitian metric and fundamental 2-form are
\EQ{
g=g_{\xii\xjj}dz^{\xii} \,d\bar z^{\xjj}\ ,\qquad
\omega=ig_{\xii\xjj}dz^{\xii} \wedge d\bar z^{\xjj}\ .
}

A K\"ahler manifold is one for which the fundamental 2-form is closed,
in which case the latter is called the {\it K\"ahler form\/}.
The closure of the K\"ahler form implies (and follows from) the complex
structure $\BI$ is covariantly constant
\EQ{
\nabla_{\mu}\BI=0\ ,
}
with respect to the
Levi-Civita connection associated to $g$. In addition, in
holomorphic coordinates, the metric is given in terms
of a {\it K\"ahler potential\/} $K$ by
\EQ{
g_{\xii\xjj}=\frac{\partial}{\partial z^{\xii}}
\frac{\partial}{\partial \bar z^{\xjj}}K\ .
}
Another way to characterize a K\"ahler manifold is via its holonomy
group. Generically, on a manifold of dimension $2n$ this would be
$\O(2n)$; however, since parallel transport with the connection
$\nabla_{\mu}$ respects the holomorphic structure (does not mix
holomorphic and anti-holomorphic indices) the holonomy of a K\"ahler
manifold is contained in the subgroup $\U(n)\subset\O(2n)$.

We now proceed to define a hyper-K\"ahler manifold. Such a space
$\ms$ admits three independent complex structures
$\BI^{(c)}$, $c=1,2,3$, (which we often represent
as a three-vector $\vec\BI$), satisfying the algebra
\EQ{
\BI^{(c)}\BI^{(d)}=-\delta^{cd}+\epsilon^{cde}\BI^{(e)}\ ,
}
which are covariantly constant with respect to the Levi-Civita
connection of a metric $g$ which is
Hermitian \eqref{meti} with respect to each of the $\vec\BI$.
There are three K\"ahler forms $\vec\omega$ each related to
the metric as in \eqref{orm}. The space $\ms$ is necessarily $4n$
dimensional. As before, we will take $x^{\mu}$, $\mu=1,\ldots,4n$,
to be a set of real local coordinates for $\ms$. There are holomorphic
coordinates with respect to each of the complex structures, but
generally there are no local coordinates which simultaneously represent the
action of all the $\vec\BI$; consequently we will have to work in
terms of non-coordinate bases.
The tangent space $T\ms$ of a hyper-K\"ahler manifold admits the following
$\Sp(n)\times\SU(2)$ description.
There is a vielbein $h_{\xii\aD}{}^{\mu}$ and an inverse
$h^{\xii\aD}{}_{\mu}$.
The composite index $\xii\aD$, $\xii=1,\ldots,2n$ and $\aD=1,2$,
manifests the fact that the vielbein picks out an
$\Sp(n)\times\SU(2)$ structure in $T\ms$.\footnote{The group
$\Sp(n)$, which has rank $n$,
is often denoted as $\USp(2n)$.} The metric takes the form
\EQ{
g=h^{\xii\aD}{}_{\mu}h^{\xjj\bD}
{}_{\nu}\Omega_{\xii\xjj}\epsilon_{\aD\bD}dx^{\mu}\,
dx^{\nu}\ .
\elabel{metsp}
}
Here, $\Omega_{\xii\xjj}$, $\xii=1,\ldots,2n$, is an $\Sp(n)$ 2-form.
The index $\aD=1,2$ is an $\SU(2)$ spinor index which can be raised and
lowered in the usual way by the
$\epsilon$ tensor associated to $\SU(2)$ spinors (see Appendix \ref{app:A1}).
The complex structures act on
the $\SU(2)$ indices of the vielbeins in a simple way:
\EQ{
(\BI^{(c)}\cdot h)^{\xii\aD}{}_\mu
=-i\vec\tau^{c\aD}{}_{\bD}h^{\xii\bD}{}_\mu\ ,
}
where $\tau^{c}$ are the three Pauli matrices. Notice that the three
complex structures can be rotated as a three-vector under action of
the $\SU(2)$, although this action is not generally an isometry of $\ms$.
The three K\"ahler forms corresponding to the three complex structures
are
\SP{
\vec\omega=i\Omega_{\xii\xjj}\epsilon_{\aD\bD}
\vec\tau^\aD{}_{\gD}h^{\xii\bD}{}_{\mu}h^{\xjj\gD}{}_{\nu}
dx^{\mu}\wedge dx^{\nu}\ .
}

On a hyper-K\"ahler manifold, since the tangent space admits an
$\Sp(n)\times\SU(2)$ structure, one can define the notion of
{\it symplectic tangent vectors\/}. These are quantities, like
$\CM^\xii$, which carry $\Sp(n)$ indices only and which have a symplectic inner
product
\EQ{
\Omega(\CM,\CN)=\Omega_{\xii\xjj}\CM^\xii\CN^\xjj\ .
}
For a hyper-K\"ahler manifold the $\SU(2)$ part of the spin connection
$\omega_{\mu}{}^{\xii\aD}{}_{\xjj\bD}$ (not to be confused with the K\"ahler
forms) vanishes. This means
\EQ{
\omega_{\mu}{}^{\xii\aD}{}_{\xjj\bD}=
\omega_{\mu}{}^{\xii}{}_{\xjj}\delta^\aD{}_{\bD}\ .
}
The tensor $\Omega_{\xii\xjj}$ is covariantly constant with respect to
the $\Sp(n)$ part of the connection.
Similarly, the only non-vanishing components of the Riemann tensor are
\EQ{
R_{\xii\aD\,\xjj\bD\,\xkk\gD\,\xll\dD}=
R_{\xii\xjj\xkk\xll}\epsilon_{\aD\bD}\epsilon_{\gD\dD}\
.
\elabel{rtusp}
}
This is a symptom of the fact that the
holonomy group of a hyper-K\"ahler manifold is restricted to
$\Sp(n)\subset\U(2n)\subset\O(4n)$. The $\Sp(n)$ curvature
$R_{\xii\xjj\xkk\xll}$ is a totally symmetric tensor in all its indices.

After the brief introduction to complex geometry and, in particular,
hyper-K\"ahler spaces,
let us now turn to a way of constructing such spaces known as
the {\it quotient construction\/}. We begin, initially, with the simpler
K\"ahler case.
The quotient construction is a way of constructing a K\"ahler
manifold $\ms$ in terms of some larger dimensional K\"ahler manifold,
the ``mother'' space,
$\tilde\ms$ which admits some group of isometries $G$ which preserves both
the metric and complex structure. We will be interested in the case when
$G$ is some compact Lie group and we will denote the Hermitian
generators $T^r$, $r=1,\ldots,\text{dim}\,G$.\footnote{We will take an
inner product on the Lie algebra of $G$ with ${\rm
Tr}\,T^rT^s=\delta^{rs}$ in some matrix representation.}
Each generator $T^r$ of $G$ defines a
vector field $X_r$ over $\tilde\ms$ in the usual way.
Since the group action preserves the metric and complex structure
the Lie derivatives ${\cal L}_{X_r}\tilde\BI={\cal
L}_{X_r}\tilde g=0$. This implies that $X_r$ is a
{\it holomorphic Killing vector\/}. The two conditions imply a third; namely,
${\cal L}_{X_r}\tilde \omega=0$. Since
\EQ{
{\cal L}_{X_r}\tilde \omega=\imath_{X_r}d\tilde\omega+d(
\imath_{X_r}\tilde\omega)\equiv
d(\imath_{X_r}\tilde\omega)\ ,}
there exits a {\it Hamiltonian\/} function $\mu^{X_r}$ where
$d\mu^{X_r}=\imath_{X_r}\tilde\omega$.\footnote{This reasoning assumes
that the first cohomology group
$H^1(\tilde\ms,{\mathbb R})$ is trivial.}
In fact this only defines $\mu^{X_r}$ up to a constant which can be
fixed, up to the abelian factors in $G$, by requiring that $\mu^X$
is {\it equivariant\/}: $X\mu^Y=\mu^{[X,Y]}$. In this case, the Hamiltonians
are known as {\it moment maps\/}. As $T^r$ varies in the Lie algebra
of $G$ we can define
the function $\mu$ from $\tilde\ms$ to the Lie algebra
\EQ{
\mu=\sum_{r=1}^{\text{dim}\,G}\mu^{X_r}T^r\ .
}

The K\"ahler quotient is then
\EQ{
\ms=\mu^{-1}(0)/G\ ,
}
in other words the quotient of the subspace $\ns\subset\tilde\ms$ on which the
moment map vanishes, the so-called {\it level set\/},
by the group $G$. The daughter $\ms$ is of dimension
\EQ{
{\rm dim}\,\ms={\rm
dim}\,\tilde\ms-2\,{\rm dim}\,G
}
and inherits a
complex structure, metric and connection from $\tilde\ms$.
In order for the quotient to be well defined,
we require that $G$, at least generically, acts freely on $\tilde\ms$. There
may be points at which the action fails to be free, in which case there
will be orbifold-type singularities in the quotient. Since,
\EQ{
\tilde g(\tilde\BI X_r,X_s)=\tilde\omega(X_r,X_s)=-\langle
X_r,d\mu^{X_s}\rangle=-\mu^{[X_r,X_s]}=0\ ,
\elabel{tty}
}
using the equivariance property, a basis for the
vectors in the tangent space normal to the level set is given by
$\tilde\BI X_r$, $r=1,\ldots,\text{dim}\,G$.

Before we discuss the induced metric, complex structure and
connection, let us say a little more about the
$G$-quotient part of the construction.
It is useful to think of $\ns$ as a principal $G$-bundle over the quotient
$\ms=\ns/G$. Hence there is a projection $p:\,\ns\to\ms$. The
$G$-action picks out a subspace of $
{\EuScript V}\subset T\ns$ called the {\it vertical\/}
space spanned by the vector fields $X_r$. The tangent space $T\ms$ is
then identified with the quotient vector space $T\ns/{\EuScript
V}$. Let the {\it
horizontal\/} space, ${\EuScript H}\subset T\ns$
be the subspace of vectors orthogonal to ${\EuScript V}$, {\it
i.e.\/}~$\tilde g(X,X_r)=0$ for all $r$, so that
\EQ{
T\ns={\EuScript H}\oplus{\EuScript V}\ .
}
A tangent vector $X\in T\ms$ has a
unique horizontal lift to $\EuScript H$ (which by a slight abuse of
notation we denote with the same letter). From these definition it
follows that a
vector $X\in{\EuScript H}\subset T\tilde\ms$ satisfies
\EQ{
\tilde g(X,\tilde\BI X_r)=\tilde g(X,X_r)=0\ .
\elabel{uoo}
}

The complex structure $\tilde\BI$ in the mother space naturally
induces one in the daughter. To see this it is enough to notice from
\eqref{uoo} that $\tilde\BI$ preserves $\EuScript H$. Consequently, the lift
of $\BI$ is simply equal to $\tilde\BI$ acting on $\EuScript H$ and the
integrability condition \eqref{xmk} is automatically
satisfied. The induced Riemannian metric on $\ms$ is obtained
in a analogous way. For two vectors $X,Y\in T\ms$ (with lifts to
$\EuScript H$ denoted by the same letter), $g(X,Y)=\tilde g(X,Y)$.
Finally a Levi-Civita connection $\tilde\nabla$ on $\tilde\ms$
induces such a connection $\nabla$ on the daughter in the following way.
First of all,
the projection of the connection $\tilde\nabla$
to the tangent space $T\ns\subset T\tilde\ms$ defines a
connection on $\ns$. It is easy to prove that the
connection on $\ns$ is of Levi-Civita type.
This is then identified with the pull-back of the
Levi-Civita connection $\nabla$ on $\ms$ to $\ns$, under the projection
$p$.

Of particular importance for the instanton calculus is the volume form
induced on the daughter space (in the hyper-K\"ahler case discussed below).
It will be convenient for applications to the instanton
calculus to define a volume form which, using physicists'
language, is not gauge fixed. This means a
$G$-invariant volume form $\Bomega$ on the level
set $\ns$ divided by the volume of the $G$-orbit at that point. A
{\it bone-fide\/} volume form on $\ms$ could then be obtained by a
gauge fixing procedure, but we shall not necessarily 
need to perform this operation.
{}From what we have said above,
it is straightforward to see that the volume form we are after is
\EQ{
\int_{\ms}\Bomega\,
\overset{\rm def}=\,\int_{\tilde\ms}
\tilde\Bomega\ \frac{J(x)}{\text{Vol}_G(x)}\
\prod_{r=1}^{\text{dim}\,G}\delta(\mu^{X_r})\ .
}
Here, $\tilde\Bomega$ is the canonical volume form on $\ms$,
$\text{Vol}_G(x)$ is the volume of the $G$-orbit through a point
$x\in\tilde\ms$ and $J(x)$ is a Jacobian factor that arises when the
integral over $\tilde\ms$ is
restricted to $\ns$ by the explicit
$\delta$-functions. Geometrically, $J(x)$ is the square root of the
determinant of the matrix of inner products of the normal vectors to
the level set. Since a basis of such vectors is provided by $\tilde\BI X_r$,
the Jacobian is simply
\EQ{
J(x)=\big|\det\,\BL\big|^{1/2}\ ,
}
where $\BL$ is the $\text{dim}\,G\times\text{dim}\,G$ matrix
of inner products with elements
\EQ{
\BL_{rs}\equiv \tilde g(\tilde\BI X_r,\tilde\BI X_s)=\tilde g(X_r,X_s)\ ,
\elabel{mip}
}
Since $G$ generically acts freely on $\tilde\ms$, each orbit
is, up to a scale factor, a copy of $G$ itself. Hence, the volume of
the orbit through a point on the level set is
\EQ{
\text{Vol}_G(x)=\big|\det\,\BL\big|^{1/2}\ {\rm Vol}\,G\ ,
\elabel{volg}
}
Here, ${\rm Vol}\,G$ a constant, the volume of the group
in some canonical normalization. Notice that the factors of the
determinant of $\BL$ cancel to leave
\EQ{
\int_{\ms}\Bomega=\frac1{{\rm Vol}\,G}\int_{\tilde\ms}
\tilde\Bomega\ \prod_{r=1}^{\text{dim}\,G}\delta(\mu^{X_r})\ .
\elabel{mkq}
}

The hyper-K\"ahler quotient construction is an obvious generalization
of the K\"ahler quotient
construction described above. One starts with a hyper-K\"ahler space
$\tilde\ms$ admitting a
group action $G$ which preserves the metric and three complex
structures. The isometries correspond to vector fields which are {\it
tri-holomorphic Killing vectors\/}, that is (i) holomorphic
with respect to each of the three complex structures ${\cal
L}_X\tilde\omega^{(c)}=0$ and (ii) preserving
the metric ${\cal L}_X\tilde g=0$.
Associated to each complex structure and K\"ahler form there is a
moment map, defined as in the K\"ahler case above,
which we can assembled into the triplet
$\vec\mu$. In an identical way to the K\"ahler case,
one can prove that the quotient
\EQ{
\ms=\vec\mu^{-1}(0)/G
}
is a hyper-K\"ahler space of dimension ${\rm dim}\,\tilde\ms-4\,{\rm
dim}\,G$. In the hyper-K\"ahler
case, a basis of vectors normal to the level set is provided by the
$3\,\text{dim}\,G$ vectors $\tilde\BI^{(c)}X_r$. As in the, K\"ahler case a
natural metric $g(X,Y)$
is induced on $\ms$ by taking $\tilde g(X,Y)$, with $X$ and
$Y$  (denoted by the same symbol) being the unique lifts of $X$ and $Y$
to $\EuScript H$. The local condition for a vector
$X\in T\tilde\ms$ to be in $\EuScript H$ is
\EQ{
\tilde g(X,\tilde\BI^{(c)}X_r)=\tilde g(X,X_r)=0\ ,
\elabel{uooo}
}
generalizing \eqref{uoo}. In
the $\SU(2)\times\Sp(n)$ basis this condition is
$X^{\ii\aD}\tilde\Omega_{\ii\,\jj}\epsilon_{\aD\bD}X_r^{\jj\gD}\vec\tau^\bD{}_\gD=
X^{\ii\aD}\tilde\Omega_{\ii\,\jj}\epsilon_{\aD\bD}X_r^{\jj\bD}=0$, or equivalently
\EQ{
X^{\ii\aD}\tilde\Omega_{\ii\,\jj}X_r^{j\bD}=0\ .
\elabel{hah}
}
The fact that
the $\SU(2)$ indices of the tangent vector are un-summed,
allows us to identify symplectic tangent vectors to the quotient space.
Their lifts are simply symplectic tangent vectors of $\tilde\ms$
subject to the projection
\eqref{hah}:
\EQ{
\CM^{\ii} \tilde\Omega_{\ii\,\jj}X_r^{\jj\bD}=0\ .
\elabel{yrr}
}

Of particular importance to the instanton calculus
is the volume form on the quotient space,
generalizing the expression \eqref{mkq} in the K\"ahler case.
The Jacobian factor is, as before, the determinant of the matrix of
inner products of the basis vectors $\tilde\BI^{(c)}X_r$ normal to the level
set. We have,
\EQ{
\tilde g(\tilde\BI^{(c)}X_r,\tilde\BI^{(d)}X_s)=\delta^{cd}\tilde g(X_r,X_s)
-\epsilon^{cde}\tilde g(X_r,\tilde\BI^{(e)}X_s)
=\begin{cases} \tilde g(X_r,X_s) & c=d\ ,\\ 0 & c\neq d\ ,\end{cases}
\elabel{uay}
}
where we have used \eqref{tty}. This means that
\EQ{
J(x)=\big|\det\,\BL\big|^{3/2}\ ,
}
where the matrix $\BL$ is defined in \eqref{mip}.
As previously the volume of the gauge orbit through a point of the
level set is \eqref{volg}.
Hence the $G$-invariant volume form is
\EQ{
\int_{\ms}\Bomega=\frac1{{\rm Vol}\,G}\int_{\tilde\ms}\tilde\Bomega\
\big|\det\,\BL\big|\ \prod_{r=1}^{\text{dim}\,G}\prod_{c=1}^3
\delta(\mu^{(c)X_r})\ .
\elabel{measq}
}
Note the factor of $|\det\,\BL|$ as compared with the case of the
K\"ahler quotient \eqref{mkq}.

As an example of hyper-K\"ahler quotient construction which is
directly relevant to
the ADHM construction, let us start from flat Euclidean
space $\tilde\ms={\mathbb R}^{4n}$. In this case there are local
coordinates $z^{\ii\aD}$ for which the one forms
$h^{\ii\aD}=dz^{\ii\aD}$. The symplectic matrix is simply
\EQ{
\tilde\Omega=\MAT{0 & 1_{\sst[n]\times[n]}\\ -1_{\sst[n]\times[n]} & 0}\ .
}
and the flat metric is
\EQ{
\tilde g=4\pi^2
\tilde\Omega_{\ii\,\jj}\epsilon_{\aD\bD}dz^{\ii\aD}\,dz^{\jj\bD}\ .
\elabel{mothmet}
}
In this case, there exists a hyper-K\"ahler potential
\EQ{
\chi=
\tilde\Omega_{\ii\,\jj}\epsilon_{\aD\bD}z^{\ii\aD}z^{\jj\bD}\ .
}

We now consider the hyper-K\"ahler quotient of ${\mathbb R}^{4n}$
by some compact group $G$ of tri-holomorphic
isometries. The most general isometries of this type are generated by
the vector fields
\EQ{
X_r=iT_{\ii\,\jj}^rz^{\jj\aD}\PD{}{z^{\ii\aD}}\ ,
\elabel{defxr}
}
where $T^r$ generators in some $2n$-dimensional representation of $G$
which satisfies
\EQ{
(\tilde\Omega T^r)^t=\tilde\Omega T^r\ .
\elabel{fftt}
}
The moment maps are
\EQ{
\vec\mu^{X_r}=-\tfrac i2z^{\ii\aD}
(\tilde\Omega T^r)_{\ii\,\jj}z^{\jj\bD}\epsilon_{\aD\gD}
\vec\tau^\gD{}_\bD-\vec\zeta^r\ ,
\elabel{mommp}
}
where $\sum\vec\zeta^rT^r$,
are arbitrary constant elements in the abelian component of the
Lie algebra of $G$.

The (gauge un-fixed)
volume form on the quotient follows from the general formula
\eqref{measq}:
\EQ{
\int_{\ms}\Bomega=\frac1{{\rm Vol}\,G}\int
\Bigg\{\prod_{\aD=1,2}\prod_{\ii=1}^{2n}dz^{\ii\aD}
\Bigg\}\  \big|\det\,\BL\big|\ \prod_{r=1}^{\text{dim}\,G}\prod_{c=1}^3
\delta\Big(\tfrac i2z^{\ii\aD}
(\tilde\Omega T^r)_{\ii\,\jj}z^{\jj\bD}\epsilon_{\aD\gD}
\tau^{c\gD}{}_\bD-\vec\zeta^r\mu^{(c)X_r}\Big)\ ,
\elabel{measqex}
}
where, as previously,
\EQ{
\BL_{rs}=\tilde g(X_r,X_s)=-
\tilde\Omega_{\ii\,\jj}\epsilon_{\aD\bD}T^r_{\ii\,\kk}z^{\kk\aD}
T^s_{\jj\,\llr}z^{\llr\bD}\ .
\elabel{lso}
}

For this example, we now discuss more concretely the connection on the
quotient and find an explicit expression for the Riemann
tensor. First the connection. As we have explained previously, the
prescription is simple: lift tangent vectors to
$\ms$ to $\EuScript H$
(denoted by the same symbol). The Levi-Civita connection is then obtained by
projection to $\EuScript H$ from that on $\tilde\ms$.
Locally, the tangent space of the horizontal subspace consists of
vectors $X$ that are orthogonal to the $4\,\text{dim}\,G$ vectors
$\{X_r,\tilde\BI^{(c)} X_r\}$, the conditions \eqref{uooo}; hence, for
two vectors $X,Y\in T\ms(\simeq{\EuScript H})$
\EQ{
\nabla_{X}Y=\big(\tilde\nabla_XY\big)^\parallel=
\tilde \nabla_XY-\big(\tilde\nabla_XY\big)^\perp\ ,
}
where
\EQ{
\big(\tilde\nabla_XY\big)^\perp
=\sum_{rs}\BL^{-1}_{rs}\Big\{\tilde g(\tilde\nabla_XY,X_s)X_r+
\sum_{c=1}^3g(\tilde\nabla_XY,\tilde\BI^{(c)}X_s)\tilde\BI^{(c)}X_r\Big\}\ .
\elabel{kjh}
}
We can make the inner product more explicit by using
$(\tilde\nabla_XY)^{i\aD}
=X^{\jj\bD}\partial Y^{\ii\aD}/\partial z^{\jj\bD}$ and the
expression for $X_r$ in \eqref{defxr}
\SP{
&\tilde g(\tilde\nabla_XY,X_r)
=-iX^{\ii\aD}(\tilde\Omega T^r)_{\ii\,\jj}Y^{\jj\bD}
\epsilon_{\bD\aD}\ ,\\
&\tilde g(\tilde\nabla_XY,\tilde\BI^{(c)} X_r)=
-X^{\ii\aD}(\tilde\Omega
T^r)_{\ii\,\jj}Y^{\jj\bD}\epsilon_{\bD\gD}\tau^{c\gD}{}_{\aD}\ .
}
In deriving these expressions we used the facts
$\tilde g(Y,X_r)=\tilde
g(Y,\BI^{(c)} X_r)=0$ and, from \eqref{fftt}, $\tilde\Omega T^r$ is
a symmetric matrix.

The Riemann tensor of $\ms$ can be expressed using the standard
formula in terms of the lifts $W,X,Y,Z\in{\EuScript H}$:
\SP{
&R(X,Y,W,Z)
=\tilde g\big(W,\tilde\nabla_X(\tilde\nabla_YZ)^\parallel-
\tilde\nabla_Y(\tilde\nabla_XZ)^\parallel-\tilde\nabla_{[X,Y]^\parallel}
Z\big)\\
&=\tilde g\big(W,\tilde\nabla_X(\tilde\nabla_YZ)^\parallel-
\tilde\nabla_Y(\tilde\nabla_XZ)^\parallel-\tilde\nabla_Z(\tilde
\nabla_XY)^\parallel
+\tilde\nabla_Z(\tilde\nabla_YX)^\parallel-[Z,[X,Y]]\big)\ .
\elabel{tttp}
}
Notice that some of the projections are unnecessary here, since
$W\in{\EuScript H}$ and
$[Z,K]\in{\EuScript V}$ for any $K\in{\EuScript V}$. Using the fact that
$\tilde g(W,\tilde\nabla_X(\tilde\nabla_YZ)^\perp)=-
\tilde g(\tilde\nabla_XW,(\tilde\nabla_YZ)^\perp)$ we
can write \eqref{tttp} as
\SP{
R(X,Y,W,Z)=&\tilde R(X,Y,W,Z)+\tilde
g\big((\tilde\nabla_XZ)^\perp,(\tilde\nabla_YW)^\perp\big)
-\tilde g\big((\tilde\nabla_XW)^\perp,(\tilde\nabla_YZ)^\perp\big)\\
&\qquad-\tilde g\big((\tilde\nabla_XY)^\perp,(\tilde\nabla_WZ)^\perp\big)
+\tilde g\big((\tilde\nabla_YX)^\perp,(\tilde\nabla_WZ)^\perp\big)\ ,
}
where $\tilde
R$ is the Riemann tensor of $\tilde\ms$. In the example relevant to
the ADHM construction this vanishes since
the mother space is flat.

We now extract the components of the Riemann tensor in the $z^{\ii\aD}$
basis by choosing
$X=\partial/\partial
z^{\ii\aD}$, $Y=\partial/\partial z^{\jj\bD}$, $W=\partial/\partial
z^{\kk\gD}$ and $Z=\partial/\partial z^{\llr\dD}$. One finds
\SP{
&R_{\ii\aD,\jj\bD,\kk\gD,\llr\dD}\\
&=2\epsilon_{\aD\bD}\epsilon_{\gD\dD}\sum_{rs}\Big\{(\tilde\Omega
T^r)_{\ii\,\jj}\BL^{-1}_{rs}(\tilde\Omega T^s)_{\kk\,\llr}+
(\tilde\Omega T^r)_{\ii\,\llr}\BL^{-1}_{rs}(\tilde\Omega T^s)_{\jj\,\kk}+
(\tilde\Omega T^r)_{\ii\,\kk}\BL^{-1}_{rs}(\tilde\Omega T^s)_{\jj\,\llr}\Big\}
\ .
}
This forms reflects the decomposition of the Riemann tensor
in \eqref{rtusp}
and so we can extract the symplectic curvature of the quotient:
\EQ{
R_{\ii\,\jj\,\kk\,\llr}=2\sum_{rs}\Big\{(\tilde\Omega
T^r)_{\ii\,\jj}\BL^{-1}_{rs}
(\tilde\Omega T^s)_{\kk\,\llr}+(\tilde\Omega
T^r)_{\ii\,\llr}\BL^{-1}_{rs}
(\tilde\Omega T^s)_{\jj\,\kk}+(\tilde\Omega T^r)_{\ii\,\kk}\BL^{-1}_{rs}
(\tilde\Omega T^s)_{\jj\,\llr}\Big\}\ .
\elabel{gju}
}
Since, from Eq.~\eqref{fftt},
$\tilde\Omega T^r$ is a symmetric matrix, it is apparent that
the symplectic curvature is totally symmetric in all its indices.

\rsen\Appendix{ADHM Algebra}\elabel{app:A4}

In this appendix, we collect together most the pieces of ADHM algebra
that we need in the text. Performing ADHM algebra is rather an art,
requiring a good deal of chicanery and experience. The ADHM constraints and
identities \eqref{dbd}, \eqref{ADHMbi},  \eqref{bbident},
\eqref{mdid} and \eqref{zmconb}:
\SP{
&\bar\Delta^\aD\Delta_\bD=\delta^\aD{}_\bD f^{-1}\ ,\qquad\bar\Delta^\aD\bar
b^\alpha=\bar b^\alpha\Delta^\aD\ ,
\qquad\bar b_\alpha b^\beta=\delta_\alpha{}^\beta1_{\sst[k]\times[k]}\\
&\bar\Delta^\aD\CM+\bar\CM\Delta^\aD=0\ ,\qquad
\bar\CM b^\alpha=\bar b^\alpha\CM
\elabel{collect}
}
are used pervasively.
The following differentiation formulae which are proved using the
definitions \eqref{dbd} and \eqref{cmpl}, are particularly useful:
\AL{
\partial_nf &=
-f\partial_n(\tfrac12\bar\Delta^\aD\Delta_\aD)f =
\begin{cases}
-f\sigmabar_n^{\aD\alpha}\bar b_\alpha\Delta_\aD f & \\
-f\bar\Delta^\aD b^\alpha\sigma_{n\alpha\aD}f & \end{cases}
\elabel{fderivs}\\
\square f&=-4 f\bar b_\alpha {\cal P} b^\alpha f\ ,\elabel{ddf}\\
\partial_n {\cal P}&=-\Delta_\aD f\bar\sigma^{\aD\alpha}_n\bar b_\alpha
{\cal P}-{\cal P} b^\alpha\sigma_{n\alpha\aD} f\bar\Delta^\aD\ ,\elabel{sdp}\\
\square{\cal P}&= -4\{{\cal P},b^\alpha f\bar b_\alpha\}
+4\Delta_\aD f\bar b_\alpha {\cal P} b^\alpha f\bar\Delta^\aD\ .
\elabel{ddP}
}
To complete this compendium of differentiation formulae, using the
expression for the gauge potential \eqref{vdef} and the ADHM identity
\eqref{uan} along with \eqref{collect}, for any ${\cal J}(x)$, we have
\AL{
\D_n(\bar U\J U)&=\partial_n(\bar U\J U)+[A_n,\bar U\J U]\notag \\
&= \bar U\,\partial_n\J\,U-\bar Ub^\alpha\sigma_{n\alpha\aD}
f\bar\Delta^\aD\J U-\bar U\J\Delta_\aD f\bar\sigma_n^{\aD\alpha}\bar
b_\alpha U\ ,\elabel{genderiv}\\
{\cal D}^2(\bar U\J U)&=-4\bar U\{b^\alpha f\bar b_\alpha,\J\}U
+4\bar U b^\alpha f\bar\Delta^\aD\J\Delta_\aD f\bar b_\alpha U\notag \\
&+\bar U\partial^2\J U-2\bar Ub^\alpha
f\sigma_{n\alpha\aD}\bar\Delta^\aD
\partial_n\J U-2\bar
U\partial_n\J\Delta_\aD\bar\sigma^{\aD\alpha}_nf\bar b_\alpha U\ .\elabel{gdd}
}

Finally, there are various other tricks that we will explain {\it in
situ\/}; however, there are particularly useful ones involving
quantities of the form
\EQ{
{\cal J}=\MAT{B_{\sst[N]\times[N]}&0_{\sst[N]\times[2k]}\\
0_{\sst[2k]\times[N]}&C_{\sst[k]\times[k]}1_{\sst[2]\times[2]}}\ .
}
The pair of identities is
\EQ{
\bar\Delta^\aD{\cal J}U=\big(\bar a^\aD{\cal J}-
C a_\aD\big)U\ ,\qquad
\bar U{\cal J}\Delta_\aD=\bar U\big({\cal J}a_\aD-a_\aD C
\big)\ .
\elabel{trick}
}
The first of these is proved by expanding $\bar\Delta=\bar
a+\bar x\bar b$, as in
\eqref{del}. Then with $\bar b$ assuming the canonical form
\eqref{aad} we can commute $\bar x\bar b$
it through ${\cal J}$, picking out the component $C$, and then
use to \eqref{uan} to re-write $\bar x\bar bU=-\bar aU$. The other
identity is proved in a similar way.

Note that where possible we suppress indices. However, in most situations
the spinor indices $\alpha$ and $\aD$ need to be written explicitly
because they
are often not contracted in an obvious way.

{\it Osborn's formula}

This identity reads \cite{OSB}
\EQ{
\TrN\,F_{mn}^2=-g^{-2}\,\square^2\,{\rm tr}_k\,\log\,f\ .
\elabel{oruf}
}

We start by expanding out the left-hand side using the ADHM form for
the field strength \eqref{sdu}:
\EQ{
\TrN\,F_{mn}^2=-16g^{-2}{\rm tr}_k\big(\bar b_\alpha{\cal P}b^\beta f\bar
b^\alpha{\cal P}b_\alpha f+\bar b_\alpha{\cal P}b^\alpha f\bar
b_\beta{\cal P}b^\beta f\big)\ .
\elabel{ffadhm}
}
Now the right-hand side. Firstly,
\SP{
\square\,{\rm tr}_k\,\log\,f&={\rm tr}_k\big(f^{-1}\square
f-f^{-1}\,\partial_n f\,f^{-1}\,\partial_nf\big)\\
&=-2{\rm tr}_k\big(\bar b_\alpha{\cal P}b^\alpha f+2f\big)\ ,
}
using the differentiation formulae \eqref{fderivs} and \eqref{ddf}
along with the expression for ${\cal P}$ in \eqref{cmpl}. Then
employing the differentiation formulae \eqref{fderivs}-\eqref{ddP} once
again
\SP{
\square^2\,{\rm tr}_k\,\log\,f&=8{\rm tr}_k\,\big(\bar b_\alpha\{{\cal
P},b^\beta f\bar b_\beta\}b^\alpha f-\bar b_\alpha\Delta_\aD f\bar
b_\beta{\cal P}b^\beta f\bar\Delta^\aD b^\alpha f\\
&-\bar b_\alpha\Delta_\aD f\bar b_\beta{\cal P}b^\alpha
f\bar\Delta^\aD b^\beta f-\bar b_\alpha{\cal P}b^\beta f\bar\Delta^\aD
b^\alpha f\bar b_\beta\Delta_\aD f\\
&+\bar b_+\alpha{\cal P}b^\alpha f\bar b_\beta{\cal P}b^\beta f+
2f\bar b_\alpha{\cal P}b^\alpha f\big)\ .
}
Then using the ADHM identity \eqref{ADHMbi}, the
definition of ${\cal P}$ in \eqref{cmpl} along with \eqref{bbident},
this becomes
\EQ{
\square^2\,{\rm tr}_k\,\log\,f=
16{\rm tr}_k\big(\bar b_\alpha{\cal P}b^\beta f\bar
b^\alpha{\cal P}b_\alpha f+\bar b_\alpha{\cal P}b^\alpha f\bar
b_\beta{\cal P}b^\beta f\big)\ ,
}
precisely $-g^2$ times \eqref{ffadhm}.

{\it Zero modes}

We prove that the quantity
\EQ{
\Lambda_\alpha(C)\equiv\bar U Cf\bar b_\alpha U
-\bar U b_\alpha f\bar CU
\elabel{dzmm}
}
satisfies the zero mode condition
\EQ{
\bar{\cal D}^{\aD\alpha}\Lambda_\alpha(C)=0\ .
\elabel{bvz2}
}

{}From Eqs.~\eqref{dzmm}, \eqref{fderivs} and \eqref{genderiv} we calculate:
\begin{equation}\bar D^{\aD\alpha}
\Lambda_\alpha(C)=
2\bar U b^\alpha f\big(\bar\Delta^{\aD} C+
\bar C\Delta^\aD\big)f\bar b_\alpha U\ .
\elabel{zmid}\end{equation}
Hence the condition for a zero mode is
\EQ{
\bar\Delta^{\aD}C+\bar C\Delta^\aD=0
\ .
\elabel{hap}
}
Expanding $\Delta(x)$ as in \eqref{del}, we have
\SP{
\bar C_{i}^\lambda a_{\lambda j\aD}&=-\bar a_{i\aD}^\lambda C_{\lambda j}\\
\bar C_{i}^\lambda b^\alpha_{\lambda j}&=\bar
b_{i}^{\alpha\lambda} C_{\lambda j}\ .
\elabel{hbp}
}

{\it Derivative of the gauge field by a collective
coordinate}

We now prove that
\EQ{
g\PD{A_n}{X^\mu}=-
{\cal D}_n\big(\PD{\bar U}{X^{\mu}}U\Big)+\bar U\PD a{X^{\mu}}
f\bar\sigma_n\bar bU-\bar U b\sigma_n f\PD{\bar a}{X^{\mu}}U\ .
\elabel{wwtpt}
}

Firstly, using $\bar UU=1$,
\EQ{
g\frac{\partial A_n}{\partial X^{\mu}}=\PD{\bar U}{X^{\mu}}\partial_nU+
\bar U\partial_n\PD U{X^{\mu}}=
\PD{\bar U}{X^{\mu}}\partial_nU+\partial_n\Big(\bar U\PD
U{X^\mu}\Big)-(\partial_n\bar U)\PD U{X^{\mu}}\ .
}
Next we insert $1\equiv U\bar U+\Delta f\bar\Delta$ into the middle of the
$1^{\rm th}$ and $3^{\rm th}$ terms, and then use the fact that $A_n=g^{-1}\bar
U\partial_n U$, to arrive at
\EQ{
g\PD{A_n}{X^\mu}=
-D_n\Big(\PD{\bar U}{X^{\mu}}U\Big)+\PD{\bar U}{X^{\mu}}\Delta
f\bar\Delta\partial_nU-(\partial_n\bar U)\Delta f\bar\Delta\PD U{X^{\mu}}\ .
}
{}From the ADHM identity \eqref{uan}, we have
\SP{
&\bar\Delta\partial_nU=-(\partial_n\bar\Delta)U=-\bar\sigma_n\bar bU\ ,\qquad
(\partial_n\bar U)\Delta=-\bar U\partial_n\Delta=-\bar Ub\sigma_n\
,\\
&\bar\Delta\PD UX=-\PD{\bar\Delta}XU=-\PD{\bar a}XU\ ,\qquad
\PD{\bar U}X\Delta=-\bar U\PD\Delta X=-\bar U\PD aX\ ,
\elabel{silid}
}
from which \eqref{wwtpt} follows.

{\it Corrigan's inner product formula}

The expression to be proved reads:
\begin{equation}
\int d^4x\,{\rm tr}_N\,\Lambda(C) \Lambda(C')= -{\pi^2\over2}
{\rm tr}_k\big[\bar C (\Pinfty+1){C'}-
\bar C'(\Pinfty+1)C\big]\ .
\elabel{corrigan}
\end{equation}
Here
\EQ{
\Pinfty=
\underset{x\rightarrow\infty}{\rm lim}\ {\cal P} =
1-b\bbar=\MAT{1_{\sst[N]\times[N]}&0_{\sst[N]\times[2k]}\\
0_{\sst[2k]\times[N]} & 0_{\sst[2k]\times[2k]}}\ ,
\elabel{Pinftydef2}
}
as per Eqs.~\eqref{cmpl} and \eqref{asymadhm}.

The strategy of the proof
is to show that the integrand is actually a total derivative,
\begin{equation}
{\rm tr}_N\,\Lambda(C) \Lambda(C')
 =\eighth\square\,{\rm tr}_k\big[\bar C({\cal P}+1)C' f-
\bar C'({\cal P}+1) C f\big]\ ,
\elabel{express2}
\end{equation}
after which Eq.~\eqref{corrigan} follows from Gauss' Theorem, together with the
asymptotic formulae of \S\ref{sec:S12}.
To verify this, let us first write out the
left-hand side of Eq.~\eqref{express2}:\footnote{Here, and in the
following, the trace on the right-hand side is either over instanton
or ADHM indices, depending on the context.}
\begin{equation}
{\rm tr}\big[\,(\bar C^{\prime} {\cal P} C-\bar C {\cal P} C^{\prime})f\bar b_\alpha {\cal P} b^\alpha
f
-\bar C {\cal P} b^\alpha f\bar C^{\prime} {\cal P} b_\alpha f-{\cal P} C f\bar b_\alpha {\cal P} C^{\prime}
f\bar b^\alpha\,\big].
\elabel{lamzeta}
\end{equation}
We have used the cyclicity of the trace, together with the definition
\eqref{cmpl} for the projector ${\cal P}.$ Turning to the right-hand side of
Eq.~\eqref{express2}, one calculates:
\SP{
\eighth\square\,{\rm tr}\big[\bar C({\cal P}+1) C^{\prime} f\big]
&=
\fourth{\rm tr}\big[\,-2\bar C\{{\cal P},b^\alpha f\bar b_\alpha\} C^{\prime} f
+2\bar C\Delta_\aD f\bar b_\alpha {\cal P} b^\alpha f\bar\Delta^\aD C^{\prime} f
\\
&\qquad\qquad-2\bar C({\cal P}+1) C^{\prime} f\bar b_\alpha {\cal P} b^\alpha f
+\bar  C\Delta_\aD f \bar\sigma^{n\aD\alpha}\bar b_\alpha
{\cal P} C^{\prime}\partial_nf
\\&\qquad\qquad
-\bar C {\cal P} b^\alpha\sigma_{n\alpha\aD}f\bar\Delta^\aD C^{\prime}\partial_nf\,
\big]
\\
&=
\hf{\rm tr}\big[\, C f\bar b_\alpha {\cal P} b^\alpha\bar C^{\prime}({\cal P}-1)
-\bar C({\cal P}+1) C^{\prime} f\bar b_\alpha {\cal P} b^\alpha f
\\
&\qquad\qquad
+ C f\bar b^\alpha {\cal P} C^{\prime} f\bar b_\alpha {\cal P}
+\bar C {\cal P} b_\alpha f\bar C^{\prime} {\cal P} b^\alpha f\,\big]
\\
&=\
\hf{\rm tr}_N\Lambda(C) \Lambda(C')
-\hf{\rm tr}\big[\,(\bar C C^{\prime}+\bar C^{\prime} C)f\bar b_\alpha {\cal P} b^\alpha
f\,\big]\ .
\elabel{mess}
}
Here the expressions on the right-hand sides follow from the
differentiation formulae \eqref{fderivs}-\eqref{ddP}.
We have also invoked the relations \eqref{zmcona}, \eqref{zmconb} and
\eqref{cmpl} and, once again, cyclicity under the trace. From the
final rewrite in Eq.~\eqref{mess},
 the desired result \eqref{express2}
follows by inspection upon anti-symmetrization in $C$ and $C'$.

{\it Covariant Laplace equation with bi-fermion source}

We now prove that for an adjoint-valued scalar field the solution of
the covariant Laplace equation with a bi-fermion source:
\EQ{
{\cal
D}^2\phi=\Lambda(C)\Lambda(C')
\elabel{uyt}
}
and boundary condition $\lim_{x\to\infty}\phi(x)=\phi^0$, is
\EQ{
\phi=-\tfrac14\bar UCf\bar C'U+\bar U\MAT{\phi^0 & 0\\
0&\varphi1_{\sst[2]\times[2]}}U\ ,
\elabel{vxx}
}
where
\EQ{
\varphi=\BL^{-1}\big(\bar w^\aD\phi^0w_\aD+\tfrac14\bar CC'\big)\ .
}
is a $k\times k$ matrix.

First of all, using the asymptotic formulae \eqref{asymadhm} it is easy to see
that \eqref{vxx} has the correct boundary condition at infinity. Next
we employ the differentiation formula \eqref{gdd}
with $\J=\tfrac14Cf\bar C'$ and compare with
\SP{
\Lambda(C)\Lambda(C')&=-\bar UCf\bar b_\alpha{\cal P} C'f\bar
b^\alpha U+\bar UCf\bar b_\alpha{\cal P} b^\alpha f\bar C'U\\
&+\bar Ub_\alpha f\bar C{\cal P} C'f\bar b^\alpha U-\bar Ub_\alpha f\bar
C{\cal P} b^\alpha
f\bar C'U\ .
\elabel{bxx}
}
Using the differentiation formula \eqref{ddf} we find that
the 3rd term in \eqref{gdd} matches the 2nd term in \eqref{bxx}. Then
writing ${\cal P}=1-\Delta_\aD f\bar\Delta^\aD$ in the 1st and 4th terms of
\eqref{bxx} and using Eqs.~\eqref{hap} and \eqref{hbp}, $\bar\Delta_\aD C=-\bar
C\Delta_\aD$ and $\bar b_\alpha C=\bar Cb_\alpha$, and the
differentiation formula \eqref{fderivs}, we find that these
terms match the 1st, 4th and 5th terms in \eqref{gdd}. This leaves
the 3nd term in \eqref{bxx} which matches the 2rd term in \eqref{gdd}
apart from the fact that ${\cal P}$ is replaced by ${\cal P}-1$. Hence
\EQ{
{\cal D}^2(\tfrac14\bar UCf\bar
C'U)=\Lambda(C)\Lambda(C')-\bar Ub^\alpha f\bar CC'f\bar
b_\alpha U\ .
\elabel{gaq}
}

Now consider \eqref{gdd} with
\EQ{
\J=\MAT{\phi^0 & 0\\ 0
&\varphi1_{\sst[2]\times[2]}}\ ,\qquad \partial_n\varphi=0\ .
}
We find
\EQ{
{\cal D}^2(\bar U\J U)=4\bar Ub^\alpha\bigg\{-\{f,\varphi\}
+f\bar\Delta^\aD\MAT{\phi^0&0\\ 0&\varphi}\Delta_\aD
f\bigg\}\bar b_\alpha U\ .
}
Then
\EQ{
\bar\Delta^\aD\MAT{\phi^0&0\\ 0&\varphi}\Delta_\aD
=\bar w^\aD\phi^0w_\aD-\BL\varphi+\{\varphi,f^{-1}\}\ ,
\elabel{funiwl}
}
where $\BL$ is defined in \eqref{vvxx}. Putting this together with
\eqref{gaq} we have solved \eqref{uyt} if
\EQ{
\BL\varphi=\bar w^\aD\phi_0w_\aD+\tfrac14\bar CC'\ .
}

{\it Anti-fermion source}

We now prove that
\begin{equation}
\bar\Sigma_{aAB}[\phi_a,\Lambda(\CM^B)]=
{\cal D}_{\alpha\dalpha}\psibar^{\dalpha}_A+\Lambda(\CN_A)\ .
\elabel{psibardef}\end{equation}
Here, $\phi_a$ is the solution to \eqref{uyt} given in \eqref{vxx}.
The collective coordinate matrix $\N_A$ (which will be seen to depend in a
nontrivial way on the original collective coordinates $\{a,{\cal M}^A\}$) is
subject to the usual zero mode conditions \eqref{zmcona}-\eqref{zmconb}.

We now solve \eqref{psibardef} for
$\psibar^{\dalpha A}$ and $\CN_{A}$. First of all, from
\eqref{ssdd} and \eqref{smid}
\EQ{
\bar\Sigma_{aAB}\phi_a=-\tfrac12\epsilon_{ABCD}
\bar U\CM^Cf\bar\CM^DU+
\bar\Sigma_{aAB}\bar U\MAT{\phi_a^0 & 0\\
0&\varphi_a1_{\sst[2]\times[2]}}U\ ,
\elabel{hji}
}
where $\varphi_a$ is defined in \eqref{dxx}.
\EQ{
\varphi_a=\BL^{-1}\big(\tfrac14\bar\Sigma_{aAB}\bar\CM^A\CM^B+\bar
w^\aD\phi_a^0w_\aD\big)\ .
}
As usual in ADHM calculus, some
educated guesswork is required. To this end we expand the left-hand
side of \eqref{psibardef}, using Eqs.~\eqref{lam} and \eqref{hji}:
\begin{equation}\begin{split}
\bar\Sigma_{aAB}[\phi_a,\Lambda(\CM^B)]=&
\tfrac12\epsilon_{ABCD}\,\bar U\bigg\{
{\cal M}^Bf\bbar_\alpha{\cal P}{\cal M}^Cf\Mbar^D
-b_\alpha f\Mbar^B{\cal P}{\cal M}^Cf\Mbar^D
\\
&-\CM^Cf\bar\CM^D{\cal P}\CM^Bf\bar b_\alpha+
\CM^Cf\bar\CM^D{\cal P} b_\alpha f\bar\CM^B\bigg\}U\\
&+\bar\Sigma_{aAB}\bar U\bigg\{-{\cal M}^Bf\bbar_\alpha{\cal P}
\begin{pmatrix}\phi_a^0 &0 \\
0 &\varphi_a\end{pmatrix}
+b_\alpha f\Mbar^B{\cal P}
\begin{pmatrix}\phi_a^0 &0 \\
0 &\varphi_a\end{pmatrix}\\
&+\begin{pmatrix}\phi_a^0 &0 \\
0 &\varphi_a\end{pmatrix} {\cal P}\CM f\bar b_\alpha
-\begin{pmatrix}\phi_a^0 &0 \\
0 &\varphi_a\end{pmatrix} {\cal P} b_\alpha f\bar\CM^B\,\bigg\}U\ .
\elabel{writeout}\end{split}\end{equation}
Here $\cal P$ is the projection operator defined in Eq.~\eqref{cmpl}
above.
Since $\partial_n\Delta=b\sigma_n,$
a comparison of Eqs.~\eqref{writeout} and
\eqref{genderiv} motivates the ansatz:
\def\one{{(1)}}
\def\two{{(2)}}
\def\three{{(3)}}
\begin{equation}
\psibar_A=\psibar^\one_A+
\psibar^\two_A+\psibar^\three_A
\elabel{psiansatz}\end{equation}
where
\begin{subequations}\begin{align}
\psibar^\one_{\dalpha A}&=
-\tfrac14\epsilon_{ABCD}
\Ubar{\cal M}^Bf\Deltabar_\dalpha {\cal M}^C f\Mbar^D U\ ,\elabel{psionedef}\\
\psibar^\two_{\dalpha A}&
=\tfrac12\bar\Sigma_{aAB}
\Ubar\bigg\{{\cal M}^Bf\Deltabar_\dalpha
\begin{pmatrix}\phi_a^0 &0 \\
0 &\varphi_a\end{pmatrix}+\begin{pmatrix}\phi_a^0 &0 \\
0 &\varphi_a\end{pmatrix}\Delta_\aD f\bar\CM^B\Bigg\}U\ ,
\elabel{psitwodef}\\
\psibar^\three_{\dalpha A}&=
\Ubar
\begin{pmatrix}0 &0 \\ 0 & \G_{\dalpha A}
\end{pmatrix}  U\ ,\quad\bar\G_{\dalpha A}=-\G_{\dalpha A}
\ ,\quad
\partial_n\G_{\dalpha A}=0\ .
\elabel{psithreedef}\end{align}
\end{subequations}
We expect $\psibar^\one_{A}$ to account (more or less) for the
first four terms on the right-hand side of
Eq.~\eqref{writeout}, and $\psibar^\two_{A}$ to
account (more or less) for the final four. The presence of
$\psibar^\three_{A}$, while less obviously motivated at this
stage, will be needed to ensure that the quantity $\CN_{A}$ defined in
Eq.~\eqref{psibardef} obeys the zero-mode constraints
\eqref{zmcona}-\eqref{zmconb}.

By explicit calculation using Eqs.~\eqref{fderivs}, \eqref{zmcona},
\eqref{zmconb}, \eqref{genderiv} and \eqref{psionedef}, one finds:
\SP{
{\cal D}_{\alpha\dalpha}\psibar^{\one\dalpha}_A =&
\tfrac12\epsilon_{ABCD}\,\bar U\bigg\{
\,{\cal M}^Bf\bbar_\alpha{\cal P}{\cal M}^C
f\Mbar^D-
b_\alpha f\Mbar^B({\cal P}-1){\cal M}^Cf\Mbar^D\\
&+\CM^C f\bar\CM^D{\cal P} b_\alpha f\bar\CM^B
-\CM^C f\bar\CM^D({\cal P}-1)\CM^B f\bar b_\alpha\bigg\}U\ .
\elabel{writeone}
}
Except for the ``$-1$'' in 2nd and 4th terms, this  reproduces
the first four terms
of Eq.~\eqref{writeout}, as expected. Similarly one calculates:
\begin{equation}\begin{split}
{\cal D}_{\alpha\dalpha}\psibar^{\two\dalpha}_A &=
-\bar\Sigma_{aAB}\bar U\bigg\{
{\cal M}^Bf\bbar_\alpha({\cal P}+1) \begin{pmatrix}\phi_a^0 &0 \\
0 &\varphi_a\end{pmatrix}
-b_\alpha f\Mbar^B({\cal P}-1)
\begin{pmatrix}\phi_a^0 &0 \\
0 &\varphi_a\end{pmatrix}
\\
&\qquad+\begin{pmatrix}\phi_a^0 &0 \\
0 &\varphi_a\end{pmatrix} ({\cal P}+1)b_\alpha f\bar\CM^B
-\begin{pmatrix}\phi_a^0 &0 \\
0 &\varphi_a\end{pmatrix} ({\cal P}-1)\CM^B f \bar b_\alpha\\
&\qquad-b_\alpha f\Deltabar^\aD
\begin{pmatrix}\phi_a^0 &0 \\
0 &\varphi_a\end{pmatrix}\Delta_\aD f\Mbar^B
+\CM^B f \bar\Delta^\aD\begin{pmatrix}\phi_a^0 &0 \\
0 &\varphi_a\end{pmatrix}\Delta_aD f\bar b_\alpha\bigg\}U
\\
&=\bar\Sigma_{aAB}\bar U\bigg\{
{\cal M}^Bf\bbar_\alpha{\cal P}
\begin{pmatrix}\phi_a^0 &0 \\
0 &\varphi_a\end{pmatrix}
-b_\alpha f\Mbar^B{\cal P}
\begin{pmatrix}\phi_a^0 &0 \\
0 &\varphi_a\end{pmatrix}
\\
&\qquad+\begin{pmatrix}\phi_a^0 &0 \\
0 &\varphi_a\end{pmatrix} {\cal P} b_\alpha f\bar\CM^B
-\begin{pmatrix}\phi_a^0 &0 \\
0 &\varphi_a\end{pmatrix} \CM^B fb_\alpha\\
&\qquad+b_\alpha f\bigg[\Mbar^B
\begin{pmatrix}\phi_a^0 &0 \\
0 &\varphi_a\end{pmatrix}-\varphi_a\Mbar^B\bigg]
+\bigg[
\begin{pmatrix}\phi_a &0 \\
0 &\varphi_a\end{pmatrix} \CM^B-\CM^B\varphi_a\bigg]f\bar b_\alpha
\bigg\}\\
&\qquad-\tfrac12\epsilon_{ABCD}U\Big\{
b_\alpha f\Mbar^B{\cal M}^C f\Mbar^D+\CM^C f\bar\CM^D\CM^B f\bar b_\alpha
\Big\}U
\elabel{writetwo}\end{split}\end{equation}
so that the last four lines of Eq.~\eqref{writeout} are accounted for,
as well as the ``$-1$'' terms in \eqref{writeone}.
Here the final equality follows from the commutator identity \eqref{funiwl}:
\begin{equation}\begin{split}
\Deltabar^\aD
\begin{pmatrix}\phi_a^0 &0 \\
0 &\varphi_a\end{pmatrix}\Delta_\aD
&=\bar w^\aD\phi_a^0w_\aD
-\BL\cdot\varphi_a+ \{\varphi_a\,,\,f^{-1}\}
\\&=
-\tfrac14\bar\Sigma_{aAB}\Mbar^A{\cal M}^B
+\{\varphi_a\,,\,f^{-1}\}
\elabel{comid}\end{split}\end{equation}
implied by Eqs.~\eqref{dbd}, \eqref{vvxx} and
\eqref{dxx}.  Finally,
\begin{equation}\begin{split}
{\cal D}_{\alpha\dalpha}\psibar^{\three\dalpha}_A
&=
2\Ubar\bigg\{ b_\alpha f\Deltabar_\dalpha
\begin{pmatrix}0 &0 \\ 0 &\G^\dalpha_A\end{pmatrix}
+\begin{pmatrix}0 &0 \\ 0 &\G^\dalpha_A\end{pmatrix}
\Delta_\aD f\bar b_\alpha\bigg\}U
\\
&=2\Ubar\bigg\{ b_\alpha f\bigg[\abar_\dalpha
\begin{pmatrix}0 &0 \\ 0 &\G^\dalpha_A\end{pmatrix}-
\G^\dalpha_A\abar_\dalpha\bigg]+\bigg[
\begin{pmatrix}0 &0 \\ 0 &\G^\dalpha_A\end{pmatrix}  a_\aD-
a_\aD\G^\dalpha_A\bigg]f \bar b_\alpha\bigg\}U\ ,
\elabel{psithreeb}\end{split}\end{equation}
where to obtain the final equality we used the moves summarized in
Eq.~\eqref{trick}.

Next we sum the expressions \eqref{writeone}, \eqref{writetwo} and
\eqref{psithreeb}, and compare to the right-hand side of
Eq.~\eqref{writeout}. By inspection, we confirm the ans\"atze
\eqref{psiansatz}, where the Grassmann zero mode matrix
has the form
\begin{equation}\N_A=-\bar\Sigma_{aAB}\bigg[\,
\begin{pmatrix}\phi_a^0&0\\0&\varphi_a
\end{pmatrix}{\cal M}^B-{\cal M}^B\varphi_a\,\bigg]
+2\begin{pmatrix}0&0\\0&\G^\dalpha_A\end{pmatrix}a_\dalpha-
2a_\dalpha\G^\dalpha_A\ .
\elabel{calNdef}\end{equation}
Up till this point we have yet to solve for $\G^\dalpha_A$. This is
accomplished by inserting $\N_A$ into the fermionic constraints
\eqref{zmcona}-\eqref{zmconb}.  One finds that
Eq.~\eqref{zmconb} is satisfied automatically by the expression
\eqref{calNdef}.  In contrast, Eq.~\eqref{zmcona} amounts to $2k^2$
independent real linear constraints, which is precisely the number
required to fix the anti-Hermitian $k\times k$ matrices $\G^\dalpha_A$
completely (the explicit form for $\G^\dalpha_A$ is not required).

{\it Supersymmetry transformations of the fermion zero
modes}

{}From the supersymmetry transformation of the fermion zero modes
\EQ{
\delta\Lambda(\CM^A)=-i\Sigma_a^{AB}(\Dslash\phi_a)\bar\xi_B\ ,
\elabel{susst}
}
and the expression \eqref{ssdd} for $\phi_a$ in the instanton background,
we now extract the transformation of the Grassmann collective
coordinates
\EQ{
\delta\CM^A=2i\Sigma_a^{AB}{\cal C}_{a\aD}\bar\xi_{B}^\aD\ ,\qquad
\delta\bar\CM^A=2i\Sigma_a^{AB}\bar\xi_{\aD B}\bar{\cal C}_a^\aD\ ,
}
where
\EQ{
{\cal C}_{a\aD}=\MAT{\phi^0_a& 0\\ 0 &\varphi_a}a_\aD-a_\aD\varphi_a\
,\qquad \bar{\cal C}_{a}^\aD=\bar a^\aD
\MAT{\phi^0_a& 0\\ 0 &\varphi_a}-\varphi_a\bar a^\aD\ .
}

The proof begins by expressing $\delta\lambda$ in terms of a variation
of the $c$-number collective coordinates, as in \eqref{uiui}, and a
variation of the Grassmann collective coordinates that must be
determined. Up to a gauge transformation, one finds
\SP{
\delta\Lambda_\alpha(\CM^A)&=\Lambda_\alpha(\delta\M^A)+
i\xi_{\aD B}\bar U\bigg\{-\CM^Bf\bar\Delta^\aD\CM^Af\bar b_\alpha-\CM
f\bar b_\alpha\Delta^\aD f\bar\CM^B+\CM^B f\bar\Delta^\aD b_\alpha
f\bar\CM^A\\
&+b_\alpha f\bar\CM^A\Delta^\aD f\bar\CM^B+\CM^Af\bar\CM^B\Delta^\aD
f\bar b_\alpha+b_\alpha f\bar\Delta^\aD\CM^B f\bar\CM^A\bigg\}U\ .
\elabel{hhes}
}
In order to derive this we used
\SP{
&\delta a_\aD=i\xi_{\aD A}\CM^A\ ,\qquad\delta\bar a_\aD=i\bar\xi_{\aD
A}\bar\CM^A\ ,\\
&\delta\bar U=-(\bar U\delta U)\bar U-\bar U\delta a_\aD f\bar\Delta^\aD
\ ,\qquad\delta U=U(\bar U\delta U)-\Delta_\aD f\delta\bar a^\aD U\
,\\
&\delta f=-\tfrac12f\big(\delta\bar
a^\aD\Delta_\aD+\bar\Delta^\aD\delta a_\aD\big)f\ .
}
Notice that only the last term in \eqref{hhes} depends on the variation
of the Grassmann collective coordinates.

{}From the supersymmetry transformation rule,
the right-hand side of \eqref{hhes} must be equated with the
right-hand side of \eqref{susst}
where $\phi_a$ assumes its value in the instanton background as in
\eqref{ssdd}.
Using the derivative identity \eqref{genderiv} and properties of the
$\Sigma$-matrices, one finds
\SP{
-i\Sigma^{AB}_a\Dslash\phi_a\bar\xi_B
=&-i\bar\xi_{\aD B}\bar U\bigg\{\CM^A f\bar\Delta^\aD b_\alpha f\bar\CM^B
+b_\alpha f\bar\Delta^\aD\CM^Af\bar\CM^B+\CM^A f\bar\CM^B\Delta^\aD
f\bar b_\alpha\\
&-\CM^B f\bar\Delta^\aD b_\alpha f\bar\CM^A-b_\alpha
f\bar\Delta^\aD\CM^B f\bar\CM^A-\CM^Bf\bar\CM^A\Delta^\aD f\bar
b_\alpha\\
&-2i\Sigma_a^{AB}b_\alpha f\bar\Delta^\aD
 \MAT{\phi^0_a&0\\0&\varphi_a}
-2i\Sigma_a^{AB}
\MAT{\phi^0_a&0\\0&\varphi_a} \Delta^\aD f\bar b_\alpha\bigg\}U\ .
\elabel{yyqq}
}
One can verify using the fermionic ADHM constraints \eqref{fadhm} that the
first six terms in the above are equal to the first six terms in
\eqref{hhes}. This means that the variation
$\Lambda_\alpha(\delta\CM^A)$ is then equated with the final two terms
in \eqref{yyqq}. These terms are not quite in the right form due to
the presence of the $x$-dependent $\Delta$ and $\bar\Delta$
terms. However, this can
easily be removed by using the tricks \eqref{trick}.
When this has been done one
extracts the variations of the Grassmann collective coordinates given
in \eqref{vivi}.

{\it Variation of the fermion zero modes}

We now prove that under a variation by a collective coordinate $X^\mu$
\EQ{
\PD{\Lambda(\CM)}{X^\mu}+
[\Omega_\mu,\Lambda(\CM)]=\Dslash\bar\varrho_\mu+\Lambda
\big(\partial\CM/\partial X^\mu\big) ,
\elabel{toproo}
}
where
\EQ{
\bar\varrho^{\aD}_\mu=\tfrac14\bar U\PD{a^\aD}{X^\mu}f\bar{\cal M}
U\ .
\elabel{vxxmm}
}

Using \eqref{lam}, \eqref{cgt} and \eqref{silid}, the left-hand side of
\eqref{toproo} is
\SP{
&\bar U\bigg\{\PD{a_\aD}{X^\mu}f\bar\Delta^\aD\big(b_\alpha
f\bar\CM-\CM f\bar b_\alpha\big)
+\big(b_\alpha f\bar\CM
-\CM f\bar b_\alpha\big)\Delta_\aD f\PD{\bar a^\aD}{X^\mu}
+\PD{\CM}{X^\mu}f\bar b_\alpha-b_\alpha f\PD{\bar\CM}{X^\mu}\\
&+\tfrac12\CM f\Big(\PD{\bar a^\aD}{X^\mu}\Delta_\aD+\bar\Delta^\aD
\PD{a_\aD}{X^\mu}\Big)f\bar b_\alpha-
\tfrac12 b_\alpha f\Big(\PD{\bar a^\aD}{X^\mu}\Delta_\aD+\bar\Delta^\aD
\PD{a_\aD}{X^\mu}\Big)f\bar\CM\bigg\}U\ .
\elabel{twed}
}
Using \eqref{genderiv}, the right-hand side of \eqref{toproo} is
\SP{
&-\bar U\bigg\{\PD{a^\aD}{X^\mu}f\bar b_\alpha\Delta_\aD f\bar\CM+
\CM f\bar b_\alpha\Delta_\aD f\PD{\bar a^\aD}{X^\mu}
+b_\alpha
f\bar\Delta_\aD\PD{a^\aD}{X^\mu}f\bar\CM+\PD{\CM}{X^\mu}f\bar
b_\alpha-b_\alpha f\PD{\bar\CM}{X^\mu}\\
&+\PD{a^\aD}{X^\mu}f\bar\CM\Delta_\aD f\bar b_\alpha+
b_\alpha f\bar\Delta_\aD\CM f\PD{\bar a^\aD}{X^mu}+
\CM f\PD{\bar a^\aD}{X^mu}\Delta_\aD f\bar b_\alpha\bigg\}U\ .
\elabel{twfd}
}
The difference of \eqref{twed} and \eqref{twfd} is
\EQ{
\tfrac12\bar U\bigg\{\CM f\Big(\PD{\bar a^\aD}{X^\mu}\Delta_\aD+
\bar\Delta_\aD\PD{a^\aD}{X^\mu}\Big)f\bar b_\alpha
-b_\alpha f\Big(\PD{\bar a^\aD}{X^\mu}\Delta_\aD+
\bar\Delta_\aD\PD{a^\aD}{X^\mu}\Big)f\bar\CM\bigg\}U\ .
}
But this vanishes by virtue of the constraints \eqref{opa} satisfied
by $C_\aD=\partial a_\aD/\partial X^\mu$.


\begin{thebibliography}{99}

{\small


\bibitem{BPST}A. Belavin, A. Polyakov, A.Schwartz and Y. Tyupkin,
Phys. Lett. {\bf B59} (1975) 85.

\bibitem{tHooft}G.~'t~Hooft, Phys. Rev. {\bf D14} (1976) 3432; ibid.
(E) {\bf D18} (1978) 2199.


\bibitem{CDG}
C.~G.~Callan, R.~F.~Dashen and D.~J.~Gross,
Phys.\ Rev.\ {\bf D17} (1978) 2717.

\bibitem{ABC}
A.I.~Vainshtein, V.I.~Zakharov
V.A.~Novikov and M.A.~Shifman,
Sov. Phys. Usp. 25 (1982) 195 and in ``ABC of Instantons,''
{\it Instantons in Gauge Theories},
ed. M. Shifman, World Scientific (1994).



\bibitem{SSI}E. Shuryak and T. Schafer,
Nucl. Phys. Proc. Suppl. {\bf53} (1997) 472.


\bibitem{DonaldsonKron}S.K.~Donaldson and P.B~Kronheimer, ``The
Geometry of Four Manifolds,'' Oxford University Press (1990).


\bibitem{ADHM} M.~Atiyah, V.~Drinfeld, N.~Hitchin and
Yu.~Manin, Phys.\ Lett.\ {\bf A65} (1978) 185.



\bibitem{OSB}H. Osborn, Ann. of Phys. {\bf135} (1981) 373.


\bibitem{Coleman}
S. Coleman, ``Uses of Instantons,'' in {\it Aspects of Symmetry},
Cambridge University Press (1985) 265.

\bibitem{Amati:1988ft}
D.~Amati, K. Konishi, Y. Meurice, G.C. Rossi and G. Veneziano,
Phys. Rept. {\bf 162} (1988) 169.

\bibitem{Shifman:1994ee}
M.~A.~Shifman, ``Instantons in gauge theories,''
{\it  World Scientific\/} (1994).


\bibitem{Shifman:1999mv}
M.~A.~Shifman and A.~I.~Vainshtein,
{\tt arXiv:hep-th/9902018}.

\bibitem{Belitsky:2000ws}
A.~V.~Belitsky, S.~Vandoren and P.~van Nieuwenhuizen,
Class.\ Quant.\ Grav.\  {\bf 17} (2000) 3521
{\tt[arXiv:hep-th/0004186]}.



\bibitem{AtiyahSinger}
M.~F.~Atiyah and I.~M.~Singer,
Annals Math.\  {\bf 87} (1968) 484; Annals Math.\  {\bf 87} (1968) 531;
Annals Math.\  {\bf 87} (1968) 546; Annals Math.\  {\bf 93} (1971)
119; Annals Math.\  {\bf 93} (1971) 139.



\bibitem{BCGW}C. Bernard, N.H. Christ, A. Guth and E.J. Weiberg,
Phys. Rev. {\bf D16} (1977) 2967.



\bibitem{Maciocia:1991ph}
A.~Maciocia,
Commun.\ Math.\ Phys.\  {\bf 135} (1991) 467.


\bibitem{Corrigan:1979pr}
E.~F.~Corrigan and P.~Goddard,
``Some Aspects Of Instantons,''
in {\it C79-09-3.2.3}
DAMTP-79-18,
{\it Lectures given at Canadian Mathematical Society Meeting, 
Montreal, Canada, Sep 3-8\/} (1979).


\bibitem{CGTone}E. Corrigan, D. Fairlie, P. Goddard and S. Templeton,
Nucl. Phys. {\bf B140} (1978) 31.\\
E. Corrigan, P. Goddard and S. Templeton,
Nucl. Phys. {\bf B151} (1979) 93.


\bibitem{CWS}{N.H. Christ, E.J. Weinberg and N.K. Stanton,
Phys. Rev. {\bf D18} (1978) 2013.}


\bibitem{KMS}
V.~V.~Khoze, M.~P.~Mattis and M.~J.~Slater,
Nucl.\ Phys.\ {\bf B536} (1998) 69
{\tt[arXiv:hep-th/9804009]}.


\bibitem{MO3}
N.~Dorey, T.J.~Hollowood, V.V.~Khoze, M.P.~Mattis and S.~Vandoren,
Nucl.\ Phys.\ {\bf B552} (1999) 88
{\tt [arXiv:hep-th/9901128]}.


\bibitem{Hitchin:1987ea}
N.~J.~Hitchin, A.~Karlhede, U.~Lindstrom and M.~Rocek,
Commun.\ Math.\ Phys.\  {\bf 108} (1987) 535.

\bibitem{Bernard}{C.~Bernard, Phys. Rev. {\bf D19} (1979) 3013.}


\bibitem{Corrup}{E. Corrigan, {\it unpublished}.}


\bibitem{MO-I}
N.~Dorey, V.~V.~Khoze and M.~P.~Mattis,
Phys.\ Rev.\ {\bf D54} (1996) 2921
{\tt[arXiv:hep-th/9603136]}.


\bibitem{Gibbons:1998xa}
G.~W.~Gibbons and P.~Rychenkova,
Phys.\ Lett.\  {\bf B443} (1998) 138
{\tt[arXiv:hep-th/9809158]}.

\bibitem{Boyer:1998sf}
C.~P.~Boyer and K.~Galicki,
{\tt arXiv:hep-th/9810250}.

\bibitem{Vandoren:2000qr}
S.~Vandoren,
{\tt arXiv:hep-th/0009150}.

\bibitem{Uhlenbeck:1982zm}
K.~K.~Uhlenbeck,
Commun.\ Math.\ Phys.\  {\bf 83} (1982) 11.


\bibitem{measure1}
N.~Dorey, V.~V.~Khoze and M.~P.~Mattis,
Nucl.\ Phys.\ {\bf B513} (1998) 681
{\tt[arXiv:hep-th/9708036]}.

\bibitem{DHKM}
N.~Dorey, T.~J.~Hollowood, V.~V.~Khoze and M.~P.~Mattis,
Nucl.\ Phys.\ {\bf B519} (1998) 470
{\tt[arXiv:hep-th/9709072]}.


\bibitem{Bruzzo:2001di}
U.~Bruzzo, F.~Fucito, A.~Tanzini and G.~Travaglini,
Nucl.\ Phys.\ {\bf B611} (2001) 205
{\tt[arXiv:hep-th/0008225]}.


\bibitem{Mehta}
M.L.~Mehta, ``Random Matrices,'' Academic Press (1991).


\bibitem{D'Adda:1978ur}
A.~D'Adda and P.~Di Vecchia,
Phys. Lett. {\bf B73} (1978) 162.

\bibitem{Berg:1979ku}
B.~Berg and M.~Luscher,
Nucl. Phys. {\bf B160} (1979) 281.

\bibitem{Corrigan:1979di}
E.~Corrigan, P.~Goddard, H.~Osborn and S.~Templeton,
Nucl. Phys. {\bf B159} (1979) 469.

\bibitem{Jack:1980rn}
I.~Jack,
Nucl. Phys. {\bf B174} (1980) 526.


\bibitem{NSVZ}V.A. Novikov, M.A. Shifman, A.I. Vainshtein and
V.I. Zakharov, Nucl. Phys. {\bf B229} (1983) 394;
Nucl. Phys. {\bf B229} (1983) 407;
 Nucl. Phys. {\bf B260} (1985) 157.


\bibitem{Belitsky:2000ii}
A.~V.~Belitsky, S.~Vandoren and P.~van Nieuwenhuizen,
Phys.\ Lett.\ {\bf B477} (2000) 335
{\tt [arXiv:hep-th/0001010]}.


\bibitem{Zumino:1977yh}
B.~Zumino,
Phys.\ Lett.\ {\bf B69} (1977) 369.


\bibitem{Nicolai:1978vc}
H.~Nicolai,
Nucl.\ Phys.\ {\bf B140} (1978) 294.


\bibitem{vanNieuwenhuizen:1996tv}
P.~van Nieuwenhuizen and A.~Waldron,
Phys.\ Lett.\ {\bf B389} (1996) 29
{\tt[arXiv:hep-th/9608174]}.


\bibitem{Blau:1997pp}
M.~Blau and G.~Thompson,
Phys.\ Lett.\ {\bf B415} (1997) 242
{\tt[arXiv:hep-th/9706225]}.

\bibitem{Acharya:1998jn}
B.~S.~Acharya, J.~M.~Figueroa-O'Farrill, B.~Spence and M.~O'Loughlin,
Nucl.\ Phys.\ {\bf B514} (1998) 583
{\tt[arXiv:hep-th/9707118]}.


\bibitem{Tong1}
D.~Tong,
Phys.\ Lett.\ {\bf B460} (1999) 295
{\tt[arXiv:hep-th/9902005]}.



\bibitem{Tong2}
N.~D.~Lambert and D.~Tong,
Phys.\ Lett.\ {\bf B462} (1999) 89
{\tt[arXiv:hep-th/9907014]}. 



\bibitem{WB}{J.~Wess and J.~Bagger, ``Supersymmetry and
Supergravity,'' Princeton University Press (1992).}


\bibitem{MO-II}
N.~Dorey, V.~V.~Khoze and M.~P.~Mattis,
Phys.\ Rev.\ {\bf D54} (1996) 7832
{\tt [arXiv:hep-th/9607202]}.

\bibitem{Affleck}{I. Affleck, Nucl. Phys. {\bf B191} (1981) 429.}

\bibitem{ADS}
I.~Affleck, M.~Dine and N.~Seiberg,
Phys. Rev. Lett. {\bf 51} (1983) 1026;
Nucl. Phys. {\bf B241} (1984) 493.

\bibitem{Derrick}G.H.~Derrick,  J. Math.~Phys.~{\bf 5} (1964) 1252\\
R.~Hobart,  Proc.~Royal.~Soc. London  {\bf 82} (1963) 201.


\bibitem{deWit:2000fp}
B.~de Wit, B.~Kleijn and S.~Vandoren,
Nucl.\ Phys.\ {\bf B568} (2000) 475
{\tt[arXiv:hep-th/9909228]}.


\bibitem{weinb}S.~Weinberg, Phys. Lett. {\bf B91} (1980) 51\newline
L.~Hall, Nucl. Phys. {\bf B178} (1981) 75.


\bibitem{FP}D. Finnell and P. Pouliot,
Nucl. Phys. {\bf B453} (1995) 225 {\tt [arXiv:hep-th/9503115]}.


\bibitem{GILMORE}R. Gilmore, ``Lie groups, Lie algebras and some of
their applications,'' Wiley-Interscience (1974).


\bibitem{Corrigan:1978ce}
E.F.~Corrigan, D.B.~Fairlie, S.~Templeton and P.~Goddard,
Nucl. Phys. {\bf B140} (1978) 31.

\bibitem{Christ:1978jy}
N.H.~Christ, E.J.~Weinberg and N.K.~Stanton,
Phys. Rev. {\bf D18} (1978) 2013.

\bibitem{DKMn4}
N.~Dorey, V.~V.~Khoze and M.~P.~Mattis,
Phys.\ Lett.\  {\bf B396} (1997) 141
{\tt[arXiv:hep-th/9612231]}.

\bibitem{IS} K. Intriligator and N. Seiberg,
{\tt arXiv:hep-th/9509066}.


\bibitem{Shifman}  M. Shifman,
Prog. Part. Nucl. Phys. {\bf 39} (1997) 1 {\tt [arXiv:hep-th/9704114]}.

\bibitem{AOYAMA}H. Aoyama, T. Harano, M. Sato and S. Wada,
Phys. Lett. {\bf B388} (1996) 331 {\tt [arXiv:hep-th/9607076]}.



\bibitem{Rossi:1984bu}
G.C.~Rossi and G.~Veneziano,
Phys. Lett. {\bf B138} (1984) 195.

\bibitem{Amati:1985uz}
D.~Amati, G.C.~Rossi and G.~Veneziano,
Nucl. Phys. {\bf B249} (1985) 1.


\bibitem{Fuchs:1986ft}
J.~Fuchs and M.G.~Schmidt,
Z. Phys. {\bf C30} (1986) 161.

\bibitem{Novikov:1985ic}
V.A.~Novikov, M.A.~Shifman, A.I.~Vainshtein and V.I.~Zakharov,
Nucl. Phys. {\bf B260} (1985) 157.

\bibitem{Shifman:1988ia}
M.A.~Shifman and A.I.~Vainshtein,
Nucl. Phys. {\bf B296} (1988) 445.

\bibitem{Hollowood:2000qn}
T.~J.~Hollowood, V.~V.~Khoze, W.~Lee and M.~P.~Mattis,
Nucl.\ Phys.\ {\bf B570} (2000) 241
{\tt[arXiv:hep-th/9904116]}.


\bibitem{SeibWitt}N. Seiberg and E. Witten, Nucl. Phys. {\bf B426} (1994) 19,
(E) {\bf B430} (1994) 485 {\tt [arXiv:hep-th/9407087]}.


\bibitem{Glumag}
N.~M.~Davies, T.~J.~Hollowood, V.~V.~Khoze and M.~P.~Mattis,
Nucl.\ Phys.\ B {\bf 559} (1999) 123
{\tt[arXiv:hep-th/9905015]}.


\bibitem{Davies:2000nw}
N.~M.~Davies, T.~J.~Hollowood and V.~V.~Khoze,
{\tt arXiv:hep-th/0006011}.


\bibitem{Fuchs:1987ae}
J.~Fuchs,
Nucl.\ Phys.\ {\bf B282} (1987) 437.


\bibitem{Novikov:1983uc}
V.A.~Novikov, M.A.~Shifman, A.I.~Vainshtein and V.I.~Zakharov,
Nucl.\ Phys.\ {\bf B229} (1983) 381.

\bibitem{Peskin:1997qi}
M.E.~Peskin,
{\tt arXiv:hep-th/9702094}.

\bibitem{Cordes}
S.F.~Cordes,
Nucl. Phys. {\bf B273} (1986) 629.

\bibitem{Weinberg}S. Weinberg,
Phys. Lett. {\bf B91} (1980) 51.

\bibitem{Ritz:2000mq}
A.~Ritz and A.~I.~Vainshtein,
Nucl.\ Phys.\ {\bf B566} (2000) 311
{\tt[arXiv:hep-th/9909073]}.

\bibitem{SWtwo}N. Seiberg and E. Witten, Nucl. Phys. {\bf B431} (1994) 484
{\tt [arXiv:hep-th/9408099]}.

\bibitem{test2}
N.~Dorey, V.~V.~Khoze and M.~P.~Mattis,
Phys.\ Lett.\ {\bf B388} (1996) 324
{\tt[arXiv:hep-th/9607066]}.

\bibitem{mybps}
T.~J.~Hollowood,
Adv.\ Theor.\ Math.\ Phys.\  {\bf 2} (1998) 335
{\tt[arXiv:hep-th/9710073]}.


\bibitem{Matone:1995rx}
M.~Matone,
Phys.\ Lett.\ {\bf B357} (1995) 342
{\tt[arXiv:hep-th/9506102]}.


\bibitem{Fucito:1997ua}
F.~Fucito and G.~Travaglini,
Phys.\ Rev.\ {\bf D55} (1997) 1099
{\tt[arXiv:hep-th/9605215]}.

\bibitem{dkmmatone}
N.~Dorey, V.~V.~Khoze and M.~P.~Mattis,
Phys.\ Lett.\ {\bf B390} (1997) 205
{\tt[arXiv:hep-th/9606199]}.


\bibitem{Ito:1996qj}
K.~Ito and N.~Sasakura,
Phys.\ Lett.\ {\bf B382} (1996) 95
{\tt[arXiv:hep-th/9602073].}

\bibitem{Ito:1997fd}
K.~Ito and N.~Sasakura,
Mod.\ Phys.\ Lett.\ {\bf A12} (1997) 205
{\tt[arXiv:hep-th/9609104]}.


\bibitem{Klemm:1995qs}
A.~Klemm, W.~Lerche, S.~Yankielowicz and S.~Theisen,
Phys.\ Lett.\ {\bf B344} (1995) 169
{\tt[arXiv:hep-th/9411048]}.

\bibitem{Argyres:1995xh}
P.~C.~Argyres and A.~E.~Faraggi,
Phys.\ Rev.\ Lett.\  {\bf 74} (1995) 3931
{\tt[arXiv:hep-th/9411057]}.

\bibitem{Argyres:1995wt}
P.~C.~Argyres, M.~R.~Plesser and A.~D.~Shapere,
Phys.\ Rev.\ Lett.\  {\bf 75} (1995) 1699
{\tt[arXiv:hep-th/9505100]}.

\bibitem{Hanany:1995na}
A.~Hanany and Y.~Oz,
Nucl.\ Phys.\ {\bf B452} (1995) 283
{\tt[arXiv:hep-th/9505075]}.

\bibitem{Minahan:1996er}
J.~A.~Minahan and D.~Nemeschansky,
Nucl.\ Phys.\ {\bf B464} (1996) 3
{\tt[arXiv:hep-th/9507032]}.

\bibitem{D'Hoker:1997nv}
E.~D'Hoker, I.~M.~Krichever and D.~H.~Phong,
Nucl.\ Phys.\ {\bf B489} (1997) 179
{\tt[arXiv:hep-th/9609041]}.




\bibitem{Dorey:1997bn}
N.~Dorey, V.~V.~Khoze and M.~P.~Mattis,
Nucl.\ Phys.\ {\bf B492} (1997) 607
{\tt[arXiv:hep-th/9611016]}.

\bibitem{Argyres:2000ty}
P.~C.~Argyres and S.~Pelland,
JHEP {\bf 0003} (2000) 014
{\tt[arXiv:hep-th/9911255]}.


\bibitem{Slater:1997zn}
M.~J.~Slater,
Phys.\ Lett.\ {\bf B403} (1997) 57
{\tt[arXiv:hep-th/9701170]}.

\bibitem{Pomeroy}
N.~B.~Pomeroy,
Phys.\ Lett.\ {\bf B510} (2001) 305
{\tt[arXiv:hep-th/0103181]}.


\bibitem{local}
T.~J.~Hollowood,
{\tt arXiv:hep-th/0201075}.

\bibitem{locn4}
T.~J.~Hollowood,
{\tt arXiv:hep-th/0202197}.


\bibitem{Edelstein:1999sp}
J.~D.~Edelstein, M.~Marino and J.~Mas,
Nucl.\ Phys.\ {\bf B541} (1999) 671
{\tt[arXiv:hep-th/9805172]}.


\bibitem{Edelstein:1999dd}
J.~D.~Edelstein, M.~Gomez-Reino and J.~Mas,
Nucl.\ Phys.\ {\bf B561} (1999) 273
{\tt[arXiv:hep-th/9904087]}.


\bibitem{Edelstein:2000xk}
J.~D.~Edelstein, M.~Gomez-Reino, M.~Marino and J.~Mas,
Nucl.\ Phys.\ {\bf B574} (2000) 587
{\tt[arXiv:hep-th/9911115]}.


\bibitem{Chan:2000gj}
G.~Chan and E.~D'Hoker,
Nucl.\ Phys.\ {\bf B564} (2000) 503
{\tt[arXiv:hep-th/9906193]}.


\bibitem{Eguchi:1996jh}
T.~Eguchi and S.~K.~Yang,
Mod.\ Phys.\ Lett.\ {\bf A11} (1996) 131
{\tt[arXiv:hep-th/9510183]}.


\bibitem{D'Hoker:1997ph}
E.~D'Hoker, I.~M.~Krichever and D.~H.~Phong,
Nucl.\ Phys.\ {\bf B494} (1997) 89
{\tt[arXiv:hep-th/9610156]}.



\bibitem{Sonnenschein:1996hv}
J.~Sonnenschein, S.~Theisen and S.~Yankielowicz,
Phys.\ Lett.\ {\bf B367} (1996) 145
{\tt[arXiv:hep-th/9510129]}.


\bibitem{GMO}P.~Goddard, P.~Mansfield and H.~Osborn,
Phys.\ Lett.\ {\bf B98} (1981) 59.

\bibitem{Mansfield:1981sk}
P.~Mansfield,
Nucl.\ Phys.\ {\bf B186} (1981) 287.




\bibitem{LETT}
N.~Dorey, T.~J.~Hollowood, V.~V.~Khoze, M.~P.~Mattis and S.~Vandoren,
JHEP {\bf 9906} (1999) 023
{\tt[arXiv:hep-th/9810243]}.


\bibitem{Hollowood:1999sk}
T.~J.~Hollowood, V.~V.~Khoze and M.~P.~Mattis,
JHEP {\bf 9910} (1999) 019
{\tt [arXiv:hep-th/9905209]}.



\bibitem{Hollowood:1999ev}
T.~J.~Hollowood, V.~V.~Khoze and M.~P.~Mattis,
Adv.\ Theor.\ Math.\ Phys.\ {\bf 4} (2000) 545
{\tt[arXiv:hep-th/9910118]}.


\bibitem{Gava:2000ky}
E.~Gava, K.~S.~Narain and M.~H.~Sarmadi,
Nucl.\ Phys.\ {\bf B569} (2000) 183
{\tt[arXiv:hep-th/9908125]}.

\bibitem{Hollowood:1999nq}
T.~J.~Hollowood,
JHEP {\bf 9911} (1999) 012
{\tt[arXiv:hep-th/9908201]}.

\bibitem{Hollowood:2000bm}
T.~J.~Hollowood and V.~V.~Khoze,
Nucl.\ Phys.\  {\bf B575} (2000) 78
{\tt[arXiv:hep-th/9908035]}.

\bibitem{MAL}{J.~Maldacena,
Adv. Theor. Math. Phys. {\bf 2} (1998) 231 {\tt [arXiv:hep-th/9711200]}.}

\bibitem{Aharony:2000ti}
O.~Aharony, S.~S.~Gubser, J.~Maldacena, H.~Ooguri and Y.~Oz,
Phys.\ Rept.\  {\bf 323} (2000) 183
{\tt[arXiv:hep-th/9905111]}.

\bibitem{GN} D. J. Gross and A. Neveu, {  Phys. Rev.} {\bf D10}
(1974) 3235.\\
K. Wilson, Phys. Rev. {\bf D7} (1973) 2911.

\bibitem{MNS}{G.~Moore, N.~Nekrasov and S.~Shatashvili,
{\tt arXiv:hep-th/9803265}.}

\bibitem{KNS}{W.~Krauth, H.~Nicolai and M.~Staudacher,
Phys. Lett. {\bf B431} (1998) 31 {\tt [arXiv:hep-th/9803117]}.\\
W.~Krauth and M.~Staudacher, Phys. Lett. {\bf B435} (1998) 350  {\tt[arXiv:hep-th/9804199]}.}

\bibitem{BGKR}{M.~Bianchi, M.B.~Green, S.~Kovacs and G.~Rossi, JHEP
{\bf9808:013} (1998) {\tt [arXiv:hep-th/9807033]}.}

\bibitem{BG}{T.~Banks and M.B.~Green, JHEP {\bf05:002}
(1998) {\tt [arXiv:hep-th/9804170]}.}

\bibitem{WIT150}E.~Witten,
Adv. Theor. Math. Phys. {\bf 2:253} (1998)
{\tt [arXiv:hep-th/9802150]}.

\bibitem{GKP}
S.~Gubser, I.~Klebanov and A.~Polyakov,
Phys. Lett. {\bf B428} (1998) 105 {\tt [arXiv:hep-th/9802109]}.

\bibitem{HW}S. Howe and P.C. West,
{\tt arXiv:hep-th/9509140};
Phys. Lett. {\bf B400} (1997) 307 {\tt [arXiv:hep-th/9611075]}.

\bibitem{AFGJ}D. Anselmi, D. Freedman, M. Grisaru and A. Johansen,
Phys. Lett. {\bf B394} (1997) 329 {\tt [arXiv:hep-th/9708042]}.

\bibitem{FMMR}
D.~Z.~Freedman, S.~D.~Mathur, A.~Matusis and L.~Rastelli,
Nucl.\ Phys.\ {\bf B546} (1999) 96
{\tt[arXiv:hep-th/9804058]};
Phys.\ Lett.\  {\bf B452} (1999) 61
{\tt[arXiv:hep-th/9808006]}.

\bibitem{EFS}
E.~D'Hoker, D.~Z.~Freedman and W.~Skiba,
Phys.\ Rev.\  {\bf D59} (1999) 045008
{\tt[arXiv:hep-th/9807098]}.


\bibitem{LMRS}
S.~M.~Lee, S.~Minwalla, M.~Rangamani and N.~Seiberg,
Adv.\ Theor.\ Math.\ Phys.\  {\bf 2} (1998) 697
{\tt[arXiv:hep-th/9806074]}.


\bibitem{CNSS}
G.~Chalmers, H.~Nastase, K.~Schalm and R.~Siebelink,
Nucl.\ Phys.\  {\bf B540} (1999) 247
{\tt[arXiv:hep-th/9805105]}.

\bibitem{LT}
H.~Liu and A.~A.~Tseytlin,
Phys.\ Rev.\  {\bf D59} (1999) 086002
{\tt[arXiv:hep-th/9807097]}.

\bibitem{MV}W. Muck and K. S. Viswanathan, Phys.
Rev. {\bf D58} (1998) 041901 {\tt [arXiv:hep-th/9804035]};
Phys. Rev. {\bf D58} (1998) 106006 {\tt [arXiv:hep-th/9805145]}


\bibitem{BGut}
J.~H.~Brodie and M.~Gutperle,
Phys.\ Lett.\ {\bf B445} (1999) 296
{\tt[arXiv:hep-th/9809067]}.

\bibitem{AF1}
G.~E.~Arutyunov and S.~A.~Frolov,
Nucl.\ Phys.\ {\bf B544} (1999) 576
{\tt[arXiv:hep-th/9806216]}.


\bibitem{SOLOD}S.N. Solodukhin, Nucl. Phys. {\bf B539} (1999) 403
{\tt [arXiv:hep-th/9806004]}.


\bibitem{GHEZ}A.M. Ghezelbash, Phys. Lett. {\bf B435} (1998) 291
{\tt [arXiv:hep-th/9805162]}.

\bibitem{GG1}M.B.~Green and M.~Gutperle,
Nucl. Phys. {\bf B498} (1997) 195 {\tt [arXiv:hep-th/9701093]}.

\bibitem{GGK}M.B.~Green, M.~Gutperle and H. Kwon,
Phys. Lett. {\bf B421} (1998) 149
{\tt [arXiv:hep-th/9710151]}.

\bibitem{KP2}A. Kehagias and H. Partouche, Int. J. Mod. Phys. {\bf
A13} (1998) 5075
{\tt [arXiv:hep-th/9712164]}.

\bibitem{KP1}A. Kehagias and H. Partouche, Phys. Lett. {\bf B422}
(1998) 109 {\tt[arXiv:hep-th/9710023]}.

\bibitem{GSI}
M.~B.~Green and S.~Sethi,
Phys.\ Rev.\ {\bf D59} (1999) 046006
{\tt[arXiv:hep-th/9808061]}.



\bibitem{GG2}
M.~B.~Green and M.~Gutperle,
JHEP {\bf 9801} (1998) 005
{\tt[arXiv:hep-th/9711107]}.


\bibitem{HS1}M. Henningson and K. Sfetsos, Phys. Lett. {\bf B431}
(1998) 63 {\tt arXiv:hep-th/9803251}.

\bibitem{DKMV}
N.~Dorey, V.~V.~Khoze, M.~P.~Mattis and S.~Vandoren,
Phys.\ Lett.\ {\bf B442} (1998) 145
{\tt[arXiv:hep-th/9808157]}.

\bibitem{Gopakumar:1999xh}
R.~Gopakumar and M.~B.~Green,
JHEP {\bf 9912} (1999) 015
{\tt[arXiv:hep-th/9908020]}.


\bibitem{Gauntlett:1994sh}
J.~P.~Gauntlett,
Nucl.\ Phys.\  {\bf B411} (1994) 443
{\tt[arXiv:hep-th/9305068]}.

\bibitem{Bak:2000vd}
D.~Bak, K.~Lee and P.~Yi,
Phys.\ Rev.\ {\bf D62} (2000) 025009
{\tt[arXiv:hep-th/9912083]}.

\bibitem{Gauntlett:2001ks}
J.~P.~Gauntlett, C.~Kim, K.~Lee and P.~Yi,
Phys.\ Rev.\ {\bf D63} (2001) 065020
{\tt[arXiv:hep-th/0008031]}.

\bibitem{Manton:1982mp}
N.~S.~Manton,
Phys.\ Lett.\  {\bf B110} (1982) 54.

\bibitem{Ward:1985ij}
R.~S.~Ward,
Phys.\ Lett.\  {\bf B158} (1985) 424.


\bibitem{Alvarez-Gaume:1981hm}
L.~Alvarez-Gaume and D.~Z.~Freedman,
Commun.\ Math.\ Phys.\  {\bf 80} (1981) 443.


\bibitem{Howe:1987qv}
P.~S.~Howe and G.~Papadopoulos,
Nucl.\ Phys.\ {\bf B289} (1987) 264.


\bibitem{Howe:1988cj}
P.~S.~Howe and G.~Papadopoulos,
Class.\ Quant.\ Grav.\  {\bf 5} (1988) 1647.


\bibitem{Alvarez-Gaume:1983ab}
L.~Alvarez-Gaume and D.~Z.~Freedman,
Commun.\ Math.\ Phys.\  {\bf 91} (1983) 87.


\bibitem{Gauntlett:2001bd}
J.~P.~Gauntlett, D.~Tong and P.~K.~Townsend,
Phys.\ Rev.\  {\bf D63} (2001) 085001
{\tt[arXiv:hep-th/0007124]}.

\bibitem{Scherk:1979zr}
J.~Scherk and J.~H.~Schwarz,
Nucl.\ Phys.\ B {\bf 153} (1979) 61.

\bibitem{Howe:1983fr}
P.~S.~Howe, G.~Sierra and P.~K.~Townsend,
Nucl.\ Phys.\ {\bf B221} (1983) 331.

\bibitem{Sierra:1983cc}
G.~Sierra and P.~K.~Townsend,
LPTENS-83-26
{\it Lectures given at the 19th Karpacz Winter School on Theoretical Physics, Karpacz, Poland, Feb 14-28\/} (1983).


\bibitem{W1} E, Witten, Nucl. Phys. {\bf B460} (1996) 541
{\tt [arXiv:hep-th/9511030]}.

\bibitem{D2}
M.~R.~Douglas,
J.\ Geom.\ Phys.\  {\bf 28} (1998) 255
{\tt[arXiv:hep-th/9604198]}.

\bibitem{D1} M. Douglas,
{\tt arXiv:hep-th/9512077}.

\bibitem{Witten:1995tz}
E.~Witten,
J.\ Geom.\ Phys.\  {\bf 15} (1995) 215
{\tt[arXiv:hep-th/9410052]}.


\bibitem{Barbon:1998ak}
J.~L.~Barbon and A.~Pasquinucci,
Nucl.\ Phys.\ {\bf B517} (1998) 125
{\tt[arXiv:hep-th/9708041]}.


\bibitem{Barbon:1999fx}
J.~L.~Barbon and A.~Pasquinucci,
Fortsch.\ Phys.\  {\bf 47} (1999) 255
{\tt[arXiv:hep-th/9712135]}.



\bibitem{P}
J. Polchinski,
{\tt arXiv:hep-th/9602052}.


\bibitem{Johnson:2000ch}
C.~V.~Johnson,
{\tt arXiv:hep-th/0007170}.


\bibitem{W135} 
E, Witten, Nucl. Phys. {\bf B460} (1996) 335
{\tt [arXiv:hep-th/9510135]}.

\bibitem{GG3}
M.~B.~Green and M.~Gutperle,
Phys.\ Rev.\ {\bf D58} (1998) 046007
{\tt[arXiv:hep-th/9804123]}.

\bibitem{GG4}M.B. Green and M. Gutperle, Phys. Lett. {\bf B398} (1997)
69 {\tt [arXiv:hep-th/9612127]}.


\bibitem{CHS} C. Callan, J. Harvey and A. Strominger,
{  Nucl. Phys.} {\bf B367} (1991) 60.

\bibitem{Fre:1996dw}
P.~Fre,
Nucl.\ Phys.\ Proc.\ Suppl.\  {\bf 45BC} (1996) 59
{\tt[arXiv:hep-th/9512043]}.

\bibitem{Antoniadis:1997ra}
I.~Antoniadis and B.~Pioline,
Int.\ J.\ Mod.\ Phys.\ {\bf A12} (1997) 4907
{\tt[arXiv:hep-th/9607058]}.

\bibitem{A} O. Aharony, M. Berkooz, S. Kachru, N. Seiberg and E. Silverstein,
Adv. Theor. Math. Phys. {\bf 1} (1998) 148 {\tt [arXiv:hep-th/9707079]}.

\bibitem{W2} E. Witten, J. High Energy Phys. {\bf 07} (1997) 003
{\tt [arXiv:hep-th/9707093]}.

\bibitem{Douglas:1996sw}
M.~R.~Douglas and G.~W.~Moore,
{\tt arXiv:hep-th/9603167}.

\bibitem{Douglas:1997xg}
M.~R.~Douglas,
JHEP {\bf 9707} (1997) 004
{\tt[arXiv:hep-th/9612126]}.

\bibitem{Diaconescu:1998br}
D.~Diaconescu, M.~R.~Douglas and J.~Gomis,
JHEP {\bf 9802} (1998) 013
{\tt[arXiv:hep-th/9712230]}.




\bibitem{Myers:1999ps}
R.~C.~Myers,
JHEP {\bf 9912} (1999) 022
{\tt[arXiv:hep-th/9910053]}.


\bibitem{Millar:2000ib}
K.~Millar, W.~Taylor and M.~V.~Raamsdonk,
{\tt arXiv:hep-th/0007157}.

\bibitem{Dorey:2001zq}
N.~Dorey, T.~J.~Hollowood and V.~V.~Khoze,
JHEP {\bf 0103} (2001) 040
{\tt[arXiv:hep-th/0011247]}.


\bibitem{DHull}
M.~R.~Douglas and C.~Hull,
JHEP {\bf 9802} (1998) 008
{\tt [arXiv:hep-th/9711165]}.

\bibitem{Chu:1999qz}
C.~Chu and P.~Ho,
Nucl.\ Phys.\ {\bf B550}  (1999) 151
{\tt [arXiv:hep-th/9812219]}.

\bibitem{Schomerus:1999ug}
V.~Schomerus,
JHEP {\bf 9906} (1999) 030
{\tt [arXiv:hep-th/9903205]}.

\bibitem{SWnc}
N. Seiberg and E. Witten,
 JHEP {\bf 9909} (1999) 032
{\tt [arXiv:hep-th/9908142]}.
{DHull,Chu:1999qz,Schomerus:1999ug,SWnc}

\bibitem{DNrev}
M.~R.~Douglas and N.~A.~Nekrasov,
{\tt arXiv:hep-th/0106048}.


\bibitem{Nakajima:1993jg}
H.~Nakajima,
``Resolution of moduli spaces of ideal instantons on ${\mathbb R}^4$,''
{\it  Sanda , Topology, geometry and field theory\/} (1993) 129.


\bibitem{Nakajima:1996ka}
H.~Nakajima,
Nucl.\ Phys.\ Proc.\ Suppl.\  {\bf 46} (1996) 154
{\tt[arXiv:alg-geom/9510003]}.


\bibitem{NS}
N.~Nekrasov and A.~Schwarz,
Commun.\ Math.\ Phys.\  {\bf 198} (1998) 689
{\tt[arXiv:hep-th/9802068]}.

\bibitem{Furuuchi123}
K.~Furuuchi,
Prog.\ Theor.\ Phys.\  {\bf 103} (2000) 1043
{\tt[arXiv:hep-th/9912047]};
{\tt [arXiv:hep-th/0010006]};
JHEP {\bf 0103} (2001) 033
{\tt [arXiv:hep-th/0010119]}.

\bibitem{LTY}
K.~Lee, D.~Tong and S.~Yi,
Phys.\ Rev.\  {\bf D63} (2001) 065017
{\tt[arXiv:hep-th/0008092]}.




\bibitem{KLY12}
K.~Kim, B.~Lee and H.~S.~Yang,
{\tt arXiv:hep-th/0003093}; {\tt arXiv:hep-th/0109121}.

\bibitem{Nekrasov:2000zz}
N.~A.~Nekrasov,
{\tt[arXiv:hep-th/0010017]}.

\bibitem{Schwarz:2001ru}
A.~Schwarz,
Commun.\ Math.\ Phys.\  {\bf 221} (2001) 433
{\tt [arXiv:hep-th/0102182]}.


\bibitem{Chu:2001cx}
C.~S.~Chu, V.~V.~Khoze and G.~Travaglini,
{\tt arXiv:hep-th/0108007}.

\bibitem{Hamanaka:2001dr}
M.~Hamanaka,
{\tt arXiv:hep-th/0109070}.




\bibitem{Hollowood:2001ng}
T.~J.~Hollowood, V.~V.~Khoze and G.~Travaglini,
JHEP {\bf 0105} (2001) 051
{\tt [arXiv:hep-th/0102045]}.


\bibitem{Armoni:2001br}
A.~Armoni, R.~Minasian and S.~Theisen,
{\tt arXiv:hep-th/0102007}.


\bibitem{Blau:1993pm}
M.~Blau,
J.\ Geom.\ Phys.\  {\bf 11} (1993) 95
{\tt[arXiv:hep-th/9203026]}.

\bibitem{Niemi:1994ej}
A.~J.~Niemi and K.~Palo,
{\tt arXiv:hep-th/9406068}.

\bibitem{Blau:1995rs}
M.~Blau and G.~Thompson,
theories,''
J.\ Math.\ Phys.\  {\bf 36} (1995) 2192
{\tt[arXiv:hep-th/9501075]}.

\bibitem{Schwarz:1997dg}
A.~Schwarz and O.~Zaboronsky,
Commun.\ Math.\ Phys.\  {\bf 183} (1997) 463
{\tt[arXiv:hep-th/9511112]}.


\bibitem{berline}
N.~Berline, E.~Getzler and M.~Vergne, ``Heat Kernals and Dirac
Operators,'' Springer Verlag (1991).


\bibitem{Fucito:2001ha}
F.~Fucito, J.~F.~Morales and A.~Tanzini,
JHEP {\bf 0107} (2001) 012
{\tt[arXiv:hep-th/0106061]}.


\bibitem{Bellisai:2000tn}
D.~Bellisai, F.~Fucito, A.~Tanzini and G.~Travaglini,
Phys.\ Lett.\ {\bf B480} (2000) 365
{\tt[arXiv:hep-th/0002110]}.

\bibitem{Bellisai:2000bc}
D.~Bellisai, F.~Fucito, A.~Tanzini and G.~Travaglini,
JHEP {\bf 0007} (2000) 017
{\tt[arXiv:hep-th/0003272]}.



\bibitem{Flume:2001nc}
R.~Flume, R.~Poghossian and H.~Storch,
{\tt arXiv:hep-th/0110240}; {\tt arXiv:hep-th/0112211}.


\bibitem{EGH}
T.~Eguchi, P.~B.~Gilkey and A.~J.~Hanson,
Phys.\ Rept.\  {\bf 66} (1980) 213.



}

\end{thebibliography}
\end{document}